\documentclass{article}

\usepackage[usenames,dvipsnames]{xcolor}     
\usepackage{verbatim} 
\usepackage{mathrsfs} 
\usepackage{amsmath, amssymb} 
\usepackage{xspace}
\usepackage{epubtk}
\usepackage{textcomp}
\usepackage{caption}
\usepackage{subcaption}
\usepackage[nottoc]{tocbibind}
\usepackage{lineno}

\usepackage[T1]{fontenc}
\usepackage[utf8]{inputenc}
\usepackage{lmodern} 
\input{glyphtounicode}  
\showlistoftablesfalse

\usepackage{graphicx}
\usepackage{pgf}
\usepackage{tikz}
\usetikzlibrary{arrows,decorations.pathmorphing,shapes,backgrounds}

\definecolor{linkcolor}{rgb}{.8,0,0}
\definecolor{urlcolor}{rgb}{0,0,.7}
\definecolor{citecolor}{rgb}{0,.5,0}
\definecolor{acrocolor}{rgb}{0,0,.7}

\hypersetup{bookmarksopen,colorlinks=true}
\hypersetup{pdfstartview=FitH}
\hypersetup{linktocpage=true,bookmarksnumbered=true}
\hypersetup{plainpages=false,breaklinks=true}
\hypersetup{linkcolor=linkcolor,citecolor=citecolor,urlcolor=urlcolor}

\hypersetup{pdfinfo={
Title={Interferometer Techniques for Gravitational-Wave Detection},
Author={Andreas Freise, Charlotte Bond, Daniel Brown,
         Kenneth Strain},
Subject={Review article},
Keywords={interferometer;simulation;gravitational wave; optics}
}}

\newif\iffinal 
\newcommand{\I}{\ensuremath{\mathrm{i}\,}}
\newcommand{\myRe}[1]{\ensuremath{{\Re}\left\{ #1 \right\} }}
\newcommand{\myIm}[1]{\ensuremath{{\Im}\left\{{#1}\right\}}}
\newcommand{\mEx}[1]{\ensuremath\exp{\left(#1\right)}}
\newcommand{\mExB}[1]{\ensuremath\exp{\Bigl(#1\Bigr)}}

\newcommand{\mCos}[1]{\ensuremath\cos{\left(#1\right)}}
\newcommand{\mSin}[1]{\ensuremath\sin{\left(#1\right)}}
\newcommand{\HG}{Hermite--Gauss\xspace}
\newcommand{\zr}{\ensuremath{z_{\mathrm{R}}}}
\newcommand{\w}{\omega}
\newcommand{\T}{\ensuremath{\,t}}
\newcommand{\WT}[1]{\ensuremath{\widetilde{#1}}}

\newcommand{\Tun}{\ensuremath \phi}
\newcommand{\sollgleich}{\ensuremath{\stackrel{!}{=}}}
\newcommand{\roc}{\ensuremath{R_{\mathrm{C}}}}

\newcommand{\SSm}[1]{\scriptscriptstyle\mathrm{#1}}
\newcommand{\OmMod}{\ensuremath{\Omega}}
\newcommand{\Finesse}{\textsc{Finesse}\xspace}
\newcommand\dif{\mathop{}\!\mathrm{d}}

\newcommand{\mum}{\textmu\hspace{-0.2pt}m}
\newcommand{\x}{\textit{x}\xspace}
\newcommand{\y}{\textit{y}\xspace}
\newcommand{\z}{\textit{z}\xspace}

\numberwithin{equation}{section}
\numberwithin{figure}{section}

\begin{document}

\title{Interferometer Techniques for Gravitational-Wave Detection}

\author{%
\epubtkAuthorData{Charlotte Bond}{%
School of Physics and Astronomy\\
University of Birmingham\\
Birmingham, B15 2TT, UK}{%
czb@star.sr.bham.ac.uk}%
{}%
\\
\and
\epubtkAuthorData{Daniel Brown}{%
School of Physics and Astronomy\\
University of Birmingham\\
Birmingham, B15 2TT, UK}{%
ddb@star.sr.bham.ac.uk}%
{}%
\\
\and \\
\epubtkAuthorData{Andreas Freise}{%
School of Physics and Astronomy\\
University of Birmingham\\
Birmingham, B15 2TT, UK}{%
adf@star.sr.bham.ac.uk}%
{http://www.gwoptics.org}
\\
\and \\
\epubtkAuthorData{Kenneth Strain}{%
School of Physics and Astronomy\\
University of Glasgow\\
Glasgow G12 8QQ, UK}{%
kenneth.strain@glasgow.ac.uk}%
{}%
}

\date{\today}
\maketitle

\begin{abstract}
Several km-scale gravitational-wave detectors have been constructed
world wide. These instruments combine a number of advanced
technologies to push the limits of precision length measurement. The
core devices are laser interferometers of a new kind; developed from
the classical Michelson topology these interferometers integrate
additional optical elements, which significantly change the properties
of the optical system. Much of the design and analysis of these laser
interferometers can be performed using well-known classical optical
techniques; however, the complex optical layouts provide a new
challenge. In this review we give a textbook-style introduction to the
optical science required for the understanding of modern gravitational
wave detectors, as well as other high-precision laser
interferometers. In addition, we provide a number of examples for a
freely available interferometer simulation software and encourage the
reader to use these examples to gain hands-on experience with the
discussed optical methods.
\end{abstract}

\epubtkKeywords{Gravitational waves, Gravitational wave detectors,
  Laser interferometry, Optics, Simulations, Finesse}

\newpage
\tableofcontents  
\newpage

\section{Introduction}
\label{sec:intro}

\subsection{The scope and style of the review}
The historical development of laser interferometers for application as
gravitational-wave detectors~\cite{lrr-rowan-hough} has involved the
combination of relatively simple optical subsystems into more and more
complex assemblies.  The individual elements that compose the
interferometers, including mirrors, beam splitters, lasers,
modulators, various polarising optics, photo detectors and so forth,
are individually well described by relatively simple, mostly-classical
physics. Complexity arises from the combination of multiple mirrors,
beam splitters etc.\ into optical cavity systems that have narrow
resonant features, and the consequent requirement to stabilise
relative separations of the various components to sub-wavelength
accuracy, and indeed in many cases to very small fractions of a
wavelength.

Thus, classical physics describes the interferometer techniques and the
operation of current gravitational-wave detectors. However, we note
that at signal frequencies above a couple of hundreds of Hertz, the
sensitivity of current detectors is limited by the photon counting
noise at the interferometer readout, also called shot-noise. The next
generation systems such as Advanced LIGO~\cite{Fritschel_adligo03,
  advanced_ligo}, Advanced Virgo~\cite{AdvancedVirgo15} and
KAGRA~\cite{KAGRA13} are expected to operate in a regime where the
quantum physics of both light and mirror motion couple to each
other. Then, a rigorous quantum-mechanical description is certainly
required. Sensitivity improvements beyond these `Advanced' detectors
necessitate the development of \emph{non-classical} techniques; a
comprehensive discussion of such techniques is provided in~\cite{Danilishin12}.
This review provides a brief introduction to quantum noise in
Section~\ref{sec:quantum_noise}
but otherwise focusses on the non-quantum aspects of interferometry
that play an important role in overcoming other limits to
current detectors, due to, for example, thermal effects and feedback
control systems. At the same time these classical techniques will
provide the means for implementing new, non-classical schemes and
just remain as important as ever.

The optical components employed tend to behave in a linear fashion with
respect to the optical field, i.e.~nonlinear optical effects need
hardly be considered. Indeed, almost all aspects of the design of laser
interferometers are dealt with in the linear regime.  Therefore the underlying
mathematics is relatively simple and many standard
techniques are available, including those that naturally allow
numerical solution by computer models. Such computer models are in
fact necessary as the exact solutions can become quite complicated
even for systems of a few components. In practice, workers in the field
rarely calculate the behaviour of the optical systems from first
principles, but instead rely on various well-established numerical
modelling techniques. An example of software that enables modelling of
interferometers and their component systems is \Finesse~\cite{Freise04,
  finesse_webpage}. This was developed by some of us (AF, DB), has been
validated in a wide range of situations, and was used to prepare the
examples included in the present review.

The target readership we have in mind is the student or researcher who
desires to get to grips with practical issues in the design of
interferometers or component parts thereof. For that reason, this review
consists of sections covering the basic physics and approaches to
simulation, intermixed with some practical examples. To make this as
useful as possible, the examples are intended to be realistic with
sensible parameters reflecting typical application in gravitational
wave detectors. The examples, prepared using \Finesse, are designed to
illustrate the methods typically applied in designing gravitational
wave detectors. We encourage the reader to obtain \Finesse and to
follow the examples (see Appendix~\ref{sec:finesse}).

\subsection{Overview of the goals of interferometer design}
\label{sec:intro_design}
Gravitational-wave detectors strive to pick out signals carried by
passing gravitational waves from a background of self-generated
noise. The principles of operation are set out at various points in
the review, but in essence, the goal has been to prepare many photons,
stored for as long as practical in the `arms' of a laser
interferometer (traditionally the two arms are at right angles), so
that tiny phase shifts induced by the gravitational waves yield the largest possible effect,
when the light leaving the appropriate
`port' of the interferometer is detected and the resulting signal
analysed.

The evolution of gravitational-wave detectors can be seen by following
their development from prototypes and early observing systems towards
the so-called `Advanced detectors', which are currently under construction, or in the case of
Advanced LIGO, in the first phase of scientific observing (as of late 2015).
Starting from the simplest Michelson interferometer~\cite{forward78},
then by the application of techniques to increase the number of
photons stored in the arms: delay lines~\cite{herriott64},
Fabry-Perot arm cavities~\cite{fabry_perot, Fattaccioli86} and
power recycling~\cite{billing83, drever_pr}. The final step in the development of
classical interferometry was the inclusion of signal
recycling~\cite{meers88, heinzel98}, which, among other effects,
allows the signal from a gravitational-wave signal of
approximately-known spectrum to be enhanced above the noise.

Reading out a signal from even the most basic interferometer requires
minimising the coupling of local environmental effects to the detected
output. Thus, the relative positions of all the components must be
stabilised. This is commonly achieved by suspending the mirrors etc.\
as pendulums, often multi-stage pendulums in series, and then applying
closed-loop control to maintain the desired operating condition. The
careful engineering required to provide low-noise suspensions with the
correct vibration isolation, and also low-noise actuation, is
described in many works, for example, \cite{braccini96, plissi2000, barriga2009, Aston2012}.

As the interferometer optics become more complicated, the resonance
conditions, i.e.~the allowed combinations of inter-component path
lengths required to allow the photon number in the interferometer arms
to reach maximum, become more narrowly defined. It is likewise
necessary to maintain angular alignment of all components, such that
beams required to interfere are correctly co-aligned. Typically the
beams need to be aligned within a small fraction, and sometimes a very
small fraction, of the far-field diffraction angle: the
requirement can be in the low nano-radian range for km-scale
detectors~\cite{Morrison1994, Freise07}. Therefore, for each optical
component there is typically one longitudinal, i.e.~along the direction
of light propagation, plus two angular degrees of freedom: pitch and
yaw about the longitudinal axis. A complex interferometer consists of up to
around seven highly sensitive components and so there
can be of order 20 degrees of freedom to be measured and
controlled~\cite{virgo_align06, Winkler07}.

Although the light fields are linear in their behaviour, the coupling between the
position of a mirror and the complex amplitude of the detected light field
typically shows strongly nonlinear dependence on mirror positions due
to the sharp resonance features exhibited by cavity systems. The
fields do vary linearly or at least smoothly close to the
desired operating point, however. So, while well-understood linear control
theory suffices to design the control system needed to maintain the
optical configuration at its operating point, bringing the system to
that operating condition is often a separate and more challenging
nonlinear problem. In the current version of this work we consider
only the linear aspects of sensing and control.

Control systems require actuators, and those employed are typically
electrical-force transducers that act on the suspended optical
components, either directly or -- to provide enhanced noise rejection
-- at upper stages of multi-stage suspensions. The transducers are
normally coil-magnet actuators, with the magnets on the moving part,
or, less frequently, electrostatic actuators of varying design.  The
actuators are frequently regarded as part of the mirror suspension
subsystem and are not discussed in the current work.

To give order to our review we consider the main physics describing
the operation of the basic optical components: mirrors, beam splitters,
modulators, etc., required to construct interferometers.
Although all of the relevant physics is generally well known and not
new, we take it as a starting point that permits the introduction of
notation and conventions. It is also true that the interferometry
employed for gravitational-wave detection has a different emphasis
than other interferometer applications. As a consequence, descriptions or
examples of a number of crucial optical properties for gravitational
wave detectors cannot be found in the literature.

The purpose of this review is especially to provide a coherent theoretical
framework for describing such effects. With the basics
established, it can be seen that the interferometer configurations that
have been employed in gravitational-wave detection may be built up and
simulated in a relatively straightforward manner.

\subsection{Plane-wave analysis}
\label{sec:fields}
The main optical systems of interferometric gravitational-wave
detectors are designed such that all system parameters are well known
and stable over time. The stability is achieved through a mixture of
passive isolation systems and active feedback control.
In particular, the light sources are some of the most stable,
low-noise continuous-wave laser systems so that electromagnetic fields
can be assumed to be essentially monochromatic. Additional frequency
components can be modelled as small modulations in amplitude or
phase. The laser beams are well collimated, propagate along a
well-defined optical axis and remain always very much smaller than the
optical elements they interact with. Therefore, these beams can be
described as \emph{paraxial} and the well-known paraxial
approximations can be applied.

It is useful to first derive a  mathematical model based on
monochromatic, scalar, plane waves. As it turns out, a more detailed
model including the polarisation and the shape of the laser beam as
well as multiple frequency components, can be derived as an extension
to the plane-wave model.
A plane electromagnetic wave is typically described by its electric
field component:

\epubtkImage{field1-external.png}{%
  \begin{figure}[h!]
    \centerline{\includegraphics[width=0.7\textwidth]{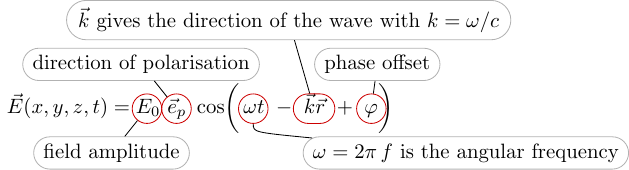}}
    \caption{}
    \label{eq:field_parameters}
\end{figure}}
\noindent
with $E_0$ as the (constant) field amplitude in V/m, $ \vec{e}_p$ the
unit vector in the direction of polarisation, such as, for example,
$\vec{e}_y$ for $\mathscr{S}$-polarised light, $\omega$ the angular
oscillation frequency of the wave, and $\vec{k}=\vec{e}_k \omega/c$
the wave vector pointing in the direction of propagation. The
absolute phase $\varphi$ only becomes meaningful when the field is
superposed with other light fields.

In this document we will consider waves propagating along the optical
axis given by the \z-axis, so that $\vec{k}\vec{r}=kz$. For the moment
we will ignore the polarisation and use scalar waves, which can be
written as
\begin{equation}
E(z,t)=E_0 \cos(\omega t - kz +\varphi).
\end{equation}
Further, in this document we use complex notation, i.e.~
\begin{equation}
E=\myRe{E'} \qquad \text{with} \qquad E'= E'_0 \exp\big(\I (\omega t - kz)\big).
\end{equation}
This has the advantage that the scalar amplitude and the phase
$\varphi$ can be given by one, now complex, amplitude $E'_0=E_0
\exp(\I \varphi)$. We will use this notation with complex numbers
throughout. For clarity we will simply use the unprimed letters for
the auxiliary field. In particular, we will use the letter $E$ and
also $a$ and $b$ to denote complex electric-field amplitudes. But
remember that, for example, in $E=E_0 \exp(-\I kz)$ neither $E$ nor
$E_0$ are physical quantities. Only the real part of $E$ exists and
deserves the name field amplitude.

\subsection{Frequency domain analysis}
In most cases we are either interested in the fields \emph{at one
  particular location}, for example, on the surface of an optical
element, or we want to know the fields at all places in the
interferometer but \emph{at one particular point in time}. The latter
is usually true for the \emph{steady state} approach: assuming that
the interferometer is in a steady state, all solutions must be
independent of time so that we can perform all computations at $t=0$
without loss of generality. In that case, the scalar plane wave can be
written as
\begin{equation}
E=E_0 \exp(-\I kz).
\end{equation}
The frequency domain is of special interest as numerical models of
gravitational-wave detectors tend to be much faster to compute in the
frequency domain than in the time domain.


\newpage
\section{Optical Components: Coupling of Field Amplitudes}
\label{sec:components}
When an electromagnetic wave interacts with an optical system, all of
its parameters can be changed as a result. Typically optical
components are designed such that, ideally, they only affect one of
the parameters, i.e.~either the amplitude \emph{or} the polarisation
\emph{or} the shape. Therefore, it is convenient to derive separate
descriptions concerning each parameter. This section introduces the
coupling of the complex field amplitude at optical
components. Typically, the optical components are described in the
simplest possible way, as illustrated by the use of abstract
schematics such as those shown in Figure~\ref{fig:ecoupling}.

\epubtkImage{ecoupling-external.png}{%
  \begin{figure}[htb]
    \centerline{\includegraphics{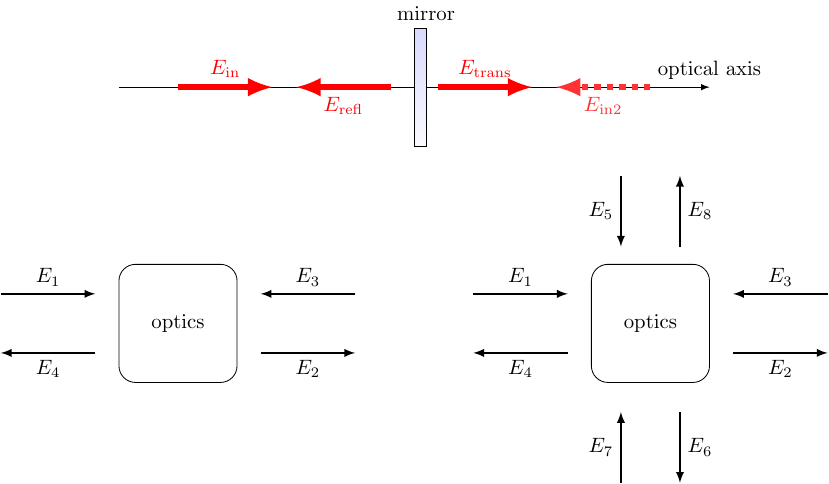}}
    \caption{This set of figures introduces an abstract form of illustration,
      which will be used in this document. The top figure shows a
      typical example taken from the analysis of an optical system: an
      incident field $E_{\mathrm{in}}$ is reflected and transmitted by a
      semi-transparent mirror; there might be the possibility of
      second incident field $E_{in2}$. The lower left figure shows the
      abstract form we choose to represent the same system. The lower
      right figure depicts how this can be extended to include a beam
      splitter object, which connects two optical axes.}
    \label{fig:ecoupling}
\end{figure}}

\subsection{Mirrors and spaces: reflection, transmission and propagation}
\label{sec:mirrors_spaces}
The core optical systems of current interferometric gravitational
interferometers are composed of two building blocks: a)~resonant
optical cavities, such as Fabry-Perot resonators, and b)~beam
splitters, as in a Michelson interferometer. In other words, the laser
beam is either propagated through a vacuum system or interacts with a
partially-reflecting optical surface.

The term \emph{optical surface} generally refers to a boundary between
two media with possibly different indices of refraction $n$, for
example, the boundary between air and glass or between two types of
glass.  A real fused silica mirror in an interferometer features two
surfaces, which interact with a reflected or transmitted laser
beam. However, in some cases, one of these surfaces has been treated
with an anti-reflection (AR) coating to minimise the effect on the
transmitted beam.

The terms \emph{mirror} and \emph{beam splitter} are sometimes used to
describe a (theoretical) optical surface in a model.
We define real \emph{amplitude coefficients} for reflection and
transmission $r$ and $t$, with $0\leq r,t \leq 1$, so that the field
amplitudes can be written as

\epubtkImage{mirror_coupling-external.png}{%
  \begin{figure}[h!]
    \centerline{\includegraphics[width=0.9\textwidth]{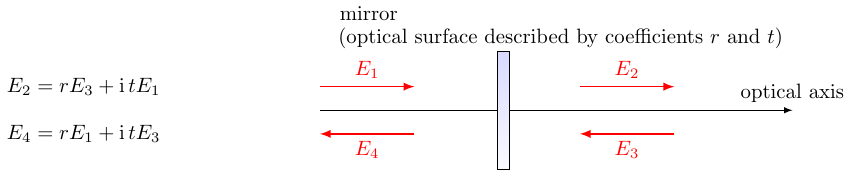}}
    \caption{}
    \label{eq:mirror_coupling}
\end{figure}}
\noindent
The $\pi/2$ phase shift upon transmission (here given by the factor
$\I$) refers to a phase convention explained in
Section~\ref{sec:bsphase}.

The free propagation of a distance $D$ through a medium with index of
refraction $n$ can be described with the following set of equations:

\epubtkImage{space_coupling-external.png}{%
  \begin{figure}[h!]
    \centerline{\includegraphics[width=0.9\textwidth]{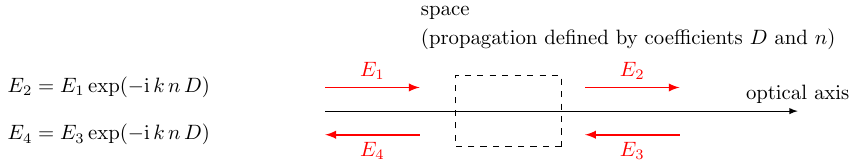}}
    \caption{}
    \label{eq:space_coupling}
\end{figure}}
\noindent
In the following we use $n=1$ for simplicity.

Note that we use above relations to demonstrate various mathematical
methods for the analysis of optical systems. However, refined versions
of the coupling equations for optical components, including those for
spaces and mirrors, are also required, see, for example,
Section~\ref{sec:mirrors_spaces_2}.

\subsection{The two-mirror resonator}
\label{sec:two_mirror}
The linear optical resonator, also called a \emph{cavity} is formed by
two partially-transparent mirrors, arranged in parallel as shown in
Figure~\ref{fig:two_mirror1}. This simple setup makes a very good
example with which to illustrate how a mathematical model of an
interferometer can be derived, using the equations introduced in
Section~\ref{sec:mirrors_spaces}. A more detailed description of the
two-mirror cavity is provided in Section~\ref{sec:two_mirror2}.

\epubtkImage{two_mirror1-external.png}{%
  \begin{figure}[h!]
    \centerline{\includegraphics{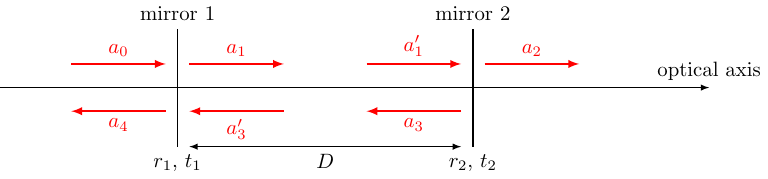}}
    \caption{Simplified schematic of a two mirror cavity. The two
      mirrors are defined by the amplitude coefficients for reflection
      and transmission. Further, the resulting cavity is characterised
      by its length $D$. Light field amplitudes are shown and
      identified by a variable name, where necessary to permit their
      mutual coupling to be computed.}
    \label{fig:two_mirror1}
\end{figure}}

The cavity is defined by a propagation length $D$ (in vacuum), the
amplitude reflectivities $r_1$, $r_2$ and the amplitude transmittances
$t_1$, $t_2$. The amplitude at each point in the cavity can be
computed simply as the superposition of fields. The entire set of
equations can be written as
\begin{equation}
\label{eq:two_mirror_set}
\begin{array}{ll}
a_1  &= \I t_1 a_0 + r_1 a_3'\\
a_1' &= \exp(-\I k D)~ a_1\\
a_2  &= \I t_2 a_1'\\
a_3  &= r_2 a_1'\\
a_3' &= \exp(-\I k D) ~a_3\\
a_4  &= r_1 a_0 + \I t_1 a_3'
\end{array}
\end{equation}
The circulating field impinging on the first mirror (surface) $a_3'$
can now be computed as
\begin{equation}
\begin{array}{ll}
a_3'&=\exp(-\I k D) ~a_3=\exp(-\I k D)~ r_2 a_1' =\exp(-\I 2 k D)~ r_2 a_1 \\
&=\exp(-\I 2 k D)~ r_2~ (\I t_1 a_0 + r_1 a_3').
\end{array}
\end{equation}
This then yields
\begin{equation}
a_3'=a_0\frac{\I r_2 t_1 \exp(-\I 2 k D)}{1-r_1r_2\exp(-\I 2 k D)}.
\label{eq:a3}
\end{equation}
We can directly compute the reflected field to be
\begin{equation}
\label{eq:cav_refl}
a_4=a_0\left(r_1 - \frac{r_2 t_1^2 \exp(-\I 2 k D)}{1-r_1 r_2 \exp(- \I 2 k D)}\right)=
a_0\left(\frac{r_1-r_2(r_1^2+t_1^2)\exp(-\I 2 k D)}{1-r_1 r_2 \exp(- \I 2 k D)}\right),
\end{equation}
while the transmitted field becomes
\begin{equation}
\label{eq:cav_trans}
a_2=a_0 \frac{-t_1 t_2 \exp(-\I k D)}{1-r_1 r_2 \exp(- \I 2 k D)}.
\end{equation}
The properties of two mirror cavities will be discussed in more detail
in Section~\ref{sec:two_mirror2}.

\subsection{Coupling matrices}
\label{sec:coupling_matrices}
Computations that involve sets of linear equations as shown in
Section~\ref{sec:two_mirror} can often be done or written efficiently
with matrices. Two methods of applying matrices to coupling field
amplitudes are demonstrated below, using again the example of a two
mirror cavity. First of all, we can rewrite the coupling equations in
matrix form. The mirror coupling as given in
Figure~\ref{eq:mirror_coupling} becomes

\epubtkImage{mirror_coupling_matrix1-external.png}{%
  \begin{figure}[h!]
    \centerline{\includegraphics[width=0.9\textwidth]{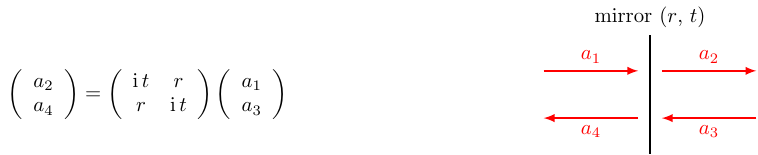}}
    \caption{}
    \label{eq:mirror_coupling_matrix1}
\end{figure}}

\newpage
\noindent
and the amplitude coupling at a `space', as given in
Figure~\ref{eq:space_coupling}, can be written as

\epubtkImage{space_coupling_matrix1-external.png}{%
  \begin{figure}[h!]
    \centerline{\includegraphics[width=0.9\textwidth]{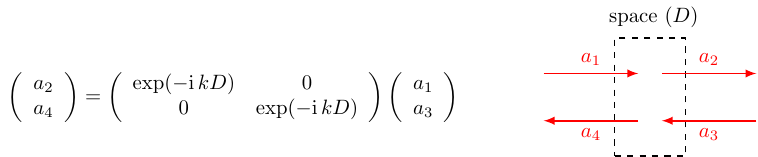}}
    \caption{}
    \label{eq:space_coupling_matrix1}
\end{figure}}
\noindent
In these examples the matrix simply transforms the `known' impinging
amplitudes into the `unknown' outgoing amplitudes.

\subsubsection*{Coupling matrices for numerical computations}
\label{sec:Coupling_matrices}
An obvious application of the matrices introduced above would be to
construct a large matrix for an extended optical system appropriate
for computerisation. A very flexible method is to setup one equation
for each field amplitude. The set of linear equations for a mirror
would expand to
\begin{equation}
\left(
\begin{array}{cccc}
 1 &0 & 0 & 0 \\
 -\I t & 1& -r & 0\\
0 &0 & 1 & 0\\
-r & 0 & -\I t & 1
\end{array}
\right)
\left(
\begin{array}{c} a_1\\ a_2\\a_3 \\ a_4\end{array}
\right)
=
\left(
\begin{array}{c} a_1\\ 0\\ a_3 \\ 0\end{array}
\right) =M_{\mathrm{system}}~\vec{a}_{\mathrm{sol}}\,=\,\vec{a}_{\mathrm{input}},
\end{equation}
where the input vector\epubtkFootnote{In many implementations of
  numerical matrix solvers the input vector is also called the
  \emph{right-hand side} vector.} $\vec{a}_{\mathrm{input}}$ has
non-zero values for the impinging fields and $\vec{a}_{\mathrm{sol}}$
is the `solution' vector, i.e.~after solving the system of equations
the amplitudes of the impinging as well as those of the outgoing
fields are stored in that vector.

As an example we apply this method to the two mirror cavity. The
system matrix for the optical setup shown in
Figure~\ref{fig:two_mirror1} becomes
\begin{equation}
\left(
\begin{array}{ccccccc}
 1       & 0            & 0 & 0       & 0       & 0    & 0\\
 -\I t_1 & 1            & 0 & -r_1    & 0       & 0    & 0\\
 -r_1    & 0            & 1 & -\I t_1 & 0       & 0    & 0\\
 0       & 0            & 0 & 1       & 0       & 0    & -e^{-\I k D}\\
 0       & -e^{-\I k D} & 0 & 0       & 1       & 0    & 0\\
 0       & 0            & 0 & 0       & -\I t_2 & 1    & 0\\
 0       & 0            & 0 & 0       & -r_2    & 0 & 1
\end{array}
\right)
\left(
\begin{array}{c}
a_0\\ a_1\\a_4 \\ a'_3\\a'_1\\a_2\\a_3
\end{array}
\right)
=
\left(
\begin{array}{c}
a_0\\0\\0 \\ 0\\0\\0\\0
\end{array}
\right)
\end{equation}
This is a \emph{sparse} matrix.  Sparse matrices are an important
subclass of linear algebra problems and many efficient numerical
algorithms for solving sparse matrices are freely available (see, for
example, \cite{DavisKLU2006}). The advantage of this method of
constructing a single matrix for an entire optical system is the
direct access to all field amplitudes. It also stores each coupling
coefficient in one or more dedicated matrix elements, so that
numerical values for each parameter can be read out or changed after
the matrix has been constructed and, for example, stored in computer
memory. The obvious disadvantage is that the size of the matrix
quickly grows with the number of optical elements (and with the
degrees of freedom of the system, see, for example,
Section~\ref{sec:beamshapes}).

\subsubsection*{Coupling matrices for a compact system descriptions}
The following method is probably most useful for analytic
computations, or for optimisation aspects of a numerical
computation. The idea behind the scheme, which is used for computing
the characteristics of dielectric coatings \cite{Hecht-english,
  matuschek97} and has been demonstrated for analysing gravitational
wave detectors \cite{MY99}, is to rearrange equations as in
Figure~\ref{eq:mirror_coupling_matrix1} and
Figure~\ref{eq:space_coupling_matrix1} such that the overall matrix
describing a series of components can be obtained by multiplication of
the component matrices. In order to achieve this, the coupling
equations have to be re-ordered so that the input vector consists of
two field amplitudes \emph{at one side of the component}. For the
mirror, this gives a coupling matrix of
\begin{equation}
\label{eq:mirror_coupling_matrix1b}
\left(
\begin{array}{c} a_1\\a_4\end{array}
\right)=\frac{\I}{t}
\left(
\begin{array}{cc}
-1&  r \\
-r &  r^2+t^2
\end{array}
\right)
\left(
\begin{array}{c} a_2\\a_3\end{array}
\right).
\end{equation}
In the special case of the lossless mirror this matrix simplifies as
we have $r^2+t^2=R+T=1$. The space component would be described by the
following matrix:
\begin{equation}
\left(
\begin{array}{c} a_1\\a_4\end{array}
\right)=
\left(
\begin{array}{cc} \exp(\I k D)  & 0\\ 0 & \exp(-\I k D)\end{array}
\right)
\left(
\begin{array}{c} a_2\\a_3\end{array}
\right).
\end{equation}
With these matrices we can very easily compute a matrix for the cavity
with two lossless mirrors as
\begin{eqnarray}
M_{\mathrm{cav}}&=&M_{\mathrm{mirror1}}\times M_{\mathrm{space}}\times M_{\mathrm{mirror2}}\\
&=&\frac{-1}{t_1 t_2}\left(\begin{array}{cc}
e^{+}-r_1 r_2 e^{-} & -r_2e^{+}+ r_1 e^{-} \\
 -r_2e^{-}+ r_1 e^{+}  & e^{-}-r_1 r_2 e^{+}
\end{array}\right),
\end{eqnarray}
with $e^{+}=\exp(\I k D)$ and $e^{-}=\exp(-\I k D)$. The system of
equation describing a cavity shown in
Equation~(\ref{eq:two_mirror_set}) can now be written more compactly
as
\begin{equation}
\left(
\begin{array}{c} a_0\\a_4\end{array}
\right)=
\frac{-1}{t_1 t_2}\left(\begin{array}{cc}
e^{+}-r_1 r_2 e^{-} & -r_2e^{+}+ r_1 e^{-} \\
 -r_2e^{-}+ r_1 e^{+}  & e^{-}-r_1 r_2 e^{+}
\end{array}\right)
\left(
\begin{array}{c} a_2\\0\end{array}
\right).
\end{equation}
This allows direct computation of the amplitude of the transmitted
field resulting in
\begin{equation}
a_2=a_0\frac{-t_1t_2\exp(-\I k D)}{1-r_1 r_2 \exp(-\I 2 k D)},
\end{equation}
which is the same as Equation~(\ref{eq:cav_trans}).

The advantage of this matrix method is that it allows compact storage
of any series of mirrors and propagations, and potentially other
optical elements, in a single 2~\texttimes~2 matrix. The disadvantage
inherent in this scheme is the lack of information about the field
amplitudes inside the group of optical elements.

\subsection{Phase relation at a mirror or beam splitter}
\label{sec:bsphase}
The magnitude and phase of reflection at a single optical surface can
be derived from Maxwell's equations and the electromagnetic boundary
conditions at the surface, and in particular the condition that the
field amplitudes tangential to the optical surface must be
continuous. The results are called \emph{Fresnel's
  equations}~\cite{kenyon08}. Thus, for a field impinging on an optical
surface under normal incidence we can give the reflection
coefficient as
\begin{equation}
r=\frac{n_1-n_2}{n_1+n_2},
\end{equation}
with $n_1$ and $n_2$ the indices of refraction of the first and
second medium, respectively. The transmission coefficient for a
lossless surface can be computed as
$t^2=1-r^2$. We note that
the phase change upon reflection is either 0 or 180\textdegree,
depending on whether the second medium is optically thinner or thicker
than the first. It is not shown here but Fresnel's equations can also
be used to show that the phase change for the transmitted light at a
lossless surface is zero. This contrasts with the definitions given in
Section~\ref{sec:mirrors_spaces} (see
Figure~(\ref{eq:mirror_coupling})ff.), where the phase shift upon any
reflection is defined as zero and the transmitted light experiences a
phase shift of $\pi/2$. The following section explains the motivation
for the latter definition having been adopted as the common notation
for the analysis of modern optical systems.

\subsubsection*{Composite optical surfaces}
\epubtkImage{mirror-details-newport-substrates.png}{%
\begin{figure}[htbp]
    \centerline{
      \includegraphics[scale=.7]{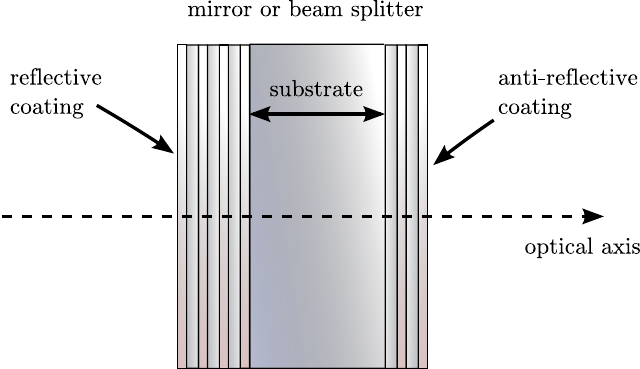}
      \includegraphics[scale=.6]{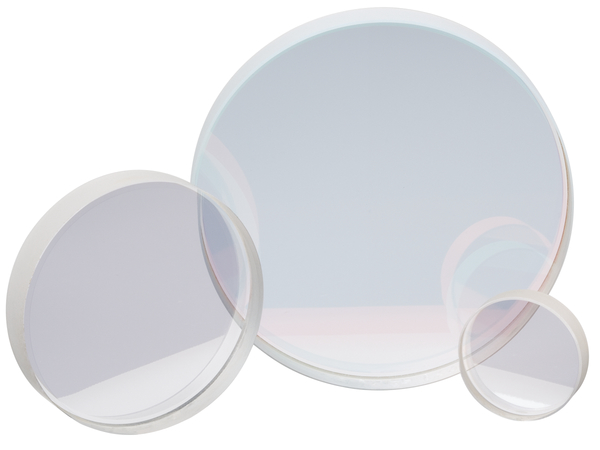}
    }
    \caption{This sketch shows a mirror or beam splitter component
      with dielectric coatings and the photograph shows some typical
      commercially available examples~\cite{newport-catalogue}. Most
      mirrors and beam splitters used in optical experiments are of
      this type: a substrate made from glass, quartz or fused silica
      is coated on both sides. The \emph{reflective coating} defines
      the overall reflectivity of the component (anything between
      $R\approx 1$ and $R\approx 0$, while the \emph{anti-reflective
      coating} is used to reduce the reflection at the second optical
      surface as much as possible so that this surface does not
      influence the light. Please note that the drawing is not to
      scale, the coatings are typically only a few microns thick on a
      several millimetre to centimetre thick substrate.
    }
    \label{fig:mirror-details}
\end{figure}}

Modern mirrors and beam splitters that make use of dielectric coatings
are complex optical systems, see Figure~\ref{fig:mirror-details} whose
reflectivity and transmission depend on the multiple interference
inside the coating layers and thus on microscopic parameters. The
phase change upon transmission or reflection depends on the details of
the applied coating and is typically not known. In any case, the
knowledge of an absolute value of a phase change is typically not of
interest in laser interferometers because the absolute positions of
the optical components are not known to sub-wavelength
precision. Instead the \emph{relative} phase between the incoming and
outgoing beams is of importance. In the following we demonstrate how
constraints on these relative phases, i.e.~the phase relation between
the beams, can be derived from the fundamental principle of power
conservation. To do this we consider a Michelson interferometer, as
shown in Figure~\ref{fig:bs_phase}, with perfectly-reflecting
mirrors. The beam splitter of the Michelson interferometer is the
object under test. We assume that the magnitude of the reflection $r$
and transmission $t$ are known. The phase changes upon transmission
and reflection are unknown. Due to symmetry we can say that the phase
change upon transmission $\varphi_t$ should be the same in both
directions. However, the phase change on reflection might be different
for either direction, thus, we write $\varphi_{r1}$ for the reflection
at the front and $\varphi_{r2}$ for the reflection at the back of the
beam splitter.

\epubtkImage{bs_phase01.png}{%
  \begin{figure}[htb]
    \centerline{\includegraphics[scale=1.2]{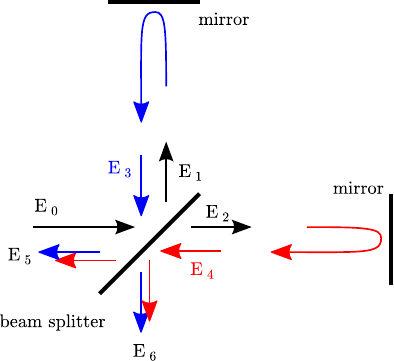}}
    \caption{The relation between the phase of the light field
	amplitudes at a beam splitter can be computed assuming a
	Michelson interferometer, with arbitrary arm length but
	perfectly-reflecting mirrors. The incoming field $E_0$ is
	split into two fields $E_1$ and $E_2$ which are reflected
	atthe end mirrors and return to the beam splitter, as $E_3$
	and $E_4$, to be recombined into two outgoing fields. These
	outgoing fields $E_5$ and $E_6$ are depicted by two arrows to
	highlight that these are the sum of the transmitted and
	reflected components of the returning fields. We can derive
	constraints for the phase of $E_1$ and $E_2$ with respect to
	the input field $E_0$ from the conservation of energy:
	$|E_0|^2=|E_5|^2+|E_6|^2$.}
      \label{fig:bs_phase}
\end{figure}}

Then the electric fields can be computed as
\begin{equation}
E_1=r~ E_0~ e^{\I \varphi_{r1}}~;\qquad E_2=t~ E_0~ e^{\I \varphi_{t}}.
\end{equation}
We do not know the length of the interferometer arms. Thus, we
introduce two further unknown phases: $\Phi_1$ for the total phase
accumulated by the field in the vertical arm and $\Phi_2$ for the
total phase accumulated in the horizontal arm. The fields impinging on
the beam splitter compute as
\begin{equation}
E_3=r~ E_0~ e^{\I (\varphi_{r1}+\Phi_1)}~;\qquad E_4=t~ E_0~ e^{\I (\varphi_{t}+\Phi_2)}.
\end{equation}
The outgoing fields are computed as the sums of the reflected and
transmitted components:
{\renewcommand{\arraystretch}{1.5}
\begin{equation}
\label{eq:MI_bsphase}
\begin{array}{ll}
E_5&=E_0\left(R~e^{\I (2\varphi_{r1}+\Phi_1)}~+~T~e^{\I (2\varphi_{t}+\Phi_2)}\right)\\
E_6&=E_0~rt\left(e^{\I (\varphi_{t}+\varphi_{r1}+\Phi_1)}~+~e^{\I (\varphi_{t}+\varphi_{r2}+\Phi_2)}\right),
\end{array}
\end{equation}
}
with $R=r^2$ and $T=t^2$.

It will be convenient to separate the phase factors into common and
differential ones. We can write
\begin{equation}
E_5=E_0~e^{\I \alpha_{+}}\left(R~e^{\I \alpha_{-}}~+~T~e^{-\I \alpha_{-}}\right),
\end{equation}
with
\begin{equation}
\alpha_{+}=\varphi_{r1}+\varphi_t+\frac12\left(\Phi_1+\Phi_2\right)~;\qquad
\alpha_{-}=\varphi_{r1}-\varphi_t+\frac12\left(\Phi_1-\Phi_2\right),
\end{equation}
and similarly
\begin{equation}
E_6=E_0~rt~e^{\I \beta_{+}}~2\cos(\beta_{-}),
\end{equation}
with
\begin{equation}
\beta_{+}=\varphi_t+\frac12\left(\varphi_{r1}+\varphi_{r2}+\Phi_1+\Phi_2\right)~;\qquad
\beta_{-}=\frac12\left(\varphi_{r1}-\varphi_{r2}+\Phi_1-\Phi_2\right).
\end{equation}
For simplicity we now limit the discussion to a 50:50 beam splitter
with $r=t=1/\sqrt{2}$, for which we can simplify the field expressions
even further:
\begin{equation}
E_5=E_0~e^{\I \alpha_{+}}\cos(\alpha_{-})~;\qquad E_6=E_0~e^{\I \beta_{+}}~\cos(\beta_{-}).
\end{equation}
Conservation of energy requires that $|E_0|^2=|E_5|^2+|E_6|^2$, which
in turn requires
\begin{equation}
\cos^2(\alpha_{-})+\cos^2(\beta_{-})=1,
\end{equation}
which is only true if
\begin{equation}
\alpha_{-}-\beta_{-}=(2N+1)\frac{\pi}{2},
\end{equation}
with $N$ as in integer (positive, negative or zero). This gives the
following constraint on the phase factors
\begin{equation}
\label{eq:phase_relation}
\frac12\left(\varphi_{r1}+\varphi_{r2}\right)-\varphi_{t}=(2N+1)\frac{\pi}{2}.
\end{equation}
One can show that exactly the same condition results in the case of
arbitrary (lossless) reflectivity of the beam
splitter~\cite{Ruediger98}.

We can test whether two known examples fulfil this condition. If the
beam-splitting surface is the front of a glass plate we know that
$\varphi_t=0$, $\varphi_{r1}=\pi$, $\varphi_{r2}=0$, which conforms
with Equation~(\ref{eq:phase_relation}). A second example is the
two-mirror resonator, see Section~\ref{sec:two_mirror}. If we consider
the cavity as an optical `black box', it also splits any incoming beam
into a reflected and transmitted component, like a mirror or beam
splitter. Further we know that a symmetric resonator must give the
same results for fields injected from the left or from the right. Thus,
the phase factors upon reflection must be equal
$\varphi_r=\varphi_{r1}=\varphi_{r2}$. The reflection and transmission
coefficients are given by Equations~(\ref{eq:cav_refl}) and
(\ref{eq:cav_trans}) as
\begin{equation}
r_{\mathrm{cav}}=\left(r_1 - \frac{r_2 t_1^2 \exp(-\I 2 k D)}{1-r_1 r_2 \exp(- \I 2 k D)}\right),
\end{equation}
and
\begin{equation}
t_{\mathrm{cav}}=\frac{-t_1 t_2 \exp(-\I k D)}{1-r_1 r_2 \exp(- \I 2 k D)}.
\end{equation}
We demonstrate a simple case by putting the cavity on resonance ($k
D=N\pi$). This yields
\begin{equation}
r_{\mathrm{cav}}=\left(r_1 - \frac{r_2 t_1^2}{1-r_1 r_2}\right)~;\qquad t_{\mathrm{cav}}=\frac{\I~t_1 t_2}{1-r_1 r_2},
\end{equation}
with $r_{\mathrm{cav}}$ being purely real and $t_{\mathrm{cav}}$
imaginary and thus $\varphi_t=\pi/2$ and  $\varphi_r=0$ which also
agrees with Equation~(\ref{eq:phase_relation}).

In most cases we neither know nor care about the exact phase
factors. Instead we can pick any set which fulfils
Equation~(\ref{eq:phase_relation}). For this document we have chosen
to use phase factors equal to those of the cavity, i.e.~
$\varphi_t=\pi/2$ and  $\varphi_r=0$, which is why we write the
reflection and transmission at a mirror or beam splitter as
\begin{equation}
\label{eq:rt_convention}
E_{\mathrm{refl}}=r ~E_0\qquad\mathrm{and}\qquad E_{\mathrm{trans}}=\I~t~ E_0.
\end{equation}
In this definition $r$ and $t$ are positive real numbers satisfying
$r^2+t^2=1$ for the lossless case.

Please note that we only have the freedom to chose convenient phase
factors when we do not know or do not care about the details of the
optical system, which performs the beam splitting. If instead the
details are important, for example, when computing the properties of a
thin coating layer, such as anti-reflex coatings, the proper phase
factors for the respective interfaces must be computed and used.

\subsection{Lengths and tunings: numerical accuracy of distances}
\label{sec:tuning}
The resonance condition inside an optical cavity and the operating
point of an interferometer depends on the optical path lengths modulo
the laser wavelength, i.e.~for light from an Nd:YAG laser length
differences of less than 1~\mum\ are of interest, not the full
magnitude of the distances between optics. On the other hand, several
parameters describing the general properties of an optical system,
like the finesse or free spectral range of a cavity (see
Section~\ref{sec:two_mirror2}) depend on the macroscopic distance and
do not change significantly when the distance is changed on the order
of a wavelength. This illustrates that the distance between optical
components might not be the best parameter to use for the analysis of
optical systems. Furthermore, it turns out that in numerical
algorithms the distance may suffer from rounding errors. Let us use
the Virgo~\cite{virgo_web} arm cavities as an example to illustrate
this. The cavity length is approximately 3~km, the wavelength is on
the order of 1~\mum, the mirror positions are actively controlled with
a precision of 1~pm and the detector sensitivity can be as good as
10\super{-18}~m, measured on $\sim$~10~ms timescales (i.e.~many
samples of the data acquisition rate). The floating point accuracy of
common, fast numerical algorithms is typically not better than
10\super{-15}. If we were to store the distance between the cavity
mirrors as such a floating point number, the accuracy would be limited
to 3~pm, which does not even cover the accuracy of the control
systems, let alone the sensitivity.

\epubtkImage{length-tuning.png}{%
  \begin{figure}[htbp]
    \centerline{\includegraphics[scale=1]{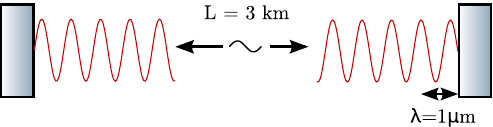}}
      \caption{Illustration of an arm cavity of the Virgo
	gravitational-wave detector~\cite{virgo_web}: the macroscopic
	length $L$ of the cavity is approximately 3~km, while the
	wavelength of the Nd:YAG laser is $\lambda\approx
	1\,\mu\mathrm{m}$. The resonance condition is only affected by
	the microscopic position of the wave nodes with respect to the
	mirror surfaces and not by the macroscopic length, i.e.~displacement of one mirror by $\Delta x=\lambda/2$ re-creates
	exactly the same condition. However, other parameters of the
	cavity, such as the finesse, only depend on the macroscopic
	length $L$ and not on the microscopic tuning.}
    \label{fig:length-tuning}
\end{figure}}

A simple and elegant solution to this problem is to split a distance
$D$ between two optical components into two
parameters~\cite{Heinzel99}: one is the macroscopic `length' $L$,
defined as the multiple of a constant wavelength $\lambda_0$ yielding
the smallest difference to $D$. The second parameter is the
microscopic \emph{tuning} $T$ that is defined as the remaining
difference between $L$ and $D$, i.e.~\ $D=L+T$.
Typically, $\lambda_0$ can be understood as the wavelength of the
laser in vacuum, however, if the laser frequency changes during the
experiment or multiple light fields with different frequencies are
used simultaneously, a default constant wavelength must be chosen
arbitrarily. Please note that usually the term $\lambda$ in any
equation refers to the actual wavelength at the respective location as
$\lambda=\lambda_0/n$ with $n$ the index of refraction at the local
medium.

We have seen in Section~\ref{sec:mirrors_spaces} that distances appear
in the expressions for electromagnetic waves in connection with the
wavenumber, for example,
\begin{equation}
E_2=E_1~\exp(-\I k z).
\end{equation}
Thus, the difference in phase between the field at $z=z_1$ and
$z=z_1+D$ is given as
\begin{equation}
\label{eq:phaseD}
\varphi=- k D.
\end{equation}
We recall that $k=2\pi/\lambda=\omega/c$. We can define
$\omega_0=2\pi~c/\lambda_0$ and $k_0=\omega_0/c$. For any given
wavelength $\lambda$ we can write the corresponding frequency as a sum
of the default frequency and a difference frequency
$\omega=\omega_0+\Delta\omega$. Using these definitions, we can
rewrite Equation~(\ref{eq:phaseD}) with length and tuning as
\begin{equation}
-\varphi= k D = \frac{\omega_0 L}{c} + \frac{\Delta \omega L}{c}  + \frac{\omega_0 T}{c}+ \frac{\Delta \omega
T}{c}.
\end{equation}
The first term of the sum is always a multiple of $2\pi$, which is
equivalent to zero. The last term of the sum is the smallest,
approximately of the order $\Delta \omega \cdot 10^{-14}$. For typical
values of $L\approx 1\mathrm{\ m}$, $T<1\ \mu\mathrm{m}$ and $\Delta
\omega<2\pi\cdot100\mathrm{\ MHz}$ we find that
\begin{equation}
\label{eq:lengths_tuns}
\frac{\omega_0 L}{c} =0,\quad \frac{\Delta \omega L}{c}\lessapprox 2,\quad \frac{\omega_0
T}{c}\lessapprox 6,\quad \frac{\Delta \omega T}{c} \lessapprox 2~10^{-6},
\end{equation}
which shows that the last term can often be ignored.

We can also write the tuning directly as a phase. We define as the
dimensionless tuning
\begin{equation}
\Tun=\omega_0 T/c.
\end{equation}
This yields
\begin{equation}
\exp\left(\I \frac{\omega}{c}T\right)=\exp\left( \I \frac{\omega_0}{c}T\frac{\omega}{\omega_0}\right)=
\exp\left(\I \frac{\omega}{\omega_0} \Tun\right).
\end{equation}
The tuning $\Tun$ is given in radian with $2\pi$ referring to a
microscopic distance of one wavelength\epubtkFootnote{Note that in
  other publications the tuning or equivalent microscopic
  displacements are sometimes defined via an optical path-length
  difference. In that case, a tuning of $2\pi$ is used to refer to the
  change of the optical path length of one wavelength, which, for
  example, if the reflection at a mirror is described, corresponds to
  a change of the mirror's position of $\lambda_0/2$.} $\lambda_0$.

Finally, we can write the following expression for the phase difference
between the light field taken at the end points of a distance $D$:
\begin{equation}
\varphi=- k D =-\left(\frac{\Delta \omega L}{c} + \Tun\frac{\omega}{\omega_0}\right),
\end{equation}
or if we neglect the last term from Equation~(\ref{eq:lengths_tuns})
we can approximate ($\omega/\omega_0\approx1$) to obtain
\begin{equation}
\varphi\approx -\left(\frac{\Delta \omega L}{c} + \Tun\right).
\end{equation}
This convention provides two parameters $L$ and $\Tun$, that can
describe distances with a markedly improved numerical accuracy. In
addition, this definition often allows simplification of the algebraic
notation of interferometer signals. By convention we associate a
length $L$ with the propagation through free space, whereas the tuning
will be treated as a parameter of the optical components. Effectively
the tuning then represents a microscopic \emph{displacement} of the
respective component. If, for example, a cavity is to be resonant to
the laser light, the tunings of the mirrors have to be the same
whereas the length of the space in between can be arbitrary.

\clearpage
\subsection{Revised coupling matrices for space and mirrors}
\label{sec:mirrors_spaces_2}

Using the definitions for length and tunings we can rewrite the
coupling equations for mirrors and spaces introduced in
Section~\ref{sec:mirrors_spaces} as follows. The mirror coupling
becomes

\epubtkImage{mirror_coupling_deltax-external.png}{%
  \begin{figure}[h!]
    \centerline{\includegraphics[width=0.9\textwidth]{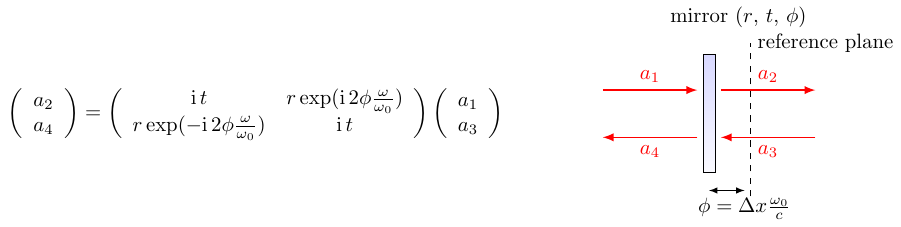}}
    \caption{}
    \label{eq:mirror_coupling_matrix2}
\end{figure}}
\noindent
(compare this to Figure~\ref{eq:mirror_coupling_matrix1}), and the
amplitude coupling for a `space', formally written as in
Figure~\ref{eq:space_coupling_matrix1}, is now written as

\epubtkImage{space_coupling_matrix2-external.png}{%
  \begin{figure}[h!]
    \centerline{\includegraphics[width=0.9\textwidth]{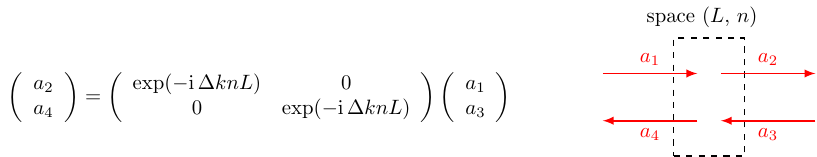}}
    \caption{}
    \label{eq:space_coupling_matrix2}
\end{figure}}

\subsection{\Finesse examples}
\subsubsection{Mirror reflectivity and transmittance}

\epubtkImage{fexample_rt_mirror.png}{%
  \begin{figure}[htbp]
    \centerline{\includegraphics[width=0.85\textwidth]{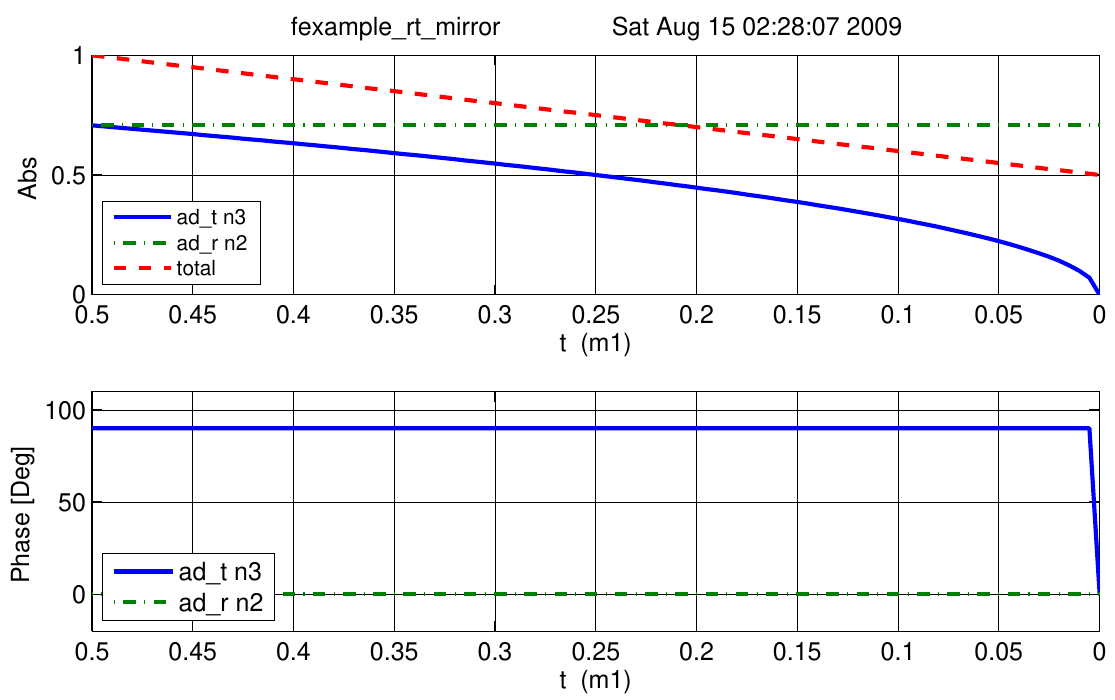}}
    \caption{\Finesse example: Mirror reflectivity and transmittance.}
    \label{fig:fexample_rt_mirror}
\end{figure}}

\noindent
We use \Finesse to plot the amplitudes of the light fields transmitted
and reflected by a mirror (given by a single surface). Initially, the
mirror has a power reflectance and transmittance of $R=T=0.5$ and is,
thus, lossless. For the plot in Figure~\ref{fig:fexample_rt_mirror} we
tune the transmittance from 0.5 to 0. Since we do not explicitly
change the reflectivity, $R$ remains at 0.5 and the mirror loss
increases instead, which is shown by the trace labelled `total'
corresponding to the sum of the reflected and transmitted light
power. The plot also shows the phase convention of a 90\textdegree\
phase shift  for the transmitted light.

\vspace{3mm}\noindent
{\small
\textbf{Finesse input file for `Mirror reflectivity and transmittance'}
{\renewcommand{\baselinestretch}{.8}

\nopagebreak
\tt
\noindent
\mbox{} \\
\mbox{}\textbf{\textcolor{RoyalBlue}{laser}}\ \ l1\ \textcolor{Purple}{1}\ \textcolor{Purple}{0}\ n1\ \ \textcolor{Gray}{\%\ laser\ with\ P=1W\ at\ the\ default\ frequency} \\
\mbox{}\textbf{\textcolor{RoyalBlue}{space}}\ \ s1\ \textcolor{Purple}{1}\ n1\ n2\ \textcolor{Gray}{\%\ space\ of\ 1m\ length} \\
\mbox{}\textbf{\textcolor{RoyalBlue}{mirror}}\ m1\ \textcolor{Purple}{0.5}\ \textcolor{Purple}{0.5}\ \textcolor{Purple}{0}\ n2\ n3\ \textcolor{Gray}{\%\ mirror\ with\ T=R=0.5\ at\ zero\ tuning} \\
\mbox{}\textbf{\textcolor{RoyalBlue}{ad}}\ ad\_t\ \textcolor{Purple}{0}\ n3\ \ \ \ \ \ \textcolor{Gray}{\%\ an\ `amplitude'\ detector\ for\ transmitted\ light} \\
\mbox{}\textbf{\textcolor{RoyalBlue}{ad}}\ ad\_r\ \textcolor{Purple}{0}\ n2\ \ \ \ \ \ \textcolor{Gray}{\%\ an\ `amplitude'\ detector\ for\ reflected\ light} \\
\mbox{}\textbf{\textcolor{Red}{set}}\textcolor{ForestGreen}{\ t}\ ad\_t\ abs \\
\mbox{}\textbf{\textcolor{Red}{set}}\textcolor{ForestGreen}{\ r}\ ad\_r\ abs \\
\mbox{}\textbf{\textcolor{Red}{func}}\textcolor{ForestGreen}{\ total}\ \textcolor{BrickRed}{=}\ \textcolor{ForestGreen}{\$r}\textcolor{BrickRed}{\textasciicircum{}}\textcolor{Purple}{2}\ \textcolor{BrickRed}{+}\ \textcolor{ForestGreen}{\$t}\textcolor{BrickRed}{\textasciicircum{}}\textcolor{Purple}{2}\ \textcolor{Gray}{\%\ computing\ the\ sum\ of\ the\ reflected\ and\ transmitted\ power} \\
\mbox{} \\
\mbox{}\textbf{\textcolor{Red}{xaxis}}\ m1\ t\ lin\ \textcolor{Purple}{0.5}\ \textcolor{Purple}{0}\ \textcolor{Purple}{100}\ \textcolor{Gray}{\%\ changing\ the\ transmittance\ of\ the\ mirror\ `m1'} \\
\mbox{}\textbf{\textcolor{Red}{yaxis}}\ abs\textcolor{BrickRed}{:}deg\ \ \ \ \ \textcolor{Gray}{\%\ plotting\ amplitude\ and\ phase\ of\ the\ results} \\
\mbox{}

}}

\subsubsection{Length and tunings}

\epubtkImage{fexample_length_tunings.png}{%
  \begin{figure}[ht]
    \centerline{\includegraphics[width=0.9\textwidth]{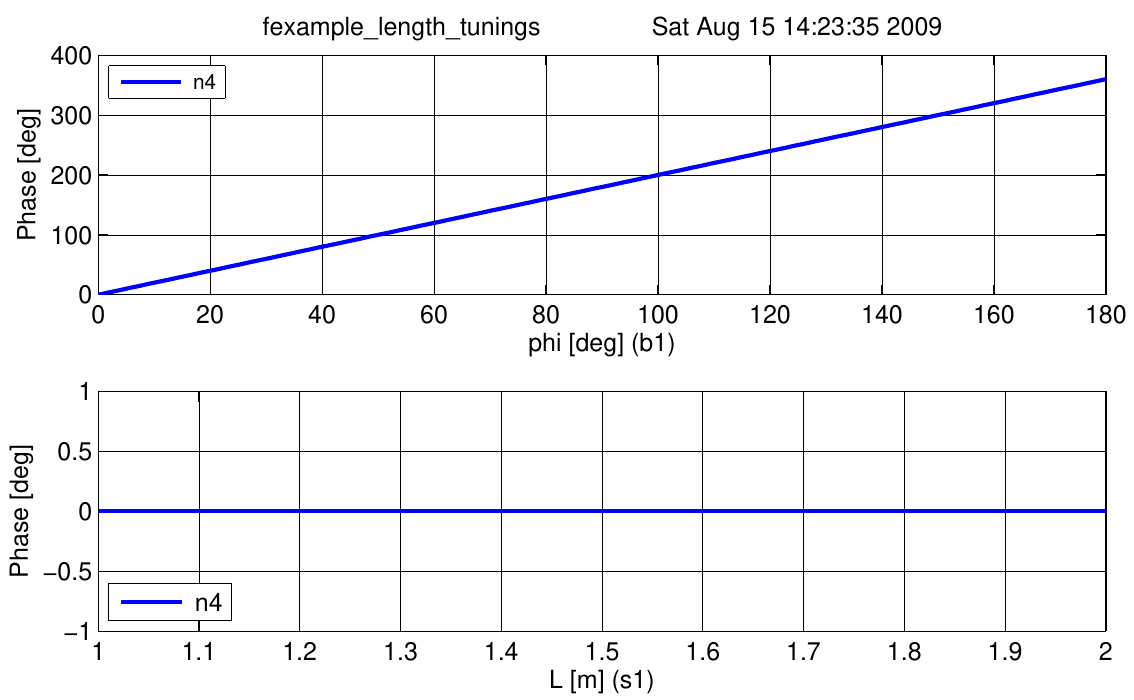}}
    \caption{\Finesse example: Length and tunings.}
    \label{fig:fexample_length_tunings}
\end{figure}}

These \Finesse files demonstrate the conventions for lengths and
microscopic positions introduced in Section~\ref{sec:tuning}. The top
trace in Figure~\ref{fig:fexample_length_tunings} depicts the phase
change of a beam reflected by a beam splitter as the function of the
beam splitter tuning. By changing the tuning from 0 to 180\textdegree\
the beam splitter is moved forward and shortens the path length by one
wavelength, which by convention increases the light phase by
360\textdegree.  On the other hand, if a length of a space is changed,
the phase of the transmitted light is unchanged (for the default
wavelength $\Delta k=0$), as shown in the lower trace.

\vspace{3mm}\noindent
{\small
\textbf{Finesse input files for `Length and tunings'}
{\renewcommand{\baselinestretch}{.8}\\

\nopagebreak
\noindent
File for top trace:
{\tt
\noindent
\mbox{} \\
\mbox{}\textbf{\textcolor{RoyalBlue}{laser}}\ \ l1\ \textcolor{Purple}{1}\ \textcolor{Purple}{0}\ n1\ \ \ \ \textcolor{Gray}{\%\ laser\ with\ P=1W\ at\ the\ default\ frequency} \\
\mbox{}\textbf{\textcolor{RoyalBlue}{space}}\ \ s1\ \textcolor{Purple}{1}\ \textcolor{Purple}{1}\ n1\ n2\ \textcolor{Gray}{\%\ space\ of\ 1m\ length} \\
\mbox{}\textbf{\textcolor{RoyalBlue}{bs}}\ \ \ \ \ b1\ \textcolor{Purple}{1}\ \textcolor{Purple}{0}\ \textcolor{Purple}{0}\ \textcolor{Purple}{0}\ n2\ n3\ dump\ dump\ \textcolor{Gray}{\%\ beam\ splitter\ as\ `turning\ mirror',\ \ normal\ incidence} \\
\mbox{}\textbf{\textcolor{RoyalBlue}{space}}\ \ s2\ \textcolor{Purple}{1}\ \textcolor{Purple}{1}\ n3\ n4\ \textcolor{Gray}{\%\ another\ space\ of\ 1m\ length} \\
\mbox{}\textbf{\textcolor{RoyalBlue}{ad}}\ \ \ \ \ ad1\ \textcolor{Purple}{0}\ n4\ \ \ \ \ \textcolor{Gray}{\%\ amplitude\ detector\ } \\
\mbox{}\textcolor{Gray}{\%\ 1)\ first\ trace:\ change\ microscopic\ position\ of\ beamsplitter} \\
\mbox{}\textbf{\textcolor{Red}{xaxis}}\ b1\ phi\ lin\ \textcolor{Purple}{0}\ \textcolor{Purple}{180}\ \textcolor{Purple}{100}\  \\
\mbox{}\textbf{\textcolor{Red}{yaxis}}\ deg\ \ \ \ \ \textcolor{Gray}{\%\ plotting\ the\ phase\ of\ the\ results} \\
\mbox{}
}

\noindent
File for bottom trace:
{\tt
\noindent
\mbox{} \\
\mbox{}\textbf{\textcolor{RoyalBlue}{laser}}\ \ l1\ \textcolor{Purple}{1}\ \textcolor{Purple}{0}\ n1\ \ \ \ \textcolor{Gray}{\%\ laser\ with\ P=1W\ at\ the\ default\ frequency} \\
\mbox{}\textbf{\textcolor{RoyalBlue}{space}}\ \ s1\ \textcolor{Purple}{1}\ \textcolor{Purple}{1}\ n1\ n2\ \textcolor{Gray}{\%\ space\ of\ 1m\ length} \\
\mbox{}\textbf{\textcolor{RoyalBlue}{bs}}\ \ \ \ \ b1\ \textcolor{Purple}{1}\ \textcolor{Purple}{0}\ \textcolor{Purple}{0}\ \textcolor{Purple}{0}\ n2\ n3\ dump\ dump\ \textcolor{Gray}{\%\ beam\ splitter\ as\ `turning\ mirror',\ \ normal\ incidence} \\
\mbox{}\textbf{\textcolor{RoyalBlue}{space}}\ \ s2\ \textcolor{Purple}{1}\ \textcolor{Purple}{1}\ n3\ n4\ \textcolor{Gray}{\%\ another\ space\ of\ 1m\ length} \\
\mbox{}\textbf{\textcolor{RoyalBlue}{ad}}\ \ \ \ \ ad1\ \textcolor{Purple}{0}\ n4\ \ \ \ \ \textcolor{Gray}{\%\ amplitude\ detector\ } \\
\mbox{}\textcolor{Gray}{\%\ second\ trace:\ change\ length\ of\ space\ s1} \\
\mbox{}\textbf{\textcolor{Red}{xaxis}}\ s1\ L\ lin\ \textcolor{Purple}{1}\ \textcolor{Purple}{2}\ \textcolor{Purple}{100}\  \\
\mbox{}\textbf{\textcolor{Red}{yaxis}}\ deg\ \ \ \ \ \textcolor{Gray}{\%\ plotting\ the\ phase\ of\ the\ results} \\
\mbox{}
}
}}

\newpage

\section{Light with Multiple Frequency Components}
\label{sec:multi-frequency}

So far we have considered the electromagnetic field to be
monochromatic. This has allowed us to compute light-field amplitudes
in a quasi-static optical setup. In this section, we introduce the
frequency of the light as a new degree of freedom. In fact, we
consider a field consisting of a finite and discrete number of
frequency components. We write this as
\begin{equation}
\label{eq:mf_intro}
E(t,z)=\sum_{j}~a_{j}~\mEx{\I(\w_j \T -k_jz)},
\end{equation}
with complex amplitude factors $a_{j}$, $\w_j$ as the angular
frequency of the light field and $k_j=\w_j/c$. In many cases the
analysis compares different fields at one specific location only, in
which case we can set $z=0$ and write
\begin{equation}
E(t)=\sum_{j}~a_{j}~\mEx{\I \w_j \T}.
\end{equation}
In the following sections the concept of light modulation is
introduced. As this inherently involves light fields with multiple
frequency components, it makes use of this type of field
description. Again we start with the two-mirror cavity to illustrate
how the concept of modulation can be used to model the effect of
mirror motion.

\epubtkImage{modulation-external.png}{%
  \begin{figure}[htbp]
    \centerline{\includegraphics[width=0.9\textwidth]{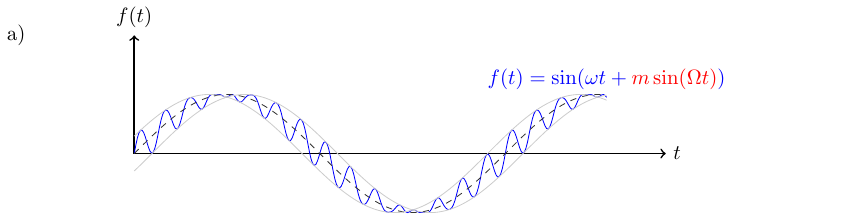}}
    \centerline{\includegraphics[width=0.9\textwidth]{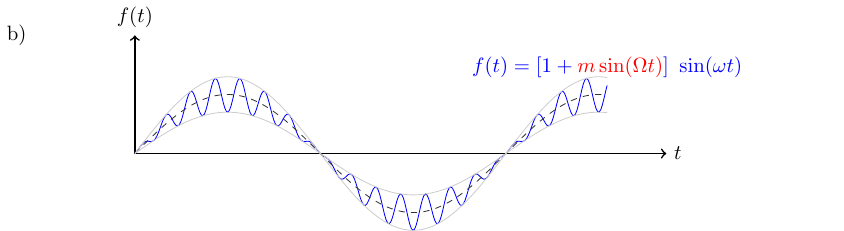}}
    \caption{Example traces for phase and amplitude modulation: the
      upper plot a) shows a phase-modulated sine wave and the lower
      plot b) depicts an amplitude-modulated sine wave. Phase
      modulation is characterised by the fact that it mostly affects
      the zero crossings of the sine wave. Amplitude modulation
      affects mostly the maximum amplitude of the wave. The equations
      show the modulation terms in red with $m$ the modulation index
      and $\OmMod$ the modulation frequency.}
    \label{fig:modulation}
\end{figure}}

\subsection{Modulation of light fields}
\label{sec:mod}

Laser interferometers typically use three different types of light
fields: the laser with a frequency of, for example,
$f\approx2.8\cdot10^{14}\mathrm{\ Hz}$, \emph{radio frequency} (RF)
sidebands used for interferometer control with frequencies (offset to
the laser frequency) of $f\approx1\cdot10^{6}$ to
$150\cdot10^{6}\mathrm{\ Hz}$, and the \emph{signal} sidebands at
frequencies of 1 to 10,000~Hz\epubtkFootnote{The signal sidebands
  are sometimes also called \emph{audio sidebands} because of their
  frequency range.}. As these modulations usually have as their origin
a change in optical path length, they are often phase modulations of
the laser frequency, the RF sidebands are utilised for optical readout
purposes, while the signal sidebands carry the signal to be measured
(the gravitational-wave signal plus noise created in the interferometer).

Figure~\ref{fig:modulation} shows a \emph{time domain} representation of
an electromagnetic wave of frequency $\omega_0$, whose amplitude or
phase is modulated at a frequency $\OmMod$. One can easily see some
characteristics of these two types of modulation, for example, that
amplitude modulation leaves the zero crossing of the wave unchanged
whereas with phase modulation the maximum and minimum amplitude of the
wave remains the same.
In the \emph{frequency domain} in which a modulated field is expanded
into several unmodulated field components, the interpretation of
modulation becomes even easier: any sinusoidal modulation of amplitude
or phase generates new field components, which are shifted in frequency
with respect to the initial field. Basically, light power is shifted
from one frequency component, the \emph{carrier}, to several others,
the \emph{sidebands}. The relative amplitudes and phases of these
sidebands differ for different types of modulation and different
modulation strengths. This section demonstrates how to compute the
sideband components for amplitude, phase and frequency modulation.

\subsection{Phase modulation}
\label{sec:phasemod}

Phase modulation can create a large number of sidebands. The number of
sidebands with noticeable power depends on the modulation strength (or
depth) given by the \emph{modulation index} $m$. Assuming an input
field
\begin{equation}
E_{\mathrm{in}}=E_0~\mEx{\I\w_0 \T},
\end{equation}
a sinusoidal phase modulation of the field can be described as
\begin{equation}
\label{eq:mod1}
E=E_0~\mExB{\I(\w_0 \T + m \mCos{\OmMod \T})}.
\end{equation}
This equation can be expanded using the identity~\cite{Gradstein8511}
%
\begin{equation}
\exp(\I z \cos\varphi)=\sum_{k=-\infty}^\infty \I^kJ_k(z)\exp(\I k \varphi),
\end{equation}
with \emph{Bessel functions of the first kind} $J_k(m)$. We can write
\begin{equation}
\label{eq:bessel0}
E=E_0~\mEx{\I\w_0 \T}~\sum_{k=-\infty}^{\infty}\I^{\,k}~J_k(m)~\mEx{\I k
\OmMod\T}.
\end{equation}
The field for $k=0$, oscillating with the frequency of the input field
$\w_0$, represents the carrier. The sidebands can be divided into
\emph{upper} ($k>0$) and \emph{lower} ($k<0$) sidebands. These
sidebands are light fields that have been shifted in frequency by $k\,
\OmMod$. The upper and lower sidebands with the same absolute value of
$k$ are called a pair of sidebands of order
$k$. Equation~(\ref{eq:bessel0}) shows that the carrier is surrounded
by an infinite number of sidebands.  However, for small modulation
indices ($m<1$) the Bessel functions rapidly decrease with increasing
$k$ (the lowest orders of the Bessel functions are shown in
Figure~\ref{fig:bessel_functions}). For small modulation indices we
can use the approximation~\cite{Abramowitz65}
\begin{equation}
J_k(m)~=\left(\frac{m}{2}\right)^k\sum_{n=0}^\infty\frac{\left(-\frac{m^2}{4}\right)^n}
{n! (k+n)!}=\frac{1}{k!}\left(\frac{m}{2}\right)^k+O\left(m^{k+2}\right).
\end{equation}
In which case, only a few sidebands have to be taken into account. For
$m\ll1$ we can write
\begin{equation}
{\renewcommand{\arraystretch}{1.5}
\begin{array}{lcl}
E&=&E_0~\mEx{\I\w_0 \T}\\
& & \times\Bigl(J_0(m)-\I J_{-1}(m)~\mEx{-\I \OmMod\T}+\I J_{1}(m)~\mEx{\I
\OmMod\T}\Bigr),
\end{array}}
\end{equation}
and with
\begin{equation}
J_{-k}(m)=(-1)^kJ_k(m),
\end{equation}
we obtain
\begin{equation}
E=E_0~\mEx{\I\w_0 \T}~\left(1+\I\frac{m}{2}\Bigl(\mEx{-\I
\OmMod\T}+\mEx{\I \OmMod\T}\Bigr)\right),
\label{eq:phase_mod}
\end{equation}
as the first-order approximation in $m$. In the above equation the
carrier field remains unchanged by the modulation, therefore this
approximation is not the most intuitive. It is clearer if the
approximation up to the second order in $m$ is given:
\begin{equation}
E=E_0~\mEx{\I\w_0 \T}~\left(1-\frac{m^2}{4}+\I\frac{m}{2}\Bigl(\mEx{-\I
\OmMod\T}+\mEx{\I \OmMod\T}\Bigr)\right),
\end{equation}
which shows that power is transferred from the carrier to the sideband fields.

Higher-order expansions in $m$ can be performed simply by specifying
the highest order of Bessel function, which is to be used in the sum in
Equation~(\ref{eq:bessel0}), i.e.~
\begin{equation}
E=E_0~\mEx{\I\w_0 \T}~\sum_{k=-order}^{order}i^{\,k}~J_k(m)~\mEx{\I k \OmMod\T}.
\label{eq:bessel1}
\end{equation}

\epubtkImage{bessel_function.png}{%
  \begin{figure}[htbp]
    \centerline{\includegraphics[width=0.9\textwidth]{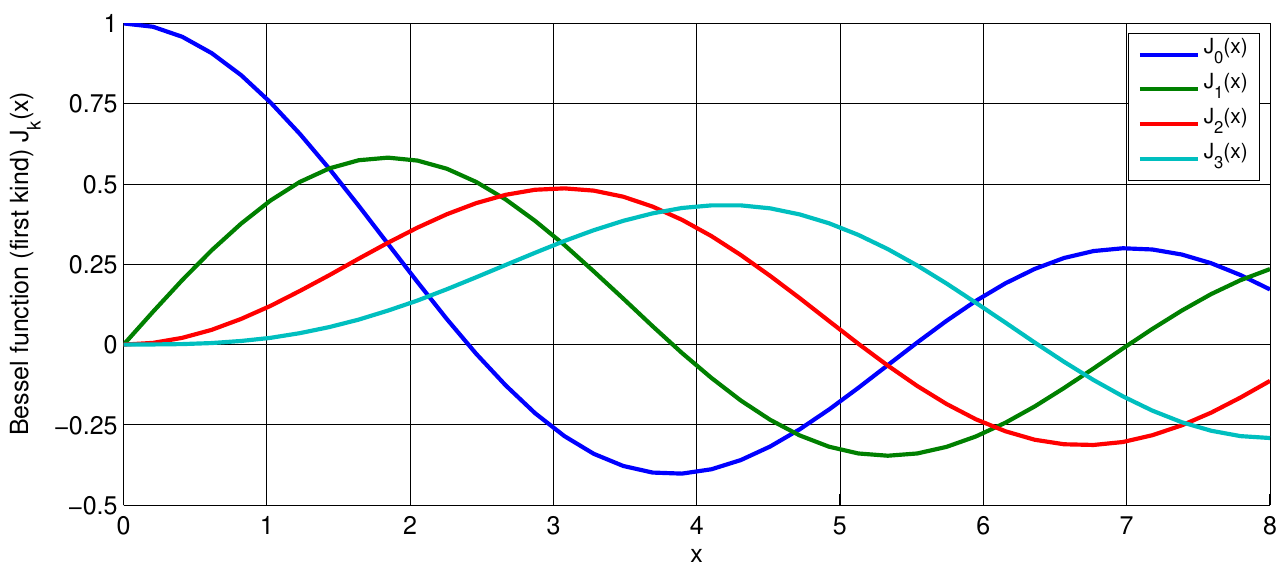}}
    \caption{Some of the lowest-order Bessel functions $J_k(x)$ of the
    first kind. For small $x$ the expansion shows a simple $x^k$
    dependency and higher-order functions can often be neglected.}
    \label{fig:bessel_functions}
\end{figure}}

\subsection{Frequency modulation}

For small modulation, indices, phase modulation and frequency
modulation can be understood as different descriptions of the same
effect~\cite{Heinzel99}. Following the same spirit as above we would
assume a modulated frequency to be given by
\begin{equation}
\label{eq:freqmod}
\omega=\omega_0+m'\mCos{\OmMod\T},
\end{equation}
and then we might be tempted to write
\begin{equation}
E=E_0~\mExB{\I(\w_0 + m' \mCos{\OmMod \T})\T},
\end{equation}
which would be wrong. The frequency of a wave is actually defined as
$\omega/(2\pi)=f= d\varphi/dt$. Thus, to obtain the frequency given in
Equation~(\ref{eq:freqmod}), we need to have a phase of
\begin{equation}
\w_0\T + \frac{m'}{\OmMod} \mSin{\OmMod \T}.
\end{equation}
For consistency with the notation for phase modulation, we define the
modulation index to be
\begin{equation}
m=\frac{m'}{\OmMod}=\frac{\Delta \omega}{\OmMod},
\end{equation}
with $\Delta\w$ as the frequency swing -- how \emph{far} the frequency
is shifted by the modulation -- and $\OmMod$ the modulation frequency
-- how \emph{fast} the frequency is shifted. Thus, a sinusoidal
frequency modulation can be written as
\begin{equation}
E=E_0\mEx{\I\varphi}=E_0~\mEx{\I\left(\w_0\T + \frac{\Delta\w}{\OmMod} \mCos{\OmMod \T}\right)},
\end{equation}
which is exactly the same expression as Equation~(\ref{eq:mod1}) for
phase modulation. The practical difference is the typical size of the
modulation index, with phase modulation having a modulation index of
$m<10$, while for frequency modulation, typical numbers might be
$m>10^4$. Thus, in the case of frequency modulation, the
approximations for small $m$ are not valid. The series expansion using
Bessel functions, as in Equation~(\ref{eq:bessel0}), can still be
performed; however, very many terms of the resulting sum need to be
taken into account.

\subsection{Amplitude modulation}

In contrast to phase modulation, (sinusoidal) amplitude modulation
always generates exactly two sidebands.  Furthermore, a natural
maximum modulation index exists: the modulation index is defined to be
one ($m=1$) when the amplitude is modulated between zero and the
amplitude of the unmodulated field.

If the amplitude modulation is performed by an active element, for
example by modulating the current of a laser diode, the following
equation can be used to describe the output field:
\begin{equation}
{\renewcommand{\arraystretch}{1.5}
\begin{array}{lcl}
E&=&E_0~\mEx{\I\w_0 \T}~\Bigl(1+m\mCos{\OmMod \T}\Bigr)\\
&=&E_0~\mEx{\I\w_0 \T}~\Bigl(1+\frac{m}{2}~\mEx{\I \OmMod
\T}+\frac{m}{2}~\mEx{-\I \OmMod \T}\Bigr).
\end{array}}
\end{equation}
However, passive amplitude modulators (like acousto-optic modulators or
electro-optic modulators with polarisers) can only reduce the amplitude.
In these cases, the following equation is more useful:
\begin{equation}
{\renewcommand{\arraystretch}{1.5}
\begin{array}{lcl}
E&=&E_0~\mEx{\I\w_0 \T}~\left(1-\frac{m}{2}\Bigl(1-\mCos{\OmMod
\T}\Bigr)\right)\\
&=&E_0~\mEx{\I\w_0 \T}~\Bigl(1-\frac{m}{2}+\frac{m}{4}~\mEx{\I \OmMod
\T}+\frac{m}{4}~\mEx{-\I \OmMod \T}\Bigr).
\end{array}}
\end{equation}

\epubtkImage{rotating-frame-adf.png}{%
  \begin{figure}[htbp]
    \centerline{\includegraphics[viewport=18 23 190 90,scale=1.5]{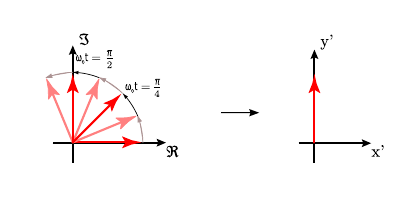}}
    \caption{Electric field vector $E_0\exp(\I \omega_0 t)$ depicted
      in the complex plane and in a rotating frame ($x^{\prime}$,
      $y^{\prime}$) rotating at $\omega_0$ so that the field vector
      appears stationary.}
    \label{fig:rotatingframe}
\end{figure}}

\subsection{Sidebands as phasors in a rotating frame}

A common method of visualising the behaviour of sideband fields
in interferometers is to use \emph{phase diagrams} in which each
field amplitude is represented by an arrow in the complex plane.

\epubtkImage{Amplitude-Phase-Modulation-MK4-adf.png}{%
  \begin{figure}[htbp]
    \centerline{\includegraphics[scale=.4, angle=0]{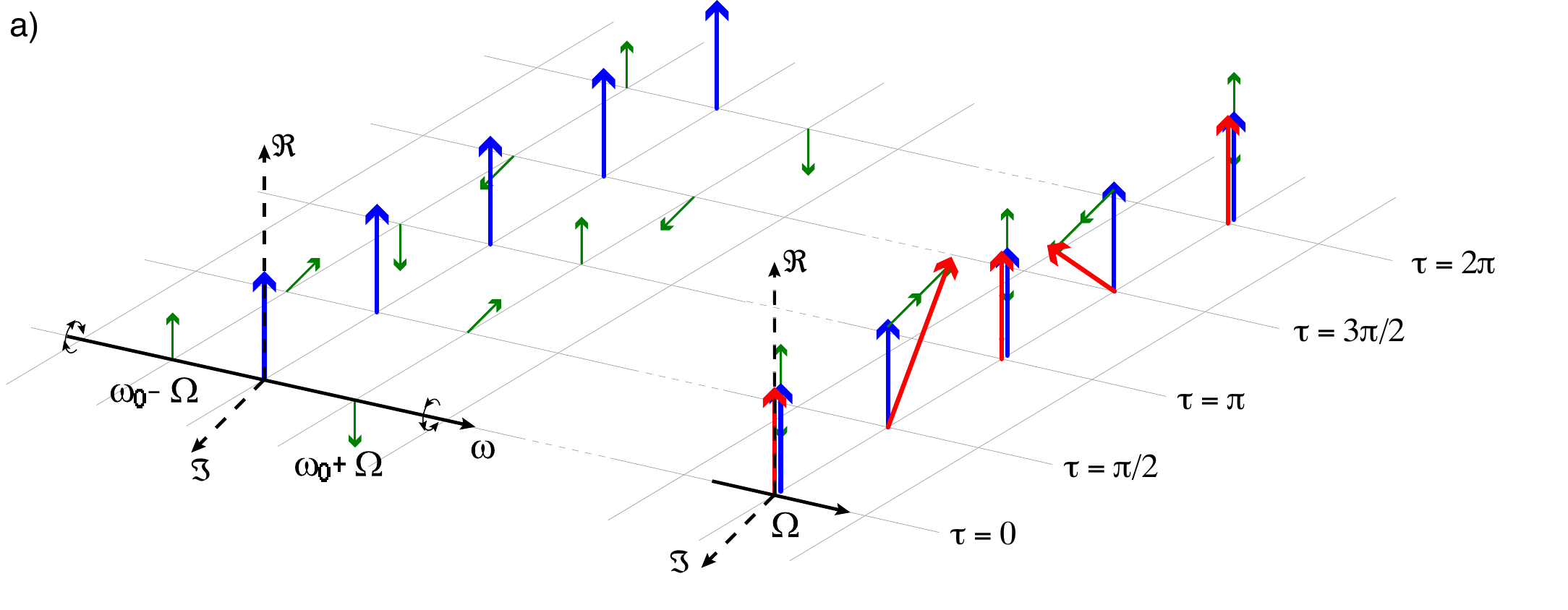}}
    \vspace{5mm}
    \centerline{\includegraphics[scale=.4, angle=0]{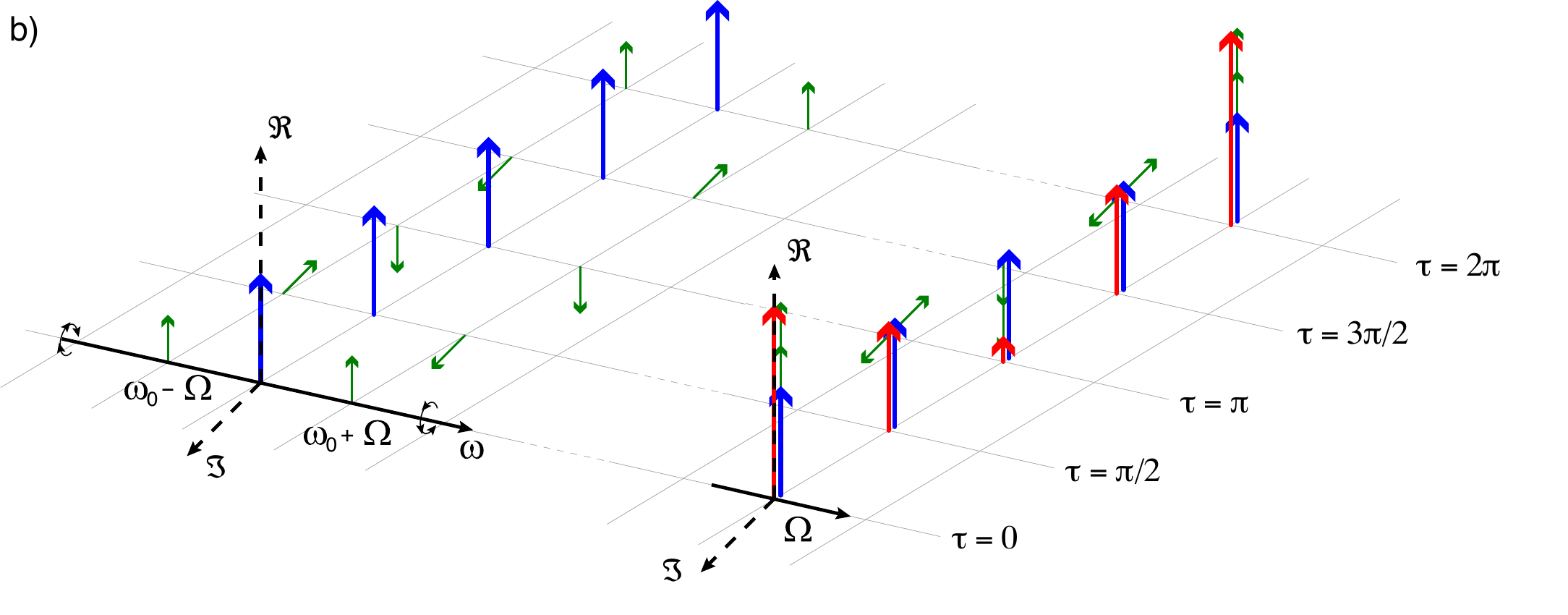}}
    \caption{Amplitude and phase modulation in the `phasor'
      picture. The upper plots a) illustrate how a phasor diagram can
      be used to describe phase modulation, while the lower plots
      b) do the same for amplitude modulation. In both cases the left hand
      plot shows the carrier in blue and the modulation sidebands in
      green as snapshots at certain time intervals. One can see
      clearly that the upper sideband ($\omega_0+\Omega$) rotates
      faster than the carrier, while the lower sideband rotates
      slower. The right plot in both cases shows how the total field
      vector at any given time can be constructed by adding the three
      field vectors of the carrier and sidebands. [Drawing courtesy of
      Simon Chelkowski]}
    \label{fig:phasors}
\end{figure}}

We can think of the electric field amplitude $E_0\exp(\I\omega_0 t)$
as a vector in the complex plane, rotating around the origin with
angular velocity $\omega_0$. To illustrate or to help visualise the
addition of several light fields it can be useful to look at this
problem using a \emph{rotating reference frame}, defined as follows. A
complex number shall be defined as $z=x+\I y$ so that the real part is
plotted along the \x-axis, while the \y-axis is used for the imaginary
part. We want to construct a new coordinate system ($x^{\prime}$,
$y^{\prime}$) in which the field vector is at a constant
position. This can be achieved by defining
\begin{equation}
  \begin{array}{ll}
    x &= x^{\prime}\ \cos\omega_0 t - y^{\prime}\ \sin\omega_0 t\\
    y &=  x^{\prime}\ \sin\omega_0 t + y^{\prime}\ \cos\omega_0 t,
  \end{array}
\end{equation}
or
\begin{equation}
  \begin{array}{ll}
    x^{\prime} &= x\ \cos\left(-\omega_0 t\right) - y\ \sin\left( -\omega_0 t \right)\\
    y^{\prime} &=  x\ \sin\left( -\omega_0 t \right) + y\ \cos\left( -\omega_0 t \right).
  \end{array}
\end{equation}
Figure~\ref{fig:rotatingframe} illustrates how the transition into the
rotating frame makes the field vector to appear stationary. The angle
of the field vector in a rotating frame depicts the phase offset of
the field. Therefore these vectors are also called \emph{phasors} and
the illustrations using phasors are called \emph{phasor diagrams}. Two
more complex examples of how phasor diagrams can be employed is shown
in Figure~\ref{fig:phasors}~\cite{phd.Chelkowski}.

Phasor diagrams can be especially useful to see how frequency coupling
of light field amplitudes can change the type of modulation, for
example, to turn phase modulation into amplitude modulation. An
extensive introduction to this type of phasor diagram can be found
in~\cite{phd.Malec}.

\subsection{Phase modulation through a moving mirror}

Several optical components can modulate transmitted or reflected light
fields. In this section we discuss in detail the example of phase
modulation by a moving mirror. Mirror motion does not change the
transmitted light; however, the phase of the reflected light will be
changed as shown in Equation~(\ref{eq:mirror_coupling_matrix2}).

\epubtkImage{mirror_coupling_modx-external.png}{%
  \begin{figure}[htbp]
    \centerline{\includegraphics[width=0.6\textwidth]{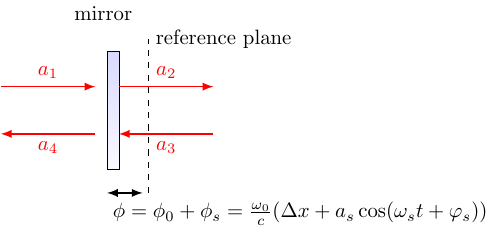}}
    \caption{A sinusoidal signal with amplitude $a_s$ frequency
      $\omega_s$ and phase offset $\varphi_s$ is applied to a mirror
      position, or to be precise, to the mirror tuning. The equation
      given for the tuning $\Tun$ assumes that $\omega_s/\omega_0 \ll
      1$, see Section~\ref{sec:tuning}.}
    \label{fig:mirror_mod}
\end{figure}}

We assume sinusoidal change of the mirror's tuning as shown in
Figure~\ref{fig:mirror_mod}. The position modulation is given as
$x_{\mathrm{m}}=a_{\mathrm{s}}\cos(\w_{\mathrm{s}}t+\varphi_{\mathrm{s}})$,
and thus the reflected field at the mirror becomes (assuming $a_4=0$)
\begin{equation}
a_3=r\, a_1 \exp(-\I 2\phi_0)\,\mEx{\I2k x_{\mathrm{m}}}~\approx r a_1 \exp(-\I 2\phi_0)\,\mExB{\I2k_{\mathrm{0}} a_{\mathrm{s}}\cos(\w_{\mathrm{s}}t+\varphi_{\mathrm{s}})},
\end{equation}
setting $m=2k_{\mathrm{0}} a_{\mathrm{s}}$. This can be expressed as
\begin{equation}
{\renewcommand{\arraystretch}{1.5}
\begin{array}{rcl}
a_3&=&r a_1 \exp(-\I 2\phi_0)\,\Bigl(1+\I\frac{m}{2}\mExB{-\I(\w_{\mathrm{s}}t+\varphi_{\mathrm{s}})}+
\I\frac{m}{2}\mExB{\I(\w_{\mathrm{s}}t+\varphi_{\mathrm{s}})}\Bigr)\\
&=&r a_1 \exp(-\I 2\phi_0)\,\Bigl(1+\frac{m}{2}\mExB{-\I(\w_{\mathrm{s}}t+\varphi_{\mathrm{s}}-\pi/2)}~\Bigr.\\
& &+\Bigl.\frac{m}{2}\mExB{\I(\w_{\mathrm{s}}t+\varphi_{\mathrm{s}}+\pi/2)}\Bigr).
\end{array}}
\end{equation}
%

\subsection{Coupling matrices for beams with multiple frequency
  components}

The coupling between electromagnetic fields at optical components
introduced in Section~\ref{sec:components} referred only to the
amplitude and phase of a simplified monochromatic field, ignoring all
the other parameters of the electric field of the beam given in
Equation~(\ref{eq:field_parameters}). However, this mathematical
concept can be extended to include other parameters provided that we
can find a way to describe the total electric field as a sum of
components, each of which is characterised by a discrete value of the
related parameters. In the case of the frequency of the light field,
this means we have to describe the field as a sum of monochromatic
components. In the previous sections we have shown how this could be
done in the special case of an initial monochromatic field that is
subject to modulation: if the modulation index is small enough we can
limit the number of frequency components that we need to consider. In
many cases it is actually sufficient to describe a modulation only by
the interaction of the carrier at $\omega_0$ (the unmodulated field)
and two sidebands with a frequency offset of $\pm \omega_m$ to the
carrier. A beam given by the sum of three such components can be
described by a complex vector:
\begin{equation}
\label{eq:freq_amp_vector}
\vec{a}=
\left(\begin{array}{c}
a(\omega_0)\\
a( \omega_0-\omega_m)\\
a( \omega_0+\omega_m)\\
\end{array}\right)=
\left(\begin{array}{c}
a_{\omega0}\\
a_{\omega1}\\
a_{\omega2}\\
\end{array}\right)
\end{equation}
with $\omega_0=\omega0$, $\omega_0-\omega_m=\omega1$ and
$\omega_0+\omega_m=\omega2$. In the case of a phase modulator that
applies a modulation of small modulation index $m$ to an incoming
light field  $\vec{a}_1$, we can describe the coupling of the
frequency component as follows:
\begin{equation}
\label{eq:freq_amp_coupling1}
\begin{array}{l}
a_{2,\omega0}=J_0(m) a_{1,\omega0} +J_{1}(m) a_{1,\omega1} +J_{-1}(m) a_{1,\omega2}\\
a_{2,\omega1}=J_{0}(m)a_{1,\omega1}+J_{-1}(m) a_{1,\omega0} \\
 a_{2,\omega2}=J_{0}(m)a_{1,\omega2} + J_1(m) a_{1,\omega0 },\\
\end{array}
\end{equation}
which can be written in matrix form:
\begin{equation}
\label{eq:freq_amp_coupling2}
\vec{a}_2=\left(
\begin{array}{ccc}
J_0(m)    & J_{1}(m) & J_{-1}(m)\\
J_{-1}(m) & J_0(m)   & 0 \\
J_{1}(m)  & 0        & J_0(m) \\
\end{array}\right)\vec{a}_1 .
\end{equation}
And similarly, we can write the complete coupling matrix for the
modulator component, for example, as
\begin{equation}
\label{eq:freq_amp_coupling3}
\left(\begin{array}{c}
a_{2,w0}\\
a_{2,w1}\\
a_{2,w2}\\
a_{4,w0}\\
a_{4,w1}\\
a_{4,w2}\\
\end{array}\right)
\left(
\begin{array}{cccccc}
J_0(m)    & J_{1}(m) & J_{-1}(m) & 0         & 0        & 0\\
J_{-1}(m) & J_0(m)   & 0         & 0         & 0        & 0\\
J_{1}(m)  & 0        & J_0(m)    & 0         & 0        & 0\\
0         & 0        & 0         & J_0(m)    & J_{1}(m) & J_{-1}(m)\\
0         & 0        & 0         & J_{-1}(m) & J_0(m)   & 0 \\
0         & 0        & 0         & J_{1}(m)  & 0        & J_0(m)\\
\end{array}\right)
\left(\begin{array}{c}
a_{1,w0}\\
a_{1,w1}\\
a_{1,w2}\\
a_{3,w0}\\
a_{3,w1}\\
a_{3,w2}\\
\end{array}\right)
\end{equation}

\subsection{\Finesse examples}

\subsubsection{Modulation index}

\epubtkImage{fexample_bessel.png}{%
  \begin{figure}[htbp]
    \centerline{\includegraphics[width=0.9\textwidth]{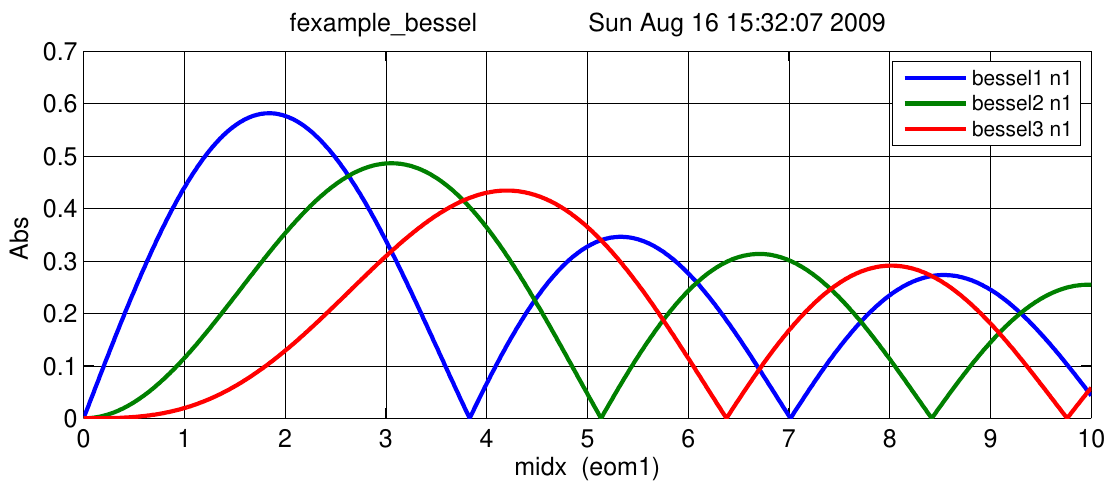}}
    \caption{\Finesse example: Modulation index.}
    \label{fig:fexample_bessel}
\end{figure}}

\noindent
This file demonstrates the use of a modulator. Phase modulation (with
up to five higher harmonics is applied to a laser beam and amplitude
detectors are used to measure the field at the first three
harmonics. Compare this to Figure~\ref{fig:bessel_functions} as well.

\vspace{3mm}\noindent
{\small
\textbf{Finesse input file for `Modulation index'}
{\renewcommand{\baselinestretch}{.8}

\nopagebreak
\tt
\noindent
\mbox{}\ \ \ \ \ \ \ \ \ \ \ \ \ \ \ \ \ \ \ \ \ \ \ \ \ \ \ \ \ \ \ \ \ \ \ \ \ \ \ \ \ \ \ \ \ \ \ \ \ \ \ \ \ \ \ \ \ \ \ \ \ \ \ \ \ \  \\
\mbox{}\textbf{\textcolor{RoyalBlue}{laser}}\ i1\ \textcolor{Purple}{1}\ \textcolor{Purple}{0}\ n0\ \ \ \ \ \ \ \ \ \ \ \ \ \ \ \textcolor{Gray}{\%\ laser\ P=1W\ f\_offset=0Hz\ \ \ \ \ \ \ \ \ \ } \\
\mbox{}\textbf{\textcolor{RoyalBlue}{mod}}\ \ \ eom1\ 40k\ \textcolor{BrickRed}{.}\textcolor{Purple}{05}\ \textcolor{Purple}{5}\ pm\ n0\ n1\ \textcolor{Gray}{\%\ phase\ modulator\ f\_mod=40kHz,\ modulation\ index=0.05} \\
\mbox{}\textbf{\textcolor{RoyalBlue}{ad}}\ \ \ \ bessel1\ 40k\ n1\ \ \ \ \ \ \ \ \ \ \textcolor{Gray}{\%\ amplitude\ detector\ f=40kHz\ \ \ \ \ \ \ \ \ \ } \\
\mbox{}\textbf{\textcolor{RoyalBlue}{ad}}\ \ \ \ bessel2\ 80k\ n1\ \ \ \ \ \ \ \ \ \ \textcolor{Gray}{\%\ amplitude\ detector\ f=80kHz\ \ \ \ \ \ \ \ \ \ } \\
\mbox{}\textbf{\textcolor{RoyalBlue}{ad}}\ \ \ \ bessel3\ 120k\ n1\ \ \ \ \ \ \ \ \ \textcolor{Gray}{\%\ amplitude\ detector\ f=120kHz\ \ \ \ \ \ \ \ \ \ } \\
\mbox{}\textbf{\textcolor{Red}{xaxis}}\ eom1\ midx\ lin\ \textcolor{Purple}{0}\ \textcolor{Purple}{10}\ \textcolor{Purple}{1000}\ \textcolor{Gray}{\%\ x-axis:\ modulation\ index\ of\ eom1\ } \\
\mbox{} \\
\mbox{}\textbf{\textcolor{Red}{yaxis}}\ abs\ \ \ \ \ \ \ \ \ \ \ \ \ \ \ \ \ \ \ \ \ \textcolor{Gray}{\%\ y-axis:\ plot\ `absolute'\ amplitude\ \ } \\
\mbox{} \\
\mbox{} \\
\mbox{}

}}

\vspace{-0.8cm}
\subsubsection{Mirror modulation}
\label{sec:mirror_mod}
\epubtkImage{fexample_mirror_modulation.png}{%
  \begin{figure}[htbp]
    \centerline{\includegraphics[width=0.9\textwidth]{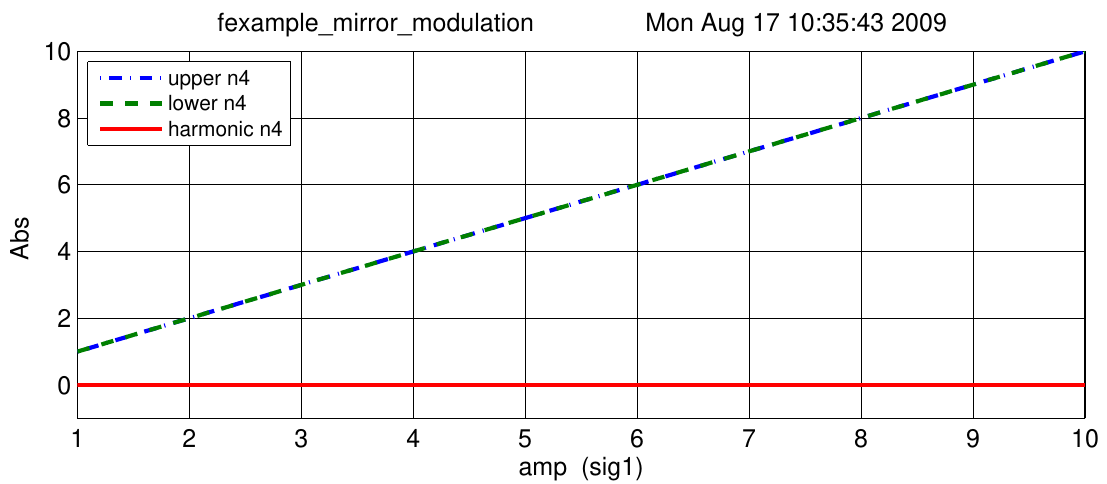}}
    \caption{\Finesse example: Mirror modulation.}
    \label{fig:fexample_mirror_modulation}
\end{figure}}

\noindent
\Finesse offers two different types of modulators: the `modulator'
component shown in the example above, and the `fsig' command, which can
be used to apply a \emph{signal modulation} to existing optical
components. The main difference is that `fsig' is meant to be used for
transfer function computations. Consequently \Finesse discards all
nonlinear terms, which means that the sideband amplitude is
proportional to the signal amplitude and harmonics are not created.

\vspace{3mm}\noindent
{\small
\textbf{Finesse input file for `Mirror modulation'}
{\renewcommand{\baselinestretch}{.8}

\nopagebreak
\tt
\noindent
\mbox{}\ \ \ \ \ \ \ \ \ \ \ \ \ \ \ \ \ \ \ \ \ \ \ \ \ \ \ \ \ \ \ \ \ \ \ \ \ \ \ \ \ \ \ \ \ \ \ \ \ \ \ \ \ \ \ \ \ \ \ \ \ \ \ \ \ \  \\
\mbox{}\textbf{\textcolor{RoyalBlue}{laser}}\ i1\ \textcolor{Purple}{1}\ \textcolor{Purple}{0}\ n1\ \ \ \ \ \ \ \ \ \ \ \ \ \ \ \textcolor{Gray}{\%\ laser\ P=1W\ f$\_$offset=0Hz\ \ \ } \\
\mbox{}\textbf{\textcolor{RoyalBlue}{space}}\ \ s1\ \textcolor{Purple}{1}\ \textcolor{Purple}{1}\ n1\ n2\ \ \ \ \ \ \ \ \ \ \ \textcolor{Gray}{\%\ space\ of\ 1m\ length} \\
\mbox{}\textbf{\textcolor{RoyalBlue}{bs}}\ \ \ \ \ b1\ \textcolor{Purple}{1}\ \textcolor{Purple}{0}\ \textcolor{Purple}{0}\ \textcolor{Purple}{0}\ n2\ n3\ dump\ dump\ \textcolor{Gray}{\%\ beam\ splitter\ as\ `turning\ mirror',\ \ normal\ incidence} \\
\mbox{}\textbf{\textcolor{RoyalBlue}{space}}\ \ s2\ \textcolor{Purple}{1}\ \textcolor{Purple}{1}\ n3\ n4\ \ \ \ \ \ \ \ \ \ \ \textcolor{Gray}{\%\ another\ space\ of\ 1m\ length} \\
\mbox{}\textbf{\textcolor{RoyalBlue}{fsig}}\ sig1\ b1\ 40k\ \textcolor{Purple}{1}\ \textcolor{Purple}{0}\ \ \ \ \ \ \ \ \ \ \textcolor{Gray}{\%\ signal\ modulation\ applied\ to\ beam\ splitter\ b1} \\
\mbox{}\textbf{\textcolor{RoyalBlue}{ad}}\ \ \ \ upper\ \ \ \ 40k\ n4\ \ \ \ \ \ \ \ \ \textcolor{Gray}{\%\ amplitude\ detector\ f=40kHz\ \ \ \ \ \ \ \ } \\
\mbox{}\textbf{\textcolor{RoyalBlue}{ad}}\ \ \ \ lower\ \ \ \textcolor{BrickRed}{-}40k\ n4\ \ \ \ \ \ \ \ \ \textcolor{Gray}{\%\ amplitude\ detector\ f=-40kHz\ \ \ \ \ \ \ \ } \\
\mbox{}\textbf{\textcolor{RoyalBlue}{ad}}\ \ \ \ harmonic\ 80k\ n4\ \ \ \ \ \ \ \ \ \textcolor{Gray}{\%\ amplitude\ detector\ f=80kHz\ \ \ \ \ \ \ \ } \\
\mbox{}\textbf{\textcolor{Red}{xaxis}}\ sig1\ amp\ lin\ \textcolor{Purple}{1}\ \textcolor{Purple}{10}\ \textcolor{Purple}{100}\ \ \ \textcolor{Gray}{\%\ x-axis:\ amplitude\ of\ signal\ modulation\ } \\
\mbox{}\textbf{\textcolor{Red}{yaxis}}\ abs\ \ \ \ \ \ \ \ \ \ \ \ \ \ \ \ \ \ \ \ \ \textcolor{Gray}{\%\ y-axis:\ plot\ `absolute'\ amplitude\ \ } \\
\mbox{}

}}


\newpage

\section{Optical Readout}

\label{sec:detection}
In previous sections we have dealt with the amplitude of light fields directly and
also used the \emph{amplitude detector} in the \Finesse examples.
This is the advantage of a mathematical analysis versus experimental tests, in which
only light intensity or light power can be measured directly. This section gives the
mathematical details for modelling photo detectors.

The intensity of a field impinging on a photo detector is given as the magnitude
of the Poynting vector, with the Poynting vector given as~\cite{Yariv}
\begin{equation}
\label{eq:poynting}
\vec{S}=\vec{E}\times\vec{H}=\frac{1}{\mu_0}\vec{E}\times\vec{B}.
\end{equation}
Inserting the electric and magnetic components of a plane wave, we obtain
%
\begin{equation}
\label{eq:intensity}
|\vec{S}|=\frac{1}{\mu_0 c}E^2=c\epsilon_0E^2_0\cos^2(\omega t)=\frac{c\epsilon_0}{2}
E_0^2\left(1+\cos(2\omega t)\right),
\end{equation}
with $\epsilon_0$ the electric permeability of vacuum and $c$ the speed
of light.


The response of a photo detector is given by the total flux of
effective radiation\epubtkFootnote{The term \emph{effective} refers to
that amount of incident light, which is converted into
photo-electrons that are then usefully extracted from the junction
(i.e.~do not recombine within the device). This fraction is usually
 referred to as \emph{quantum efficiency} $\eta$ of the photodiode.}
during the response time of the detector. For example, in a photodiode a photon will release a charge in the n-p junction. The response
time is given by the time it takes for the charge to travel through
the detector (and further time may be taken up in the electronic
processing of the signal). The size of the photodiode and the applied
bias voltage determine the travel time of the charges with typical
values of approximately 10\,ns.  Thus, frequency components faster than
perhaps 100~MHz are not resolved by a standard photodiode. For example,
a laser beam with a wavelength of $\lambda$~=~1064~nm has a frequency of
$f=c/\lambda\approx282~10^{12}\mathrm{\ Hz}=282\mathrm{\ THz}$. Thus,
the $2\omega$ component is much too fast for the photo detector;
instead, it returns the average power
\begin{equation}
|\overline{\vec{S}}|=\frac{c\epsilon_0}{2} E_0^2.
\end{equation}
In complex notation we can write
\begin{equation}
|\overline{\vec{S}}|=\frac{c\epsilon_0}{2}E E^*.
\end{equation}
%
However, for more intuitive results the light fields can be given in
converted units, so that the light power can be computed as the square
of the light field amplitudes.  Unless otherwise noted, throughout
this work the unit of light field amplitudes is
$\sqrt{\mathrm{watt}}$. Thus, the notation used in this document to
describe the computation of the light power of a laser beam is
\begin{equation}
P=E E^*.
\end{equation}

\epubtkImage{opticalbeat-external.png}{%
  \begin{figure}[htbp]
    \centerline{\includegraphics[width=0.9\textwidth]{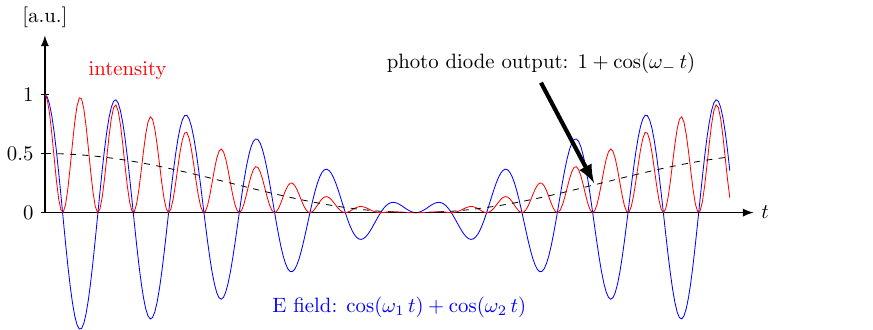}}
    \caption{A beam with two frequency components hits the photo
      diode. Shown in this plot are the field amplitude, the
      corresponding intensity and the electrical output of the photodiode.}
    \label{fig:opticalbeat}
\end{figure}}

\subsection{Detection of optical beats}
\label{sec:beats}

What is usually called an \emph{optical beat} or simply a \emph{beat}
is the sinusoidal behaviour of the intensity of two overlapping and
coherent fields. For example, if we superpose two fields of slightly
different frequency, we obtain
\begin{equation}\begin{array}{ll}
E=&E_0\cos(\omega_1 t)+ E_0 \cos(\omega_2 t)\\
\ \\
P=&E^2=E_0^2\left(\cos^2(\omega_1 t) + \cos^2(\omega_2 t) + 2\cos(\omega_{1} t)\cos(\omega_{2} t)\right)\\
&=E_0^2\left(\cos^2(\omega_1 t) + \cos^2(\omega_2 t) + \cos(\omega_{+} t) + \cos(\omega_{-} t)\right),
\end{array}
\end{equation}
with $\omega_{+}=\omega_{1}+\omega_{2}$ and
$\omega_{-}=\omega_{1}-\omega_{2}$. In this equation the frequency
$\omega_{-}$ can be very small and can then be detected with the
photodiode as illustrated in Figure~\ref{fig:opticalbeat}.
\begin{equation}
P_{\mathrm{diode}}=E_0^2\left(1+\cos(\omega_{-} t)\right)
\end{equation}
Using the same example photodiode as before: in order to be able to
detect an optical beat $\omega_{-}$ would need to be smaller than
100~MHz. If we take two, sightly detuned Nd:YAG lasers with
$f$~=~282~THz, this means that the relative detuning of these lasers
must be smaller than 10\super{-7}.

In general, for a field with several frequency components, the
photodiode signal can be written as
\begin{equation}
\begin{array}{rcl}
|E|^2&=&E\cdot E^*=\sum\limits_{i=0}^N\sum\limits_{j=0}^N
a_ia_j^*~e^{\I(\w_i-\w_j)\T}.\\
\end{array}
\label{eq:intensity2}
\end{equation}
For example, if the photodiode signal is filtered with a low-pass
filter, such that only the DC part remains, we can compute the
resulting signal by looking for all components without frequency
dependence. The frequency dependence vanishes when the frequency
becomes zero, i.e.~in all parts of Equation~(\ref{eq:intensity2}) with
$\w_i = \w_j$.  The output is a real number, calculated like this:
\begin{equation}
\label{eq:dc_det}
x=\sum\limits_i\sum\limits_j a_ia_j^*\quad\mathrm{with}\quad
\{i,j~|~i,j\in\{0,\dots,N\}~\wedge~\w_i=\w_j\}.
\end{equation}

\subsection{Signal demodulation}
\label{sec:demodulation}

A typical application of light modulation, is its use in a
modulation-demodulation scheme, which applies an electronic
demodulation to a photodiode signal. A `demodulation' of a photodiode
signal at a user-defined frequency $\omega_{x}$, performed by an
electronic mixer and a low-pass filter, produces a signal, which is
proportional to the amplitude of the photo current at DC and at the
frequency $\omega_0\pm\omega_x$. Interestingly, by using two mixers with
different phase offsets one can also reconstruct the phase of the
signal, or to be precise the phase difference of the light at
$\omega_0 \pm \omega_x$ with respect to the carrier light. This
feature can be very powerful for generating interferometer control
signals.

Mathematically, the demodulation process can be described by a
multiplication of the output with a cosine:
$\cos(\w_x+\varphi_x)$, where $\varphi_x$ is
the \emph{demodulation phase}. This cosine is also called the `local oscillator'.
After the multiplication was performed only the \textbf{DC} part of the
result is taken into account. The signal is
\begin{equation}
S_0=|E|^2=E\cdot E^*=\sum\limits_{i=0}^N\sum\limits_{j=0}^N
a_ia_j^*~e^{\I(\w_i-\w_j)\T}.
\end{equation}
Multiplied with the local oscillator it becomes
\begin{equation}
\begin{array}{rcl}
S_{1}&=&S_0\cdot
\cos(\w_xt+\varphi_x)=S_0\frac12\left(e^{\I(\w_xt+\varphi_x)} +
e^{-\I(\w_xt+\varphi_x)}\right)\\
&=&\frac12\sum\limits_{i=0}^N\sum\limits_{j=0}^N
a_ia_j^*~e^{\I(\w_i-\w_j)\T}\cdot\left(e^{\I(\w_xt+\varphi_x)} +
e^{-\I(\w_xt+\varphi_x)}\right).
\end{array}
\end{equation}
With $A_{ij}=a_ia_j^*$ and $e^{\I\w_{ij}\T}=e^{\I(\w_i-\w_j)\T}$ we can
write
\begin{equation}
S_{1}=\frac12\left(\sum\limits_{i=0}^NA_{ii}+\sum\limits_{i=0}^N
\sum\limits_{j=i+1}^N (A_{ij}~e^{\I\w_{ij}\T}+A_{ij}^*~e^{-\I\w_{ij}\T})\right)\cdot
\left(e^{\I(\w_xt+\varphi_x)}+e^{-\I(\w_xt+\varphi_x)}\right).
\end{equation}
When looking for the DC components of $S_1$ we get the
following~\cite{phd.Freise}:
\begin{equation}
  \begin{array}{lll}
    S_{\mathrm{1,DC}}&=&\sum\limits_{ij}
    \frac12(A_{ij}~e^{-\I\varphi_x}+A_{ij}^*~e^{\I\varphi_x})\quad\mathrm{with}\quad
    \{i,j~|~i,j\in\{0,\dots,N\}~\wedge~\w_{ij}=\w_x\}\\
    &=&\sum\limits_{ij}\myRe{A_{ij}~e^{-\I\varphi_x}}.
  \end{array}
\end{equation}
This would be the output of a mixer and a subsequent low-pass
filter. The results for $\varphi_x=0$ and $\varphi_x=\pi/2$ are called
\emph{in-phase} and \emph{in-quadrature}, respectively (or also
\emph{first} and \emph{second quadrature}). They are given by
\begin{equation}
  \begin{array}{rcl}
    S_{\mathrm{1,DC,phase}}&=&\sum\limits_{ij}\myRe{A_{ij}},\\
    S_{\mathrm{1,DC,quad}}&=&\sum\limits_{ij}\myIm{A_{ij}}.
  \end{array}
\end{equation}
If only one mixer is used, the output is always real and is determined
by the demodulation phase. However, with two mixers generating the
in-phase and in-quadrature signals, it is possible to construct a
complex number representing the signal amplitude and phase:
\begin{equation}
\label{eq:single_demod}
z=\sum\limits_{ij}a_{i}a^*_{j}\quad\mathrm{with}\quad
\{i,j~|~i,j\in\{0,\dots,N\}~\wedge~\w_{ij}=\w_x\}.
\end{equation}

Often several sequential demodulations are applied in order to measure
very specific phase information. For example, a double demodulation can be
described as two sequential multiplications of the signal with two
local oscillators and taking the DC component of the result. First
looking at the whole signal, we can write:
\begin{equation}
S_{2}=S_0\cdot \cos(\w_x t +\varphi_x)\cos(\w_y t +\varphi_y).
\end{equation}
This can be written as
\begin{equation}
\begin{array}{rcl}
S_{2}&=&S_0\frac12(
\cos(\w_y t +\w_x t +\varphi_y+\varphi_x)+\cos(\w_y t -\w_x t +\varphi_y-\varphi_x))\\
&=&S_0\frac12( \cos(\w_+ t +\varphi_+)+\cos(\w_- t +\varphi_-)),
\end{array}
\end{equation}
and thus reduced to two single demodulations. Since we now only care for the
DC component we can use the expression from above
(Equation~(\ref{eq:single_demod})). These two demodulations give two complex
numbers:
\begin{equation}
  \begin{array}{rcl}
    z1&=&\sum\limits_{ij}A_{ij}\quad\mathrm{with}\quad
    \{i,j~|~i,j\in\{0,\dots,N\}~\wedge~\w_i-\w_j=\w_+\},\\
    z2&=&\sum\limits_{ij}A_{kl}\quad\mathrm{with}\quad
    \{k,l~|~k,l\in\{0,\dots,N\}~\wedge~\w_k-\w_l=\w_-\}.\\
    \end{array}
\end{equation}
The demodulation phases are applied as follows to get a real output
(two sequential mixers)
\begin{equation}
x=\myRe{(z_1~e^{-\I\varphi_x}+z_2~e^{\I\varphi_x})~e^{-\I\varphi_y}}.
\end{equation}
In a typical setup, a user-defined demodulation phase for the first
frequency (here $\varphi_x$) is given. If two mixers are used for the
second demodulation, we can reconstruct the complex number
\begin{equation}
z=z_1~e^{-\I\varphi_x}+z_2~e^{\I\varphi_x}.
\end{equation}
More demodulations can also be reduced to single demodulations as
above.


\subsection{\Finesse examples}

\subsubsection{Optical beat}

\epubtkImage{fexample_optical_beat.png}{%
  \begin{figure}[htbp]
    \centerline{\includegraphics[width=0.8\textwidth]{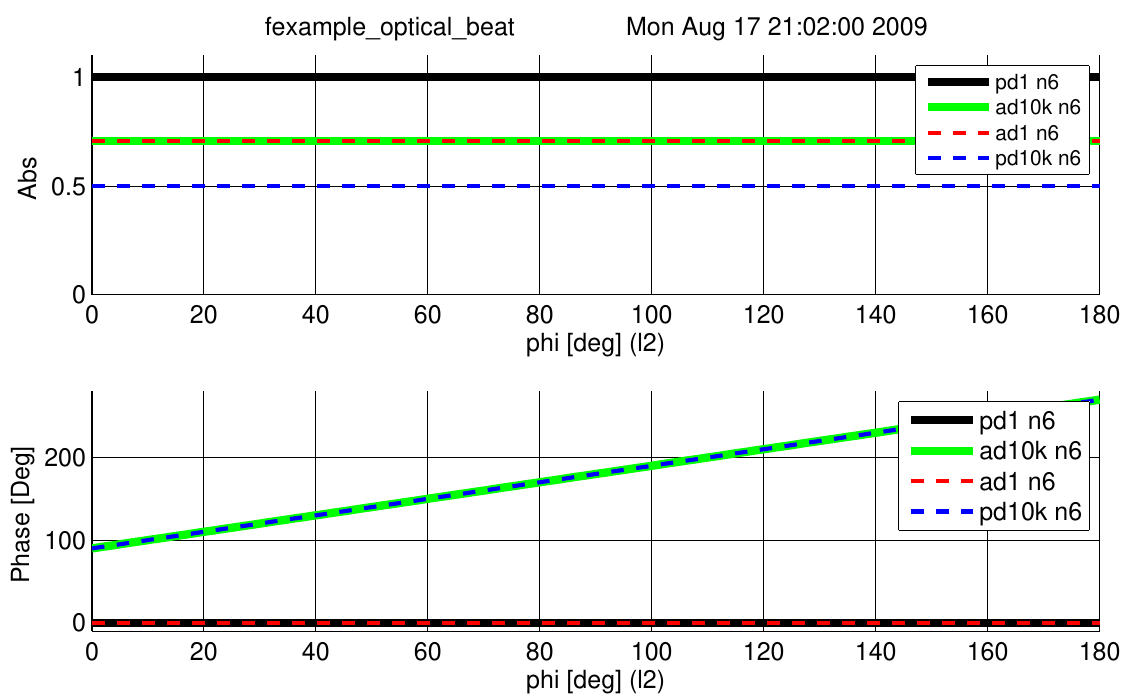}}
    \caption{\Finesse example: Optical beat.}
    \label{fig:fexample_optica_beat}
\end{figure}}

\noindent
In this example two laser beams are superimposed at a 50:50 beam
splitter. The beams have a slightly different frequency: the second
beam has a 10~kHz offset with respect to the first (and to the default
laser frequency). The plot illustrates the output of four different
detectors in one of the beam splitter output ports, while the phase of
the second beam is tuned from 0\textdegree\ to 180\textdegree. The
photodiode `pd1' shows the total power remaining constant at a value of 1. The
amplitude detectors `ad1' and `ad10k' detect the laser light at 0~Hz
(default frequency) and 10~kHz respectively. Both show a constant
absolute of $\sqrt{1/2}$ and the detector `ad10k' tracks the tuning of
the phase of the second laser beam. Finally, the detector `pd10k'
resembles a photodiode with demodulation at 10~kHz. In fact, this
represents a photodiode and two mixers used to reconstruct a complex
number as shown in Equation~(\ref{eq:single_demod}). One can see that
the phase of the resulting electronic signal also directly follows the
phase difference between the two laser beams.

\vspace{3mm}\noindent
{\small
\textbf{Finesse input file for `Optical beat'}
{\renewcommand{\baselinestretch}{.8}

\nopagebreak
\tt
\noindent
\mbox{} \\
\mbox{}\textbf{\textcolor{Red}{const}}\textcolor{ForestGreen}{\ \ freq}\ 10k\ \ \ \ \ \ \ \textcolor{Gray}{\%\ creating\ a\ constant\ for\ the\ frequency\ offset} \\
\mbox{}\textbf{\textcolor{RoyalBlue}{laser}}\ \ l1\ \textcolor{Purple}{1}\ \textcolor{Purple}{0}\ \ n1\ \ \ \ \ \textcolor{Gray}{\%\ laser\ with\ P=1W\ at\ the\ default\ frequency} \\
\mbox{}\textbf{\textcolor{RoyalBlue}{space}}\ \ s1\ 1n\ \textcolor{Purple}{1}\ n1\ n2\ \ \textcolor{Gray}{\%\ space\ of\ 1nm\ length} \\
\mbox{}\textbf{\textcolor{RoyalBlue}{laser}}\ \ l2\ \textcolor{Purple}{1}\ \textcolor{ForestGreen}{\$freq}\ \ n3\ \textcolor{Gray}{\%\ a\ second\ laser\ with\ f=10kHz\ frequency\ offset} \\
\mbox{}\textbf{\textcolor{RoyalBlue}{space}}\ \ s2\ 1n\ \textcolor{Purple}{1}\ n3\ n4\ \ \textcolor{Gray}{\%\ another\ space\ of\ 1nm\ length} \\
\mbox{}\textbf{\textcolor{RoyalBlue}{bs}}\ \ \ \ \ b1\ \textcolor{Purple}{0.5}\ \textcolor{Purple}{0.5}\ \textcolor{Purple}{0}\ \textcolor{Purple}{0}\ n2\ n5\ dump\ n4\ \textcolor{Gray}{\%\ 50:50\ beam\ splitter\ } \\
\mbox{}\textbf{\textcolor{RoyalBlue}{space}}\ \ s3\ 1n\ \textcolor{Purple}{1}\ n5\ n6\ \ \textcolor{Gray}{\%\ another\ space\ of\ 1nm\ length} \\
\mbox{}\textbf{\textcolor{RoyalBlue}{ad}}\ \ \ \ \ ad0\ \textcolor{Purple}{0}\ n6\ \ \ \ \ \ \ \textcolor{Gray}{\%\ amplitude\ detector\ at\ f=0Hz} \\
\mbox{}\textbf{\textcolor{RoyalBlue}{ad}}\ \ \ \ \ ad10k\ \textcolor{ForestGreen}{\$freq}\ n6\ \textcolor{Gray}{\%\ amplitude\ detector\ at\ f=10kHz} \\
\mbox{}\textbf{\textcolor{RoyalBlue}{pd}}\ \ \ \ \ pd1\ n6\ \ \ \ \ \ \ \ \ \textcolor{Gray}{\%\ simple\ photo\ detector} \\
\mbox{}\textbf{\textcolor{RoyalBlue}{pd}}\textcolor{Purple}{1}\ \ \ \ pd10k\ \textcolor{ForestGreen}{\$freq}\ n6\ \textcolor{Gray}{\%\ photo\ detector\ with\ demodulation\ at\ 10kHz} \\
\mbox{} \\
\mbox{}\textbf{\textcolor{Red}{xaxis}}\ l2\ phi\ lin\ \textcolor{Purple}{0}\ \textcolor{Purple}{180}\ \textcolor{Purple}{100}\ \textcolor{Gray}{\%\ changing\ the\ phase\ of\ the\ l2-beam} \\
\mbox{} \\
\mbox{}\textbf{\textcolor{Red}{yaxis}}\ abs\textcolor{BrickRed}{:}deg\ \ \ \ \ \ \ \ \ \textcolor{Gray}{\%\ plotting\ amplitude\ and\ phase\ } \\
\mbox{}

}}


\newpage

\section{Basic Interferometers}
\label{sec:interferometers}

The large interferometric gravitational-wave detectors currently in
operation are based on two fundamental interferometer topologies: the
\emph{Fabry-Perot} interferometer and the \emph{Michelson}  interferometer.
The main instrument is very similar to the original interferometer
concept used in the famous experiment by Michelson and Morley,
published in 1887~\cite{michelson1887}. The main difference is that
modern instruments use laser light to illuminate the interferometer to
achieve much higher accuracy. Already the first prototype by Forward
and Weiss has thus achieved a sensitivity a million times better than
Michelson's original instrument~\cite{forward78}.
In addition, the Michelson interferometer used in current
gravitational-wave detectors has been enhanced by resonant cavities,
which in turn have been derived
from the original idea for a spectroscopy standard published by Fabry
and Perot in 1899~\cite{fabry_perot}. The following section will
describe the fundamental properties of the Fabry-Perot
interferometer and the Michelson interferometer. A thorough
understanding of these basic instruments is essential for the study of
the high-precision interferometers used for gravitational-wave
detection.

\subsection{The two-mirror cavity: a Fabry-Perot interferometer}
\label{sec:two_mirror2}
We have computed the field amplitudes in a linear two-mirror cavity,
also called a \emph{Fabry-Perot} interferometer, in
Section~\ref{sec:two_mirror}. In order to understand the features of
this optical instrument it is interesting to have a closer look at the
power circulating in the cavity. A typical optical layout is shown in
Figure~\ref{fig:cavity_layout}; two parallel mirrors form the
Fabry-Perot cavity. A laser beam is injected through the first
mirror (at normal incidence).

\epubtkImage{fabry-perot01.png}{%
  \begin{figure}[htbp]
    \centerline{\includegraphics{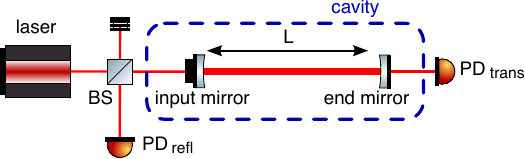}}
   \caption{Typical optical layout of a two-mirror cavity, also called a
   Fabry-Perot interferometer. Two mirrors form the Fabry-Perot
   interferometer, a laser beam is injected through one of the mirrors
   and the reflected and transmitted light can be detected by photo
   detectors.}
   \label{fig:cavity_layout}
\end{figure}}

The behaviour of the (ideal) cavity is determined by the length of the
cavity $L$, the wavelength of the laser $\lambda$ and the reflectivity
and transmittance of the mirrors. Using the mathematical description
introduced in Section~\ref{sec:two_mirror} and assuming an input power of
$|a_0|^2=1$, we obtain the following equation for the circulating power:
\begin{equation}
\label{eq:cav-power}
P_1=|a_1|^2
=\frac{T_1}{1+R_1R_2-2r_1r_2\cos\left(2 k L\right)},
\end{equation}
with $k=2\pi/\lambda$, $P$, $T=t^2$ and $R=r^2$, as defined in Section~\ref{sec:fields}.
Similarly we could compute the transmission of the optical system
as the input-output ratio of the field amplitudes. For example, with
$a_0$ the field injected into the cavity and $a_2$ the field
transmitted by the cavity,
\begin{equation}
\frac{a_2}{a_0}=\frac{-t_1 t_2 \exp(-\I k L)}{1-r_1 r_2 \exp(- \I 2 k L)}
\end{equation}
is the frequency-dependent transfer function of the cavity in
transmission (the frequency dependence is hidden inside the $k=2\pi
f/c$).

\epubtkImage{example_cavity_power.png}{%
  \begin{figure}[htbp]
    \centerline{\includegraphics[viewport=0 0 710 410,width=13cm]{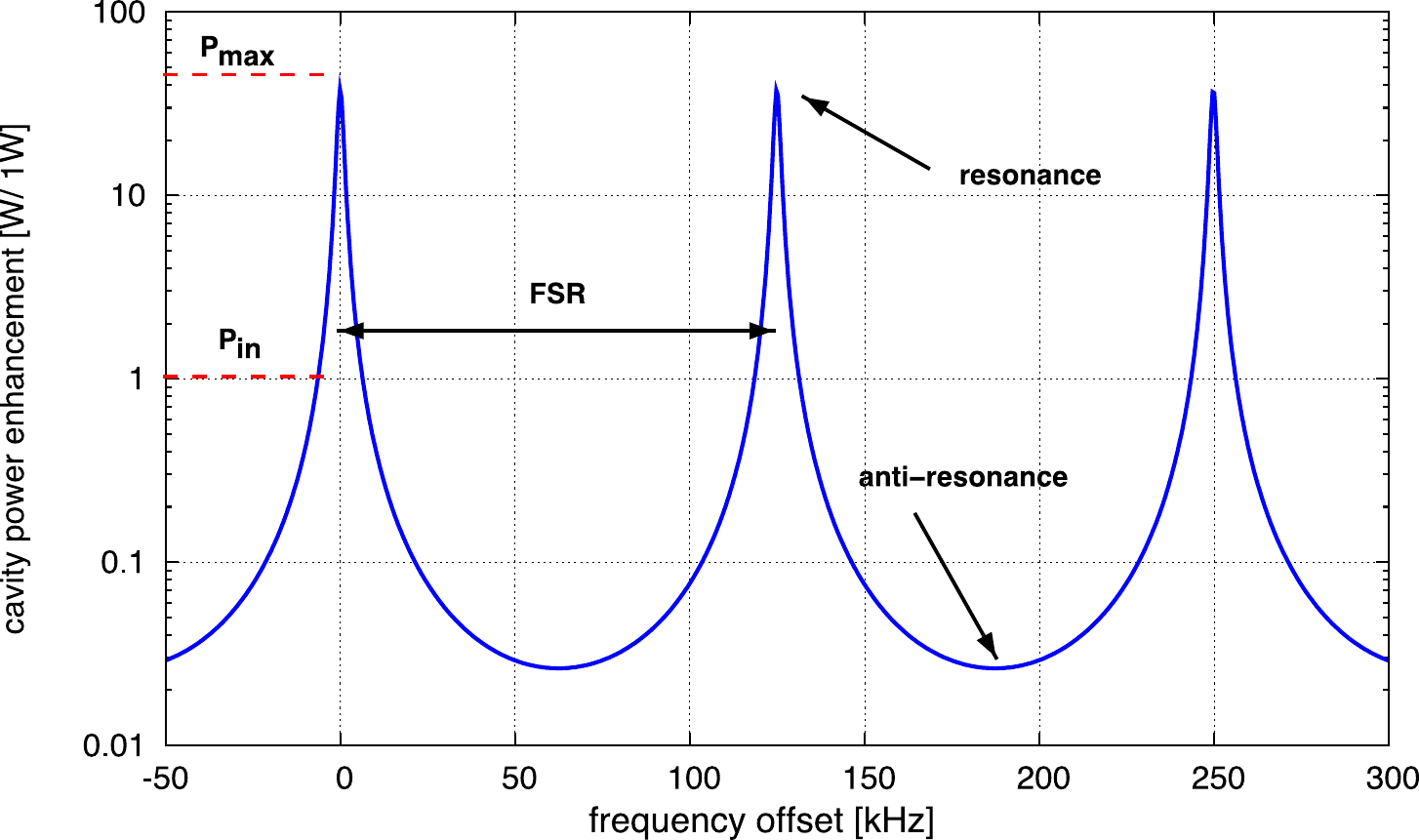}}
    \caption{Power enhancement in a two-mirror cavity as a function of
      the laser-light frequency. The peaks marks the resonances of the
      cavity, i.e.~modes of operation in which the injected light is
      resonantly enhanced. The frequency distance between two peaks is
      called \emph{free-spectral range} (FSR).}
    \label{fig:cav-powerenhance}
\end{figure}}

Figure~\ref{fig:cav-powerenhance} shows a plot of the circulating
light power $P_1$ over the laser frequency. The maximum power is
reached when the cosine function in the denominator becomes equal to
one, i.e.~at $k L = N \pi$ with $N$ an integer. This occurs when
the round-trip length is an integer multiple of the wavelength of the
injected light: $2L =  N 2\pi/k=N \lambda$. This is called the
cavity \textit{resonance}. The lowest power values are reached at
\textit{anti-resonance} when $k L = (N +1/2)\pi$. We can also rewrite
\begin{equation}
2 k L=\omega \frac{2L}{c} = 2 \pi f \frac{2 L}{c} = \frac{2 \pi f} {\mathrm{FSR}},
\end{equation}
with FSR being the \textit{free-spectral range} of the cavity as shown
in Figure~\ref{fig:cav-powerenhance}. Thus, it becomes clear that
resonance is reached for laser frequencies
\begin{equation}
f_{r}= N \cdot \mathrm{FSR},
\end{equation}
where $N$ is an integer.

\epubtkImage{cavity_phase.png}{%
  \begin{figure}[htbp]
    \centerline{\includegraphics[width=12.5cm]{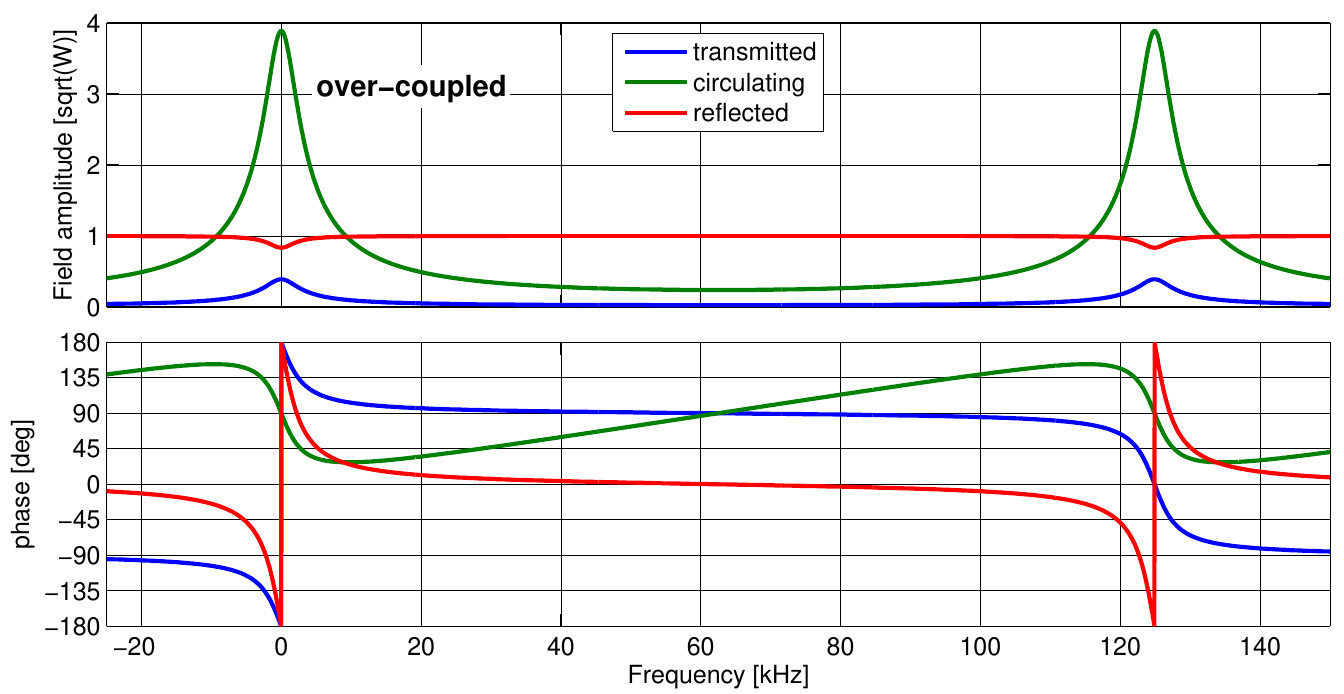}}
    \centerline{\includegraphics[width=12.5cm]{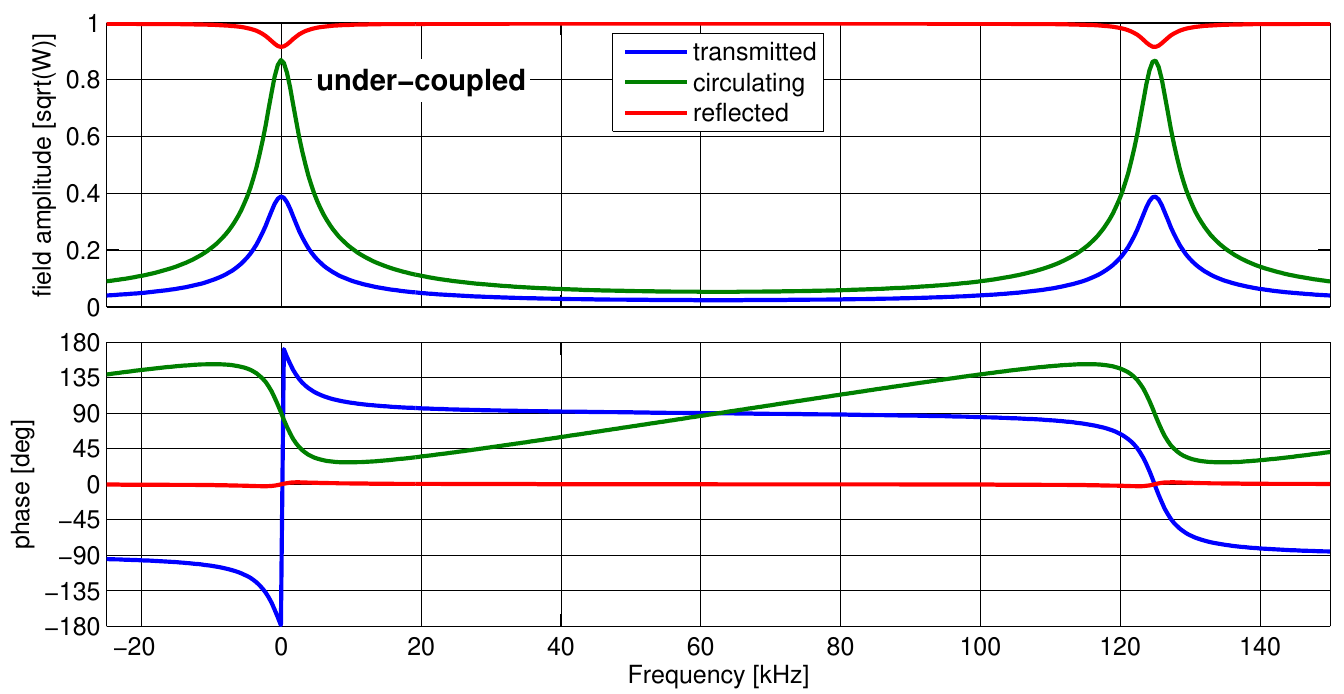}}
    \centerline{\includegraphics[width=12.5cm]{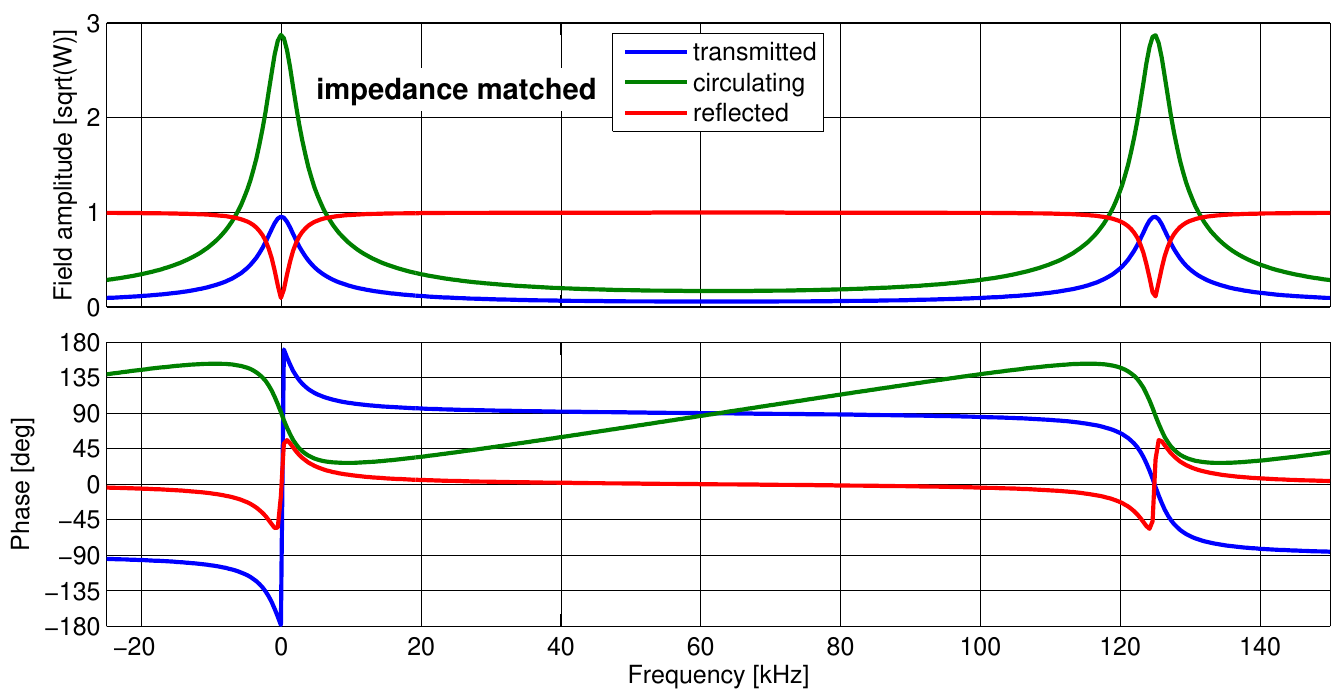}}
    \caption{This figure compares the fields reflected by, transmitted
      by and circulating in a Fabry-Perot cavity for the three
      different cases: over-coupled, under-coupled and impedance
      matched cavity (in all cases $T_1+T_2=0.2$ and the round-trip
      loss is 1\%). The traces show the phase and amplitude of the
      electric field as a function of laser frequency detuning.}
    \label{fig:cav-phase}
\end{figure}}

Another characteristic parameter of a cavity is its linewidth, usually
given as its \textit{full width at half maximum} (FWHM) or its
\textit{pole frequency}, $f_p$. In order to compute the linewidth we
have to ask at which frequency the circulating power becomes half the
maximum:
\begin{equation}
\begin{array}{l}
|a_1(f_p)|^2 \sollgleich \frac12 |a_{1,\mathrm{max}}|^2.\\
\end{array}
\end{equation}
This results in the following expression for the full linewidth:
\begin{equation}
\mathrm{FWHM}=2f_p=\frac{2 \mathrm{FSR}}{\pi}\arcsin\left(\frac{1-r_1r_2}{2\sqrt{r_1r_2}}\right).
\end{equation}
The ratio of the linewidth to the free spectral range is called the \textit{finesse} of a cavity:
\begin{equation}
F=\frac{\mathrm{FSR}}{\mathrm{FWHM}}=\frac{\pi}{2\arcsin\left(\frac{1-r_1r_2}{2\sqrt{r_1r_2}}\right)}.
\end{equation}
In the case of high finesse, i.e.~when $r_1$ and $r_2$ are close to $1$, we
can use the fact that the argument of the $\arcsin$ function is small
and make the approximation
\begin{equation}
F\approx\frac{\pi\sqrt{r_1r_2}}{1-r_1r_2}\approx\frac{\pi}{1-r_1r_2}.
\end{equation}

\epubtkImage{cavity-coupled.png}{%
  \begin{figure}[htbp]
    \centerline{\includegraphics[width=13cm]{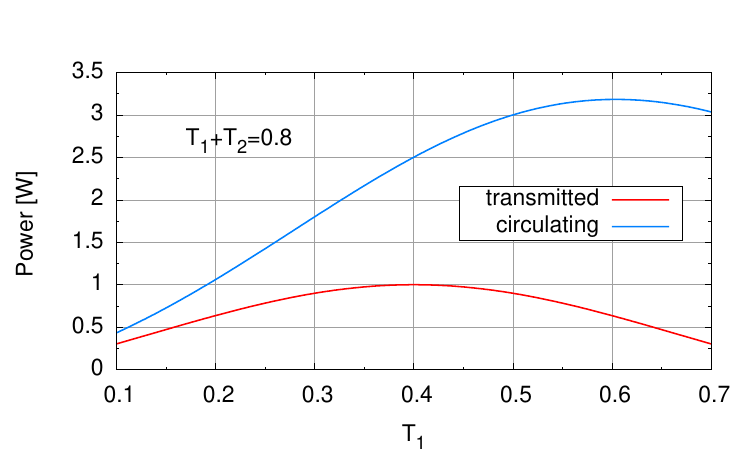}}
    \caption{Power transmitted and circulating in a two mirror cavity
    with input power 1~W. The mirror transmissions are set such that
    $T_1+T_2=0.8$ and the reflectivities of both mirrors are set as
    $R=1-T$. The cavity is undercoupled for $T_1<0.4$, impedance
    matched at $T_1=T_2=0.4$, and overcoupled for $T_1>0.4$. The
    transmission is maximised in the impedance-matched case and falls
    similarly for over or undercoupled settings. However, the
    circulating power (and any resonance performance of the cavity) is
    much larger in the overcoupled case.}
    \label{fig:cav-coupled}
\end{figure}}

The behaviour of a two mirror cavity depends on the length of the
cavity (with respect to the frequency of the laser) and on the
reflectivities of the mirrors. Regarding the mirror parameters, one
distinguishes three cases\epubtkFootnote{Please note that in the
  presence of losses the coupling is defined with respect to the
  transmission and losses.  In particular, the impedance-matched case
  is defined as $T_1=T_2\times\mathrm{Loss}$, so that the input power
  transmission exactly matches the light power lost in one
  round-trip.}:
\begin{itemize}
\item when $T_1<T_2$ the cavity is \textit{undercoupled}
\item when $T_1=T_2$ the cavity is \textit{impedance matched}
\item when $T_1>T_2$ the cavity is \textit{overcoupled}
\end{itemize}
The differences between these three cases can seem subtle mathematically but
have a strong impact on the application of cavities in laser systems. One of the main differences
is the phase evolution of the light fields, as shown in Figure~\ref{fig:cav-phase}.
The circulating power shows that the resonance effect is better used in over-coupled
cavities; this is illustrated in
Figure~\ref{fig:cav-coupled}, which shows the transmitted and circulating power
for the three different cases.
Only in the impedance-matched case can the cavity transmit (on resonance) \emph{all} the
incident power. Given the same total transmission $T_1+T_2$, the overcoupled
case allows for the largest circulating power and thus a stronger
`resonance effect' of the cavity, which is useful, for example, when the cavity is used as
a mode filter. Hence, most commonly used cavities are impedance matched or overcoupled.

\subsection{Michelson interferometer}
\label{sec:Michelson}
We came across the Michelson interferometer in
Section~\ref{sec:bsphase} when we discussed the phase relation at a
beam splitter. The typical optical layout of the Michelson
interferometer is shown again in Figure~\ref{fig:michelson_layout}, a
laser beam is split by a beam splitter and sent along two
perpendicular \emph{interferometer arms}. The four directions seen
from the beam splitter are often labelled North, East, West and
South. Another common naming scheme, also shown in
Figure~\ref{fig:michelson_layout}
refers to the interferometer arms as X and Y; the two outputs are
labelled as the symmetric port (towards the laser input) and anti-symmetric
port respectively. Both conventions are common in the literature and
we will make use of both in this article.

The ends
of the interferometer arms (North and East or Y and X) are marked by highly reflective
\emph{end mirrors}, sometimes called \emph{end test masses} (ETM),
The laser beams are reflected by the end mirrors and then recombined at
the central beam splitter. Generally, the
Michelson interferometer has two outputs, namely the so far unused
beam splitter port (\textit{South port} or \textit{anti-symmetric port}) and the input port
(\textit{West port } or \textit{symmetric port}). Both output
ports can be used to obtain interferometer signals; however most
setups are designed such that the main signals are detected in the
South port\footnote{The term 'main signals' refers to the optical
  signal providing the readout of the interferometric measurement, for
  example, of a position
  or length change. In addition, other output signals exist: for
  example, the light power reflected back into the West port can be
  recorded for monitoring the interferometer status.}.

\epubtkImage{michelson01.png}{%
  \begin{figure}[htbp]
    \centerline{\hspace{1em}\includegraphics{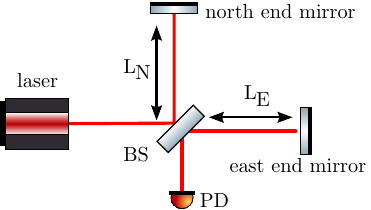}\hfill\includegraphics{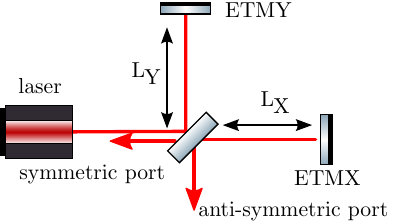}\hspace{1em}}
   \caption{Optical layout and two common naming convention sfor a Michelson interferometer: a
     laser beam is split into two and sent along two perpendicular
     interferometer arms. We will sometimes label the directions in a Michelson interferometer
     as North, East, West and South, as shown in the
     left plot. The end mirror,s or end test masses (ETMs),
     reflect the beams towards the beam
     splitter, where they recombine. The South and West ports of the beam splitter are
     possible output port; however in many cases only the South port
     is used. The plot on the right shows an alternative naming scheme
   commonly used, in which the two arms are labelled X and Y,
   the output towards the laser is called the
   \textit{symmetric port} and the other output is referred to as the
   \textit{anti-symmetric port}.}
   \label{fig:michelson_layout}
\end{figure}}

The Michelson interferometer output signal is determined by the laser
wavelength $\lambda$, the reflectivity and transmittance of the beam
splitter and the end mirrors, and the relative length of the interferometer
arms. In many cases the end mirrors are highly reflective and the beam
splitter is ideally a 50:50 beam splitter. In this case, we can compute
the output for a monochromatic field as shown in
Section~\ref{sec:bsphase}. Using Equation~(\ref{eq:MI_bsphase}) we can
write the field in the South port as
\begin{equation}
E_S=E_0~\frac{\I}{2}\left(e^{\I 2 k L_N}~+~e^{\I 2 k L_E}\right).
\end{equation}
We define the common and differential arm lengths as
\begin{equation}
\label{eq:armlength}
\begin{array}{lcl}
\bar{L}&=&\frac{L_N+L_E}{2}\\
\Delta L&=&L_N-L_E,\\
\end{array}
\end{equation}
which yield $2L_N=2\bar{L}+\Delta L$ and $2L_E=2\bar{L}-\Delta L$.
Thus, we can further simplify to get
\begin{equation}
E_S=E_0~\frac{\I}{2} e^{\I 2 k \bar{L}} \left(e^{\I k \Delta
  L}~+~e^{-\I k \Delta L}\right)=E_0~\I e^{\I 2 k \bar{L}} \cos(k
\Delta L).\label{eq:michelson_south_port_carrier}
\end{equation}
The photo detector then produces a signal proportional to
\begin{equation}
S=E_SE^*_S=P_0 \cos^2(k \Delta L)=P_0 \cos^2(2\pi \Delta L/\lambda).
\end{equation}
This signal is depicted in Figure~\ref{fig:mi_output}; it shows that
the power in the South port changes between zero and the input power
with a period of $\Delta L/\lambda=0.5$. The tuning at which the
output power drops to zero is called the \emph{dark fringe}. Current
interferometric gravitational-wave detectors operate their Michelson
interferometer at or near the dark fringe.

\epubtkImage{michelson_south.png}{%
  \begin{figure}[htbp]
    \centerline{\includegraphics[width=0.85\textwidth]{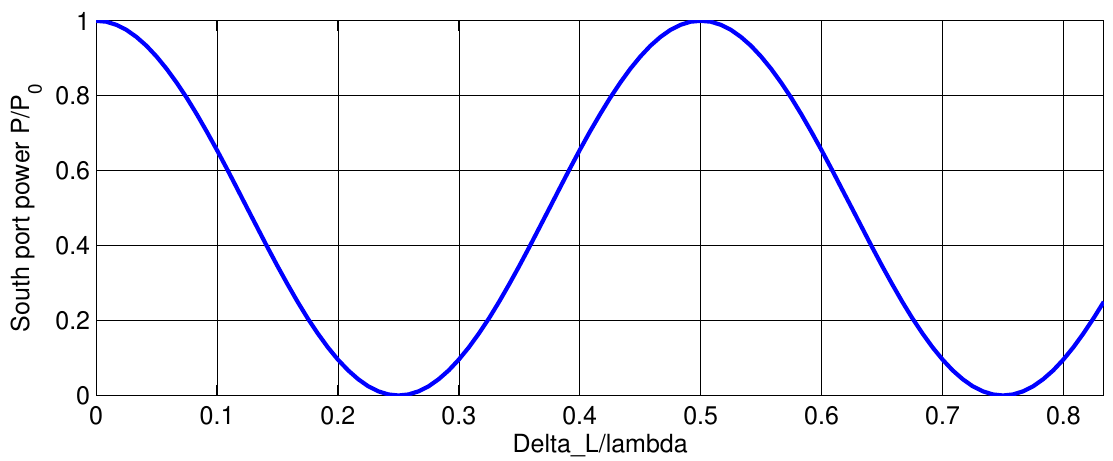}}
   \caption{Power in the South port of a symmetric Michelson
   interferometer as a function of the arm length difference $\Delta
   L$. When the interferometer is set to $\Delta L/\lambda=0.25$ the
   input light is not transmitted into the South port: this condition is
   called the \textit{dark fringe}.}
   \label{fig:mi_output}
\end{figure}}

The above seems to indicate that the macroscopic arm-length difference
plays no role in the Michelson output signal. However, this is only
correct for a monochromatic laser beam with infinite coherence
length. In real interferometers care must be taken that the arm-length
difference is well below the coherence length of the light source. In
gravitational-wave detectors the macroscopic arm-length difference is
an important design feature; it is kept very small in order to reduce
coupling of laser noise into the output but needs to retain a finite
size to allow the transfer of phase modulation sidebands from the
input to the output port; this is illustrated in the \Finesse example
below and will be covered in detail in Section~\ref{sec:mi_control}.

\subsection{Michelson interferometer and the sideband picture}
\label{sec:MIandSB}
In the context of gravitational wave detection the Michelson
interferometer is used for measuring a very small differential change
in the length of one arm versus the other.
The very small amplitude of gravitational waves, or the equivalent
small differential change of the arm lengths,
requires additional optical techniques to increase the sensitivity
of the interferometer.  In this section we briefly
introduce the interferometer configurations and review their effect on the
detector sensitivity.

The Michelson interferometer can achieve its best sensitivity when
operated in a quasi stationary mode, i.e.~when the positions of mirrors
and beamsplitters are carefully controlled so that the key
parameters, for example the light power inside the interferometer and
at the output ports, are nearly constant. We call such an interferometer
state, described by a unique set of the key parameters,
an \textit{operating point} of the interferometer
(see Section~\ref{sec:control} for a discussion of the control systems
involved to reach and maintain an operating point). For an
interferometer in a steady state it is possible to describe and
analyse the behaviour using a \textit{steady state model}, describing
the light field coupling in the frequency domain and making use of the
previously introduced concept of sidebands, see Section~\ref{sec:mod}.

\epubtkImage{MIsidebands.png}{%
  \begin{figure}[htbp]
    \centerline{\includegraphics{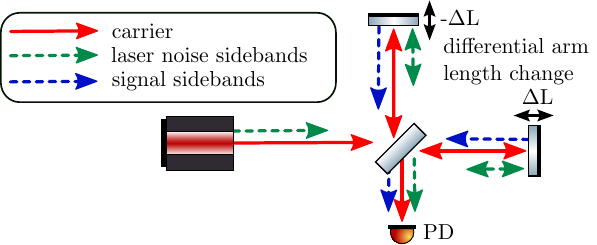}}
    \vspace{3mm}
    \centerline{\includegraphics[width=0.9\textwidth]{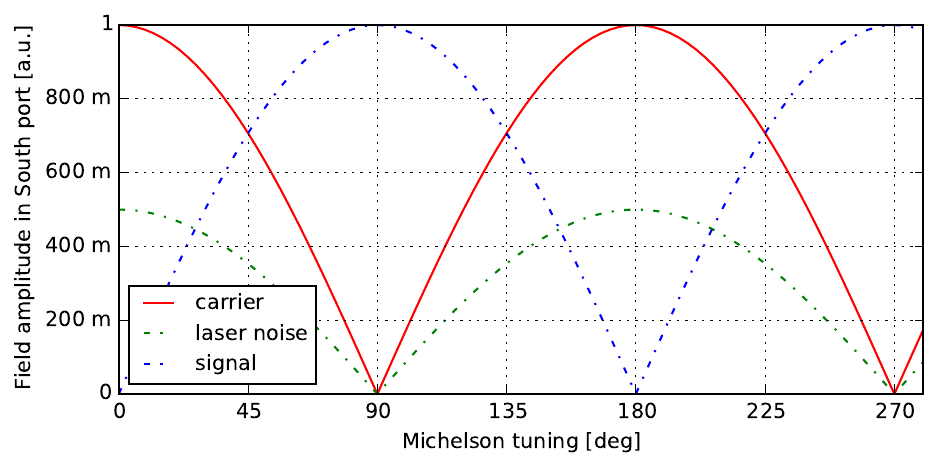}}
    \caption[MIchelson with signal sidebands]{A Michelson
      interferometer shown with
three types of light field: the `carrier', representing the
undistorted laser input field, `laser phase noise sidebands', which
enter the interferometer with the carrier, and  `signal sidebands',
which are phase modulation sidebands caused by differential arm length
motion. All
   three fields leave the interferometer through both output
   ports (here only the detector in the South port is shown). The
   graph shows the amplitude of the three light fields in the
   South port as a function of the Michelson tuning (differential arm
   length change).  At 0 degrees the
   Michelson is on a \textit{bright fringe} and at 90 degrees on a
   \textit{dark fringe}.}
    \label{fig:MIsidebands}
\end{figure}}

Consider a Michelson interferometer which is to be used to measure
a differential arm length change. As an example for a signal to noise
comparison we consider the phase
noise of the injected laser light.
For this example the noise can be represented by a sinusoidal modulation with a small
amplitude at a single frequency, say 100\,Hz. Therefore we can describe
the phase noise of the laser by a pair of sidebands superimposed on
the main carrier light field entering the Michelson interferometer.
Equally the change of an interferometer arm represents a phase
modulation of the light reflected back from the end mirrors and the
generated optical signal can be represented by a pair of phase modulation
sidebands, see Section~\ref{sec:response_gravwave}.

In order to get an estimation of the signal to noise
ratio we can trace the individual sidebands through the interferometer
and compute their amplitude in the output port.
Figure~\ref{fig:MIsidebands} shows the setup of a basic Michelson interferometer,
indicating the insertion of the noise and signal sidebands. It also
provides a plot of the sideband amplitude in the South output
port as a function of the differential arm length of the Michelson interferometer. We can see that
a tuning of 90\,degrees corresponds to the dark fringe, the state of
the interferometer in which the injected light (the carrier
and laser noise) is reflected back
towards the laser and is not transmitted into the South port.
The plot reveals two advantages of the dark fringe as an operating point:
first of all the transmission of the signal sidebands to the photo
detector is maximised while the laser phase noise is minimised.
More generally at the dark fringe, all common mode effects, such
as laser noise, or common length changes of the arms, produce a
minimal optical signal at the output port, whereas differential
effects in the arms are maximised. Furthermore at the dark fringe
the least amount of carrier light is transmitted to the photo
detector. This is an advantage because it is technically often easier
to make an accurate light power measurement when the total
detected power is low.

It should be noted that typically one set of sidebands alone does not create
a strong signal during detection.  In the case of gravitational wave
detection these sidebands are many orders of magnitude smaller than the amplitude of
the carrier. Instead we require a beat between the
signal sidebands and another field, a so-called
\textit{local oscillator}, to generate a strong electronic signal
proportional to the amplitude of the signal sidebands. The local oscillator
can be created in different ways, the most common are:
\begin{itemize}
\item Apply an RF modulation to the laser beam, either before
  injecting it into the interferometer or inside the interferometer.
  A small macroscopic length asymmetry between the two
  arms (Schnupp asymmetry, see Section~\ref{sec:Schnupp}) allows
  a significant amount of the RF sidebands to
  reach the South port when the interferometer is operating
  on the dark fringe for the carrier.  The RF sideband fields
  can be used as a local oscillator.
\item Set the Michelson such that it is close to, but not exactly on,
  the dark fringe. The carrier leaking into the South port can thus be
  used as a local oscillator. This scheme preserves the advantages of the dark fringe
  but relies on very good power stability of the carrier light.
\item Superimpose an auxiliary beam onto the output before the
  photodetector. For example, a pick-off beam from the main laser can
  be used for this. The main disadvantage of this concept is that it
  requires a very stable auxiliary beam (in phase as
  well as position) thus creating new control problems.
\end{itemize}

\subsection{Michelson interferometer signal readout with DC offset,
 or RF modulation}
\label{sec:signal_readout}
As discussed in Section \ref{sec:beats},
one method for providing a local oscillator is to use a small
microscopic DC offset to tune the Michelson interferometer slightly away from the dark fringe.
This allows a small amount of carrier to leak
through to the output port to beat with the signal sidebands.
The differential arm length difference required is
\begin{equation}
    \Delta L = \frac{\pi}{2 k_0} + \delta_{\mathrm{off}},\label{eq:dc_offset_phase}
\end{equation}
where $k_0 = \omega_0/c$ is the wavenumber of the carrier field and
the DC offset is $\delta_{\mathrm{off}} \ll 1$.
The field at the output port of a Michelson (as shown in figure \ref{fig:bs_phase})
for a single carrier field and one pair of signal sidebands is:
\begin{eqnarray}
    E_{6} &=& itr E_0 e^{-i2k_0\bar{L}}\left( 2\cos(k_0 \Delta L) + s^+ + s^-\right) e^{\I \omega_0 t},\nonumber \\
    &=& itr E_0 e^{-i2k_0\bar{L}}\left( 2\cos\left(\frac{\pi}{2} + k_0\delta_{\mathrm{off}}\right) + s^+ + s^-\right) e^{\I \omega_0 t}, \nonumber \\
    &=& itr E_0 e^{-i2k_0\bar{L}}\left( 2\sin\left(k_0\delta_{\mathrm{off}}\right) + s^+ + s^-\right) e^{\I \omega_0 t},
    \label{eq:dc_offset_amp}
\end{eqnarray}
where $s^\pm$ are the complex amplitudes (magnitude and phase) of the upper and lower sidebands
that reach the output port, for example, sidebands generated by a gravitational wave signal
or via the modulation of a mirror position.
The power in this field as measured by a photodiode will then contain
the beats between the carrier and both sidebands.
As the magnitude of any signal sideband is assumed to be very
small, $|s^\pm| \ll 1$, we only need to consider terms linear in
$s^\pm$.
The DC power and terms linear in the $s^\pm$ are then given by:
\begin{eqnarray}
    E_{6}E^\ast_{6} &=& TR |E_0|^2 \left( 4\sin^2\left(k_0\delta_{\mathrm{off}}\right) + 2\sin\left(k_0\delta_{\mathrm{off}}\right)(s^+ + s^-) + O(s^2) \right).
    \label{eq:dc_offset}
\end{eqnarray}
As expected the signal sideband terms are not visible in the power if
$\sin(k_0\delta_{\mathrm{off}}) = 0$,
because, if we operate purely at the dark fringe for the
carrier field, no local oscillator is present to beat with the signal.
The signal amplitude and phase can then be read out by demodulating the
photocurrent at the signal frequency.
In practice the choice of $\delta_{\mathrm{off}}$
depends on a number of technical issues, in particular the
laser power in the main output port and the transfer of common mode
noise into the output.

Another option for providing a local oscillator is by phase modulating the input laser
light, which is typically done at radio-frequencies (RF).
This method of readout is also referred to as a \textit{heterodyne readout scheme}.
The RF sidebands will have a different interference condition at the
beam  splitter compared to the carrier, and the inteferometer can be setup so
that the RF sidebands are present at the output port, to  be used as a local
oscillator, whilst the carrier field is at a dark fringe.

Consider a phase modulated beam with modulation index $b$ and
modulation frequency $\omega_b$, the input field will be:
\begin{eqnarray}
    E_0 = E'_{0}e^{\I\omega_0 t}(1 + \I b (e^{\I \omega_{b}t} + e^{-\I \omega_{b}t})).
\end{eqnarray}
The propagation of these three input fields to the output port can be treated separately and
is similar to equation \ref{eq:dc_offset_amp}, except that we must
keep track of their different frequencies: $k_0 = \omega_0/c$ and $k_{b} = \omega_b/c$
for the upper and lower RF sidebands. Ignoring the signal sidebands the fields present at
the output port are
\begin{eqnarray}
    E_6 &=& E'_6(\omega_0) + E'_6(\omega_0+\omega_b) + E'_6(\omega_0-\omega_b) \nonumber \\
    E'_6(\omega_0) &=& \I 2 rt E'_0 \cos\left(k_0 \Delta L \right) \nonumber \\
    E'_6(\omega_0\pm\omega_b) &=& \I 2 b E'_{0}e^{-\I2(k_0 \pm k_b) \bar{L}} \cos\left((k_0 \pm k_b)\Delta L\right) e^{\I(\omega_0\pm \omega_{b})t}
\end{eqnarray}
For using an RF readout scheme we want to
set the Michelson to be on the dark fringe for $E'_6(\omega_0) = 0$.
This is done by using a differential arm length difference of
$\Delta L = (2N+1)\frac{\pi}{2 k_0}$ so that
$\cos\left(k_0 \Delta L \right)=0$, where $N$ is any integer.
The condition for the RF sidebands is now:
\begin{eqnarray}
    \cos\left((k_0 \pm k_b)\Delta L\right) &=& \cos\left((k_0 \pm k_b)(2N+1)\frac{\pi}{2 k_0}\right) \nonumber \\
    &=& \cos\left(\frac{\pi}{2} + N\pi \pm (2N+1)\frac{\pi k_b}{2 k_0}\right)\nonumber \\
    &=& \sin\left(N\pi \pm (2N+1)\frac{\pi k_b}{2 k_0}\right) \nonumber \\
    &=& \pm(-1)^N\sin\left(k_b \Delta L_N\right) \\
    \Delta L_N &\equiv& (2N+1)\frac{\lambda_0}{4},
\end{eqnarray}
where $\lambda_0$ is the wavelength of the carrier light field.
Thus the $\sin\left(k_b \Delta L_N\right)$
term now determines the amplitude of the RF sidebands
that will be present at the output port, where $N$ is our free variable to choose.
Although $\Delta L_0$ is a microscopic distance the actual differential arm length
difference required to allow a reasonable amount of sidebands through requires a
large choice of $N$ as $k_b \Delta L_N \ll 1$ for radio frequency modulations. For example,
the GEO\,600 detector, which uses such an RF modulation scheme, operates with
$\Delta L = 13.5$cm~\cite{lueck2010}.
The final step of including the signal sidebands is not
elaborated on here but can be included with some careful algebra, remembering
that there will be signal sidebands created around the
carrier and both RF sidebands that could be present at the output port.

See Section~\ref{sec:Schnupp} for an more detailed comparison of the
DC and RF techniques to produce control signals and
Section~\ref{sec:dc} for detailed arguments for the advantages and
disadvantages of both techniques.

\subsection{Response of the Michelson interferometer to a
  gravitational waves signal}
\label{sec:response_gravwave}

In this section we derive how the sideband picture can be used
to decribe how the length modulation caused by a gravitational wave
affects a laser beam travelling through space.
This method can then be applied to any interferometer setup,
for example to compute how the signal readout
of a Michelson interferometer when using a DC offset.
Modulating a space of proper length $L$ will induce a phase modulation
to any laser beam travelling along it. The phase such a beam accumulates along
a path modulated by a gravitational wave signal $h(t)$ is~\cite{phd.Mizuno}
\begin{equation}
\varphi=-k_0 L \mp \frac{\omega_0}{2}\int_{t-L/c}^{t} h(t) =
-k_0 L\mp \delta \varphi,
\end{equation}
with $k_{c} = \omega_{0}/c$ being the wavenumber of the light field and
$\delta \varphi$ being the additional phase accumulated due to the modulation
of the path. For our analysis here we can assume the gravitational wave signal
is a simple sinusoidal function
\begin{equation}
h(t)=h_0\cos\left( \omega_{gw} t + \varphi_{gw}\right),
\end{equation}
where $\omega_{gw}$ and $\varphi_{gw}$ are the frequency and phase of the gravitational wave.
The phase accumulated from propagating along the space is
then\footnote{Derivations of the accumulated phase can be found in many works, a simple
  example is presented in ~\cite{phd.bond2014}.}
\begin{equation}
    \delta \varphi = \frac{\omega_0 h_0}{\w_{gw}}\cos\left(\w_{gw} t +\varphi_{gw} -\omega_{gw}\frac{L}{2c}\right)\sin\left(\omega_{gw}\frac{L}{2c}\right).
\end{equation}
Thus an oscillating, time dependent phase is present in the light fields travelling along the space.
Section~\ref{sec:phasemod} describes how such a modulation generates
sideband fields; the respective modulation index and phase are
\begin{eqnarray}
m &=& -\frac{\omega_0 h_0}{\omega_{gw}} \sin{\left(\frac{k_{gw} L}{2}\right)}, \\
\varphi &=& - \frac{k_{gw} L}{2} + \varphi_{gw},
\end{eqnarray}
with $k_{gw} = \omega_{gw}/c$ being the wavenumber for the
gravitational wave  signal sidebands.
Using equation \ref{eq:phase_mod} the unscaled amplitude and phase of the upper, $\alpha^+_{gw}$,
and lower, $\alpha^-_{gw}$, sidebands generated
by a gravitational wave are then
\begin{eqnarray}
A_{gw} &=& - \frac{w_0 h_0}{2\omega_{gw}} \sin{\left(\frac{k_{gw} L}{2}\right)}, \\
\Phi_{gw}^\pm &=& \frac{\pi}{2} - L\left(k_0 \pm \frac{k_{gw}}{2}\right) \pm \varphi_{gw}, \\
\alpha^\pm_{gw} &=& A_{gw} e^{i\Phi^\pm_{gw}} e^{\pm i\omega_{gw}t}.\label{eq:sbamp}
\end{eqnarray}
Note that $\alpha^\pm_{gw}$ must be scaled by the carrier field that is propagating into the space
for the complete sideband amplitude.

\epubtkImage{single_reflection.png}{%
  \begin{figure}[htb]
    \centerline{\includegraphics[width=0.4\textwidth]{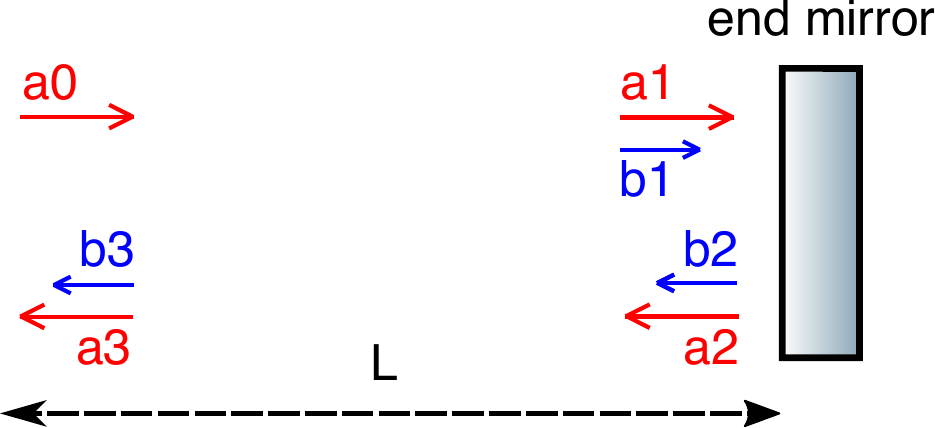}}
    \caption{Simplified sketch of a single arm of the Michelson
      interferometer, the arrows show the carrier fields, denoted by
      $a$, and the signal sidebands $b$ at the locations where they
      are computed.}
    \label{fig:single_reflection}
\end{figure}}

To compute how a Michelson responds to a gravitational wave  we must
first consider the modulation of the carrier field travelling in both
directions along the arms.
Both the carrier and the created signal sideband fields propagate along each arm, as shown in Figure~\ref{fig:single_reflection}, and are reflected by a mirror with amplitude reflectivity $r_{etm}$.
The relevant carrier fields are
\begin{eqnarray}
    a_3 &=& a_2 \exp{(-\I k_0 L)},   \\
    a_2 &=& r_{etm} a_1,                 \\
    a_1 &=& a_0 \exp{(-\I k_0 L)}.
\end{eqnarray}
The sidebands that are generated along such an arm are
\begin{eqnarray}
    b^\pm_1 &=& a_0 \alpha_{gw}^{\pm}, \nonumber\\
    b^\pm_2 &=& r_{etm} b^\pm_1, \nonumber\\
    b^\pm_3 &=& b^\pm_2 \exp{(-\I(k_0\pm k_{gw})L)} + a_2 \alpha_{gw}^{\pm}, \nonumber\\
    &=&  2 r_{etm} a_0 \alpha_{gw}^{\pm} \exp\left(-\I k_0 L\mp \I \frac{k_{gw}L}{2}\right)\cos\left(\mp \I \frac{k_{gw}L}{2}\right)
\end{eqnarray}
and by substituting the sideband amplitude $\alpha_{gw}^{\pm}$, see Equation~\ref{eq:sbamp}, we find
that:
\begin{eqnarray}
    b^\pm_3 &=&  - \I\frac{r_{etm} a_0 w_0 h_0}{2\omega_{gw}}\sin\left(k_{gw}L\right) \exp\left(-\I 2k_0 L\right)\exp\left(\pm(\omega_{gw}t - \I k_{gw}L + \varphi_{gw})\right)\label{eq:gw_sidebands}.
\end{eqnarray}
These are the sidebands that will leave the arm due to some gravitational
wave modulating the space of an arm. One point to note is that
the gravitational wave induced sidebands can cancel themselves out
for frequencies $f_{gw} = \frac{Nc}{2L}$.

Now we assume the Michelson interferometer is operated with a DC
offset for the signal readout, see Section~\ref{sec:signal_readout}.
For such a setup the field at the output port
is given by equation \ref{eq:dc_offset_amp}
which when applied here gives:
\begin{eqnarray}
    E_{out} &=& \I 2 r t E_0 \cos(k_0\Delta L) + b^+_{N} + b^-_{N} + b^+_{E} + b^-_{E}
\end{eqnarray}
The gravitational wave signal sidebands created in the North
and East arms with perfect end mirrors, $r_{etm}=1$, is given
by \ref{eq:gw_sidebands} where care should be taken to use the
correct lengths and carrier term: $b^\pm_N \equiv b^\pm_3$ with $L=L_N,\,a_0=rE_0$ and
$b^\pm_E \equiv b^\pm_3$ with $L=L_E,\,a_0=itE_0$. These sidebands at the output port, once transmitted
or reflected at the central beam splitter again, are
\begin{eqnarray}
    b^\pm_N &=&  \frac{rt E_0 w_0 h_0}{2\omega_{gw}}\sin\left(k_{gw}L_N\right) \exp\left(-\I 2k_0 L_N\right)\exp\left(\pm(\omega_{gw}t - \I k_{gw}L_N + \varphi_{gw})\right),\\
    b^\pm_E &=&  -\frac{rt E_0 w_0 h_0}{2\omega_{gw}}\sin\left(k_{gw}L_E\right) \exp\left(-\I 2k_0 L_E\right)\exp\left(\pm(\omega_{gw}t - \I k_{gw}L_E + \varphi_{gw})\right).
\end{eqnarray}
Note that an extra minus sign is included for
the East-arm sidebands because the gravitational wave modulate
the North and East arms differentially. Next we will write the arm lengths in
terms of a macroscopic differential $\Delta L$, and common mode $\bar{L}$,
lengths: $L_N = \bar{L} + \Delta L/2$ and $L_E = \bar{L} - \Delta L/2$.
Along with this we also assume that the central beam splitter has a
50:50 splitting ration $r = t = 1/\sqrt{2}$,
that the common mode length is
an integer number of wavelengths for the carrier light $\exp(\I k_0\bar{L}) = 1$,
that $\bar{L} \gg \Delta L$, and that the gravitational wave's
wavelength is much larger than $\Delta L$, so $k_{gw}(\bar{L} + \Delta L/2) \approx k_{gw}\bar{L}$.
Taking these assumptions into account the sideband terms become
\begin{eqnarray}
    b^\pm_N &=&   \frac{E_0 w_0 h_0}{4\omega_{gw}}\sin\left(k_{gw}\bar{L}\right) \exp\left(-\I k_0\Delta L\right)\exp\left(\pm(\omega_{gw}t - \I k_{gw}\bar{L} + \varphi_{gw})\right),\\
    b^\pm_E &=&  -\frac{E_0 w_0 h_0}{4\omega_{gw}}\sin\left(k_{gw}\bar{L}\right) \exp\left( \I k_0\Delta L\right)\exp\left(\pm(\omega_{gw}t - \I k_{gw}\bar{L} + \varphi_{gw})\right).
\end{eqnarray}
Finally the sum of the sidebands at the output is
\begin{eqnarray}
    b^+_{N} + b^-_{N} + b^+_{E} + b^-_{E} = \frac{\I E_0 w_0 h_0}{\omega_{gw}}\sin\left(k_{gw}\bar{L}\right)\sin\left(k_0\Delta L\right) \cos\left(\omega_{gw}t - k_{gw}\bar{L} + \varphi_{gw}\right).
\end{eqnarray}
Now that we know the signal sideband fields at the output port,
we can combine them with the carrier field that is also present:
\begin{eqnarray}
    E_{out} &=& \I E_0 \cos(k_0\Delta L) + b^+_{N} + b^-_{N} + b^+_{E} + b^-_{E} \nonumber \\
    &=& \I E_0 \left[\cos(k_0\Delta L)  + \frac{w_0 h_0}{\omega_{gw}}\sin\left(k_{gw}\bar{L}\right)\sin\left(k_0\Delta L\right) \cos\left(\omega_{gw}t - k_{gw}\bar{L} + \varphi_{gw}\right) \right].
\end{eqnarray}
A photodiode placed at the output of the Michelson will then measure the power in this beam
from which  we want to extract the gravitational wave
amplitude, $h_0$ and phase, $\varphi_{gw}$. The power in the beam
contains multiple beat frequencies between all the carrier and signal
sidebands, with the terms oscillating at the frequency $\omega_{gw}$,
are those linearly proportional to $h_0$:
\begin{eqnarray}
    P_{gw} &=& |E_0|^2\frac{w_0 h_0}{\omega_{gw}}\sin\left(k_{gw}\bar{L}\right)\sin\left(2k_0\Delta L\right) \cos\left(\omega_{gw}t - k_{gw}\bar{L} + \varphi_{gw}\right).
\end{eqnarray}

As we are using DC readout, the differential arm length is chosen to operate slightly
away from the dark fringe of the carrier field $\Delta L = \frac{\pi}{2 k_0} + \delta_{\mathrm{off}}$
as discussed in Section~\ref{sec:signal_readout}. The choice of DC offset is typically $\delta_{\mathrm{off}} \ll \lambda_0$,
the wavelength of the carrier light. So for small DC offset the power signal can be approximated as
\begin{eqnarray}
    P_{gw} &\approx& 2k_0\delta_{\mathrm{off}}|E_0|^2\frac{w_0 h_0}{\omega_{gw}}\sin\left(k_{gw}\bar{L}\right)\cos\left(\omega_{gw}t - k_{gw}\bar{L} + \varphi_{gw}\right).
    \label{eq:michelson_dc_gw_signal}
\end{eqnarray}
As described before, we now see that some DC offset is required to measure
the signal; the DC offset provides the local oscillator field for
the signal sidebands to beat with.
Finally, the transfer function from a gravitational wave signal
to the output photodiode, $T_{gw\rightarrow P}$, shows that the diode
measures $T_{gw\rightarrow P}$ Watts per unit $h_0$ at frequency $\omega_{gw}$:
\begin{eqnarray}
    T_{gw\rightarrow P}(\omega_{gw}) &\approx& k_0\delta_{\mathrm{off}}|E_0|^2\frac{w_0}{\omega_{gw}}\sin\left(k_{gw}\bar{L}\right) e^{ - \I k_{gw}\bar{L}}.
    \label{eq:michelson_dc_gw_TF}
\end{eqnarray}
For an example on how to model the response of a Michelson to a gravitational wave
modelled using \Finesse see Section~\ref{sec:fexample_mich_gw_sig}.


\subsection{\Finesse examples}


\subsubsection{Cavity power}
\label{sec:fexample_cavity_power}
\epubtkImage{fexample_cavity_power.png}{%
  \begin{figure}[htbp]
    \centerline{\includegraphics[width=0.9\textwidth]{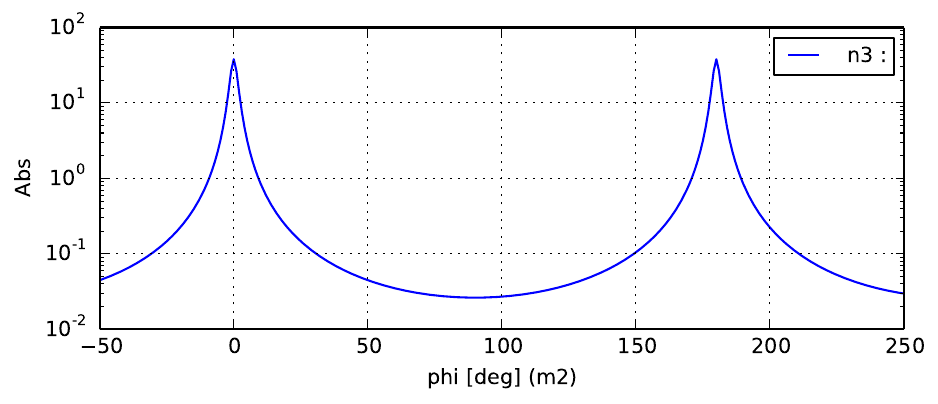}}
    \caption{\Finesse example: Cavity power.}
    \label{fig:fexample_cavity_power}
\end{figure}}

\noindent
This is  a simple \Finesse example showing the power enhancement in a
two-mirror cavity as a function of the microscopic tuning of a mirror
position (the position is given in degrees with 360 degrees referring
to a change of longitudinal position by one wavelength). Compare this plot to the one shown in
Figure~\ref{fig:cav-powerenhance}, which instead shows the power
enhancement as a function of the laser frequency detuning.

\vspace{3mm}\noindent
{\small
\textbf{Finesse input file for `Cavity power'}
{\renewcommand{\baselinestretch}{.8}

\nopagebreak
\tt
\noindent
\mbox{} \\
\mbox{}\textbf{\textcolor{RoyalBlue}{laser}}\ \ l1\ \textcolor{Purple}{1}\ \textcolor{Purple}{0}\ n1\ \ \ \ \textcolor{Gray}{\%\ laser\ with\ P=1W\ at\ the\ default\ frequency} \\
\mbox{}\textbf{\textcolor{RoyalBlue}{space}}\ \ s1\ \textcolor{Purple}{1}\ \textcolor{Purple}{1}\ n1\ n2\ \textcolor{Gray}{\%\ space\ of\ 1m\ length} \\
\mbox{}\textbf{\textcolor{RoyalBlue}{mirror}}\ m1\ \textcolor{Purple}{0.9}\ \textcolor{Purple}{0.1}\ \textcolor{Purple}{0}\ n2\ n3\ \ \ \textcolor{Gray}{\%\ cavity\ input\ mirror} \\
\mbox{}\textbf{\textcolor{RoyalBlue}{space}}\ \ L\ \textcolor{Purple}{1200}\ \textcolor{Purple}{1}\ n3\ n4\ \ \ \ \ \ \ \textcolor{Gray}{\%\ cavity\ length\ of\ 1200m} \\
\mbox{}\textbf{\textcolor{RoyalBlue}{mirror}}\ m2\ \textcolor{Purple}{1.0}\ \textcolor{Purple}{0.0}\ \textcolor{Purple}{0}\ n4\ dump\ \textcolor{Gray}{\%\ cavity\ output\ mirror} \\
\mbox{}\textbf{\textcolor{RoyalBlue}{pd}}\ \ \ \ \ P\ n3\ \ \ \ \ \ \ \ \ \textcolor{Gray}{\%\ photo\ diode\ measuring\ the\ intra-cavity\ power\ } \\
\mbox{} \\
\mbox{}\textbf{\textcolor{Red}{yaxis}}\ log\ abs \\
\mbox{}\textbf{\textcolor{Red}{xaxis}}\ m2\ phi\ lin\ \textcolor{BrickRed}{-}\textcolor{Purple}{50}\ \textcolor{Purple}{250}\ \textcolor{Purple}{300}\ \textcolor{Gray}{\%\ changing\ the\ microscopic\ tuning\ of\ mirror\ m2\ } \\
\mbox{} \\
\mbox{}

}}

\subsubsection{Michelson power}

\epubtkImage{fexample_michelson_power.png}{%
  \begin{figure}[htbp]
    \centerline{\includegraphics[width=0.9\textwidth]{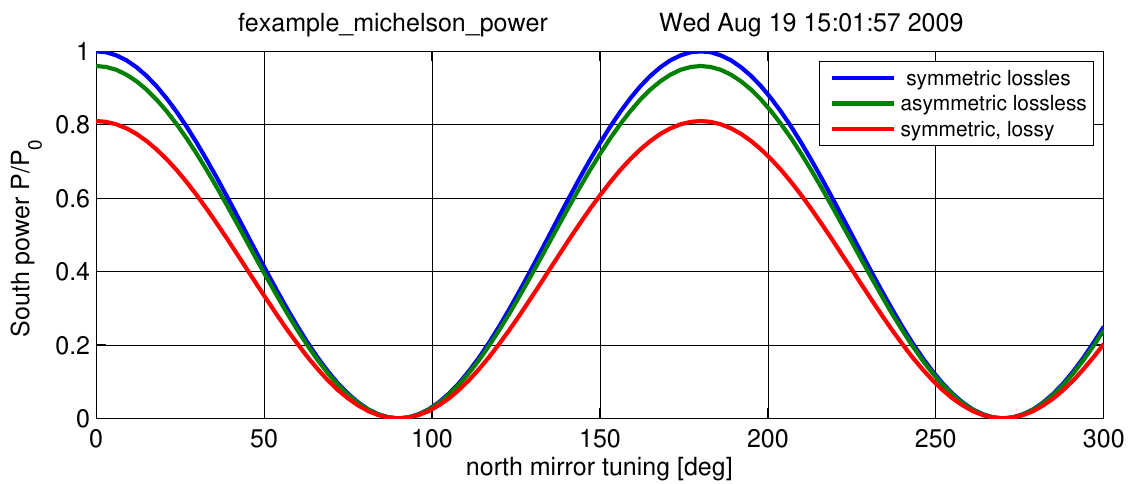}}
    \caption{\Finesse example: Michelson power.}
    \label{fig:fexample_michelson_power}
\end{figure}}

\noindent
The power in the South port of a Michelson detector varies as the
cosine squared of the microscopic arm length difference. The maximum
output can be equal to the input power, but only if the Michelson
interferometer is symmetric and lossless. The tuning for which the
South port power is zero is referred to as the \emph{dark fringe}.

\vspace{3mm}\noindent
{\small
\textbf{Finesse input file for `Michelson power'}
{\renewcommand{\baselinestretch}{.8}

\nopagebreak
\tt
\noindent
\mbox{} \\
\mbox{}\textbf{\textcolor{RoyalBlue}{laser}}\ l1\ \textcolor{Purple}{1}\ \textcolor{Purple}{0}\ \ n1\ \ \ \ \ \ \textcolor{Gray}{\%\ laser\ with\ P=1W\ at\ the\ default\ frequency} \\
\mbox{}\textbf{\textcolor{RoyalBlue}{space}}\ s1\ \textcolor{Purple}{1}\ \textcolor{Purple}{1}\ n1\ n2\ \ \ \ \textcolor{Gray}{\%\ space\ of\ 1m\ length} \\
\mbox{}\textcolor{Gray}{\%\ first\ trace:\ symmetric\ BS} \\
\mbox{}\textbf{\textcolor{RoyalBlue}{bs}}\ b1\ \textcolor{Purple}{0.5}\ \textcolor{Purple}{0.5}\ \textcolor{Purple}{0}\ \textcolor{Purple}{0}\ n2\ nN1\ nE1\ nS1\ \textcolor{Gray}{\%\ 50:50\ beam\ splitter\ } \\
\mbox{}\textcolor{Gray}{\%\ second\ trace:} \\
\mbox{}\textcolor{Gray}{\%bs\ b1\ 0.4\ 0.6\ 0\ 0\ n2\ nN1\ nE1\ nS1\ \%\ 40:60\ beam\ splitter\ } \\
\mbox{}\textcolor{Gray}{\%\ third\ trace:} \\
\mbox{}\textcolor{Gray}{\%bs\ b1\ 0.45\ 0.45\ 0\ 0\ n2\ nN1\ nE1\ nS1\ \%\ 45:45\ beam\ splitter\ } \\
\mbox{}\textbf{\textcolor{RoyalBlue}{space}}\ \ LN\ \textcolor{Purple}{1}\ \textcolor{Purple}{1}\ nN1\ nN2\ \ \ \ \textcolor{Gray}{\%\ north\ arm} \\
\mbox{}\textbf{\textcolor{RoyalBlue}{space}}\ \ LE\ \textcolor{Purple}{1}\ \textcolor{Purple}{1}\ nE1\ nE2\ \ \ \ \textcolor{Gray}{\%\ east\ arm} \\
\mbox{}\textbf{\textcolor{RoyalBlue}{mirror}}\ mN\ \textcolor{Purple}{1}\ \textcolor{Purple}{0}\ \textcolor{Purple}{0}\ nN2\ dump\ \textcolor{Gray}{\%\ north\ end\ mirror,\ lossless} \\
\mbox{}\textbf{\textcolor{RoyalBlue}{mirror}}\ mE\ \textcolor{Purple}{1}\ \textcolor{Purple}{0}\ \textcolor{Purple}{0}\ nE2\ dump\ \textcolor{Gray}{\%\ east\ end\ mirror,\ lossless} \\
\mbox{}\textbf{\textcolor{RoyalBlue}{space}}\ \ s2\ \textcolor{Purple}{1}\ \textcolor{Purple}{1}\ nS1\ nout\  \\
\mbox{}\textbf{\textcolor{RoyalBlue}{pd}}\ South\ nout\ \ \ \ \ \ \ \ \ \ \ \ \textcolor{Gray}{\%\ photo\ detector\ in\ South\ port} \\
\mbox{}\textbf{\textcolor{Red}{xaxis}}\ mN\ phi\ lin\ \textcolor{Purple}{0}\ \textcolor{Purple}{300}\ \textcolor{Purple}{100}\ \textcolor{Gray}{\%\ changing\ the\ microscopic\ position\ of\ mN} \\
\mbox{} \\
\mbox{}

}}

\subsubsection{Michelson gravitational wave response}
\label{sec:fexample_mich_gw_sig}
\epubtkImage{fexample_mich_gw_sig.png}{%
  \begin{figure}[htp]
    \centerline{\includegraphics[width=0.9\textwidth]{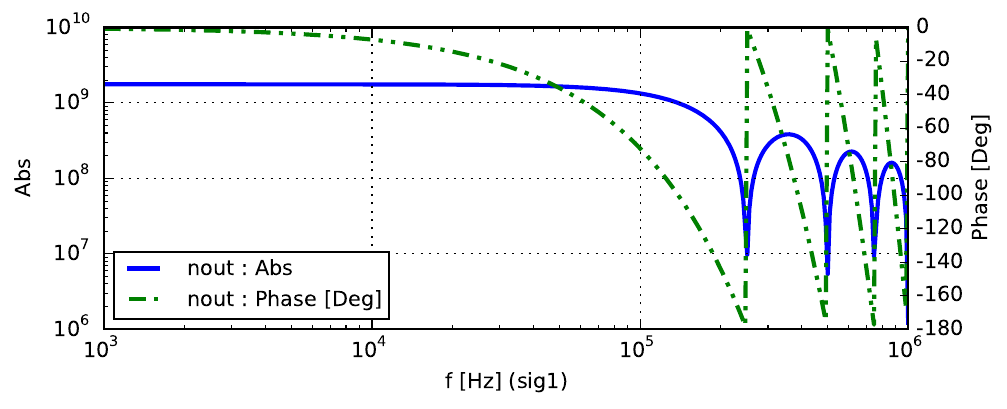}}
    \caption{\Finesse example: Michelson sideband output from a gravitational wave}
    \label{fig:fexample_mich_gw_sig}
\end{figure}}

\noindent
This is  a simple \Finesse example showing how the arm
spaces can be modulated to produce the effect a gravitational
wave would have on it. It outputs the amplitude and phase of the
upper sideband that reaches the output port.

\vspace{3mm}\noindent
{\small
\textbf{Finesse input file for `Michelson gravitational wave response'}
{\renewcommand{\baselinestretch}{.8}

\nopagebreak
\tt
\noindent
\mbox{}\textbf{\textcolor{RoyalBlue}{laser}}\ l1\ \textcolor{Purple}{1}\ \textcolor{Purple}{0}\ nin \\
\mbox{}\textbf{\textcolor{RoyalBlue}{space}}\ s0\ \textcolor{Purple}{1}\ nin\ n1 \\
\mbox{}\textbf{\textcolor{RoyalBlue}{bs}}\ BS\ \textcolor{Purple}{0.5}\ \textcolor{Purple}{0.5}\ \textcolor{Purple}{0}\ \textcolor{Purple}{45}\ n1\ ny1\ nx1\ nout \\
\mbox{}\textbf{\textcolor{RoyalBlue}{s}}\ syarm\ \textcolor{Purple}{600}\ ny1\ ny2 \\
\mbox{}\textbf{\textcolor{RoyalBlue}{m1}}\ ETMy\ \textcolor{Purple}{100e-6}\ \textcolor{Purple}{0}\ \textcolor{Purple}{0}\ ny2\ ny3 \\
\mbox{}\textbf{\textcolor{RoyalBlue}{s}}\ sxarm\ \textcolor{Purple}{600}\ nx1\ nx2 \\
\mbox{}\textbf{\textcolor{RoyalBlue}{m1}}\ ETMx\ \textcolor{Purple}{100e-6}\ \textcolor{Purple}{0}\ \textcolor{Purple}{90}\ nx2\ nx3 \\
\mbox{}\textcolor{Gray}{\#\ Apply\ a\ signal\ each\ arm\ with\ 180\ phase\ difference\ between\ them} \\
\mbox{}\textbf{\textcolor{RoyalBlue}{fsig}}\ sig1\ syarm\ \textcolor{Purple}{1}\ \textcolor{Purple}{180} \\
\mbox{}\textbf{\textcolor{RoyalBlue}{fsig}}\ sig1\ sxarm\ \textcolor{Purple}{1}\ \textcolor{Purple}{0} \\
\mbox{}\textcolor{Gray}{\#\ output\ the\ upper\ sideband} \\
\mbox{}\textbf{\textcolor{RoyalBlue}{ad}}\ upper\ \textcolor{Purple}{0}\ nout \\
\mbox{}\textbf{\textcolor{Red}{xaxis}}\ sig1\ f\ log\ 1k\ 1M\ \textcolor{Purple}{400} \\
\mbox{}\textcolor{Gray}{\#\ sets\ the\ ad\ detector\ frequency\ to\ the\ upper\ sideband\ freq.\ } \\
\mbox{}\textbf{\textcolor{Red}{put}}\ upper\ f\ \textcolor{ForestGreen}{\$x1} \\
\mbox{}\textbf{\textcolor{Red}{yaxis}}\ log\ abs\textcolor{BrickRed}{:}deg \\
\mbox{}

}}


\newpage
\section{Radiation pressure and quantum fluctuations of light}
\label{sec:quantum_noise}
Once classical noise sources 
are sufficiently reduced, the quantum fluctuations of light become one of the
limiting noise sources for interferometric
gravitational-wave detectors \cite{Braginskii1975, jaekel90,
  Meers1991, Niebauer91}. To reduce this quantum noise the basic Michelson
interferometer has been significantly altered over time, as we discuss
in Section~\ref{sec:advanced}. This section aims to outline what
quantum noise is and how its effects can be calculated.

The coupling
of the quantum fluctuations of light into the output signal of the
detector has traditionally been described as two separate
effects: shot noise in the output current of the photodiodes and
radiation pressure effects due to the use of suspended optics. Caves
has shown that both noise components can be understood as originating from
vacuum fluctuations coupling into the dark port of the Michelson
interferometer~\cite{Caves81} and the two-photon formalism suggested
by Caves and Schumaker~\cite{Caves85} has led to a large body of
work towards understanding and reducing quantum noise in gravitational
wave interferometers~\cite{Miao2013, McClelland2011, Chen10,
  et_mueller_ebhard09, CCM05, Buonanno03b}.

In the following we outline a method to compute quantum noise in
interferometer output ports using
sidebands and the classical framework presented in
Sections~\ref{sec:components}, \ref{sec:multi-frequency} and
\ref{sec:detection}.
We apply this method  to investigate the quantum noise limits of
several interferometer readout schemes and finally discuss
how suspended optics effect the quantum noise.

The interested reader can explore this topic further
with a modern and comprehensive treatment of quantum noise in
the review provided in~\cite{Danilishin12} and the following references:
the standard quantum limit~\cite{Caves81, jaekel90}
squeezing~\cite{loudon87, vahlbruch07} and quantum non-demolition
interferometry~\cite{braginsky2000, giovannetti04}.

\subsection{Quantum noise sidebands}
The two quadratures of the light field, its amplitude and phase, form an
observable conjugate pair thus both cannot be measured simultaneously
without some uncertainty in the result~\cite{Caves85}.
This quantum noise of a single mode laser can be depicted as a
phasor with the coherent carrier field and the addition of some
stochastic Gaussian-distributed noise which affects both its phase and
amplitude~\cite{bachor90,Meers1991}. The quantities $\sigma^2_{\phi}$ and
$\sigma^2_{a}$ are the
variances that characterise fluctuations in phase and amplitude
respectively. The noise present in a light field with an equal,
minimum $\sigma_\phi$ and $\sigma_a$ is known as \textit{vacuum
  fluctuations} or \textit{vacuum noise}. Vacuum noise can be understood
as the  photon at all frequencies being incoherently
created and annihilated. Therefore vacuum noise is all-pervasive,
existing at all locations in space, at every frequency and in every spatial mode. Such
photons also enter our interferometer and
limit the sensitivity of any measurement of a field's amplitude or
phase.

\begin{figure}[htb]
    \centering
    \includegraphics[width=0.3\textwidth]{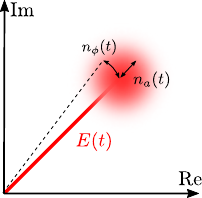}
    \caption{Phasor diagram of equation \ref{eq:carrier_noise} depicting the
    Gaussian random amplitude and phase fluctuations due to vacuum noise. Here $n_{a,\phi}(t)$ are
    random gaussian noises in either the phase or amplitude of the carrier. Shown is only
    the positive frequency part of the carrier field, as $E(t)$ is real a conjugate negative
    frequency term also exists.}
    \label{fig:quant_field}
\end{figure}

Consider a carrier field at one location with
amplitude $a_0$ and frequency $\omega_0$ along with a continuum of
noise fields (the positive frequency spectrum):
\begin{equation}
    E(t) =  \frac{a_0}{2} e^{\I\omega_0 t} + \frac{1}{2}\int_0^\infty q(\omega)e^{\I\omega t}\,d\omega + {\rm c.c}. \label{eq:carrier_noise_sb}
\end{equation}
where $q(\omega)$ is the Fourier component of a stochastic
process, representing the vacuum fluctuation of the electric field.

We can rewrite the continuum of noise in reference to the carrier field frequency:
\begin{equation}
    E(t) =  \frac{a_0}{2} e^{\I\omega_0 t} + \frac{e^{\I\omega_0
        t}}{2}\int_{-\omega_0}^\infty q(\omega_0+\omega)e^{\I\omega t}
    \,d\omega + {\rm c.c}.
\end{equation}
where we can view our quantum noise fields as sidebands of the carrier
instead. For 
gravitational wave detectors the bandwidth $B$ of the signals
induced by a gravitational wave is of the order of several kHz and
thus $B \ll \omega_0$.
Hence, we can focus on a small range of the noise sidebands that will actually
affect our sensitivity:
\begin{equation}
    E(t) =  \frac{1}{2}\left[a_0 + \int_{-B}^B q(\omega_0+\Omega)e^{\I\Omega t} \,d\Omega\right]e^{\I\omega_0 t} + {\rm c.c}.
\end{equation}
Here $\Omega$ will be used in notation to refer to frequencies in the
signal bandwidth with $-B < \Omega \leq B \ll \omega_0$.
We can also represent the quantum fluctuations
as noise in both amplitude and phase:
\begin{equation}
    E(t) = [a_0 + n_a(t)] e^{\I\omega_0 t + n_\phi(t)/a_0} + {\rm c.c}=
    [a_0 + n_a(t) + \I n_\phi(t)]e^{\I\omega_0 t} + {\rm c.c}.
\label{eq:carrier_noise},
\end{equation}
With $n_a$, $n_\phi$ being real amplitudes of the amplitude and phase
fluctuations (of the stochastic process)
with $n_a$, $n_\phi \ll 1$. This equation is represented in
the phasor diagram in Figure~\ref{fig:quant_field}.


We can now related the amplitude and phase fluctuation to the complex
quantum noise $q(\omega)$:
\begin{equation}
    q(\omega) = n_a(\omega) + \I n_\phi(\omega)
\end{equation}
Both $n_{a,\phi}(\omega)$ of a vacuum noise sideband
are characterised by a Gaussian probability density function
with a mean $\mu_{a,\phi}=0$ and variance $\sigma^2_{a,\phi}$.
Note that that the sidebands for the quantum noise are not
representing a coherent and deterministic signal.
This semi-classical approach is sufficient to motivate the
design choices in laser interferometers for gravitational wave detection.
A rigorous approach would require to use operators instead of sidebands.
This approach is beyond the scope of this article, and instead fully
covered in the review article~\cite{Danilishin12}.

The variances $n_{a,\phi}(\omega)$  are limited by the minimum uncertainty in the
relation
\begin{eqnarray}
   \sigma_\phi \sigma_a \geq \frac{\hbar \omega}{2}, \label{eq:quant_uncertainty}
\end{eqnarray}
which gives for an integration time of one second, $\sigma^2_{\phi} = \sigma^2_{a} = \hbar\omega/2$.
As the phase and amplitude of $q(\omega)$ is random we can only
compute its \textit{expected value} or \textit{ensemble value} at a
particular frequency:
\begin{eqnarray}
    \left< q(\omega) \right> = \left<\mu_a\right> + \I \left<\mu_\phi\right> = 0,
\end{eqnarray}
which is zero as the mean of the noise is zero, hence on average no
sidebands are actually observed. We can also consider the covariance
between any two sidebands at frequency $\omega$ and $\omega'$. As
$q(\omega)$ is a complex value there are multiple
ways the covariance can be taken when considering the conjugates of
either sideband, for example $\left< q(\omega)q^\ast(-\omega') \right>$,
$\left< q^\ast(\omega)q(-\omega') \right>$, etc..
However as the fluctuations in amplitude and phase at different frequencies
are independent, the covariance between any two vacuum noise sidebands is:
\begin{eqnarray}
    \left< q(\omega)q^\ast(\omega') \right> &=& \frac{\hbar \omega}{2}\delta(\omega-\omega') \label{eq:qnoise_autocovariance},\\
    \left< q(\omega)q(\omega') \right> &=& 0 \label{eq:qnoise_covariance},
\end{eqnarray}
The delta function in the covariance signifies there is no correlation between different frequencies.
The auto-covariance is then $\left< q(\omega)q^\ast(\omega) \right> \propto \delta(\omega-\omega) = \infty$,
which may seem odd at a first glance. However, this can be better understood in the time domain picture, as we are measuring our signal
over an idealistic infinite time span and as our noise is Markovian (and therefore also
ergodic), the time average of the power of a signal will be infinitely large.

\epubtkImage{quantum_spectrum.png}{%
\begin{figure}[htb]
    \centering
    \includegraphics[width=0.6\textwidth]{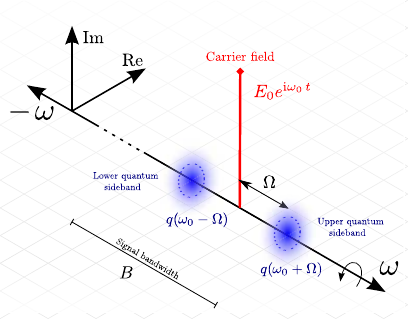}
    \caption{This diagram depicts a carrier field as shown in Figure~\ref{fig:quant_field}
    but expanded to show the vacuum noise sideband phasors that contribute towards the noise.
    The amplitude and phase of each sideband is a stochastic Gaussian
    noise so that its real and imaginary parts are described by some
    probability distribution depicted by the blue faded region, the
    dashed circle represents the standard deviation of such fluctuations.
    The signal bandwidth $B$ can be imagined as containing
    an infinite number of such vacuum noise sidebands, each oscillating with a random
    phase and amplitude. Pictured are two upper and lower sidebands selected from this
    continuum of vacuum noise.  The negative frequency phasors are not shown, they would be the
    mirrored conjugate versions of the positive phasors.}
    \label{fig:quantum_spectrum}
\end{figure}}

\subsection*{Noise power spectral densities}
Noise, i.e. a random signal, can be quantified using a \textit{power spectral
density} (PSD) which is a measure of the power in a signal per frequency.
The definition of a single-sided PSD of some frequency domain value $x(\omega)$ is:
\begin{eqnarray}
    S_{xx}(\omega)\delta(\omega-\omega') &=& 2\left< x(\omega)x^\ast(\omega') \right>, \label{eq:psd}
\end{eqnarray}
with units $[x]^2/{\rm Hz}$. The cross-spectral-density
between two values $x(\omega)$ and $y(\omega)$ is similarly:
\begin{eqnarray}
    S_{xy}(\omega)\delta(\omega-\omega') &=& 2\left< x(\omega)y^\ast(\omega') \right> \label{eq:csd}.
\end{eqnarray}
The eventual physical noise we wish to compute is the noise in the demodulated photocurrent
of the photodiode measuring the interferometer output signal, here we
will consider only photodiodes with 100\% quantum efficiency\footnote{
    It is proportional to a factor $\chi$, the photodiodes quantum efficiency, which states how many Amps per Watt
	of incident power is output by the photodiode. We will assume here the efficiency is perfect, $\chi=1$, for simplicity.}.
The photocurrent $I$ is proportional to the detected light power
$I(t)\sim P(t)$ and the PSD of the noise in the photocurrent is:
\begin{eqnarray}
    S_I(\omega)\delta(\omega-\omega') = 2\left< I(\omega)I^\ast(\omega') \right>\label{eq:qnoie_psd}
\end{eqnarray}
The DC and $\omega\pm\Omega$ terms of the power on a photodiode for a
single carrier with quantum noise sidebands is:
\begin{eqnarray}
    P(t) = E(t)E^\ast(t) &=&  |a_0|^2 + a_0^\ast \int_{-B}^{B} q(\omega_0+\Omega) e^{\I\Omega t} \,d\Omega \nonumber \\
    && + a_0 \int_{-B}^{B} q^{\ast}(\omega_0+\Omega) e^{-\I\Omega t} \,d\Omega + O(q^2),
\end{eqnarray}
terms of the order $q^2$ are assumed to be a negligibly small contribution.
The positive half of the photocurrent spectrum for $0<\Omega \leq B$
is given by its Fourier transform:
\begin{eqnarray}
    I(\Omega) &\equiv& \mathcal{F}[I(t)]= a_0^\ast q(\omega_0+\Omega) + a_0 q^\ast(\omega_0-\Omega). \label{eq:photocurrent}
\end{eqnarray}
The spectrum for frequencies in the signal bandwidth is thus
defined by just quantum noise scaled by the carrier field.
From this point on for the sake of brevity we will define the following notation
without the carrier frequency, as we are only using a single carrier
for this derivation:
\begin{eqnarray}
    q(\omega_0\pm\Omega) \Rightarrow q_\pm\,\,\, {\rm and}\,\,\,q(\omega_0\pm\Omega') \Rightarrow q'_\pm .
\end{eqnarray}
Using equations \ref{eq:qnoie_psd} and \ref{eq:photocurrent},
the PSD of the photocurrent is:
\begin{eqnarray}
    S_I(\Omega)\delta(\Omega-\Omega') &=& 2P_0\bigl(\bigl< q_+q'^\ast_+ \bigr> + \bigl< q_-q'^\ast_- \bigr>\bigr)  + 2a_0^2\bigl<q_-q'_+ \bigr>^\ast + 2{a_0^2}^\ast \bigl<q_+q'_- \bigr>\label{eq:photocurrent_psd}.
\end{eqnarray}
Now applying equations \ref{eq:qnoise_autocovariance} and \ref{eq:qnoise_covariance} in equation \ref{eq:photocurrent_psd}
the noise PSD for a single carrier field with vacuum noise is:
\begin{eqnarray}
    S_I(\omega_0\pm\Omega)\delta(\Omega-\Omega') &=& 2 P_0 \left(\bigl<q_+ q_+'^\ast\bigr> + \bigl<q_- q_-'^\ast\bigr>\right), \nonumber \\
    &=& P_0 \left(\hbar(\omega_0+\Omega) + \hbar(\omega_0-\Omega)\right)\delta(\Omega-\Omega') \nonumber \\
    S_I(\omega_0\pm\Omega) &=&  2 P_0 \hbar \omega_0 . \label{eq:quantum_shottky}
\end{eqnarray}
Here we see that the quantum noise of a single carrier field does not depend on the
sideband frequency $\Omega$. The vacuum fluctuations interfering with our carrier field produces a
broadband frequency-independent noise source proportional to the carrier power and frequency.
It should also be noted that equation~\ref{eq:quantum_shottky} is the same result
as the semi-classical Schottky shot-noise equation, equation~\ref{eq:classical_shottky}.
An interesting aspect to note here are the differing reasons for the presence of this quantum or shot noise.
The Schottky formula derives this noise from the
Poisson statistics of electrons generated in the photocurrent due to the light field power.
Whereas the quantum approach reasons that such fluctuations in the photocurrent
are in fact due to vacuum noise superimposing itself onto our
light fields introducing a noise into our measurements.

The description of quantum noise with semi-classical sidebands has the
advantage that the propagation of a stochastic signal through a linear system is described
by the same transfer functions as for a deterministic signal. Therefor
we can use the classical model of the optical system to compute the
propagation of the quantum noise as well as any signal.

\subsection{Vacuum noise and gravitational wave detector readout schemes}
\label{sec:readout}
\begin{figure}[htb]
   \centering
   \begin{subfigure}[b]{0.45\textwidth}
    \includegraphics[width=\textwidth]{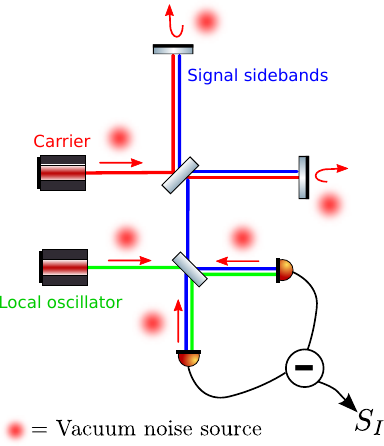}
    \caption{Balanced homodyne}
    \label{fig:quant_michelson_balanced}
   \end{subfigure}
   \hfill
   \begin{subfigure}[b]{0.45\textwidth}
    \includegraphics[width=\textwidth]{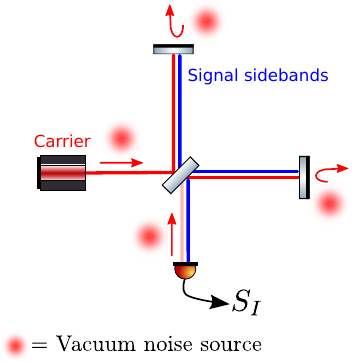}
    \caption{DC offset}
    \label{fig:quant_michelson_DC}
   \end{subfigure}
   \caption{Shown are two possible readout schemes that can be used to extract signal sideband
   information from a Michelson along with the various classical fields and sources of vacuum noise.}
   \label{fig:quant_michelson}
\end{figure}

Let us now consider a Michelson interferometer, as described in
Section~\ref{sec:Michelson}, and the limiting sensitivity due to vacuum
noise leaking into the detector. Figure~\ref{fig:quant_michelson}
depicts two exemplary readout schemes for measuring
the gravitational wave signal. For both schemes we can identify the sources
of vacuum noise that will enter the interferometer. The input is assumed
to be a perfect single-mode laser whose noise is purely vacuum noise. The end mirrors
in the arms are taken to be perfectly reflective thus no vacuum noise
enters through them; however if $r<1$ any vacuum noise leaking out would be replaced with
an equal amount of  uncorrected noise injected back in.
The output port is fully open and thus allow vacuum noise to enter into the
system proving the primary contribution of noise in Michelson setup used
for gravitational wave detectors. This is due to the fact that such
the Michelson is operated on the dark fringe
for the input carrier, meaning any the laser noise will leave the
system back towards the laser, whereas the noise entering through
the output port will return to the output port.

When no non-linear optical effects (effects proportional to the beam's
power) are present in an interferometer and the only quantum noise
present is uncorrelated vacuum noise, there will always be the same
amount of vacuum noise incident on any photodiode. This is
irrespective of the topology of the interferometer or components used
because noise can never be effectively lost from the system;  an
equivalent amount of uncorrelated noise is always injected back in.
In such cases propagation of the noise sidebands through the
interferometer do not need to be computed.
Instead, when computing $S_I$ at any of the photodiodes shown in
Figure~\ref{fig:quant_michelson} we only need to consider  pure vacuum noise sidebands and
the local oscillator field, $E_{LO}$; the source of location of the vacuum
noise sources is not of importance. This is why for early generation
gravitational wave detectors, which had negligible non-linear optical
effects, the semi-classical Schottky expression could be used to estimate
the quantum noise correctly.

The detailed computation of quantum noise limited sensitivity of a detector
depends on the readout scheme used. Early generations of gravitational
wave detectors such as LIGO, Virgo, GEO\,600 and TAMA300 used
heterodyne readout schemes, where RF modulation sidebands applied
to the input field are used as local oscillators at the output (see
Sections~\ref{sec:signal_readout}). However, such schemes
included some technical challenges, the oscillator noise of
the RF modulator being one of them, and also
increase the shot-noise level when demodulating the
photocurrent~\cite{Meers1991, Niebauer91, Rakhmanov:01,Buonanno03b}. Thus
the next generation of detectors opted for a DC readout
scheme~\cite{Fricke12, Hild09b}, see Section~\ref{sec:dc}.
Both schemes depicted in Figure~\ref{fig:quant_michelson} use a form
of DC readout, which we will analyse in more detail in the following
sections. We do not cover the computation of  quantum noise with
RF modulation readout schemes, the interested reader should
see~\cite{Harms06,Rakhmanov:01,Buonanno03b}.

\subsubsection*{Noise-to-signal ratio for DC offset}

\epubtkImage{shot_noise_sens.png}{%
  \begin{figure}[htbp]
    \centerline{\includegraphics[width=1\textwidth]{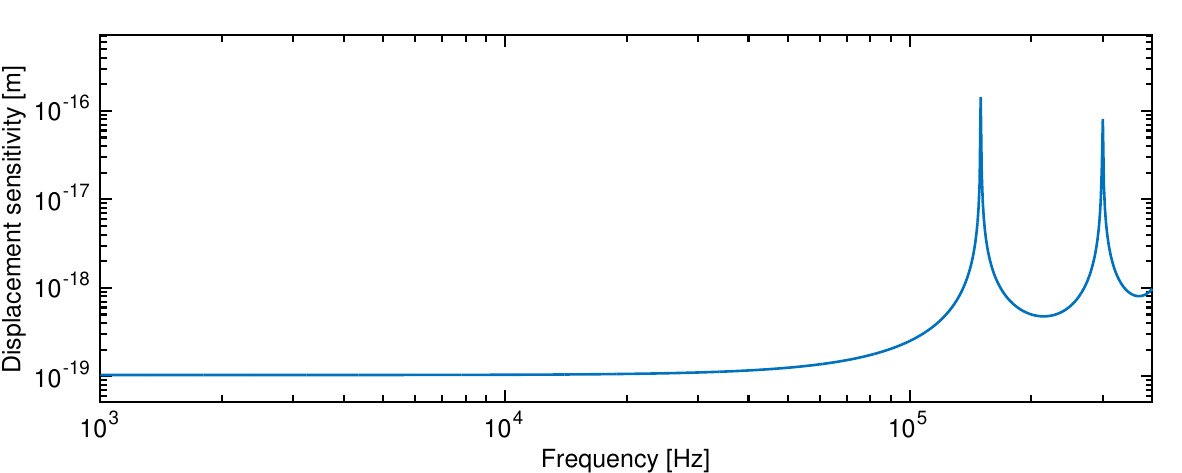}}
   \caption{Shot-noise limited sensitivity of a Michelson, see
     Equation~\ref{eq:mich_shot_noise_NSR}, with $\bar{L}=1000$\,m, $R=T=0.5$
     and  $P_0=1$\,W.}
   \label{fig:shot_noise_sens}
\end{figure}}

\noindent
A DC offset in the main Michelson interferometer (see Section~\ref{sec:signal_readout}) provides
a local oscillator by making the interferometer operate slightly
away from the dark fringe for the carrier, and hence allowing
some to leak through to the output port along with any signal sideband fields.
The sources of vacuum noise that will contribute to the quantum noise
are shown in Figure~\ref{fig:quant_michelson_DC};
however, as stated previously the total amount of noise present at the photodiode
will be just pure vacuum noise as it is assumed that there are no non-linear optical effects. The
local oscillator field at the output is
given by Equation~\ref{eq:michelson_south_port_carrier} and along
with the vacuum noise sidebands the output field is:
\begin{eqnarray}
    E_{out} &=&  \left[\I E_0 e^{-i2k\bar{L}}\sin(k_0 \delta_{\mathrm{off}}) + q_+e^{\I\Omega t} + q_-e^{-\I\Omega t}\right]e^{\I\omega_0 t}.
\end{eqnarray}
where we have used the dark fringe offset as stated in Equation~\ref{eq:dc_offset_phase}.
The quantum noise PSD when using a DC offset is now essentially the
same scenario as when deriving Equation~\ref{eq:quantum_shottky},
where a single carrier and noise sidebands were considered; except that the carrier
power now depends on $\delta_{\mathrm{off}}$:
\begin{eqnarray}
    S_{P,DC} &=& \left<|P_{out}(\Omega)|^2\right> = 2 P_{0}\sin^2\left(k_0\delta_{\mathrm{off}}\right) \hbar \omega_0 \approx 2P_{0}(k_0\delta_{\mathrm{off}})^2 \hbar \omega_0 ,
\end{eqnarray}
where $P_0$ is the power of the laser injected into the Michelson and $k_0\delta_{\mathrm{off}} \ll 1$.

To compute the noise-to-signal (NSR) ratio, which is used to describe  the
sensitivity of our Michelson, the transfer function from a signal we
want to measure to the photodiode output is required. Here we will use
the gravitational wave signal transfer function from
equation~\ref{eq:michelson_dc_gw_TF} which describes the Watts of power per unit of
strain, $h$, at the output detector
\begin{eqnarray}
    \left|T_{gw\rightarrow P}(\omega_{gw})\right| &\approx& k_0\delta_{\mathrm{off}}P_0\frac{w_0}{\omega_{gw}}\sin\left(\frac{\omega_{gw}\bar{L}}{c}\right)  \, {\rm W}/{\rm h}.
\end{eqnarray}
We note that $T_{gw\rightarrow P}$ refers to the
\textit{amplitude} of the differential length modulation that the arms experience.
Thus the NSR should be computed with  the \textit{amplitude spectral
  density} given as
${\rm ASD} = \sqrt{{\rm PSD}}$:
\begin{eqnarray}
    {\rm NSR} = \frac{\sqrt{S_{P,DC}}}{T_{gw\rightarrow P}} = \sqrt{\frac{2\hbar}{P_0 \omega_0}} \frac{\omega_{gw}}{\sin(\omega_{gw}\bar{L}/c)} \,\, \frac{{\rm h}}{\sqrt{\rm Hz}}.
    \label{eq:mich_shot_noise_NSR}
\end{eqnarray}
The displacement sensitivity does not depend on the DC offset and can
be improved, for example, by increasing the laser power. Eventually, building a more powerful laser is not possible without
sacrificing stability in power and frequency. Instead we
can also use Fabry-Perot cavities to increase the effective
power inside the interferometer, see Section~\ref{sec:advanced}.

\subsubsection*{Noise-to-signal ratio for balanced homodyne}

Balanced homodyne readout involves the use of an external local
oscillator whose optical frequency is the same as the
main carrier light in the interferometer. The main Michelson
interferometer is operated on the dark
fringe for the carrier so no carrier light  is present at the output
port. This local oscillator
is mixed with the signal sidebands using a beam splitter,
such a setup is depicted in Figure~\ref{fig:quant_michelson_balanced}
and in more detail for the readout in Figure~\ref{fig:quant_output_fields_balanced}.
As the signal sidebands are now split into two optical paths we
require two photodiodes to measure the signal, otherwise half the
signal will be lost instantly. The balanced aspect of this readout
scheme refers to the fact that the two photocurrents $I_a$ and $I_b$
are combined in such a way that the noise from either the local
oscillator port or the signal port can be completely removed from the
measurement.

No current generation gravitational-wave detector uses this form of
homodyne readout for extracting gravitational wave signals. This has
been due to the additional technical challenges which are not present
when using DC readout. It is however
used extensively for quantum noise measurements when non-vacuum states
are injected into interferometers~\cite{Stefszky12, Chua14}
and offers potential benefits over DC readout if the technical challenges
can be overcome, as we show later in this section.
Although not currently used, such a readout scheme
is a current topic of investigation for future generations of
detectors for extracting gravitational wave signals~\cite{Fritschel14}.

\epubtkImage{quant_output_fields_balanced.png}{%
\begin{figure}[htb]
   \centering
    \includegraphics[width=0.7\textwidth]{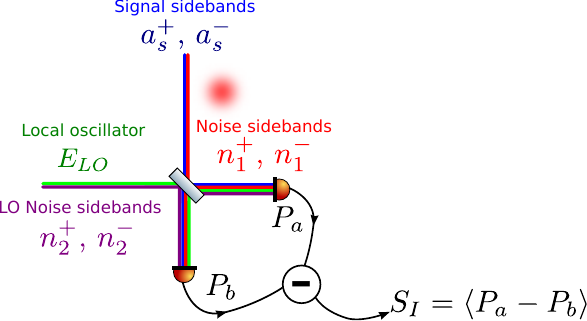}
    \caption{The signal and noise fields in the homodyne detector as
      used in the balanced homodyne readout scheme.}
    \label{fig:quant_output_fields_balanced}
\end{figure}}

There are two possible sources for the local oscillator field
when using balanced homodyne detection: a separate laser system or
a pick-off of the same carrier field used in the interferometer.
The former is technically challenging as the separate system must
be locked to the input laser to ensure temporal coherence when beating
with the signal sidebands. The latter option of using a pick-off beam
does not have this issue as it is from the same laser. Other technical
challenges that exist for both options are~\cite{McKenzie07} that the beam splitter is
exactly 50:50; that the signal sidebands and local oscillator fields
have a particularly good spatial overlap, also referred to as \textit{mode-matching}
and that the local oscillator does not back-scatter into the output
port of the interferometer.

Assuming a perfect 50:50 beam splitter and a coherent local oscillator
which is well aligned to some signal beam we want to measure, both the incoming
signal and local oscillator include vacuum noise.
To calculate the photocurrent noise $S_I$ we describe the  noise
sidebands are at as shown in Figure~\ref{fig:quant_output_fields_balanced}:
\begin{eqnarray}
    E_a = \left[ r(n_1^+e^{\I\Omega t} + n_1^-e^{-\I\Omega t}) + \I t(E_{LO} + n_2^+e^{\I\Omega t} + n_2^-e^{-\I\Omega t})\right]e^{\I\omega_0 t} \nonumber \\
    E_b = \left[ r (E_{LO} + n_2^+e^{\I\Omega t} + n_2^-e^{-\I\Omega t}) + \I t(n_1^+e^{\I\Omega t} + n_1^-e^{-\I\Omega t})\right]e^{\I\omega_0 t}
\end{eqnarray}
The photocurrent noise PSD is then proportional to:
\begin{eqnarray}
    S_I \propto \left< \left|E_aE^\ast_a - E_bE^\ast_b\right|^2 \right>
\end{eqnarray}
The for the incident power on each photodiode we ignore noise terms
that are not scaled by
the local oscillator  as negligible:
\begin{eqnarray}
    P_a(t) &=& -\I t E^\ast_{LO} \left[r(n_1^+e^{\I\Omega t} + n_1^-e^{-\I\Omega t}) + \I t(n_2^+e^{\I\Omega t} + n_2^-e^{-\I\Omega t})\right] +{\rm c.c}  \nonumber \\
    P_b(t) &=& r E^\ast_{LO} \left[\I t (n_1^+e^{\I\Omega t} + n_1^-e^{-\I\Omega t}) + r (n_2^+e^{\I\Omega t} + n_2^-e^{-\I\Omega t})\right]  +{\rm c.c}
\end{eqnarray}
Assuming that each photodiode is identical in its response to the power,
the photocurrents proportional to these two powers can then be subtracted or
summed:
\begin{eqnarray}
    P_a(t) \pm P_b(t) &=& E^\ast_{LO}\left[ \I rt (n_1^+e^{\I\Omega t} + n_1^-e^{-\I\Omega t})(-1\pm 1) + (n_2^+e^{\I\Omega t} + n_2^-e^{-\I\Omega t})(T\pm R) \right]  +{\rm c.c} .
\end{eqnarray}
This shows that either the noise from the local oscillator, $n_2^\pm$,
or that coming along with the signal, $n_1^\pm$, can be removed. Typically
the local oscillator noise will be larger than that accompanying the signal
thus we can compute $P_a - P_b$ to remove it. It can also be seen here if
the beam splitter is not 50:50, $R \neq T$, the local oscillator noise cannot
be fully removed. Finally the subtracted photocurrent for the sideband
frequency $\Omega$ is:
\begin{eqnarray}
    P_{a-b}(\Omega) \equiv \mathcal{F}[P_a(t) - P_b(t)](\Omega) &=& -\I E^\ast_{LO} (n_1^+ + {n_1^-}^\ast),
\end{eqnarray}
where $r=t=1/\sqrt{2}$. For pure vacuum noise, $n_{1}^\pm \Rightarrow
q_\pm$, the resulting photocurrent noise PSD for this is that given by
Equation~\ref{eq:quantum_shottky}:
\begin{eqnarray}
    S_I &=& \left<\left| P_{a-b}(\Omega) \right|^2\right> = 2P_{LO} \left(\left<q_+ q_+^\ast\right> + \left<q_-q_-^\ast\right>\right),\nonumber \\
    &=& 2 P_{LO} \hbar \omega_{0}
\end{eqnarray}
Therefore if correctly balanced the quantum noise is no greater than
what is present for a DC offset readout.
If the local oscillator power $P_{LO}$ is identical to the carrier
power in the DC offset scheme, then
the sensitivity for the balanced homodyne detection is
the same as that for the DC offset detection, stated in Equation~\ref{eq:mich_shot_noise_NSR},
because the transfer function from signal to the two photodiodes is
essentially the same. One aspect where it differs however is the phase
of the local oscillator relative to the signal sidebands which is now a free parameter,
which is known as the \textit{homodyne angle} or the \textit{readout phase}.
When using a DC offset readout scheme the readout phase is essentially fixed.
Having the ability to vary this readout phase
provides an extra degree of freedom for optimising the quantum noise-to-signal
ratio for gravitational wave signals. This is an assumed feature in some
\textit{quantum non-demolition} schemes~\cite{Braginsky01081980,RevModPhys.68.1}
which introduce new methods for reducing the quantum
noise~\cite{Purdue02,Kimble02,Khalili96,Chen03,Chen10}.

\subsection{Quantum noise with non-linear optical effects or squeezed states}
\label{sec:quantfluc_rp}
\epubtkImage{qn_squeezing.png}{%
\begin{figure}[htb]
    \centering
    \includegraphics[width=\textwidth]{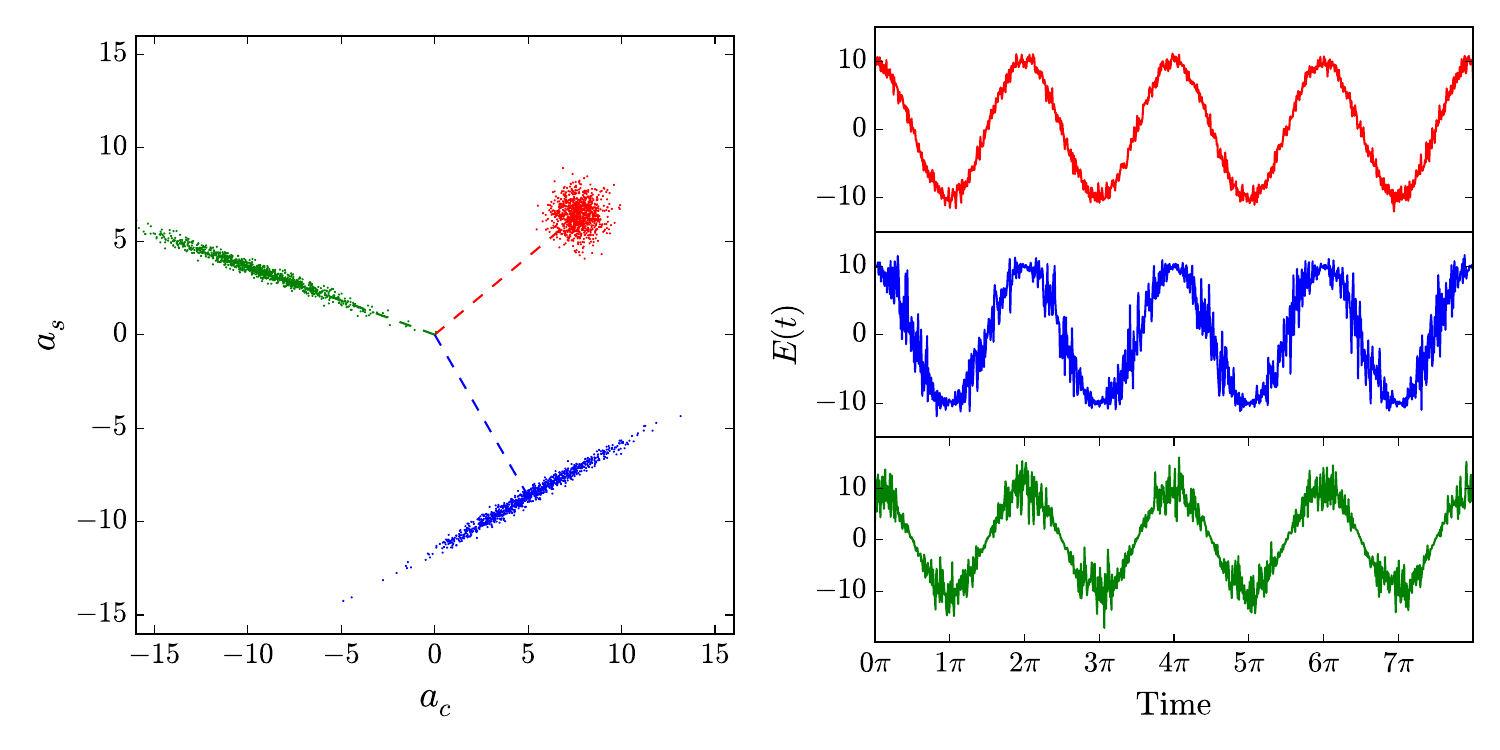}
    \caption{Depicted are the phasor and the time signals of pure vacuum
    noise (Red), amplitude-squeezed (Blue) and phase-squeezed (Green)
    noise. 
    The effect here is greatly exaggerated to produce a visible noise thus
    the scaling on the axes do not represent any realistic values. The phasor
    diagram shows $E(t)$ at some arbitrary time value and each point is a sample
    retrieved from the probability density function of the noise. The squeezed
    states show clearly a correlation between phase and amplitude fluctuations; the
    area of each state is equal representing the minimum given by the uncertainty
    relation \ref{eq:quant_uncertainty}.}
    \label{fig:quant_quadrature}
\end{figure}}%
Up to this point we have only considered pure vacuum noise and linear optical
effects, both of which are valid approximations for previous generations of gravitational
wave detectors. As the effective laser power in the interferometer is increased
to reduce shot noise, the radiation pressure exerted on suspended optics by the
circulating laser beams will increase another noise significantly, the
\textit{radiation pressure noise}. As the suspended mirrors are free to move under the influence of radiation pressure, any fluctuation in the laser's
power will couple back into itself as phase modulation. This is a non-linear process as
the amplitude of the motion is proportional to the power in the beam,
which leads to the upper and lower sidebands becoming correlated
with one another.
As explained in Section~\ref{sec:quant_radp}, such noise is prominent at low frequencies.

Then there is also the possibility of \textit{squeezing} the vacuum
noise, whereby still satisfying the
relationship~\ref{eq:quant_uncertainty}, the uncertainty in either the
phase or amplitude is increased whilst being and decreased in the
other. This squeezed noise can be represented by correlated noise
sidebands~\cite{Caves85, Danilishin12},
Figure~\ref{fig:quant_quadrature} shows qualitatively the effect of
squeezed vacuum noise on a coherent field. Injecting squeezed noise
into the output port of the interferometer can thus
reduce the dominant vacuum noise. Upon returning to the output port
the noise should still be squeezed, but to a slightly lesser degree
due to various optical losses which degrade the amount of
squeezing. If squeezing is implemented effectively, the noise can be reduced below the
typical shot-noise level of $2P_0\hbar\omega_0$, thus providing a
broad improvement in shot-noise limited regions of the detectors
sensitivity~\cite{Caves81}. Although we will not cover squeezing in
detail in this article, squeezed light injection has been used
routinely by the GEO\,600 since 2010~\cite{Abadie2011}, and further upgrades to
advanced gravitational wave detectors  using squeezed light sources
are actively being developed~\cite{oelker14}.

One important aspect to note here is that when either non-linear
optical effects or non-vacuum states of light are significant, the
correlations introduced in the propagation of light fields through the interferometer
have to be considered. This is due to the fact that the noise sidebands
will be altered in amplitude and phase and the correlation
between sidebands introduced as they propagate becomes an
important feature. Such a calculation involves constructing the full interferometer
matrix, see Section~\ref{sec:Coupling_matrices}, for the noise
sideband frequencies and including, if so required, the radiation
pressure coupling at suspended mirrors as discussed in the next
section.

\subsection{Radiation pressure coupling at a suspended mirror}
\label{sec:quant_radp}
As the laser power is increased to reduce the shot-noise, the higher
power results in a significant radiation pressure force being exerted
on the interferometer mirrors. The frequency spectrum of the force exerted
on a perfectly reflectivity mirror by a single beam with power $P(\Omega)$
is given by~\cite{Meystre:85}
\begin{equation}
    F_{rp}(\Omega) = \frac{2 P(\Omega)}{c}.\label{eq:Frp}
\end{equation}
In order to attenuate seismic vibrations and produce free-masses as
probes for gravitational waves, the mirrors in gravitational wave detectors are
suspended via a series of active and passive suspension systems.
At frequencies well above the resonances of the suspension systems the
mirrors can be considered to be free (or quasi-free). Any fluctuation in the light power induces a motion in
the suspended mirrors. This process converts power fluctuations into phase
fluctuations, and this coupling can lead to opto-mechanical effects such
as \textit{optical springs}, which couple the motion of multiple
suspended optics together~\cite{sheard04, aspelmeyer14}.

The induced longitudinal motion of a suspended mirror due to $N_f$
separate forces being applied to it is:
\begin{equation}
    \delta z(\Omega) = H(\Omega) \sum_{n=0}^{N_f} F_n(\Omega),
    \label{eq:zf}
\end{equation}
where $H(\Omega)$ is the \textit{mechanical susceptibility} or
\textit{mechanical transfer function} from a force applied to motion
parallel to the mirrors surface normal. Similar relationships are
possible for rotational motions considering torques applied to the
mirror.

\subsubsection*{Mechanical transfer functions}
The transfer function $H(\Omega)$ is determined by the specific setup
of the suspension systems.
The various resonances and features of the system can be represented
with an expansion into \textit{poles and zeros}:
\begin{eqnarray}
    H(\Omega) &=& -\frac{1}{M\Omega^2} \frac{\prod_{k=1}^{N_z} (\I\Omega-\mathcal{Z}_k)(\I\Omega-\mathcal{Z}_k^\ast)}{\prod_{j=1}^{N_p}(\I\Omega-\mathcal{P}_j)(\I\Omega-\mathcal{P}_j^\ast)} \nonumber\\
    \mathcal{P}_k &=& \Omega_{p,k} \left(\I-\frac{1}{2 Q_{p,k}}\right), \, \mathcal{Z}_j = \Omega_{z,k} \left(\I-\frac{1}{2 Q_{z,k}}\right)
\end{eqnarray}
where $M$ is the mass of the mirror in kg,
$\{\Omega_{p/z,i}\}_{i=1}^{N_{p/z}}$ is a set of frequencies for each
pole and zero and $\{Q_{p/z,i}\}_{i=1}^{N_{p/z}}$ the respective
quality factors.
When the frequencies of interest (signal frequencies) are much
higher than any pole or zero frequency, $\Omega \gg \Omega_{p/z}$, we
can assume a \textit{free mass}, $N_z = 0$ and $N_p = 0$:
\begin{equation}
    H(\Omega) = -\frac{1}{M \Omega^2}.
\end{equation}

\subsubsection*{Approximations for radiation pressure}
With the mirror position change being proportional to the laser power,
$\delta z \propto P$, the problem is non-linear in terms of the
complex field amplitudes. Solving such a problem in a complex
interferometer setup is challenging and not possible using the
methods outlined in Section~\ref{sec:components}, as the frequency
domain model is assuming a linear system. However, for
gravitational wave detectors we can make some assumptions about the
system:
\begin{itemize}
\item the motion of any optic is small, $|\delta z| \ll \lambda$,
  when the interferometer is controlled and well-behaved, and we can
  linearise equations in $\delta z$,
\item any high-frequency fluctuations in the beam are negligible
  due to $H(\Omega) \propto 1/\Omega^2$ and we ignore the effects of
  RF sidebands on the optics,
\item any low-frequency fluctuations are very small, such that
  the magnitude of any sidebands is much less than the magnitude of
  its carrier field, which allows us to identify a well defined
  carrier field in our calculations.
\end{itemize}
These are all valid assumptions for gravitational wave detectors
once they are operating in a steady state and have well controlled
optics.
For a single carrier with amplitude $E_0$ and frequency
$\omega_0$ and noise sidebands at frequency $\Omega$, the
incident field on a suspended mirror is:
\begin{eqnarray}
E_i &=& (E_0 + q_+e^{\I\Omega t}+q_-e^{-\I\Omega t}) e^{\I\omega_0 t} + {\rm c.c}.
\end{eqnarray}
As with the approximations listed above we can assume $|q_\pm| \ll
|E_0|$ and $\omega_0 \gg \Omega$. The 
fluctuation in the beam power is then given by:
\begin{eqnarray}
P(\Omega) &=& q_+ E^\ast_0 + q_-^\ast E_0
\label{eq:Ps}
\end{eqnarray}
where we only consider sideband-carrier product terms and those with a
frequency $\Omega$. Substituting the fluctuating power~\ref{eq:Ps} into
the radiation pressure force~\ref{eq:Frp} to compute the displacement~\ref{eq:zf}
the motion of the mirror can be found.
The amplitude of the motion at frequency $\Omega$ induced via radiation
pressure for a perfectly reflective, free-mass mirror is
\begin{equation}
    \delta z = -\frac{2}{Mc\Omega^2} \left(q_+ E^\ast_{0} + q_-^\ast E_{0}\right).
    \label{eq:zf_rp}
\end{equation}
Such a moving mirror, as discussed in Section~\ref{sec:phasemod},
creates phase modulation sidebands around any carrier that is reflected from
it. The reflected field, using Equation~\ref{eq:phase_mod}, is:
\begin{equation}
E_r= E_i\left(1+\frac{\I k_0}{2}\Bigl(\delta z^{+} e^{-\I\Omega t} + \delta z^{-} e^{\I \Omega t}\Bigr)\right)e^{\I\omega_0 t},\label{eq:refl_rp}
\end{equation}
where to keep notation simpler, $\delta z^{+}\equiv \delta z$ and $\delta z^{-} \equiv \delta z^\ast$.

Take the simple example of vacuum noise and a single carrier,
with amplitude $E_0$, incident on a free mass mirror of mass $M$ and calculate the
noise after being reflected. The amplitude of the reflected upper and lower
noise sidebands, $q_{r,\pm}$, using Equations~\ref{eq:zf} and~\ref{eq:refl_rp}, are:
\begin{eqnarray}
    q_{r,\pm} &=& v_\pm + \I E_0 k_0\frac{\delta z^\pm}{2}, \nonumber  \\
&=& v^\pm - \I E_0 k_0\frac{v_{\pm} E^\ast_0 + v_{\mp}^\ast E_0}{Mc\Omega^2}.\label{eq:q_rp_refl}
\end{eqnarray}
where $v_\pm$ are the incident pure vacuum noise sidebands.
\eqref{eq:q_rp_refl} shows that the reflected upper and lower noise sidebands are now a
mix of the incident upper and lower sidebands, i.e.~after reflection they are correlated.
There is additional phase noise present, and it scales as $\propto|E_0|/\Omega^2$.
Thus, this noise is relevant at low frequencies and
when the beam power to mass ratio is significant.

The noise PSD for the reflected beam is computed using equation~\ref{eq:photocurrent_psd}.
This requires computing the various covariance and auto-covariances of the reflected noise sidebands~\ref{eq:q_rp_refl}
and their beating with the carrier field:
\begin{eqnarray}
        \left<q_{r,\pm}q_{r,\mp}'\right> &=& -\dfrac{\hbar\omega_0 E_0^2}{M\Omega^2 c}\left( k_0 + \dfrac{|E_0|^2 k_0^2}{M \Omega^2 c} \right)\\
        \left<q_{r,\pm}q_{r,\pm}'^\ast\right> &=& \dfrac{\hbar(\omega_0\pm\Omega)}{2} + \dfrac{\hbar\omega_0 |E_0|^4 k_0^2}{M^2 c^2 \Omega^4}.
\end{eqnarray}
To simplify the above we will also assume the carrier has zero phase, $E_0^\ast = E_0$ and that $E_0 = \sqrt(P_0)$.
The power noise PSD using equation~\ref{eq:photocurrent_psd} is then
\begin{eqnarray}
    S_I(\Omega) &=& 2P_0\bigl(\bigl< q_{r,+}q'^\ast_{r,+} \bigr> + \bigl< q_{r,-}q'^\ast_{r,-} \bigr>\bigr)  + 2P_0\bigl<q_{r,-}q'_{r,+} \bigr>^\ast + 2P_0^\ast \bigl<q_{r,+}q'_{r,-} \bigr> \nonumber \\
    &=& 2 \hbar\omega_0 P_0.
\end{eqnarray}
Thus the noise is still just a flat shot noise limit, as expected. However, this only
shows the amplitude noise in the beam, not any phase noise. To compute the phase quadrature
the local oscillator must have an additional $\pi/2$ phase relative to the sidebands.
Experimentally this could be achieved using the balanaced homodyne readout as mentioned in previous sections.
Here though we can cheat by simply adding an additional phase to the carrier to the beam \textit{after} reflection,
i.e.~compute the PSD of a power fluctuation
\begin{equation}
	P(\Omega) = q_+ E^\ast_0 e^{-\I\phi} + q_-^\ast E_0 e^{\I\phi}
\end{equation}
where $\phi$ is our additional homodyne phase. Computing the PSD of this with $\phi=\pi/2$ to
compute the phase fluctuations we see:
\begin{eqnarray}
    S_\phi(\Omega) &=& 2 P_0 \hbar \omega_0 + \dfrac{8\hbar P_0^3	\omega_0 k_0^2}{M^2 c^2 \Omega^4}.
\end{eqnarray}
Here there is a flat shot noise fluctuation plus additional phase noise due to the vacuum noise
perturbing the mirror. In the limit of an infinitely heavy mirror we can see this radiation pressure
noise is removed and we are left with the vacuum noise fluctuations in phase. It is these phase fluctuations
that are converted from phase to amplitude noise at the Michelson dark port that then lead to
quantum noise limited sensitivity of the detector at the output photodiode.

\subsection{Semi-classical Schottky shot-noise formula}
\label{sec:shottky}
Shot noise historically has been described as the noise arising from the statistical
distribution of electrons in photo detectors. The Schottky formula for the
(single-sided) power spectral density of the fluctuation of the photocurrent
for a given mean current $\bar{I}$ is:
\begin{equation}
S_I(f) =2\,e\,\bar{I},
\end{equation}
with $e$ the electron charge.  Here $S_X (f)$ denotes the single-sided power
spectral density of $X$ over the Fourier frequency $f$.
The link between (mean) photocurrent $\bar{I}$ and (mean) light power
$\bar{P}$ is given by the relation:
\begin{equation}
\bar{I}= e N= \frac{e~\eta~\lambda }{\hbar 2 \pi c} \bar{P},
\end{equation}
with $N$ as the number of photons and $\eta$ the quantum efficiency of the
diode.  Instead of Planck's constant we write $\hbar\,2\,\pi$ to avoid
confusion with the typical use of $h(t)$ for the strain of a gravitational
wave. We can now give a power spectral density for the fluctuations of the
photocurrent:
\begin{equation}
S_P(f)=2\,\frac{2\pi\,\hbar\,c~}{\lambda}\bar{P}.\label{eq:classical_shottky}
\end{equation}
As stated above this equation estimates the shot noise correctly when
the interferometer does not contain any non-linear effects or squeezed
input fields.

\subsection{Optical springs}\label{sec:optical_springs}

Optical springs are a result of a feedback loop being created
by the optical field scattered by a suspended
mirror being fed back in to itself.
Using the properties of the opto-mechanical coupling introduced in the previous section it will be shown how this
feedback process introduces a force that is analogous to a damped spring being attached to the mirror.
This optical spring will having a particular resonance frequency and damping coefficient that
depends on the optical and mechanical properties of the interferometer.

\begin{figure}[htb]
\centering
\includegraphics{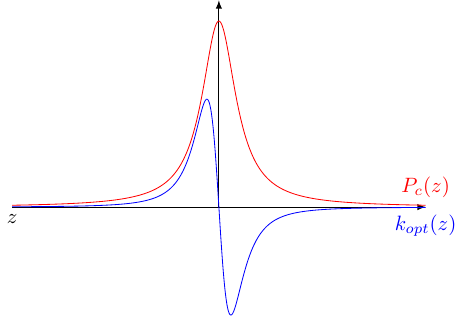}
\caption{Illustrative example of the circulating power (red) in a Fabry-Perot cavity. The blue line
	shows the spring constant (blue). Not to scale.}
\label{fig:fp_circ}
\end{figure}

\subsubsection*{Adiabatic optical spring}
A Fabry-Perot cavity with a suspended end mirror is the simplest system which can feedback
the sidebands created to the mirror. Firstly the case when the mirror is moving slowly
compared to the round-trip time of the cavity is considered.
In this situation the optical response to a mirror moving is effectively instantaneous throughout the interferometer.
The power circulating in a Fabry-Perot cavity, hence the power incident on the suspended mirror,
as a function of a cavity length change $z$ in meters is:
\begin{equation}
P_c(z) = \dfrac{P_0\,T_1}{1 + R_1 R_2 - 2 r_1 r_2 \cos(2k z)}
\label{eq:fabry_perot_Pcirc}
\end{equation}
and shown in figure~\ref{fig:fp_circ}.
As the radiation pressure force is $\propto P_c$
the force varies with respect to the end mirror's position.
A position dependent force is the definition of a spring constant, thus
for our optical spring we find:
\begin{equation}
\mathit{k}_{opt} = -\dfrac{\dif F(z)}{\dif z} = \dfrac{\dif}{\dif z} \left[\dfrac{-2 P_c(z)}{c}\right] = \dfrac{-8 P_0 r_1 r_2 k T_1 \sin(2k z)}{c(1+R_1 R_2-2r_1 r_2 \cos(2 k z))^2}.
\end{equation}
Plotting the $k_{opt}$ in figure~\ref{fig:fp_circ},
when the cavity is perfectly resonant for the carrier field there is no optical
spring, for positive $\delta z$ we have a restoring force, $k_{opt} < 0$, and
anti-restoring force, $k_{opt} > 0$, with negative detunings.

\begin{figure}[htb]
	\centering
\includegraphics{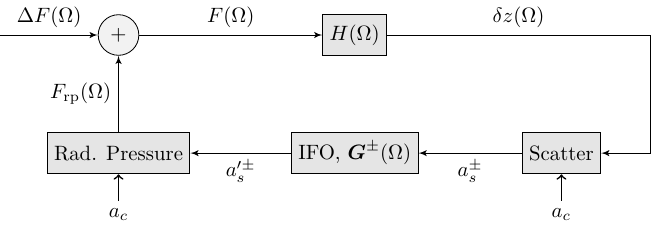}
\caption{A generic view of a closed-loop opto-mechanical transfer function for a suspended mirror
	with mechanical susceptibility $H(\Omega)$. Due to some motion $\delta z(\Omega)$ of a mirror the light is scattered from the carrier.
	The IFO plant describes the optical transfer function of the sidebands propagating through the interferometer and back
	to the mirror in question. The the interference of these then creates some radiation pressure force which is fed back
	into the mirror. Here ${a}_c$ is the carrier field at the mirror in question. }
\label{fig:os_block}
\end{figure}

\subsubsection*{Full steady state optical spring}
To compute the full response of a suspended mirror
we have to consider the propagation
of the sidebands generated when the mirror moves,
through the rest of the interferometer, and back to the mirror. This
process of scattering and feedback
is represented by the block diagram in figure~\ref{fig:os_block}.
Here we have some force $F(\Omega)$ acting on a
mirror with mechanical susceptibility $H(\Omega)$.
The motion $\delta z(\Omega)$ combined with the incident
carrier field ${a}_c$ scatters light into
the sidebands ${a}_s^\pm$. The IFO plant is the optical
transfer functions from the port leaving the suspended
mirror to the incoming port. Lastly these transformed
sidebands, ${a}'^\pm_s = G^\pm(\Omega) {a}_s^\pm$, are
combined again with the carrier field to compute the
radiation pressure force $F_\mathrm{rp}(\Omega)$ along with an external
excitation $\Delta F(\Omega)$ to feedback into the mirror.

To illustrate this in more detail only a single carrier with a pair
of sidebands describing a modulation at a
frequency $\Omega$ will be considered. This optical field is
incident on a perfectly reflective suspended mirror.
The incident and reflected optical fields are then:
\begin{align}
a^{'\pm}_s &= G^\pm(\Omega) a_s^\pm \label{eq:os_a'}, \\
a^\pm_s &= (a'^\pm_s + \I k Z^\pm a_c) \label{eq:os_a}.
\end{align}
Here the mirror displacement is $\delta z(\pm \Omega) \equiv Z^+ \equiv Z^{-\ast}$.
Removing the circular dependence of the fields with \ref{eq:os_a'} and \ref{eq:os_a}
\begin{equation}
a^{'\pm}_s = \dfrac{\I k G^\pm Z^\pm a_c}{1 - G^\pm}. \label{eq:inc_field}
\end{equation}
The radiation pressure force exerted on the mirror is
\begin{equation}
F_\mathrm{rp}(\Omega) = \dfrac{4}{c}\left[a_s^{'+}a_c^\ast + a_s^{'-\ast} a_c\right] \label{eq:os_force},
\end{equation}
substituting into~\ref{eq:inc_field} the force is seen to be
\begin{align}
F_\mathrm{rp}(\Omega) &= \dfrac{4\I k P_c}{c}  \left[ \dfrac{G^+ -
                        G^{-\ast}}{ 1 - r G^+ - r G^{-\ast} + R G^+
                        G^{-\ast}} \right] z \equiv \kappa(\Omega) z. \label{eq:opt_spring_force}
\end{align}
This shows that the radiation pressure force is linearly dependent on
$z$ for an arbitrary interferometer layout described by $G^\pm$.
The complex valued scaling factor, $\kappa(\Omega)$,
represents how the dynamic response of the suspended mirror is altered.
Those terms independent of $\Omega$ define the stiffness of the optical spring.
Terms $\propto \Omega$ describe any damping, $c_\mathrm{opt}$ being the optical damping coefficient:
\begin{equation}
\kappa(\Omega) = k_\mathrm{opt} + \I \Omega c_\mathrm{opt}(\Omega) + \mathcal{O}(\Omega^2). \label{eq:kappa}
\end{equation}
Higher order terms can also be significant, depending on the optical feedback, and can
alter the inertial behaviour by introducing terms $\propto \Omega^2$. Such manipulation of
the opto-mechanical coupling here can be exploited to improve the sensitivity
of gravitational wave detectors~\cite{PhysRevLett.113.151102}.

The above analysis is applicable in the case of a single optical field. If there are multiple optical fields of comparable amplitude, the sum of the multiple radiation pressure forces must be considered to compute the overall value of $\kappa$.
The result \eqref{eq:opt_spring_force} is only applicable
when fields are incident on a single side of a perfectly reflective mirror.
As can be imagined, analytical calculation of the spring dynamics in more detailed cases
such as a 50:50 suspended beam splitter, multiple suspended optics, or multiple carrier frequencies
with higher order spatial modes. Tools such as \Finesse take all these effects into account to ease studying such systems.

\subsubsection*{Optical spring in a cavity}

\begin{figure}[htb]
\centering
\includegraphics{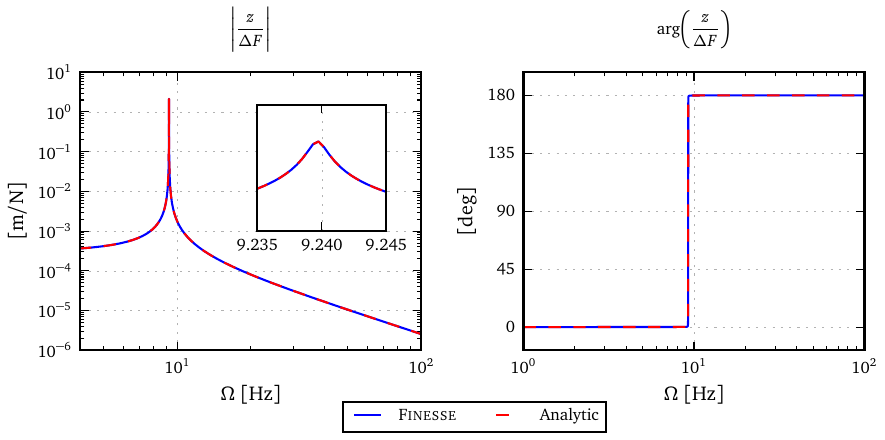}
\caption{
	Analytical v.s. numerical (\Finesse) comparison of an optical spring. This is for a fixed
	input field and suspended (free-mass) end mirror. A force is applied to the end mirror and shown
	is the force-to-displacement of the end mirror transfer function. Inset plot shows zoomed
	region around the peak which shows a good agreement with the peak shape and position.
}
\label{fig:os_comparison}
\end{figure}

The simplest case of an optical spring we can consider is that of an two-mirror optical cavity
with a single suspended mirror. The optical spring constant for this setup can be determined using equation~\ref{eq:opt_spring_force}.
Here $G^\pm$ are the optical transfer functions for the upper and lower sidebands through one round-trip of the cavity
after being created at the suspended mirror. For this example we will take the input mirror to be fixed and the end
mirror to be a suspended free mass.
We determine $G^\pm$ by starting from where the sidebands are created at the suspended mirror, whose reflectivity is $r_2$.
These then propagate along the cavity length $L$ twice with a reflection from the fixed input mirror, with reflectivity $r_1$,
before returning to the end mirror. In total the propagation is
\begin{equation}
G^\pm = r_1 e^{\mp\I 2 \frac{\Omega}{c}L} e^{\I 2\phi}.
\end{equation}
Here $\phi$ is some detuning of the input mirror position.
Substituting this into \eqref{eq:opt_spring_force} we find:
\begin{equation}
\dfrac{G^+ - G^{-\ast}}{ 1 - r_2 G^+ - r_2 G^{-\ast} + R_2 G^+ G^{-\ast}} = \dfrac{\I 2 r_1 e^{-\I 2 \frac{\Omega}{c}L} \sin(2\phi)}{1 + R_1 R_2 e^{-\I 4 \frac{\Omega}{c}L} - 2 r_2 r_1 e^{-\I 2 \frac{\Omega}{c}L} \cos(2\phi)}
\end{equation}
and:
\begin{equation}
\kappa(\Omega) = -\dfrac{8 k P_c r_1 r_2 \sin(2\phi)}{c} \dfrac{e^{-\I 2 \frac{\Omega}{c}L}}{1 + R_1 R_2 e^{-\I 4 \frac{\Omega}{c}L} - 2 r_2 r_1 e^{-\I 2 \frac{\Omega}{c}L} \cos(2\phi)} \label{eq:k_opt_fp}.
\end{equation}
When the cavity is on resonance, $\phi=0$, we see no optical spring when there is no power.
Likewise, in the DC limit $\Omega, \rightarrow 0$, we find an agreement with \eqref{eq:fabry_perot_Pcirc}.
Shown in figure~\ref{fig:os_comparison} is and example of a force-to-displacement
transfer function for the suspended end mirror when a force is applied.

\subsection{Finesse examples}

\subsubsection{Optical Spring}

\epubtkImage{fexample_optical_sprinf.png}{%
  \begin{figure}[htbp]
    \centerline{\includegraphics[width=0.9\textwidth]{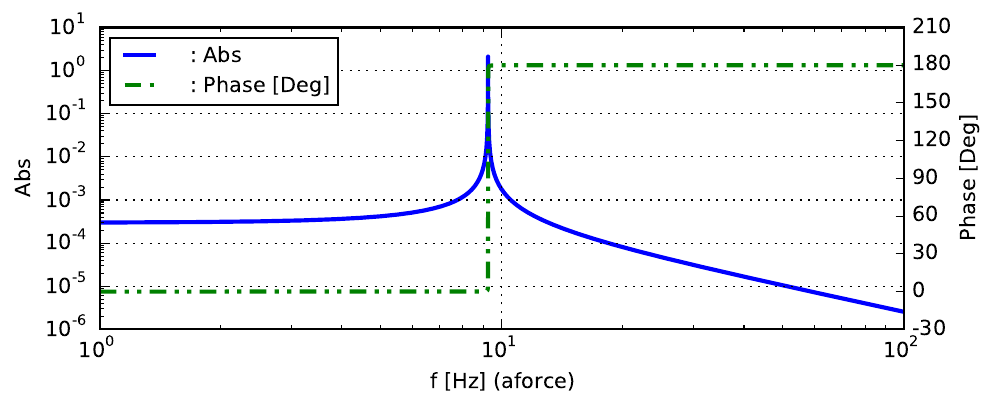}}
    \caption{\Finesse example: Fabry-Perot cavity with an optical
      spring. The two traces show the amplitude and phase of the
      mechanical response of one cavity mirror to an exciting
      force. The resonance feature close to 10\,Hz is the result of
      the opto-mechanical coupling of the mirror with the cavity field.}
    \label{fig:fexample_optical_spring}
\end{figure}}

\noindent
A simple example for an optical spring in a two mirror cavity with
suspended mirrors. The cavity is slightly detuned which is required
for creating the spring. The output is the motion of a mirror while
it is excited with a force of constant amplitude. The resonance
feature shown in Figure~\ref{fig:fexample_optical_spring} is the
result of the opto-mechanical coupling of the mirror with the
cavity field.

\vspace{3mm}\noindent
{\small
\textbf{Finesse input file for `optical spring'}
{\renewcommand{\baselinestretch}{.8}

\nopagebreak
\tt
\noindent
\mbox{}\textbf{\textcolor{RoyalBlue}{l}}\ l1\ \textcolor{Purple}{1}\ \textcolor{Purple}{0}\ n1 \\
\mbox{}\textbf{\textcolor{RoyalBlue}{m}}\ ITM\ \textcolor{Purple}{0.99}\ \textcolor{Purple}{0.01}\ \textcolor{Purple}{0}\ n1\ n2 \\
\mbox{}\textbf{\textcolor{RoyalBlue}{s}}\ s1\ \textcolor{Purple}{1}\ n2\ n3 \\
\mbox{}\textbf{\textcolor{RoyalBlue}{m}}\ ETM\ \textcolor{Purple}{1}\ \textcolor{Purple}{0}\ \textcolor{BrickRed}{-}\textcolor{Purple}{0.048}\ n3\ n4 \\
\mbox{}\textcolor{Gray}{\%\ Set\ the\ mass\ of\ the\ ETM\ in\ kg} \\
\mbox{}\textcolor{Blue}{attr}\ ETM\ mass\ \textcolor{Purple}{1}\  \\
\mbox{}\textcolor{Gray}{\%\ Here\ we\ are\ applying\ a\ force\ to\ the\ end\ mirror} \\
\mbox{}\textbf{\textcolor{RoyalBlue}{fsig}}\ aforce\ ETM\ Fz\ \textcolor{Purple}{1}\ \textcolor{Purple}{0}\ \textcolor{Purple}{1} \\
\mbox{}\textcolor{Gray}{\%\ A\ detector\ for\ force-to-motion\ transfer\ function} \\
\mbox{}xd\ ETMz\ ETM\ z \\
\mbox{}\textcolor{Gray}{\%\ scanning\ the\ frequency\ of\ the\ force\ } \\
\mbox{}\textbf{\textcolor{Red}{xaxis}}\ aforce\ f\ log\ \textcolor{Purple}{1}\ \textcolor{Purple}{100}\ \textcolor{Purple}{100000} \\
\mbox{}\textbf{\textcolor{Red}{yaxis}}\ log\ abs\textcolor{BrickRed}{:}deg \\
\mbox{}

}}

\subsubsection{Homodyne detector and squeezed light}

\epubtkImage{fexample_homodyne.png}{%
  \begin{figure}[htbp]
    \centerline{\includegraphics[width=0.9\textwidth]{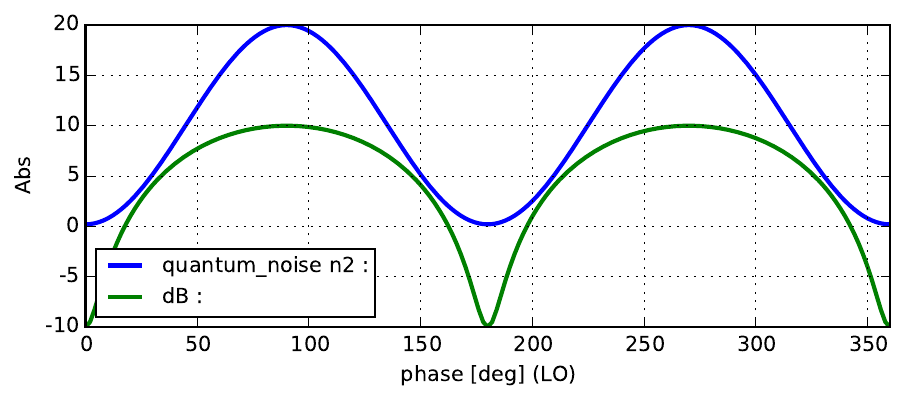}}
    \caption{\Finesse example: Homodyne detector with a squeezed light
      input. The blue trace shows the quantum noise in units of $\hbar
    f$ with a range of 0.2 to 20, compared to a quantum noise of $2
    \hbar f$ for an unsqueezed source. The green trace instead has the
  units 'dB' and shows an effective squeezing level, inferred from
  the detected quantum noise.}
    \label{fig:fexample_homodyne}
\end{figure}}

\noindent
A laser and a squeezed light source are mixed with a beam splitter and
then detected with a homodyne detector. The nominal quantum noise
of an un-squeezed light field in the units of the blue trace are $2 \hbar
f$. The squeezing level of the squeezed light source is $10$\,dB,
which means that the noise in one quadrature is 10 times lower than
this whereas the other quadrature should be 10 times higher. With the
phase of the local oscillator the homodyne detector can be tuned to
measure the different quadratures.
The green trace shows a computation of an effective squeezing level
from the detected quantum noise using the Schottky equation.

\vspace{3mm}\noindent
{\small
\textbf{Finesse input file for `homodyne detector and squeezed light'}
{\renewcommand{\baselinestretch}{.8}

\nopagebreak
\tt
\noindent
\mbox{}\textbf{\textcolor{RoyalBlue}{l}}\ LO\ \textcolor{Purple}{1}\ \textcolor{Purple}{0}\ n1 \\
\mbox{}\textbf{\textcolor{RoyalBlue}{s}}q\ sqz\ \textcolor{Purple}{0}\ \textcolor{Purple}{10}\ \textcolor{Purple}{0}\ n4 \\
\mbox{}\textbf{\textcolor{RoyalBlue}{bs}}\ bs1\ \textcolor{Purple}{0.5}\ \textcolor{Purple}{0.5}\ \textcolor{Purple}{0}\ \textcolor{Purple}{0}\ n1\ n2\ n3\ n4 \\
\mbox{}\textcolor{Gray}{\%\ homodyne\ detector\ attached\ to\ the\ two\ bs\ outputs} \\
\mbox{}qhd\ quantum$\_$noise\ \textcolor{Purple}{180}\ n2\ n3 \\
\mbox{}\textcolor{Gray}{\%\ what\ noise\ frequency\ (Omega)\ do\ we\ plot} \\
\mbox{}\textbf{\textcolor{RoyalBlue}{fsig}}\ noise\ \textcolor{Purple}{1} \\
\mbox{}\textcolor{Gray}{\%\ Output\ noise\ in\ units\ of\ hbar*f} \\
\mbox{}\textbf{\textcolor{Red}{scale}}\ PSD$\_$hf\ quantum$\_$noise \\
\mbox{}\textcolor{Gray}{\%\ varying\ the\ LO\ phase\ } \\
\mbox{}\textbf{\textcolor{Red}{xaxis}}\ LO\ phase\ lin\ \textcolor{Purple}{0}\ \textcolor{Purple}{360}\ \textcolor{Purple}{180} \\
\mbox{}\textcolor{Gray}{\%\ compute\ the\ squeezing\ level\ } \\
\mbox{}\textbf{\textcolor{Red}{set}}\textcolor{ForestGreen}{\ QN$\_$re}\ quantum$\_$noise\ re \\
\mbox{}\textbf{\textcolor{Red}{func}}\textcolor{ForestGreen}{\ dB}\ \textcolor{BrickRed}{=}\ \textcolor{Purple}{10}\ \textcolor{BrickRed}{*}\ log10\textcolor{BrickRed}{((}\textcolor{ForestGreen}{\$QN$\_$re}\ \textcolor{BrickRed}{+}\ \textcolor{Purple}{1E-20}\textcolor{BrickRed}{)/}\textcolor{Purple}{2}\textcolor{BrickRed}{)} \\
\mbox{}

}}

\subsubsection{Quantum-noise limited interferometer sensitivity}

\epubtkImage{fexample_sensitivity.png}{%
  \begin{figure}[htbp]
    \centerline{\includegraphics[width=0.9\textwidth]{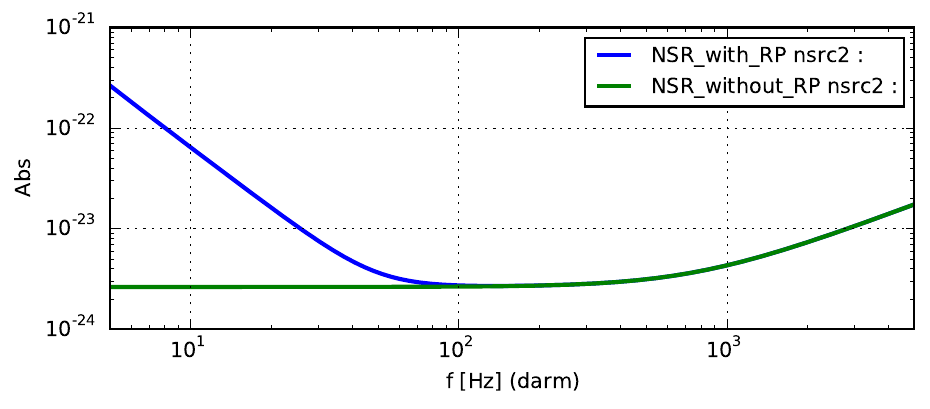}}
    \caption{\Finesse example: Quantum limited sensitivity of a simplified model of an
      Advanced LIGO interferometer. The interferometer setup is
      similar to a broadband RSE configuration of LIGO. The blue trace shows the
      full quantum-noise-limited sensitivity. For comparison the green
    trace shows the shot-noise-limited sensitivity.}
    \label{fig:fexample_sensitivity}
\end{figure}}

\noindent
This example shows the quantum-noise limited sensitivity of an
advanced detectors. See Appendix~\ref{sec:ALOL} for the optical
layout of the detector and Section~\ref{sec:DRFPMI} for more details about the
interferometer operation.
The model is loosely based on the Advanced LIGO design file and thus
we expect to see the peak sensitivity around $100$\,Hz at a sensitivity
of about $10^{-23}/\sqrt{\rm Hz}$. We can see the both the `qnoised' and `qshot'
detectors agree at high frequencies, where the sensitivity is purely
limited by shot noise. At low frequencies the two traces differ
because only `qnoised' takes into account the radiation pressure effects.

The \Finesse input file for this example is more complex than for
other examples because it contains a more complex interferometer setup
and uses relatively advanced concepts such as setting mechanical
transfer function. See Appendix~\ref{sec:finesse} for more information on
\Finesse and where to find the documentation, such as the syntax
reference, required to follow this example.

\vspace{3mm}\noindent
{\small
\textbf{Finesse input file for `quantum-limited interferometer sensitivity'}
{\renewcommand{\baselinestretch}{.8}

\nopagebreak
\tt
\noindent
\mbox{}\textbf{\textcolor{RoyalBlue}{l}}\ l1\ \textcolor{ForestGreen}{\$Pin}\ \textcolor{Purple}{0}\ nin \\
\mbox{}\textbf{\textcolor{RoyalBlue}{s}}\ s1\ \textcolor{Purple}{0}\ nin\ nprc1 \\
\mbox{}\textcolor{Gray}{\#\ Power\ recycling\ mirror} \\
\mbox{}\textbf{\textcolor{RoyalBlue}{m1}}\ prm\ \textcolor{ForestGreen}{\$prmT}\ \textcolor{Purple}{37.5u}\ \textcolor{Purple}{90}\ nprc1\ nprc2 \\
\mbox{}\textbf{\textcolor{RoyalBlue}{s}}\ \ prc\ \textcolor{ForestGreen}{\$lprc}\ nprc2\ nbsin \\
\mbox{}\textcolor{Gray}{\#\ Central\ beamsplitter} \\
\mbox{}\textbf{\textcolor{RoyalBlue}{bs}}\ bs1\ \textcolor{BrickRed}{.}\textcolor{Purple}{5}\ \textcolor{BrickRed}{.}\textcolor{Purple}{5}\ \textcolor{Purple}{0}\ \textcolor{Purple}{45}\ nbsin\ n0y\ n0x\ nbsout \\
\mbox{}\textcolor{Gray}{\#\ X-arm} \\
\mbox{}\textbf{\textcolor{RoyalBlue}{s}}\ ichx\ \textcolor{ForestGreen}{\$lmichx}\ n0x\ n1x \\
\mbox{}\textbf{\textcolor{RoyalBlue}{m1}}\ itmx\ \textcolor{ForestGreen}{\$itmT}\ \textcolor{Purple}{37.5u}\ \textcolor{Purple}{90}\ n1x\ n2x \\
\mbox{}\textbf{\textcolor{RoyalBlue}{s}}\ armx\ \textcolor{ForestGreen}{\$Larm}\ n2x\ n3x \\
\mbox{}\textbf{\textcolor{RoyalBlue}{m1}}\ etmx\ \textcolor{Purple}{5u}\ \textcolor{Purple}{37.5u}\ \textcolor{Purple}{89.999875}\ n3x\ n4x \\
\mbox{}\textcolor{Blue}{attr}\ itmx\ mass\ \textcolor{ForestGreen}{\$Mtm}\ zmech\ sus1 \\
\mbox{}\textcolor{Blue}{attr}\ etmx\ mass\ \textcolor{ForestGreen}{\$Mtm}\ zmech\ sus1 \\
\mbox{}\textcolor{Gray}{\#\ Y-arm} \\
\mbox{}\textbf{\textcolor{RoyalBlue}{s}}\ \ ichy\ \textcolor{ForestGreen}{\$lmichy}\ n0y\ n1y \\
\mbox{}\textbf{\textcolor{RoyalBlue}{m1}}\ itmy\ \textcolor{ForestGreen}{\$itmT}\ \textcolor{Purple}{37.5u}\ \textcolor{ForestGreen}{\$michy$\_$phi}\ n1y\ n2y \\
\mbox{}\textbf{\textcolor{RoyalBlue}{s}}\ \ army\ \textcolor{ForestGreen}{\$Larm}\ n2y\ n3y \\
\mbox{}\textbf{\textcolor{RoyalBlue}{m1}}\ etmy\ \textcolor{Purple}{5u}\ \textcolor{Purple}{37.5u}\ \textcolor{Purple}{0.000125}\ n3y\ n4y \\
\mbox{}\textcolor{Blue}{attr}\ itmy\ mass\ \textcolor{ForestGreen}{\$Mtm}\ zmech\ sus1 \\
\mbox{}\textcolor{Blue}{attr}\ etmy\ mass\ \textcolor{ForestGreen}{\$Mtm}\ zmech\ sus1 \\
\mbox{}\textcolor{Gray}{\#\ Signal\ recycling\ mirror} \\
\mbox{}\textbf{\textcolor{RoyalBlue}{s}}\ \ src\ \textcolor{ForestGreen}{\$lsrc}\ nbsout\ nsrc1 \\
\mbox{}\textbf{\textcolor{RoyalBlue}{m1}}\ srm\ \textcolor{ForestGreen}{\$srmT}\ \textcolor{Purple}{37.5u}\ \textcolor{ForestGreen}{\$srm$\_$phi}\ nsrc1\ nsrc2 \\
\mbox{}\textcolor{Gray}{\#\ Force-to-position\ transfer\ function\ for\ longitudinal\ } \\
\mbox{}\textcolor{Gray}{\#\ motions\ of\ test\ masses} \\
\mbox{}tf\ sus1\ \textcolor{Purple}{1}\ \textcolor{Purple}{0}\ p\ \textcolor{ForestGreen}{\$mech$\_$freq}\ \textcolor{ForestGreen}{\$mech$\_$Q} \\
\mbox{}\textbf{\textcolor{Red}{const}}\textcolor{ForestGreen}{\ mech$\_$freq}\ \textcolor{Purple}{1}\ \  \\
\mbox{}\textbf{\textcolor{Red}{const}}\textcolor{ForestGreen}{\ mech$\_$Q}\ \ \ \ 1M\ \textcolor{Gray}{\#\ Guess\ for\ suspension\ Q\ factor} \\
\mbox{}\textcolor{Gray}{\#\ offsets\ for\ DC\ readout:\ 100mW\ =\ michy$\_$phi\ 0.07\ OR\ darm$\_$phi\ .00025} \\
\mbox{}\textbf{\textcolor{Red}{const}}\textcolor{ForestGreen}{\ michy$\_$phi}\ \textcolor{Purple}{0}\ \  \\
\mbox{}\textbf{\textcolor{Red}{const}}\textcolor{ForestGreen}{\ darm$\_$phi}\ \ \textcolor{BrickRed}{.}\textcolor{Purple}{00025} \\
\mbox{}\textbf{\textcolor{Red}{const}}\textcolor{ForestGreen}{\ Larm}\ \textcolor{Purple}{3995} \\
\mbox{}\textbf{\textcolor{Red}{const}}\textcolor{ForestGreen}{\ itmT}\ \textcolor{Purple}{0.014} \\
\mbox{}\textbf{\textcolor{Red}{const}}\textcolor{ForestGreen}{\ srmT}\ \textcolor{Purple}{0.2} \\
\mbox{}\textbf{\textcolor{Red}{const}}\textcolor{ForestGreen}{\ prmT}\ \textcolor{Purple}{0.03} \\
\mbox{}\textbf{\textcolor{Red}{const}}\textcolor{ForestGreen}{\ Pin}\ \ \textcolor{Purple}{125}\  \\
\mbox{}\textbf{\textcolor{Red}{const}}\textcolor{ForestGreen}{\ Mtm}\ \ \textcolor{Purple}{40} \\
\mbox{}\textbf{\textcolor{Red}{const}}\textcolor{ForestGreen}{\ srm$\_$phi}\ \textcolor{BrickRed}{-}\textcolor{Purple}{90}\  \\
\mbox{}\textbf{\textcolor{Red}{const}}\textcolor{ForestGreen}{\ lmichx}\ \textcolor{Purple}{4.5} \\
\mbox{}\textbf{\textcolor{Red}{const}}\textcolor{ForestGreen}{\ lmichy}\ \textcolor{Purple}{4.45} \\
\mbox{}\textbf{\textcolor{Red}{const}}\textcolor{ForestGreen}{\ lprc}\ \ \ \textcolor{Purple}{53} \\
\mbox{}\textbf{\textcolor{Red}{const}}\textcolor{ForestGreen}{\ lsrc}\ \ \ \textcolor{Purple}{50.525}\  \\
\mbox{}\textcolor{Gray}{\#\ A\ squeezed\ source\ could\ be\ injected\ into\ the\ dark\ port} \\
\mbox{}\textbf{\textcolor{RoyalBlue}{s}}q\ sq1\ \textcolor{Purple}{0}\ \textcolor{Purple}{0}\ \textcolor{Purple}{90}\ nsrc2 \\
\mbox{}\textcolor{Gray}{\#\ Differentially\ modulate\ the\ arm\ lengths} \\
\mbox{}\textbf{\textcolor{RoyalBlue}{fsig}}\ darm\ \ armx\ \textcolor{Purple}{1}\ \textcolor{Purple}{0} \\
\mbox{}\textbf{\textcolor{RoyalBlue}{fsig}}\ darm2\ army\ \textcolor{Purple}{1}\ \textcolor{Purple}{180} \\
\mbox{}\textcolor{Gray}{\#\ Output\ the\ full\ quantum\ noise\ limited\ sensitivity} \\
\mbox{}qnoisedS\ NSR$\_$with$\_$RP\ \ \ \ \textcolor{Purple}{1}\ \textcolor{ForestGreen}{\$fs}\ nsrc2 \\
\mbox{}\textcolor{Gray}{\#\ Output\ just\ the\ shot\ noise\ limited\ sensitivity} \\
\mbox{}qshotS\ \ \ NSR$\_$without$\_$RP\ \textcolor{Purple}{1}\ \textcolor{ForestGreen}{\$fs}\ nsrc2 \\
\mbox{}\textcolor{Gray}{\#\ We\ could\ also\ display\ the\ quantum\ noise\ and\ the\ signal\ } \\
\mbox{}\textcolor{Gray}{\#\ separately\ by\ uncommenting\ these\ two\ lines.} \\
\mbox{}\textcolor{Gray}{\#\ qnoised\ noise\ \ \ \ \$fs\ nsrc2} \\
\mbox{}\textcolor{Gray}{\#\ pd1\ \ \ \ \ signal\ 1\ \$fs\ nsrc2} \\
\mbox{}\textbf{\textcolor{Red}{xaxis}}\ darm\ f\ log\ \textcolor{Purple}{5}\ 5k\ \textcolor{Purple}{1000} \\
\mbox{}\textbf{\textcolor{Red}{yaxis}}\ log\ abs \\
\mbox{}

}}

\newpage
\section{Advancing the interferometer layout}
\label{sec:advanced}
The first generation of interferometric gravitational wave detectors
was limited in the upper-frequency band by shot noise, one
manifestation of the quantum noise of the laser light, see
Section~\ref{sec:quantum_noise}. To improve
the ratio between gravitational-wave signal and shot noise we must
increase the light energy stored in the interferometer arms. This
can be achieved in several ways, for example, by increasing the
arms' length or by increasing the injected laser power. The lengths
of the arms is typically limited by the associated costs of the
building the infrastructure. High-power lasers are used; however do not
come near the power levels required for the anticipated sensitivity.
For example, the design sensitivity of Advanced LIGO requires a light
power or several hundred kilowatts in  the interferometer arms. The
Advanced LIGO laser can provide up to 200\,W of power, and
represents a state of the art system (for a CW laser with the required
stability in frequency, amplitude and beam profile)~\cite{kwee12}.

In order to increase the laser power inside the arms further
we can utilise the concept of resonant light enhancement in the
Fabry-Perot cavity: so-called advanced interferometer topologies
are created by introducing optical cavities to the
Michelson interferometer. In the following we will briefly introduce
the most common concepts, which are used by modern
gravitational wave detectors today.

We have shown in Section~\ref{sec:Michelson} how the dark fringe
operating point allows to maximise the throughput of differential
signals (with respect to common mode noise), using the sideband
picture. Similarly we can compute the transfer functions of the
signal sidebands to illustrate the concepts behind the advanced
interferometer layout. The motivation for all the advanced concepts
shown below is
the improvement of the ratio between signal and shot noise. However,
we will ignore here the detailed computation of the shot noise and
quantum noise discussed in Section~\ref{sec:quantum_noise}. Instead
we will compute only the transfer functions of the signal to the
photo detector using the sideband picture. We will ignore radiation
pressure noise and shot-noise contributions from any light field but
the local oscillator. Thus the amplitude of the signal sidebands
in the detection port give a good figure of merit for the shot-noise
limited sensitivity of the detector.

\epubtkImage{michelsonPR01.png}{%
  \begin{figure}[htb]
    \centerline{\includegraphics{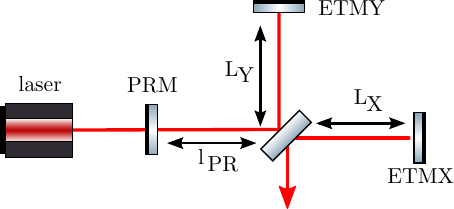}}
   \caption{Optical layout of a Michelson interferometer with arm
     power recycling.}
   \label{fig:michelsonPR_layout}
\end{figure}}
\subsection{Michelson interferometers with power recycling}
\label{sec:power_recycling}
The Michelson interferometer, when held on the dark fringe and
ignoring internal losses, reflects all the incoming light back into
the laser port; seen from the laser it acts like a highly reflective
mirror. It was soon realised we can utilise this fact to increase the light power
inside the interferometer: an additional mirror inter the input port,
the so-called power-recycling mirror (PRM), will generate an optical
cavity with the Michelson interferometer acting as a second `mirror'.
This scheme which is now called \textit{power recycling} was
first proposed in 1983 independently by
Schilling~\cite{billing83} and Drever~\cite{drever_pr}.
The newly formed cavity is often called \textit{power-recycling cavity}.
The optical layout
of a power-recycled Michelson interferometer is shown in
Figure~\ref{fig:michelsonPR_layout}. Figure~\ref{fig:MIPR_sidebands}
shows the amplitude of signal sidebands for different levels of power
recycling, as a function of the frequency of the signal. We will
compare this to similar plots for other techniques described below.
\epubtkImage{fexample_MIPR_sidebands.png}{%
  \begin{figure}[htb]
    \centerline{\includegraphics[width=0.9\textwidth]{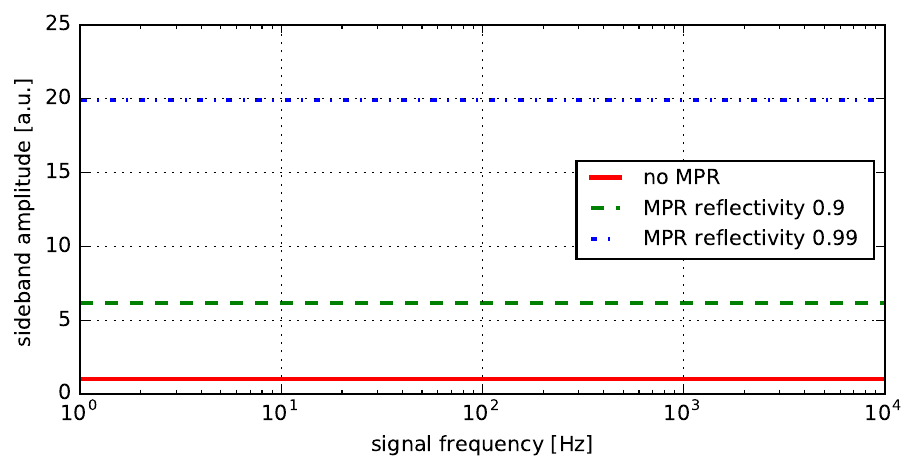}}
   \caption{This graph shows the signal sideband amplitude for a
     differential arm length change, as detected in the anti-symmetric
     output port, as a function of the frequency of the signal. The
     solid red trace at an amplitude of $1$ refers to the case without
   power recycling. The other two traces show the increased amplitude
   for different reflectivity's of the power-recoiling mirror. Compare
 this plot also with Figures~\ref{fig:MIcav_sidebands} and~\ref{fig:MISR_sidebands} below.}
   \label{fig:MIPR_sidebands}
\end{figure}}

As we have discussed in Section~\ref{sec:two_mirror2}, the power
circulating inside a cavity can be much higher than the injected
light power. The power enhancement is given by the finesse of the
cavity which is given by the optical losses in the interferometer
and the reflectivity of the power-recycling mirror. When the losses
inside the Michelson interferometer are negligible the cavity formed
by the Michelson and the power-recycling mirror is over-coupled and
the power enhancement in the interferometer arms, also called
\textit{power-recycling gain} computes as
\begin{equation}
G_{\rm PR}=\frac{4}{T_{\rm PRM}}\approx\frac{2\mathcal{F}}{\pi}
\end{equation}
with $\mathcal{F}$ the finesse of the power-recycling cavity.

When the optical losses can not be ignored the maximum power-recycling
gain can be reached by impedance matching, i.e.~setting the
transmission of the power-recycling mirror equal to the round trip
losses of the power-recycling cavity and the gain becomes
\begin{equation}
G_{\rm PR}=\frac{1}{T_{\rm PRM}}\approx\frac{\mathcal{F}}{\pi}
\end{equation}

The power in the signal sidebands proportional to the carrier power
and thus scales with the power-recycling gain as well. The amplitudes
plotted in Figure~\ref{fig:MIPR_sidebands} thus show values of
$\sqrt{4/0.1}\approx 6.32$ and $\sqrt{4/0.01}=20$.

Power-recycling has further advantages: the cavity effect can be used
to reduce beam jitter and to filter laser frequency noise. The
disadvantage is that another mirror position needs to be carefully
maintained by a feedback control system. In addition, the increase in
circulating power also increases the laser power with  the substrate
of the beam splitter which can cause thermal distortions which lead to
higher-optical losses. In practise this often limites the
achievable power-recycling gain.

\epubtkImage{michelson_cavities01.png}{%
  \begin{figure}[htbp]
    \centerline{\includegraphics{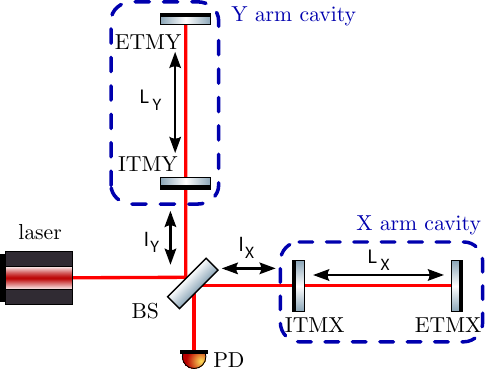}}
   \caption{Optical layout of a Michelson interferometer with arm cavities.}
   \label{fig:michelson_cavities_layout}
\end{figure}}

\subsection{Michelson interferometers with arm cavities}
\label{sec:MIFP}

Another way to employ cavities to enhance the light power circulating
in the interferometer arms is to place optical cavities into these
arms, as so-called \textit{arm cavities}, as shown in
Figure~\ref{fig:michelson_cavities_layout}.
This optical configuration sometimes referred to as
\textit{Fabry-Perot Michelson interferometer}. Similar to power-recycling
the finesse of the cavity determines the enhancement of the light.

\epubtkImage{fexample_MIcav_sidebands.png}{%
  \begin{figure}[htbp]
    \centerline{\includegraphics[width=0.9\textwidth]{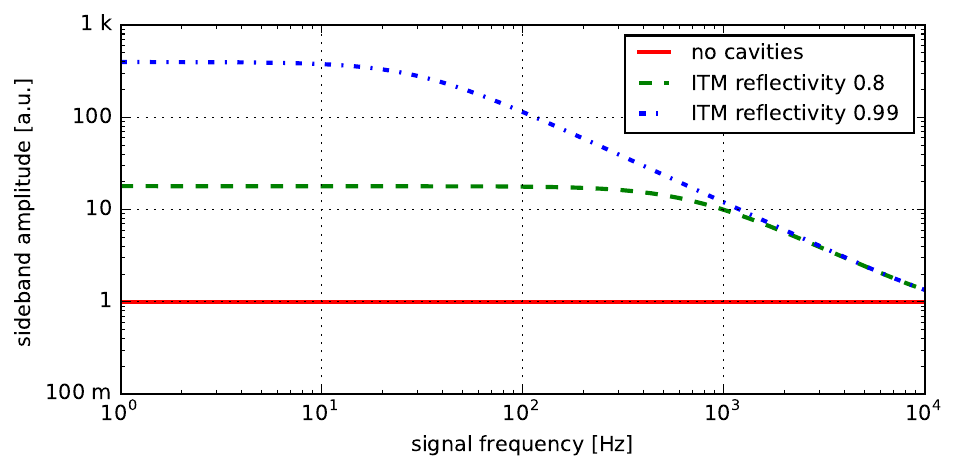}}
   \caption{ This graph shows the signal sideband amplitude for a
     Fabry-Perot Michelson interferometer. The signal is a
     differential arm length change detected in the anti-symmetric
     output port, as a function of the frequency of the signal. The
     solid red trace at an amplitude of $1$ refers to the case without
   arm cavities. The other two traces show the increased amplitude
   for different reflectivities of the cavities' input mirrors.
   Compare this also with Figures~\ref{fig:MIPR_sidebands} and~\ref{fig:MISR_sidebands}.}
   \label{fig:MIcav_sidebands}
\end{figure}}
The arm cavities have another effect on the detector sensitivity: they
affect not only the power of the circulating carrier field, but also
that sidebands generated by a length change. This results in a further
increase of the sensitivity for signals with a frequency within the
linewidth of the arm cavities but to a decrease in sensitivity regarding
signals with frequencies that fall outside the linewidth of the cavities.
This can be shown again very clearly with the sideband amplitudes
detected at the interferometer output as shown in
Figure~\ref{fig:MIcav_sidebands}. We can compare this results to the
power-recycling case (Figure~\ref{fig:MIPR_sidebands}): when the
reflectivity of the PRM and ITMs is set to $R=0.99$, the expected gain
for the carrier field inside the cavities must be the same and equal
to 400, assuming an over-coupled case. At low frequencies the signal
sidebands will experience the same enhancement, namely by a factor of
400 in power. Thus the total enhancement for the signal sidebands in
the Michelson with arm cavities is 16000, which gives the amplitude
of 400 shown for sideband amplitude in Figure~\ref{fig:MIcav_sidebands}.
Therefore the arm cavities also change the detector response function
in a way that limits the possible sensitivity increase.

The limited bandwidth of the arm cavities is a disadvantage when
compared to the power-recycling technique; however, the arm cavities have the
significant advantage of not increasing the light power in the beam
splitter substrate. In practise the two techniques are commonly used
together, with the finesse of the arm cavities and the reflectivity of
the power-recycling mirror the result of a trade-off analysis between
the bandwidth reduction of the arm cavities and the light power
increase in the beam splitter substrate. Such an optical layout is also
called \textit{power-recycled Fabry-Perot Michelson interferometer}.

\epubtkImage{michelson_cavities01.png}{%
  \begin{figure}[htbp]
    \centerline{\includegraphics{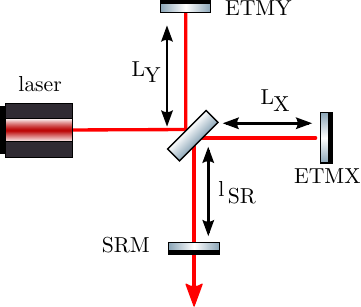}}
   \caption{Optical layout of a Michelson interferometer with signal recycling.}
   \label{fig:michelson_SR_layout}
\end{figure}}

\subsection{Signal recycling, dual recycling and
  resonant sideband extraction}
\label{sec:SRDRRSE}
Soon after the development of power recycling in which an additional
mirror is used to `recycle' the laser light leaving the Michelson
interferometer through the symmetric port, Brian Meers recognised that
it would be of interest to employ a similar technique employed in
the anti-symmetric port. In the ideal Michelson interferometer on the
dark fringe, the carrier light and the signal sidebands become
separated at the central beam splitter and leave the interferometer
though different ports. Meers suggested~\cite{meers88} the addition of a
\textit{signal-recycling mirror} at the anti-symmetric port, to
form a signal-recycling cavity with the Michelson
interferometer. In a similar manner to the power-recycling cavity the
signal-recycling cavity could resonantly enhance the light circulating
within, i.e.~the signal sidebands. The optical layout of a
signal-recycled Michelson interferometer is shown in
Figure~\ref{fig:michelson_SR_layout}.

It is somewhat counterintuitive
that placing a highly-reflective mirror in front of the photo detector
would increase the power detected on same photo detector. This
is because the signal sidebands are created within the interferometer,
and thus within the signal recycling cavity, by a parametric effect,
in which light is transferred from much larger reservoir, the carrier field.
Gerhard Heinzel provides, in Appendix D of his
thesis~\cite{Heinzel99}, a clear and compact mathematical overview of a
two-mirror cavity including this effect.

\epubtkImage{fexample_MISR_sidebands.png}{%
  \begin{figure}[htbp]
    \centerline{\includegraphics[width=0.9\textwidth]{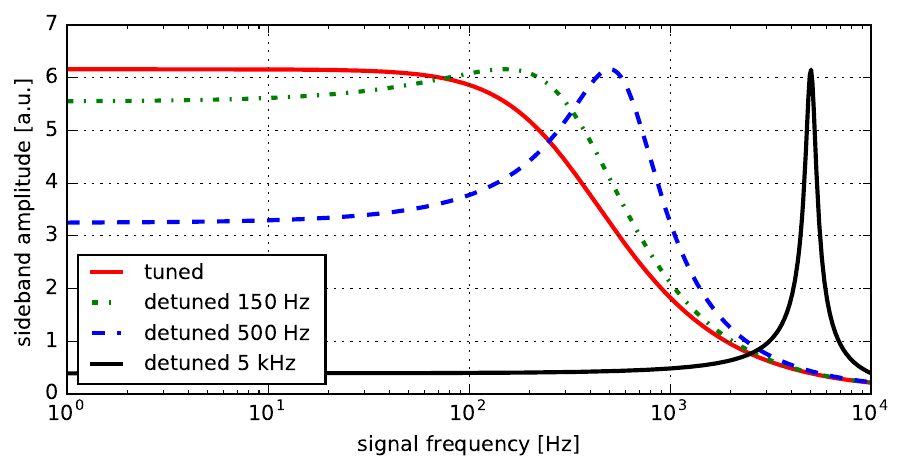}}
   \caption{
This graph shows the signal sideband amplitude for a Michelson
interferometer with different signal-recycling configurations.
for all 4 traces the reflectivity of the signal-recycling mirror was
set to $R=0.9$. The red trace shows the \textit{tuned} case in which
the signal-recycling cavity is resonant for the carrier light and
thus maximises signals around DC. The other red traces show different
detunings, microscopic offsets to the longitudinal positions of the
signal-recalling mirror. The maximum amplitude and bandwidth of the
trace is the same in all four cases, just the frequency of the peak
sensitivity is shifted by the detuning. Compare this the plots for
arm cavities in Figures~\ref{fig:MIcav_sidebands} and power
recycling, Figure~\ref{fig:MIPR_sidebands}.}
   \label{fig:MISR_sidebands}
\end{figure}}

When both recycling techniques are used together, power recycling for
enhancing the carrier power and signal recycling
for increasing the signal interaction time, the combination of the two
methods is called \textit{dual recycling}. It was actually the concept
of dual recycling which Meers proposed in 1988~\cite{meers88},
and this was demonstrated first as a table-top experiment by the
Glasgow group in 1991~\cite{Strain91}.

The combination of arm cavities and a signal-recycling mirror is
sometimes also called \textit{resonant sideband
  extraction}~\cite{Mizuno93}. The difference between signal-recycling
and resonant sideband extraction is that in the latter case
the arm cavities have a very high finesse and the signal-recycling
mirror is tuned to or near the anti-resonant
operating point, thus effectively increasing the bandwidth of the
detector for the signal sidebands. An analysis of the different
techniques can be found in the thesis of Jun
Mizuno~\cite{phd.Mizuno}. It is interesting to note that for all
variants of the signal recycling the total integrated gain remains
constant. For example, the areas under curves for the different
detunings shown in  Figure~\ref{fig:MISR_sidebands} are
constant\footnote{The tuned case is slightly special, because the
  integrated area is half compared to the others, because the plot
shows only the positive  half of the total linewidth seen by signal sidebands.}.
This means that signal-recycling is used to \textit{shape} the
response function of the detector with respect to the
signal-to-shot-noise ratio.

The main interferometer of an Advanced LIGO detector is based on a
Michelson interferometer with arm cavities plus power and signal
recycling. This configuration is most commonly called \textit{dual-recycled
  Fabry-Perot Michelson interferometer} even though the signal
recycling mirror is here used in the resonant sideband extraction mode, see
Figure~\ref{fig:DRFPMI_fields} for a schematic of this layout.

\epubtkImage{sagnac.png}{%
  \begin{figure}[htb]
    \centerline{\hspace{1em}\includegraphics{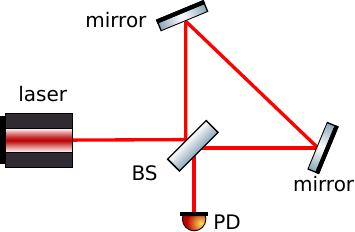}\hfill\includegraphics{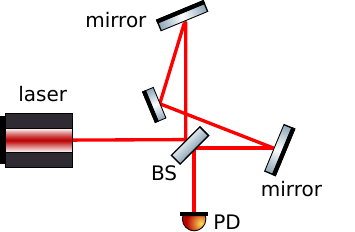}\hspace{1em}}
   \caption{The left sketch shows a typical layout of the original
     Sagnac interferometer: similar to the Michelson interferometer
     the injected light is split and recombined at a central beam
     splitter. However, unlike the Michelson the Sagnac has not two
     different interferometer arms, but the two split beams travel
     along the same path, in different direction. This makes the
     Sagnac interferometer insensitive to the actual path length,
     instead is sensitivity to rotation of the whole interferometer.
The sketch on the right shows a so-called \textit{zero-area} Sagnac
interferometer: an additional mirror is used so that the beam path is
folded reducing the effective circulated area (see text).}
   \label{fig:sagnac_layout}
\end{figure}}
\subsection{Sagnac interferometer}
\label{sec:sagnac}
Another interferometer type which has a similar-looking optical layout
to the Michelson interferometer is the \textit{Sagnac} interferometer,
see Figure~\ref{fig:sagnac_layout}. Originally proposed by Sagnac for
measuring rotation~\cite{Sagnac1913a, Sagnac1913b} it became of interested to the
gravitational-wave community as a possible alternative to the
Michelson interferometer: in 1995 successful experimental tests of a
zero-area Sagnac demonstrated a different mode of operation, in which
it becomes insensitive to rotation but sensitive to mirror
motion~\cite{sun96}.
Further investigations into the performance and technical limitations
of a Sagnac interferometric gravitational wave detector have been
undertaken~\cite{Mizuno97, Petrovichev98,Beyersdorf02} and
the community interest was renewed after understanding that the Sagnac
topology can be used as a speed-meter with the potential to suppress
radiation pressure noise in future detectors~\cite{Chen03,
  Danilishin04, wang13, voronchev14, danilishin15}.
\epubtkImage{example_sag_vs_mich_TF.png}{%
  \begin{figure}[htb]
    \centerline{\includegraphics[width=0.7\textwidth]{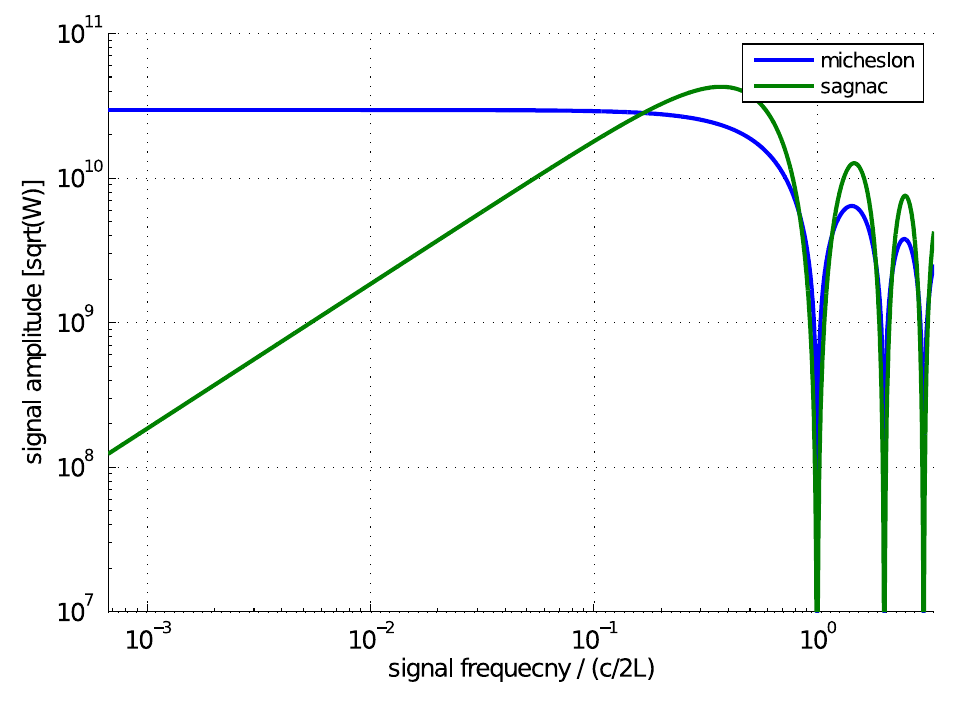}}
   \caption{Transfer function of a Michelson and Sagnac interferometer
     for a gravitational-wave signal to the main output channel. It
     can be seen that the Sagnac response falls off for lower
     frequencies and reached twice the peak response; however at
     relatively large frequency, in the case of Advanced LIGO the peak
   would be at f=$c/2L\approx37.5$\,kHz, i.e.~above the measurement window.}
   \label{fig:sag_v_mich}
\end{figure}}

\subsection{\Finesse examples}

\subsubsection{Michelson interferometer with arm cavities}

\epubtkImage{fexample_MIcav.png}{%
  \begin{figure}[htb]
    \centerline{\includegraphics[width=1\textwidth]{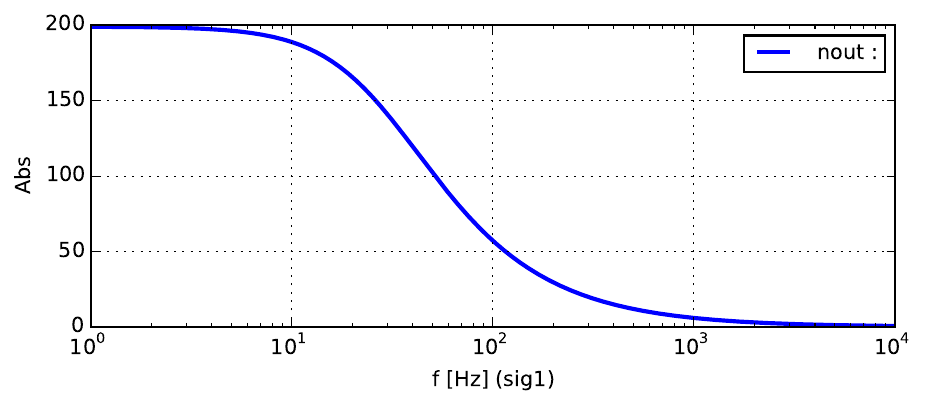}}
    \caption{\Finesse example: Michelson interferometer with arm
      cavities. The trace shows the signal sideband amplitude in the
      anti-symmetric port as a function of signal frequency.}
    \label{fig:fexample_MIcav}
\end{figure}}

\noindent
This example shows how to setup a Michelson interferometer, tune it to
the dark fringe and compute a transfer function from the differential
length change to the output signal, using the sideband amplitude for simplicity.

\vspace{3mm}\noindent
{\small
\textbf{Finesse input file for `Michelson interferometer with arm cavities'}
{\renewcommand{\baselinestretch}{.8}

\nopagebreak
\tt
\noindent
\mbox{} \\
\mbox{}\textbf{\textcolor{RoyalBlue}{laser}}\ l1\ \textcolor{Purple}{1}\ \textcolor{Purple}{0}\ \ n1\ \ \ \ \ \ \textcolor{Gray}{\%\ laser\ with\ P=1W\ at\ the\ default\ frequency} \\
\mbox{}\textbf{\textcolor{RoyalBlue}{space}}\ si1\ \textcolor{Purple}{1}\ n1\ n2\ \ \ \ \textcolor{Gray}{\%\ space\ of\ 1m\ length} \\
\mbox{}\textbf{\textcolor{RoyalBlue}{mirror}}\ MPR\ \textcolor{Purple}{0}\ \textcolor{Purple}{1}\ \textcolor{Purple}{0}\ n2\ n3 \\
\mbox{}\textbf{\textcolor{RoyalBlue}{space}}\ si2\ \textcolor{Purple}{50}\ n3\ n4\ \ \ \  \\
\mbox{} \\
\mbox{}\textbf{\textcolor{RoyalBlue}{bs}}\ b1\ \textcolor{Purple}{0.5}\ \textcolor{Purple}{0.5}\ \textcolor{Purple}{0}\ \textcolor{Purple}{0}\ n4\ nY1\ nX1\ nO1\ \textcolor{Gray}{\%\ 50:50\ beam\ splitter\ } \\
\mbox{}\textbf{\textcolor{RoyalBlue}{space}}\ \ slY\ \textcolor{Purple}{7}\ nY1\ nY2\ \ \  \\
\mbox{}\textbf{\textcolor{RoyalBlue}{space}}\ \ slX\ \textcolor{Purple}{7}\ nX1\ nX2\ \ \  \\
\mbox{}\textbf{\textcolor{RoyalBlue}{mirror}}\ ITMY\ \textcolor{Purple}{0.99}\ \textcolor{Purple}{0.01}\ \textcolor{Purple}{0}\ nY2\ nY3\ \textcolor{Gray}{\%\ Y\ input\ mirror,\ lossless} \\
\mbox{}\textbf{\textcolor{RoyalBlue}{mirror}}\ ITMX\ \textcolor{Purple}{0.99}\ \textcolor{Purple}{0.01}\ \textcolor{Purple}{90}\ nX2\ nX3\ \textcolor{Gray}{\%\ X\ input\ mirror,\ lossless} \\
\mbox{}\textbf{\textcolor{RoyalBlue}{space}}\ \ LY\ 4k\ nY3\ nY4\ \ \ \ \textcolor{Gray}{\%\ Y\ arm} \\
\mbox{}\textbf{\textcolor{RoyalBlue}{space}}\ \ LX\ 4k\ nX3\ nX4\ \ \ \ \textcolor{Gray}{\%\ X\ arm} \\
\mbox{}\textbf{\textcolor{RoyalBlue}{mirror}}\ ETMY\ \textcolor{Purple}{1}\ \textcolor{Purple}{0}\ \textcolor{Purple}{0}\ nY4\ dump\ \textcolor{Gray}{\%\ Y\ end\ mirror,\ lossless} \\
\mbox{}\textbf{\textcolor{RoyalBlue}{mirror}}\ ETMX\ \textcolor{Purple}{1}\ \textcolor{Purple}{0}\ \textcolor{Purple}{90}\ nX4\ dump\ \textcolor{Gray}{\%\ X\ end\ mirror,\ lossless,\ tuned\ for\ dark\ fringe} \\
\mbox{}\textbf{\textcolor{RoyalBlue}{space}}\ \ so1\ \textcolor{Purple}{1}\ nO1\ nout\  \\
\mbox{} \\
\mbox{}\textbf{\textcolor{RoyalBlue}{fsig}}\ sig1\ ETMX\ \textcolor{Purple}{100}\ \textcolor{Purple}{0} \\
\mbox{}\textbf{\textcolor{RoyalBlue}{fsig}}\ sig2\ ETMY\ \textcolor{Purple}{100}\ \textcolor{Purple}{180} \\
\mbox{}\textbf{\textcolor{RoyalBlue}{ad}}\ signal\ \textcolor{Purple}{100}\ nout\ \ \ \ \ \ \ \ \ \ \ \  \\
\mbox{} \\
\mbox{}\textbf{\textcolor{Red}{xaxis}}\ sig1\ f\ log\ \textcolor{Purple}{1}\ 10k\ \textcolor{Purple}{300} \\
\mbox{}\textbf{\textcolor{Red}{put}}\ signal\ f\ \textcolor{ForestGreen}{\$x1} \\
\mbox{}

}}

\subsubsection{Michelson interferometer with signal recycling}

\epubtkImage{fexample_MISR_sidebands.png}{%
  \begin{figure}[htbp]
    \centerline{\includegraphics[width=0.9\textwidth]{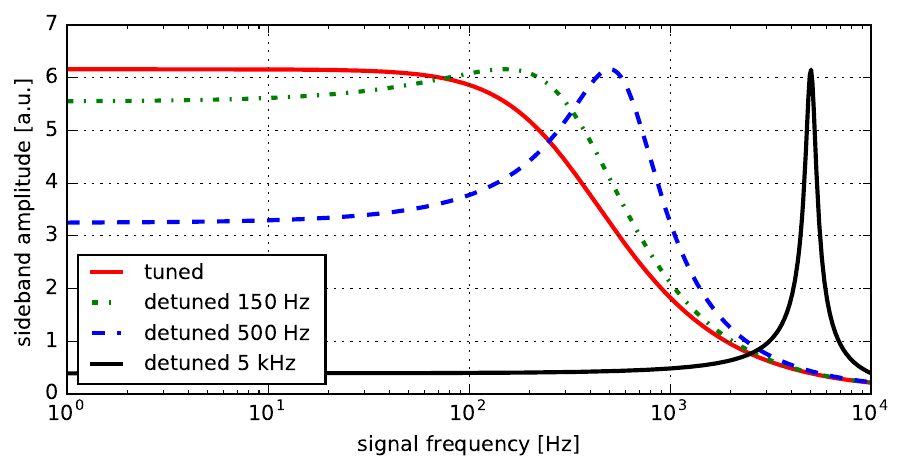}}
    \caption{\Finesse example: This graph shows the signal sideband
      amplitude for a Michelson interferometer with different
      signal-recycling configurations, see Figure~\ref{fig:MISR_sidebands}.}
    \label{fig:fexample_michelson_SR_sidebands}
\end{figure}}

\noindent
This example recreates the plot shown in
Figure~\ref{fig:MISR_sidebands}, the four traces show the transfer
function for a Michelson interferometer with different signal
recycling tunings.

\vspace{3mm}\noindent
{\small
\textbf{Finesse input file for `Michelson with signal recycling'}
{\renewcommand{\baselinestretch}{.8}

\nopagebreak
\tt
\noindent
\mbox{}\textbf{\textcolor{RoyalBlue}{laser}}\ l1\ \textcolor{Purple}{1}\ \textcolor{Purple}{0}\ \ n1\ \ \ \ \ \textcolor{Gray}{\%\ laser\ with\ P=1W\ at\ the\ default\ frequency} \\
\mbox{}\textbf{\textcolor{RoyalBlue}{space}}\ si1\ \textcolor{Purple}{1}\ n1\ n2\ \ \ \ \textcolor{Gray}{\%\ space\ of\ 1m\ length} \\
\mbox{}\textbf{\textcolor{RoyalBlue}{mirror}}\ MPR\ \textcolor{Purple}{0}\ \textcolor{Purple}{1}\ \textcolor{Purple}{0}\ n2\ n3 \\
\mbox{}\textbf{\textcolor{RoyalBlue}{space}}\ si2\ \textcolor{Purple}{50}\ n3\ n4\ \ \  \\
\mbox{} \\
\mbox{}\textbf{\textcolor{RoyalBlue}{bs}}\ b1\ \textcolor{Purple}{0.5}\ \textcolor{Purple}{0.5}\ \textcolor{Purple}{0}\ \textcolor{Purple}{0}\ n4\ nY1\ nX1\ nO1\ \textcolor{Gray}{\%\ 50:50\ beam\ splitter\ } \\
\mbox{}\textbf{\textcolor{RoyalBlue}{space}}\ \ slY\ \textcolor{Purple}{7}\ nY1\ nY2\ \ \  \\
\mbox{}\textbf{\textcolor{RoyalBlue}{space}}\ \ slX\ \textcolor{Purple}{7}\ nX1\ nX2\ \ \  \\
\mbox{}\textbf{\textcolor{RoyalBlue}{mirror}}\ ITMY\ \textcolor{Purple}{0}\ \textcolor{Purple}{1}\ \textcolor{Purple}{0}\ nY2\ nY3\ \textcolor{Gray}{\%\ Y\ input\ mirror,\ lossless} \\
\mbox{}\textbf{\textcolor{RoyalBlue}{mirror}}\ ITMX\ \textcolor{Purple}{0}\ \textcolor{Purple}{1}\ \textcolor{Purple}{90}\ nX2\ nX3\ \textcolor{Gray}{\%\ X\ input\ mirror,\ lossless} \\
\mbox{}\textbf{\textcolor{RoyalBlue}{space}}\ \ LY\ 4k\ nY3\ nY4\ \ \ \ \textcolor{Gray}{\%\ Y\ arm} \\
\mbox{}\textbf{\textcolor{RoyalBlue}{space}}\ \ LX\ 4k\ nX3\ nX4\ \ \ \ \textcolor{Gray}{\%\ X\ arm} \\
\mbox{}\textbf{\textcolor{RoyalBlue}{mirror}}\ ETMY\ \textcolor{Purple}{1}\ \textcolor{Purple}{0}\ \textcolor{Purple}{0}\ nY4\ dump\ \textcolor{Gray}{\%\ Y\ end\ mirror,\ lossless} \\
\mbox{}\textbf{\textcolor{RoyalBlue}{mirror}}\ ETMX\ \textcolor{Purple}{1}\ \textcolor{Purple}{0}\ \textcolor{Purple}{90}\ nX4\ dump\ \textcolor{Gray}{\%\ X\ end\ mirror,\ lossless,\ tuned\ for\ dark\ fringe} \\
\mbox{}\textbf{\textcolor{RoyalBlue}{space}}\ \ so1\ \textcolor{Purple}{50}\ nO1\ nO2 \\
\mbox{}\textcolor{Gray}{\%\ change\ the\ signal\ recycling\ mirror\ for\ th\ 4\ traces} \\
\mbox{}\textbf{\textcolor{RoyalBlue}{mirror}}\ MSR\ \textcolor{Purple}{0.9}\ \textcolor{Purple}{0.1}\ \textcolor{Purple}{0}\ nO2\ nO3\ \textcolor{Gray}{\%\ tuned} \\
\mbox{}\textcolor{Gray}{\%mirror\ MSR\ 0.9\ 0.1\ 0.73026\ nO2\ nO3\ \%\ detuned\ to\ 150\ Hz} \\
\mbox{}\textcolor{Gray}{\%mirror\ MSR\ 0.9\ 0.1\ 2.4342\ nO2\ nO3\ \%\ detuned\ to\ 500\ Hz} \\
\mbox{}\textcolor{Gray}{\%mirror\ MSR\ 0.9\ 0.1\ 24.342\ nO2\ nO3\ \%\ detuned\ to\ 5\ kHz} \\
\mbox{}\textbf{\textcolor{RoyalBlue}{space}}\ so2\ \textcolor{Purple}{1}\ nO3\ nout\ \ \  \\
\mbox{} \\
\mbox{}\textbf{\textcolor{RoyalBlue}{fsig}}\ sig1\ ETMX\ \textcolor{Purple}{100}\ \textcolor{Purple}{0} \\
\mbox{}\textbf{\textcolor{RoyalBlue}{fsig}}\ sig2\ ETMY\ \textcolor{Purple}{100}\ \textcolor{Purple}{180} \\
\mbox{}\textbf{\textcolor{RoyalBlue}{ad}}\ signal\ \textcolor{Purple}{100}\ nout\ \ \ \ \ \ \ \ \ \ \ \  \\
\mbox{} \\
\mbox{}\textbf{\textcolor{Red}{xaxis}}\ sig1\ f\ log\ \textcolor{Purple}{1}\ 10k\ \textcolor{Purple}{600} \\
\mbox{}\textbf{\textcolor{Red}{put}}\ signal\ f\ \textcolor{ForestGreen}{\$x1} \\
\mbox{}

}}

\subsubsection{Sagnac interferometer}

\epubtkImage{fexample_Sagnac_sidebands.png}{%
  \begin{figure}[htbp]
    \centerline{\includegraphics[width=0.9\textwidth]{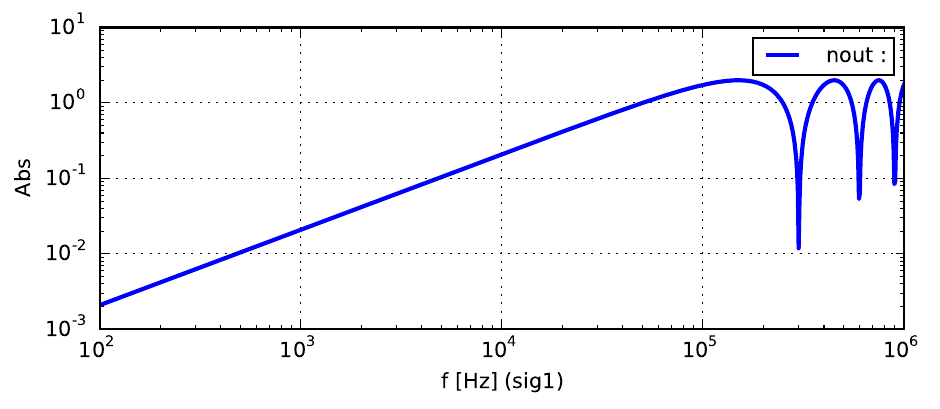}}
    \caption{\Finesse example: Frequency response of a Sagnac
      interferometer: transfer function from differential mirror
      position change to signal sideband amplitude in the main output port.}
    \label{fig:fexample_sagnac_sidebands}
\end{figure}}

\noindent
This example demonstrates how compute the frequency response of a
simple Sagnac interferometer.

\vspace{3mm}\noindent
{\small
\textbf{Finesse input file for Sagnac interferometer}
{\renewcommand{\baselinestretch}{.8}

\nopagebreak
\tt
\noindent
\mbox{} \\
\mbox{}\textbf{\textcolor{RoyalBlue}{laser}}\ l1\ \textcolor{Purple}{1}\ \textcolor{Purple}{0}\ \ n1\ \ \ \ \  \\
\mbox{}\textbf{\textcolor{RoyalBlue}{space}}\ s1\ \textcolor{Purple}{1}\ n1\ n2\ \ \ \  \\
\mbox{}\textbf{\textcolor{RoyalBlue}{bs}}\ b1\ \textcolor{Purple}{0.5}\ \textcolor{Purple}{0.5}\ \textcolor{Purple}{0}\ \textcolor{Purple}{0}\ n2\ nN1\ nE1\ nS1\ \textcolor{Gray}{\%\ 50:50\ beam\ splitter\ } \\
\mbox{}\textbf{\textcolor{RoyalBlue}{space}}\ \ LN\ 1k\ nN1\ nN2\ \ \ \ \textcolor{Gray}{\%\ north\ arm} \\
\mbox{}\textbf{\textcolor{RoyalBlue}{space}}\ \ LE\ 1k\ nE1\ nE2\ \ \ \ \textcolor{Gray}{\%\ east\ arm} \\
\mbox{}\textbf{\textcolor{RoyalBlue}{bs}}\ mN\ \textcolor{Purple}{1}\ \textcolor{Purple}{0}\ \textcolor{Purple}{0}\ \textcolor{Purple}{0}\ nN2\ nN3\ dump\ dump\ \textcolor{Gray}{\%\ north\ end\ mirror} \\
\mbox{}\textbf{\textcolor{RoyalBlue}{space}}\ \ link\ 1k\ nN3\ nE3\ \ \ \textcolor{Gray}{\%\ third\ arm} \\
\mbox{}\textbf{\textcolor{RoyalBlue}{bs}}\ mE\ \textcolor{Purple}{1}\ \textcolor{Purple}{0}\ \textcolor{Purple}{0}\ \textcolor{Purple}{0}\ nE2\ nE3\ dump\ dump\ \textcolor{Gray}{\%\ east\ end\ mirror} \\
\mbox{}\textbf{\textcolor{RoyalBlue}{space}}\ \ s2\ \textcolor{Purple}{1}\ nS1\ nout\  \\
\mbox{} \\
\mbox{}\textbf{\textcolor{RoyalBlue}{fsig}}\ sig1\ mE\ \textcolor{Purple}{100}\ \textcolor{Purple}{0} \\
\mbox{}\textbf{\textcolor{RoyalBlue}{fsig}}\ sig2\ mN\ \textcolor{Purple}{100}\ \textcolor{Purple}{180} \\
\mbox{}\textbf{\textcolor{RoyalBlue}{ad}}\ signal\ \textcolor{Purple}{100}\ nout\ \ \ \ \ \ \ \ \ \ \ \  \\
\mbox{} \\
\mbox{}\textbf{\textcolor{Red}{xaxis}}\ sig1\ f\ log\ \textcolor{Purple}{100}\ 1M\ \textcolor{Purple}{1000} \\
\mbox{}\textbf{\textcolor{Red}{put}}\ signal\ f\ \textcolor{ForestGreen}{\$x1} \\
\mbox{}\textbf{\textcolor{Red}{yaxis}}\ log\ abs \\
\mbox{}

}}

\newpage
\section{Interferometric Length Sensing and Control}
\label{sec:control}
In this section we introduce interferometers as length sensing
devices. In particular, we explain how the Fabry-Perot
interferometer and the Michelson interferometer can be used for
high-precision measurements and that both require a careful control of
the base length (which is to be measured) in order to yield their
large sensitivity. In addition, we briefly introduce the general
concepts of \emph{error signals}, \emph{transfer functions} and
relevant elements of \emph{control theory}, which are used to describe
most essential features of length sensing and control.

In addition to sensing and controlling the distances between the
components of an interferometer, alignment sensing and control is
required for correct operation.  While we do not deal with this aspect
in detail, all of the ideas we develop for length sensing and control
can be applied.  The essential differences are that split
photo-detectors are required to sense the relative angles of optical
wavefronts, and control is be means of actuators that are able to
adjust the angles of optical components.  For an introduction to the
essential ideas see Section~\ref{sec:HOM} for an introduction to the
relevant theory and~\cite{Morrison1994}, for details of a practical
implementation.

\subsection{An overview of the control problem}
\label{sec:control1}
A complete interferometer can have a large number of control loops for
the various mirrors and beam splitters, their suspension systems and
many other components, such as the laser, active vibration-isolation systems
etc.  For practical purposes these are usually divided into two broad
classes that are often considered separately in the design
process. These divisions reflect a degree of independence of the
various categories of control and simplify the design process by
allowing the problem to be split into a number of more easily
tractable design elements.

The set of control loops that obtain signals from the detection of
interference conditions or other properties of the light within the
interferometer, and act on the major optical components of the
interferometer to control those properties, is usually called {\em
  global control}. As an example, a description of the global control system
  of Virgo can be found in~\cite{arnaud05}.
On the other hand, loops that sense properties
associated with a single component, and act on that component are
called {\em local} loops.  A good example of local control is the
system employed to damp the rigid-body modes of a mirror suspension, for an
example from Advanced LIGO see~\cite{Aston2012}.
By `cooling' or quieting the motion of individual mirrors, the task
faced by the global control system can be simplified.
Further division of global interferometer control is frequently made
between systems that control longitudinal degrees of freedom, i.e.~relative
positions of the mirrors and beam splitters along the direction of
propagation of the light, and angular (alignment) control systems that
are designed to stabilise the pointing of components.

Due to the presence of strong nonlinearity throughout much of the
phase-space volume, there has been no attempt thus far to solve the
multi-dimensional control problem as a whole. At least up to the
present, the problem has been divided into several smaller parts, with
methods developed to deal with the particular details of each facet of
the system, and each stage of operation from completely uncontrolled
to held at the operating point -- a condition that is called `locked'.

This leads to yet another division: it is normal to separate the start-up phase i.e.~the
process called {\em acquisition of lock} from the stable running
condition (`in lock').  This split is motivated, at least in part, by
the consideration that signal sizes can differ greatly between the two
stages.  During acquisition electronic signals tend to be large --
corresponding to adjusting mirror positions by of order wavelengths,
or more.  By contrast, in operation the signals representing residual
motion in the sensitive frequency band may be 12 or more orders of magnitude smaller than the wavelength of the light, to
deliver the required measurement sensitivity\footnote{The typical light wavelength
is $\sim 10^{-6}\,$m while all ground-based interferometers built or planned
have target displacement noise spectral densities below $\sim 10^{-19}\,{\mathrm m}/\sqrt{\mathrm Hz}$ in a frequency
band of order 100\,Hz wide.}.

The jump in signal size between these two states is often dealt with by
switching gain levels or even substituting large parts of the control
system: starting with large-range but noisy methods for acquisition
and switching to low-noise, but small-range controls for operation.  A
good example of a more substantial transition is the {\em arm  length
  stabilization} (ALS) scheme in Advanced LIGO, which employs
additional lasers, mirror coatings and interferometric methods to
provide wavelength-range sensing during the acquisition
phase~\cite{Evans2009}.  When the long cavities (see Appendix~\ref{sec:ALOL}) are locked,
the control systems are switched over to the high-sensitivity, low-noise signals
derived from the main interferometer systems.

During acquisition of lock, the instantaneous operating point
frequently lies in a non-linear region of the control space.  Several
methods have been developed to cope with this problem.

The simplest approach, employed in the early interferometer prototypes, was to wait for a random
co-incidence of suitable values to occur then to catch the system
quickly enough to hold it in the desired state.  As the complexity of the interferometer topologies increased, and with that the number of degrees of freedom, the  probability of the desired state occurring in a conveniently short time became rare.  This led to the development of more sophisticated
techniques for the first long-baseline
interferometers.

As a first step it was realised that digital logic, implemented directly in electronics or as software, could be employed to identify when one or more degrees of freedom happened to fall close to the desired operating point, and to activate the relevant control loop.  This prevents false signals, frequently present in regions of phase space close to the desired operating point, being fed back to the actuators and perturbing the system. In Pound-Drever-Hall locking, for example, when the phase modulation sidebands pass through resonance in the cavity, the error signal has the opposite sign from that produced by a carrier resonance.  The acquisition process can be improved by enabling the control system only when the circulating power within the cavity has exceeded the maximum possible power that a sideband can produce.  By this means the control system is only activated close to the desired operating point, improving the chances of a successful lock.

A second way to improve matters is to linearise the behaviour, and so to increase the capture range: i.e.~the volume of
phase space within which the various control signals are valid.  This improvement can be accomplished by
normalising the relevant error signal according to some estimate of its slope, as measured by
another signal such as the circulating optical power.  As an example, the linear range in the Pound-Drever-Hall signal for locking a cavity may be extended by normalising with respect to the power within the cavity, as measured
by probing the light transmitted by the cavity.

Another approach, is to arrange for the first first locking to be in a region of phase space that is relatively smooth, compared to the
region in phase space surrounding the final operating point.  This was an enabling technique for
GEO\,600 when it was first operated in dual-recycled mode.
It was found that by locking with signal recycling detuned by a few kHz, an initial lock was possible. The tuning was then stepped towards the target value, in steps chosen to be small enough to avoid perturbing the lock.  By this means it was possible to reach a location in phase space which would essentially never have occurred by chance~\cite{phd.Grote}.

After lock has been achieved by one or more of the above means, the control task is generally managed by
linear control systems that may be analysed using standard {\em linear
  time invariant} (LTI) control theory. Two generic approaches have
been employed with success. In one approach there is a set of separate
single-input single output (SISO) controllers, one for each degree of
freedom. The alternative is to deal with several degrees of freedom in
a single multi-input multi-output (MIMO) controller.  Recently, since
the advent of computer-based, digital control systems, the MIMO
approach has become much more practical than it would be if
implemented in analogue electronics. An important difference between
the SISO and MIMO approaches concerns how cross-coupling between the
degrees of freedoms can dealt with.

Cross-coupling is commonly seen in both sensing and actuation, and
considerable effort is needed to develop control systems that operate
correctly in the presence of undesired mixing of signals.  The main
approaches to solving these problems with MIMO controllers is
described in Section~\ref{sec:GHCM}.  Otherwise we discuss SISO
controllers to provide illustrative examples of control in idealised
interferometers where there is no mixing of degrees of freedom at the
point of sensing or actuation.

We introduce standard terminology from control theory.  In each
control loop a point of reference is taken, called the {\em error
  point} at which we measure how the gain of the loop acts to suppress
deviations from the desired operating point. Since a loop has no end,
the selection of this point is somewhat arbitrary, but it is usually
convenient to take the output from a photo-detector or its associated
demodulator.

 Likewise, we choose an
actuation or {\em feedback point} at the interface between the control
electronics and the interferometer -- again the precise division is
somewhat arbitrary, but the electronic signal input to an actuator is
frequently employed as the point of reference.

With these points
defined, the part of the loop from error-point to feedback-point is
called the {\em controller} or just the feedback, and the rest of the
loop from feedback point to the error point, in the causal direction,
is called the {\em plant}.

Before a loop is activated, the signal that would be measured at the
error point is called the {\em error signal}.  In interferometry this
is usually derived as an output from the optical system and its
photo-detectors, as explained in Section~\ref{sec:ES_TF}.

\subsection{Linear time-invariant control theory -- introductory concepts}
\label{sec:LTI}
A full description of linear time-invariant (LTI) theory is beyond the
scope of this article, therefore we restrict our description to a
short summary of the essential concepts, with some relevant examples
presented in the following sections.

In LTI models the superposition principle applies, frequencies do not
mix and it is possible to represent any physical time-domain signals
in the frequency domain through their Fourier transforms.  The
time-invariance means that the response of a system to an input does
not depend on the time at which that input is applied.  This implies
that the differential equations describing the system are linear and
homogeneous with coefficients that are constant in time. In this case
it is common to solve these equations by employing methods based on
the Laplace transform. The response of the system is represented by
its {\em transfer function} which is the Laplace transform of its {\em
  impulse response} -- i.e.~the output produced in the time domain
when the input is a Dirac-delta function.  We look at transfer
functions in mode detail in Section~\ref{sec:ES_TF}.

For LTI systems the eigenfunctions or basis functions of the solution
are the complex exponentials.  Consider an input of the form
\begin{equation}
S_I=Ae^{st},
\end{equation}
 where $A$ is a complex factor, $s$ the Laplace transform
variable and $t$ is time, to a particular system.  If the output is
\begin{equation}
S_O=Be^{st},
\end{equation}
with different complex factor $B$, the system is described in
full by the transfer function
\begin{equation}
T=\frac{S_O}{S_I}=\frac{Be^{st}}{Ae^{st}}=\frac{B}{A},
\end{equation}
which is not a
function of time.

By implication, if the input to the system is a
sinusoid (a single Fourier component) the output is also a sinusoid
with, generally, different amplitude and phase as described by the
transfer function.  If the system consists of a series of optical,
electronic and mechanical stages or sub-systems, the overall transfer
function is the product of the individual transfer functions.  If
there is feedback, then {\em loop-algebra} can be applied, and this
represents summation of the signal from one point in a loop into an
earlier point (feedback), or in the case of feed-forward to a later
point. This is also a linear operation, and so the system with
feedback or feed-forward retains its LTI property.  The most
complicated case that need be considered is when there are two or more
nested feedback loops (or equivalently two or more actuators that
adjust one degree of freedom~\cite{Shapiro2014}), in this case there
are loop-algebraic techniques that allow the system to be reduced to a
single equivalent loop, such that nested loops of any practical
topology may be considered step by step.

In modelling LTI systems it is common to pick a {\em representation}
from one of a mathematically-equivalent set of options. The
alternatives are described in, e.g.~\cite{Franklin1998}. To give
some examples, individual loops or blocks with one input and one
output (called SISO -- single input, single output systems), are
commonly represented by their transfer functions.

For LTI systems, the transfer functions can be written as rational polynomials in the
complex frequency variable $s$.  The polynomials on both the numerator and denominator
 of the transfer function can be factorised.  This leads to an equivalent mathematical
 representation of the transfer function as a list of zeros (zeros of the numerator) and poles (zeros
 of the denominator). In addition an overall gain factor, independent of frequency, is usually
 required.   The poles and zeros may be represented by their coordinates in the
complex ($s$)-plane, or more phenomenologically by their resonant
frequencies and damping factors. This leads to the so-called  $(z, p,k)$-notation, where $z$ represents one or
more zeros, $p$ the poles, and  $k$ represents an overall frequency-independent gain factor.

For MIMO systems arrays of transfer functions can be employed to describe all possible input-output
relations, but it is more common to use the
{\em state-space} representation.  Here a single set of matrices
encapsulates the behaviour of the entire system. A description of this
important method is beyond the scope, but full detail is given in~\cite{Franklin1998}.

In briefest summary, the state-space method involves writing the set
of $N$ second order differential equations representing the internal
dynamics of a system which has $N$ degrees of freedom.  These are
reduced to $2N$ first-order equations by introducing the time
derivatives, i.e.~generalised velocities, of the displacement-like coordinates. The
solution of the resulting set of equations is then usually carried out
numerically, using matrix methods.

\epubtkImage{dsp1.png}{
  \begin{figure}[htb]    
    \centerline{\includegraphics[width=0.9\textwidth]{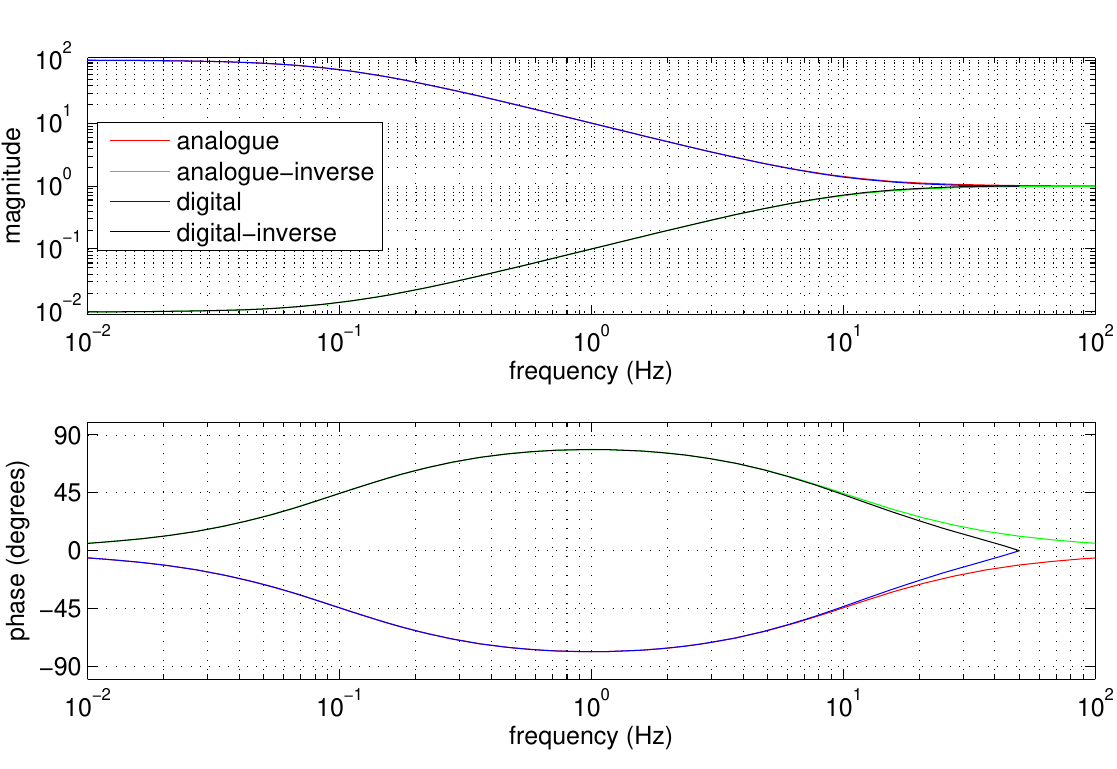}}
    \caption{Bode plots to compare a digital representations with a first-order analogue transfer function and its inverse.  The analogue system has a single real pole at 0.1\,Hz and single real zero at 10\,Hz. When the transfer function is inverted the pole becomes a zero, and the zero a pole.  The bilinear transform is employed to produce the digital equivalents, with a sampling frequency of 100\,Hz.  The highest frequency that can be represented in this case is 50\,Hz and it can be seen that the digital response becomes a poor approximation to the analogue one at frequencies approaching this limit. Note that, in a practical digital system, there would be a finite time delay and corresponding phase-lag, not included here, and that further delays may be present from anti-aliasing and anti-imaging filters -- see text.}
    \label{fig:dsp1}
\end{figure}}

\subsection{Digital signal processing for control}
\label{sec:DSP}
In the past two decades digital control systems have been introduced
into the control of interferometers, in modern instruments the
majority of control systems contain digital processing elements,
although the interfaces with the interferometer remain analogue in
almost all cases.  The essential principles remain the same as in the continuous-time
systems, and a common
approach to the  design of digital control systems starts by designing and
simulating a continuous-time analogue model.  When this model operates as
required in simulation, the result is transformed to the
discrete-time mathematics of digital control.  The resulting filters
are then  implemented in a combination of software and hardware.   In discrete-time
models only a finite set of frequencies exist, limited at high
frequencies by the Nyquist frequency, i.e.~half of the sampling rate
in the digital system.

Digital models also have finite amplitude resolution, with the
practical resolution limits occurring at the analogue-digital and
digital-analogue interfaces (ADC, DAC respectively), rather than in
the digital signal processing.  These limitations are generally
handled by {\em whitening} the signal at the input and {\em
  de-whitening} at the output.  For example, input signals with
predominant low-frequency content may be high-pass filtered to render
their spectral content relatively uniform, i.e.~white, before sampling
to make best use of the available resolution.  The converse process
can be applied at the output, with sufficient low-pass filtering
applied to ensure that white noise resulting from the output
conversion (DAC) is suppressed within the gravitational wave band.

In the description of digital controllers, the discrete mathematics of
the $z$-plane replaces the continuous nature of the
$s$-plane~\cite{Franklin1998}. Transforming from one space to the other is something of an
art, mainly due to the consequences of finite precision in the associated
calculations. A bilinear transformation is commonly employed.  To avoid problems
of numerical accuracy in the associated calculations, complicated systems are
broken down in to a series of second-order sections.  These subsystems have up to two poles
and two zeros, i.e.~the transfer functions have no higher order than quadratic numerator and denominator. Such subsystems
can be transformed more reliably.

The finite time-steps in digital processing limit the filter transfer functions that
may be produced at frequencies approaching the Nyquist limit -- see Figures~\ref{fig:dsp1} and~\ref{fig:dsp2}.  Note however that
with modern computers the sampling rate is often limited by the analogue interfaces rather than
the speed of processing.  Where this is true the data stream can be up-sampled (interpolated) to a higher sampling
rate for filtering, and then down-sampled (decimated) to the original sampling rate before conversion back
to the analogue domain. This process can be used to improve the high-frequency response of digital filters.

\epubtkImage{dsp2.png}{
  \begin{figure}[htb]    
    \centerline{\includegraphics[width=0.9\textwidth]{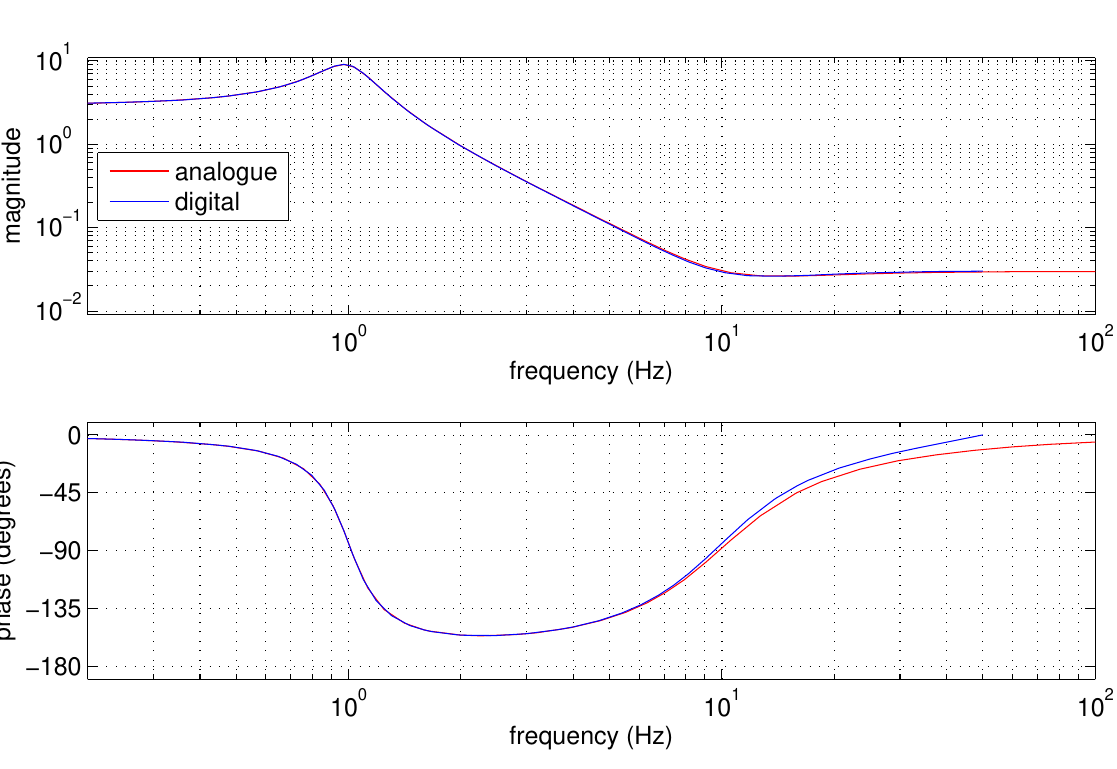}}
    \caption{Bode plots for a second-order analogue system, and a digital approximation thereto. Here there is a complex pair of poles at 1\,Hz with a resonance quality factor $Q=3$, and a pair of zeros at  10\,Hz, $Q=1$. There is an overall gain factor of 3.     The digital sampling frequency is 100\,Hz, so signal frequencies are limted to 50\,Hz. }
    \label{fig:dsp2}
\end{figure}}

One further consequence of discrete time is that there is a finite delay associated with the analogue to digital, signal processing and digital to analogue steps.  This must be considered in the development of feedback loops based on digital controls.  In practice, even more severe limits to high-frequency performance often arise from anti-aliasing or anti-imaging filters that may be required on the analogue input and outputs, respectively.

Aliasing occurs when the ingoing signal contains significant amplitude components at Fourier frequencies above the Nyquist limit.  If these are not filtered out, they are incorrectly recorded their beat frequencies with the nearest harmonic of the sampling frequency.  At the output, the digital signal has discrete steps from one sample to the next. To properly reconstruct the required analogue signal these steps require to be removed by the low-pass action of an anti-imaging (or reconstruction) filter.  Further detail of the sampling and reconstruction processes is found in~\cite{Franklin1998}.

\subsection{Degrees of freedom and operating points}
\label{sec:DoF}

We consider the optical components to be rigid bodies, each with six
degrees of freedom.  With practical, high-quality spherical surfaces,
only three degrees of freedom per component are important: position
along the direction of propagation of the light, referred to as the
{\em longitudinal} coordinate, the {\em yaw} angle with respect to
that direction, i.e.~in the horizontal plane, and the {\em pitch}
angle in the vertical plane. The other three degrees of freedom
(vertical, horizontal normal to the beam and roll around the axis of
the beam) may be important with respect to noise coupling into the
length measurement in the case of imperfect mirrors. As discussed in Section~\ref{sec:characterising},
the mirrors typically have only small deviations from ideal spheres, so the coupling factors are
small and do not significantly affect the control of the interferometer.

In the interferometer as a whole, one component, for example, a mirror or
beam splitter, may be chosen as the origin for the coordinate system. This allows
one position and a pair of angles to be pre-defined.  The
positions and angles of the other components may then be described
with respect to this origin. Note, however, that the longitudinal
degrees of freedom are measured with an optical `ruler' that is based
on the wavelength of the light, and so the wavelength should be
counted as one longitudinal degree of freedom in the system as a whole (in the sense that the light has the same frequency everywhere which is usually true to a good approximation in the ultra-stable environment of a gravitational wave detector). Similarly the direction of the light beam entering the interferometer defines two angles.

To take an example, a simple cavity that is to be held on resonance with in-going light has two meaningful longitudinal degrees of freedom.  For a cavity in isolation it would be usual to consider the the position of one mirror relative to the other and the frequency, or wavelength, of the light as the important parameters.  Mathematically there are other equivalent choices, but in the control and operation of interferometers the point is to find a convenient set of control variables.

Similarly, a simple Michelson interferometer has three components and three longitudinal degrees of freedom.  Again it would be usual to consider one component as a reference.  If the beam splitter is fixed, the three degrees of freedom are the two arm-lengths and the optical wavelength, or frequency.

If a pair of  cavities were to be placed, one each, into the arms of the simple Michelson, the single degree of freedom of each mirror is replaced with the two of the cavity, for a total of five: once again, the same as the number of components.  Fixing the beam splitter, these are the laser wavelength, the two distances from the beam splitter to the near mirrors of the cavities and the two lengths of the cavities.

An example of the degrees of freedom relevant to longitudinal control of a more complex system is shown in Section~\ref{sec:DRFPMI}.

The choice between employing absolute or relative coordinates for the positions (and angles) of interferometer components is reflected in differences of approach in the available modelling software. In a \Finesse
model of a two-mirror cavity, for example, the longitudinal positions
of the two mirrors are specified, and adjusting either of them changes
the resonant condition of the cavity (see, for example,
Section~\ref{sec:fexample_cavity_power}).  Likewise, adjusting the
position of the input mirror changes the phase of both the light in
the cavity, and the light reflected from the cavity.  See
Section~\ref{sec:mirrors_spaces} for a discussion of this point.

An operating interferometer requires various interference conditions
to be maintained, e.g.~cavities should be kept on resonance, the
dark-fringe condition in a Michelson interferometer must be met, and
so forth.  For each degree of freedom this implies that there is an
optimum value for best sensitivity or an {\em operating point} in the
multi-dimensional space representing the degrees of freedom. This
point is not usually unique: for example, signals repeat modulo one
round-trip wavelength, see below.

As most or all degrees of freedom are subject to movement or drift,
they must be controlled, generally by designing and implementing a
separate control loop for each one. These
loops must be designed to hold the value of the degree of freedom
close to such an operating point, where `close' is determined by
tolerance bounds that must be determined by calculation.

In most cases it is possible to evaluate a tolerance interval around
the operating point.  The limits usually arise in the consideration of
the coupling of some kind of noise into the sensitive measurement
(frequency noise, power noise, beam direction noise, etc.).  For
example, in the case of the dark fringe, sensitivity to laser power
noise is at a minimum at the perfectly dark condition, and the
tolerable increase in this coupling may be used to set bounds on
deviations from the operating point.

\epubtkImage{two_dofs_two_plots.png}{%
  \begin{figure}[htbp]
   \centerline{\includegraphics[width=1\textwidth]{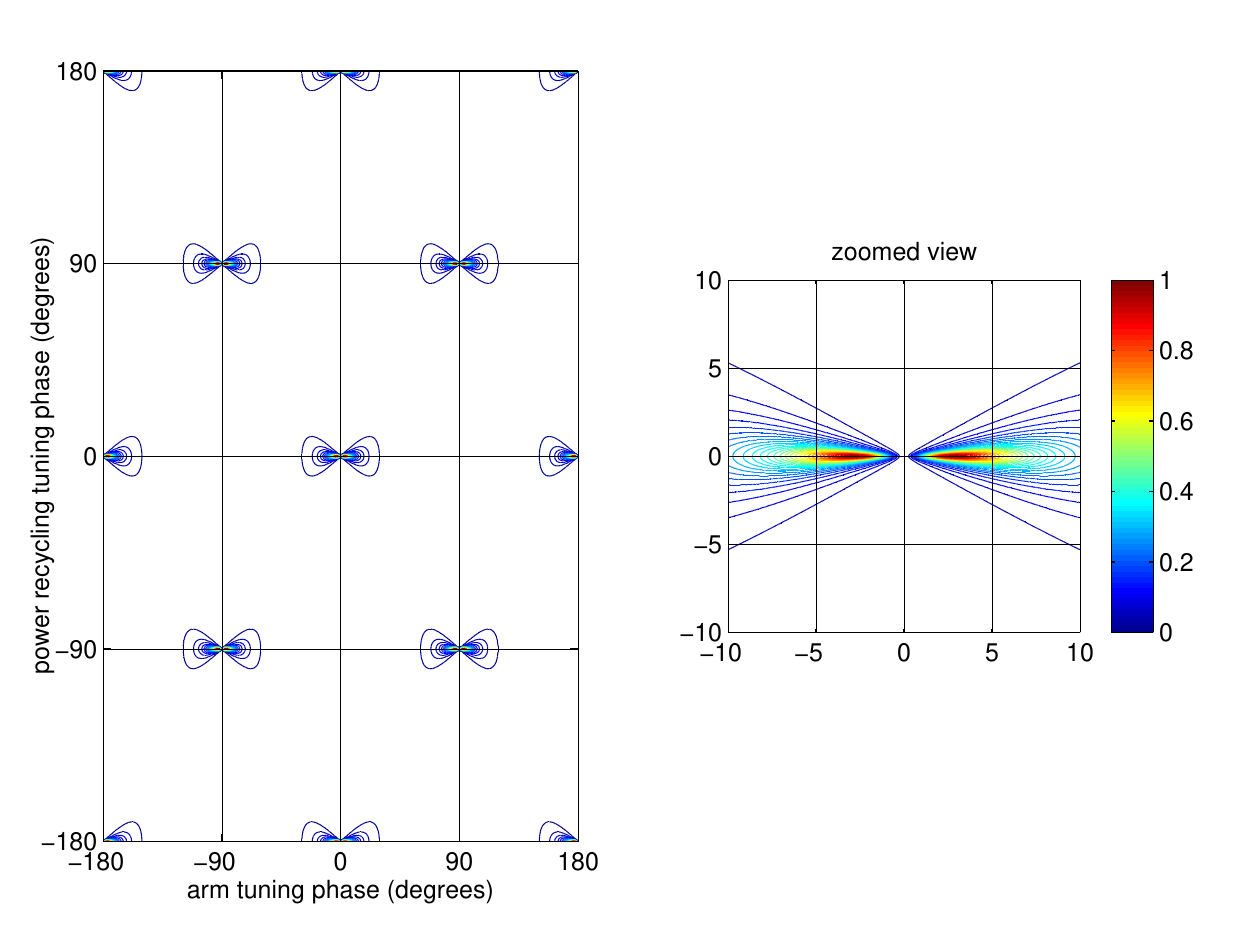}}
     \caption[Varying two degrees of freedom]{Simplified example illustrating the restricted phase space
    volume in which interferometer control signals are expected to achieve significant magnitudes.  In the left panel we
    show the modulus of the light amplitude emerging at the detection/output port of a power recycled Michelson interferometer as a function of both the length difference between the arms (in the plot, the two end mirrors have their tuning phase shifted by the amount shown, but in opposition) and the tuning of the power recycling (the mirror has its tuning phase shifted by the amount shown). The normalised amplitude is shown on a linear scale from zero to one. These two degrees of freedom are swept over two cycles from the nominal operating point at (0,0).  The other possible longitudinal degrees of freedom, namely the
    common mode arm length and the optical frequency are kept constant. The size of a signal designed to sense the differential degree
    of freedom, and allow the interferometer to be locked at the operating point, would have significant magnitude only in the region of the
    features like the one near (0,0) where there is a significant gradient in the horizontal direction in the plot.
    In this
    example, the power recycling mirror has a transmission of $1\%$, and other components have no loss. In
    a practical detector there would be several other degrees of freedom.  Typically there would also be cavities of higher finesse, leading to even narrower features. The right panel shows a magnified view of one of the `islands' of useful signal, the operating point is a small region at coordinates (0,0) where the light power is low.}
    \label{fig:two_dofs}
\end{figure}}

Bounds may also be set by
considering the required linearity of signals. Non-linearity can lead
to beating, which mixes noise into the measurement band. For example
if there is a narrow spectral feature or `line', such as a calibration line that may be applied to monitor instrumental sensitivity, or a suspension violin mode\footnote{To minimise thermal noise in suspensions, low-loss materials and techniques are employed to avoid dissipation. The resonant modes of these suspensions are seen in the frequency domain as narrow spectral features, or lines.  The violin modes are transverse oscillations of the stretched suspension fibres that support the mirrors, which vibrate much like a violin string. The frequencies of these modes typically lie between 300 and 800\,Hz, and they are often conspicuous in the spectra of signals from gravitational wave detectors.} in the measurement band, beating this with low frequency motion
of suspended components will produce sidebands on either side of the
narrow feature, and these may be of higher amplitude than the noise
background at the frequencies of interest near the line. Non-linear
operation may also cause problems for control systems, as its presence
implies that the gain of control loops will depend on the magnitude of
deviations from the nominal operating point.  In the following
(Section~\ref{sec:ES_TF}) it will be seen that the normal process of sensing the
length of a cavity is reasonably linear only within a very narrow
range, in comparison to the wavelength of light, around the operating
point, at least for a cavity of high finesse  -- a range of distance
of order $\lambda/{\mathit F}$ (see Section~\ref{sec:two_mirror2}), or
smaller.

In designing a control system for an interferometer, one can in
principle consider the space of possible values of all the degrees of
freedom in an interferometer, but it is more usual to work with a
sub-space, e.g.~only the longitudinal degrees. In the angular case
there is usually a unique operating point per degree of freedom
corresponding to one optimal alignment, but in the longitudinal case
operating points are repeated as the relevant round-trip phase change
steps in multiples of $2\pi$, i.e.~one wavelength change in round-trip
optical path distance. In this case it is usual to consider the
(hyper-)volume containing one repeat in each dimension: for km-long
interferometers there is very little difference between adjacent
volumes. In a typical interferometer designed for gravitational wave
detection, the number of degrees of freedom in combination with
requirements on noise coupling and linearity mean that only a
vanishingly small volume within the phase space corresponds to the
acceptable region around the desired operating point.  This suggests
one of the important questions in operating a gravitational wave
detector: how to bring every degree of freedom to the desired
operating point.

The diminutive scale of the useful volume in phase space can be illustrated by means of
a simplified example.  Here we consider the case of two degrees of freedom in a power
recycled Michelson interferometer.  Even fixing the location of the beam splitter, there are three degrees of
freedom (i.e.~common arm length, differential arm length and longitudinal position of the
power recycling mirror.  However, we choose to produce a contour plot showing signal sizes as a function
of just two degrees of freedom, Figure~\ref{fig:two_dofs}. Here we vary the difference in the lengths of the two arms while keeping the average (or common mode) arm length fixed, and also to vary the
 position of the power recycling mirror.
In a practical interferometer there would be several other degrees of freedom associated with, for example, arm cavities, signal recycling and control of the common-mode arm length (or laser frequency), and in most cases the cavities
would be of higher finesse producing even narrower features
 -- see, for example, the parameters for Advanced LIGO in Appendix~\ref{sec:ALOL}.

The complexity of sensing and control becomes apparent when one
considers that, in the common case of the freely-suspended optical
components in a ground-based interferometric gravitational wave
detector, the initial condition, at the point of `switching on' the
controls can be any random point within the space, with -- in addition
-- a wide range of initial velocities associated with each degree of
freedom: up to perhaps of order one wavelength per second, in
a typical ground-based instrument.  How this is dealt with is
summarised in Section~\ref{sec:control1}.

\subsection{Error signals and transfer functions}
\label{sec:ES_TF}

In general, we will call an \emph{error signal} any measured signal
suitable for stabilising a certain experimental parameter $p$ with a
servo loop. The aim is to maintain the variable $p$ at a user-defined
value, the \emph{operating point}, $p_0$. Therefore, the error signal
must be a function of the parameter $p$. In most cases it is
preferable to have a bipolar signal with a zero crossing at the
operating point.  The slope of the error signal at the operating point
is a measure of the `gain' of the sensor, which in the case of
interferometers is a combination of optics and electronics.

\emph{Transfer functions} describe the propagation of a
periodic signal through a \emph{plant} and are usually given as plots
of amplitude and phase over frequency, e.g.~as Bode plots (see the following section).
By definition a transfer
function describes only the linear coupling of signals inside a
system. This means a transfer function is independent of the actual
signal size. For small signals or small deviations, most systems can
be linearised and correctly described by transfer functions.

Experimentally, network analysers are commonly used to measure a
transfer function: one connects a periodic signal (the \emph{source})
to an actuator of the plant (which is to be analysed) and to an input
of the analyser. A signal from a sensor that monitors a certain
parameter of the plant is connected to the second analyser input. By
mixing the source with the sensor signal the analyser can determine
the amplitude and phase of the input signal with respect to the source
(amplitude equals one and the phase equals zero when both signals are
identical).

Mathematically, transfer functions can be modelled similarly: applying a
sinusoidal signal $~\sin(\omega_s t)$ to the interferometer, e.g., as a
position modulation of a cavity mirror, will create phase modulation
sidebands with a frequency offset of $\pm \omega_s$ to the carrier
light. If such light is detected in the right way by a photodiode, it
will include a signal at the frequency component $\omega_s$, which can
be extracted, for example, by means of demodulation (see
Section~\ref{sec:demodulation}).

Transfer functions are of particular interest in relation to error
signals. Typically a transfer function of the error signal is required
for the design of the respective electronic servo. A `transfer
function of the error signal' usually refers to a very specific setup:
the system is held at its operating point, such that, on average,
$\bar{p}=p_0$. A signal is applied to the system in the form of a very
small sinusoidal disturbance of $p$. The transfer function is then
constructed by computing for each signal frequency the ratio of the
error signal and the injected signal. Figure~\ref{fig:slope1} shows an
example of an error signal and its corresponding transfer
function. The operating point shall be at
\begin{equation}
x_{\mathrm{d}} = 0 \qquad \text{and} \qquad x_{\SSm{EP}}(x_{\mathrm{d}}=0) = 0.
\end{equation}
The optical transfer function $T_{\mathrm{opt, x_{\mathrm{d}}}}$ with
respect to this error signal is defined by
\begin{equation}
\WT{x}_{\SSm{EP}}(f) = T_{\mathrm{opt, x_{\mathrm{d}}}} T_{\mathrm{det}} \WT{x}_d(f),
\end{equation}
with $T_{\mathrm{det}}$ as the transfer function of the sensor. In the
following, $T_{\mathrm{det}}$ is assumed to be unity.  At the zero
crossing the slope of the error signal represents the magnitude of the
transfer function for low frequencies:
\begin{equation}
\label{eq:slope_tf}
\left|\frac{d x_{\SSm{EP}}}{d
  x_{\mathrm{d}}}\right|_{~\bigl|x_{\mathrm{d}}=0\bigr.}=|T_{\mathrm{opt, x_{\mathrm{d}}}}|_{~\bigl|f\rightarrow 0\bigr.}
\end{equation}
The quantity above will be called the \emph{error-signal slope} in the
following text. It is proportional to the \emph{optical gain}
$|T_{\mathrm{opt}, x_{\mathrm{d}}}|$, which describes the
amplification of the gravitational-wave signal by the optical instrument.

\epubtkImage{slope1-slope2.png}{%
  \begin{figure}[htbp]
    \centerline{\includegraphics[viewport=0 0 360 220, width=0.8\textwidth]{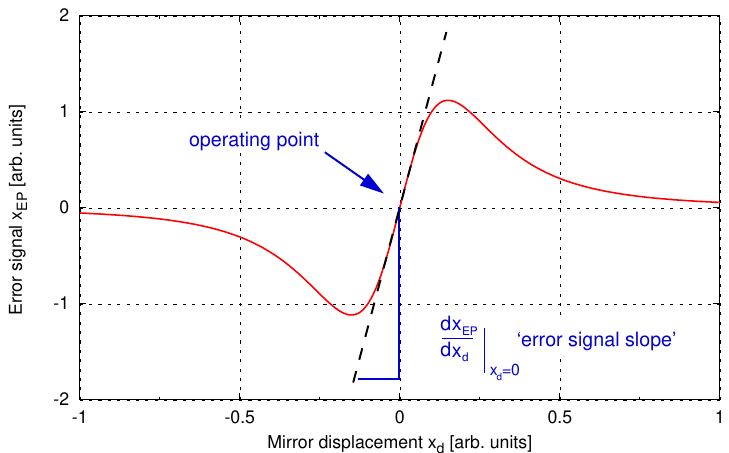}}
    \vspace{5mm}
    \centerline{\includegraphics[viewport=0 0 360 220, width=0.8\textwidth]{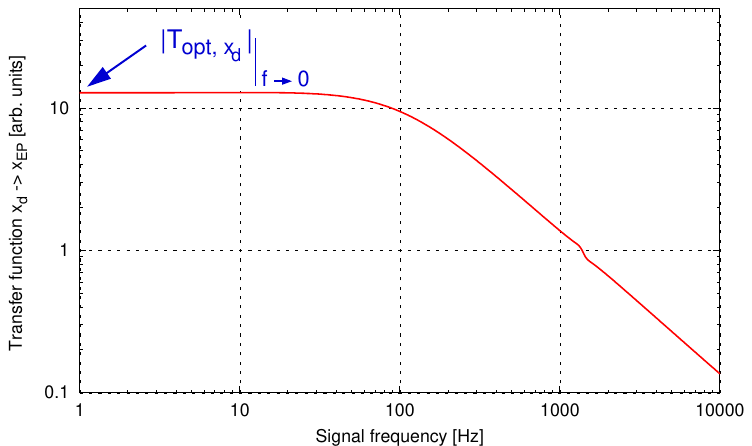}}
    \caption[Example error-signal slope]{Example of an error signal:
      the top graph shows the electronic interferometer output signal
      as a function of mirror displacement. The operating point is
      given as the zero crossing, and the \emph{error-signal slope} is
      defined as the slope at the operating point.  The right graph
      shows the magnitude of the transfer function \emph{mirror
      displacement $\rightarrow$ error signal}. The slope of the error
      signal (left graph) is equal to the low frequency limit of the
      transfer function magnitude (see Equation~(\ref{eq:slope_tf})).}
    \label{fig:slope1}
\end{figure}}

\subsection{Bode plots -- traditional control theory for SISO loops}
\label{sec:Bode}
An essential feature of a control system is stability, i.e.~for a
finite input the output should always be bounded.  This is equivalent
to requiring all of the transfer function poles to correspond to
decaying exponentials, so their real parts must be strictly negative.

Prior to the routine application of computers, a number of tools (plots)
were developed to facilitate control system design. Although the root-locus, Nyquist and Bode plots continue to be applied,  computer models remove the practical (calculational) advantages of one over another. All of these methods present essentially equivalent information, and the choice of one over another is a matter of convenience or familiarity.  Since Bode plots provide a complete description of single-input single-output (SISO) LTI control loops we choose to describe that approach as an example.

 Throughout this section, the system is described in continuous time, i.e.~as an analogue model.
When the technique is applied to digital control systems discrete-time models are needed as discussed in Section~\ref{sec:DSP}.

A Bode plot of a system shows its transfer function in the form of log-magnitude and linear-phase
graphs against a logarithmic frequency axis -- conventionally in vertically-stacked plots
with matching, aligned frequency axes. In the context of the design of
complete negative feedback loops, it is common, though not universal,
to add $\pi$ to the phase to represent the overall negative sign --
this convention is assumed here. The standard procedure starts with
consideration of the open-loop Bode plot.  In this, the loop is broken
(in the model) at a convenient point, and the transfer function from
there back to just before the break is calculated and plotted.
Remember that the total transfer function is computed as the product
of individual transfer functions of parts of the loop that are
connected in series. Particular attention is paid to the regions close
to points where the transfer function magnitude crosses unity (i.e.~zero
on the log scale), called the unity gain point(s), and where the phase
crosses $-\pi$ in absolute terms,  {\em not} modulo $2\pi$. The
transfer function is then characterised by the phase margin and the
gain margin.  The phase margin is the phase of the transfer function
plus $\pi$ at the frequency where the gain is unity, and the gain
margin is the inverse of the gain where the phase is $-\pi$ (or the
negative of the log gain). If there are multiple unity gain points the
smallest phase margin, and the smallest gain margin, dominate.  If the
smallest gain and phase margins are both positive, the system is
stable.  Note that if there are multiple paths or `loops', these are dealt
with by applying loop algebra to reduce the system to a single
feedback loop without subsidiary loops.

 \epubtkImage{bode_simple_pp.png}{%
  \begin{figure}[htbp]
    \centerline{\includegraphics[width=0.9\textwidth]{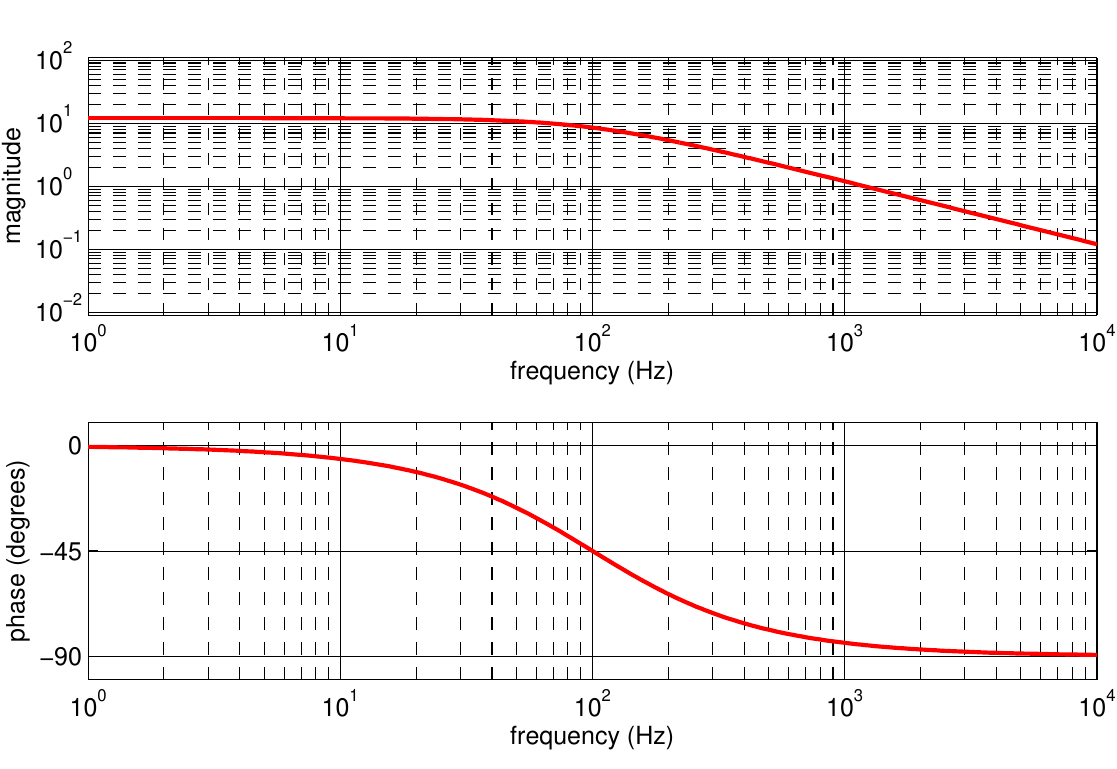}}
       \caption[Example Bode Plot]{Example Bode plot for a simple first order low-pass
    transfer function with a gain of 12 at low-frequency and a single pole at 100\,Hz.
      The upper panel shows the magnitude of the response, presented on a logarithmic
      scale, as a function of frequency (here in Hz, may also be radian/s). The lower panel shows the corresponding phase in degrees. As noted in the text, for minimum-phase systems such as this, the magnitude and phase are not independent, but both provide useful information in the design of control loops. Where the response is flat with frequency, the phase is asymptotically zero; where the response varies as $1/f$, there is a phase lag approaching 90$^{\circ}$. As explained in the text, if this plot were to represent the open-loop
      transfer function of a negative feedback loop, important properties of the loop can be
      read-off by inspection.  For example, stability can be assessed by checking the phase
      margin at the unity gain frequency (here the gain is unity at 1200\,Hz).  At this frequency the phase is about $-85^{\circ}$ and the phase margin is given by the difference of this  from $-180^{\circ}$,
      or $95^{\circ}$. This is far from zero, and so the closed-loop response is predicted to be
      stable.}
    \label{fig:bode_simple}
\end{figure}}

Traditionally these methods were extended to reveal properties of the
closed-loop system, i.e.~of the original model without any break. This
was done because, for transfer functions of low order (one, two or three
poles), there are simple expressions that relate the phase margin to
the ringing, or equivalently damping, of the closed-loop response.
When computer models are employed for systems of greater complexity
there is no need for these rules or guidelines and it is common to
transform back to the time domain, calculate the impulse response of
the closed loop system, and characterise its resonant frequencies and
damping without reliance on rules.

 As a concrete example of a Bode plot, we include one representing a system that approximates the transfer function shown in Figure~\ref{fig:slope1}).  The system  consists of a gain factor (12) and a single pole at 100\,Hz (or $2\pi\times100$ radian/s). The transfer function may be written
\begin{equation}
   G(s) = \frac{12}{s+200\pi},
\end{equation}
such that there is a single real pole at complex frequency $s=-200\pi$.
Further explanation of the mathematics of transfer functions is given
in Section~\ref{sec:LTI}, while Bode plots of
higher order systems in both continuous and discrete time are found in Section~\ref{sec:DSP}.

The construction and utility of the Bode plot originates in part from
the properties of a common subset of transfer functions that
represents stable, causal systems.  Such systems are called {\em
  minimum phase} as a consequence of the locations of their zeros in
the $s$-plane.  In a causal system the output lags the input. Stable,
causal LTI systems are also invertible, i.e.~the transfer function
numerator and denominator can be swapped, or equivalently all the
poles and zeros may be exchanged resulting in another stable, causal
system.  For this to work the zeros of the system must have negative
real parts, so that when they become poles in the inverse system they
are damped. It can be shown that in such a system there is a strict
relationship between the phase and the slope of the log-magnitude, as
shown on a Bode plot -- one is a Hilbert transform of the other. In
practice this is equivalent to writing that the when the magnitude
graph has a slope of $f^{-n}$, where $f$ is the frequency, the phase
approaches $-n\pi/2$. This method allows the loop to be designed to meet
various goals that are usually expressed in terms of gain (or
attenuation) that must be achieved in one or more range of
frequencies, with stability checked by reading off the phase and gain
margins.

In interferometry the {\em optical transfer function} is usually a
significant aspect of control loops.  Such transfer functions may be
measured or found by calculation (e.g.~with \Finesse).  The corresponding transfer functions can be
found by applying the techniques described in Section~\ref{sec:ES_TF}.

\epubtkImage{cavity_power.png}{%
  \begin{figure}[htbp]
    \centerline{\includegraphics[width=0.9\textwidth]{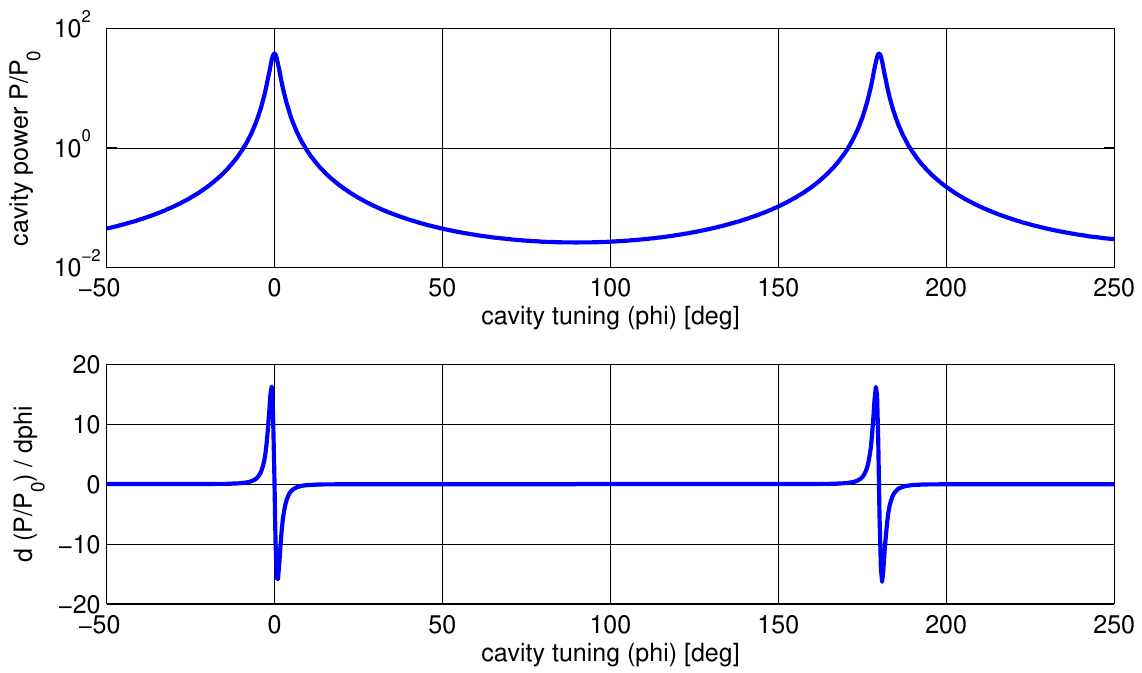}}
    \caption{The top plot shows the cavity power as a function of the
      cavity tuning. A tuning of 360\textdegree\ refers to a change in the
      cavity length by one laser wavelength. The bottom plot shows
      the differentiation of the upper trace. This illustrates that
      near resonance the cavity power changes very rapidly when the
      cavity length changes. However, for most tunings the cavity
      seems not sensitive at all.}
    \label{fig:cav_slope}
\end{figure}}

\subsection{Separating mixtures of the degrees of freedom: control matrices}
\label{sec:GHCM}

In practice, each error signal intended to represent a particular
degree of freedom of the optical arrangement also contains some information about other degrees of freedom. To give a simple example of the mixing that may occur, any motion that leads to a change in the circulating light
power in a cavity is likely to couple, at some level, to every signal
that depends on the intra-cavity light, unless the signal is precisely zero.

In most cases such mixing is undesirable as it is easier to design control systems
to deal with one degree of freedom in isolation.
In the worst case, if the mixing, or cross-coupling is strong, it can lead to
the formation of unintended feedback paths.  If the transfer function of such loops has
a magnitude exceeding unity, there is a chance that the loop may be unstable.  A common cause
of such instability is a resonance in the unintended or `parasitic' loop.
At such a resonance high gain is
typically accompanied by a phase lag of $-\pi$ which will tend to be unstable unless some compensation
is included, e.g.~in the form of a notch filter to cancel the resonance.

Unwanted mixing of signals can also occur at the point of actuation.
For example, a mirror may be common to two degrees of freedom of an
interferometer. In an interferometer with arm cavities, the cavity mirrors
  closest to the beam splitter behave in this way.
Moving such a mirror must then affect at least two
length degrees of freedom.  This can be seen in Figure~\ref{fig:michelson_cavities_layout} where
motion of either of the two mirrors labelled ITMX and ITMY affects the phase of the light in the respective arm cavity and also the interference condition of the Michelson interferometer.  In contrast, the end mirrors (ETMX, ETMY) each affect only one longitudinal degree of freedom.

A further possible source of mixing between degrees of freedom arises at the point of actuation. Feedback to control a mirror is
often carried out in practice using an array of actuators, such as coil-magnet
pairs, that push on the mirror at various points on its surface. For example, it is common to employ a square-array of four magnets attached to the rear surface of the
mirror, as these allow longitudinal, pitch and yaw adjustment.  If they are mounted close to the perimeter of the rear surface they may be out of the way of a transmitted light beam.  With such an arrangement,  each individual actuator causes changes to a mixture of
angular and longitudinal degrees of freedom.  If the actuators are not
of precisely uniform strength and alignment, this leads to unintended components in the
resultant force produced by the array.  An {\em actuation
  matrix}, with frequency-dependent elements where necessary,
can be employed to orthogonalise the response of the system to
commands from the controller, at least to some degree of precision.

The elements of actuation and sensing matrices are typically determined as  a result of simulation and measurement.  Modelling may yield a set of starting values that suffice to allow the interferometer to operate.  When operational residual mixing is normally determined by carrying out all possible transfer function measurements.  The measurements allow coupling matrices to be determined, and inverting the coupling matrix provides the appropriate matrix necessary to remove unwanted mixing. This process is somewhat involved and benefits from automation.

\subsection{Modern control methods in gravitational wave detectors}

During the past few decades new methods of designing sophisticated
controllers based on digital signal processing have emerged.
A major benefit of the resulting `digital controls' is that the response of
a control filter can be adjusted by changing filter coefficients, this can even be achieved
while the controller is operating, if that is required.

Digital control facilitates
the application of so-called modern control methods in which
optimisation methods are employed.  As an indication of the possible
advantages that may arise from this, we briefly mention two approaches
to modern control of application in interferometry.  For a relevant
description of these see, e.g.~\cite{Franklin1998}.

In the first approach, we consider the generation of an optimal filter
with fixed coefficients (gain, poles and zeros).  In such a case, the
plant to be controlled is characterised by some means, and the results
are used in the design of an optimal filter.  For example, if it can
be assumed that a measurement produces an estimate of the system
contaminated by noise, and a model of the system with the correct
number of degrees of freedom exists, a Wiener filter may be formed as
a result of least-squares fitting the model to the data.  If the
result is to be inverted to provide a compensating filter in a control
system, then the fit must be constrained produce a causal filter (with
all poles and zeros having negative real parts, in an analogue
model). The are standard methods by which this may be accomplished.

The next step up in sophistication is to find a controller that
remains optimal even if the underlying plant changes (or if its
parameters cannot be measured accurately before the controller is put
into operation).  Such an {\em adaptive}  controller, employs a Kalman
filter -- also called a Linear Quadratic Estimator.  This is
implemented as an algorithm that operates on a series of measurements
taken over time.  These measurements are assumed to be contaminated with noise.
The algorithm operates recursively to produce an optimal estimate of
the state of the physical system.  During this process a model of the
system, i.e.~a representation of the equations of motion, with
relevant coefficients available to be adjusted,  is iteratively
updated.  The model is assumed to have errors either as a result of
poor starting estimates or due to drifting of parameters over time.
A weighting function, also called a {\em cost function}, is applied to
the measured data to allow less noisy or otherwise more important
aspects of the data to have a stronger influence on the outcome.  At
each iteration the model is employed to predict the current state,
this is then compared with the actual state and the results of the
comparison are used to refine and update the model.  When this method
is made to operate, the model of the underlying system converges to an
optimal solution for the given weighting function.

\subsection{Fabry-Perot length sensing}
\label{sec:fp_control}


In Figure~\ref{fig:cav-powerenhance} we have plotted the circulating
power in a Fabry-Perot cavity as a function of the laser
frequency. The steep features in this plot indicate that such a cavity
can be used to measure changes in the laser frequency. From the
equation for the circulating power (see Equation~(\ref{eq:cav-power})),
\begin{equation}
\label{eq:cav_power2}
P_1/P_0=\frac{T_1}{1+R_1R_2-2r_1r_2\cos\left(2 k L\right)}=\frac{T_1}{d},
\end{equation}
we can see that the actual frequency dependence is given by the
$\cos(2 k L)$ term. Writing this term as
\begin{equation}
 \cos(2 k L)=\cos\left(2\pi \frac{Lf}{c}\right),
\end{equation}
we can highlight the fact that the cavity is in fact a reference for
the laser frequency in relation to the cavity length. If we know the
cavity length very well, a cavity should be a good instrument to
measure the frequency of a laser beam. However, if we know the laser
frequency very accurately, we can use an optical cavity to measure a
length. In the following we will detail the optical setup and
behaviour of a cavity used for a length measurement. The same
reasoning applies for frequency measurements. If we make use of the
resonant power enhancement of the cavity to measure the cavity length,
we can derive the sensitivity of the cavity from the differentiation
of Equation~(\ref{eq:cav-power}), which gives the slope of the trace
shown in Figure~\ref{fig:cav-powerenhance},
\begin{equation}
\frac{d\,P_1/P_0}{d\,L}=\frac{-4 T_1 r_1 r_2 k \sin(2 k L)}{d^2},
\end{equation}
with $d$ as defined in Equation~(\ref{eq:cav_power2}). This is plotted
in Figure~\ref{fig:cav_slope} together with the cavity power as a
function of the cavity tuning. From  Figure~\ref{fig:cav_slope} we can
deduce a few key features of the cavity:
\begin{itemize}
  \item The cavity must be held as near as possible to the resonance
  for maximum sensitivity. This is the reason that active servo control
  systems play an important role in modern laser interferometers.
  \item If we want to use the power directly as an error signal for
    the length, we cannot use the cavity directly on resonance because
    there the optical gain is zero. A suitable error signal (i.e.~a
    bipolar signal) can be constructed by adding an offset to the
    light power signal. A control system utilising this method is
    often called \emph{DC-lock} or \emph{offset-lock}. However, we
    show below that more elegant alternative methods for generating
    error signals exist.
  \item The differentiation of the cavity power looks like a perfect
  error signal for holding the cavity on resonance. A signal
  proportional to such differentiation can be achieved with a
  modulation-demodulation technique.
\end{itemize}

\epubtkImage{pound_drever_hall01.png}{%
  \begin{figure}[htb]
    \centerline{\includegraphics{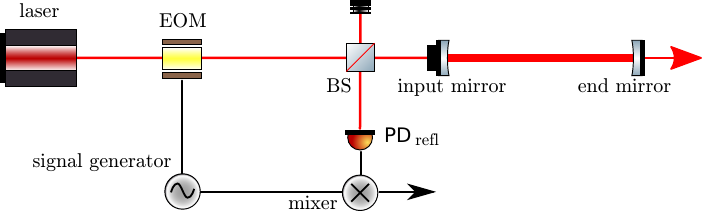}}
   \caption{Typical setup for using the Pound--Drever--Hall scheme for
     length sensing and with a two-mirror cavity: the laser beam is
     phase modulated with an electro-optical modulator (EOM). The
     modulation frequency is often in the radio frequency range. The
     photodiode signal in reflection is then electrically demodulated
     at the same frequency.}
   \label{fig:pound_drever_hall}
\end{figure}}

\subsection{The Pound--Drever--Hall length sensing scheme}
\label{sec:pdh}
This scheme for stabilising the frequency of a light field to the
length of a cavity, or vice versa, is based on much older techniques
for performing very similar actions with microwaves and microwave
resonators~\cite{Pound1946}. Drever and Hall have adapted such techniques for use
in the optical regime~\cite{drever83} and today what is now called the
\emph{Pound--Drever--Hall} technique can be found in a great number of
different types of optical setups. An example layout of this scheme is
shown in Figure~\ref{fig:pound_drever_hall}, in this case for
generating a length (or frequency) signal of a two-mirror cavity. The
laser is passed through an electro-optical modulator, which applies a
periodic phase modulation at a fixed frequency. In many cases the
modulation frequency is chosen such that it resides in the radio
frequency band for which low-cost, low-noise electronic components are
available. The phase modulated light is then injected into the
cavity. However, from the frequency domain analysis introduced in
Section~\ref{sec:interferometers}, we know that in most cases not all
the light can be injected into the cavity. Let's consider the example
of an over-coupled cavity with the reflectivity of the end mirror
$R_2<1$. Such a cavity would have a frequency response as shown in the
top traces of Figure~\ref{fig:cav-phase} (recall that the origin of
the frequency axis refers to an arbitrarily chosen default frequency,
which for this figure has been selected to be a resonance frequency of
the cavity). If the cavity is held on resonance for the unmodulated
carrier field, this field enters the cavity, gets resonantly enhanced
and a substantial fraction is transmitted. If the frequency offset of
the modulation sidebands is chosen such that it does not coincide with
(or is near to) an integer multiple of the cavity's free spectral
range, the modulation sidebands are mostly reflected by the cavity and
will not be influenced as much by the resonance condition of the
cavity as the carrier. The photodiode measuring the reflected light
will see the optical beat between the carrier field and the modulation
sidebands. This includes a component at the modulation frequency which
is a measure of the phase difference between the carrier field and the
sidebands (given the setup as described above). Any slight change of
the cavity length would introduce a proportional change in the phase
of the carrier field and no change in the sideband fields.  Thus the
photodiode signal can be used to measure the length changes of the
cavity. One of the advantages of this method is the fact that the
so-generated signal is bipolar with a zero crossing and steep slope
exactly at the cavity's resonance, see
Figure~\ref{fig:fexample_cavity_pdh01}.
\epubtkImage{fexample_cavity_pdh.png}{%
  \begin{figure}[htbp]
    \centerline{\includegraphics[width=0.9\textwidth]{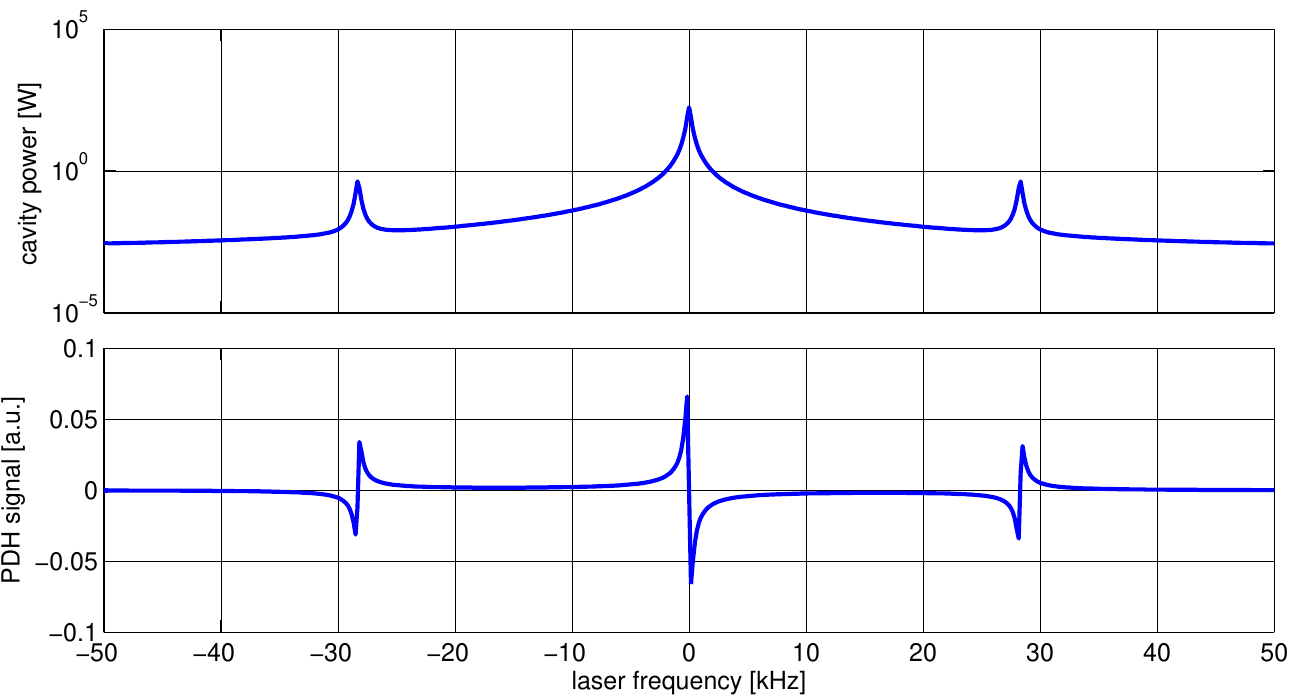}}
    \caption{This figure shows an example of a Pound--Drever--Hall
      (PDH) signal of a two-mirror cavity. The plots refer to a setup
      in which the cavity mirrors are stationary and the frequency of
      the input laser is tuned linearly. The upper trace shows the
      light power circulating in the cavity. The three peaks
      correspond to the frequency tunings for which the carrier (main
      central peak) or the modulation sidebands (smaller side peaks)
      are resonant in the cavity. The lower trace shows the PDH signal
      for the same frequency tuning. Coincident with the peaks in the
      upper trace are bipolar structures in the lower trace. Each of
      the bipolar structures would be suitable as a length-sensing
      signal. In most cases the central structure is used, as
      experimentally it can be easily identified because its slope has
      a different sign compared to the sideband structures.}
    \label{fig:fexample_cavity_pdh01}
\end{figure}}


\subsection{Michelson length sensing}
\label{sec:mi_control}

Similarly to the two-mirror cavity, we can start to understand the
length-sensing capabilities of the Michelson interferometer by looking
at the output light power as a function of a mirror movement, as shown
in Figure~\ref{fig:mi_output}. The power changes as sine squared
with the maximum slope at the point when the output power (in what we
call the South port) is half the input power. The slope of the output
power, which is the \emph{optical gain} of the instrument for
detecting a differential arm-length change $\Delta L$ with a photo
detector in the South port can be written as
\begin{equation}
\frac{d\,S}{d\,\Delta L}=\frac{2\pi P_0}{\lambda}\sin\left(\frac{4\pi}{\lambda}\Delta L\right)
\end{equation}
and is shown in Figure~\ref{fig:mi_slope}. The most notable difference
of the optical gain of the Michelson interferometer with respect to the
Fabry-Perot interferometer (see Figure~\ref{fig:cav_slope}) is the
wider, more smooth distribution of the gain. This is due to the fact that
the cavity example is based on a high-finesse cavity in which the
optical resonance effect is dominant. In a basic Michelson
interferometer such resonance enhancement is not present.

\epubtkImage{example_michelson_slope.png}{%
  \begin{figure}[htbp]
    \centerline{\includegraphics[width=0.9\textwidth]{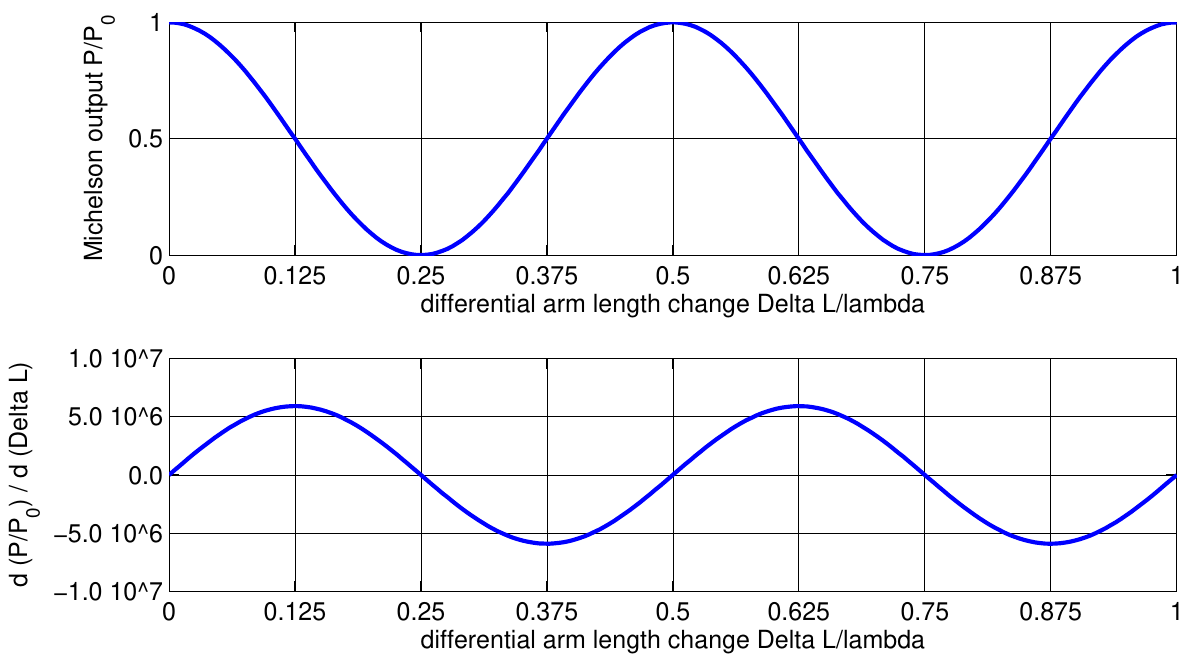}}
    \caption{Power and slope of a Michelson interferometer. The upper
      plot shows the output power of a Michelson interferometer as
      detected in the South port (as already shown in
      Figure~\ref{fig:mi_output}). The lower plot shows the optical
      gain of the instrument as given by the slope of the upper plot.}
    \label{fig:mi_slope}
\end{figure}}

However, the main difference is that the measurement is made
differentially by comparing two lengths. This allows one to separate a
larger number of possible noise contributions, for example noise in
the laser light source, such as amplitude or frequency noise. This is
why the main instrument for gravitational-wave measurements is a
Michelson interferometer. However, the resonant enhancement of light
power can be added to the Michelson, for example, by using
Fabry-Perot cavities within the Michelson as introduced in Section~\ref{sec:MIFP}.
This construction of new topologies by combining Michelson and
Fabry-Perot interferometers has culminated in the dual-recycled Fabry-Perot Michelson
configuration that is the subject of the following section.

The Michelson interferometer has two longitudinal degrees of
freedom (setting aside the optical wavelength as a third degree of freedom).
These can be represented by the positions (along the optical
axes) of the end mirrors. However, it is more efficient to use proper
linear combinations of these and describe the Michelson interferometer
length or position information by the \emph{common} and
\emph{differential arm length}, as introduced in
Equation~(\ref{eq:armlength}):
\begin{equation*}
  \begin{array}{l}
    \bar{L}=\frac{L_N+L_E}{2}\\
    \Delta L=L_N-L_E.\\
  \end{array}
\end{equation*}
The Michelson interferometer is intrinsically insensitive to the
common arm length $\bar{L}$.

\subsection{Advanced LIGO: an example of a complex interferometer}
\label{sec:DRFPMI}
In this section we present a simplified overview of the dual-recycled
Fabry-Perot Michelson interferometer ({\bf DRFPMI}) topology, as
exemplified by the Advanced LIGO detectors~\cite{Harry2010_aligo}.  At
this level of detail, the description applies equally to Advanced
Virgo~\cite{AdvancedVirgo15}.

Our description builds on the ideas presented in
Section~\ref{sec:advanced}.  The DRFPMI configuration is built around
a Michelson interferometer, with 4 cavities added to modify the
behaviour of the system.  As shown in Figure~\ref{fig:DR-FPMI_fields},
there are two Fabry-Perot arm cavities that extend the light path in
the arms of the interferometer to enhance the signal due to
gravitational waves.
\epubtkImage{DR-FPMI_fields.png}{%
  \begin{figure}[htbp]
    \centerline{\includegraphics[width=0.9\textwidth]{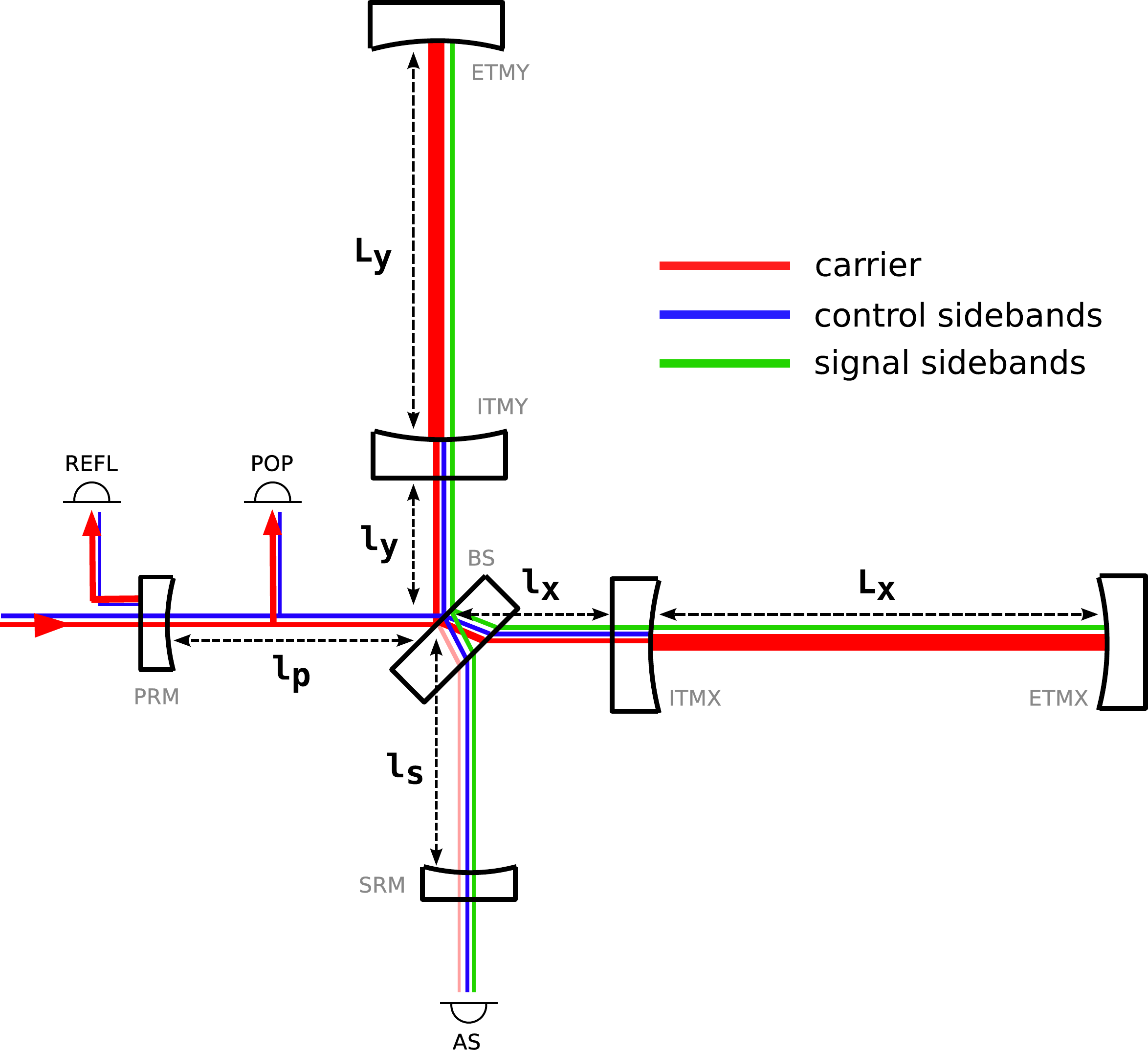}}
   \caption{Schematic illustration of the dual-recycled Fabry-Perot Michelson configuration showing the main optical components (i.e.~6 mirrors and the beam splitter), components of the light field in different regions of the interferometer, photo-detectors and one possible representation of the degrees of freedom. The system is controlled by signals obtained from three photo-detectors: REFL, short for reflected port, and POP, short for pick-off-port detect aspects of the light reflected by the Michelson, while the transmitted light is detected at the anti-symmetric port (AS). The degrees of freedom are indicated by the various lengths `L' and `l' with subscripts described in the text of  the current section.  Note that the lengths marked with capital `L's involve the long arms of the interferometer, while the others involve the short distances from the beam splitter to the nearby components. Further detail of the sensing and control is discussed in Appendix~\ref{sec:ALOL}.}
   \label{fig:DR-FPMI_fields}
\end{figure}}The Michelson is operated at, or very close to, a dark fringe so that, apart from losses, most of the light is reflected back in the direction towards laser and injection optics, hence this input port is also called the `bright' port of the interferometer.  A partially transmitting power-recycling mirror, placed at the bright port and adjusted to resonate the light, allows the power circulating within the interferometer to build up (ideally by a factor of 1/loss), reducing the requirement for input light power.

The final cavity is formed by placing a partially transmitting mirror between the output or `dark' port of the Michelson and the detection optics (consisting of a photo-detector, and perhaps some other components). This mirror recycles light that carries signal information to the photo-detector, and is called the signal recycling mirror -- see Section~\ref{sec:SRDRRSE} for an introduction to this aspect of the interferometer configuration.

The idea of a bright fringe or port and dark fringe or port can be extended to form one of the central concepts in the control of complex interferometers.  In the condition described, with the input or power recycling port maintained in the bright state, and the output or signal recycling port held in the dark state, there is a separation of light-field components to one or other port according to their relative state in the interferometer arms.  Here `component' means light at a single frequency, i.e.~a carrier or a sideband, and in a single optical mode (for a discussion of spatial modes, see Section~\ref{sec:beamshapes}).  Such light-field components, which have spatial and temporal coherence, can interfere. If they have the same phase in the two arms they interfere constructively at the bright port.  If they have the opposite phase in the two arms they interfere constructively at the dark port.  Note that this arises because of the choice of interference of the carrier light to create the bright and dark ports.

In the same way that the carrier light which has a common phase in the two arms appears at the bright port,  any perturbation of the interferometer that is common to the two arms generates higher order modes and/or sidebands that have the same phase in the two arms and thus causes an effect on the optical field at the bright port.  Examples of this would be in-phase arm-length changes, or the addition of the same amount of optical loss in the two arms.  On the other hand, perturbations that are exactly out-of-phase between the two arms have an effect on the light field at the dark port. An example would be that gravitational waves produce differential phase modulation sidebands that have opposite phase in the two arms, and these interfere constructively at the output port.

 The distinction between effects that are either in-phase or have  opposing  phases is frequently important in the control of interferometers.  As noted in the previous section for the case of the simple Michelson, it has become standard to consider the two physical degrees of freedom associated with the arms of an interferometer in logical-combination as the \emph{common mode} and the \emph{differential mode}.  For the same reasons, the bright port is also called the \emph{symmetric port} and the dark port is called the \emph{anti-symmetric port}.

The advantage of the choice of common and differential modes may be seen in consideration of control loops to deal with laser-frequency fluctuations and to keep the interferometer locked at the dark fringe, to give but two examples.  In a nearly-symmetrical interferometer a fluctuation of the frequency of the in-going light will lead to a primarily common-mode effect, and it makes sense to stabilise the laser frequency with respect to the common mode of the two arms.  Similarly the gravitational wave signal may be read-out as part of the error signal of a control loop for the dark fringe. Such a loop should act on the differential mode, rather than on the length of one arm cavity, or the other.

Referring to Figure~\ref{fig:DR-FPMI_fields}, the physical optical path-lengths shown on the diagram may be related to the logical degrees of freedom applied in interferometer control in the following way.  The solution presented here is not unique, but is intended as an example of one way to approach the problem.  We deal with the degrees of freedom in turn,  take the beam splitter as a point of reference for the small 'l' lengths (as suggested on the figure) and assume the laser frequency is fixed.  This leaves 5 degrees of freedom to be controlled:

\begin{itemize}
\item {\bf CARM}  Common-mode arm length, CARM $=L_x+L_y$.
This corresponds to the average length
of the arm cavities and is adjusted to keep both arm-cavities on resonance.
\item {\bf DARM} Differential arm length, DARM $=L_x-L_y$.  This corresponds to the difference
in length of the two arm cavities and is used to maximise the constructive interference,  at the output port, of sidebands resulting from differential arm-length changes (this degree of freedom is therefore the source of the  gravitational wave channel).
\item {\bf MICH} Michelson arm length difference, MICH $=l_x-l_y$.  MICH corresponds to the difference
in length of the short
arms of the Michelson, between the ITMs and the beam splitter,  and determines the state of interference
at the output port.  In Advanced LIGO, the Michelson is operated close to the dark fringe.
\item {\bf PRCL} Power recycling cavity length, PRCL $=L_p +\frac{l_x+l_y}{2}$.  The power recycling cavity
is operated on resonance to maximise the power coupled into the central
interferometer.
\item {\bf SRCL} Signal recycling cavity length, SRCL $=L_s+\frac{l_x+l_y}{2}$.  This corresponds to the
resonance condition of the signal recycling cavity.  The operating point of SRCL depends
on the mode of operation of the interferometer.  It can be tuned for a particular frequency
of gravitational wave or for broadband operation.
\end{itemize}

As a reminder, we restate that in a gravitational wave detector, we are concerned with microscopic variations of path lengths that may be up to several km.

In the following sections we discuss, in general terms, how the error signals can be extracted from the optical system, combined and processed to provide signals representing the degrees of freedom to be controlled, and how the resulting signals can be fed-back to force the optical system into the desired condition.


\subsection{The Schnupp modulation scheme}
\label{sec:Schnupp}
In this and the following three sections (\ref{sec:extpdh}, \ref{sec:intextdither}, \ref{sec:dc}), we introduce techniques for reading out signals from interferometers.  These approaches complement and extend the Pound-Drever-Hall method for readout from Fabry-Perot cavities presented in Section~\ref{sec:pdh}.

Similar to the Fabry-Perot cavity, the Michelson
interferometer is also often used to set an operating point where the
optical gain of a direct light power detection is zero. This operating
point, given by $\Delta L/\lambda=(2N+1)\cdot0.25$ with $N$ a
non-negative integer, is called \emph{dark fringe}. This operating
point has several advantages, the most important being the low
(ideally zero) light power on the diode. Highly efficient and
low-noise photodiodes usually use a small detector area and thus are
typically not able to detect large power levels. By using the dark
fringe operating point, the Michelson interferometer can be used as a
\emph{null instrument} or \emph{null measurement}, which generally is a
good method to reduce systematic errors~\cite{Saulson_book}.

One approach to make use of the advantages of the dark fringe
operating point is to use an operating point very close to the dark
fringe at which the optical gain is not yet zero. In such a scenario a
careful trade-off calculation can be done by computing the
signal-to-noise with noises that must be suppressed, such as the laser
amplitude noise. This type of operation is usually referred to as
\emph{DC control} or \emph{offset control} and is very similar to the
similarly-named mechanism used with Fabry-Perot cavities.

\epubtkImage{michelson_schnupp01.png}{%
  \begin{figure}[htbp]
    \centerline{\includegraphics{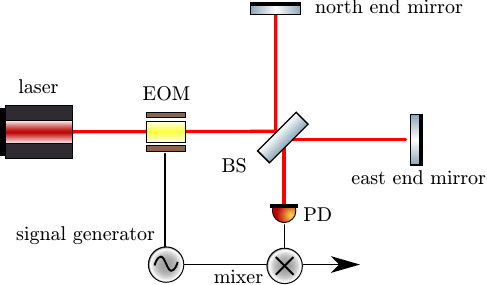}}
    \caption{This length sensing scheme is often referred to as
      \emph{frontal} or \emph{Schnupp modulation}: an EOM is used to
      phase modulate the laser beam before entering the Michelson
      interferometer. The signal of the photodiode in the South port
      is then demodulated at the same frequency used for the
      modulation.}
    \label{fig:michelson_schnupp}
\end{figure}}

Another option is to employ phase modulated light, similar to the
Pound--Drever--Hall scheme described in Section~\ref{sec:pdh}. The
optical layout of such a scheme is depicted in
Figure~\ref{fig:michelson_schnupp}: an electro-optical modulator is
used to apply a phase modulation at a fixed frequency, usually in the RF range,
to the monochromatic laser light before it enters the
interferometer. The photodiode signal from the interferometer output
is then demodulated at the same frequency.  This scheme allows one to
operate the interferometer precisely on the dark fringe. The method
originally proposed by Lise Schnupp is also sometimes referred to as
\emph{frontal modulation}.

The optical gain of a Michelson interferometer with Schnupp modulation
is shown in Figure~\ref{fig:fexample_michelson_schnupp} in
Section~\ref{sec:finesse_lengthsensing}.


\subsection{Extending the Pound-Drever-Hall technique to more complicated optical systems}
\label{sec:extpdh}
To recap Section~\ref{sec:pdh}, in the Pound-Drever-Hall (PDH) or
RF-reflection locking technique sinusoidal radio-frequency phase
modulation is applied to the light to produce phase modulation
sidebands.  With phase modulation, higher order sidebands are imposed
on the light, though the beats due to these are generally not employed
in the normal implementation of the Pound-Drever-Hall technique. The light is then incident on the
cavity that is to be controlled.  The signal is obtained by detecting
the reflected light on a photo-detector which has a square-law
response to the light amplitude, and analysing the resulting beats.
The important beats are between the carrier and the first order RF
sidebands.  The electronic signal from the photodiode is filtered to
pass the beats in a frequency range around the modulation frequency,
and multiplied or `mixed' with an electronic signal at the modulation
frequency: an electronic local oscillator.  The
output from the mixer is then low-pass filtered to remove oscillations
at harmonics of the modulation frequency.  The useful signal is in one
quadrature of the output from the photo-detector at the modulation
frequency.  The phase of the local oscillator is chosen to select the
required quadrature.

During the 1980s and 1990s, the question arose of how to obtain
control signals for systems of coupled cavities and systems with
combination of cavities in a Michelson interferometer. A good example
is the power-recycled Fabry-Perot Michelson interferometer
configuration as employed in initial LIGO and Virgo. In such a system,
one possibility is to add pick-offs (low-reflectivity beam splitters)
to remove some of the light reflected from each arm cavity for
detection.  This approach introduces a conflict between efficient
power recycling that requires low loss, and the generation of a
low-noise control signal, which argues for more highly reflecting
beam splitters.  It is of interest to identify other approaches that
do not require additional detection ports. With this restriction, the
problem becomes one of sensing all internal degrees of freedom by
analysing light fields reflected from or transmitted by the entire
interferometer.  This has been accomplished for the dual-recycled
Michelson topology of GEO\,600~\cite{phd.Grote}, for the dual-recycled
Fabry-Perot Michelson configuration, e.g.~Advanced LIGO -- see Appendix~\ref{sec:ALOL}
for a short description of the sensing scheme.

The scope of this section permits discussion only of design
principles.  It is worth noting, however, some practical matters that
constrain the acceptable solutions.  For example, the choice of
modulation frequencies is usually restricted.  One limit is the speed
of photo-detectors, and in particular quadrant photo-detectors, for
alignment sensing.  This restriction makes the use of modulation
frequencies above of order 100\,MHz highly challenging.  Another
limitation results from the presence of mode-cleaning cavities in the
path from the laser to the interferometer. Due to practical
difficulties in the design of in-vacuum modulators, modulation is
usually applied prior to the light passing the mode-cleaner.  In this
case the only available modulation frequencies are whole-multiples of
the free-spectral-range of the mode-cleaner.  Note however that
in-vacuum modulation is possible, and has been applied in GEO\,600
allowing a relatively free choice of modulation frequency which is
important for the method of locking the dual recycled
system~\cite{phd.Grote}.

The essence of the Pound-Drever-Hall method is that the light field is divided,
according to frequency, into a component that suffers a phase change
in response to variation of the target degree of freedom for
measurement, and a component that does not.  Therefore, a starting
point in the discovery of alternatives is to create circumstances in
which different light components, distinguished by frequency, resonate
in different locations.  Secondly, to produce a useful error signal,
the output from the detection process should contain a dominant linear
component in terms of its magnitude as a function of the target degree
of freedom.  Although it is desirable that the signal crosses zero at
the operating point, it may be necessary and acceptable to subtract a
(hopefully steady) offset to obtain the required result.  These
aspects are dealt with in turn.

First we consider how zero-crossing signals may be obtained from beats. The desired
zero-crossing linear slope is achieved most directly if the components
of the light are in quadrature, as is the case in Pound-Drever-Hall sensing: see
sections~\ref{sec:phasemod} and~\ref{sec:pdh}. This ensures that the
measurement depends on the relative phase of the optical field
components, rather than their amplitudes.  As an example of an
alternative, quadrature is also achieved in the case of beating
amplitude modulation sidebands against phase modulation sidebands.

In cases like this, where beats are obtained
between various sidebands, rather than by beating with the carrier,
the demodulated signal may either be obtained directly by mixing the
electronic signal with a local oscillator at the beat frequency, or by
employing double demodulation.  A description of this process is
shown in Section~\ref{sec:demodulation}.

The condition for quadrature requires pairs of sidebands to be
symmetrical so that they represent either pure phase modulation or pure
amplitude modulation.  In either of these cases, their resultant sum maintains a
constant phase over time. If there is an imbalance of the amplitude of
the lower and upper sidebands, the phase of the resultant must
oscillate. This is equivalent to saying that the sidebands represent a
mixture of amplitude and phase modulation, or equivalently, that there
is an unbalanced single-sideband component.  Extraction of useful
error signals is still possible, but it is to be expected that there
will be an offset in the demodulated signal, rather than a
zero-crossing at the desired resonance condition.

Such sideband imbalance arises naturally in interferometers with
detuned signal recycling, see Section~\ref{sec:SRDRRSE}.
In these interferometers, the resonance of the signal
recycling cavity is not centred on the carrier and so the response to
upper and lower modulation sidebands can be expected to be
asymmetrical.  The beats produced on detection of the unbalanced
sidebands may still produce a useful linear component, corresponding
to the part of the amplitude that is in the appropriate quadrature.

As an example of obtaining signals from beats between sidebands, we
cite the important method called {\em third-harmonic demodulation},
introduced and explained in detail in~\cite{Arai2000}. In brief
summary, this technique exploits the natural presence of higher
harmonics in phase modulation for moderate to large modulation
indices, e.g.~0.8\,rad in the cited work.  As noted above, such
harmonics are passed by a mode-cleaner that is resonant at the first
harmonic, and depending on the design of the interferometer, at least
some can be expected to be resonant in the power recycling cavity (the
odd members of the series in the scheme described by Arai~{\em et
al.} in~\cite{Arai2000}. By combining this method of demodulation with the introduction
of asymmetry in the geometry of the interferometer, as described in
the following section, it is possible to construct a sensing system
that provides well separated readout of the various degrees of
freedom.  In the cited scheme, neither the first or third order
sidebands are strongly affected by the phase of the arm cavities (when
the carrier is on resonance), and the method allows relatively
independent control of the other degrees of freedom.

The third harmonic demodulation approach has been extended, with
results proven in a series of investigations on prototype
interferometers, including a 4\,m interferometer with resonant
sideband enhancement~\cite{Kawazoe2006}, and experiments on the
CalTech 40\,m apparatus~\cite{Miyakawa2006} as part of the development of
control systems for Advanced LIGO, in which third-harmonic
demodulation is employed -- see Appendix~\ref{sec:ALOL}.

Next we return to the question of how sideband fields may be separated by
breaking the symmetry of the interferometer.  To reduce noise couplings, interferometers are usually designed and
built to be as symmetrical as possible.  For instance, an
interferometer with perfectly matched arms is insensitive to the
frequency of the light. In the design process it is usually assumed
that the long arms of the interferometer must be kept as symmetrical
as can be arranged in practice, but that controlled amounts of
asymmetry can be introduced in the paths from the beam splitter to the
arm cavities or recycling mirrors as appropriate to facilitate the
design of sensing schemes.

The methods discussed in this section stem from the Schnupp modulation
technique described in Section~\ref{sec:Schnupp}.  In the unmodified
Michelson interferometer, shown in Figure~\ref{fig:michelson_schnupp},
the asymmetry required to maximise the strength of the sidebands at
the output, with modulation frequencies in the usual range (typically
10 to 100\,MHz) is one quarter of the RF wavelength.  The addition of
power-recycling lowers the required asymmetry because in this case
optimum transfer of sideband power occurs when the asymmetry leads to
an out-coupling of equal strength to the transmission of the power
recycling mirror.  This is in direct analogy with the transmission of
light through an equal-mirror Fabry-Perot cavity.

An example of this `classical' application of Schnupp modulation is
found in GEO\,600.  Here the approximately 1200\,m (optical path) arms
are adjusted to differ in length by about 10\,cm, and this provides
efficient transfer of $\approx 15\,$MHz sidebands to the output
port. The approach is described in~\cite{phd.Grote}.

The idea of Schnupp modulation influenced the development of Advanced
LIGO see, for example,~\cite{Strain2003}. It had been
decided that phase modulation would be applied prior to the in-vacuum
mode-cleaner, thus constraining the modulation sidebands to fall in a
harmonic series. A detailed description of these methods is beyond the
scope of this review, but some important features are
described below.

The objective is always to cause distinct modulation sidebands to
resonate in different physical regions within the interferometer.  In
a dual-recycling Fabry-Perot Michelson configuration, it is necessary
to control the (inner) Michelson, the power recycling cavity and the
signal recycling cavity.  Controlling the arm cavities may be achieved
by beating the carrier with suitable sidebands, the hard part of the
problem is to remove the influence of arm cavities on signals for the
other degrees of freedom.  For control of the signal recycling cavity,
for example, at least one sideband must be directed towards the signal
recycling mirror.  This can be accomplished by choosing a difference
in the lengths of the two arms of the Michelson to arrange that one
sideband is on a bright fringe, and therefore strongly directed
towards the signal recycling mirror. For further detail of this
aspect of interferometer sensing, see~\cite{Strain2003}
and Appendix~\ref{sec:ALOL}.

One last design ingredient is that, in a `closed' configuration like
the dual-recycling Fabry-Perot Michelson, light travelling back from
one of the arms `sees' another (effective) Michelson interferometer
formed by the beam splitter and the two recycling mirrors.  A
variation of the Schnupp technique can also be applied in that case,
by adjusting the optical paths from the beam splitter to the recycling
mirrors to be unequal.  This provides further control over sideband
resonance conditions in the various parts of the interferometer.

It can be appreciated that the design problem rapidly becomes too
complex for a full description in this review, but all of the main
principles are included, and numerical calculation allows these
principles to be developed into a complete sensing scheme.

\subsection{Complementary techniques: internal modulation, external modulation and dithering}
\label{sec:intextdither}
For completeness we review a range of methods that have been applied
in interferometry for gravitational wave detection.  The ideas follow on from the
basic RF heterodyne methods introduced in Section~\ref{sec:signal_readout}.
The first RF-modulation based signal readout scheme for a Michelson
interferometer involved generating the RF sidebands in phase
modulators placed into the arms of the interferometer, as shown in Figure~\ref{fig:michelson_internal}.

\epubtkImage{michelson_internal.png}{%
  \begin{figure}[htbp]    
    \centerline{\includegraphics[width=0.6\textwidth]{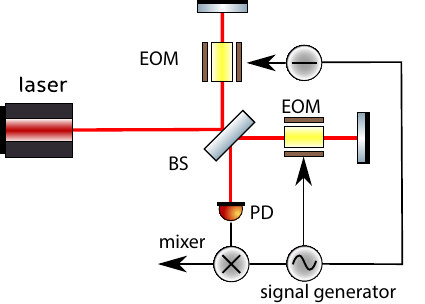}}
    \caption{Michelson interferometer with internal modulation.  Phase modulators are placed in the arms
of the interferometer and driven sinusoidally in opposing phase at a radio-frequency.  The strength of the
modulation is chosen such that the light field at the output of the interferometer, at the dark fringe is strongly
dominated by the modulation sidebands.  Since the sidebands are applied differentially, they appear predominantly
at the anti-symmetric port when the interferometer output is at the dark fringe for the carrier light. If, as shown here,
the light passes the modulators in both directions, the position of the modulators and the frequency chosen must be
taken into account to avoid unwanted cancellation or enhancement of the effect. }
    \label{fig:michelson_internal}
\end{figure}}

Although this
technique, called {\em internal modulation} was shown to be successful
in interferometry up to the late 1980s~\cite{Shoemaker1988}, it has not been
possible to devise an implementation that operates with the low noise
levels required for modern detectors.  A related concept, dithering of
interferometer mirrors to phase modulate the light within the
interferometer, is described below.  See also
Section~\ref{sec:mirror_mod}.

\epubtkImage{michelson_external.png}{%
  \begin{figure}[htbp]    
    \centerline{\includegraphics[width=0.6\textwidth]{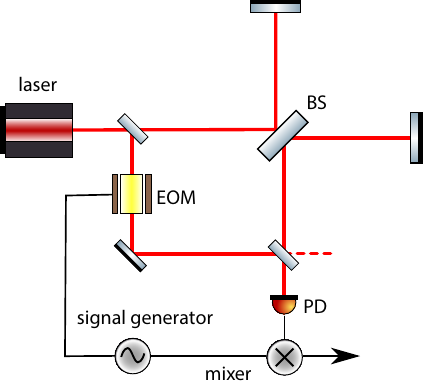}}
    \caption{Michelson interferometer with external modulation.  In this version of external modulation, a sample of the in-going  light is picked off, phase modulated and recombined with the light emerging from the anti-symmetric port in a Mach-Zehnder arrangement.  In an interferometer with power recycling, the light to be modulated may be extracted from within the power-recycling cavity, where the filtering action of the cavity may render it more stable.  As with internal modulation, the sidebands should dominate the detected light.  In that case to improve efficiency and minimise the amount of light that is extracted from the power recycling cavity, detectors may be placed at both ports of the Mach-Zehnder interferometer, and the resulting signals subtracted prior to demodulation.  In the case of external modulation, the path-lengths involved are normally small compared to the RF wavelength. }
    \label{fig:michelson_external}
\end{figure}}
In the technique of {\em external modulation}~\cite{Man1990}, a phase modulated field
is derived from the common mode light within the interferometer as shown in Figure~\ref{fig:michelson_external}.
Light picked-off from a convenient location, usually close to the
beam splitter or even at its imperfectly anti-reflection coated rear
surface, is phase modulated and recombined with the main output field,
by means of a second beam splitter.  This Mach-Zehnder interferometer
geometry is distinguished from general heterodyne methods in that,
when power recycling is present, the modulated field is obtained from
within the power recycling cavity, where the light field may be more
stable than the ingoing light, due to the passive filtering provided
by the power recycling cavity. External modulation adds significant
complexity to the output optics of an interferometer, and is
disfavoured in advanced interferometers where the application of
squeezed light is considered.

Another approach to the generation of suitable signals is {\em
dithering}, this is, effectively, the application of phase modulation
sidebands by modulating parameters of the system, usually the positions or angles of mirrors, rather than
modulating the ingoing light.  In principle, dithering could be
applied at distinct frequencies to as many components of the system
as there are degrees of freedom requiring to be controlled.

There are
practical limitations that restrict the application of dithering, and it is normally applied to lock
auxiliary degrees of freedom where the signal to noise requirements
are less severe.  The limitations arise because dithering is commonly applied by mechanical means,
resulting in restricted actuation force (to avoid either causing
damage or adding noise due from powerful actuators).   This imposes a limit to the product of imposed
displacement and (dither-) frequency-squared, resulting in typical dither frequencies that do not
exceed a few kHz.  Dithering is, therefore, typically employed to monitor and
control slowly varying aspects of the interferometer.  A relatively recent application
of dithering is in locking an output mode-cleaner for use with DC
readout.  This is discussed in the following section and in
\cite{Ward2008}.

\subsection{Circumstances in which offset locking is favoured over modulation\-based techniques}
\label{sec:dc}
As mentioned in Sections~\ref{sec:signal_readout} and~\ref{sec:Schnupp}, the idea of
offset-locking of Michelson interferometers to produce a zero-crossing
error signal for the differential displacement arises naturally.  There are, however,
disadvantages associated with this method of readout, and it has only become favoured
over heterodyne methods due to particular circumstances that associated with recently
developed interferometer designs, as explained below.

In a simple Michelson
interferometer, the steepest gradient in the length to intensity
transfer function occurs half-way-up the fringe. However, operating in this
condition has two disadvantages: half of the light is directed back
towards the laser and sensitivity to laser power fluctuations is
maximised.  The latter problem can be ameliorated by symmetrising the
readout through the addition a photo-detector for the reflected light.
On subtracting the signals from the detectors at the two ports of the
interferometer, the displacement signals add while laser power
fluctuations cancel, to the extent that balance is achieved. In this
case, however, all the light is detected and there is no possibility to
take advantage of low-loss optics by adding power recycling.

A further problem when a simple Michelson is offset-locked is that
the optical local oscillator for the measurement is a relatively noisy
component of the light field.  Indeed this last concern led to the
choice of radio frequency modulation in the Pound-Drever-Hall and other techniques
described above.  In those techniques modulation frequencies are
chosen to fall at Fourier frequencies where technical laser noise is
less than shot noise in the detected light power. This is typically
true above about 10\,MHz for detection of the tens of mW of light from
the argon-ion or Nd:YAG lasers typically employed.

During the design of Enhanced and Advanced LIGO, Advanced Virgo and
GEO-HF, three motivations emerged to prompt reconsideration of
offset-locking methods.  As noted in Section~\ref{sec:mod} it had been
shown that modulation generally worsens shot-noise limited
performance, and these arguments were extended to show that it is
impractical to benefit from squeezed light in modulation based
readout~\cite{Buonanno2003}.  Secondly, it was realised that, for the
interferometer to achieve the planned sensitivity, the light within
the power recycling cavity in a system such as Advanced LIGO, must be
more stable than the best available RF oscillators, at Fourier
frequencies of interest, and so the arguments against employing this
light for signal readout scheme become moot.  Finally, whether the
detected light amplitude is shot noise limited or not depends on the
power that is detected, because the shot noise in the detection of
small light power can make technical noise unimportant.  It was
realised that, by adding a mode-cleaner on the output of the
interferometer, to pass the signal, which would predominantly be in
the $\mathrm{TEM}_{00}$ mode of the arms, but exclude other light
resulting from imperfect interference, mainly in other modes, it would
suffice to detect relatively low light power, at which level the
measurement should be shot noise limited.  See
Section~\ref{sec:imperfect} for a description of modes resulting from
imperfect interference.

In modern detectors this scheme, where signals are read out directly
in the base-band i.e.~near zero frequency or `DC', is often called
{\em DC readout}. As an example of its application, the details of the
DC readout scheme developed for Advanced LIGO are described in~\cite{Ward2008}. The
technique has also been tested on GEO\,600, where the method has been
shown to be compatible with squeezing~\cite{Abadie2011}.

It should be noted that offset locking applies to the control of one
length degree of freedom per interferometer, and the remaining degrees
of freedom are typically sensed using the modulation methods described
above.

\subsection{\Finesse examples}
\label{sec:finesse_lengthsensing}

\subsubsection{Michelson modulation}

\epubtkImage{fexample_michelson_modulation.png}{%
  \begin{figure}[htbp]
    \centerline{\includegraphics[width=0.9\textwidth]{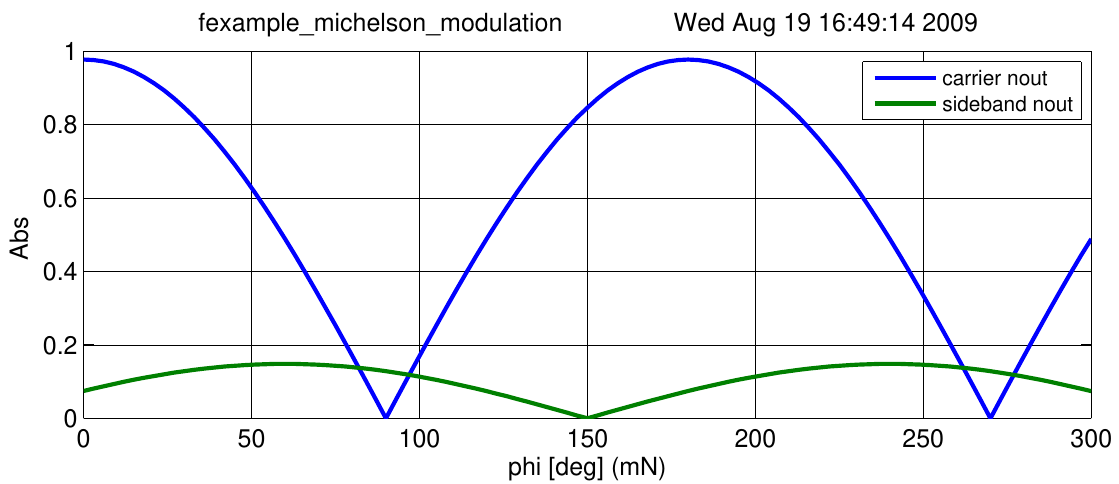}}
    \caption{\Finesse example: Michelson modulation.}
    \label{fig:fexample_michelson_modulation}
\end{figure}}

\noindent
This example demonstrates how a macroscopic arm length difference can
cause different `dark fringe' tuning for injected fields with
different frequencies. In this case, some of the 10~MHz modulation
sidebands are transmitted when the interferometer is tuned to a dark
fringe for the carrier light. This effect can be used to separate
light fields of different frequencies. It is also the cause for
transmission of laser noise (especially frequency noise) into the
Michelson output port when the interferometer is not perfectly
symmetric.

\vspace{3mm}\noindent
{\small
\textbf{Finesse input file for `Michelson modulation'}
{\renewcommand{\baselinestretch}{.8}

\nopagebreak
\tt
\noindent
\mbox{} \\
\mbox{}\textbf{\textcolor{RoyalBlue}{laser}}\ l1\ \textcolor{Purple}{1}\ \textcolor{Purple}{0}\ \ n1\ \ \ \ \ \ \textcolor{Gray}{\%\ laser\ with\ P=1W\ at\ the\ default\ frequency} \\
\mbox{}\textbf{\textcolor{RoyalBlue}{space}}\ s1\ \textcolor{Purple}{1}\ \textcolor{Purple}{1}\ n1\ n2\ \ \ \ \textcolor{Gray}{\%\ space\ of\ 1m\ length} \\
\mbox{}\textbf{\textcolor{RoyalBlue}{mod}}\ eom1\ 10M\ \textcolor{Purple}{0.3}\ \textcolor{Purple}{1}\ pm\ n2\ n3\ \textcolor{Gray}{\%\ phase\ modulation\ at\ 10\ MHz} \\
\mbox{}\textbf{\textcolor{RoyalBlue}{space}}\ s2\ \textcolor{Purple}{1}\ \textcolor{Purple}{1}\ n3\ n4\ \ \ \ \textcolor{Gray}{\%\ another\ space\ of\ 1m\ length} \\
\mbox{}\textbf{\textcolor{RoyalBlue}{bs}}\ b1\ \textcolor{Purple}{0.5}\ \textcolor{Purple}{0.5}\ \textcolor{Purple}{0}\ \textcolor{Purple}{0}\ n4\ nN1\ nE1\ nS1\ \textcolor{Gray}{\%\ 50:50\ beam\ splitter\ } \\
\mbox{}\textbf{\textcolor{RoyalBlue}{space}}\ \ LN\ \textcolor{Purple}{100}\ \textcolor{Purple}{1}\ nN1\ nN2\ \ \textcolor{Gray}{\%\ north\ arm} \\
\mbox{}\textbf{\textcolor{RoyalBlue}{space}}\ \ LE\ \textcolor{Purple}{110}\ \textcolor{Purple}{1}\ nE1\ nE2\ \ \textcolor{Gray}{\%\ east\ arm} \\
\mbox{}\textbf{\textcolor{RoyalBlue}{mirror}}\ mN\ \textcolor{Purple}{1}\ \textcolor{Purple}{0}\ \textcolor{Purple}{0}\ nN2\ dump\ \textcolor{Gray}{\%\ north\ end\ mirror,\ lossless} \\
\mbox{}\textbf{\textcolor{RoyalBlue}{mirror}}\ mE\ \textcolor{Purple}{1}\ \textcolor{Purple}{0}\ \textcolor{Purple}{0}\ nE2\ dump\ \textcolor{Gray}{\%\ east\ end\ mirror,\ lossless} \\
\mbox{}\textbf{\textcolor{RoyalBlue}{space}}\ \ s3\ \textcolor{Purple}{1}\ \textcolor{Purple}{1}\ nS1\ nout\  \\
\mbox{}\textbf{\textcolor{RoyalBlue}{ad}}\ carrier\ \textcolor{Purple}{0}\ nout\ \ \ \ \ \ \ \ \textcolor{Gray}{\%\ amplitude\ detector\ for\ carrier\ field} \\
\mbox{}\textbf{\textcolor{RoyalBlue}{ad}}\ sideband\ 10M\ nout\ \ \ \ \ \textcolor{Gray}{\%\ amplitude\ detector\ for\ +10\ MHz\ sideband} \\
\mbox{}\textbf{\textcolor{Red}{xaxis}}\ mN\ phi\ lin\ \textcolor{Purple}{0}\ \textcolor{Purple}{300}\ \textcolor{Purple}{100}\ \textcolor{Gray}{\%\ changing\ the\ microscopic\ position\ of\ mN} \\
\mbox{} \\
\mbox{}

}}

\subsubsection{Cavity power and slope}

\epubtkImage{fexample_cavity_power_slope.png}{%
  \begin{figure}[htbp]
    \centerline{\includegraphics[width=0.9\textwidth]{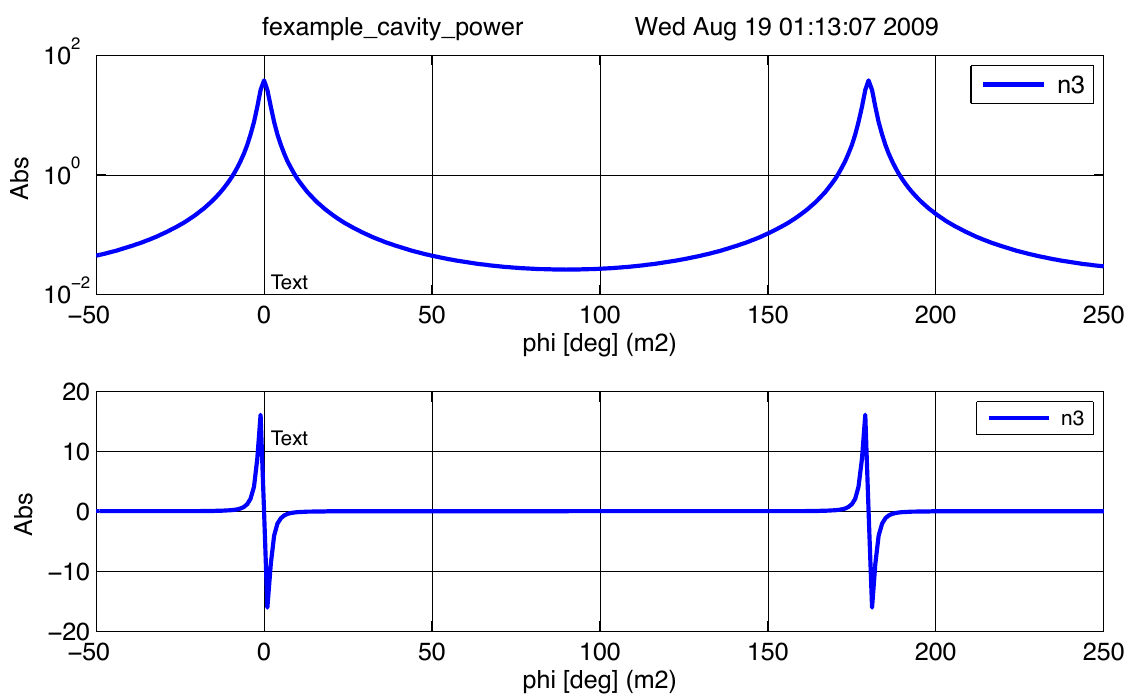}}
    \caption{\Finesse example: Cavity power and slope.}
    \label{fig:fexample_cavity_power_slope}
\end{figure}}

\noindent
Figure~\ref{fig:fexample_cavity_power_slope} (same as
Figure~\ref{fig:cav_slope}) shows a plot of the analytical functions
describing the power inside a cavity and its differentiation by the
cavity tuning. This example recreates the plot using a numerical model
in \Finesse.

\vspace{3mm}\noindent
{\small
\textbf{Finesse input file for `Cavity power and slope'}
{\renewcommand{\baselinestretch}{.8}

\nopagebreak
\tt
\noindent
\mbox{} \\
\mbox{}\textbf{\textcolor{RoyalBlue}{laser}}\ \ l1\ \textcolor{Purple}{1}\ \textcolor{Purple}{0}\ n1\ \ \ \ \textcolor{Gray}{\%\ laser\ with\ P=1W\ at\ the\ default\ frequency} \\
\mbox{}\textbf{\textcolor{RoyalBlue}{space}}\ \ s1\ \textcolor{Purple}{1}\ \textcolor{Purple}{1}\ n1\ n2\ \textcolor{Gray}{\%\ space\ of\ 1m\ length} \\
\mbox{}\textbf{\textcolor{RoyalBlue}{mirror}}\ m1\ \textcolor{Purple}{0.9}\ \textcolor{Purple}{0.1}\ \textcolor{Purple}{0}\ n2\ n3\ \ \ \textcolor{Gray}{\%\ cavity\ input\ mirror} \\
\mbox{}\textbf{\textcolor{RoyalBlue}{space}}\ \ L\ \textcolor{Purple}{1200}\ \textcolor{Purple}{1}\ n3\ n4\ \ \ \ \ \ \ \textcolor{Gray}{\%\ cavity\ length\ of\ 1200m} \\
\mbox{}\textbf{\textcolor{RoyalBlue}{mirror}}\ m2\ \textcolor{Purple}{1.0}\ \textcolor{Purple}{0.0}\ \textcolor{Purple}{0}\ n4\ dump\ \textcolor{Gray}{\%\ cavity\ output\ mirror} \\
\mbox{}\textbf{\textcolor{RoyalBlue}{pd}}\ \ \ \ \ P\ n3\ \ \ \ \ \ \ \ \ \textcolor{Gray}{\%\ photo\ diode\ measuring\ the\ intra-cavity\ power\ } \\
\mbox{} \\
\mbox{}\textcolor{Gray}{\%\ for\ the\ plot\ we\ perform\ two\ sequenctial\ runs\ of\ Finesse\ using\ `mkat'} \\
\mbox{}\textcolor{Gray}{\%\ 1)\ first\ trace:\ plot\ the\ power\ (switching\ to\ log\ plot)} \\
\mbox{}\textbf{\textcolor{Red}{yaxis}}\ log\ abs \\
\mbox{}\textcolor{Gray}{\%\ 2)\ second\ trace:\ plot\ the\ differentiation} \\
\mbox{}\textcolor{Gray}{\%diff\ m2\ phi\ } \\
\mbox{} \\
\mbox{}\textbf{\textcolor{Red}{xaxis}}\ m2\ phi\ lin\ \textcolor{BrickRed}{-}\textcolor{Purple}{50}\ \textcolor{Purple}{250}\ \textcolor{Purple}{300}\ \textcolor{Gray}{\%\ changing\ the\ microscopic\ tuning\ of\ mirror\ m2\ } \\
\mbox{} \\
\mbox{}

}}

\subsubsection{Michelson with Schnupp modulation}

\epubtkImage{fexample_michelson_schnupp.png}{%
  \begin{figure}[htbp]
    \centerline{\includegraphics[width=0.9\textwidth]{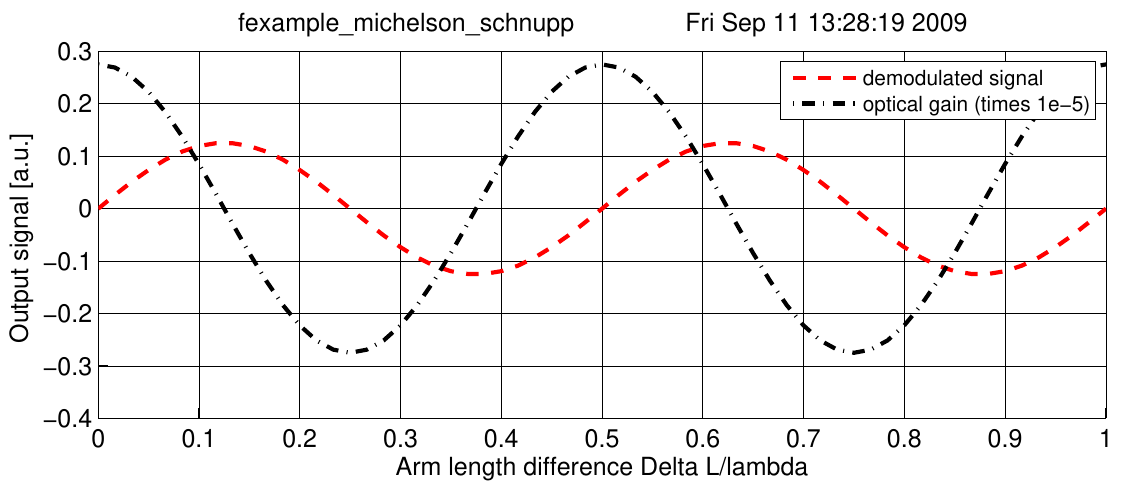}}
    \caption{\Finesse example: Michelson with Schnupp modulation.}
    \label{fig:fexample_michelson_schnupp}
\end{figure}}

\noindent
Figure~\ref{fig:fexample_michelson_schnupp} shows the demodulated
photodiode signal of a Michelson interferometer with Schnupp
modulation, as well as its differentiation, the latter being the
optical gain of the system. Comparing this figure to
Figure~\ref{fig:mi_slope}, it can be seen that with Schnupp modulation,
the optical gain at the dark fringe operating points is maximised and
a suitable error signal for these points is obtained.

\vspace{3mm}\noindent
{\small
\textbf{Finesse input file for `Michelson with Schnupp modulation'}
{\renewcommand{\baselinestretch}{.8}

\nopagebreak
\tt
\noindent
\mbox{} \\
\mbox{}\textbf{\textcolor{RoyalBlue}{laser}}\ l1\ \textcolor{Purple}{1}\ \textcolor{Purple}{0}\ \ n1\ \ \ \ \ \ \textcolor{Gray}{\%\ laser\ with\ P=1W\ at\ the\ default\ frequency} \\
\mbox{}\textbf{\textcolor{RoyalBlue}{space}}\ s1\ \textcolor{Purple}{1}\ \textcolor{Purple}{1}\ n1\ n2\ \ \ \ \textcolor{Gray}{\%\ space\ of\ 1m\ length} \\
\mbox{}\textbf{\textcolor{RoyalBlue}{mod}}\ eom1\ 10M\ \textcolor{Purple}{0.3}\ \textcolor{Purple}{1}\ pm\ n2\ n3\ \textcolor{Gray}{\%\ phase\ modulation\ at\ 10\ MHz} \\
\mbox{}\textbf{\textcolor{RoyalBlue}{space}}\ s2\ \textcolor{Purple}{1}\ \textcolor{Purple}{1}\ n3\ n4\ \ \ \ \textcolor{Gray}{\%\ another\ space\ of\ 1m\ length} \\
\mbox{}\textbf{\textcolor{RoyalBlue}{bs}}\ b1\ \textcolor{Purple}{0.5}\ \textcolor{Purple}{0.5}\ \textcolor{Purple}{0}\ \textcolor{Purple}{0}\ n4\ nN1\ nE1\ nS1\ \textcolor{Gray}{\%\ 50:50\ beam\ splitter\ } \\
\mbox{}\textbf{\textcolor{RoyalBlue}{space}}\ \ LN\ \textcolor{Purple}{100}\ \textcolor{Purple}{1}\ nN1\ nN2\ \ \textcolor{Gray}{\%\ north\ arm} \\
\mbox{}\textbf{\textcolor{RoyalBlue}{space}}\ \ LE\ \textcolor{Purple}{110}\ \textcolor{Purple}{1}\ nE1\ nE2\ \ \textcolor{Gray}{\%\ east\ arm} \\
\mbox{}\textbf{\textcolor{RoyalBlue}{mirror}}\ mN\ \textcolor{Purple}{1}\ \textcolor{Purple}{0}\ \textcolor{Purple}{22}\ \ nN2\ dump\ \textcolor{Gray}{\%\ north\ end\ mirror,\ lossless} \\
\mbox{}\textbf{\textcolor{RoyalBlue}{mirror}}\ mE\ \textcolor{Purple}{1}\ \textcolor{Purple}{0}\ \textcolor{BrickRed}{-}\textcolor{Purple}{22}\ \ nE2\ dump\ \textcolor{Gray}{\%\ east\ end\ mirror,\ lossless} \\
\mbox{}\textbf{\textcolor{RoyalBlue}{space}}\ \ s3\ \textcolor{Purple}{1}\ \textcolor{Purple}{1}\ nS1\ nout\  \\
\mbox{} \\
\mbox{}\textbf{\textcolor{RoyalBlue}{pd}}\textcolor{Purple}{1}\ South\ 10M\ \textcolor{BrickRed}{-}\textcolor{Purple}{115}\ nout\ \textcolor{Gray}{\%\ demodulated\ output\ signal} \\
\mbox{}\textcolor{Gray}{\%\ for\ the\ second\ (black)\ trace,\ we\ add\ differentiation} \\
\mbox{}\textcolor{Gray}{\%diff\ mN\ phi\ \%\ computing\ the\ slope\ of\ the\ signal} \\
\mbox{} \\
\mbox{}\textbf{\textcolor{Red}{xaxis}}\ mN\ phi\ lin\ \textcolor{Purple}{0}\ \textcolor{Purple}{300}\ \textcolor{Purple}{100}\ \textcolor{Gray}{\%\ changing\ the\ microscopic\ position\ of\ mN} \\
\mbox{}\textbf{\textcolor{Red}{put}}\ mE\ phi\ \textcolor{ForestGreen}{\$mx1}\ \ \ \ \ \ \ \ \ \ \ \ \textcolor{Gray}{\%\ moving\ mE\ as\ -mN\ to\ make\ a\ differential\ motion} \\
\mbox{}

}}

\newpage
\section{Beam Shapes: Beyond the Plane Wave Approximation}
\label{sec:beamshapes}

In previous sections we have introduced a notation for describing the
on-axis properties of electric fields. Specifically, we have described
the electric fields along an optical axis as functions of frequency
(or time) and the location \z. Models of optical systems may often use
this approach for a basic analysis even though the respective
experiments will always include fields with distinct off-axis beam
shapes. A more detailed description of such optical systems needs to
take the geometrical shape of the light field into account.
One method of treating the transverse beam geometry is to describe the
spatial properties as a sum of `spatial components' or `spatial modes'
so that the electric field can be written as a sum of the different
frequency components and of the different spatial modes. Of course,
the concept of modes is directly related to the use of a sort of
oscillator, in this case the optical cavity. Most of the work
presented here is based on the research on laser resonators reviewed
originally by Kogelnik and Li~\cite{KogelnikandLi66}. Siegman has
written a very interesting historic review of the development of
Gaussian optics~\cite{siegman2000_1, siegman2000_2} and we use
whenever possible the same notation as used in his textbook
`Lasers'~\cite{siegman}.

This section introduces the use of Gaussian modes for describing the
spatial properties along the transverse orthogonal \x and \y
directions of an optical beam. We can write
\begin{equation}
\label{eq:HG_intro1}
E(t,x,y,z)=\sum_{j}~\sum_{n,m}~a_{jnm}~u_{nm}(x,y,z)~\mEx{\I(\w_j \T -k_j z)},
\end{equation}
with $u_{nm}$ as special functions describing the spatial properties
of the beam and $a_{jnm}$ as complex amplitude factors ($\w_j$ is
again the angular frequency and $k_j=\w_j/c$). For simplicity we
restrict the following description to a single frequency component at
one moment in time ($t=0$), so
\begin{equation}
\label{eq:HG_intro2}
E(x,y,z)=\mEx{-\I k z}~\sum_{n,m}~a_{nm}~u_{nm}(x,y,z).
\end{equation}
In general, different types of spatial modes $u_{nm}$ can be used in
this context. Of particular interest are the \emph{Gaussian modes},
which will be used throughout this document. Many lasers emit light
that closely resembles a \emph{Gaussian beam}: the light mainly
propagates along one axis, is well collimated around that axis and the
cross section of the intensity perpendicular to the optical axis shows
a Gaussian distribution. The following sections provide the basic
mathematical framework for using Gaussian modes for analysing optical
systems.

\epubtkImage{laser_beam.png}{
	\begin{figure}[htb]
		\centering
		\includegraphics*[scale=0.6, viewport= 230 460 460 680]{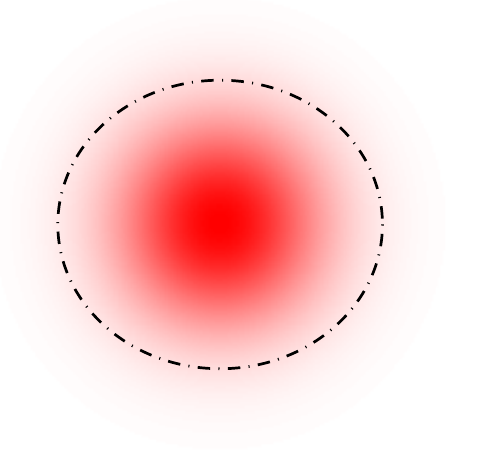}
		\includegraphics[scale=0.6, viewport= 150 440 500 600]{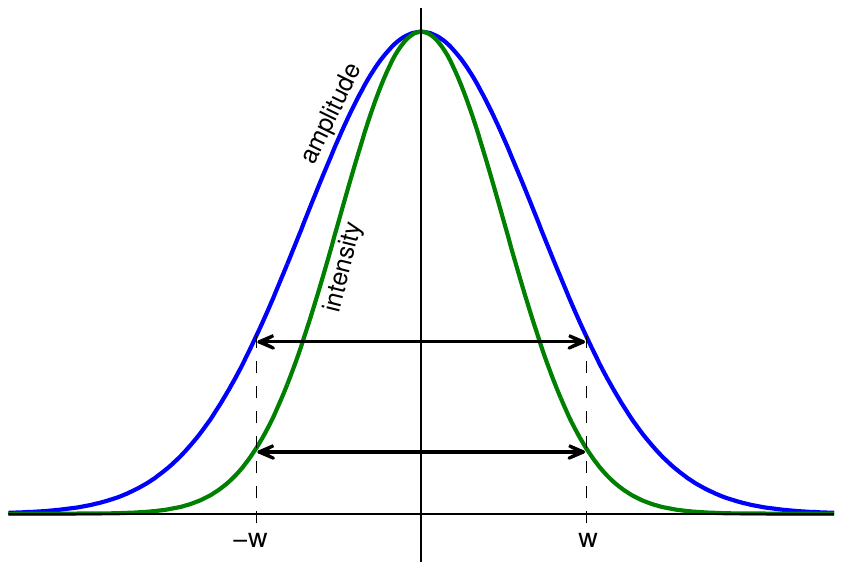}
		\caption{A typical laser beam intensity pattern (left) and the intensity
			and amplitude distributions
			of a normalised Gaussian beam (right).  A Gaussian beam is characterised by
			it's spot size, $w$, the radius at which the intensity falls to $\frac{1}{e^2}$ ($\sim
			14$\%) of the peak intensity.}
		\label{fig:laser_beam}
	\end{figure}
}

\subsection{A typical laser beam: the fundamental Gaussian mode}

The beam produced from a real laser is not a plane wave, but has some
intensity distribution.  This is typically a roughly circular beam with a peak
brightness near the centre.  The intensity pattern of a beam generated
by an ideal laser based on a stable optical cavity with spherical mirrors
would resemble a Gaussian beam.  Figure~\ref{fig:laser_beam} shows
the intensity and amplitude distribution of a typical Gaussian beam, often
characterised by the beam spot size, $w$, the radius within which $\sim86$\%
($\frac{1}{e^2}$) of the light power is contained.  As the beam propagates the beam spot size
changes slowly, which allows producing a narrow beam of light with a small diffraction
angle.

The use of cavities in interferometry provides the basis for the mathematical
description of laser beam shapes as Gaussian modes.  A well designed cavity is a perfect
optical resonator for a particular Gaussian mode.  As discussed above, the
intensity distribution can be characterised by the beam spot size, which determines
the width of the beam.  In the case of Gaussian modes
the wavefront, or phase, of the light field is curved and can be expressed with a radius
of curvature, $R_C$.  As the beam propagates the curvature of the wavefront changes.
To achieve perfect resonance in an optical cavity the curvature of the wavefront must match
the curvature of the mirrors at their positions on the optical axis.  The Gaussian beam whose
curvatures match the mirrors of a cavity is known as the cavity eigenmode,
see Figure~\ref{fig:gauss_beam_cavity_basic}.

\epubtkImage{gauss_beam_cavity_basic.png}{%
  \begin{figure}[htbp]
    \centerline{\includegraphics[width=0.4\textwidth]{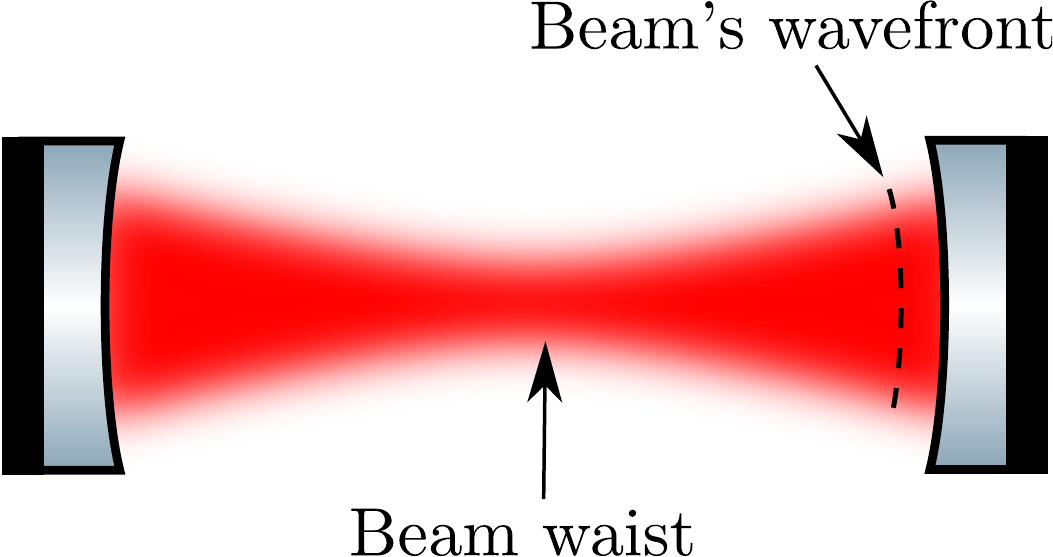}}
    \caption{Simple depiction of a cavity eigenmode. The length and curvature of the
    mirrors determine the cavity eigenmode which defined by the beam waist size
    and position relative to the mirrors.}
    \label{fig:gauss_beam_cavity_basic}
\end{figure}}

\subsection{Describing beam distortions with higher-order modes}
\label{sec:HOM}

In an ideal interferometer the laser beam would be a perfect Gaussian beam,
with wavefronts exactly matched to the shape of the mirrors.
However, in a real interferometer mismatches between the beam and mirror
curvatures, misalignments from the optical axis and deviations of the mirror
surfaces from a perfect sphere all contribute to distort the beam from the ideal
Gaussian beam.

Small distortions of the fundamental beam can be described
by the addition of \emph{higher-order modes}.
Higher-order modes have the same basic properties of the fundamental
Gaussian beam, with two exceptions: higher-order modes have different intensity patterns
from the simple spot of the fundamental mode and modes of different
order pick up an extra phase upon propagation (the Gouy phase, see Section~\ref{sec:Gouy}).

\begin{figure}[htb]
	\centering
	\includegraphics[scale=0.45, viewport= 80 400 540 730]{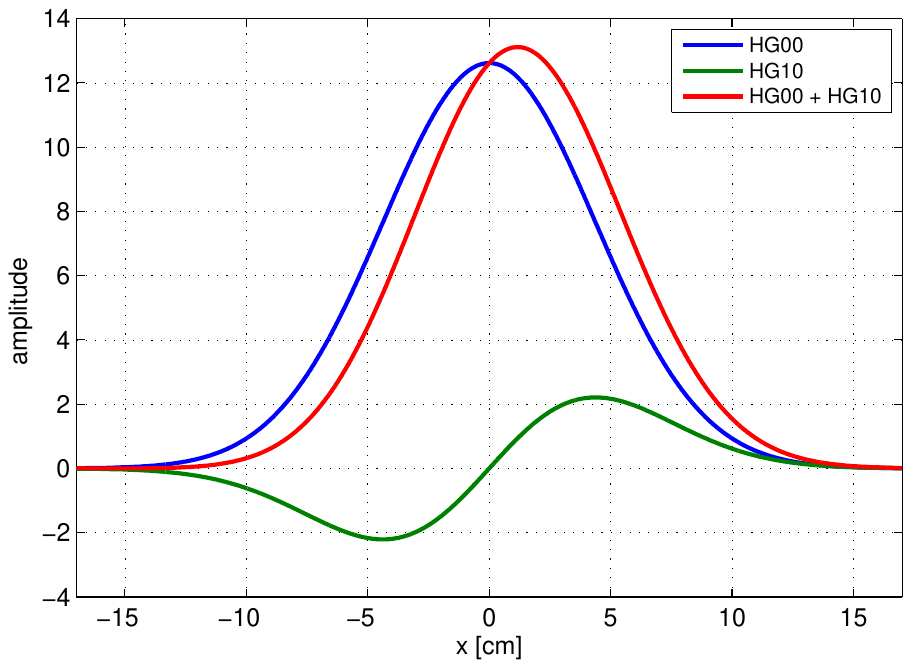}
	\includegraphics[scale=0.45, viewport= 80 400 540 730]{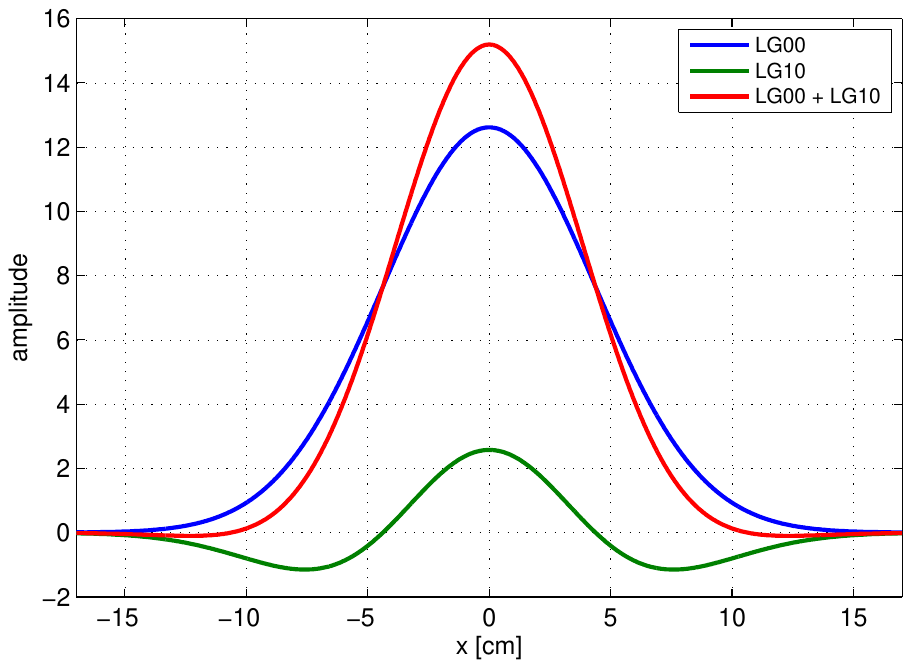}
		\caption{{\bf Left:} Amplitude distributions of a fundamental gaussian beam
			(HG$_{00}$), order 1 Hermite-Gauss beam (HG$_{10}$) and the
			sum of the two modes.
			The resulting sum is a good description of a misaligned fundamental
			beam.  The total power is 1\,W with 4\% power
			in the order 1 mode.
			{\bf Right:} Amplitude distributions of a fundamental gaussian beam (LG$_{00}$),
			order 2 Laguerre-Gauss beam (LG$_{10}$) and the sum of the two modes.
			The resulting sum is a good description of a fundamental gaussian beam with
			a smaller beam spot size.  The power in the order 2 mode is 4\% of the total 1\,W
			power.			
			}
	\label{fig:misalign_mismatch}
\end{figure}

One simple example is a misaligned beam, whose centre has been
shifted from the optical axis.  This can be described by
the addition of an order `1' Hermite-Gauss mode, HG$_{10}$ (Section~\ref{sec:fullHGmode}), as illustrated in the
left panel of Figure~\ref{fig:misalign_mismatch}.  Such a distortion is
a first order effect and, as long as the misalignment is small, can
be described with just this one additional mode.  In a similar way the second order
effect such as a mismatch in beam size can be described by the addition of a single
order `2' mode, in this case the Laguerre-Gauss mode LG$_{10}$ (Section~\ref{sec:LGmodes}).
A mismatch in beam size is illustrated in the right
panel of Figure~\ref{fig:misalign_mismatch}.

The following sections describe details of Gaussian modes and how
any paraxial laser beam with distortions can be described by a sum of Gaussian modes.

\subsection{The paraxial approximation}
\label{sec:paraxial}
All electromagnetic waves are solutions to the general wave equation
(Helmholtz equation), which in vacuum can be given as:
\begin{equation}
\label{eq:fullwave}
\Delta\vec{E}-\frac{1}{c^2}\ddot{\vec{E}}=0.
\end{equation}
Mathematically, Gaussian modes represent solutions to the
\emph{paraxial approximation} of this equation.
Laser light fields are special class of electromagnetic waves. A laser
beam will have a characteristic size $w$ describing the `width' (the
dimension of the field transverse to the main propagation axis), and a
characteristic length $l$ defining some local length along the
propagation over which the beam characteristics do not vary much. By
definition, for what we call a \emph{beam} $w$ is typically small and
$l$ large in comparison, so that $w/l$ can be considered small. In
fact, the paraxial wave equation (and its solutions) can be derived as
the first-order terms of a series expansion of
Equation~(\ref{eq:fullwave}) into orders of $w/l$~\cite{Lax75}.

A simpler approach to the paraxial-wave equation goes as follows: A
particular beam shape shall be described by a function $u(x,y,z)$ so
that we can write the electric field as
\begin{equation}
\label{eq:paraxbeam1}
E(x,y,z)=u(x,y,z)~\mEx{-\I kz}.
\end{equation}
Substituting this into the standard wave equation yields a
differential equation for $u$:
\begin{equation}
\label{eq:a_hg_wave2} \left(\partial_x^2+\partial_y^2+\partial_z^2\right)u(x,y,z)-2\I k\partial_z u(x,y,z)=0.
\end{equation}
Now we put the fact that $u(x,y,z)$ should be slowly varying with $z$
in mathematical terms. The variation of $u(x,y,z)$ with $z$ should be
small compared to its variation with $x$ or $y$. Also the second
partial derivative in $z$ should be small. This can be expressed as
\begin{equation}
\left|\partial_z^2 u(x,y,z)\right|\ll\left|2k\partial_zu(x,y,z)\right|,\left|\partial_x^2
u(x,y,z)\right|,\left|\partial_y^2 u(x,y,z)\right|.
\end{equation}
With this approximation, Equation~(\ref{eq:a_hg_wave2}) can be
simplified to the \emph{paraxial wave equation},
\begin{equation}
\label{eq:paraxial}
\left(\partial_x^2+\partial_y^2\right)u(x,y,z)-2\I k\partial_z u(x,y,z)=0.
\end{equation}
Any field $u$ that solves this equation represents a paraxial beam
shape when used in the form given in Equation~(\ref{eq:paraxbeam1}).

\subsection{Transverse electromagnetic modes}
In general, any solution $u(x,y,z)$ of the paraxial wave equation,
Equation~(\ref{eq:paraxial}), can be employed to represent the
transverse properties of a scalar electric field representing a
beam-like electro-magnetic wave. Especially useful in this respect are
special \emph{families} or \emph{sets} of functions that are solutions
of the paraxial wave equation. When such a set of functions is
complete and countable, it's called a set of \emph{transverse
  electromagnetic modes} (TEM). For instance, the set of
\emph{Hermite--Gauss modes} are exact solutions of the paraxial wave
equation. These modes are represented by an infinite, countable and
complete set of functions. The term \emph{complete} means they can be
understood as a base system of the function space defined by
\emph{all} solutions of the paraxial wave equation. In other words, we
can describe any solution of the paraxial wave equation $u'$ by a
linear superposition of \HG\ modes:
\begin{equation}
u'(x,y,z)=\sum_{n,m}~a_{jnm}~u_{nm}(x,y,z),
\end{equation}
which in turn allows us to describe any laser beam using a sum of
these modes:
\begin{equation}
\label{eq:HG_mode2}
E(t,x,y,z)=\sum_{j}~\sum_{n,m}~a_{jnm}~u_{nm}(x,y,z)~\mEx{\I(\w_j \T -k_j z)}.
\end{equation}
The \HG\ modes as given in this document (see
Section~\ref{sec:fullHGmode}) are orthonormal so that
\begin{equation}
\label{eq:HG_mode3}
\int\!\!\!\int\!dxdy~u_{n m}u^*_{n' m'}=\delta_{n n'}\delta_{m m'}=\left\{
\begin{array}{ll}
1\quad\mathrm{if}\quad n=n'\quad\mathrm{and}\quad m=m'\\
0\quad\mathrm{otherwise}
\end{array}\right\}.
\end{equation}
This means that, in the function space defined by the paraxial wave
equation, the Hermite--Gauss functions can be understood as a complete
set of \emph{unit-length} basis vectors. This fact can be utilised for
the computation of coupling factors, as shown in Section~\ref{sec:HOMcoupling}.
Furthermore, the power of a beam, as given by
Equation~(\ref{eq:HG_intro2}), being detected on a single-element
photodetector (provided that the area of the detector is large with
respect to the beam) can be computed as
\begin{equation}
EE^*=\sum_{n,m} a_{nm}a_{nm}^*,
\end{equation}
or for a beam with several frequency components (compare with Equation~(\ref{eq:dc_det})) as
\begin{equation}
\label{eq:hg_dc_det}
EE^*=\sum_{n,m}\sum\limits_i\sum\limits_j a_{inm}a_{jnm}^*\quad\mathrm{with}\quad
\{i,j~|~i,j\in\{0,\dots,N\}~\wedge~\w_i=\w_j\}.
\end{equation}

\subsection{Properties of Gaussian beams}
\label{sec:gaussian_beam_properties}

The basic or `lowest-order' Hermite--Gauss mode is equivalent to what
is usually called a \emph{Gaussian beam} and is given by
\begin{equation}
\label{eq:a_hg_base}
u(x,y,z)=\sqrt{\frac{2}{\pi}}~\frac{1}{w(z)}~\mEx{\I \Psi(z)}~ \mEx{-\I
k\frac{x^2+y^2}{2R_C(z)}-\frac{x^2+y^2}{w^2(z)}}.
\end{equation}
The parameters of this equation are explained in detail below. The
shape of a Gaussian beam is quite simple: the beam has a circular
cross section, and the radial intensity profile of a beam with total
power $P$ is given by
\begin{equation}
\label{eq:gauss_profile}
I(r)=\frac{2P}{\pi w^2(z)}\mEx{-2r^2/w^2},
\end{equation}
with $w$ the \emph{spot size}, defined as the \emph{radius} at which
the intensity is $1/e^2$ times the maximum intensity $I(0)$. This is a
Gaussian distribution, see Figure~\ref{fig:gauss_profile}, hence the
name \emph{Gaussian beam}.

\epubtkImage{gauss-profile.png}{%
  \begin{figure}[htbp]
    \begin{center}
      \begin{minipage}{.7\textwidth}
	\includegraphics[viewport=0 0 340 180]{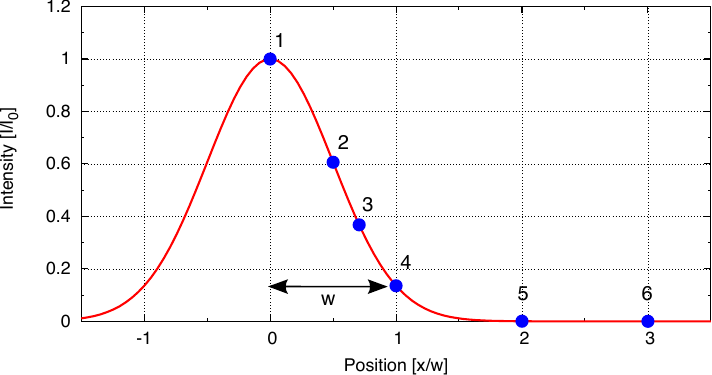}
      \end{minipage}
      \begin{minipage}{.2\textwidth}
	\vspace{-1.8cm}
	\hspace{-2.8cm}
	\begin{tabular}{ccc}
	  no. & $x/w$ & $I/I_0$ \\
	  1   & $0.0$   & $1.0$ \\
	  2   & $1/2$   & $1/\sqrt{e}$ \\
	  3   & $1/\sqrt{2}$   & $1/e$ \\
	  4   & $1.0$   & $1/e^2$ \\
	  5   & $2.0$   & $3\cdot 10^{-4}$ \\
	  6   & $3.0$   & $2\cdot 10^{-8}$ \\
	\end{tabular}
      \end{minipage}
    \end{center}
    \caption[Gaussian beam profile]{One dimensional cross-section of a
    Gaussian beam. The width of the beam is given by the radius $w$ at
    which the intensity is $1/e^2$ of the maximum intensity.}
    \label{fig:gauss_profile}
\end{figure}}

Figure~\ref{fig:beamprofile2} shows a different cross section through
a Gaussian beam: it plots the beam size as a function of the position
on the optical axis.
\epubtkImage{beam_profile_fin.png}{%
  \begin{figure}[htbp]
    \centerline{\includegraphics{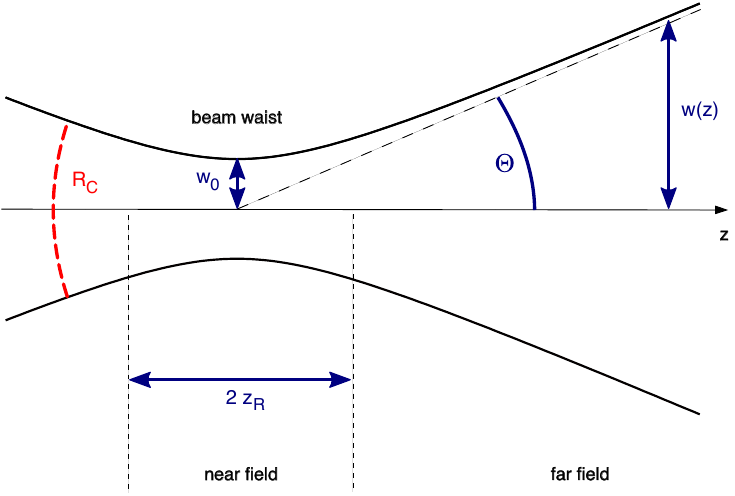}}
    \caption{Gaussian beam profile along z: this cross section along
      the \x-\z-plane illustrates how the beam size $w(z)$ of the
      Gaussian beam changes along the optical axis. The position of
      minimum beam size $w_0$ is called \emph{beam waist}. See text
      for a description of the parameters $\Theta$, $z_R$ and $R_c$.}
    \label{fig:beamprofile2}
\end{figure}}
Such a beam profile (for a beam with a given wavelength $\lambda$) can be
completely determined by two parameters: the size of the minimum spot size
$w_0$ (called \emph{beam waist}) and the position $z_0$ of the beam waist
along the \z-axis.

To characterise a Gaussian beam, some useful parameters can be derived
from $w_0$ and $z_0$. A Gaussian beam can be divided into two
different sections along the \z-axis: a \emph{near field} -- a region
around the beam waist, and a \emph{far field} -- far away from the
waist. The length of the near-field region is approximately given by
the \emph{Rayleigh range} $\zr$. The Rayleigh range and the spot size
are related by
\begin{equation}
\zr=\frac{\pi w_0^2}{\lambda}.
\end{equation}
With the Rayleigh range and the location of the beam waist, we can
usefully write
\begin{equation}
w(z)=w_0\sqrt{1+\left(\frac{z-z_0}{\zr}\right)^2}.
\end{equation}
This equation gives the size of the beam along the \z-axis. In the
far-field regime ($z\gg \zr,z_0$), it can be approximated by a linear
equation, when
\begin{equation}
w(z)\approx w_0\frac{z}{\zr}=\frac{z \lambda}{\pi w_0}.
\end{equation}

The angle $\Theta$ between the \z-axis and $w(z)$ in the far field is
called the \emph{diffraction angle}\epubtkFootnote{Also known as the
  \emph{far-field angle} or the \emph{divergence} of the beam.} and is
defined by
\begin{equation}
\Theta=\arctan\left(\frac{w_0}{\zr}\right)=\arctan\left(\frac{\lambda}{\pi
w_0}\right)\,\,\approx\frac{w_0}{\zr}.
\end{equation}

Another useful parameter is the \emph{radius of curvature} of the
wavefront at a given point \z. The radius of curvature describes the
curvature of the `phase front' of the electromagnetic wave -- a
surface across the beam with equal phase -- intersecting the optical
axis at the position \z. We obtain the radius of curvature as a
function of \z:
\begin{equation}
R_C(z)=z-z_0+\frac{\zr^2}{z-z_0}.
\end{equation}
We also find:
\begin{equation}
{\renewcommand{\arraystretch}{1.5}
  \begin{array}{lll}
    R_C\approx \infty,&  z-z_0\ll\zr\ & \qquad \text{(beam waist)}\\
    R_C\approx z, & z\gg\zr,~ z_0 & \qquad \text{(far field)}\\
    R_C=2\zr, & z-z_0=\zr & \qquad \text{(maximum curvature)}.\\
  \end{array}
}
\end{equation}

\subsection{Astigmatic beams: the tangential and sagittal plane}

If the interferometer is confined to a plane (here the \x-\z plane),
it is convenient to use projections of the three-dimensional
description into two planes~\cite{Rigrod65}: the \emph{tangential
  plane}, defined as the \x-\z plane and the \emph{sagittal plane} as
given by \y and \z.

The beam parameters can then be split into two respective parameters:
$z_{0,s}$, $w_{0,s}$ for the sagittal plane and $z_{0,t}$  and
$w_{0,t}$ for the tangential plane so that the \HG\ modes can be
written as
\begin{equation}
u_{nm}(x,y)=u_n(x,z_{0,t},w_{0,t})~u_m(y,z_{0,s},w_{0,s}).
\end{equation}
Beams with different beam waist parameters for the sagittal and
tangential plane are \emph{astigmatic}.

Remember that these Hermite--Gauss modes form a base system. This
means one can use the separation into sagittal and tangential planes
even if the actual optical system does not show this special type of
symmetry. This separation is very useful in simplifying the
mathematics. In the following, the term \emph{beam parameter}
generally refers to a simple case where $w_{0,x}=w_{0,y}$ and
$z_{0,x}=z_{0,y}$ but all the results can also be applied directly to
a pair of parameters.

\subsection{Higher-order Hermite--Gauss modes}
\label{sec:fullHGmode}

The complete set of Hermite--Gauss modes is given by an infinite
discrete set of modes $u_{\mathrm{nm}}(x,y,z)$ with the indices n and
m as \emph{mode numbers}. The sum n+m is called the \emph{order} of
the mode. The term \emph{higher-order modes} usually refers to modes
with an order $n+m>0$. The general expression for Hermite--Gauss modes
can be given as~\cite{KogelnikandLi66}
\begin{equation}
u_{\mathrm{nm}}(x,y,z)=u_{\mathrm{n}}(x,z)u_{\mathrm{m}}(y,z),
\end{equation}
with
\begin{equation}
\label{eq:HG_mode1}
{\renewcommand{\arraystretch}{1.5}
\begin{array}{rl}
u_{\mathrm{n}}(x,z)=&\left(\frac{2}{\pi}\right)^{1/4}\left(\frac{\mEx{\I(2n+1)\Psi(z)}}{2^n
n! w(z)}\right)^{1/2}~\times\\
&H_n\left(\frac{\sqrt{2}x}{w(z)}\right) \mEx{-\I\frac{kx^2}{2R_C(z)}-\frac{x^2}{w^2(z)}},
\end{array}}
\end{equation}
and $H_n(x)$ the Hermite polynomials of order n. The first Hermite
polynomials, without normalisation, can be written
\begin{equation}
\label{eq:h_poly}
\begin{array}{ll}
H_0(x)=1 & H_1(x)=2x\\
H_2(x)=4x^2-2 & H_3(x)=8x^3-12x.
\end{array}
\end{equation}
Further orders can be computed recursively since
\begin{equation}
\label{eq:h_poly_rec} H_{n+1}(x)=2xH_n(x)-2nH_{n-1}(x).
\end{equation}
For both transverse directions we can also rewrite the above to
\begin{equation}
{\renewcommand{\arraystretch}{1.5}
\begin{array}{lcl}
u_{\mathrm{nm}}(x,y,z)&=&\left(2^{n+m-1}n!m!\pi\right)^{-1/2}
\frac{1}{w(z)}~\mEx{\I(n+m+1)\Psi(z)}~\times\\
&&\qquad H_n\left(\frac{\sqrt{2}x}{w(z)}\right)H_m\left(\frac{\sqrt{2}y}{w(z)}\right)
\mEx{-\I\frac{k(x^2+y^2)}{2R_C(z)}-\frac{x^2+y^2}{w^2(z)}}.
\end{array}}
\end{equation}
The latter form has the advantage of clearly showing the extra phase shift
along the \z-axis of $(n+m+1)\Psi(z)$, called the \emph{Gouy phase}; see
Section~\ref{sec:Gouy}.

\subsection{The Gaussian beam parameter}

For a more compact description of the interaction of Gaussian modes
with optical components we will make use of the \emph{Gaussian beam
  parameter} $q$~\cite{kogelnik65}. The beam parameter is a complex
quantity defined as
\begin{equation}
\frac{1}{q(z)}=\frac{1}{R_C(z)}-\I\frac{\lambda}{\pi w^2(z)}.
\end{equation}
It can also be written as
\begin{equation}
q(z)=\I\zr +z-z_0=q_0+z-z_0\qquad\mathrm{and}\qquad q_0=\I\zr.
\end{equation}
Using this parameter, Equation~(\ref{eq:a_hg_base}) can be rewritten as
%
\begin{equation}
\label{eq:a_hg_qbase} u(x,y,z)=\sqrt{\frac{2}{\pi}}\frac{q_0}{w_0q(z)}\mEx{-\I k\frac{x^2+y^2}{2q(z)}}.
\end{equation}
%
Other parameters, like the beam size and radius of curvature, can also
be written in terms of the beam parameter $q$:
\begin{equation}
w^2(z)=\frac{\lambda}{\pi}\frac{|q|^2}{\myIm{q}},
\end{equation}
\begin{equation}
w_0^2=\frac{\myIm{q} \lambda}{\pi},
\end{equation}
\begin{equation}
\zr=\myIm{q}
\end{equation}
and
\begin{equation}
R_C(z)=\frac{|q|^2}{\myRe{q}}.
\end{equation}
The Hermite--Gauss modes can also be written using the Gaussian beam
parameter as\epubtkFootnote{Please note that this formula
  from~\cite{siegman} is very compact. Since the parameter $q$ is a
  complex number, the expression contains at least two complex square
  roots. The complex square root requires a different algebra than the
  standard square root for real numbers. Especially the third and
  fourth factors can \emph{not} be simplified in any obvious way:
$\left(\frac{q_0}{q(z)}\right)^{1/2} \left(\frac{q_0q^*(z)}{q_0^*q(z)}\right)^{n/2}
\neq\left(\frac{q_0^{n+1}{q^*}^n(z)}{q^{n+1}(z){q_0^*}^n}\right)^{1/2}$ !}
\begin{equation}
\label{eq:a_hg_qmode}
{\renewcommand{\arraystretch}{1.5}
\begin{array}{l}
u_{\mathrm{nm}}(x,y,z)=u_{\mathrm{n}}(x,z)u_{\mathrm{m}}(y,z)\qquad \text{with}\\
u_{\mathrm{n}}(x,z)=\left(\frac{2}{\pi}\right)^{1/4}\left(\frac{1}{2^nn!w_0}\right)^{1/2}
\left(\frac{q_0}{q(z)}\right)^{1/2}\left(\frac{q_0~q^*(z)}{q_0^*~q(z)}\right)^{n/2}
H_n\left(\frac{\sqrt{2}x}{w(z)}\right) \mEx{-\I\frac{kx^2}{2q(z)}}.
\end{array}}
\end{equation}

\subsection{Properties of higher-order Hermite--Gauss modes}

Some of the properties of Hermite--Gauss modes can easily be described
using cross sections of the field intensity or field
amplitude. Figure~\ref{fig:HGintensities} shows such cross sections,
i.e.~the intensity in the \x-\y plane, for a number of higher-order
modes. This shows a \x-\y symmetry for mode indices $n$ and $m$. We
can also see how the size of the intensity distribution increases with
the mode index, while the peak intensity decreases.
\epubtkImage{HGintensities.png}{%
  \begin{figure}[htbp]
    \centerline{\includegraphics[viewport=200 0 690 550, scale=.45]{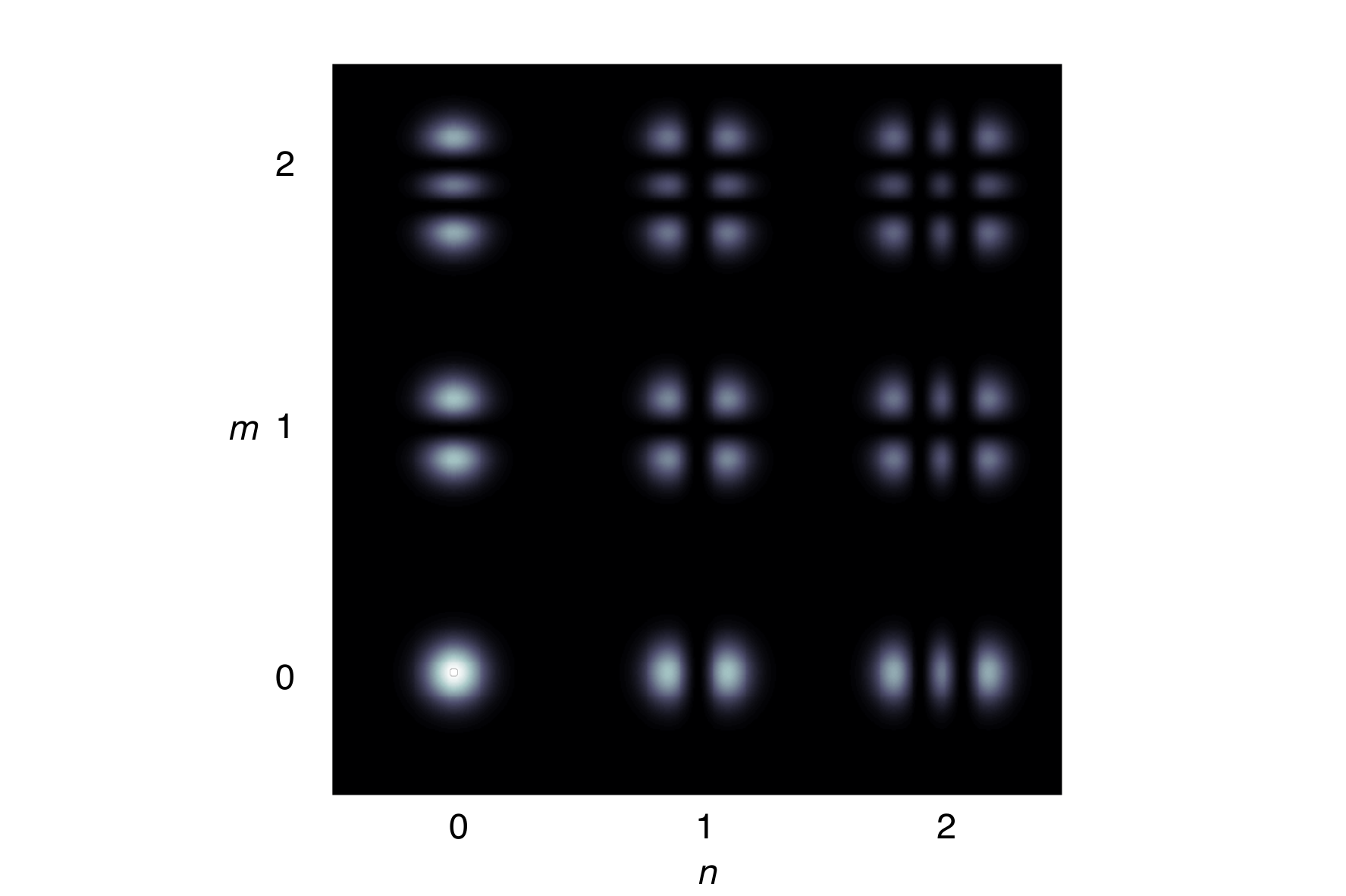}}
    \caption[Hermite--Gaussian beam shapes]{This plot shows the
      intensity distribution of Hermite--Gauss modes $u_{nm}$. One can
      see that the intensity distribution becomes wider for larger
      mode indices and the peak intensity decreases. The mode index
      defines the number of dark stripes in the respective direction.}
    \label{fig:HGintensities}
\end{figure}}
Similarly, Figure~\ref{fig:gauss_shapes} shows the amplitude and phase
distribution of several higher-order \HG\ modes. Some further features
of \HG\ modes:
\begin{itemize}
\item The size of the intensity profile of any sum of \HG\ modes
  depends on \z while its shape remains constant over propagation
  along the optical axis.
\item The phase distribution of \HG\ modes shows the curvature (or
  radius of curvature) of the beam. The curvature depends on \z but
  is equal for all higher-order modes.
\end{itemize}

Note that these are special features of Gaussian beams and not
generally true for arbitrary beam shapes. Figure~\ref{fig:triangular},
for example, shows the amplitude and phase distribution of a
triangular beam at the point where it is (mathematically) created and
after a 10~m propagation. Neither the shape is preserved nor does it
show a spherical phase distribution.

\epubtkImage{TR_s-TR_p.png}{%
  \begin{figure}[htbp]
    \centerline{
      \includegraphics[scale=.3]{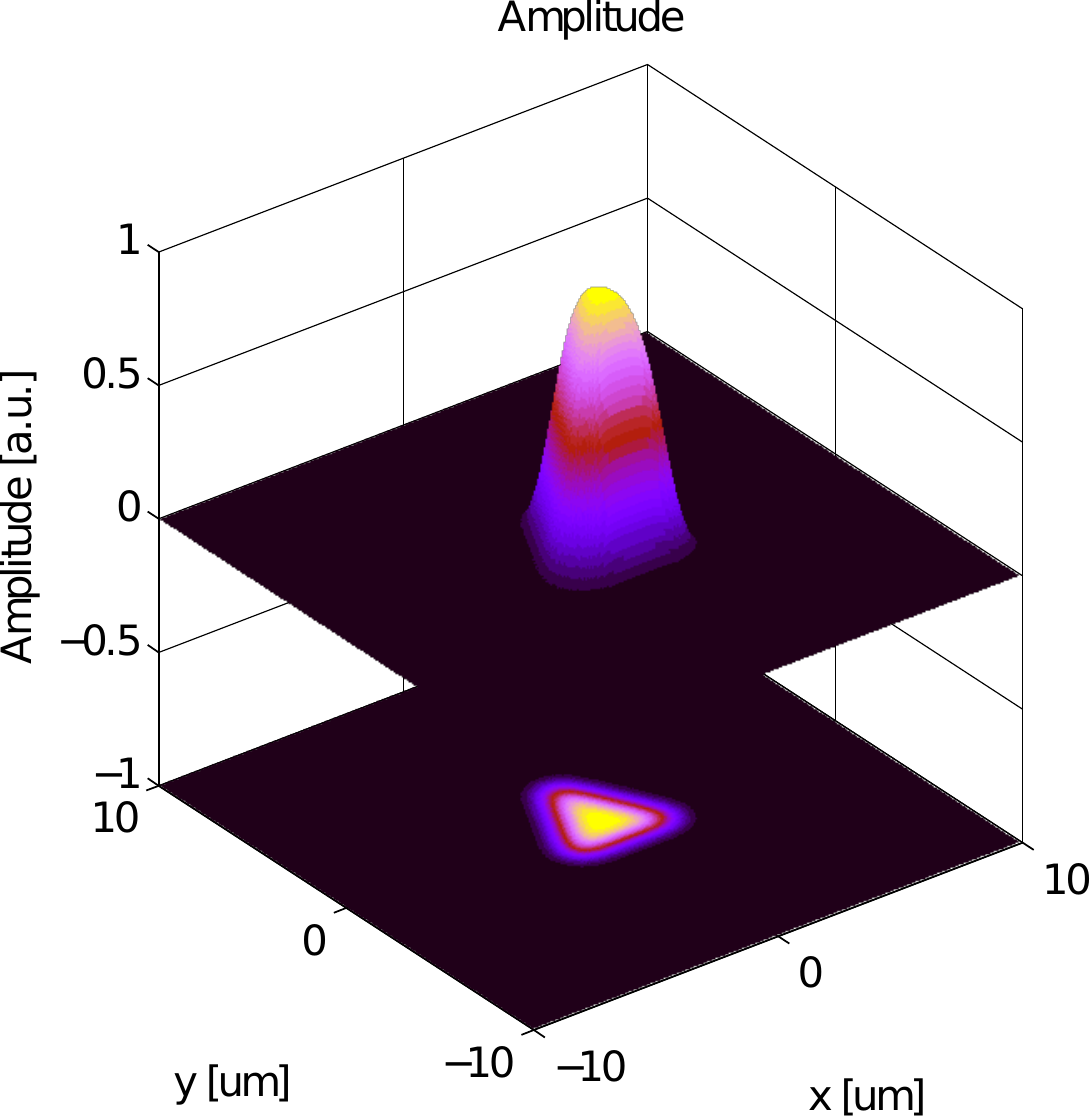}\hspace{5mm}
      \includegraphics[scale=.3]{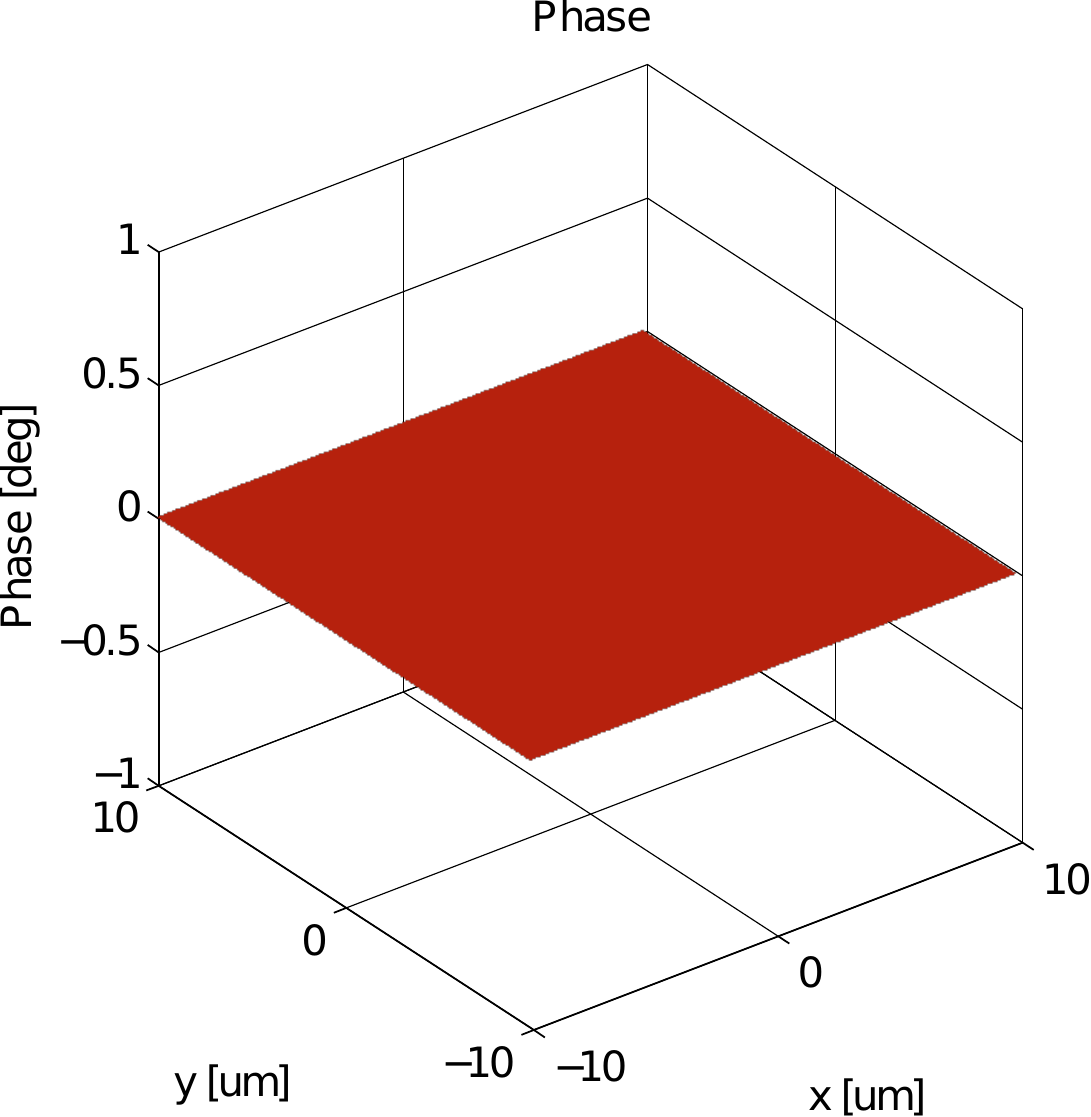}
    }
    \vspace{2mm}
    \centerline{
      \includegraphics[scale=.3]{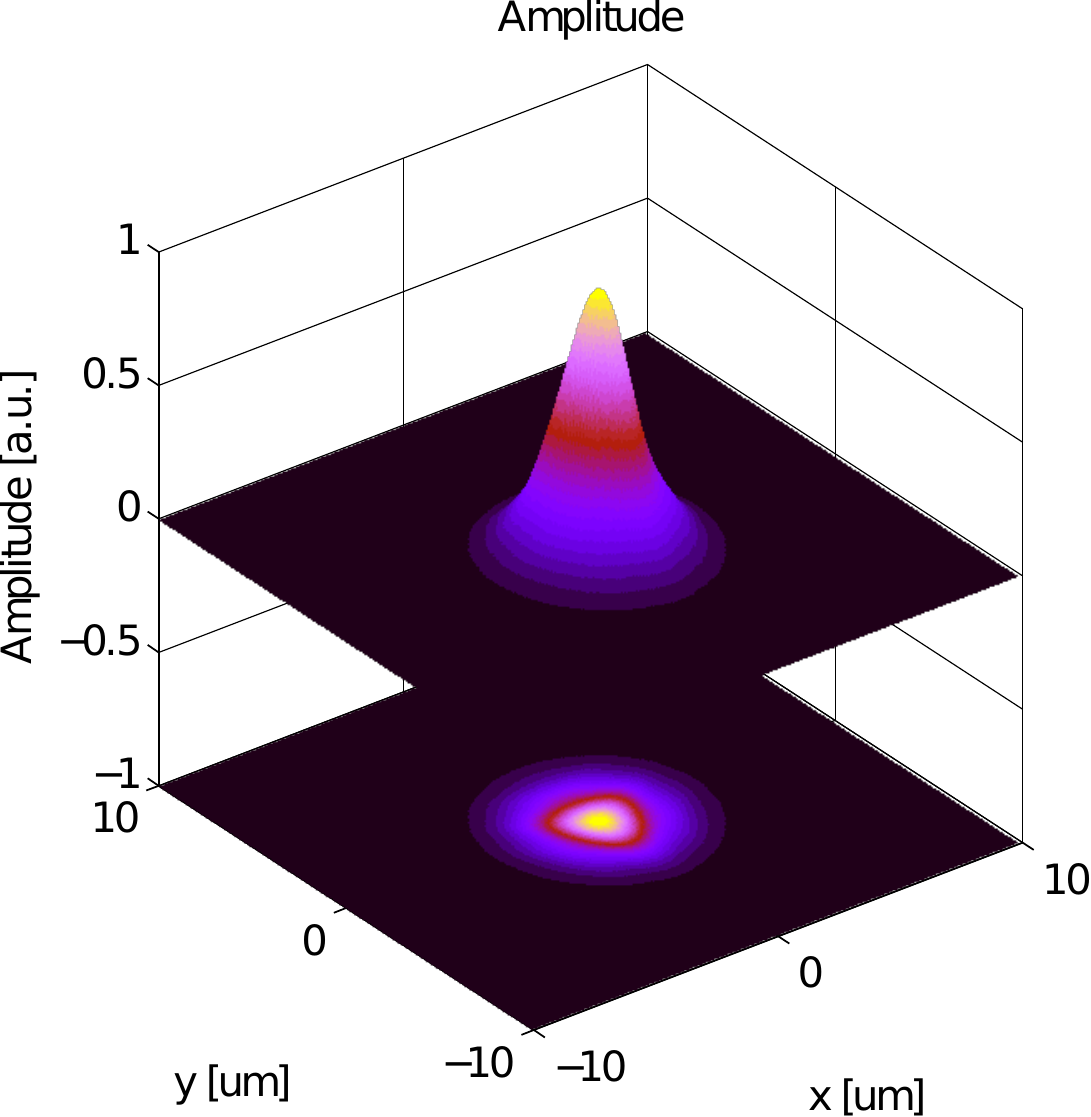}\hspace{5mm}
      \includegraphics[scale=.3]{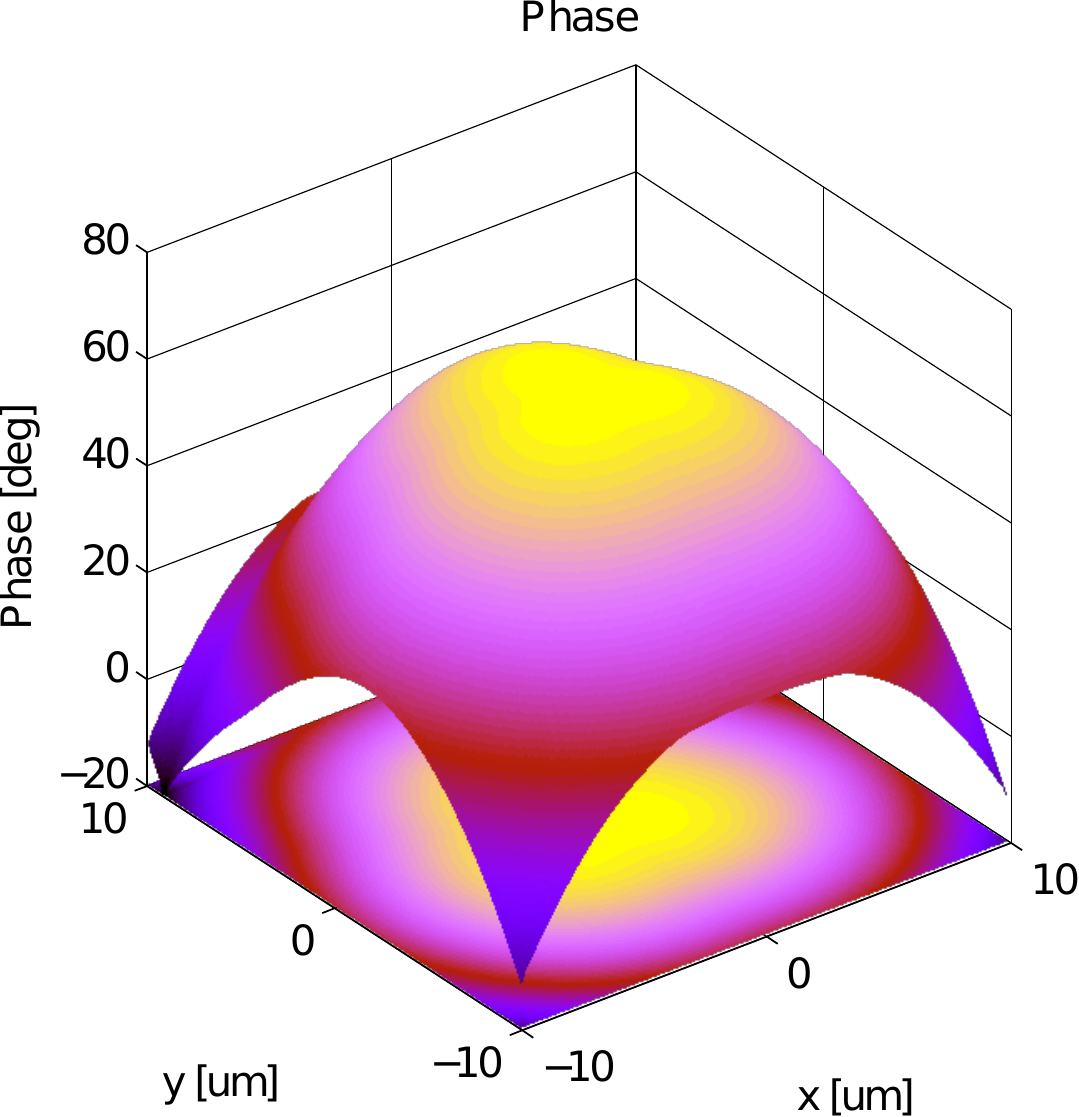}
    }
    \caption[Hermite--Gaussian beam shapes]{These top plots show a
      triangular beam shape and phase distribution and the bottom
      plots the diffraction pattern of this beam after a propagation
      of \z~=~5~m. It can be seen that the shape of the triangular
      beam is not conserved and that the phase front is not
      spherical.}
    \label{fig:triangular}
\end{figure}}

\epubtkImage{HG00-HG01-HG23.png}{%
  \begin{figure}[htbp]
    \centerline{
      \includegraphics[viewport=130 0 690 570, scale=.3]{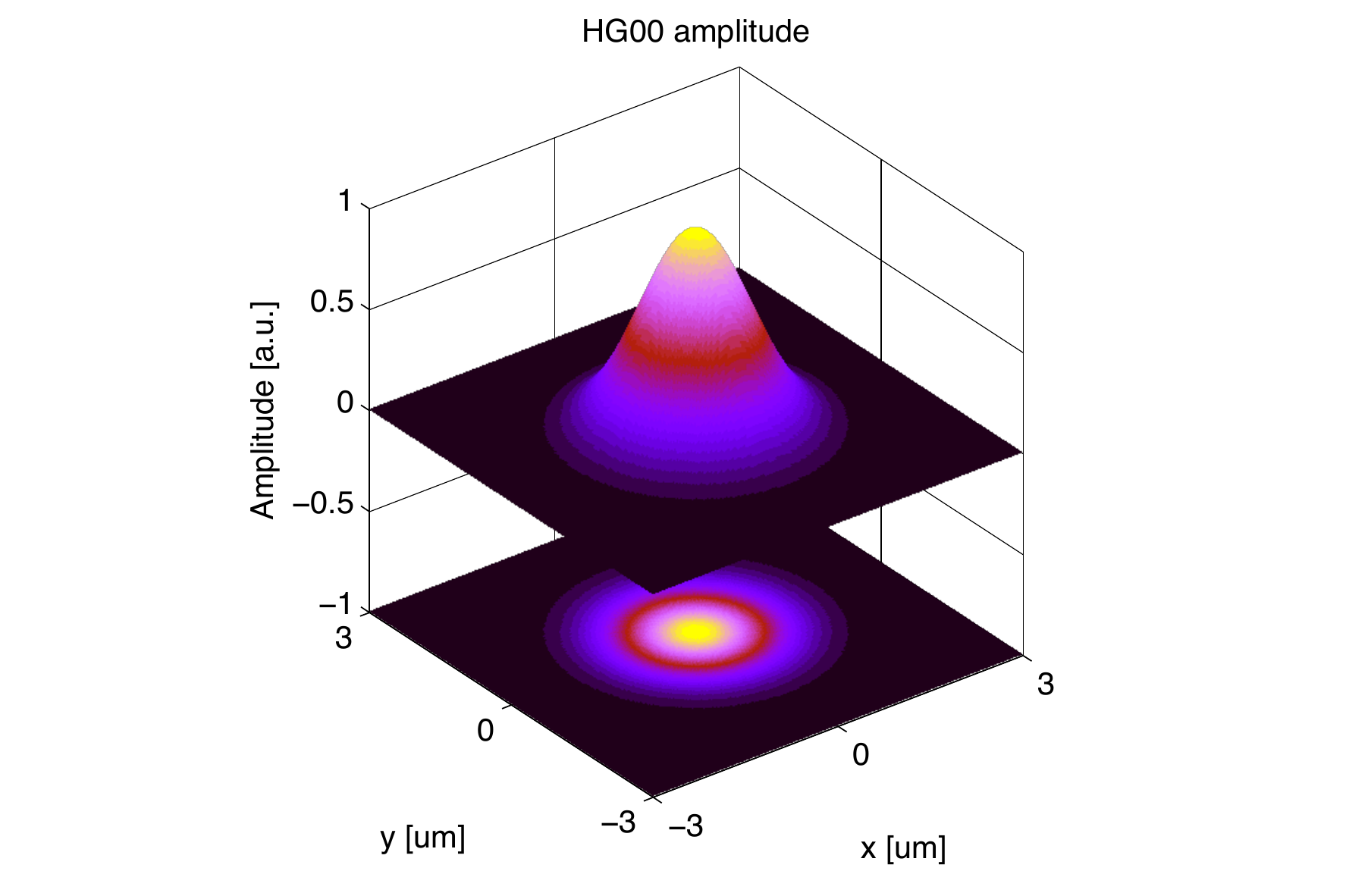}
      \includegraphics[viewport=130 0 690 570, scale=.3]{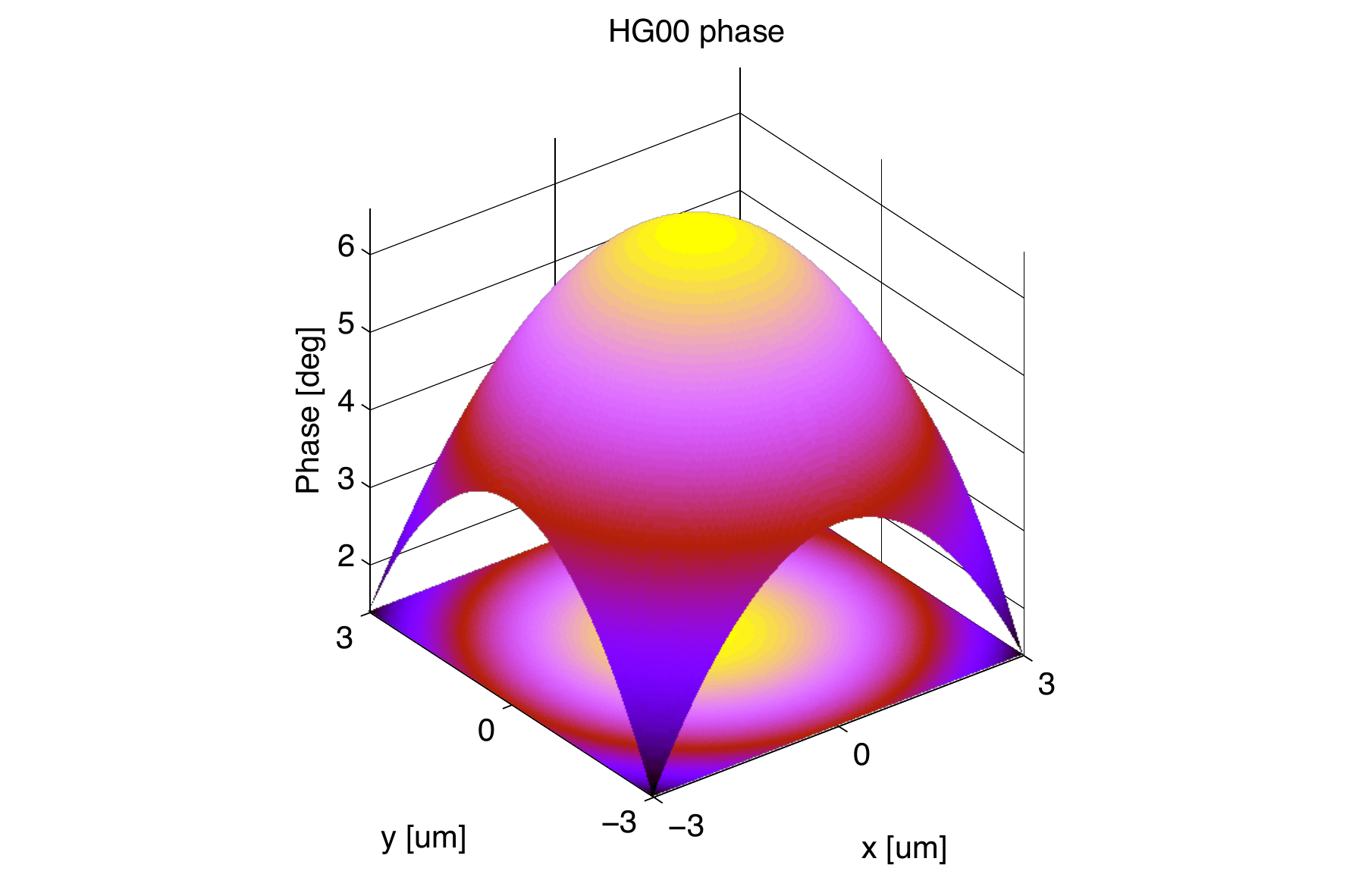}
    }
    \centerline{
      \includegraphics[viewport=130 0 690 570, scale=.3]{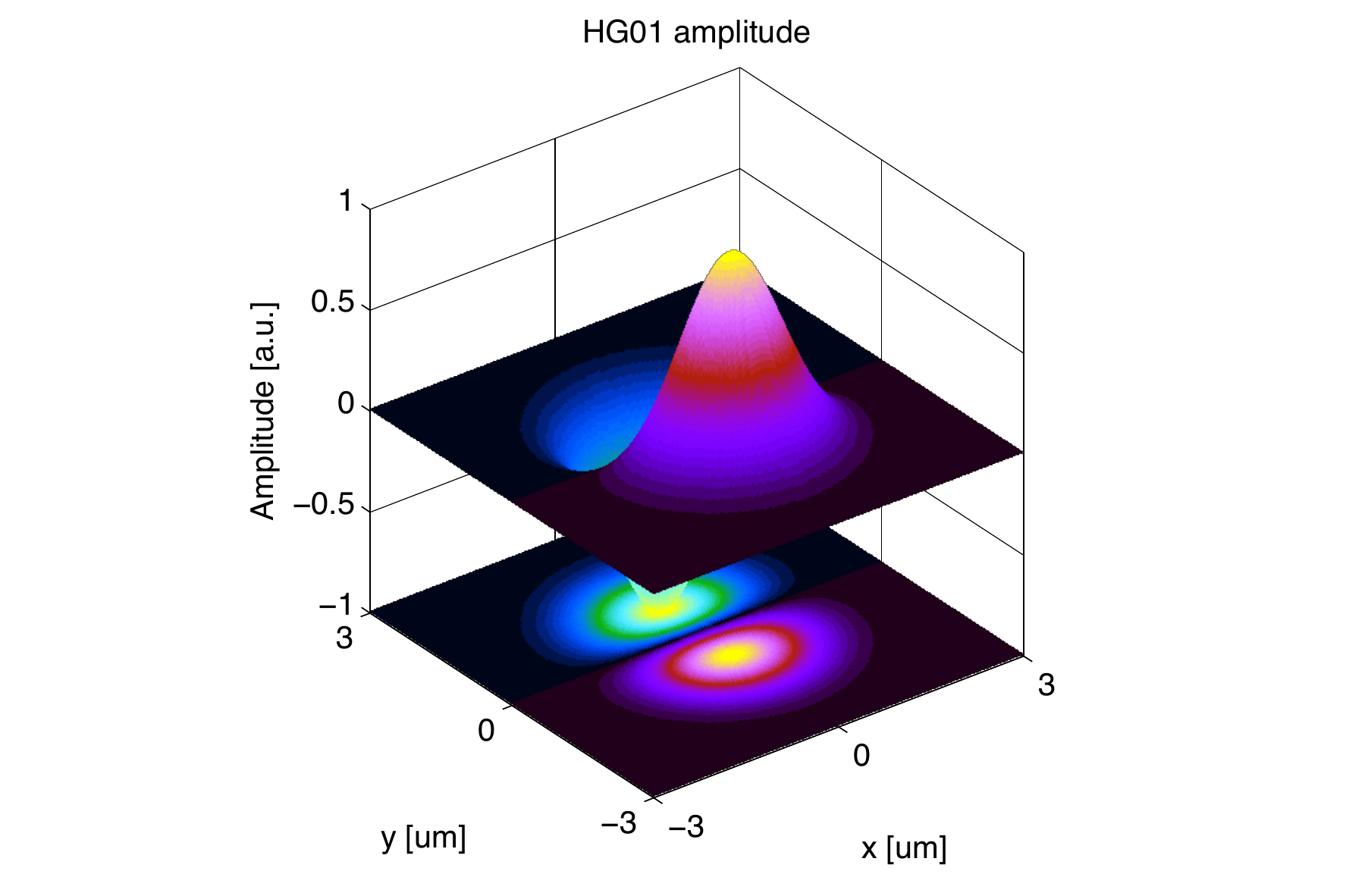}
      \includegraphics[viewport=130 0 690 570, scale=.3]{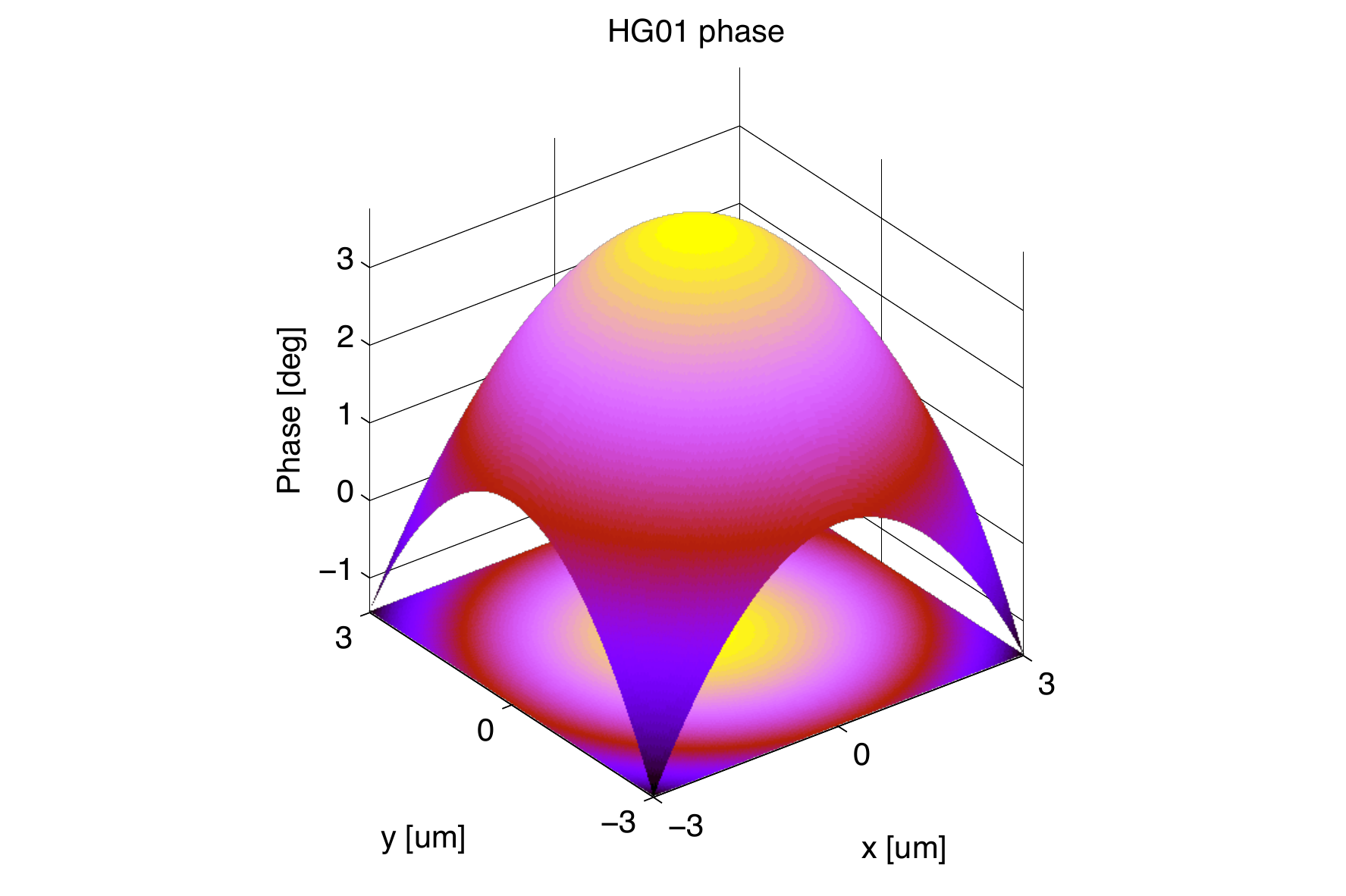}
    }
    \centerline{
      \includegraphics[viewport=130 0 690 570, scale=.3]{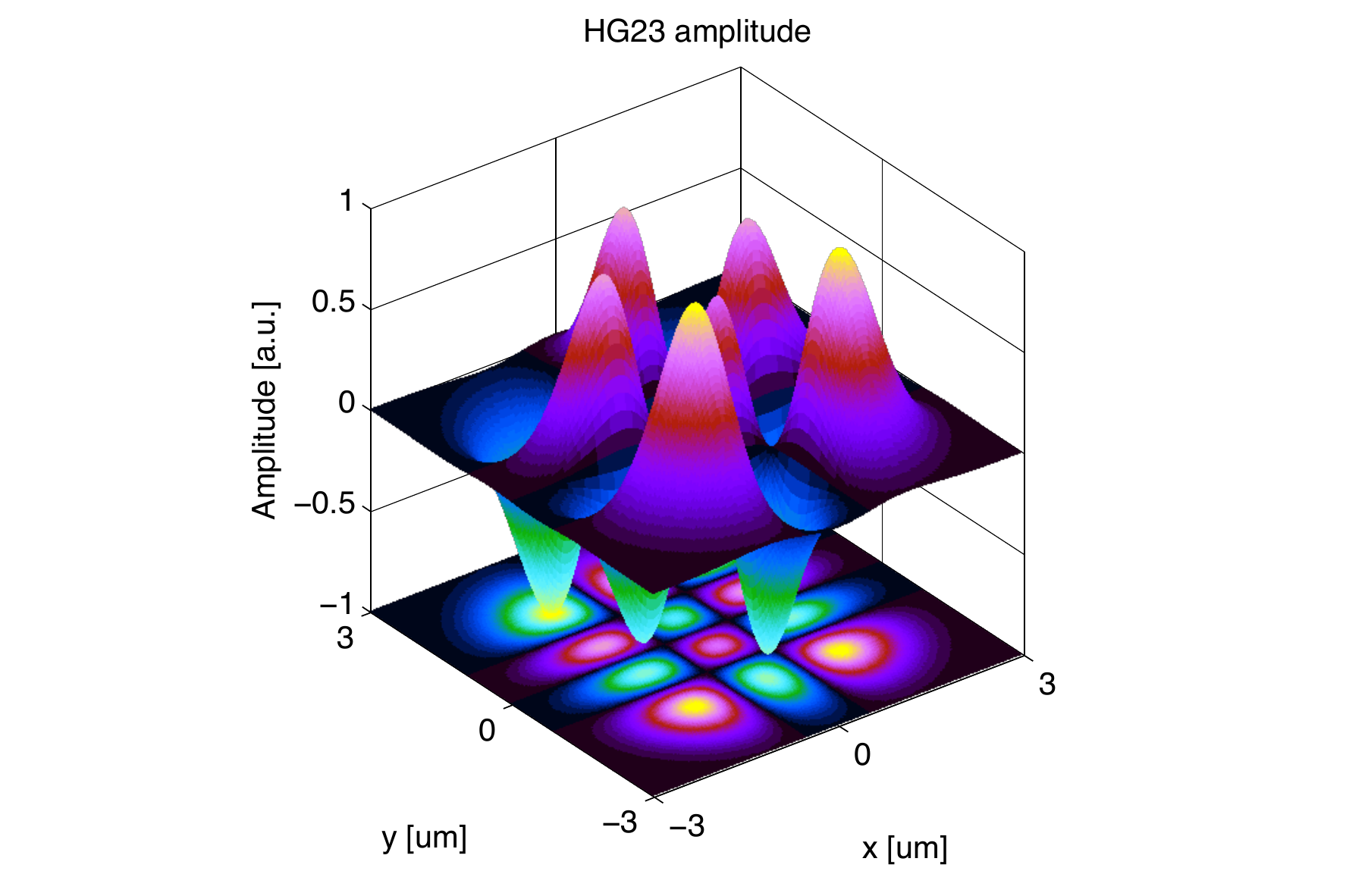}
      \includegraphics[viewport=130 0 690 570, scale=.3]{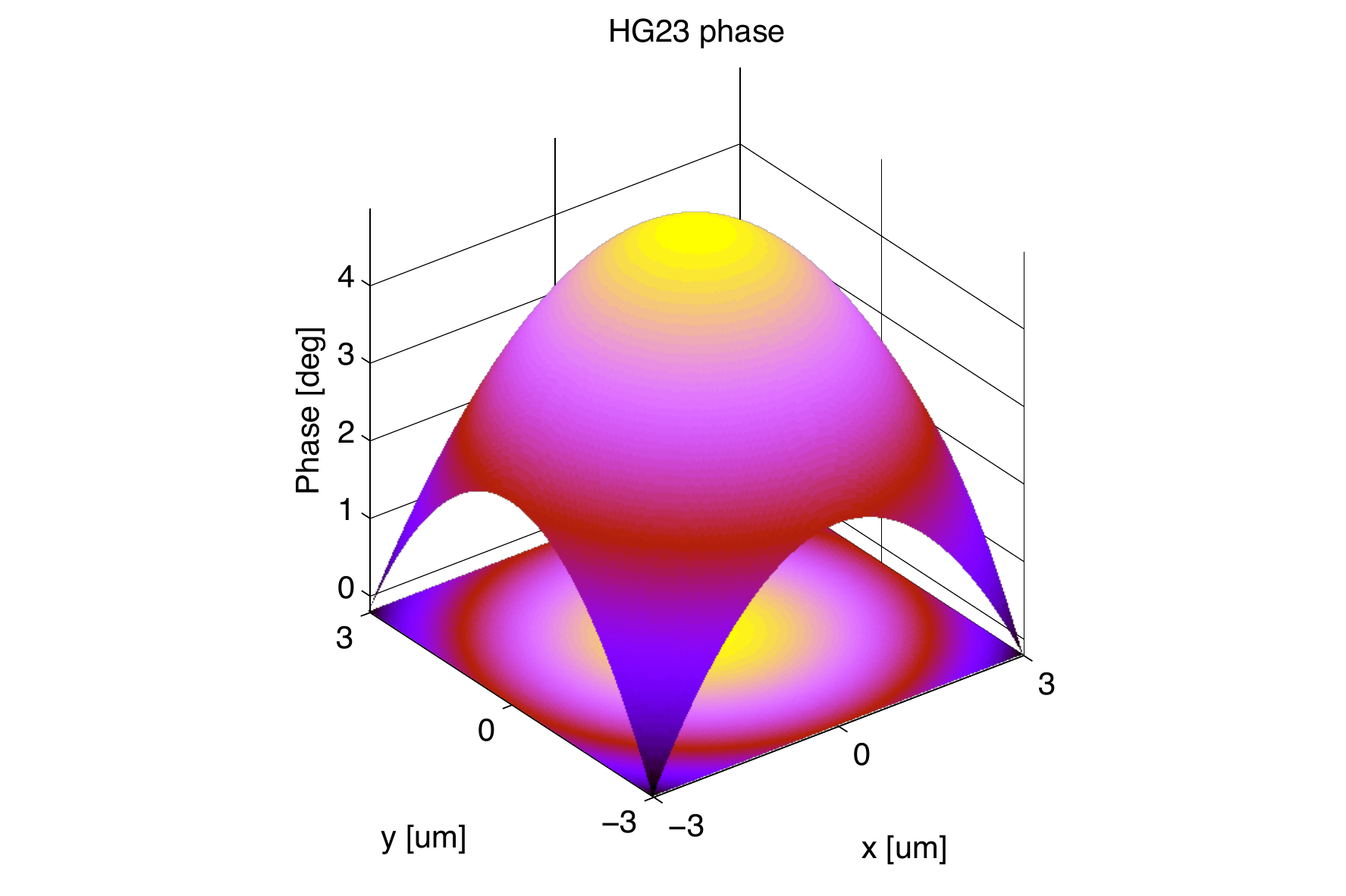}
    }
    \caption[Gaussian beam shapes]{These plots show the amplitude
      distribution and wave front (phase distribution) of
      Hermite--Gaussian modes $u_{nm}$ (labeled as HG\textit{nm} in the
      plot). All plots refer to a beam with $\lambda$~=~1~\mum,
      \textit{w}~=~1~mm and distance to waist \z~=~1~m. The mode index (in one
      direction) defines the number of zero crossings (along that
      axis) in the amplitude distribution. One can also see that the
      phase distribution is the same spherical distribution, regardless
      of the mode indices.}
    \label{fig:gauss_shapes}
\end{figure}}

\subsection{Gouy phase}
\label{sec:Gouy}

The equation for \HG\ modes shows an extra longitudinal phase
lag. This \emph{Gouy phase}~\cite{boyd79, Gouy1890, Gouy1890b}
describes the fact that, compared to a plane wave, the \HG\ modes have
a slightly slower phase velocity, especially close to the waist. The
Gouy phase can be written as
\begin{equation}
\Psi(z)=\arctan\left(\frac{z-z_0}{\zr}\right),
\end{equation}
or, using the Gaussian beam parameter,
\begin{equation}
\Psi(z)=\arctan\left(\frac{\myRe{q}}{\myIm{q}}\right).
\end{equation}
Compared to a plane wave, the phase lag $\varphi$ of a \HG\ mode is
\begin{equation}
\varphi=(n+m+1)\Psi(z).
\end{equation}
With an astigmatic beam, i.e.~different beam parameters in the
tangential and sagittal planes, this becomes
\begin{equation}
\varphi=\left(n+\frac{1}{2}\right)\Psi_t(z)+\left(m+\frac{1}{2}\right)\Psi_s(z),
\end{equation}
with
\begin{equation}
\Psi_t(z)=\arctan\left(\frac{\myRe{q_t}}{\myIm{q_t}}\right),
\end{equation}
as the Gouy phase in the tangential plane (and $\Psi_s$ is similarly
defined in the sagittal plane).

\epubtkImage{LG10-LG31-LG33.png}{%
  \begin{figure}[htbp]
    \centerline{
      \includegraphics[viewport=130 0 690 570, scale=.3]{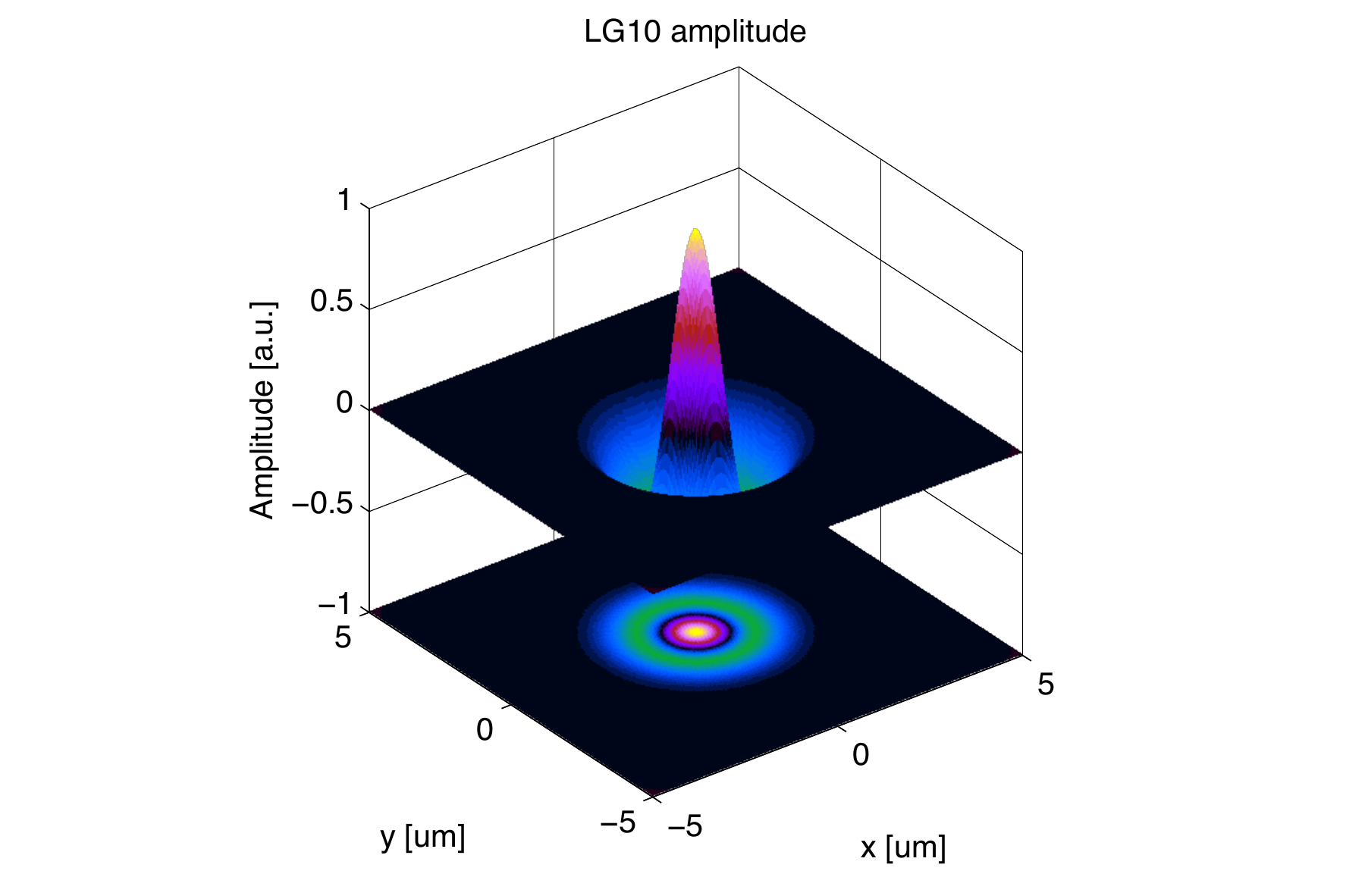}
      \includegraphics[viewport=130 0 690 570, scale=.3]{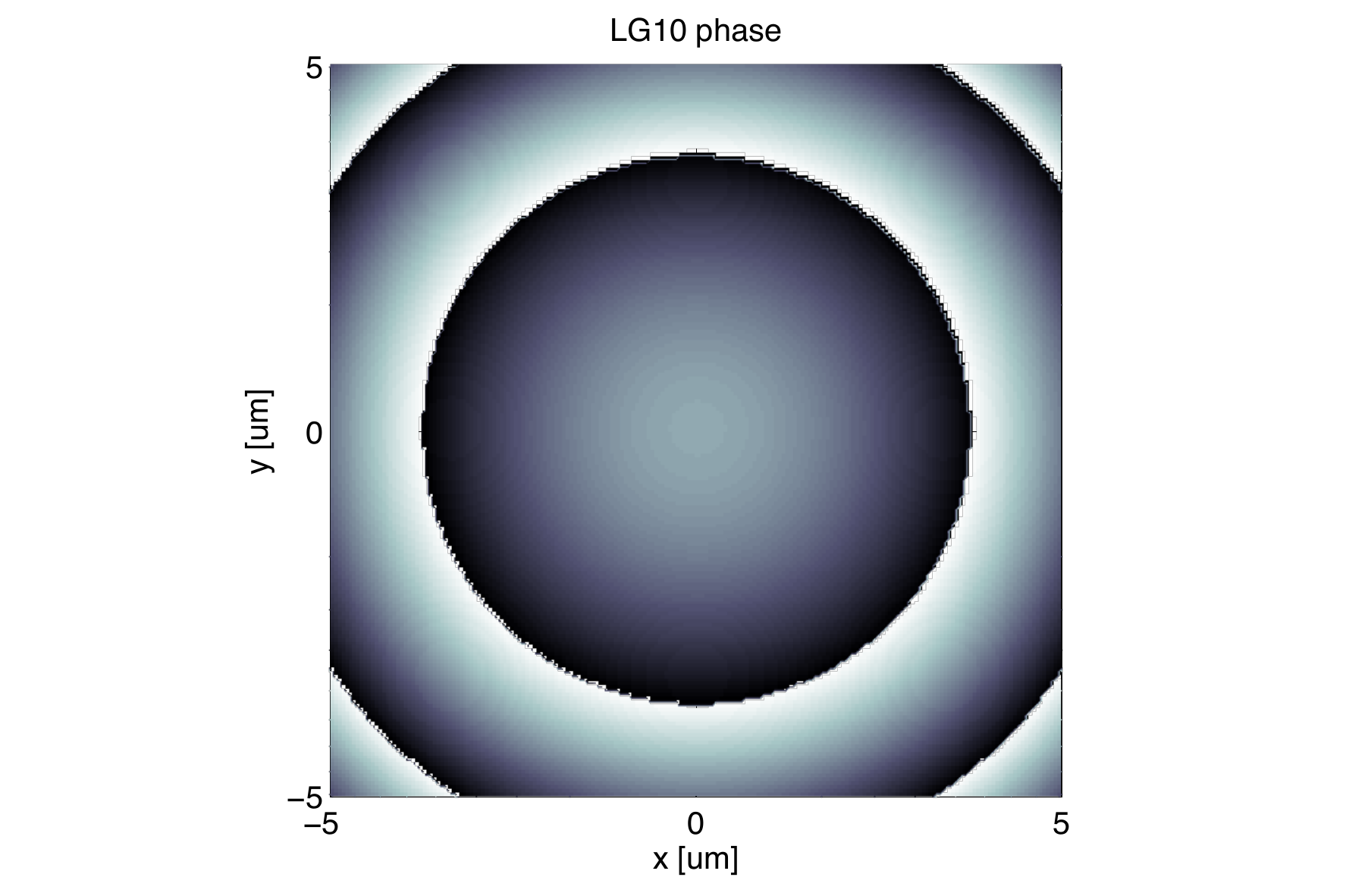}
    }
    \centerline{
      \includegraphics[viewport=130 0 690 570, scale=.3]{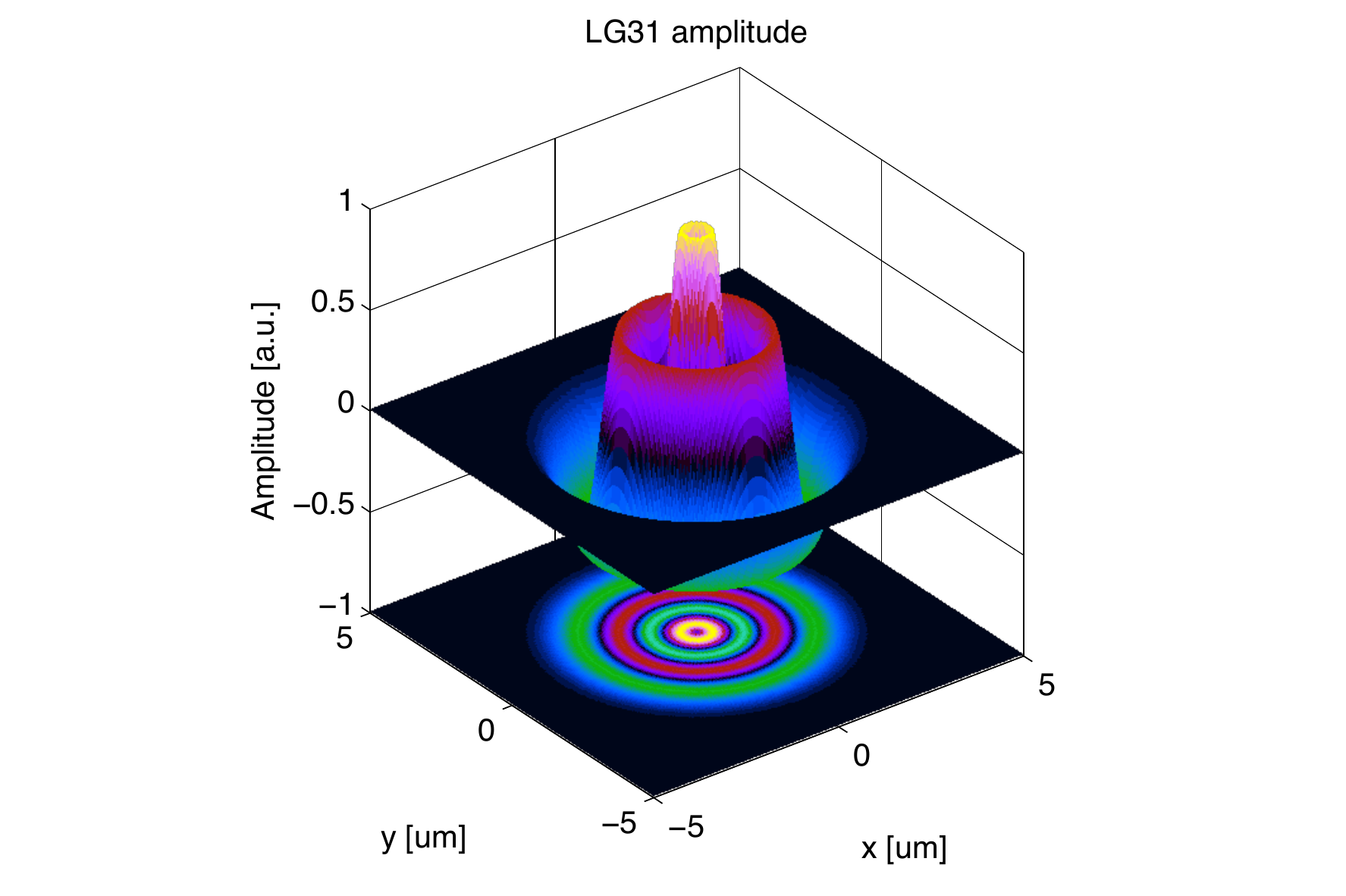}
      \includegraphics[viewport=130 0 690 570, scale=.3]{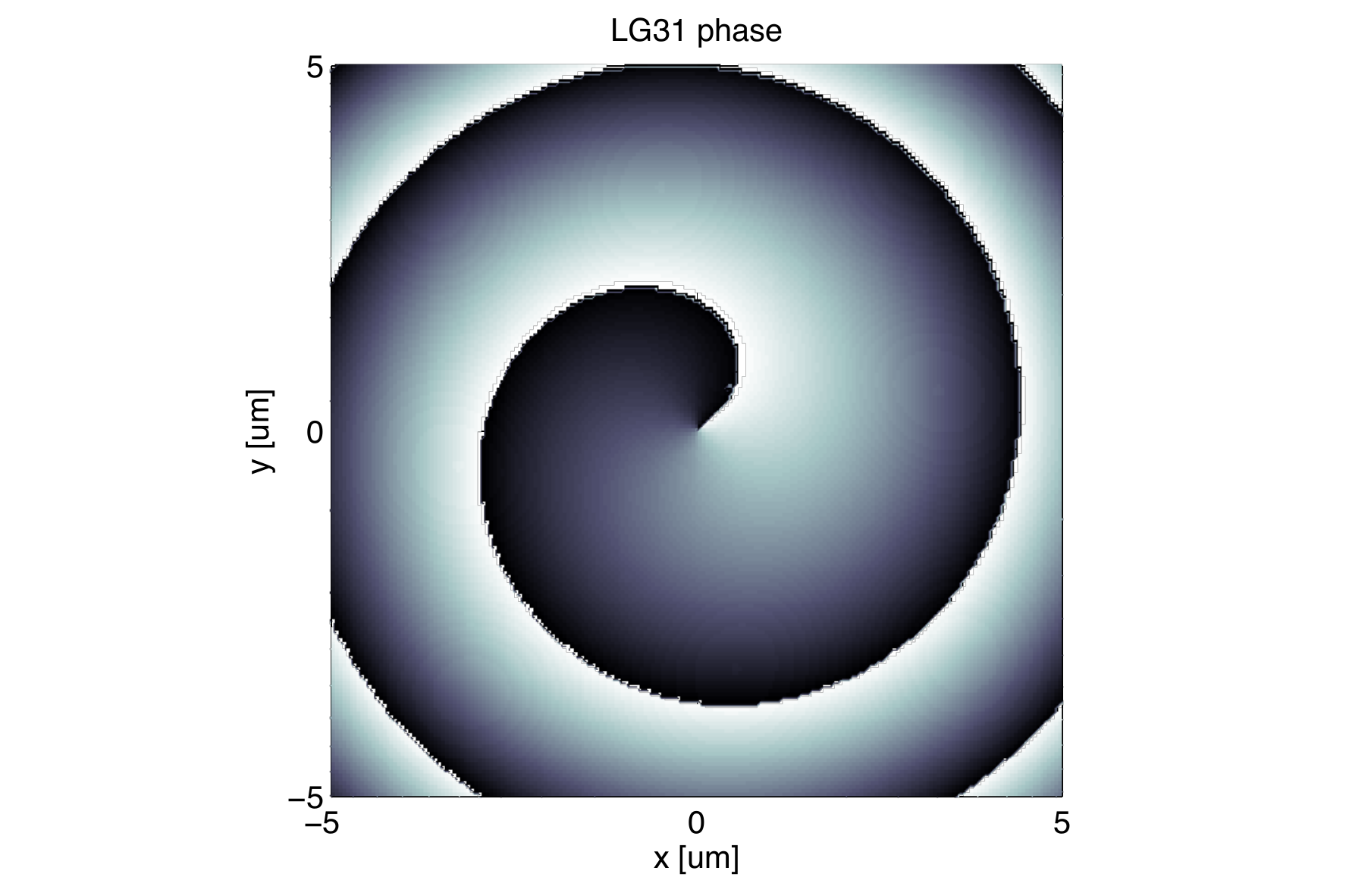}
    }
    \centerline{
      \includegraphics[viewport=130 0 690 570, scale=.3]{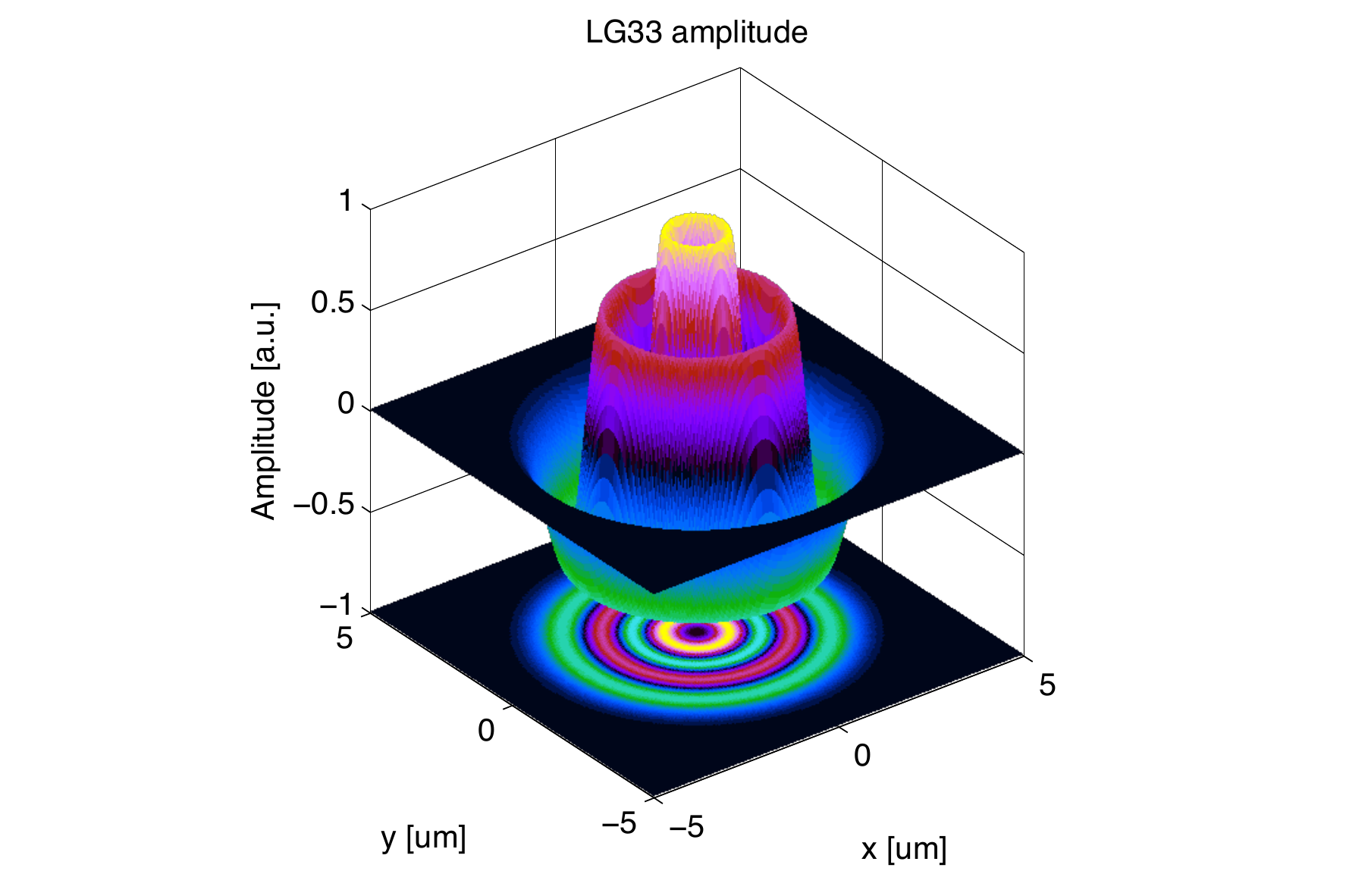}
      \includegraphics[viewport=130 0 690 570, scale=.3]{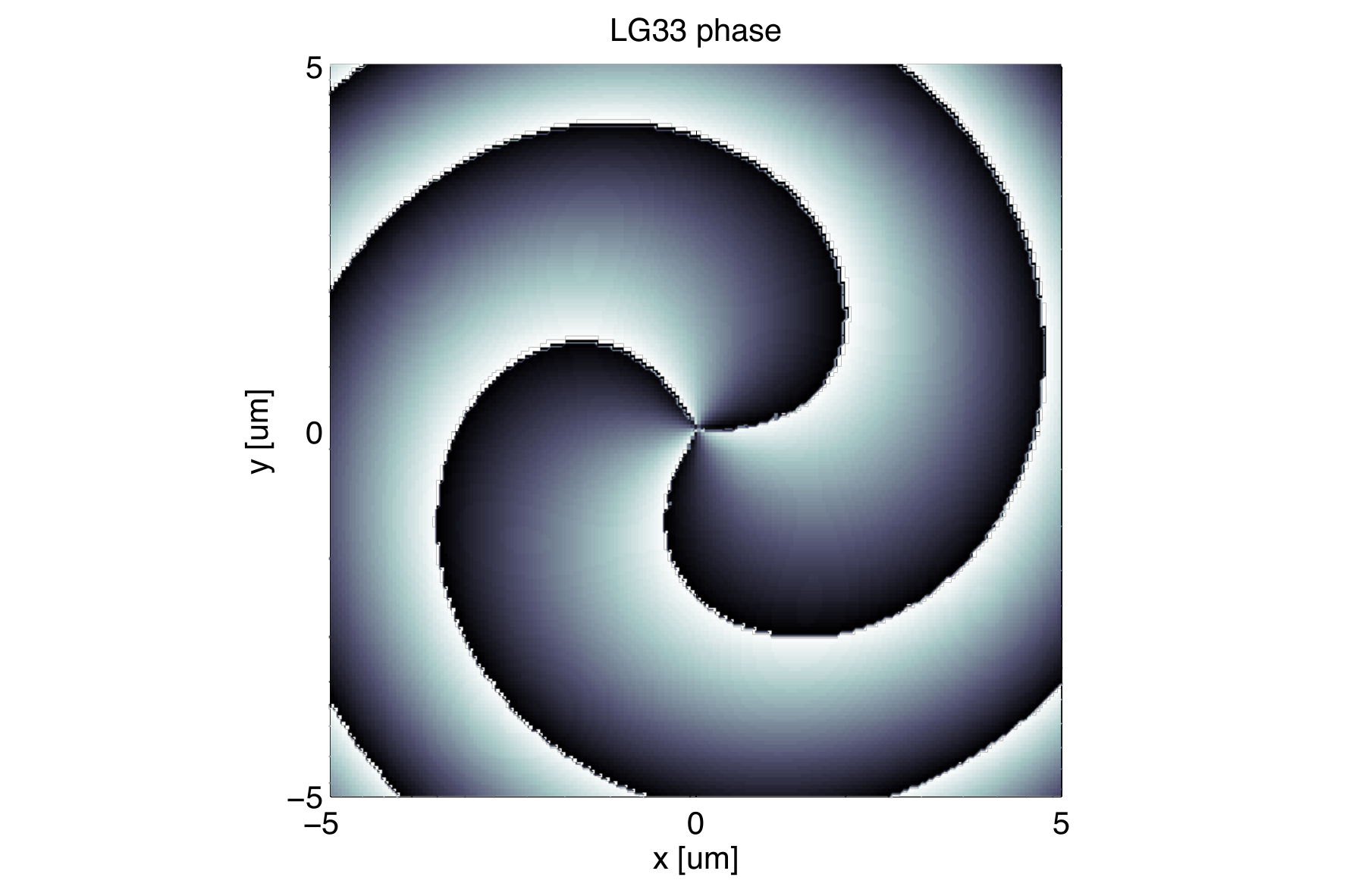}
    }
    \caption[Laguerre-Gaussian beam shapes]{These plots show the
      amplitude distribution and wave front (phase distribution) of
      helical Laguerre-Gauss modes $u_{pl}$. All plots refer to a
      beam with $\lambda$~=~1~\mum, \textit{w}~=~1~mm and distance to
      waist \z~=~1~m.
    }
    \label{fig:lg_gauss_shapes}
\end{figure}}

\epubtkImage{LGintensities.png}{%
  \begin{figure}[htbp]
    \centerline{\includegraphics[viewport=200  0 690 550, scale=.45]{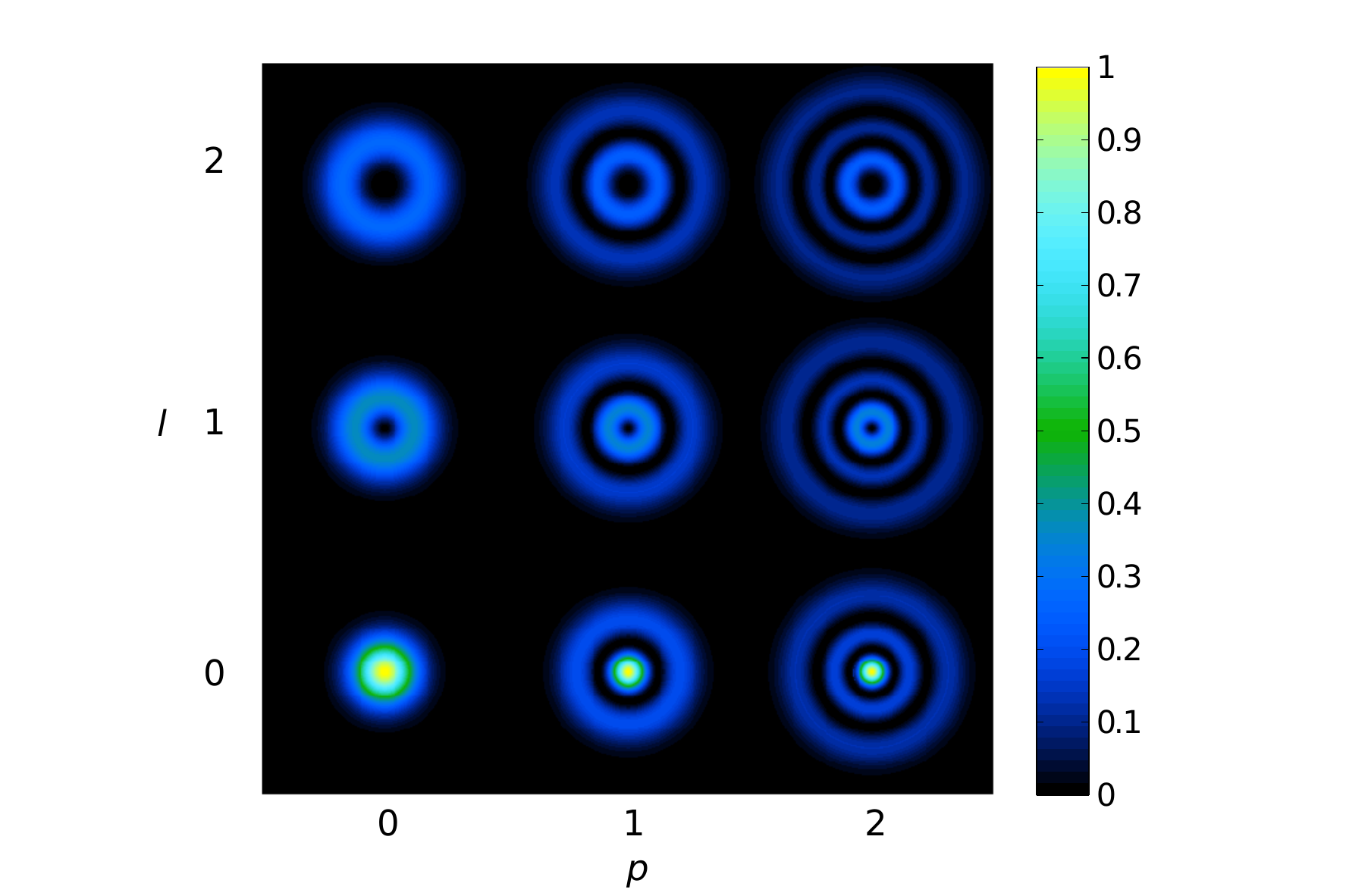}}
    \caption[Laguerre-Gaussian beam shapes]{Intensity profiles for
    helical Laguerre-Gauss modes $u_{pl}$. The $u_{00}$ mode is
    identical to the Hermite--Gauss mode of order 0. Higher-order
    modes show a widening of the intensity and decreasing peak
    intensity. The number of concentric dark rings is given by the
    radial mode index $p$.}
    \label{fig:LGintensities}
\end{figure}}

\epubtkImage{LGintensities_cos.png}{%
  \begin{figure}[htbp]
    \centerline{\includegraphics[viewport=200  0 690 550, scale=.45]{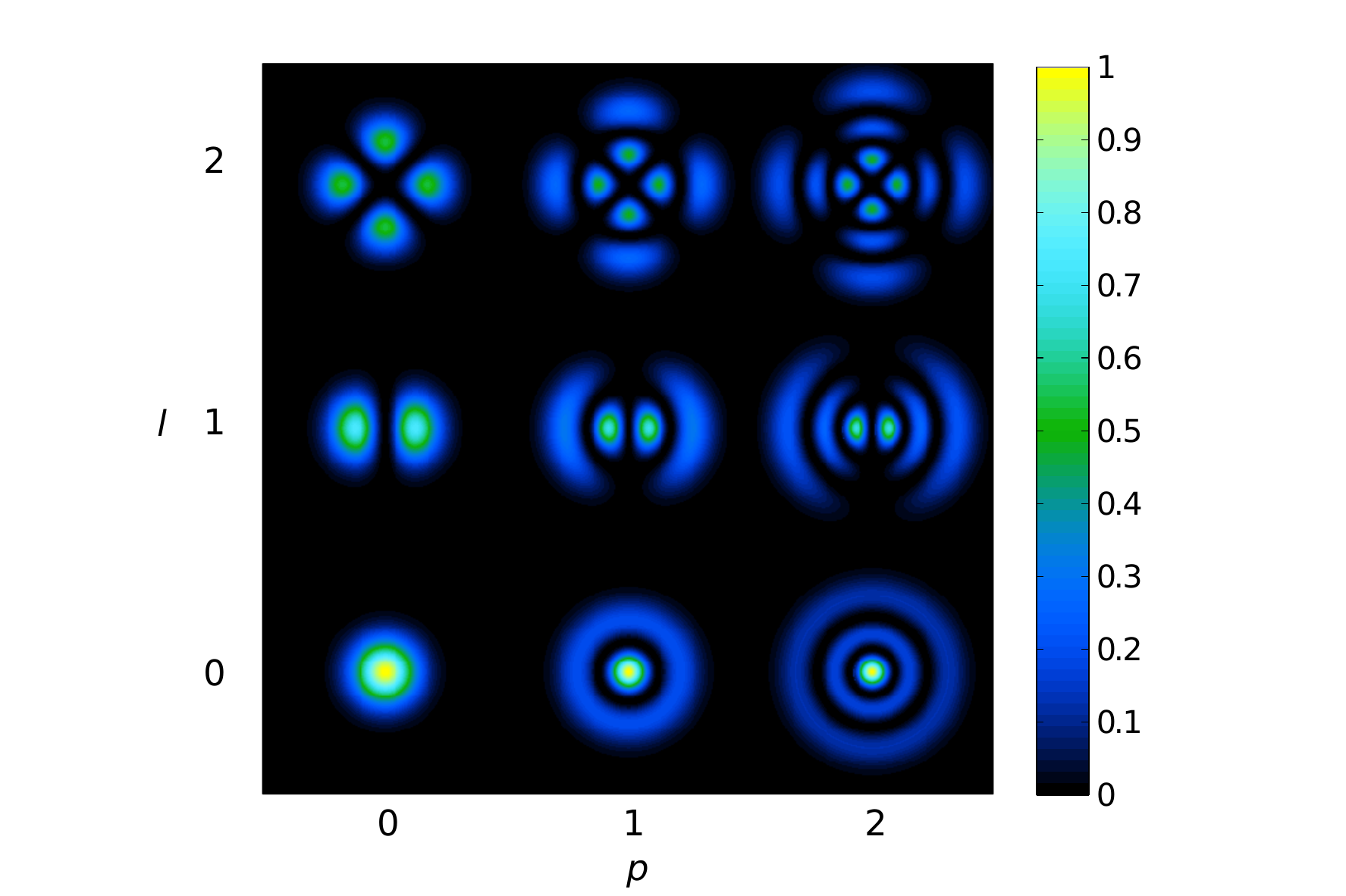}}
    \caption[Intensities of sinusoidal Laguerre--Gaussian
    modes]{Intensity profiles for sinusoidal Laguerre--Gauss modes
    $u^{\mathrm{alt}}_{pl}$. The $u_{p0}$ modes are identical to the
    helical modes. However, for azimuthal mode indices $l>0$ the
    pattern shows $l$ dark radial lines in addition to the $p$ dark
    concentric rings.}
    \label{fig:LGintensities_cos}
\end{figure}}

\subsection{Laguerre--Gauss modes}
\label{sec:LGmodes}
Laguerre--Gauss modes are another complete set of functions, which
solve the paraxial wave equation. They are defined in cylindrical
coordinates and can have advantages over Hermite--Gauss modes in the
presence of cylindrical symmetry. More recently, Laguerre--Gauss modes
are being investigated in a different context: using a pure
higher-order Laguerre--Gauss mode instead of the fundamental Gaussian
beam can significantly reduce the impact of mirror thermal noise on the
sensitivity of gravitational wave detectors~\cite{vinet09,
  Chelkowski09}. Laguerre--Gauss modes are commonly given
as~\cite{siegman}
\begin{equation}
\label{eq:uld1}
\begin{array}{rcl}
u_{p,l}(r,\phi,z)&=& \frac{1}{w(z)}\sqrt{\frac{2p!}{\pi(|l|+p)!}}\exp(\I(2p+|l|+1)\Psi(z))\\
&\times&\left(\frac{\sqrt{2}r}{w(z)}\right)^{|l|}L_p^{|l|}\left(\frac{2r^2}{w(z)^2}\right) \exp\left(-\I
k\frac{r^2}{2q(z)}+\I l \phi\right),
\end{array}
\end{equation}
with $r$, $\phi$ and $z$ as the cylindrical coordinates around the
optical axis. The letter $p$ is the radial mode index, $l$ the
azimuthal mode index\epubtkFootnote{\cite{siegman} states that the
  indices must obey the following relations: $0\leq |l|\leq p$.
  However, that is not the case.} and $L_p^{|l|}(x)$ are the associated
Laguerre polynomials:
\begin{equation}
L_p^{|l|}(x)=\frac{1}{p!}\sum_{j=0}^p\frac{p!}{j!}
\left(
\begin{array}{c}
|l|+p\\
p-j
\end{array}\right)(-x)^j .
\end{equation}
All other parameters $(w(z), q(z), \ldots)$ are defined as above for
the Hermite--Gauss modes.

The dependence of the Laguerre modes on $\phi$ as given in
Equation~(\ref{eq:uld1}) results in a spiralling phase front, while the
intensity pattern will always show unbroken concentric rings; see
Figure~\ref{fig:lg_gauss_shapes}. These modes are also called
\emph{helical} Laguerre--Gauss modes because of the their special
phase structure.

The reader might be more familiar with a slightly different type of
Laguerre modes (compare Figure~\ref{fig:LGintensities} and
Figure~\ref{fig:LGintensities_cos}) that features dark radial lines as
well as dark concentric rings. Mathematically, these can be described
simply by replacing the phase factor $\exp(\I l \phi)$ in
Equation~(\ref{eq:uld1}) by a sine or cosine function. For example, an
alternative set of Laguerre--Gauss modes is given by~\cite{VBP2}
\begin{equation}
\label{eq:uld2}
\begin{array}{rcl}
u^{\mathrm{alt}}_{p,l}(r,\phi,z)&=& \frac{2}{w(z)}\sqrt{\frac{p!}{(1+\delta_{0l}\pi(|l|+p)!}}\exp(\I(2p+|l|+1)\Psi(z))\\
&\times&\left(\frac{\sqrt{2}r}{w(z)}\right)^{|l|}L_p^{|l|}\left(\frac{2r^2}{w(z)^2}\right) \exp\left(-\I
k\frac{r^2}{2q(z)}\right)\cos(l \phi).
\end{array}
\end{equation}
This type of mode has a spherical phase front, just as the
Hermite--Gauss modes. We will refer to this set as \emph{sinusoidal}
Laguerre--Gauss modes throughout this document.

For the purposes of simulation it can be sometimes useful to decompose
Laguerre--Gauss modes into Hermite--Gauss modes. The mathematical
conversion for helical modes is given as~\cite{Beijersbergen93,
  Abramochkin91}
\begin{equation}
\label{eq:LGtoHG}
u^{LG}_{p,l}(x,y,z)=\sum_{k=0}^N (-1)^p (\mp \I)^k b(|l|+p,p,k) u^{HG}_{N-k,k}(x,y,z),
\end{equation}
where $\mp$ is negative for positive $l$ and positive for negative $l$
and with real coefficients
\begin{equation}
b(n,m,k)=\sqrt{\frac{(N-k)!k!}{2^N n!m!}} \frac{1}{k!}(\partial_t)^k[(1-t)^n(1+t)^m]_{t=0},
\end{equation}
if $N=2p+|l|$.
The coefficients $b(n,m,k)$
can be computed numerically by using Jacobi polynomials.
Jacobi polynomials can be written in various forms:
\begin{equation}
P_n^{\alpha, \beta}(x)=\frac{(-1)^n}{2^n
n!}(1-x)^{-\alpha}(1+x)^{-\beta}(\partial_x)^n(1-x)^{\alpha+n}(1+x)^{\beta+n},
\end{equation}
or
\begin{equation}
P_n^{\alpha, \beta}(x)=\frac{1}{2^n}\sum_{j=0}^n
\left(
\begin{array}{c}
n+\alpha\\
j
\end{array}\right)
\left(\begin{array}{c}
n+\beta\\
n-j
\end{array}\right)
(x-1)^{n-j}(x+1)^j ,
\end{equation}
which leads to
\begin{equation}
b(n,m,k)=\sqrt{\frac{(N-k)!k!}{2^N n!m!}}~(-2)^k P_k^{n-k,m-k}(0).
\end{equation}
%

\subsection{Tracing a Gaussian beam through an optical system}
\label{sec:trace}
Whenever Gauss modes are used to analyse an optical system, the
Gaussian beam parameters (or equivalent waist sizes and locations)
must be defined for each location at which field amplitudes are to be
computed (or at which coupling equations are to be defined). In our
experience the quality of a computation or simulation and the
correctness of the results depend critically on the choice of these
beam parameters. One might argue that the choice of a basis should not
alter the result.  This is correct, but there is a practical
limitation: the number of modes having non-negligible power might
become very large if the beam parameters are not optimised, so that in
practice a good set of beam parameters is usually required.

In general, the Gaussian beam parameter of a mode is changed at every
optical surface in a well-defined way (see Section~\ref{sec:abcd}). Thus, a possible
method of finding reasonable beam parameters for every location in
the interferometer is to first set only some specific beam parameters
at selected locations and then to derive the remaining beam parameters from
these initial ones: usually it is sensible to assume that the beam at the
laser source can be properly described by the (hopefully known) beam parameter of
the laser's output mode. In addition, in most stable cavities the light fields should
be described by using the respective cavity eigenmodes.
Then, the remaining beam
parameters can be computed by \emph{tracing} the beam through the
optical system. `Trace' in this context means that a beam starting
at a location with an already-known beam parameter is propagated mathematically through the
optical system. At every optical element along
the path the beam parameter is transformed according to the ABCD matrix of
the element (see below).

\subsection{ABCD matrices}
\label{sec:abcd}
The transformation of the beam parameter can be performed by the ABCD
matrix-formalism~\cite{kogelnik65, siegman}. When a beam passes an
optical element or freely propagates though space, the initial beam
parameter $q_1$ is transformed into $q_2$. This transformation can be
described by four real coefficients as follows:
\begin{equation}
\frac{q_2}{n_2}=\frac{A\frac{q_1}{n_1}+B}{C\frac{q_1}{n_1}+D},
\label{eq:abcd_transform}
\end{equation}
with the coefficient matrix
\begin{equation}
M=\left(
\begin{array}{cc}
 A& B \\
 C& D
\end{array}
\right),
\end{equation}
$n_1$ being the index of refraction at the beam segment defined by
$q_1$, and $n_2$ the index of refraction at the beam segment described
by $q_2$. ABCD matrices for some common optical components are given
below, for the sagittal and tangential plane.

\subsubsection*{Transmission through a mirror:}
A mirror in this context is a single, partly-reflecting surface with
an angle of incidence of 90\textdegree. The transmission is described
by

\epubtkImage{abcd_mi1-external.png}{%
  \begin{figure}[h!]
    \centerline{\includegraphics[width=0.9\textwidth]{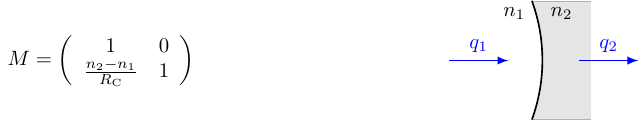}}
    \caption{}
    \label{eq:abcd-mi1}
\end{figure}}
\noindent
with $R_{\mathrm{C}}$ being the radius of curvature of the spherical
surface. The sign of the radius is defined such that $R_{\mathrm{C}}$
is negative if the centre of the sphere is located in the direction of
propagation. The curvature shown above (in Figure~\ref{eq:abcd-mi1}),
for example, is described by a positive radius. The matrix for the
transmission in the opposite direction of propagation is identical.

\clearpage
\subsubsection*{Reflection at a mirror:}
The matrix for reflection is given by%
\epubtkImage{abcd_mi2-external.png}{%
  \begin{figure}[!h]
    \centerline{\includegraphics[width=0.8\textwidth]{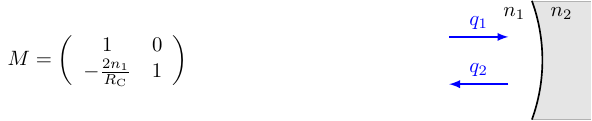}}
    \caption{}
    \label{eq:abcd-mi2}
\end{figure}}

\noindent
The reflection at the back surface can be described by the same type
of matrix by setting $C=2n_2/\roc$.

\subsubsection*{Transmission through a beam splitter:}
A beam splitter is understood as a single surface with an arbitrary
angle of incidence $\alpha_1$. The matrices for transmission and
reflection are different for the sagittal and tangential planes
($M_{\mathrm{s}}$ and $M_{\mathrm{t}}$):

\epubtkImage{abcd_bs1-external.png}{%
  \begin{figure}[h!]
    \centerline{\includegraphics[width=0.9\textwidth]{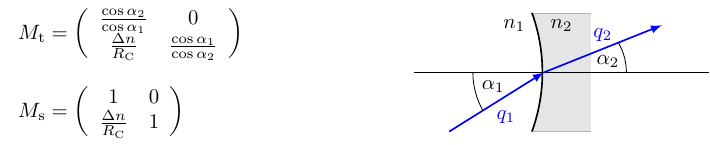}}
    \caption{}
    \label{eq:abcd_bs1}
\end{figure}}
\noindent
with $\alpha_2$ given by Snell's law:
\begin{equation}
n_1\mSin{\alpha_1}=n_2\mSin{\alpha_2},
\end{equation}
and $\Delta n$ by
\begin{equation}
\Delta
n=\frac{n_2\mCos{\alpha_2}-n_1\mCos{\alpha_1}}{\mCos{\alpha_1}\mCos{\alpha_2}}.
\end{equation}
If the direction of propagation is reversed, the matrix for the sagittal
plane is identical and the matrix for the tangential plane can be obtained
by changing the coefficients A and D as follows:
\begin{equation}
\begin{array}{l}
A\longrightarrow1/A,\\
D\longrightarrow1/D.
\end{array}
\end{equation}

\subsubsection*{Reflection at a beam splitter:}
The reflection at the front surface of a beam splitter is given by:

\epubtkImage{abcd_bs2-external.png}{%
  \begin{figure}[h!]
    \centerline{\includegraphics[width=0.9\textwidth]{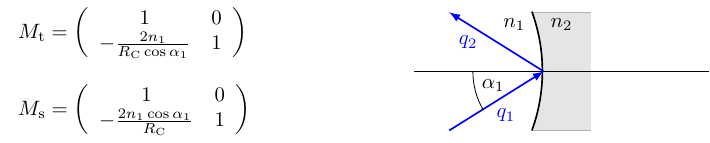}}
    \caption{}
    \label{eq:abcd_bs2}
\end{figure}}
\noindent
To describe a reflection at the back surface the matrices have to be changed
as follows:
\begin{equation}
\begin{array}{l}
R_{\mathrm{C}}\longrightarrow-R_{\mathrm{C}},\\
n_1\longrightarrow n_2,\\
\alpha_1\longrightarrow-\alpha_2.
\end{array}
\end{equation}

\subsubsection*{Transmission through a thin lens:}
A thin lens transforms the beam parameter as follows:

\epubtkImage{abcd_len-external.png}{%
  \begin{figure}[h!]
    \centerline{\includegraphics[width=0.8\textwidth]{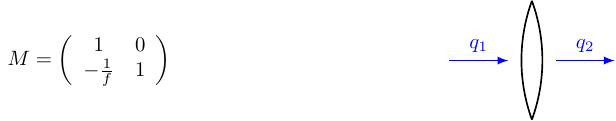}}
    \caption{}
    \label{eq:abcd_len}
\end{figure}}
\noindent
where $f$ is the focal length. The matrix for the opposite direction
of propagation is identical. Here it is assumed that the thin lens is
surrounded by `spaces' with index of refraction $n=1$.

\subsubsection*{Transmission through a free space:}
As mentioned above, the beam in free space can be described by one
base parameter $q_0$. In some cases it is convenient to use a matrix
similar to that used for the other components to describe the
$z$-dependency of $q(z)=q_0+z$. On propagation through a free space of
the length $L$ and index of refraction $n$, the beam parameter is
transformed as follows.

\epubtkImage{abcd_sp-external.png}{%
  \begin{figure}[h!]
    \centerline{\includegraphics[width=0.8\textwidth]{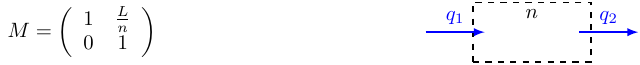}}
    \caption{}
    \label{eq:abcd-sp}
\end{figure}}
\noindent
The matrix for the opposite direction of propagation is identical.

\subsection{Computing a cavity eigenmode and stability}
\label{sec:cav_eigen}

A cavity eigenmode is defined as the optical field whose spatial
properties are such that the field after one round-trip through the cavity will
be exactly the same as the injected field. In the case of resonators
with spherical mirrors, the eigenmode will be a Gaussian mode,
defined by the Gaussian beam parameter $q_{\rm{cav}}$. For a generic
cavity (an arbitrary number of spherical mirrors or lenses) a
round-trip ABCD matrix $M_{\rm{rt}}$ can be defined and used to compute
the cavity's eigenmode.
Chapter 21 of~\cite{siegman} provides a comprehensive description of
different optical resonators including a derivation and discussion of stability
criteria. Here we provide a brief introduction focussing on the specific case
of closed and stable resonators with spherical mirrors.

The change in the $q$ parameter after one round-trip through a cavity is
given by:
\begin{eqnarray}
    \frac{A q_{1} + B}{C q_{1} + D} = q_{2} = q_{1}
\end{eqnarray}
where $A$, $B$, $C$ and $D$ are the elements of a matrix $M_{\rm{rt}}$.
If $q_1 = q_2$ then the spatial profile of the beam is recreated after
each round-trip and we have identified the cavity eigenmode. We can compute the
parameter $q_{\rm{cav}} \equiv q_1 = q_2$ by solving:
\begin{equation}
	C q_{\rm{cav}}^2+(D-A)q_{\rm{cav}} - B = 0, \label{eq:cavity_eigenmode_quadratic}
\end{equation}
For example, in the case of the  two-mirror cavity shown in
Figure~\ref{fig:cavity_eigenmode_calc} the matrix is given by:
\begin{eqnarray}
	M_{\rm{rt}} &=& M_{space}(L) \times M_{\rm{refl}}(R_2) \times M_{\rm{space}}(L)
  \times M_{\rm{refl}}(R_1),
\end{eqnarray}
with $L$ the length of the cavity, and $R_{1/2}$ the radii of curvature
of the mirrors. Now we can compute the A, B, C and D coefficients for the
round-trip matrix $M_{\rm{rt}}$ to solve
Equation~\ref{eq:cavity_eigenmode_quadratic}.
This quadratic equation generally has two solutions, one being the
complex conjugate of the other.

\epubtkImage{cavity_eigenmode_calc.png}{%
  \begin{figure}[h!]
    \centerline{\includegraphics[width=0.6\textwidth]{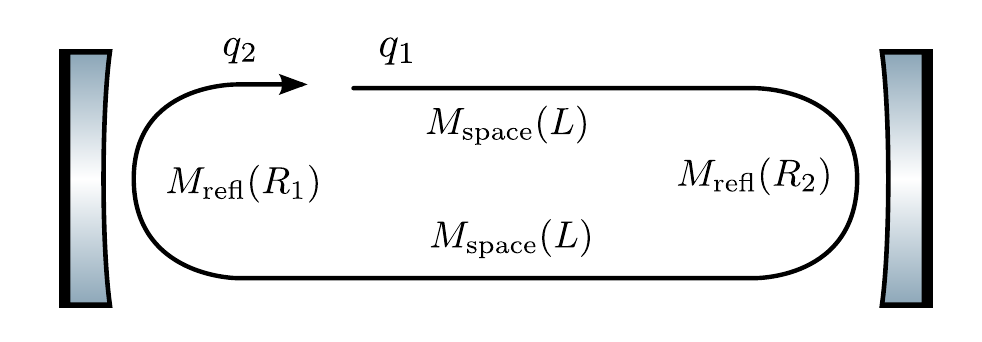}}
    \caption{Cavity round trip ABCD matrices for a 2-mirror cavity}
    \label{fig:cavity_eigenmode_calc}
\end{figure}}

When the polynomial above has a suitable solution the
optical resonator is said to be `stable'. The stability requirement can
be formulated using the Gaussian
beam parameter: a cavity is stable only when the cavity's eigenmode, $q_{\rm{cav}}$,
has a real waist size. The value for the beam waist is a real number whenever $q_{\rm{cav}}$
has a positive non-zero imaginary part, as this defines the Rayleigh range of the
beam and therefore the beam waist, $\operatorname{Im}(q_{\rm{cav}}) = \pi w_0^2/\lambda$.
A complex $q_{\rm{cav}}$ is ensured if the determinant of
equation \ref{eq:cavity_eigenmode_quadratic} is negative.
This requirement can formulated in a compact way by defining the
parameter $m$ as:
\begin{equation}
m \equiv \frac{A+D}{2},
\end{equation}
where $A$ and $B$ are the coefficients of the round-trip matrix $M_{\rm rt}$.
The stability criterion then simply becomes:
\begin{equation}
m^2 < 1.
\end{equation}

The stability of simple cavities are often described using \textit{g-factors}.
These factors are simply rescaled versions of the more generic $m$ value:
\begin{equation}
	g \equiv \dfrac{m+1}{2} = \dfrac{A+D+2}{4}, \label{eq:g-factor}
\end{equation}
For the cavity to be stable the g-factor must fulfil the requirement:
\begin{equation}
0 \leq g \leq 1
\end{equation}
The closer $g$ is to the 0 or 1, the smaller the
tolerances are for any change in the geometry before
the cavity becomes unstable.

For a simple two-mirror cavity of length $L$ and mirror radii of curvature
$R_{1,2}$, its \textit{g-factor} is
\begin{eqnarray}
	g_1 &=& 1 - \frac{L}{R_1}, \\
	g_2 &=& 1 - \frac{L}{R_2}, \\
	g &=& g_1 g_2.
\end{eqnarray}
Where $g_{1,2}$ are the individual \textit{g-factors} of the cavity
mirrors and $g$ is the g-factor of the entire cavity.

\subsection{Round-trip Gouy phase and higher-order-mode separation}

As discussed in section~\ref{sec:Gouy}, as a higher order optical mode
propagates it accumulates an additional phase, the Gouy phase, proportional
to its mode order. To determine how such a mode resonantes within an optical
cavity the accumulated Gouy phase on one round-trip through the cavity must
be included. The round-trip Gouy phase will determine which order of optical modes
are resonant within a cavity. As the resonance condition of a mode is dependent
on its order, this allows an optical setup to select particular orders of optical modes from an
incident field. This behaviour is the basis of \textit{mode-cleaner} cavities; such
as those used for the input and output light of gravitational wave detectors.

To compute the round-trip Gouy phase the evolution of the beam shape through the
cavity must first be computed. This involves computing the round-trip ABCD matrix, $M_{\rm{RT}}$,
as outlined in section~\ref{sec:cav_eigen}. With this matrix the round-trip Gouy phase
is computed using its elements~\cite{T1300189}:
\begin{eqnarray}
	\psi_{\rm{RT}} &=& 2 \rm{arccos} \left( \rm{sign}(B) \sqrt{ \dfrac{A+D+2}{4} } \right), \\
	\psi_{\rm{RT}}(g) &=& 2 \rm{arccos} \left( \rm{sign}(B) \sqrt{g} \right).
\end{eqnarray}
As can be seen, the round-trip Gouy phase is linked to the cavity's g-factor, \eqref{eq:g-factor}.
As the cavity approaches instability, $g\rightarrow 0$ or $1$, the
phase accumulated by a mode TEM\sub{nm} is then $(n+m)\psi_{\rm{RT}}(0) = (n+m)\pi/2$
or $(n+m)\psi_{\rm{RT}}(1) = 0$. In the later case all higher order optical modes---regardless of their mode order---are resonant in the cavity at the same time. In the former case
either the odd or even mode orders are resonant at once.

The effect of the round-trip Gouy phase has is often referred to
\textit{higher order mode separation frequency}. This states how
far the resonance of the next optical order is in frequency:
\begin{equation}
	\delta f = \dfrac{\psi_{\mathrm{RT}}}{2\pi} \mathrm{FSR}.
\end{equation}
For example, the Advanced LIGO arm cavities have $\delta f \approx 5$~kHz.

\subsection{Coupling of higher-order-modes}
\label{sec:kmnmn}
Now that we are able to compute the eigenmode of a particular cavity,
what happens if a beam with a slightly different eigenmode is injected into it?
The aim of this section is to outline the problem. In reality producing a
perfect Gaussian laser which matches exactly the eigenmode of a cavity is
essentially impossible, there will always be a minor difference. However
we are still able to inject lasers into a cavity and produce a resonance.
This is because as long as the eigenmode of the incoming laser is nearly the
same, the majority of the laser light will `fit' into the cavity and resonate,
the rest will be reflected from it.

Let us consider a cavity whose eigenmode is given by $q_2$
and the eigenmode of a Gaussian beam that is incident on it, $q_1$.
The incident beam has all of its power in the fundamental $00$ mode.
For a cavity with perfectly spherical mirrors there are two possible
`misconfigurations' that can take place:
\begin{itemize}
\item If the optical axes of the beam and the cavity do not
  overlap perfectly, the setup is called \emph{misaligned},
\item If the beam size or shape at cavity input does not match
  the beam shape and size of the (resonant) fundamental eigenmode
  ($q_1(z_{\mathrm{cav}})\neq q_2(z_{\mathrm{cav}})$), the beam is then not
  \emph{mode-matched} to the second cavity, i.e.~there is a
  \emph{mode mismatch}.
\end{itemize}
The coupling of a mode refers to how a spatial mode in one basis
is represented in another; e.g. which sum of modes in the cavity basis $q_2$
produces the HG\sub{00} mode in the $q_1$ basis.
\HG modes are coupled whenever a beam is not matched or aligned to a cavity or
beam segment. This
coupling is sometimes referred to as \textit{scattering} into higher-order
modes because in most cases the laser beam is a considered as a pure
HG\sub{00} mode and any mode coupling would transfer power from the
fundamental into higher-order modes. However, in general every mode
with non-zero power will transfer energy into other modes whenever
mismatch or misalignment occur, and this effect also includes the
transfer from higher orders into a low order.

To compute the amount of coupling the beam must be projected into the
base system of the cavity or beam segment it is being injected
into. This is always possible, provided that the paraxial
approximation holds, because each set of \HG modes, defined by the
beam parameter at a position \z, forms a complete set. Such a change
of the basis system results in a different distribution of light power
in the new \HG modes and can be expressed by coupling coefficients
that yield the change in the light amplitude and phase with respect to
mode number.

Let us assume that a beam described by the beam parameter $q_1$ is
injected into a segment described by the parameter $q_2$. Let the
optical axis of the beam be misaligned: the coordinate system of the
beam is given by ($x, y, z$) and the beam travels along the \z-axis.
The beam segment is parallel to the $z'$-axis and the coordinate
system ($x', y', z'$) is given by rotating the ($x, y, z$) system
around the \y-axis by the \emph{misalignment angle} $\gamma$. The
amplitude of a particular mode TEM\sub{nm} in the beam segment is then defined as:
\begin{equation}
u_{n m}(x,y;\,q_2)\mExB{\I(\w t -k z)}=\sum_{n',m'}k_{n,m,n',m'}u_{n'
m'}(x,y;\,q_1)\mExB{\I(\w t -k z')},
\end{equation}
where $u_{n' m'}(x,y;\,q_1)$ are the \HG modes used to describe the injected
beam, $u_{n m}(x,y;\,q_2)$ are the `new' modes that are used to
describe the light in the beam segment and $k_{n,m,n',m'}$ is the
\textit{coupling coefficient} from each TEM\sub{n'm'} into TEM\sub{nm}.
Note that including the plane
wave phase propagation within the definition of coupling coefficients
is important because it results in coupling coefficients that are
independent of the position on the optical axis for which the coupling
coefficients are computed.

Using the fact that the \HG modes $u_{n m}$ are orthonormal, we can
compute the coupling coefficients by the
overlap integral~\cite{bayer-helms}:
\begin{equation}
\label{eq:tem_conv}
k_{n,m,n',m'}=\mEx{\I 2 k z'
\sin^2\left(\frac{\gamma}{2}\right)}\int\!\!\!\int\!dx'dy'~
u_{n' m'}\mEx{\I k x' \sin{\gamma}}~u^*_{n m}.
\end{equation}
Since the Hermite--Gauss modes can be separated with respect to \x and
\y, the coupling coefficients can also be split into $k_{n m n'
  m'}=k_{n n'}k_{m m'}$.  These equations are very useful in the
paraxial approximation as the coupling coefficients decrease with
large mode numbers. In order to be described as paraxial, the angle
$\gamma$ must not be larger than the diffraction angle. In addition,
to obtain correct results with a finite number of modes the beam
parameters $q_1$ and $q_2$ must not differ too much.

The integral \ref{eq:tem_conv} can be computed
directly using numerical integration methods. However, this can potentially be computationally
very expensive depending on how difficult the integrand is to evaluate and complex it it.
The following part of this section is based on the work of
Bayer-Helms~\cite{bayer-helms} and provides an analytic solution to the integral.
Another description of coupling coefficients and their derivation can be found in the work of
Vinet~\cite{VBP2}. In~\cite{bayer-helms} the above integral
is partly solved and the coupling coefficients are given by multiple
sums as functions of $\gamma$ and the mode mismatch parameter $K$,
which is defined by
\begin{equation}
K=\frac12(K_0+\I K_2),
\end{equation}
where $K_0=(z_R-z_R')/z_R'$ and $K_2=((z-z_0)-(z'-z_0'))/z_R'$. This
can also be written using $q=\I\zr +z-z_0$, as
\begin{equation}
K=\frac{\I (q-q')^*}{2 \myIm{q'}}.
\end{equation}

The coupling coefficients for misalignment and mismatch (but no
lateral displacement) can then be written as
\begin{equation}
\label{eq:ccoeff}
k_{n n'}=(-1)^{n'} E^{(x)} (n!n'!)^{1/2} (1+K_0)^{n/2+1/4}
(1+K^*)^{-(n+n'+1)/2}\left\{S_g-S_u\right\},
\end{equation}
where
\begin{equation}
{\renewcommand{\arraystretch}{2.5}\begin{array}{l}
S_g=\sum\limits_{\mu=0}^{[n/2]}\sum\limits_{\mu'=0}^{[n'/2]}
\frac{(-1)^\mu \bar{X}^{n-2\mu}X^{n'-2\mu'}}{(n-2\mu)!(n'-2\mu')!}
\sum\limits_{\sigma=0}^{\min(\mu,\mu')}\frac{(-1)^\sigma
\bar{F}^{\mu-\sigma} F^{\mu'-\sigma}}
{(2\sigma)! (\mu-\sigma)! (\mu'-\sigma)!},\\
S_u=\sum\limits_{\mu=0}^{[(n-1)/2]}\sum\limits_{\mu'=0}^{[(n'-1)/2]}
\frac{(-1)^\mu \bar{X}^{n-2\mu-1}X^{n'-2\mu'-1}}{(n-2\mu-1)!(n'-2\mu'-1)!}
\sum\limits_{\sigma=0}^{\min(\mu,\mu')}\frac{(-1)^\sigma
\bar{F}^{\mu-\sigma} F^{\mu'-\sigma}}
{(2\sigma+1)! (\mu-\sigma)! (\mu'-\sigma)!}.
\end{array}}
\end{equation}
The corresponding formula for $k_{m m'}$ can be obtained by replacing
the following parameters: $n\rightarrow m$, $n'\rightarrow m'$,
$X,\bar{X}\rightarrow 0$ and $E^{(x)}\rightarrow 1$ (see below).  The
notation $[n/2]$ means
\begin{equation}
\left[\frac{m}{2}\right]=\left\{
\begin{array}{ll}
m/2 & \text{if}\ m\ \text{is even,}\\
(m-1)/2 & \text{if}\ m\ \text{is odd.}
\end{array}\right.
\end{equation}
The other abbreviations used in the above definition are
\begin{equation}
{\renewcommand{\arraystretch}{1.5}
\begin{array}{l}
\bar{X}={(\I \zr'-z')\sin{(\gamma)}}/({\sqrt{1+K^*}w_0}),\\
X={(\I \zr+z')\sin{(\gamma)}}/({\sqrt{1+K^*}w_0}),\\
F={K}/({2(1+K_0)}),\\
\bar{F}={K^*}/{2},\\
E^{(x)}=\mEx{-\frac{X\bar{X}}{2}}.
\end{array}}
\end{equation}


\epubtkImage{kmn_mi.png}{%
  \begin{figure}[htb]
    \centerline{\includegraphics[viewport= 0 0 380 120,scale=0.5]{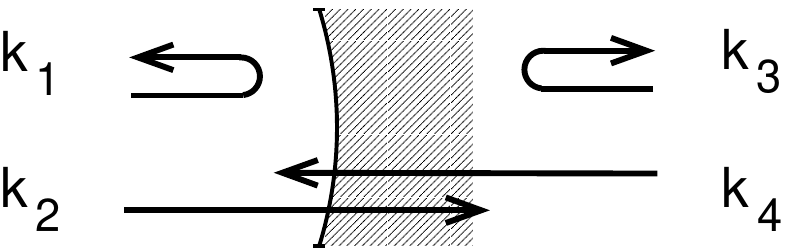}}
    \caption[Coupling coefficients for TEM modes]{Coupling
      coefficients for \HG modes: for each optical element and each
      direction of propagation complex coefficients $k$ for
      transmission and reflection have to be computed. In this figure
      $k_1$, $k_2$, $k_3$, $k_4$ each represent a matrix of
      coefficients $k_{n m n' m'}$ describing the coupling of
      $u_{n,m}$ into $u_{n',m'}$.}
    \label{fig:kmn_mi}
\end{figure}}

In general, the Gaussian beam parameter might be different for the
sagittal and tangential planes and a misalignment can be given for
both possible axes (around the \y-axis and around the \x-axis), in
this case the coupling coefficients are given by
\begin{equation}
\label{eq:knmnm}
k_{nm m'n'}=k_{n n'} k_{m m'},
\end{equation}
where $k_{n n'}$ is given above with
\begin{equation}
{\renewcommand{\arraystretch}{1}
\begin{array}{l}
q \rightarrow q_t\\
\text{and}\\
w_0\rightarrow w_{t,0} \text{, etc.}
\end{array}}
\end{equation}
and $\gamma \rightarrow \gamma_y$ is a rotation about the \y-axis. The
$k_{m m'}$ can be obtained with the same formula, with the following
substitutions:
\begin{equation}
{\renewcommand{\arraystretch}{1}
\begin{array}{l}
n \rightarrow m,\\
n' \rightarrow m', \\
q \rightarrow q_s,\\
\text{thus}\\
w_0\rightarrow w_{s,0} \text{, etc.}
\end{array}}
\end{equation}
and $\gamma \rightarrow \gamma_x$ is a rotation about the \x-axis.
At each component a matrix of coupling coefficients has to be computed
for every time a beam transfers from one eigenmode to another
for transmission and reflection as depicted in Figure~\ref{fig:kmn_mi}.

In this section we have outlined how an incoming higher-order-mode will be coupled
into an outgoing beam basis when taking into account a difference in the eigenmode
of two sections of the interferometer or misalignments. This coupling of higher-order-modes
is a very powerful tool that  is used throughout
this article, as it allows us to model
interferometers with realistic defects; like imperfect mirror surfaces or
misaligned optics. This enables us to better understand the reasons
why complex interferometers behave in certain ways
and provide solutions to combat particular problems that might arise.
The next section details how misalignments and mode-mismatching
affect the dual-recycled Fabry-Perot Michelson interferometers that were described in
Section~\ref{sec:advanced}. Then Section~\ref{sec:surface_defects} lays out the theory
behind a more general form of scattering HOMs undergo when they interact with surface
defects present on mirrors or beamsplitters.

\subsection{\Finesse examples}

\subsubsection{Beam parameter tracing}

\epubtkImage{fexample_beam_param.png}{%
  \begin{figure}[htbp]
    \centerline{\includegraphics[width=0.9\textwidth]{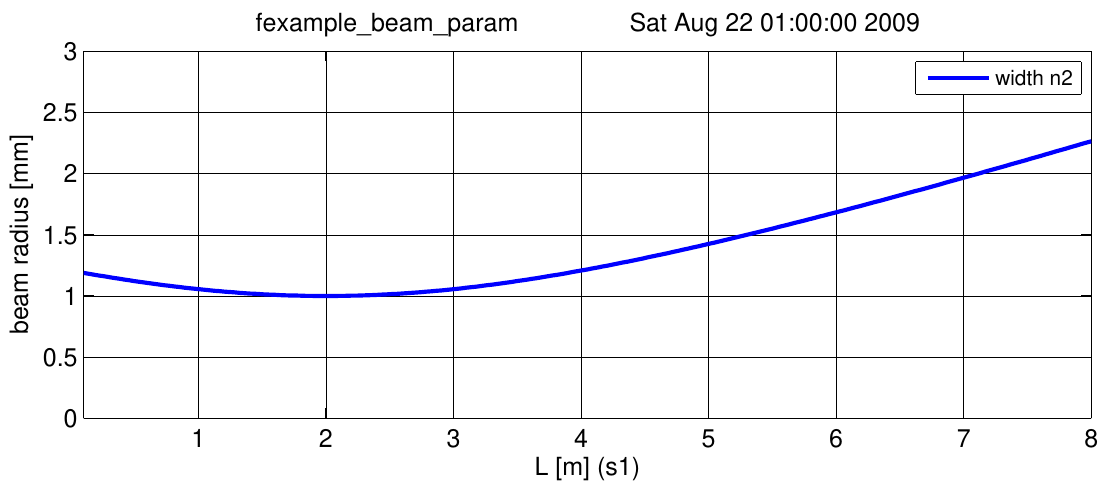}}
    \caption{\Finesse example: Beam parameter tracing}
    \label{fig:fexample_beam_parameter}
\end{figure}}

\noindent
This example illustrates a possible use of the beam parameter detector
`bp': the beam radius of the laser beam is plotted as a function of
distance to the laser. For this simulation, the interferometer matrix
does not need to be solved. `bp' merely returns the results from the
beam tracing algorithm of \Finesse.

\vspace{3mm}\noindent
{\small
\textbf{Finesse input file for `Beam parameter tracing'}
{\renewcommand{\baselinestretch}{.8}

\nopagebreak
\tt
\noindent
\mbox{} \\
\mbox{}\textbf{\textcolor{RoyalBlue}{laser}}\ i1\ \textcolor{Purple}{1}\ \textcolor{Purple}{0}\ n1\ \ \ \ \ \ \ \ \ \ \ \ \textcolor{Gray}{\%\ laser\ with\ P=1W} \\
\mbox{}\textbf{\textcolor{RoyalBlue}{gauss}}\ g1\ i1\ n1\ 1m\ \textcolor{BrickRed}{-}\textcolor{Purple}{2}\ \ \ \ \ \ \ \textcolor{Gray}{\%\ a\ dummy\ beam\ parameter} \\
\mbox{}\textbf{\textcolor{Red}{maxtem}}\ \textcolor{Purple}{0}\ \ \ \ \ \ \ \ \ \ \ \ \ \ \ \ \ \ \ \textcolor{Gray}{\%\ we\ need\ only\ the\ u$\_$00\ mode\ } \\
\mbox{}\textbf{\textcolor{RoyalBlue}{s}}\ s1\ \textcolor{Purple}{1}\ n1\ n2\ \ \ \ \ \ \ \ \ \ \ \ \ \ \ \textcolor{Gray}{\%\ a\ space\ of\ 1m\ length} \\
\mbox{}\textbf{\textcolor{RoyalBlue}{bp}}\ width\ x\ w\ n2\ \ \ \ \ \ \ \ \ \ \ \ \textcolor{Gray}{\%\ detecting\ the\ beam\ width\ (horizontal)} \\
\mbox{}\textbf{\textcolor{Red}{xaxis}}\ s1\ L\ lin\ \textcolor{Purple}{0.1}\ \textcolor{Purple}{8}\ \textcolor{Purple}{200}\ \ \ \textcolor{Gray}{\%\ tuning\ the\ length\ of\ s1}

}}

\subsubsection{Telescope and Gouy phase}

\epubtkImage{fexample_gouy_phase.png}{%
  \begin{figure}[htbp]
    \centerline{\includegraphics[width=0.9\textwidth]{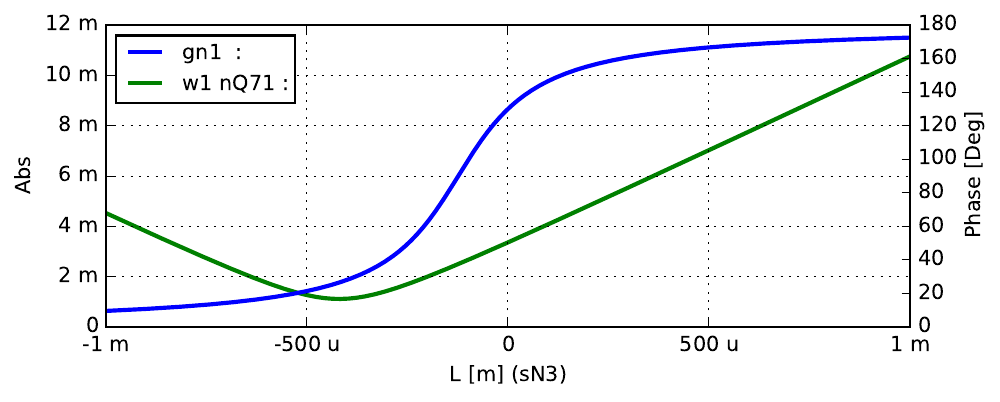}}
    \caption{\Finesse example: Telescope and Gouy phase. The blue
      trace shows the Gouy phase accumulated in the telescope, the
      green trace the beam spot size at the end of the telescope. The
      change on the xaxis represents a position tuning of the lens `L2'.}
    \label{fig:fexample_gouy_phase}
\end{figure}}

\noindent
This example shows the fine tuning of a telescope. The optical setup
is similar to the optical layout on the Virgo North-end detection
bench, resembling the telescope for the beam transmitted by the
end mirror of one arm. The purpose of the telescope is to reduce the
beam size and provide a user defined Gouy phase for a split photo
detector which is used for the alignment sensing system.

\vspace{3mm}\noindent
{\small
\textbf{Finesse input file for `Telescope and Gouy phase'}
{\renewcommand{\baselinestretch}{.8}

\nopagebreak
\tt
\noindent
\mbox{}\textbf{\textcolor{RoyalBlue}{l}}\ i1\ \textcolor{Purple}{6}\ \textcolor{Purple}{0}\ \textcolor{Purple}{0}\ nin \\
\mbox{}\textbf{\textcolor{RoyalBlue}{gauss}}\ g1\ i1\ nin\ \textcolor{Purple}{0.014852735}\ \textcolor{BrickRed}{-}\textcolor{Purple}{2}\textcolor{BrickRed}{.}2462888k \\
\mbox{}\textcolor{Gray}{\#\ multi-lens\ telescope,\ similar\ to\ the\ Virgo\ north\ end\ detction} \\
\mbox{}\textbf{\textcolor{RoyalBlue}{s}}\ sN2\ \textcolor{Purple}{1.77}\ nin\ nL1 \\
\mbox{}\textbf{\textcolor{RoyalBlue}{lens}}\ L1\ \textcolor{Purple}{1.02}\ nL1\ nL2\  \\
\mbox{}\textbf{\textcolor{RoyalBlue}{s}}\ sN3\ \textcolor{BrickRed}{.}\textcolor{Purple}{8996}\ nL2\ nL3 \\
\mbox{}\textbf{\textcolor{RoyalBlue}{lens}}\ L2\ \textcolor{BrickRed}{-.}\textcolor{Purple}{2}\ nL3\ nL4 \\
\mbox{}\textbf{\textcolor{RoyalBlue}{s}}\ sN4\ \textcolor{BrickRed}{.}\textcolor{Purple}{2146}\ nL4\ nL5 \\
\mbox{}\textbf{\textcolor{RoyalBlue}{lens}}\ L3\ \textcolor{BrickRed}{-.}\textcolor{Purple}{1}\ nL5\ nL6 \\
\mbox{}\textbf{\textcolor{RoyalBlue}{s}}\ sN5\ \textcolor{BrickRed}{.}\textcolor{Purple}{608}\ nL6\ nL12 \\
\mbox{}\textbf{\textcolor{RoyalBlue}{lens}}\ L4a\ \textcolor{BrickRed}{-.}\textcolor{Purple}{1}\ nL12\ nL13 \\
\mbox{}\textbf{\textcolor{RoyalBlue}{s}}\ sN8\ \textcolor{BrickRed}{.}\textcolor{Purple}{759}\ nL13\ nQ71 \\
\mbox{}\textcolor{Gray}{\#\ Plot\ Gouy\ phase\ from\ through\ the\ entire\ telescope} \\
\mbox{}\textbf{\textcolor{RoyalBlue}{gouy}}\ gn1\ x\ sN2\ sN3\ sN4\ sN5\  \\
\mbox{}\textcolor{Gray}{\#\ Plot\ beam\ size\ at\ end\ of\ telescope} \\
\mbox{}\textbf{\textcolor{RoyalBlue}{bp}}\ w1\ x\ w\ nQ71 \\
\mbox{}\textcolor{Gray}{\#\ Tuning\ the\ position\ of\ lens\ L2\ by\ chaning\ the\ lengths\ of\ } \\
\mbox{}\textcolor{Gray}{\#\ the\ spaces\ in\ front\ and\ behind\ the\ lens.} \\
\mbox{}\textbf{\textcolor{Red}{xaxis}}\textcolor{BrickRed}{*}\ sN3\ L\ lin\ \textcolor{BrickRed}{-}1m\ 1m\ \textcolor{Purple}{400} \\
\mbox{}\textbf{\textcolor{Red}{func}}\textcolor{ForestGreen}{\ sN4L}\ \textcolor{BrickRed}{=}\ \textcolor{Purple}{1.1142}\ \textcolor{BrickRed}{-}\ \textcolor{ForestGreen}{\$x1}\  \\
\mbox{}\textbf{\textcolor{Red}{noplot}}\ sN4L \\
\mbox{}\textbf{\textcolor{Red}{put}}\ sN4\ L\ \textcolor{ForestGreen}{\$sN4L} \\
\mbox{}

}}

\subsubsection{LG33 mode}

\epubtkImage{LG33amp-LG33phase.png}{%
  \begin{figure}[htbp]
    \centerline{
      \includegraphics[width=0.45\textwidth]{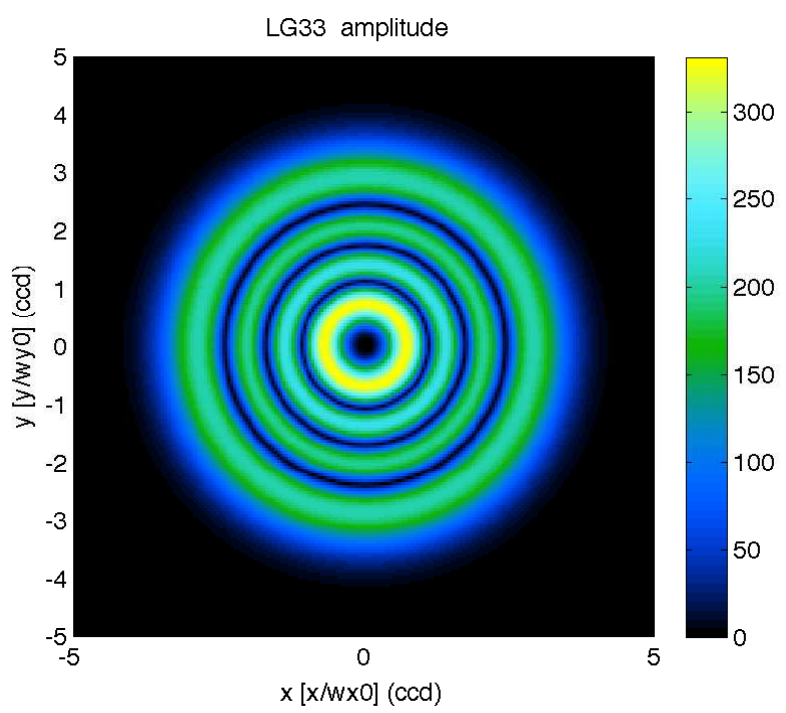}
      \includegraphics[width=0.45\textwidth]{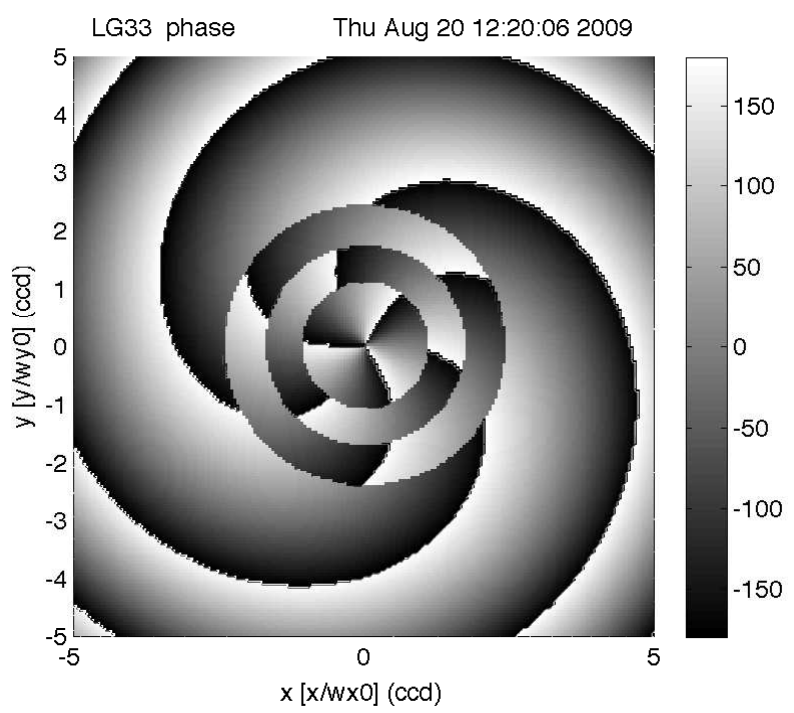}
    }
    \caption{\Finesse example: LG33 mode. The ring structure in the
      phase plot is due to phase jumps, which could be removed by
      applying a phase `unwrap'.}
    \label{fig:fexample_LG33}
\end{figure}}

\noindent
\Finesse uses the Hermite--Gauss modes as a base system for describing
the spatial properties of laser beams. However, Laguerre--Gauss modes
can be created using the coefficients given in
Equation~(\ref{eq:LGtoHG}). This example demonstrates this and the use
of a `beam' detector to plot amplitude and phase of a beam cross
section.

\vspace{3mm}\noindent
{\small
\textbf{Finesse input file for `LG33 mode'}
{\renewcommand{\baselinestretch}{.8}

\nopagebreak
\tt
\noindent
\mbox{} \\
\mbox{}\textbf{\textcolor{RoyalBlue}{laser}}\ i1\ \textcolor{Purple}{1}\ \textcolor{Purple}{0}\ n1\ \ \ \ \ \ \ \ \ \ \ \textcolor{Gray}{\%\ laser\ with\ P=1W} \\
\mbox{}\textbf{\textcolor{RoyalBlue}{gauss}}\ g1\ i1\ n1\ 1m\ \textcolor{Purple}{0}\ \ \ \ \ \ \ \textcolor{Gray}{\%\ a\ dummy\ beam\ parameter} \\
\mbox{}\textbf{\textcolor{Red}{maxtem}}\ \textcolor{Purple}{9}\ \ \ \ \ \ \ \ \ \ \ \ \ \ \ \ \ \ \textcolor{Gray}{\%\ we\ need\ modes\ up\ to\ n+m=9} \\
\mbox{}\textcolor{Blue}{tem}\ i1\ \textcolor{Purple}{0}\ \textcolor{Purple}{0}\ \textcolor{Purple}{0}\ \textcolor{Purple}{0}\ \ \ \ \ \ \ \ \ \ \ \ \textcolor{Gray}{\%\ HG\ coefficients\ to\ create\ LG33\ mode} \\
\mbox{}\textcolor{Blue}{tem}\ i1\ \textcolor{Purple}{9}\ \textcolor{Purple}{0}\ \textcolor{Purple}{0.164063}\ \textcolor{Purple}{0} \\
\mbox{}\textcolor{Blue}{tem}\ i1\ \textcolor{Purple}{8}\ \textcolor{Purple}{1}\ \textcolor{Purple}{0.164063}\ \textcolor{BrickRed}{-}\textcolor{Purple}{90} \\
\mbox{}\textcolor{Blue}{tem}\ i1\ \textcolor{Purple}{7}\ \textcolor{Purple}{2}\ \textcolor{Purple}{0}\ \textcolor{Purple}{0} \\
\mbox{}\textcolor{Blue}{tem}\ i1\ \textcolor{Purple}{6}\ \textcolor{Purple}{3}\ \textcolor{Purple}{0.125}\ \textcolor{BrickRed}{-}\textcolor{Purple}{90} \\
\mbox{}\textcolor{Blue}{tem}\ i1\ \textcolor{Purple}{5}\ \textcolor{Purple}{4}\ \textcolor{Purple}{0.046875}\ \textcolor{Purple}{180} \\
\mbox{}\textcolor{Blue}{tem}\ i1\ \textcolor{Purple}{4}\ \textcolor{Purple}{5}\ \textcolor{Purple}{0.046875}\ \textcolor{BrickRed}{-}\textcolor{Purple}{90} \\
\mbox{}\textcolor{Blue}{tem}\ i1\ \textcolor{Purple}{3}\ \textcolor{Purple}{6}\ \textcolor{Purple}{0.125}\ \textcolor{Purple}{180} \\
\mbox{}\textcolor{Blue}{tem}\ i1\ \textcolor{Purple}{2}\ \textcolor{Purple}{7}\ \textcolor{Purple}{0}\ \textcolor{Purple}{0} \\
\mbox{}\textcolor{Blue}{tem}\ i1\ \textcolor{Purple}{1}\ \textcolor{Purple}{8}\ \textcolor{Purple}{0.164063}\ \textcolor{Purple}{180} \\
\mbox{}\textcolor{Blue}{tem}\ i1\ \textcolor{Purple}{0}\ \textcolor{Purple}{9}\ \textcolor{Purple}{0.164063}\ \textcolor{Purple}{90} \\
\mbox{}\textbf{\textcolor{RoyalBlue}{s}}\ s1\ \textcolor{Purple}{1}\ n1\ n2\ \ \ \ \ \ \ \ \ \ \ \ \ \ \textcolor{Gray}{\%\ space\ of\ 1m\ lentgh} \\
\mbox{} \\
\mbox{}\textbf{\textcolor{RoyalBlue}{beam}}\ ccd\ \textcolor{Purple}{0}\ n2\ \ \textcolor{Gray}{\%\ beam\ detector\ for\ carrier\ light} \\
\mbox{}\textbf{\textcolor{Red}{xaxis}}\ ccd\ x\ lin\ \textcolor{BrickRed}{-}\textcolor{Purple}{5}\ \textcolor{Purple}{5}\ \textcolor{Purple}{200}\ \ \textcolor{Gray}{\%\ tune\ x\ position\ of\ beam\ detector} \\
\mbox{}\textbf{\textcolor{Red}{x2axis}}\ ccd\ y\ lin\ \textcolor{BrickRed}{-}\textcolor{Purple}{5}\ \textcolor{Purple}{5}\ \textcolor{Purple}{200}\ \textcolor{Gray}{\%\ tune\ y\ position\ of\ beam\ detector} \\
\mbox{}\textbf{\textcolor{Red}{yaxis}}\ abs\textcolor{BrickRed}{:}deg\ \ \textcolor{Gray}{\%\ plot\ amplitude\ and\ phase} \\
\mbox{}\textbf{\textcolor{Orange}{multi}} \\
\mbox{} \\
\mbox{}

}}

\newpage
\section{Imperfect interferometers}
\label{sec:imperfect}
Imperfections in a Michelson interferometer can refer to
any of the differences between
a real interferometer and the perfect design.
These include, but are not limited to:
deviations of the optical properties of the mirrors from the design;
the limits of longitudinal and alignment control of the mirrors;
additional noise sources not included in our models (i.e.~electronic noise);
and effects which distort the shape of the beam.
To estimate the impact of such imperfections on the Michelson's
performance is complicated
and requires substantial modelling.
The greatest impact on the sensitivity arises from asymmetries between
the two arms.  For accurate differential measurements, such as those
made in gravitational wave interferometers,
the mirrors are very carefully manufactured to make the arms as similar as possible.
Differences between the two arms, for example, imbalances in the
finesse of the two arm cavities, will couple extra light into the
anti-symmetric port of the interferometer
where it adds additional noise to the detection photodiode.

It is important to understand how imperfections in an interferometer affect the
resonating beams and impact the sensitivity of the instrument.  For this we
need accurate models which can simulate complex interferometers in
the presence of such imperfections.  This
is crucial for the design of interferometers, such as
gravitational wave detectors, and the commissioning process, in which
deviations of the interferometer behaviour from the expected design
must be diagnosed.
In this review we will consider imperfections in the form
of distortions of the beam and we discuss these
effects for gravitational wave interferometers;
firstly in terms of the behaviour of distorted optics and
how this effects the performance of different optical configurations;
and secondly in terms of solutions to these distortion problems and
implications for the design process.

\subsection{Spatial modes in optical cavities}
\label{sec:HOM_resonances}
In the previous chapter the idea of representing distortions of a beam as
higher-order Gaussian modes was introduced.
Here we use this description to investigate the behaviour of
interferometers with distorted beams.

A well designed optical cavity can act as a resonator for a particular order
of Gaussian modes, depending on its longitudinal tuning.  In
modern interferometers such cavities are operated as resonators
for the fundamental mode, filtering out unwanted spatial components
of the beam.  This is achieved as each Gauss mode is subject to
an additional phase term as it propagates.  This additional phase depends on the
mode order:
\begin{equation}
\varphi(z) = (n+m+1) \psi(z) = (n+m+1) \tan^{-1}\left(\frac{z}{z_R}\right)
\end{equation}
where $n+m$ is the mode order, $\psi$ is the Gouy phase and $z_R$
is the Rayleigh range (see
Section~\ref{sec:gaussian_beam_properties}).
This additional
phase ensures different modes are resonant in a cavity at different
longitudinal tunings.  The Gouy phase accumulated in one round trip of a cavity is
\begin{equation}
\Psi = 2(\psi_2-\psi_1) = 2\left(\tan^{-1}\left(\frac{z_2}{z_R}\right) - \tan^{-1}\left(\frac{z_1}{z_R}\right)\right)
\end{equation}
where $\psi_{1/2}$ is the Gouy phase at the input/end mirror, $z_{1/2}$ is the
distance from the waist of the input/end mirror and $z_{R}$ is the Rayleigh range,
all in terms of the cavity eigenmode.  The different resonant tunings for various
HOM in a cavity is illustrated in figure~\ref{fig:hom_scan}, where
an Advanced LIGO cavity is simulated with an input beam made up of equal parts
of 6 different order modes, orders $0$ to $5$.  Each of the
higher-order modes is resonant at a different microscopic tuning;
a cavity operated on resonance for the fundamental mode (order $0$) will suppress the power
in the higher-order modes circulating in the cavity and transmitted by the cavity,
as these are not resonant at the same tuning. In consequence the higher-order modes are
reflected by the cavity.
The parameters for the Gaussian eigenmode
of an Advanced LIGO arm cavity are summarised in table~\ref{table:aligo_params}.
This includes half the round-trip Gouy phase, in this case 24.3$^{\circ}$,
which gives the spacing between subsequent resonance peaks in terms
of cavity tuning (as seen in figure~\ref{fig:hom_scan}).

\epubtkImage{HOM_scan.png}{%
  \begin{figure}[htb]
    \centerline{\includegraphics[scale=0.7]{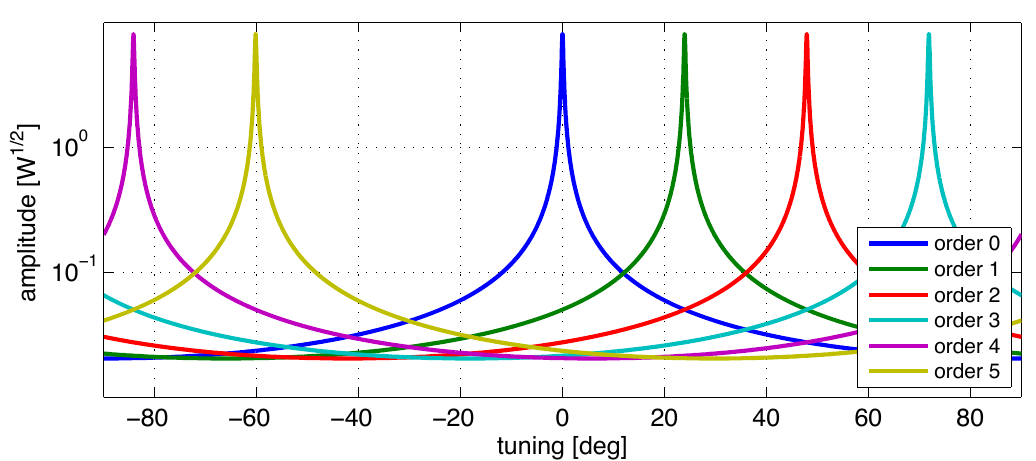}}
    \caption[]{Amplitude of 6 higher-order modes (orders 0 to 5) circulating in
    		an optical cavity as the microscopic length is tuned.  The fundamental
		mode (order 0) is resonant at 0$^{\circ}$ tuning and the mode separation
		tuning, 24$^{\circ}$ is defined by the length of the cavity an the mirror
		curvatures.
    }
    \label{fig:hom_scan}
\end{figure}}

\begin{table}[!t]
\begin{center}
\begin{tabular}{c|c|c|c|c|c|c|c|c}
		 $R_{C,1}$ [m]	& $R_{C,2}$ [m]	& $w_0$ [cm]	& $w_1$ [cm]	& $w_2$ [cm] 	& $z_1$ [m]	& $z_2$	& $z_R$ [m]	& $\frac{\Psi}{2}$ [$^{\circ}$]	 \\
\hline
		 1934			& 2245			& 1.2			& 5.3			& 6.2			& -1834	& 2160	 & 425		& 24.3			\\
\end{tabular}
\end{center}
\caption{Summary of the parameters defining the Gaussian eigenmode of an Advanced LIGO
		arm cavity.  This includes the radius of curvature, $R_C$, at the input (1) and end (2) mirrors;
		the beam spot size, $w$, at the input mirror, end mirror and at the beam waist (0);
		the distance from the waist, $z$, of the input and end mirrors; the Rayleigh range, $z_R$;
		and half the round-trip Gouy phase, $\frac{\Psi}{2}$.}
\label{table:aligo_params}
\end{table}%

This property of an optical cavity to act as a filter of spatial modes is
utilised in gravitational wave detectors.  Firstly, the input laser beam
is `cleaned' of spatial modes by passing through an \emph{input
mode cleaner}, an optical cavity carefully designed to transmit the
fundamental mode and filter out most higher-order modes before
the beam enters the main interferometer.  Within the multiple cavities
of the central interferometer careful design can take advantage of
these resonant properties to suppress distortions of the beam.
Finally, the output beam containing the gravitational wave signal
is cleaned of spatial modes and control sidebands using an
\emph{output mode cleaner}.  These design features are
discussed in grater detail in section~\ref{sec:aLIGO_design}.

\subsection{Cavity alignment in the mode picture}

In the previous example the injected beam contains
several different order modes.  This is an exaggeration of the effect
of a mode cleaner, where
a distorted beam is cleaned of unwanted spatial modes.
After the mode cleaner the input beam is well described by the
fundamental mode and higher-order modes present in interferometers
can be the result of defects in the optics and mismatches between
the incoming beam and eigenmode of the interferometer.
The simplest example of this is a misaligned 2-mirror cavity,
where the optical axis of the incoming
beam is not aligned to the optical axis of the cavity.
\epubtkImage{misaligned_cavity_axis.png}{%
  \begin{figure}[htb]
    \centerline{\includegraphics[scale=0.85]{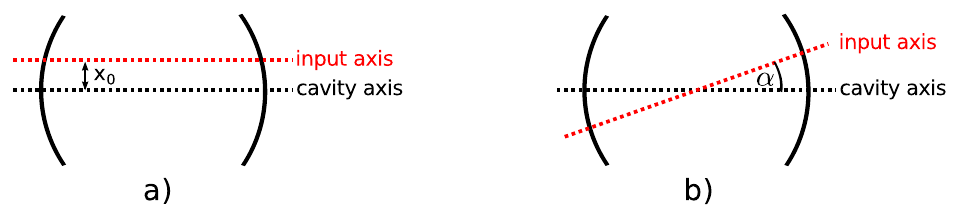}}
    \caption[]{Different possible misalignments of an optical cavity.
    	{\bf a)} Constant $x$ displacement of the input beam optical axis
	with respect to the cavity axis.
	{\bf b)} Relative tilt between the input optical axis and the
	cavity axis.
    }
    \label{fig:misalign_axis}
\end{figure}}

As discussed by Anderson~\cite{anderson84}
and illustrated in Figure~\ref{fig:misalign_axis} we can consider different possible
misalignments.  Any misalignment can be split into a displacement
in $x$ (or $y$) of the input beam axis with respect to the cavity axis
(a) and a relative tilt between the input beam and cavity axes (b).
For mathematical simplicity we consider a fundamental Gaussian
beam at the waist.  As the Hermite-Gauss modes are separable in
$x$ and $y$ we just consider the $x$ component.  The results are
equivalent for a displacement in $y$.  The fundamental mode ($n=0$)
and first order mode ($n=1$) of the cavity can be written
\begin{equation}
\begin{array}{cccc}
u_0(x) = \left(\frac{2}{\pi w_0^2}\right)^{1/4} \exp{\left(-\frac{x^2}{w_0^2}\right)}
& & &
u_1(x) = \left(\frac{2}{\pi w_0^2}\right)^{1/4} \frac{2 x}{w_0} \exp{\left(-\frac{x^2}{w_0^2}\right)}
\end{array}
\end{equation}
where $w_0$ is the beam waist size.
Assuming the input beam matches the cavity eigenmode, with the exception
of the misalignment, a displacement of the input beam (a) is
translated onto the cavity axis as
\begin{equation}
u_{\mathrm{disp.}}(x) = u_0(x-x_0) = \left(\frac{2}{\pi w_0^2}\right)^{1/4} \exp{\left(-\frac{(x-x_0)^2}{w_0^2}\right)}.
\end{equation}
As long as the displacement, $x_0$, is small compared to the beam size
($\frac{x_0}{w_0} \ll 1$) any second order terms and higher
in $\frac{x_0}{w_0}$ can be ignored and the input field is approximated
as
\begin{equation}
u_{\mathrm{disp.}}(x) \approx \left(\frac{2}{\pi w_0^2}\right)^{1/4} \left(1 + \frac{2xx_0}{w_0^2}\right)
		\exp{\left(-\frac{x^2}{w_0^2}\right)}
		= u_0(x) + \frac{x_0}{w_0} u_1(x).
\end{equation}
Thus this displacement of the input beam, with respect to the cavity, is
equivalent to the addition of a first order Hermite-Gauss mode.

Similarly a misalignment in terms of a tilt of the input axis with respect to
the cavity axis (b) can be described by the addition of a first order
mode~\cite{anderson84}.
In this case the amplitude of the input field as projected onto the
cavity axis is un-altered, for small tilts, and the
relative misalignment only effects the phase of the beam
\begin{equation}
u_{\mathrm{tilt.}}(x) = u_0(x)\exp{\left(\I k \sin{(\alpha) x}\right)} \approx u_0(x)\exp{\left(\I k \alpha x \right)}.
\end{equation}
For a small misalignment the higher-order terms
of the exponential are ignored:
\begin{equation}
u_{\mathrm{tilt.}}(x) \approx u_0(x) \left(1 +\I k \alpha x\right) = u_0(x) + \I \frac{k \alpha w_0}{2} u_1(x).
\end{equation}
The relative tilt of the input beam is expressed with
the addition of an order 1 mode, 90$^{\circ}$ out of phase
with the fundamental mode.  This 90$^{\circ}$ phase factor
is a useful feature which can be used to separate the
order 1 modes caused by a displacement of the optical
axis and those caused by a tilt of the optical axis.
A combination of these two types of misalignment can describe
any misaligned cavity.

The consequence of a misalignment of an optical cavity is the
creation of first order modes.  If we chose to describe the problem
using the Gaussian beam parameters and axis of the incoming beam as
our basis, then the incoming beam is a pure fundamental beam and a
first order modes is created when the light enters (and leaves)
the cavity. Alternatively we can using the cavity eigenmode and cavity
axis as our basis. In this case the higher-order modes are already
present in the incoming beam.
Either of these approaches is valid for such a simple distortion.

In more realistic cases the circulating field in a cavity  is not completely described by a
fundamental Gaussian beam, due to deviations of real mirrors from an
ideal sphere.  This can be modelled using the closest Gaussian eigenmode
(from now on refereed to as the eigenmode of the cavity)
superimposed with higher-order modes.  On can say that the
higher-order modes are created when the fundamental mode
interacts with the distorted mirrors.

To describe the input-output relations of a cavity for
higher-order modes it is important to know at which location
they have been created, in other words where they enter the cavity.
For higher-order modes present in the input
beam (not created inside the cavity) the amplitude of these modes
in the circulating field is given by:
\begin{equation}
a_{n,m}^{\mathrm{circ.}} = \frac{\I r_2 t_1 \exp{\left(-\I 2kL + \I (1+n+m) \Psi \right)}}{ 1-r_1r_2\exp{\left(-\I 2kL +\I(1+n+m)\Psi \right)}} a_{n,m}^{\mathrm{in}},
\end{equation}
where $a_{n,m}^{\mathrm{in}}$ is the amplitude of the HOM in the
incoming field.  The equation is very similar to that for a plane wave
(equation~\ref{eq:a3}) with the addition of the Gouy phase picked
up for different modes.  On the usual operating point of resonance
for the fundamental mode this simplifies to:
\begin{equation}
a_{n,m}^{\mathrm{circ.}} = \frac{\I r_2 t_1 \exp{\left(\I (n+m) \Psi \right)}}{ 1-r_1r_2\exp{\left(\I(n+m)\Psi \right)}} a_{n,m}^{\mathrm{in}}.
\end{equation}
A well designed cavity will have $(n+m)\Psi \neq N 2\pi$ up to
a high mode order to prevent other modes resonating.
In the case, where we consider higher-order modes
created at individual mirrors, the equations are different.
If the modes are created at the end mirror,
with no higher-order modes  present in the input beam,
the circulating field is approximated as:
\begin{equation}
a_{n,m}^{\mathrm{circ.}} \approx \frac{\I  r_2 t_1 \exp{\left(\I (n+m)\frac{\Psi}{2}\right)}}
		{(1-r_1r_2)(1-r_1r_2\exp{\left(\I(n+m)\Psi\right))}} k_{0,0,n,m} \ a_{0,0}^{\mathrm{in}},
\end{equation}
where $k_{0,0,n,m}$ is a coupling coefficient describing the phase
and amplitude of the mode HG$_{nm}$ created at the end mirror, due
to an incident HG$_{00}$ mode.  The approximation here assumes
the coupling is small and so does not include the loss of power from
the HG$_{00}$ mode and coupling from the HG$_{nm}$ mode back into HG$_{00}$.
This approximation is valid for small distortions, but including distortions
on all the optics will quickly become very complicated analytically,
and it is for such problems that simulation tools such as \Finesse
are valuable.

\epubtkImage{misaligned_cav_scan.png}{%
  \begin{figure}[htb]
    \centerline{\includegraphics[scale=0.7]{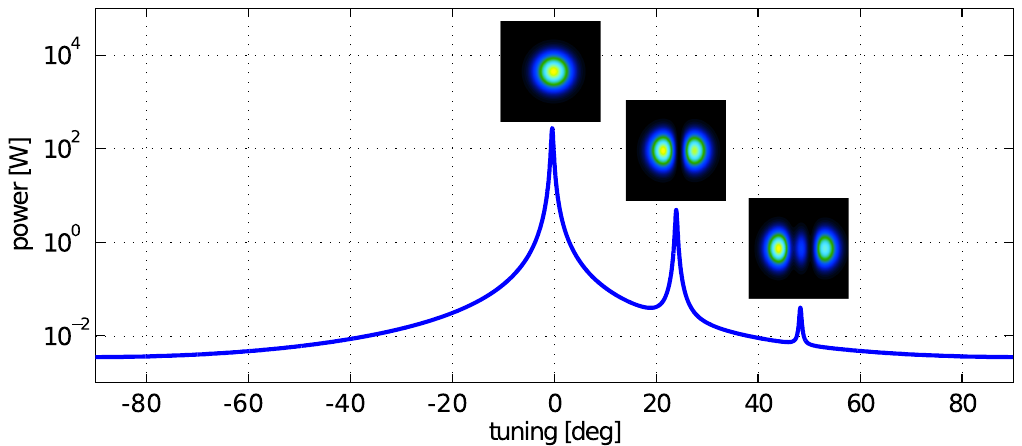}}
    \caption[]{Circulating power in an Advanced LIGO cavity with a
    	0.3\,$\mu$rad misalignment applied to the end mirror.  The cavity
	is tuned and the circulating beam is detected for each peak in
	intra-cavity power.  Most of the
	power remains in the fundamental mode, resonant at $\sim 0^{\circ}$,
	with some coupling into HG$_{10}$, resonant at $\sim 24^{\circ}$,
	and HG$_{20}$, resonant at $\sim 50^{\circ}$.
    }
    \label{fig:misalign_scan}
\end{figure}}
In Figure~\ref{fig:misalign_scan} the effects of misalignment on
intra-cavity power is illustrated.  In this example an Advanced LIGO
cavity has been modelled with a misaligned end mirror.  The circulating power
exhibits several peaks, corresponding to the resonances of the
different higher-order modes created due to the misalignment.
Most of the power remains in the fundamental mode (peak at $\sim 0^{\circ}$).
The misalignment of the cavity has induced higher-order modes, mostly the
first order HG$_{10}$ mode, whose resonance is observed at $\sim 24^{\circ}$.
In this case the extent of the misalignment also results in the creation of
the order 2 mode HG$_{20}$, resonant at $\sim 50^{\circ}$.
During operation, where the cavity is on resonance for the HG$_{00}$ mode,
the relative power in higher-order modes is suppressed.  There will still be
some higher-order modes in the beam at this tuning, which degrade the purity of
the beam transmitted and reflected from the cavity.



\subsection{Mode mismatch}
\label{sec:mismatch}
Another common defect of optical cavities with respect to an input
laser beam is known as \emph{mode mismatch}~\cite{mueller2000}.
Whilst (a small) misalignment is a first order effect described by first order
modes, a (small) mode mismatch is a second order effect,
where the wavefront curvature or beam size of the incoming
beam does not match that of the cavity eigenmode.
Such effects are described primarily by second order modes.
As with misalignment this defect can be split into two different
effects: beam size mismatch and waist position mismatch.

\epubtkImage{mismatch_cavity.png}{%
  \begin{figure}[htbp]
    \centerline{\includegraphics[scale=1.1]{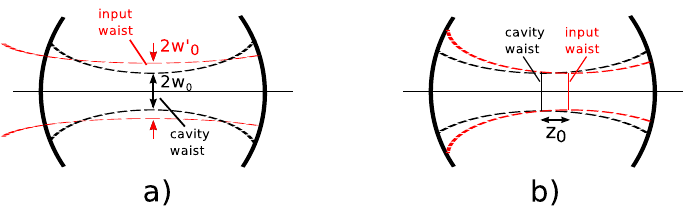}}
    \caption[]{Different possible mode mismatch between an the eigenmode of
    		an optical cavity and an injected laser beam. {\bf a)} Beam size mismatch.
		{\bf b)} Mismatch of the position of the beam waists.
    }
    \label{fig:mismatch_cav}
\end{figure}}

In the case of a pure beam size mismatch, with the
cavity and input beam waists located at the same point along
the optical axis the input beam can be described
in the cavity basis as~\cite{anderson84}:
\begin{equation}
u_{\mathrm{size.}}(r) = u_0(r) + \epsilon \ u_2(r),
\end{equation}
where $u_0$ is the fundamental cavity mode, $\epsilon$
is the fractional difference in the input beam size to the
beam size of the cavity eigenmode, $w_0' = (1+\epsilon) w_0$,
and $u_2$
is the second order Laguerre-Gauss mode of the cavity,
with no angular dependence (LG$_{10}$)
\begin{equation}
u_2(r) = \sqrt{\frac{2}{\pi}} \frac{1}{w_0} \left(1-\frac{2r^2}{w^2_0}\right) . \exp{\left(-\frac{r^2}{w_0^2}\right)}
\end{equation}
In this case the calculation is performed at the waist for simplicity.
Similarly, for a purely waist position mismatch
we have
\begin{equation}
u_{\mathrm{posit.}}(r,z) = u_0(r,z) + \I \frac{kz_0}{w_0^2} u_2(r,z),
\end{equation}
where $z_0$ is the displacement of the input beam
waist with respect to the cavity waist, and the fundamental
and second order modes are now in their more general form,
taking into account a finite radius of curvature.
As with misalignment we find that the two types of
mode mismatch result in the creation of the same mode,
with one in phase (beam size) and one with a 90$^{\circ}$
phase shift (waist position) with respect to the fundamental mode.

In figure~\ref{fig:mismatch_scan} the circulating power in a
mismatched cavity is shown.  In this case the mismatch
is a 25\% mismatch purely in beam size.
Most of the power remains in the fundamental mode (resonant at 0$^{\circ}$),
with most of the mismatch described by the order 2 mode LG$_{10}$
(resonant at $\sim 50^{\circ}$).  Such a large mismatch
results in additional modes with even mode orders: the order 4 mode LG$_{20}$
and order 6 mode LG$_{30}$.

\epubtkImage{mismatched_cav_scan.png}{%
  \begin{figure}[htb]
    \centerline{\includegraphics[scale=0.7]{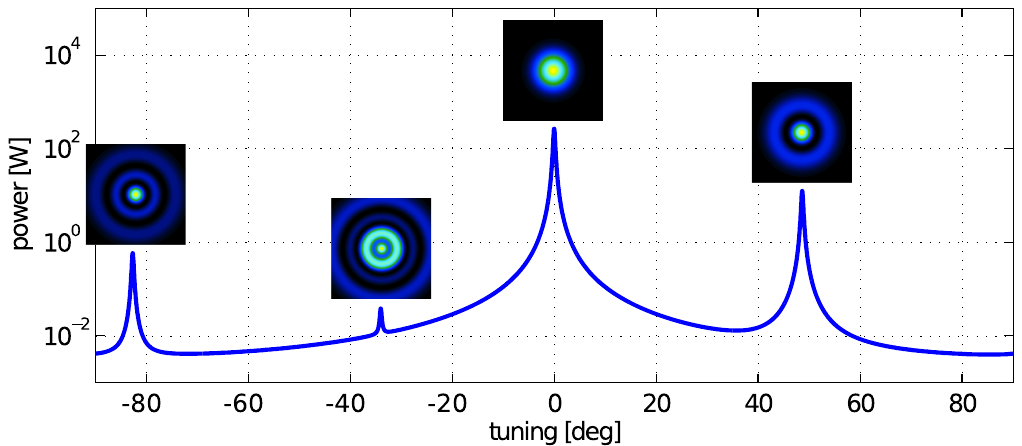}}
    \caption[]{Circulating power in a Advanced LIGO cavity with
    	a 25\% mismatch in beam size between the injected beam
	and the cavity eigenmode.  The intra-cavity power is detected as
	the longitudinal length is scanned, with the beam detected at each
	local resonance.  Most of the power is in the fundamental
	mode, resonant at $\sim 0^{\circ}$, with the mismatch represented
	by power in the order 2 mode LG$_{10}$ ($\sim 50^{\circ}$), the
	order 4 mode LG$_{20}$ ($\sim -80^{\circ}$) and the order 6 mode
	LG$_{30}$ ($\sim -35$).
    }
    \label{fig:mismatch_scan}
\end{figure}}

\subsection{Spatial defects}

Misalignment and mode mismatch are the lowest order distortions of the beam
and are well described analytically.
These low order distortions are carefully controlled in an interferometer,
using alignment control schemes and using lenses and curved optics to mode
match beams between different cavities.  Higher-order distortions
produced from more complex processes, i.e.~interaction with
distorted mirror surfaces or finite sized optics, cannot currently be controlled.
There are many different spatial defects which are likely to be
present in real interferometers.

For the design and commission of real detectors we want to represent these
more arbitrary defects, in particular the deviation of the mirror surfaces
from a perfect sphere.  In the case of interferometer design this will help
set requirements on the polishing and coating of the mirrors.  For the commissioning
process this will aide in identifying the output beam shape and other
effects associated with distortions of the beam.
In this article we will focus on
mirror surface errors and thermal effects.
The detailed mathematics of these higher-order effects are discussed in
Section~\ref{sec:surface_defects}.
 For now we just consider that higher-order modes are created when
beams are distorted.  The advantage of describing distortions of the
beam as higher-order modes is that these spatial modes are easy to trace
through the interferometer, to predict the behaviour of a distorted interferometer.

\subsection{Operating cavities at high power}
\label{sec:thermal_cavs}

Advanced gravitational wave detectors will operate with very high light power
in the arm cavities, to increase the signal compared to the shot noise.  In Advanced LIGO,
for example, the power in the arms will approach 1\,MW~\cite{aligo_design}.
Even the state of the art optics used in advanced gravitational wave detectors
will absorb a proportion of the incident power.
The mirrors produced for Advanced LIGO have requirements of
$<0.5$\,parts-per-million (ppm) power absorption.  With the expected
power in the arms this means $\sim 0.5$\,W will be absorbed in
the mirror coatings and substrates.  During operation this absorption
of power will lead to a temperature gradient evolving in the optics,
starting from a \emph{cold state} defined by the temperature of
the environment and gradually developing a temperature field
across the optic, with a hot point at the centre where the beam is
most intense.  Finally the temperature field of the mirrors will reach
a steady state, where the optic is in thermal equilibrium.
The development of such temperature gradients in the
mirrors will result in two types of mirror aberration:
\begin{enumerate}
\item A thermal lens forms within the mirror due to the temperature dependent nature
of the index of refraction of the substrate material (fused silica).
This distortion can be described mostly as
a spherical lens, with some higher-order components.
\item The mirror expands thermally, with the expansion greatest where the mirror is hottest, giving
a non-uniform expansion over the mirror surface and effectively distorting the surface
from the cold case.  This thermal distortion is primarily a change in the radius of curvature
of the mirror.
\end{enumerate}
Both these effects will impact the shape of the beam in the arms, the
thermal lens in the input mirror distorting the beam injected into the
cavities and the surface distortions of the cavity mirrors
changing the shape of the beam resonating in the arms.
These effects will be primarily second order effects, impacting
the mode matching of the beam into the arm cavities and the
resonating eigenmode.
Crucially these effects are not constant: the temperature
fields and thermal aberrations will evolve from the cold state to
thermal equilibrium, where this equilibrium state, or \emph{hot
state}, is dependent on the interferometer input power.
For example, Advanced LIGO is expected to operate within a range of input
powers up to $\sim$100\,W
\footnote{Here we refer to the input power as the power injected into the central
Michelson interferometer, after the input mode cleaner and other input optics.}.  
The transitory nature of these thermal
aberrations will be one of the key challenges for advanced
interferometers.  Effectively the input mode and
cavity eigenmodes are constantly developing and require
additional systems to control the resonating mode of the interferometer.
For a more detailed description of the evolution of these thermal effects please
refer to the Living Reviews article~\cite{vinet09}.
Here we will attempt to quantify this problem and motivate the need
for thermal compensation systems to correct the lensing and
change in curvatures of the mirrors at high power.

Firstly we consider a single arm cavity of the second generation gravitational
wave detector Advanced LIGO~\cite{Arain2011}.
The two mirrors which make up this
cavity, the input test mass (ITM) and end test mass (ETM) are separated by
$\sim 4$\,km.  The radii of curvature and optical
parameters are given in Table~\ref{table:aLIGO}.
These numbers refer to the curvature of the mirrors in the
cold state, before heating of the mirror from the laser beam.
During operation the mirrors will heat up as they absorb power from
the laser beam, creating a thermal lens in the ITM and distorting the reflective
surfaces of both mirrors.
Advanced LIGO is expected to operate within a range of input powers.
Here we will refer to the cold case (0\,W input power), low power
(12.5\,W input power) and high power (125\,W input power).
This gives a maximum of 800\,kW in the arm cavities at high power.
The thermal lensing and distortions for each case are
summarised in Table~\ref{table:thermal_abs}.
\begin{table}[!t]
\begin{center}
\begin{tabular}{l|c|c|c}
		& 	$R_C$ [m]		& 	Transmittance	& Loss [ppm]	\\
\hline
ITM		&	1934			&	1.4\%		& 37.5			\\
\hline
ETM		&	2245			&	5\,ppm		& 37.5			\\
\end{tabular}
\end{center}
\caption{The design geometric and optical parameters of an Advanced LIGO arm
	cavity.  The radius of curvature ($R_C$), proportion of power transmitted and
	proportion of power lost for the input test mass (ITM) and end test mass (ETM)
	are given.  ppm refer to parts-per-million~\cite{Arain2011}.}
\label{table:aLIGO}
\end{table}%
\begin{table}[!b]
\begin{center}
\begin{tabular}{l|c|c|c}
				& $f_{\rm{ITM}}$ [km]	& $\delta R_{C,\rm{ITM}}$	[km] & $\delta R_{C,\rm{ETM}}$ [km]	 \\
\hline
Cold case (0\,W)	& $\infty$				&	$\infty$				& $\infty$ 			\\
\hline
Low power (12.5\,W)& $50$				& $1100$					& 1600			 \\
\hline
High power (125\,W)& $5$				& 110					& 160	\\
\end{tabular}
\end{center}
\caption{The expected thermal aberration of the test masses in an Advanced LIGO arm cavity
		for 3 states of operation corresponding to different input powers.  The aberrations
		are well described by second order effects: a spherical lens in the ITM characterised
		by focal length $f_{\rm{ITM}}$ and distortions of the reflective surfaces of the mirrors
		characterised by a change in curvature $\delta R_{C,\rm{ITM}}$/$\delta R_{C,\rm{ETM}}$.
		In these cases of relatively low power absorption the distortions scale linearly with power.
		}
\label{table:thermal_abs}
\end{table}%
The thermal lens is the dominant aberration and will have a large impact on the
beam injected into the cavity. However, it will not impact the eigenmode of
the cavity, this is determined purely by the curvature of the highly reflective mirror surfaces.

Consider an individual arm cavity with an incoming laser mode
matched to the cold optics (the design curvatures).  The size of the
beam corresponding to this cold eigenmode is plotted in blue in
Figure~\ref{fig:H&C_beams} along the length of the cavity.
Next we consider the hot eigenmode of the cavity, in this case corresponding
to the curvatures of the cavity during high power operation (125\,W).
This is plotted in red.  The mirrors are less curved and the eigenmode
differs slightly from the cold case, with a slightly smaller beam size
at the input and end mirror and a larger waist.  The mismatch between
the two eigenmodes is relatively small.   The incoming beam, however,
is strongly mismatched between the possible cavity eigenmodes
(shown plotted in orange).  During
high power operation the injected beam will experience
a strong 5\,km lens in the ITM, focussing the beam and
shifting the waist closer to the ITM,
resulting in a larger beam at the ETM.

In reality the hot eigenmode and input beam will develop
over time as the mirrors heat up and the aberrations evolve.
This takes us from the cold case, where the incoming beam is well
matched to the cavity, to the hot case where there is
a strong mode mismatch.
\epubtkImage{hot_and_cold_beams.png}{%
  \begin{figure}[htb]
    \centerline{\includegraphics[scale=0.8]{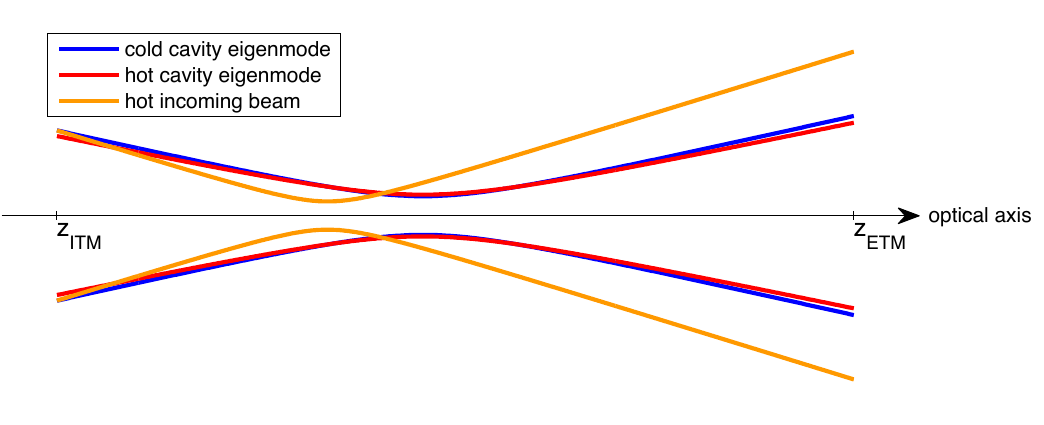}}
    \caption[]{Plots of the predicted beam sizes for different states (hot and cold)
    of an Advanced LIGO arm cavity.  In the cold case the cavity eigenmode (blue curve)
    is defined by the radius of curvature of the cold optics (0\,W input power).  We assume the incoming
    beam is mode matched to the cold cavity.  The hot case refers to high
    power operation (125\,W input power) where the mirrors absorb a proportion of the laser power.
    The hot cavity eigenmode (red curve) is determined by the curvature of the reflective surface of
    the hot optics: the mirror expands elastically, reducing the curvature.
    This changes the eigenmode of the cavity.
    In the hot case the input beam (orange curve) will no longer be matched to the cavity eigenmode,
    as it passes through the strong thermal lens ($f=5$\,km) in the ITM.
    During high power operation \emph{thermal compensation systems}
    will be used to correct the curvatures of the mirrors back to their cold
    state and compensate the lens in the ITM, with the aim of keeping the
    interferometer modes well matched and consistent for a range of
    input powers.
    }
    \label{fig:H&C_beams}
\end{figure}}
This will have a strong impact on the power injected into the cavity.
During operation the arm cavities are `locked' to the resonance of
the fundamental cavity mode.  In this state the components of
the injected beam which do not overlap with the cavity eigenmode will
be reflected.  As was discussed in section~\ref{sec:mismatch}
these will be primarily order 2 modes.

Table~\ref{table:qs} lists the beam parameters for the 3 different Gaussian
beams: the cold and hot cavity eigenmodes, calculated from the radii of
curvature of the hot and cold optics, and the hot input beam, calculated using
an ABCD matrix for a 5\,km lens (see Section~\ref{sec:abcd}).
\begin{table}[!t]
\begin{center}
\begin{tabular}{l|c|c|c|c}
		& $q$ [m]	& $w$ [cm]	& $R_C$ [m] 	& $z$ [m] \\
\hline
Cold eigenmode	& $-1834.2 + 427.8\I$ & 5.30		& $-1934$				& $-1834.2$		 \\
Hot eigenmode		& $-1832.7 + 499.0\I$ & 4.95		& $-1968.6$				& $-1832.7$		 \\
Hot input beam		& $-1356.2 + 228.1\I$ & 5.30		& $-1394.6$				& $-1356.2$		 \\

\end{tabular}
\end{center}
\caption{Beam parameter, $q$, beam size, $w$, wavefront curvature, $R_C$ and
		distance from the wasit, $z$ of 3 different Gaussian beams,
		the eigenmode of an Advanced LIGO cavity during cold operation
		(0\,W input power) and during hot operation (125\,W input power) and
		the input beam during hot operation.}
\label{table:qs}
\end{table}%
To estimate how much power will be injected into the hot cavity
we calculate the overlap between the hot cavity eigenmode and hot input beam,
i.e., we want to know how much power in the input beam is
in the 00 mode of the cavity eigenmode.  This takes the form
\begin{equation}
c = \int_S u_{\mathrm{in}} u^*_{\mathrm{cav}} \ \mbox{d}S,
\end{equation}
where $u_{\mathrm{in}}$ is the input field, $u_{\mathrm{cav}}$ is the cavity
eigenmode, $S$ is an infinite surface perpendicular to the optical axis
and the percentage of input power in the cavity eigenmode is given by
$|c|^2$.  In this case both beams are cylindrically symmetric fundamental
Gaussian beams (not astigmatic) and the overlap can be calculated as
\begin{equation}
c = \frac{2}{\pi}\frac{1}{w_{\mathrm{in}}w_{\mathrm{cav}}} \exp{\left(\I\Psi_{\mathrm{in}}-\I\Psi_{\mathrm{cav}}\right)}
\int_0^{2\pi} \int_0^{\infty} \exp{\left(-\frac{\I kr^2}{2}\left(\frac{1}{q_{\mathrm{in}}}-\frac{1}{q_{\mathrm{cav}}^*}\right)\right)} r \ \mbox{d}\phi \ \mbox{d}r .
\end{equation}
Integrating with respect to $\phi$ and changing variables to $R=r^2$
we have
\begin{equation}
c = \frac{2}{w_{\mathrm{in}}w_{\mathrm{cav}}} \exp{(\I \Psi_{\mathrm{in}} - \I \Psi_{\mathrm{cav}})}
\int_0^{\infty} \exp{\left(-\frac{\I kR}{2}\left(\frac{1}{q_{\mathrm{in}}} - \frac{1}{q_{\mathrm{cav}}^*}\right)\right)} \ \mbox{d}R .
\end{equation}
As $\myRe{\frac{\I k}{2}\left(\frac{1}{q_{\mathrm{in}}}-\frac{1}{q_{\mathrm{cav}}^*}\right)} > 0$
the solution of this equation can be written as~\cite{Gradstein8511}
\begin{equation}
c = -\frac{4 \I}{w_{\mathrm{in}}w_{\mathrm{cav}} k} \exp{(\I \Psi_{\mathrm{in}}-\I \Psi_{\mathrm{cav}})}
\frac{1}{\frac{1}{q_{\mathrm{in}}}-\frac{1}{q_{\mathrm{cav}}^*}} .
\end{equation}
Using this formula we calculate the overlap between the hot cavity eigenmode
and hot input beam as $|c|^2=52.5\%$.  Such a large mismatch between
the incoming beam and cavity would therefore result in around half the circulating power
expected from a plane wave model or with a perfectly mode-matched
beam.
This is illustrated in figure~\ref{fig:Pin_hot_cav}, which shows the power circulating
in a single arm cavity as the input power is increased from 0 to 125\,W, taking into
account the 50:50 beam splitter and assuming a power recycling gain of 45 (in reality
this will also be impacted by thermal effects).
\epubtkImage{intra_hot_cavity_power.png}{%
  \begin{figure}[htb]
    \centerline{\includegraphics[scale=0.8]{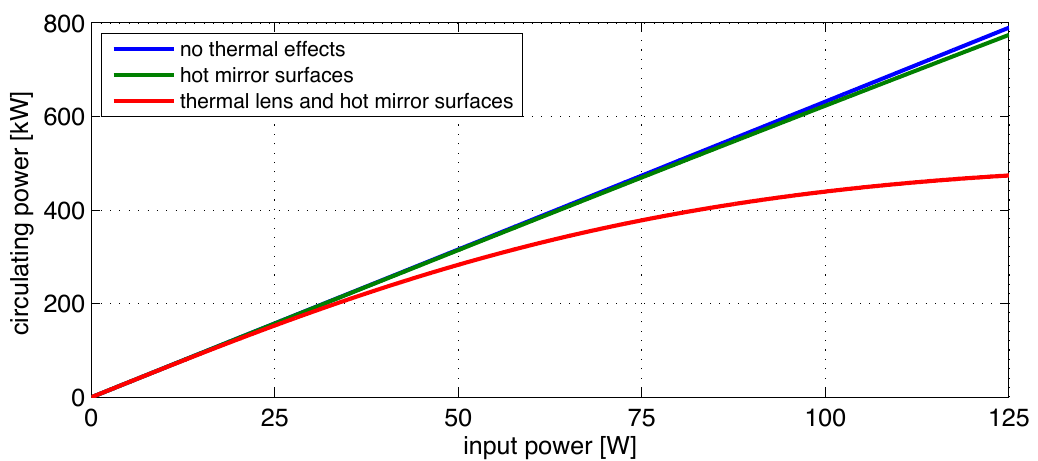}}
    \caption[]{The simulated circulating power in an Advanced LIGO
    	arm cavity vs. input laser power into the power recycled interferometer
	with linear arm cavities.  Three cases are simulated: no thermal effects,
	where the response is linear; including the change in curvature of the
	mirror surfaces due to thermal effects; and including both the change in
	curvature of the mirrors and the thermal lens induced in the input mirror due
	to power absorbed in the mirrors.  The addition of the thermal lens has the greatest
	impact, reducing the power coupled into the arm cavities.
    }
    \label{fig:Pin_hot_cav}
\end{figure}}
When no thermal effects are included
the increase in circulating power is linear.  Including the thermally induced reduction in curvature
of the mirror surfaces, with the input beam remaining mode matched to the cold
cavity, causes a small mismatch, which, at high powers can be noted in a reduction in
the gain of the cavity.  The largest effect of the internal heating is in the creation of
the thermal lens in the ITM, which induces a large mismatch between the beam injected
into the cavity and the cavity eigenmode.
This is reflected in a large reduction in the circulating power at high input
powers, around the 50\% reduction predicted by the overlap between
the hot input beam and cavity eigenmode.  In this simple model we
assume a linear scaling of the thermal distortions and lenses with input
power.  As these aberrations are determined by the circulating power,
which no longer scales linearly with input power, this is a slight over
estimation of the power loss.  In reality this process is more complicated:
as the mirrors heat and their aberrations evolve the circulating power
will decrease, but this will then reduce the thermal lensing
and distortions.  The expected stable cavity will therefore
be slightly different to the cases shown here.  However, these
plots illustrate the magnitude of the problem: such extreme mode
mismatches are unacceptably high.  We require some
method to compensate these effects, especially the thermal lensing.
Such compensation will need to be adaptive, to be applicable for
a range of input powers and for the transition from cold to hot,
controlling the mode resonating in the interferometer
as the power is increased and the mirrors reach thermal equilibrium.
In reality the thermal aberrations will differ slightly from our models
and more importantly the aberrations in each arm will
differ from each other, due to differences in absorption for
the individual mirrors.  The \emph{thermal compensation systems}
need to act on individual mirrors, incorporating sensors which
monitor the current state of the thermal aberrations
in each arm and then feed back to systems which can correct the
curvature of the mirror surfaces and the lenses in the ITMs.
These compensation systems are discussed in more detail in Section~\ref{sec:thermal}
and in the comprehensive review article~\cite{vinet09}.


\subsection{The Michelson: differential imperfections}
\label{sec:GEO_MM}
Previously we motivated operating a Michelson interferometer
on the dark fringe in order to maximise the differential gravitational
wave signal and minimise the noise at the dark port (see Section~\ref{sec:MIandSB}).
The differential degrees of freedom, the Michelson (MICH) and differential
arm length (DARM), are carefully controlled to maintain the interferometer
on the dark fringe, as discussed in Section~\ref{sec:DRFPMI}.
A well defined dark fringe relies on the
fact that the two arms are very similar and essentially the carrier
and any common mode effects cancel at the dark port.
Simply we can express the field reflected back towards the laser
at the symmetric port as
\begin{equation}
E_{\rm{sym.}} = \frac{1}{\sqrt{2}} \left(E_x+E_y\right) ,
\end{equation}
where $E_x$ and $E_y$ refer to the fields coming from the individual
arms.  The field in the asymmetric port, or output port is
\begin{equation}
E_{\rm{asym.}} = \frac{1}{\sqrt{2}}\left(E_x-E_y\right) .
\end{equation}
In the case where each arm is identical the symmetric field
is $E_{\rm{sym.}} = \sqrt{2} E_{\rm{arm}}$, where $E_{\rm{arm}}$ is
the field reflected from an individual arm, and the asymmetric field
is $E_{\rm{asym.}} = 0$.  For a generic interferometer any field
components common to both arms cancel at the
asymmetric port whilst any differential components cancel
in the symmetric port: the field reflected from the interferometer is the common mode;
the field exiting the interferometer is the differential mode.  Previously we have considered
the carrier and any noise coming from the laser to be common
mode (reflected back towards the laser) whilst the
asymmetric port is dominated by the differential gravitational
wave signal, any differential noise and potentially
a small proportion of leaked carrier light for DC readout, see Section~\ref{sec:signal_readout}.
However, a complex realistic interferometer, such as gravitational
wave detectors, contains imperfections and deviations from
specifications that lead to additional fields at the asymmetric port.
For example:
\begin{itemize}
\item Differences in the loss and finesse of each arm result
in different carrier amplitudes in each arm, degrading the
interference of the two beams at the dark fringe and leading to
additional carrier light at the output port.
\item Different resonant or interference conditions for the carrier and control sidebands.
Advanced gravitational wave detectors such as Advanced LIGO
employ a Schnupp modulation scheme, see Section~\ref{sec:Schnupp},
to control the interferometer, where by an asymmetric length
applied to the short Michelson arms ensures that the dark fringe
of the carrier is not equivalent to the dark fringe of the control
sidebands, resulting in a proportion of these radio frequency sideband fields at the dark port.
\item Spatial differences between the beams coming from
each arm.  An imperfect overlap of the spatial distribution of
these beams will degrade their interference and cause light in
higher-order modes to leak into the dark port.
\end{itemize}

In the specific examples discussed here we will focus on
the impact of higher-order modes at the detector output.
These fields do not contain the gravitational wave signal but will
carry noise to the output port.  This will not only increase the
contributions from expected differential mode noises but can couple
common mode noise, such as laser noise, into the output channel.
Such additional light fields can be refereed to as excess light,
defined as light fields exiting the interferometer which do not
contribute to the signal readout.  This should not be confused with
the local oscillator fields, such as the leaked local carrier light in
the DC readout scheme.

A figure of merit for the excess light leaving a Michelson interferometer
is the \emph{contrast defect}.  This is the ratio of the excess light
exiting through the dark port to the light circulating in the
interferometer and is calculated as
\begin{equation}
C = \dfrac{ \displaystyle\int_S  E_x-E_y \ \rm{d}S }{\displaystyle\int_S E_x +E_y \ \rm{d}S} .
\end{equation}
In a Michelson with no differential spatial effects
the contrast defect is determined by the differential
losses in the arms and at the beam splitter.  In reality slight differences
in mirror curvatures, distortions of the mirror surfaces and
limits to alignment control will result in different spatial features
in each arm, appearing as higher-order modes in the
output port and increasing the contrast defect.  The mirrors for
each arm of the detector are manufactured to be very similar
in terms of their optical properties, which will determine
differential arm losses; and their geometric and thermal properties,
which will determine the higher-order mode content
in each arm.  In Advanced LIGO the contrast defect should be lower
than a few hundred ppm ($\sim 2 \times10^{-4}$W/W)~\cite{advanced_ligo}.  In
addition the
higher-order modes can be suppressed  using an output mode cleaner.

An example of the impact of higher order modes on contrast
defect in a dual recycled Michelson have been observed at GEO\,600,
the German\,--\,British gravitational wave detector in Hannover~\cite{GEO60094,Abbott04}.
Unlike other gravitational wave detectors GEO\,600 does not include arm cavities,
but instead has folded arms to increase the effective arm length of the Michelson.
As described in~\cite{Lueck04}, during the operation of GEO\,600
it was discovered that a difference in the radii of curvature of the
folding mirrors in the $x$ and $y$ arms (687\,m in the $x$ arm, 666\,m in the $y$ arm)
was causing a significant difference in the wavefront curvatures
of the beams returning from each arm.  This mismatch between the two beams
resulted in a significant amount of power at the interferometer dark fringe:
the degrade in overlap between the two beams reduced the effective
destructive interference and consequently the output beam on the dark fringe was
dominated by the order 2 mode typical of a mode mismatch, LG$_{10}$.
The resulting loss of power into the anti symmetric port increased the effective loss
in the power recycling cavity, limiting the power build up to $\sim 200$\,W/W,
a significant reduction from the 300\,W/W predicted for this configuration.
The mismatch of the two arms also had a negative impact on the longitudinal
error signal of the Michelson, reducing the magnitude of the error signal and
increasing the susceptibility to misalignments.
\epubtkImage{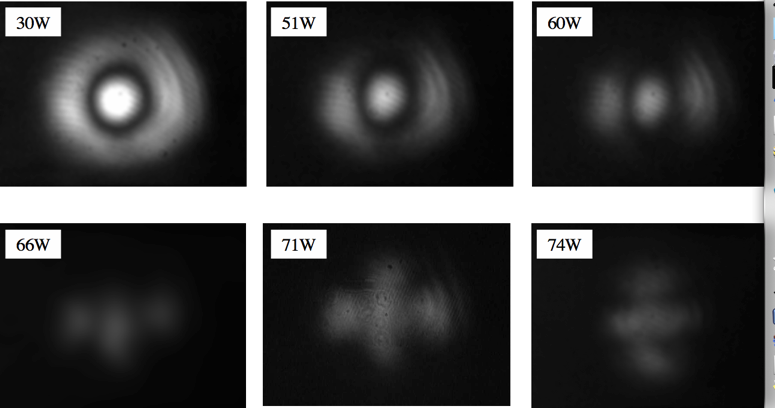}{%
  \begin{figure}[htb]
    \centerline{\includegraphics*[scale=0.8, viewport= 0 0 378 210]{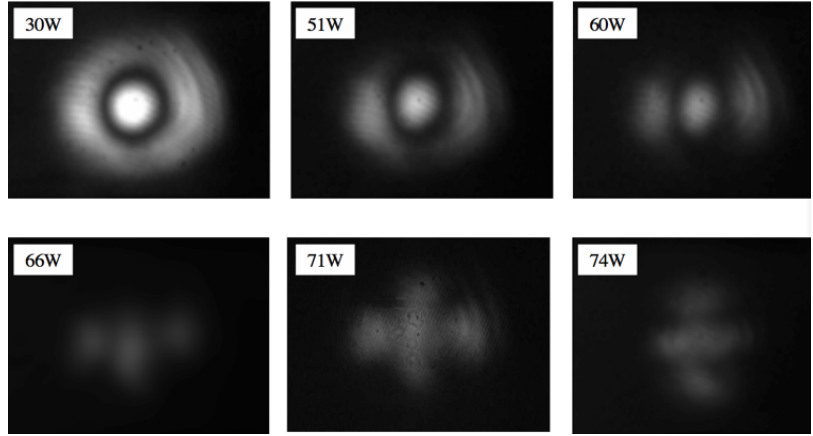}}
    \caption[]{Interference pattern at the dark port of GEO\,600, for different thermal
    compensation of the East arm folding mirrors, indicated by the different ring heater
    powers.  The brightness of the beam images in the bottom row are slightly enhanced
    for better visibility (Figure reproduced from~\cite{Lueck04}).
    }
    \label{fig:GEO_darkfringe}
\end{figure}}

To reduce this mismatch and recover the power recycling
gain the curvature of one of the folding mirrors required correcting,
to match that of the other arm.  In this case the thermal properties of
the mirrors were exploited, namely the dependence of the radius of
curvature of the mirrors to a temperature gradient, as was discussed
in Section~\ref{sec:thermal_cavs}.
In advanced detectors the temperature gradient which develops from high
powered beams incident on the mirrors is an unwanted effect which
results in the distortions of the mirror surfaces (primarily a change in
curvature) and thermal lensing.  In GEO\,600 this thermal behaviour was
manipulated to alter the curvature of the folding mirror in the East arm
($x$ arm) using a ring heater placed behind the mirror substrate to
produce an appropriate temperature gradient in the East mirror.
The extent of the change in mirror curvature is dependent on the
ring heater power, which can be gradually altered to find the
power which corresponds to the optimum curvature (i.e. $\sim 666$\,m to match the North mirror).

In Figure~\ref{fig:GEO_darkfringe} the interference pattern at the dark fringe
is shown for different ring heater powers.  For relatively low powers (30\,W)
the two arms are still not well matched and the dark port is dominated
by the typical bullseye shape of the mismatch mode, LG$_{10}$.  As the
ring heater power is increased from 30\,W to 66\,W the mode matching
between the two arms increases and the power at the
dark fringe is reduced, improving the contrast defect by an order of magnitude.
Increasing the ring heater power to 71\,W further optimised the dark fringe,
in terms of minimum power at the dark port.  At this point the limitations
of this curvature compensation are observed: the compensation is
applied as a spherical curvature correction and does not take into
account differences in curvature in the horizontal and vertical directions, i.e.
astigmatism of the mirrors or beam splitter.  For ring heater powers of 66\,--\,74\,W the output mode
is still dominated by order two modes but in this case these are the Hermite-Gauss
modes consistent with astigmatic mismatches.  At 66\,W the mismatch between
the two arms is compensated in the vertical direction whereas optimum
compensation in the horizontal direction requires a ring heater power of 74\,W.

To fully diagnose and understand the nature of this problem these measurements
were compared with \Finesse simulations of GEO\,600.  In Figure~\ref{fig:GEO_TCS}
the power circulating in the power recycling cavity is plotted against the
power at the dark fringe, showing both simulation and experimental results.
\epubtkImage{comp_fin.png}{%
  \begin{figure}[htb]
    \centerline{\includegraphics[scale=0.8]{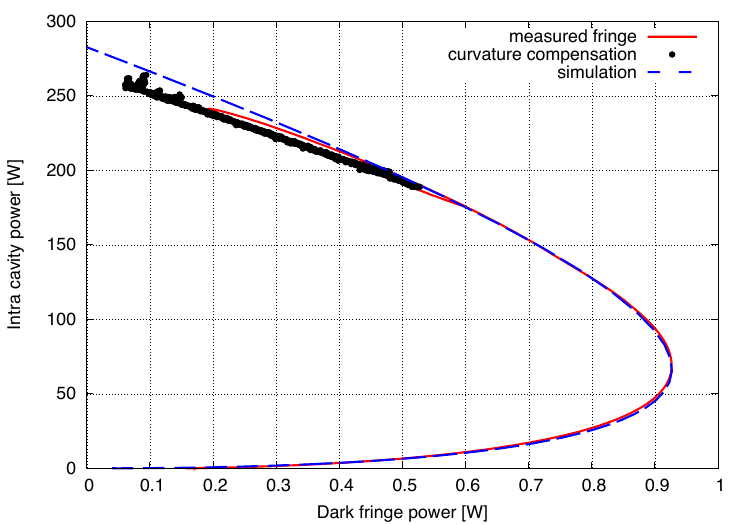}}
    \caption[]{Comparison of measurements and simulations of the circulating
    	power and dark fringe power in GEO\,600.  The measured fringe
	(red trace) and simulation (dashed blue trace) refer to the measured
	and simulated powers for GEO\,600 as the interferometer is moved through
	the dark fringe with the optimal curvature compensation
	applied to the East folding mirror (that which best matches the two arms).
	The curvature compensation (black markers) refers to measurements
	made whilst the interferometer is locked to the dark fringe and the
	curvature compensation of the East mirror is varied (Figure
  reprodcued from~\cite{Lueck04}).
    }
    \label{fig:GEO_TCS}
\end{figure}}
Two different experimental results are shown.  The first result has the optimum
curvature compensation applied with the powers measured as the interferometer
passes through the dark fringe (solid red trace).  The second result is the case where the
interferometer is locked to the dark fringe and the curvature compensation is varied (black
markers).  The experimental and simulation results are sufficiently
similar to suggest that
that our understanding of this problem is correct and that the low intra cavity power/
high contrast defect is dominated by a differential mode mismatch.  The slight differences
between the experimental results and the simulation observed at high intra cavity power
can be explained by the limits of the model: no astigmatic or higher-order spatial
effects were included in this model and hence the model represents a more
simplified system than reality. The experiment and simulation are well matched
for low intra-cavity power where the effects of the spherical
mode mismatch dominate.

This experience at GEO\,600 illustrates the need to have well matched arms.
This can be in terms of mode matching, as shown here, or in terms
of mirror surface distortions and other defects.  While low-order
aberrations such as misalignment and mode mismatch can be corrected
during operation by means of additional control systems, higher-order
effects are typically not actively controlled.  It is crucial that the impact of
higher-order modes is considered during the design of an interferometer
to avoid a large buildup of unwanted modes in the detector.

\subsection{Advanced LIGO: implications for design and commissioning}
\label{sec:aLIGO_design}

The correct modelling of the impact of beam distortions in interferometers is crucial, not only
to our understanding of the physics of real interferometers, but because it will
have implications for real experiments, in particular during the design and commissioning
of detectors.
There are many defects in an interferometer which will effect the shape of the resonating
beams.  In complex advanced interferometers, such as Advanced LIGO, additional systems
help control the shape of the beam, mitigating some higher-order mode effects.  The main
sources of higher-order modes are:
\begin{itemize}
\item {\bf Misalignment.}  Any tilt or lateral shift between the beam axis
and a cavity axis, or between the axes of the multiple interdependent cavities in advanced interferometers,
will produce higher-order modes, for small misalignments these are dominated
by first-order modes.  In modern gravitational wave detectors these effects are
carefully controlled using alignment systems to maintain consistent optical axes within the
interferometer and avoid a large amount of power in first order modes on the detection photodiode.
\item {\bf Mismatch.}  Second-order modes arise from a mismatch in beam size or
wavefront curvature between the cavity eigenmode and incoming beam, or the
multiple cavity eigenmodes in complex interferometers.  In gravitational
wave detectors mismatches are the result of second-order mirror aberrations from the manufacturing
process or environmental processes such as thermal lensing.
In Advanced interferometers {\it thermal compensation systems} will be in place to
correct the curvature of the arm cavity mirrors, to compensate any thermal
lensing and to avoid large mode mismatches.

\item {\bf Surface distortions.}  Higher-order distortions of the beam are generally the
result of higher-order mirror distortions on the highly reflective mirror surfaces.
These defects can arise during the manufacturing process (so called {\it mirror figure error})
or through environmental processes like the thermal distortion of the mirror surfaces.
Whilst first and second order distortions of the beam can be corrected
it is more difficult to actively correct modes of a higher order.  A crucial part of the
design process is to determine
the tolerances and requirements for the polishing and coating of the interferometer mirrors,
to ensure a low higher-order modes content.  This is discussed in more
detail in Section~\ref{sec:surface_defects}.

\item {\bf Apertures} Higher-order modes are also generated when the circulating beams
encounter the effective aperture caused by the finite size of optical components.
The `clipping' of the beam results in a sharp cut-off, equivalent to the addition of
high order modes.  The design of a well behaved optical setup
will ensure the size of the optics, compared to the beam, is sufficiently large such that
these higher-order effects are small and we can simple consider the effect of the aperture
as a small power loss.
\end{itemize}
In this section we consider the impact higher-order mode effects
have on the final design of an advanced interferometer.
The impact of beam distortions are carefully considered during
the design process and here we review the choices
motivated by beam shape and size for the particular case of Advanced LIGO.

\subsubsection*{The input mode cleaner}
In a gravitational wave detector use an optical cavity, called the
{\it input mode cleaner} (IMC), between the laser and the main
interferometer.  The purpose of the IMC is to produce a very pure
fundamental $\mathrm{TEM}_{00}$ Gaussian beam for the detector input,
filtering out
higher-order spatial modes. It is also used as part of the laser
frequency stabilisation system,  producing a very stable carrier
frequency.  This is motivated by the desire to avoid injecting light
fields into the interferometer which may couple additional noise to
the output photodiode.  The interferometer is tuned to the operating
point of one specific mode, the carrier $\mathrm{TEM}_{00}$ mode.  Any other fields
will propagate differently through the interferometer, for example,
most higher-order modes do not enter the  arm cavities and therefore
carry a differen phase information than the $\mathrm{TEM}_{00}$ more.

\epubtkImage{inputmodecleaner.png}{%
  \begin{figure}[htb]
    \centerline{\includegraphics[scale=1.2]{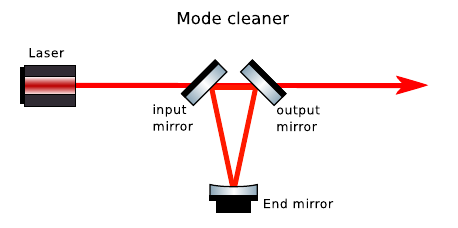}}
    \caption[]{Diagram of a 3-mirror input mode cleaner, similar to
    		that used in advanced interferometers. The cavity is impedance
		matched to ensure maximum transmission of the carrier. }
    \label{fig:IMC}
\end{figure}}

Another requirement of the IMC is to maximise transmission
of the fundamental carrier mode, whilst also transmitting the radio frequency control
sidebands applied to the beam.  For the case of Advanced LIGO
the corresponding modulation frequencies are 9\,MHz and 45\,MHz.
The requirement of high transmission for the carrier and certain sidebands
sets very specific specifications
on the length of the IMC (see Section~\ref{sec:two_mirror2}),
whereas the suppression of higher-order spatial
modes requires a choice of mirror curvatures which provide a round-trip Gouy
phase sufficient to effectively separate the resonance of the spatial modes
(see Section~\ref{sec:HOM_resonances}).
The Advanced LIGO IMC is a 3-mirror impedance matched cavity, as shown in
Figure~\ref{fig:IMC}.
It consists of two identical flat mirrors (input and output mirrors)
and one curved mirror with a very high reflectivity (the end mirror).
The final design parameters of the IMC are described in~\cite{aligo_IO_design}, and
table~\ref{table:IMC} summarises the key parameters.  The free
spectral range (FSR) is chosen to
allows transmission of the two control sidebands at
$f_1 = 1\times\mathrm{FSR}$ and $f_2 = 5\times\mathrm{FSR}$.

\begin{table}[htb]
\begin{center}
\begin{tabular}{l|r}
Parameter		 		& Value			\\
\hline
Length				& 16.473\,m		\\
Free spectral range		& 9,099,471\,Hz	\\
Input/end mirror $R_C$	& >10000\,m		\\
Input/end mirror $T$		& 0.6\%			\\
Input/end mirror $R$		& 99.4\%			\\
Input/end mirror $\alpha$  & 44.59$^{\circ}$	\\
Curved mirror $R_C$	& 27.24$\pm$0.14\,m	\\
Curved mirror $R$		& >0.9999			\\
Curved mirror $\alpha$	& 0.82$^{\circ}$		\\
Finesse				& 522			\\	
\end{tabular}
\end{center}
\caption{Summary of key design parameters of an Advanced LIGO
		input mode cleaner~\cite{aligo_IO_design}.  The cavity length, radii of curvature ($R_C$),
		reflectance ($R$), transmittance ($T$) and angle of incidence ($\alpha$)
		are all given, as well as derived parameters such as the free spectral range
		and finesse.  The cavity consists of 3 mirrors of which the input and end
		mirrors are nominally flat.}
\label{table:IMC}
\end{table}%

\subsubsection*{Recycling cavities}
\label{sec:recycling_design}
As discussed in Sections~\ref{sec:mismatch} and~\ref{sec:thermal_cavs},
it is important that any beam injected into
a cavity is well mode matched to ensure optimum coupling of the
laser beam into the cavity.  In a Michelson it is important that the
two arms are well mode matched to avoid a large amount of
power exiting the interferometer through the anti-symmetric port
(see Section~\ref{sec:GEO_MM}).
Advanced interferometers are highly complex, incorporating
a series of cavities within the general Michelson layout.
The addition of a recycling
mirror at the symmetric port (power recycling) and anti-symmetric
port (signal recycling) form recycling cavities between these mirrors
and the rest of the interferometer.  The parameters of these
cavities must be carefully
chosen to ensure a good mode match between the
eigenmodes of the recycling cavities and the arm cavities.
The following discussion of the design of the recycling cavities
refers to the most common design based on arguments
presented in~\cite{Arain08}. Note that the Advanced
Virgo project has chosen a different design
approach~\cite{AdvancedVirgo15}.

\epubtkImage{coupled_cavities.png}{%
  \begin{figure}[htb]
    \centerline{\includegraphics[scale=0.4]{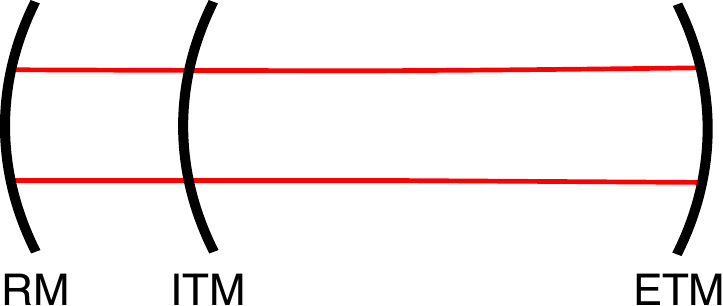}\hfill
    \includegraphics[scale=0.4]{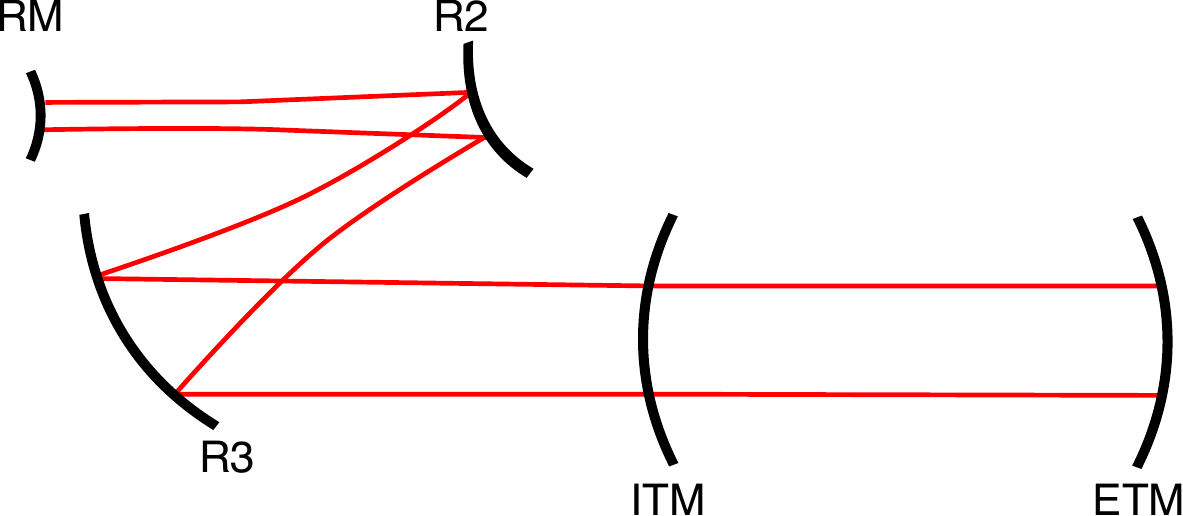}}
    \caption[]{Two examples of a coupled cavity formed between
    a recycling mirror (RM) and an Advanced LIGO arm cavity, made
    of the input and end test masses (ITM and ETM).  The diagram on
    the left has a single recycling mirror forming a cavity with the ITM.
    This is illustrative of the setup of Initial LIGO~\cite{Adhikari2006}.
    The diagram on the right illustrates a folded recycling cavity,
    where two additional mirrors in the recycling cavity reshape the beam
    between the recycling mirror and ITM.  This is the setup used in Advanced
    LIGO~\cite{Arain08}.  Illustrations are not to scale
    and in the case of LIGO the distances between the recycling optics
    is much smaller (of the order 10\,m) than the distance between the
    test masses (4\,km).
    }
    \label{fig:coupled_cavities}
\end{figure}}
For the design stage we first assume perfect matching of the arm
cavities.  We can then consider each recycling cavity acting with the
arms as a simple coupled cavity.
Two examples of a possible coupled cavity setup are shown in
figure~\ref{fig:coupled_cavities}.
The eigenmode of the arm cavities is selected to produce large beams
at the ITM (5.3\,cm) and ETM (6.2\,cm) to reduce thermal noise, with
slightly smaller beams at the ITM as the thermal noise is lower here
(fewer coating layers) and to prevent scattering into the recycling cavities.
The curvatures are also carefully selected for a specific Gouy phase to
avoid higher-order modes easily ringing up
in the arms: $R_C=1934$\,m (ITM) and $R_C=2245$\,m (ETM).
The beam parameter of the arms is therefore a fixed parameter,
and the properties of the recycling cavities should be chosen
to mode match the recycling cavity to the arms.

\epubtkImage{recycling_gouy_phase.png}{%
  \begin{figure}[htb]
    \centerline{\includegraphics[scale=0.8]{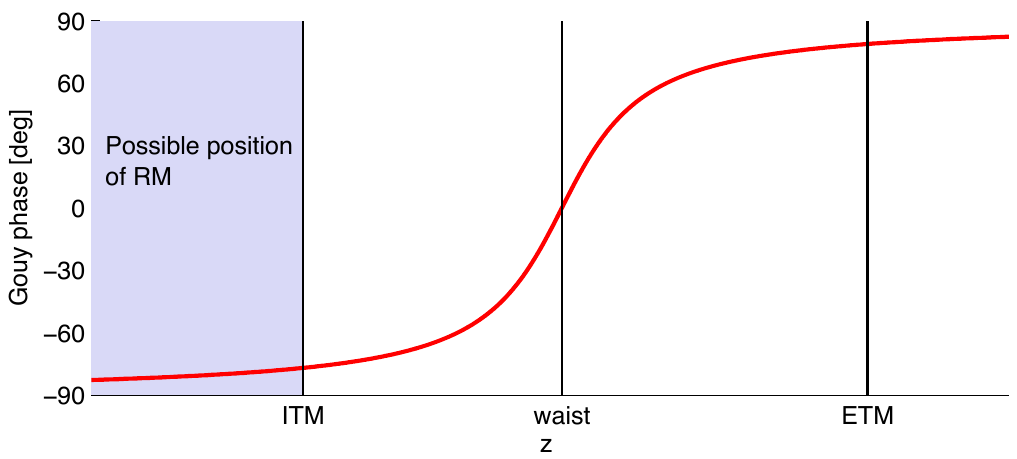}}
    \caption[]{Gouy phase as a function of position on the optical axis
    	for the Advanced LIGO arm cavity eigenmode. A single power/signal
	recycling mirror (RM) would be placed before the ITM in this
	representation~\cite{Arain08}.
    }
    \label{fig:RC_gouy}
\end{figure}}

The simplest design for the recycling cavities uses a single mirror
coupled with the arm cavities, as shown in the left diagram of
figure~\ref{fig:coupled_cavities}, where the curvature of the recycling mirror
is matched to the wavefront curvature of the arm cavity
eigenmode.  This was the layout chosen for power recycling in Initial LIGO.
In this layout the eigenmode of the arm and recycling cavity can
be matched.  However, there is another consideration for
the design of the recycling cavities: the separation of higher-order
mode resonances.  This is determined by the Gouy phase accumulated
in the recycling cavity (between RM and ITM).
In figure~\ref{fig:RC_gouy} the Gouy phase
of the eigenmode for the Advanced LIGO arms is shown at different
positions along the optical axis.  The ITM and ETM are both far from
the waist but the difference in Gouy phase (155.7$^{\circ}$,
equivalent to $-24.3^{\circ}$) is far outside the linewidth of the cavity.
With a single recycling mirror the only possible positions do not allow
for a large change in Gouy phase, as the ITM is already in the far field.
In reality there are additional limitations on the position of the
recycling mirror, such as the physical location of the vacuum chambers.

In Initial LIGO this configuration resulted in a power recycling cavity
formed in the far field where the higher-order mode
resonances were not sufficiently
separated: they fell within the linewidth of the cavity.
The
individual recycling cavity (power recycling mirror and ITM) was only
marginally stable in this setup.  When operated as a coupled cavity
the carrier $\mathrm{TEM}_{00}$ mode enters the arm cavity, whilst all higher-order modes are directly
reflected, meaning the $\mathrm{TEM}_{00}$ mode acquires 180$^{\circ}$ of phase on
reflection from the arm compared with the higher-order modes.  This
allowed stable operation of the power recycling cavity for the carrier in Initial LIGO,
as in the coupled system the HOMs are effectively anti-resonant
in the recycling cavity when the carrier is resonant.
However, as observed in LIGO~\cite{Adhikari2006}, this configuration is only
marginally stable for the control sidebands, which do not enter the
arm cavities, resulting in a near-degenerate cavity for the sidebands
with all spatial modes near resonance.
HOMs of the sidebands are easily excited through misalignment
and mode mismatch and it was only the use of thermal
compensation systems which allowed the design sensitivity of LIGO to be achieved.

In Advanced LIGO the issue of unstable recycling cavities becomes more
complex due to larger beam sizes, large thermal lensing effects and the addition of signal recycling.
Unlike the power recycling cavity the signal recycling cavity coupled
with the arm cavities will operate on anti-resonance for the carrier,
for resonant sideband extraction.  Any HOMs will be nearly resonant in
an SRC designed with a single recycling mirror.  
To avoid these problems in Advanced LIGO an alternative recycling geometry was designed.
This is shown in the right diagram of figure~\ref{fig:coupled_cavities},
adding 2 folding mirrors to the recycling cavities to alter the beam
parameter and gain significant Gouy phase between the ITM and recycling
mirror.  The curvatures of these mirrors are carefully chosen to gain this
required Gouy phase, whilst maintaining a mode matched system.
The design parameters for the power and signal recycling cavities
for Advanced LIGO are summarised in Appendix~\ref{sec:ALOL}.

Such stable recycling cavities are now installed in Advanced LIGO.
Each recycling cavity is characterised by 3 mirrors: the primary mirrors, PRM and SRM,
and two additional folding mirrors which shape and direct the beam, PR2/3 and SR2/3.
The greatest change in the beam occurs between PR2/3 (and SR2/3)
where the beam size increases by around a factor of 10 over a short distance
($\sim 16$\,m).  Any small changes in the
curvatures of the folding mirrors or the distance between them
can lead to substantially larger or smaller beams
and degrade the mode matching to the arm cavities.
This is illustrated in figure~\ref{fig:cold_MM} where the mode matching
between the power recycling cavity, arm cavity and input
beam (input mode cleaner eigenmode) is plotted against
the distance between PR2 and PR3\footnote{This shows results which are slightly different than those
  reported in~\cite{Arain08} because we have used the final design
values for this plot, not the preliminary values used in~\cite{Arain08}.}.
Two sets of results are shown, those for the nominal values
of recycling optics, and those for a slight error in the curvature
of PR3.  The mode matching between the arm and the recycling
cavity is relatively insensitive to the PR2\,--\,PR3 distance over a 100\,mm
range, whilst the mode matching between the recycling mode and input
beam falls more sharply away from the nominal value.
An error in the curvature of the recycling optics can significantly
degrade the mode matching, even pushing the recycling cavity to
instability (regions of no data).  However, the mode matching
can be recovered from any such errors by adjusting
the distance between the two folding mirrors, R2 and R3.

\epubtkImage{cold_mode_matching.png}{%
  \begin{figure}[htb]
    \centerline{\includegraphics[scale=0.8]{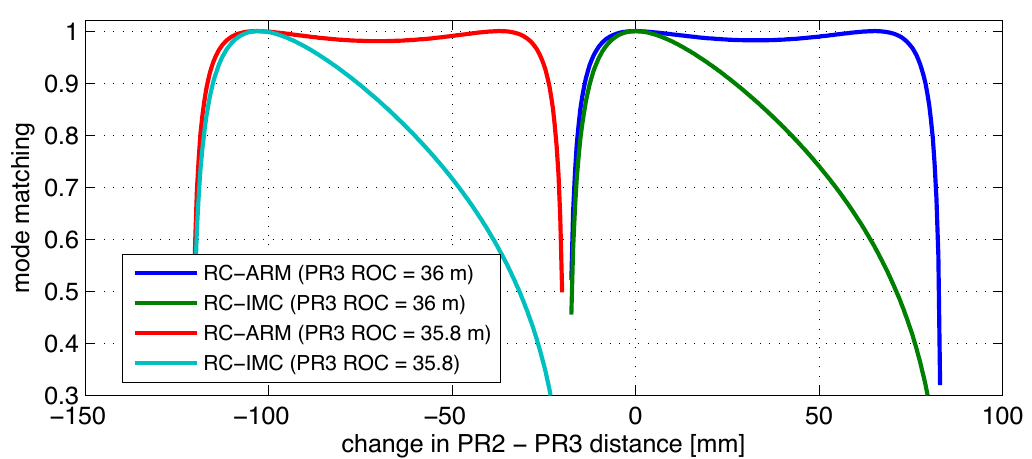}}
    \caption[]{Plots showing the mode matching between the
    	recycling cavity eigenmode, the arm cavity eigenmode and
	the incoming beam (input mode cleaner, IMC, eigenmode)
	for the Advanced LIGO design.  The mode matching is
	shown for the power recycling cavity as the distance between
	the two telescope mirrors, PR2 and PR3 (see figure~\ref{fig:coupled_cavities}),
	is adjusted from the nominal design value.  Two sets of results
	are shown, those for the design curvature of PR3 (36\,m)
	and a small error on this curvature (35.8\,m).
	Adjusting the PR2\,--\,PR3 distance recovers the mode matching
	from errors in the curvatures of the recycling optics~\cite{Arain08}.}
    \label{fig:cold_MM}
\end{figure}}

\subsubsection*{Thermal distortions}
\label{sec:thermal}

The mode matching of the beams between the recycling cavities and
arms is complicated by thermal effects, specifically thermal lensing and
the change in mirror curvatures.  Previously the need for some
thermal compensation was motivated by the behaviour of a single
cavity at high power (see Section~\ref{sec:thermal_cavs}).  For Advanced LIGO the implications
for the coupled systems of the arm and recycling cavities were considered
during the design phase~\cite{Arain2011}.  In figure~\ref{fig:MM_vs_power} the mode matching
between the recycling and arm cavities, and the recycling cavities and
the input mode is shown, as the
interferometer input power is increased.  As the thermal lens in the ITM
is by far the dominant effect (\ref{sec:thermal_cavs}) this is the only
thermal aberration included,
modelled as a simple spherical lens.
The first two traces in figure~\ref{fig:MM_vs_power} (0\,W design) show the mode matching
for the original design of the recycling cavities, where the parameters
were chosen to match the cold optics of the arm cavities.  A second design
(18\,W design) is also shown.  In this case the mode matching between
the recycling cavities, the arm cavities and the input mode was optimised
for the expected thermal lensing of 34.5\,km\footnote{34.5\,km is the
  focal length when modelled as an individual lens
in a vacuum, the approach in this document.
Sometimes quoted is 50\,km corresponding to the lens
when modelled inside the fused silica substrate of the ITM.}
at 18\,W input power. The advantages of this design
is that it gives a larger range of input power at which the interferometer is
well mode matched, without the need for thermal compensation systems.

\epubtkImage{mode_matching_at_high_power.png}{%
  \begin{figure}[htb]
    \centerline{\includegraphics[scale=0.8]{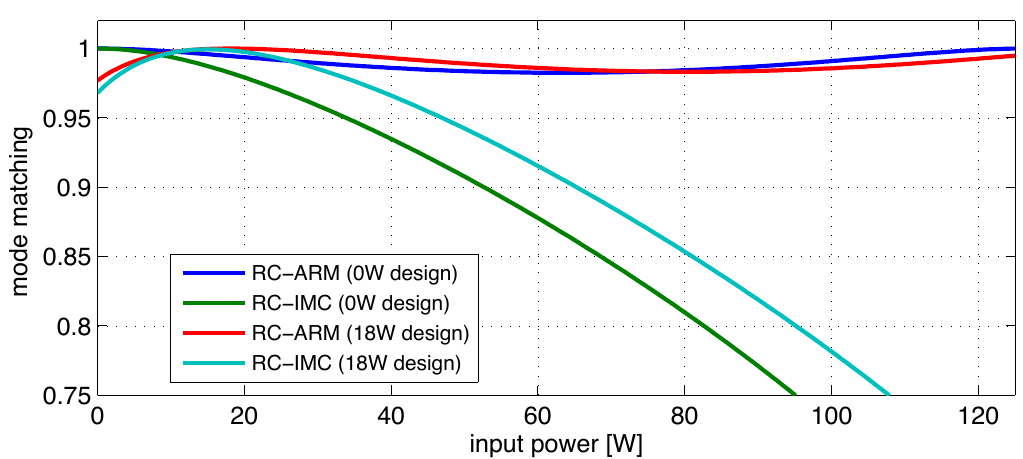}}
    \caption[]{Plots showing mode matching at different input powers between the recycling cavities
    	and the arms (RC-ARM) and recycling cavities and the input mode cleaner
	(RC-IMC) for the Advanced LIGO design.  The mode overlap is calculated
	considering the thermal lens formed in the input test masses from predicted
	absorptions in the ITMs.  Two different designs are considered, one optimised for
	mode matching at an input power of 0\,W (cold optics) and one optimised for
	18\,W, the final Advanced LIGO design. }
    \label{fig:MM_vs_power}
\end{figure}}

The plots shown in figure~\ref{fig:MM_vs_power} show that the mode matching
between the recycling cavities and arm cavities is relatively independent of
the expected thermal lensing.  Whilst the eigenmode of the arm cavity is
fixed, the recycling cavity eigenmode is affected by the thermal lens.
The recycling eigenmode curvature is fixed at the reflective ITM surface,
and the beam size at this point only varies a small amount, maintaining
the mode matching between the arm and recycling cavity.
However, the effect of the lens on the mode parameters is exaggerated
during the large divergence between the recycling mirrors R2 and R3
(see figure~\ref{fig:coupled_cavities}) and this has a large impact on the beam size
at the recycling mirror, and hence the mode matching between
the input beam and recycling cavity is significantly degraded.
As we saw previously for a single cavity, during high power
operation the power coupled into the interferometer will
be significantly reduced.

In Advanced LIGO \emph{thermal compensation systems} (TCS)
will be employed at high power, not only to ensure a large power
buildup within the interferometer but to balance the lensing and
eigenmodes of the two arms to prevent a high contrast defect
~\cite{Willems2009}.  The first is a ring-heater positioned near
the anti-reflective surface of each test mass~\cite{Arain2011}.
These are used to heat the outer edge of the mirror to produce
a curvature in the opposite direction to that from heating by the beam.
The ring heater also corrects some of the thermal lens in the ITM substrate.
An additional system is required to complete the correction of the thermal lens.
This involves a compensation plate, placed in front of the ITMs, made of the same
material (fused silica).  A heating pattern is projected onto this plate via a CO$_2$
laser.  This pattern is designed to heat the compensation plate in such a way
as to correct any thermal lensing in the ITM~\cite{Brooks2012}.

\subsubsection*{The output mode cleaner}
\label{sec:MCs}

Even with state of the art optics, alignment systems to correct any misalignments
and thermal compensation systems to correct for differential mismatches
some light at the Michelson anti-symmetric port will be in higher-order modes.
There will also be some power in the control sidebands
exiting the interferometer, as the dark fringe for the carrier
is not the dark fringe for the sidebands due to the applied Schnupp
asymmetry (see Section~\ref{sec:Schnupp} and Appendix~\ref{sec:aLIGO_params}).
The only fields which should be present on the detection photodiode
are the gravitational wave signal sidebands and the local oscillator
field, in case of Advanced LIGO this is the leaked
carrier light for DC readout (see Section~\ref{sec:readout}).
If the power in the higher-order modes and control sidebands is sufficiently
low they can be effectively stripped from the beam using
an \emph{output mode cleaner} (OMC), an optical cavity between the
Michelson interferometer and the main photodiode.

The parameters of the output mode cleaner are carefully chosen
to maximise the transmission of the gravitational wave signal and
local oscillator field, whilst sufficiently filtering
out the unwanted spatial modes and control sidebands.
The OMC design for Advanced LIGO is a Fabry-Perot cavity in a
4-mirror bow tie configuration~\cite{Arai2013}, consisting of two flat mirrors
(the input and output mirrors) and two curved mirrors (high reflectors).
To maximise the transmission
of the desired fields the cavity is impedance matched between the input
mirror and output mirrors.
Ignoring any losses, the power in an individual field transmitted by an
impedance matched cavity is
\begin{equation}
P_{\mathrm{trans.}} = \frac{T^2}{1+R^2-2R\cos{(kL_{rt} - \Psi(n+m+1))}} P_0 ,
\end{equation}
where $T$ and $R$ are the transmittance and reflectance properties of
the input and output coupler, $P_0$ is the power in the incoming field,
$k$ is the wavenumber of the field, $L_{rt}$
is the round-trip length of the cavity, $\Psi$ is the round-trip Gouy phase
and $n$ and $m$ are the higher-order mode indices of the field.
In a lossless cavity all the power in a field is transmitted on resonance.
For an Advanced LIGO OMC with realistic losses the transmission of the carrier $\mathrm{TEM}_{00}$ mode
and signal sidebands is expected to be $\sim 98\%$.
On anti-resonance the transmitted power can be approximated as
\begin{equation}
\mathrm{min}(P_{\mathrm{trans.}}) = \frac{T^2}{1+R^2+2R}P_0 \approx \frac{T^2}{4} P_0 ,
\end{equation}
as in the case of a high finesse
cavity, $R\approx1$.  In order to avoid transmitting the unwanted fields the length and curvatures
of the cavity mirrors are very carefully chosen.  The length of the cavity
must not be resonant for the 9\,MHz and 45\,MHz control sidebands.
The curvatures of the mirrors are chosen to ensure sufficient Gouy phase
to avoid the resonance of any higher-order modes entering the cavity.
This requires careful modelling and a knowledge of which higher-order
modes are expected to exit the main interferometer.  In~\cite{Arai12}
the HOM content is modelled using a power law derived from the
spectrum of higher-order modes observed in Enhanced LIGO.  This predicts
the total power in each order of modes exiting the interferometer as
\begin{equation}
P_{\mathrm{HOM}} = 1.8 \times 10^{-3} \times 10^{-\frac{N}{4.8}} ,
\end{equation}
where $N$ is the mode order and $N \leq 2$ (1st order modes are reduced
via alignment control).  Using this power law different mirror curvatures
and lengths were modelled to find the optimum design for minimum
transmission of the expected undesired fields.  This design is
presented in~\cite{Arai2013} and the key parameters are summarised in
table~\ref{table:OMC}.  With a finesse of $\sim 400$
and an expected round-trip loss of 140\,ppm the transmission of undesirable
fields is expected to be $10^{-5}$W/W, relative to the power injected into
the interferometer.  This is equivalent to $\sim 1$\,mW at high power,
compared to $\sim 100$\,mW of reference carrier light for DC readout.

\begin{table}[htb]
\begin{center}
\begin{tabular}{l|r}
Parameter		 		& Value			\\
\hline
Length				& 1.132\,m		\\
Free spectral range		& 264.8\,MHz	\\
Input/output mirror $R_C$ & $\infty$		\\
High reflectors $R_C$	& 2.5\,m		\\
Input/output mirror $T$	& 8300\,ppm	\\
High reflectors $T$		& 50\,ppm		\\
Finesse				& 390		\\
Angle of incidence		& 4$^{\circ}$	\\
\end{tabular}
\end{center}
\caption{Summary of key design parameters for the Advanced LIGO
		output mode cleaner~\cite{Arai2013}.  The length, mirror radii of curvature
		($R_C$) and transmittance ($T$) are given, as well as derived parameters
		the free spectral range and finesse.}
\label{table:OMC}
\end{table}%

\subsection{Commissioning}

Commissioning describes the process of tuning and improving a
gravitational wave detectors after it's subsystems have been installed
and before the full system. This process typically takes several years
because the interferometer couples all the subsystems in a unique and
complex way, which cannot be tested in advance.
This is particularly important for advanced detectors which employ many
cutting edge technologies which, although having been tested in the laboratory
and at prototype facilities, have never been implemented
together in interferometers of this scale. The efficiency of the
commissioning process crucial for achieving the expected sensitivity
and provide an instrument for scientific data taking in a timely manner.

Through the commissioning process we observe effects never seen before,
the interferometer will be  operated in a new regime, namely
a full scale, high power, dual recycled interferometer with arm cavities.
In this extremely sensitive configuration previously negligible effects
could have a strong impact on interferometer performance.  For example,
parametric instabilities, where higher-order mode and radiation pressure
effects couple together with the potential of ringing up high order sidebands,
will likely be a factor in this high powered
regime~\cite{BSV01, Evans2010665, gras2010}.
Subsystems of the interferometer also use cutting edge techniques which
have yet to be tested within the full framework of our advanced detectors.

During commissioning the interferometers are
assembled in increments, building towards the full dual recycled
configuration.  As the optics are installed many measurements are taken to
test the behaviour of various subsystems and finally to test the
response and noise budgets of the full interferometer.
During this process it is crucial that we have accurate models of the
interferometers.  These must include possible defects and
higher-order mode effects, typically going beyond the more simplified
models used in the design phase.
For example, in Advanced LIGO measurements of the surfaces of the
mirrors were taken prior to installation.  This surface data can be
used in simulations to model the expected distortion of the beams
within the real interferometers.  During the commissioning process
these models are used to check against experimental measurements.
In the case where a measurement is not as expected models are
used to investigate the possible causes, adding in more realistic
measurements and tuning parameters to recreate the observed
behaviour.  from such models we can then suggest solutions
in the case of underperformance.

The interested reader is directed to the following documents, which
give details on specific modelling tasks to support the commissioning of
for Advanced LIGO, particularly those which
are concerned with higher-order mode effects\footnote{Many more such
  documents exist for Advanced LIGO and other gravitational wave
  detectors. This selection is based on our
  familiarity with the described work.}:
\begin{itemize}
\item  Comparisons of alignment signals calculated for Fabry-Perot cavities
	using three methods:
	\Finesse, an analytic calculation and the FFT propagation simulation
  OSCAR \cite{ballmer14}
\item  Comparisons of the control signals and sideband build up in Advanced
	LIGO, as modelled in \Finesse and Optickle \cite{Bond14c, phd.bond2014}
\item Investigation into the effect of mode-mismatch in the control
  signals of the Advanced LIGO interferometer~\cite{Bond14b}.
\item  A dedicated commissioning investigation into power loss at the central
	beam splitter in Advanced LIGO using \Finesse \cite{Bond13c}
\item  \Finesse simulations of the alignment control signal of the Advanced
	LIGO input mode cleaner \cite{Kokeyama2013}
\end{itemize}

\subsection{Finesse examples}

\subsubsection{Higher-order mode resonances}
\epubtkImage{fexample_HOM_scan.png}{%
  \begin{figure}[htb]
    \centerline{\includegraphics[width=0.9\textwidth]{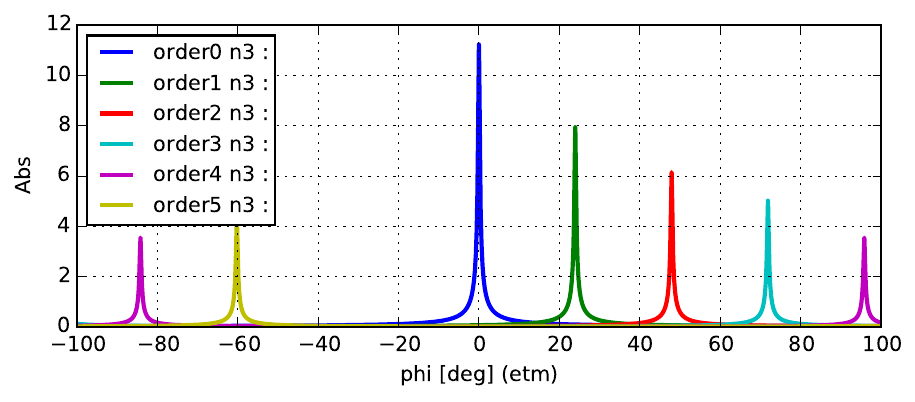}}
    \caption{\Finesse example: Higher-order mode resonances}
    \label{fig:fexample_HOM_scan}
\end{figure}}

\noindent
This example illustrates the different resonance conditions of the
higher-order modes in an optical cavity.  In the simulation an input
beam made up of 6 different order modes is generated using the
`tem' commands.  This beam is injected into an Advanced LIGO style
cavity and the cavity length is tuned.  Using amplitude detectors, `ad',
the different order modes are individually detected.

\vspace{3mm}\noindent
{\small
\textbf{Finesse input file for `Higher-order mode resonances'}
{\renewcommand{\baselinestretch}{.8}

\nopagebreak
\tt
\noindent
\mbox{}\textbf{\textcolor{RoyalBlue}{l}}\ i1\ \textcolor{Purple}{1}\ \textcolor{Purple}{0}\ nin \\
\mbox{}\textcolor{Blue}{tem}\ i1\ \textcolor{Purple}{1}\ \textcolor{Purple}{0}\ \textcolor{Purple}{0.5}\ \textcolor{Purple}{0} \\
\mbox{}\textcolor{Blue}{tem}\ i1\ \textcolor{Purple}{2}\ \textcolor{Purple}{0}\ \textcolor{Purple}{0.3}\ \textcolor{Purple}{0} \\
\mbox{}\textcolor{Blue}{tem}\ i1\ \textcolor{Purple}{3}\ \textcolor{Purple}{0}\ \textcolor{Purple}{0.2}\ \textcolor{Purple}{0} \\
\mbox{}\textcolor{Blue}{tem}\ i1\ \textcolor{Purple}{4}\ \textcolor{Purple}{0}\ \textcolor{Purple}{0.1}\ \textcolor{Purple}{0} \\
\mbox{}\textcolor{Blue}{tem}\ i1\ \textcolor{Purple}{5}\ \textcolor{Purple}{0}\ \textcolor{Purple}{0.15}\ \textcolor{Purple}{0} \\
\mbox{} \\
\mbox{}\textbf{\textcolor{RoyalBlue}{s}}\ s1\ \textcolor{Purple}{1}\ nin\ n1 \\
\mbox{}\textbf{\textcolor{RoyalBlue}{m}}\ itm\ \textcolor{Purple}{0.986}\ \textcolor{Purple}{0.014}\ \textcolor{Purple}{0}\ n1\ n2 \\
\mbox{}\textbf{\textcolor{RoyalBlue}{s}}\ scav\ \textcolor{Purple}{4000}\ n2\ n3 \\
\mbox{}\textbf{\textcolor{RoyalBlue}{m}}\ etm\ \textcolor{Purple}{1}\ \textcolor{Purple}{0}\ \textcolor{Purple}{0}\ n3\ n4 \\
\mbox{}\textbf{\textcolor{RoyalBlue}{cav}}\ arm\ itm\ n2\ etm\ n3\  \\
\mbox{}\textcolor{Blue}{attr}\ etm\ Rc\ \textcolor{Purple}{2245} \\
\mbox{}\textcolor{Blue}{attr}\ itm\ Rc\ \textcolor{BrickRed}{-}\textcolor{Purple}{1934} \\
\mbox{} \\
\mbox{}\textbf{\textcolor{RoyalBlue}{ad}}\ order0\ \textcolor{Purple}{0}\ \textcolor{Purple}{0}\ \textcolor{Purple}{0}\ n3 \\
\mbox{}\textbf{\textcolor{RoyalBlue}{ad}}\ order1\ \textcolor{Purple}{1}\ \textcolor{Purple}{0}\ \textcolor{Purple}{0}\ n3 \\
\mbox{}\textbf{\textcolor{RoyalBlue}{ad}}\ order2\ \textcolor{Purple}{2}\ \textcolor{Purple}{0}\ \textcolor{Purple}{0}\ n3 \\
\mbox{}\textbf{\textcolor{RoyalBlue}{ad}}\ order3\ \textcolor{Purple}{3}\ \textcolor{Purple}{0}\ \textcolor{Purple}{0}\ n3 \\
\mbox{}\textbf{\textcolor{RoyalBlue}{ad}}\ order4\ \textcolor{Purple}{4}\ \textcolor{Purple}{0}\ \textcolor{Purple}{0}\ n3 \\
\mbox{}\textbf{\textcolor{RoyalBlue}{ad}}\ order5\ \textcolor{Purple}{5}\ \textcolor{Purple}{0}\ \textcolor{Purple}{0}\ n3 \\
\mbox{}\textbf{\textcolor{Red}{maxtem}}\ \textcolor{Purple}{5} \\
\mbox{}\textbf{\textcolor{Red}{phase}}\ \textcolor{Purple}{2} \\
\mbox{}\textbf{\textcolor{Red}{xaxis}}\ etm\ phi\ lin\ \textcolor{BrickRed}{-}\textcolor{Purple}{90}\ \textcolor{Purple}{90}\ \textcolor{Purple}{10000} \\
\mbox{}\textbf{\textcolor{Red}{yaxis}}\ lin\ abs

}}

\subsubsection{Mode cleaner}

\epubtkImage{fexample_mc.png}{%
  \begin{figure}[htbp]
    \centerline{\includegraphics[width=0.9\textwidth]{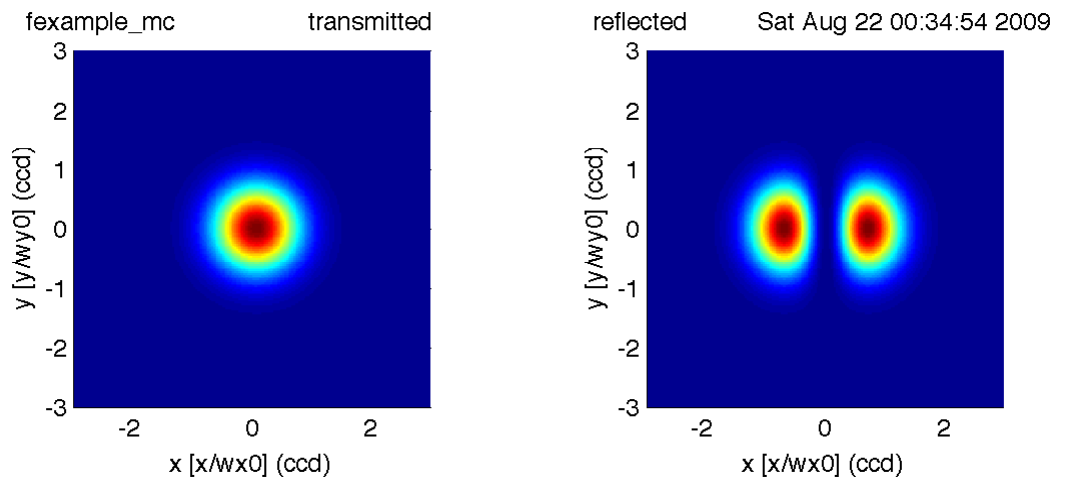}}
    \caption{\Finesse example: Mode cleaner}
    \label{fig:fexample_mc}
\end{figure}}

\noindent
This example uses the `tem' command to create a laser beam which is a
sum of equal parts in  $u_{00}$ and $u_{10}$ modes. This beam is
passed through a triangular cavity, which acts as a mode cleaner. Being
resonant for the $u_{00}$, the cavity transmits this mode and reflects
the $u_{10}$ mode as can be seen in the resulting plots.

\vspace{3mm}\noindent
{\small
\textbf{Finesse input file for `Mode cleaner'}
{\renewcommand{\baselinestretch}{.8}

\nopagebreak
\tt
\noindent
\mbox{} \\
\mbox{}\textbf{\textcolor{RoyalBlue}{laser}}\ i1\ \textcolor{Purple}{1}\ \textcolor{Purple}{0}\ n1\ \ \ \ \ \ \ \ \ \ \ \ \textcolor{Gray}{\%\ laser\ with\ P=1W} \\
\mbox{}\textbf{\textcolor{Red}{maxtem}}\ \textcolor{Purple}{1}\ \ \ \ \ \ \ \ \ \ \ \ \ \ \ \ \ \ \ \textcolor{Gray}{\%\ need\ Hermite-Gauss\ modes\ up\ to\ n+m=1} \\
\mbox{}\textcolor{Blue}{tem}\ i1\ \textcolor{Purple}{0}\ \textcolor{Purple}{0}\ \textcolor{Purple}{1}\ \textcolor{Purple}{0}\ \ \ \ \ \ \ \ \ \ \ \ \ \textcolor{Gray}{\%\ laser\ beam\ is\ a\ mix\ of\ u\_00\ and\ u\_10\ } \\
\mbox{}\textcolor{Blue}{tem}\ i1\ \textcolor{Purple}{1}\ \textcolor{Purple}{0}\ \textcolor{Purple}{1}\ \textcolor{Purple}{0} \\
\mbox{}\textbf{\textcolor{RoyalBlue}{s}}\ s1\ \textcolor{Purple}{1}\ n1\ n2\ \ \ \ \ \ \ \ \ \ \ \ \ \ \ \textcolor{Gray}{\%\ a\ space\ of\ 1m\ length} \\
\mbox{}\textcolor{Gray}{\%\ triangular\ mode\ cleaner\ cavity} \\
\mbox{}\textbf{\textcolor{RoyalBlue}{bs}}\ bs1\ \textcolor{BrickRed}{.}\textcolor{Purple}{9}\ \textcolor{BrickRed}{.}\textcolor{Purple}{1}\ \textcolor{Purple}{0}\ \textcolor{Purple}{0}\ n2\ nrefl\ n3\ n4\ \ \ \ \ \textcolor{Gray}{\%\ input\ mirror} \\
\mbox{}\textbf{\textcolor{RoyalBlue}{s}}\ sc1\ \textcolor{Purple}{2}\ n3\ n5\ \ \ \ \ \ \ \ \ \ \ \ \ \ \ \ \ \ \ \ \ \ \ \textcolor{Gray}{\%\ distance\ between\ b1\ and\ bs2} \\
\mbox{}\textbf{\textcolor{RoyalBlue}{bs}}\ bs2\ \textcolor{BrickRed}{.}\textcolor{Purple}{9}\ \textcolor{BrickRed}{.}\textcolor{Purple}{1}\ \textcolor{Purple}{0}\ \textcolor{Purple}{0}\ ntrans\ dump\ \ n5\ n6\ \textcolor{Gray}{\%\ output\ mirror} \\
\mbox{}\textbf{\textcolor{RoyalBlue}{s}}\ sc2\ \textcolor{Purple}{49}\ n4\ n7\ \ \ \ \ \ \ \ \ \ \ \ \ \ \ \ \ \ \ \ \ \ \textcolor{Gray}{\%\ distance\ between\ b1\ and\ bs3} \\
\mbox{}\textbf{\textcolor{RoyalBlue}{s}}\ sc3\ \textcolor{Purple}{49}\ n6\ n8\ \ \ \ \ \ \ \ \ \ \ \ \ \ \ \ \ \ \ \ \ \ \textcolor{Gray}{\%\ distance\ between\ b2\ and\ bs3} \\
\mbox{}\textbf{\textcolor{RoyalBlue}{bs}}\ bs3\ \textcolor{Purple}{1}\ \textcolor{Purple}{0}\ \textcolor{Purple}{0}\ \textcolor{Purple}{0}\ n7\ n8\ dump\ dump\ \ \ \ \ \ \textcolor{Gray}{\%\ end\ mirror} \\
\mbox{}\textcolor{Blue}{attr}\ bs3\ Rc\ \textcolor{Purple}{150}\ \ \ \ \ \ \ \ \ \ \ \ \ \ \ \ \ \ \ \ \ \textcolor{Gray}{\%\ Rc=150m\ for\ bs3} \\
\mbox{}\textbf{\textcolor{RoyalBlue}{cav}}\ cav1\ bs1\ n3\ bs1\ n4\ \ \ \ \ \ \ \ \ \ \ \ \ \ \textcolor{Gray}{\%\ computing\ cavity\ parameters} \\
\mbox{}run1\textcolor{BrickRed}{:}\ beam\ ccd\ ntrans\ \ \ \ \ \ \ \textcolor{Gray}{\%\ beam\ shape\ in\ transmission} \\
\mbox{}run2\textcolor{BrickRed}{:}\ beam\ ccd\ nrefl\ \ \ \ \ \ \ \ \textcolor{Gray}{\%\ beam\ shape\ in\ reflection} \\
\mbox{}\textbf{\textcolor{Red}{xaxis}}\ ccd\ x\ lin\ \textcolor{BrickRed}{-}\textcolor{Purple}{3}\ \textcolor{Purple}{3}\ \textcolor{Purple}{200}\ \ \ \ \textcolor{Gray}{\%\ tuning\ x,y\ axes\ of\ beam\ detector} \\
\mbox{}\textbf{\textcolor{Red}{x2axis}}\ ccd\ y\ lin\ \textcolor{BrickRed}{-}\textcolor{Purple}{3}\ \textcolor{Purple}{3}\ \textcolor{Purple}{200} \\
\mbox{}\textbf{\textcolor{Red}{yaxis}}\ abs\ \ \ \ \ \ \ \ \ \ \ \ \ \ \ \ \ \ \ \textcolor{Gray}{\%\ plotting\ the\ absolute\ intensity\ } \\
\mbox{} \\
\mbox{}

}}

\subsubsection{Misaligned cavity}
\epubtkImage{fexample_misaligned_cavity.png}{%
  \begin{figure}[htb]
    \centerline{\includegraphics[width=0.9\textwidth]{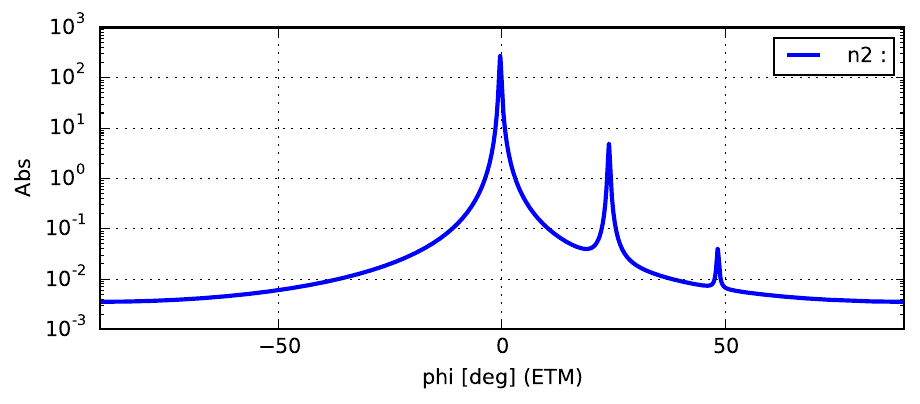}}
    \caption{\Finesse example: Misaligned cavity}
    \label{fig:fexample_misaligned_cavity}
\end{figure}}

\noindent
In this example a misaligned cavity is scanned and the circulating
power is detected.  Additional spikes in the cavity scan indicate the
higher-order modes (order one and two are visible) created by the misalignment.

\vspace{3mm}\noindent
{\small
\textbf{Finesse input file for `Misaligned cavity'}
{\renewcommand{\baselinestretch}{.8}

\nopagebreak
\tt
\noindent
\mbox{}\textbf{\textcolor{RoyalBlue}{l}}\ i1\ \textcolor{Purple}{1}\ \textcolor{Purple}{0}\ nin \\
\mbox{}\textbf{\textcolor{RoyalBlue}{s}}\ sin\ \textcolor{Purple}{1}\ nin\ n1 \\
\mbox{}\textbf{\textcolor{RoyalBlue}{m1}}\ ITM\ \textcolor{Purple}{0.014}\ \textcolor{Purple}{35u}\ \textcolor{Purple}{0}\ n1\ n2 \\
\mbox{}\textbf{\textcolor{RoyalBlue}{s}}\ scav\ \textcolor{Purple}{3994.5}\ n2\ n3 \\
\mbox{}\textbf{\textcolor{RoyalBlue}{m1}}\ ETM\ \textcolor{Purple}{5u}\ \textcolor{Purple}{35u}\ \textcolor{Purple}{0}\ n3\ n4 \\
\mbox{}\textcolor{Blue}{attr}\ ITM\ Rc\ \textcolor{BrickRed}{-}\textcolor{Purple}{1934} \\
\mbox{}\textcolor{Blue}{attr}\ ETM\ Rc\ \textcolor{Purple}{2245} \\
\mbox{}\textcolor{Blue}{attr}\ ETM\ xbeta\ \textcolor{Purple}{0.3u} \\
\mbox{}\textbf{\textcolor{RoyalBlue}{cav}}\ FP\ ITM\ n2\ ETM\ n3 \\
\mbox{}\textbf{\textcolor{Red}{maxtem}}\ \textcolor{Purple}{3} \\
\mbox{}\textbf{\textcolor{Red}{phase}}\ \textcolor{Purple}{2}\  \\
\mbox{}\textbf{\textcolor{RoyalBlue}{pd}}\ pdscan\ n2 \\
\mbox{}\textbf{\textcolor{Red}{xaxis}}\ ETM\ phi\ lin\ \textcolor{BrickRed}{-}\textcolor{Purple}{90}\ \textcolor{Purple}{90}\ \textcolor{Purple}{5000} \\
\mbox{}\textbf{\textcolor{Red}{yaxis}}\ log\ abs

}}

\subsubsection{Impact of thermal aberrations}
\epubtkImage{fexample_thermal_cavity.png}{%
  \begin{figure}[htb]
    \centerline{\includegraphics[width=0.9\textwidth]{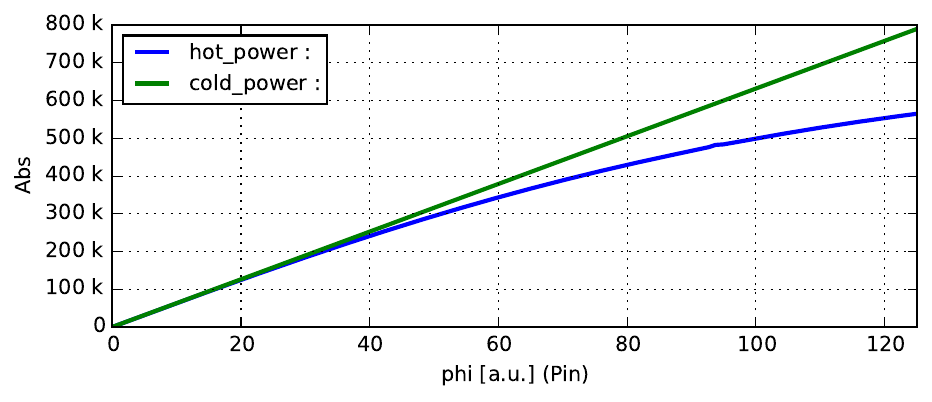}}
    \caption{\Finesse example: Thermal cavity}
    \label{fig:fexample_thermal_cavity}
\end{figure}}

\noindent
This example shows the power circulating in an Advanced LIGO
style arm cavity versus input laser power when we consider the
impact of thermal effects (lensing of the input mirror and change in
curvature of the mirror surfaces).  The mode mismatches these
aberrations cause results in less power coupled into the cavity.

We assume here that the thermal aberrations scale linearly with
power~\cite{vinet09}.  As we tune the incident laser power we also tune
the thermal changes in curvature (dRc1 and dRc2).  Here we use the
change in Rc calculated for an Advanced LIGO cavity operating at high
power (125 W) and then scale the Rcs accordingly.  The curvatures are
combined with the cold state curvatures to give the final state of the
cavity mirrors at a given laser power.  Similarly for the thermal lens
in the ITM we scale the focal length, calculated for high power, with
input power.

Finally we scale the power circulating in an individual arm cavity
(\$Pc) by the gain afforded by the power recycling cavity (45W/W) and
the beam splitter (0.5) to represent the power in an arm of the full
power recycled Michelson configuration.  We also plot the theoretical
linear circulating power, when thermal effects are not considered.
Here the 280.7W is the circulating power in a cavity simulated with no
thermal or higher order mode defects.

\vspace{3mm}\noindent
{\small
\textbf{Finesse input file for `Thermal cavity'}
{\renewcommand{\baselinestretch}{.8}

\nopagebreak
\tt
\noindent
\mbox{}\textbf{\textcolor{RoyalBlue}{l}}\ i1\ \textcolor{Purple}{1}\ \textcolor{Purple}{0}\ nin \\
\mbox{}\textbf{\textcolor{RoyalBlue}{gauss}}\ g1\ i1\ nin\ 12m\ \textcolor{BrickRed}{-}\textcolor{Purple}{1}\textcolor{BrickRed}{.}83522k \\
\mbox{}\textbf{\textcolor{RoyalBlue}{s}}\ sin\ \textcolor{Purple}{1}\ nin\ na \\
\mbox{}\textbf{\textcolor{RoyalBlue}{lens}}\ TL\ inf\ na\ nb \\
\mbox{}\textbf{\textcolor{RoyalBlue}{s}}\ sTL\ \textcolor{Purple}{0}\ nb\ n1 \\
\mbox{}\textbf{\textcolor{RoyalBlue}{m1}}\ ITM\ \textcolor{Purple}{0.014}\ \textcolor{Purple}{35u}\ \textcolor{Purple}{0}\ n1\ n2 \\
\mbox{}\textbf{\textcolor{RoyalBlue}{s}}\ scav\ \textcolor{Purple}{3994.5}\ n2\ n3 \\
\mbox{}\textbf{\textcolor{RoyalBlue}{m1}}\ ETM\ \textcolor{Purple}{5u}\ \textcolor{Purple}{35u}\ \textcolor{Purple}{0}\ n3\ n4 \\
\mbox{}\textbf{\textcolor{RoyalBlue}{cav}}\ arm\ ITM\ n2\ ETM\ n3 \\
\mbox{}\textbf{\textcolor{Red}{maxtem}}\ \textcolor{Purple}{6} \\
\mbox{}\textcolor{Blue}{attr}\ ITM\ Rc\ \textcolor{BrickRed}{-}\textcolor{Purple}{1934} \\
\mbox{}\textcolor{Blue}{attr}\ ETM\ Rc\ \textcolor{Purple}{2245} \\
\mbox{} \\
\mbox{}\textbf{\textcolor{RoyalBlue}{pd}}\ Pcirc\ n2 \\
\mbox{}\textbf{\textcolor{Red}{noplot}}\ Pcirc \\
\mbox{}\textbf{\textcolor{Red}{set}}\textcolor{ForestGreen}{\ Pc}\ Pcirc\ abs \\
\mbox{}\textbf{\textcolor{Red}{func}}\textcolor{ForestGreen}{\ hot$\_$power}\ \textcolor{BrickRed}{=}\ \textcolor{ForestGreen}{\$Pc}\ \textcolor{BrickRed}{*}\ \textcolor{ForestGreen}{\$x1}\ \textcolor{BrickRed}{*}\ \textcolor{Purple}{45}\textcolor{BrickRed}{/}\textcolor{Purple}{2} \\
\mbox{}\textbf{\textcolor{Red}{func}}\textcolor{ForestGreen}{\ cold$\_$power}\ \textcolor{BrickRed}{=}\ \textcolor{ForestGreen}{\$x1}\ \textcolor{BrickRed}{*}\ \textcolor{Purple}{280.7}\ \textcolor{BrickRed}{*}\ \textcolor{Purple}{45}\textcolor{BrickRed}{/}\textcolor{Purple}{2} \\
\mbox{}\textbf{\textcolor{Red}{phase}}\ \textcolor{Purple}{3} \\
\mbox{}variable\ Pin\ \textcolor{Purple}{0} \\
\mbox{}\textbf{\textcolor{Red}{xaxis}}\ Pin\ phi\ lin\ \textcolor{Purple}{0}\ \textcolor{Purple}{125}\ \textcolor{Purple}{500} \\
\mbox{}\textbf{\textcolor{Red}{func}}\textcolor{ForestGreen}{\ dRc1}\ \textcolor{BrickRed}{=}\ \textcolor{Purple}{0}\ \textcolor{BrickRed}{-}\ \textcolor{Purple}{1}\textcolor{BrickRed}{/(}\textcolor{Purple}{1}\textcolor{BrickRed}{/}\textcolor{Purple}{55000}\textcolor{BrickRed}{*}\textcolor{ForestGreen}{\$x1}\textcolor{BrickRed}{/}\textcolor{Purple}{125}\ \textcolor{BrickRed}{+}\ \textcolor{Purple}{1E-20}\textcolor{BrickRed}{)} \\
\mbox{}\textbf{\textcolor{Red}{noplot}}\ dRc1 \\
\mbox{}\textbf{\textcolor{Red}{func}}\textcolor{ForestGreen}{\ Rc1}\ \textcolor{BrickRed}{=}\ \textcolor{Purple}{0}\ \textcolor{BrickRed}{-}\ \textcolor{Purple}{1}\textcolor{BrickRed}{/(}\textcolor{Purple}{1}\textcolor{BrickRed}{/}\textcolor{Purple}{1934}\textcolor{BrickRed}{+}\textcolor{Purple}{1}\textcolor{BrickRed}{/}\textcolor{ForestGreen}{\$dRc1}\textcolor{BrickRed}{)} \\
\mbox{}\textbf{\textcolor{Red}{noplot}}\ Rc1 \\
\mbox{}\textbf{\textcolor{Red}{func}}\textcolor{ForestGreen}{\ dRc2}\ \textcolor{BrickRed}{=}\ \textcolor{Purple}{0}\ \textcolor{BrickRed}{-}\ \textcolor{Purple}{1}\textcolor{BrickRed}{/(}\textcolor{Purple}{1}\textcolor{BrickRed}{/}\textcolor{Purple}{80000}\textcolor{BrickRed}{*}\textcolor{ForestGreen}{\$x1}\textcolor{BrickRed}{/}\textcolor{Purple}{125}\ \textcolor{BrickRed}{+}\ \textcolor{Purple}{1E-20}\textcolor{BrickRed}{)} \\
\mbox{}\textbf{\textcolor{Red}{noplot}}\ dRc2 \\
\mbox{}\textbf{\textcolor{Red}{func}}\textcolor{ForestGreen}{\ Rc2}\ \textcolor{BrickRed}{=}\ \textcolor{Purple}{1}\textcolor{BrickRed}{/(}\textcolor{Purple}{1}\textcolor{BrickRed}{/}\textcolor{Purple}{2245}\textcolor{BrickRed}{+}\textcolor{Purple}{1}\textcolor{BrickRed}{/}\textcolor{ForestGreen}{\$dRc2}\textcolor{BrickRed}{)} \\
\mbox{}\textbf{\textcolor{Red}{noplot}}\ Rc2 \\
\mbox{}\textbf{\textcolor{Red}{func}}\textcolor{ForestGreen}{\ foc}\ \textcolor{BrickRed}{=}\ \textcolor{Purple}{6700}\textcolor{BrickRed}{*}\textcolor{Purple}{125}\textcolor{BrickRed}{/(}\textcolor{ForestGreen}{\$x1}\textcolor{BrickRed}{+}\ \textcolor{Purple}{1E-20}\textcolor{BrickRed}{)} \\
\mbox{}\textbf{\textcolor{Red}{noplot}}\ foc \\
\mbox{}\textbf{\textcolor{Red}{put}}\ TL\ f\ \textcolor{ForestGreen}{\$foc} \\
\mbox{}\textbf{\textcolor{Red}{put}}\ ITM\ Ry\ \textcolor{ForestGreen}{\$Rc1} \\
\mbox{}\textbf{\textcolor{Red}{put}}\ ITM\ Rx\ \textcolor{ForestGreen}{\$Rc1} \\
\mbox{}\textbf{\textcolor{Red}{put}}\ ETM\ Ry\ \textcolor{ForestGreen}{\$Rc2} \\
\mbox{}\textbf{\textcolor{Red}{put}}\ ETM\ Rx\ \textcolor{ForestGreen}{\$Rc2} \\
\mbox{}\textbf{\textcolor{Red}{yaxis}}\ lin\ abs

}}

\newpage


\section{Scattering into higher-order modes}
\label{sec:surface_defects}

Spatial variations in the optics that compose a laser
interferometers, such as distortions of the mirror surfaces,
will change the shape of the circulating beams.
Methods for quantifying such optical imperfections and their effects
are required during the design of an interferometer and for modelling
efforts to characterise the instrument during operation.
In particularly, this is crucial during the design phase in
order to produce, for example, polishing requirements for the mirror
surfaces.   At first glance it is not obvious how such optical defects should
be characterised and we will show that the nature of the problem
determines which approach to use.

Previously we introduced the idea that these distortions can
be described as higher-order Gaussian modes and considered
the impact of such modes on the interferometer performance.
In this section we consider the mechanisms and mathematics
of this scattering into higher-order modes, with particular
emphasis on this process for mirror surface distortions.
We will explore how different types of surface
distortions impact the beam shape and quantify which mirror shapes produce
which higher order modes. Throughout this section we use measured
data from the Advanced LIGO mirrors, kindly provided by
GariLynn Billingsley of the LIGO Laboratory~\cite{Billingsley}.

\subsection{Light scattering in interferometers}

The term `scattering' in interferometers can refer to several
different processes.  Most commonly it refers
to imperfections of high spatial
frequency that scatter light into large angles away from the optical
axis and effectively scattered out of the path of the
beam.
This is a different problem to scattering into higher order modes,
which occurs when the light is scattered back into the
path of the beam (i.e.~small angle scattering).
Light scattered at large angles has the potential to be
re-scattered back into the path of the beam by interactions
with, for example, the walls of the beam tube.  This will
couple new noises into the interferometer, from the beam tubes
into the circulating light field.  Low angle scattering into higher order modes can introduce
noise in other ways, as was discussed in Section~\ref{sec:imperfect}.
The effects of scattered light and mitigation solutions
are an ongoing research topic in the gravitational wave
community~\cite{Vinet97, Yamamoto07, virgo_scatter2010,  vander-hyde15}.

The different scattering processes require different methods for
efficient, accurate modelling. Whereas low-angle scattering can be
modelled using a paraxial approach, either via a description of
higher-order modes or using a Fourier propagation model, high-angle
scattering is outside the paraxial approximation and can require
computationally heavy numerical algorithms for accurate results.

In this review we focus on low angle scattering which manifests itself as
changes in the beam shape.  Of course low and high angle scattering are
not two separate phenomena, and we see the paraxial method
fail at scattering angles greater than $\sim20^{\circ}$.  This region
between high and low-angle scattering can be difficult to model, falling between
the two regimes.
In addition, the finite size of the mirrors in real interferometers prevents the buildup of
very high-order modes as these are wider than the mirrors and
experience significant larger losses.  In this way the finite size of the cavity mirrors
can set a limit for high angle scattering.

The two regimes of scattering provide correspond to some extend to
two commonly used categories for describing spatial surface defects:
\begin{enumerate}
\item Flatness, describing the overall shape of a mirror and its large scale, low spatial frequency features.
These defects impact the shape of the beam within the path, as can be described
with higher spatial modes.
\item Roughness, the high spatial frequency distortions of the mirror which scatter
light out of the path of the beam.
\end{enumerate}

\subsection{Mirror surface defects}\label{sec:mirrormaps}

\begin{figure}[htb]
\centering
	\includegraphics[scale=0.515]{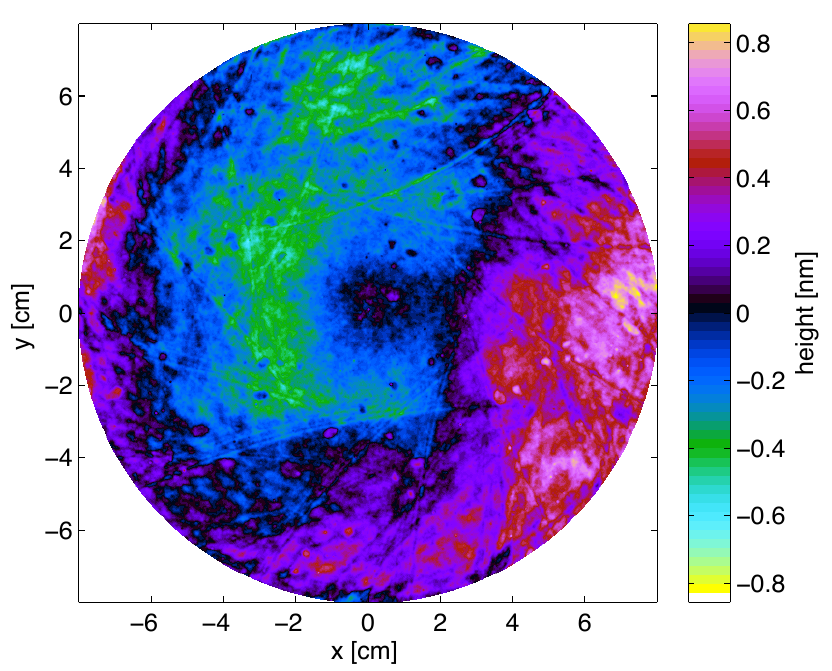} \ \
	\includegraphics[scale=0.515]{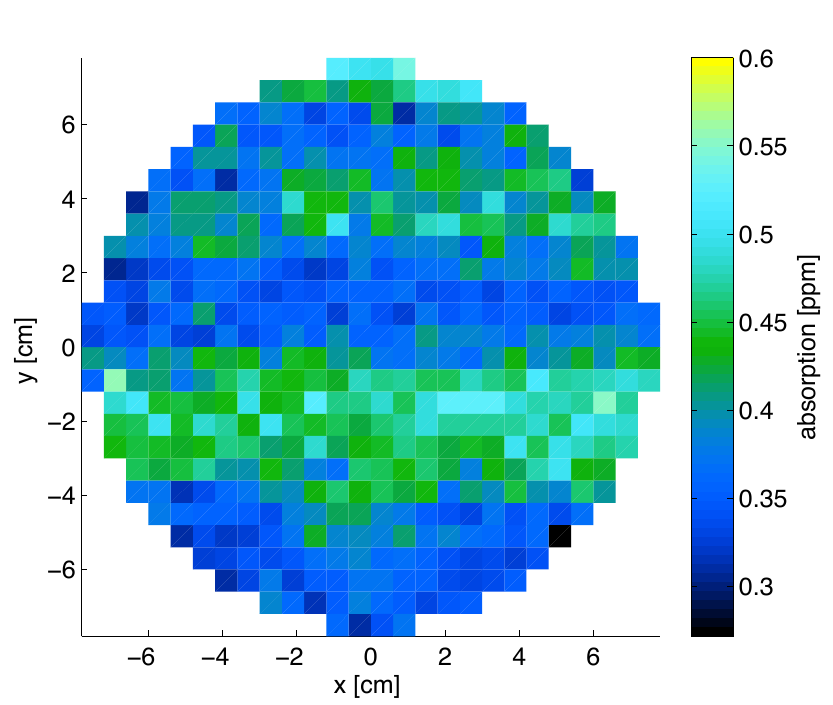}
	\includegraphics[scale=0.515]{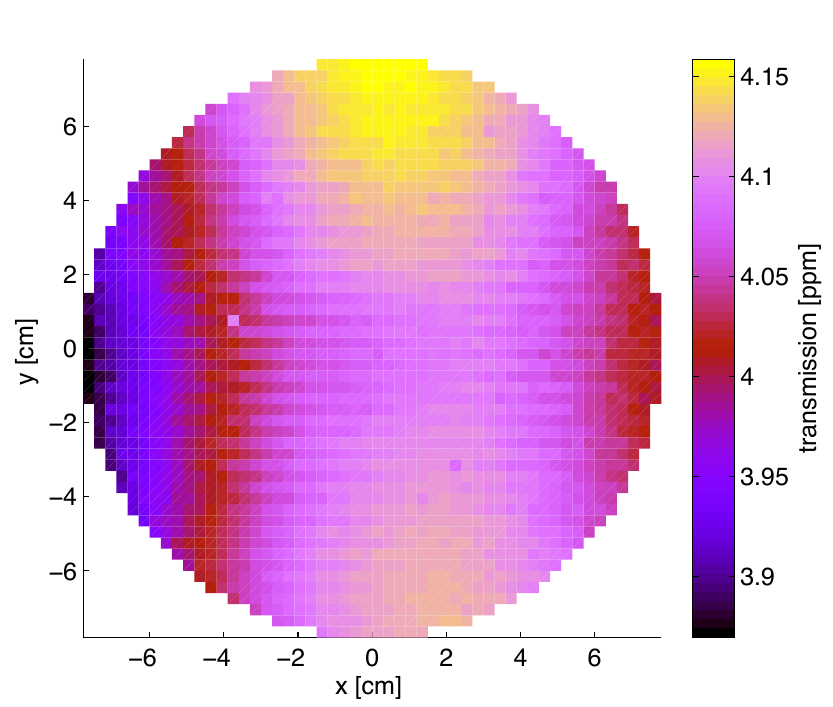} \ \
	\includegraphics[scale=0.515]{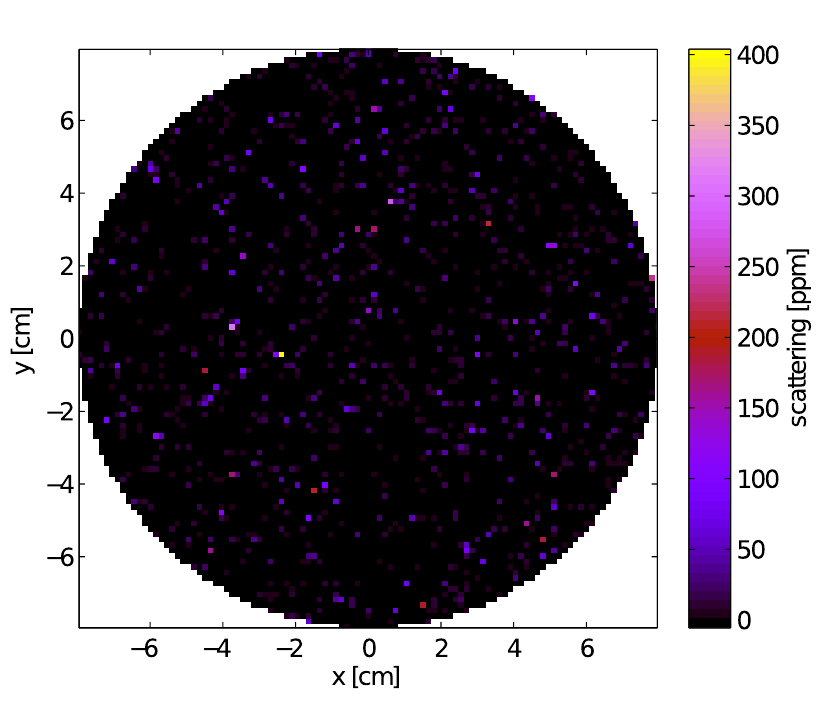}
	\caption[]
		{\emph{Maps} showing different properties of an Advanced LIGO
			mirror, measured across the optic (end test
			mass ETM08)~\cite{Billingsley}.
			{\bf Top left:} The surface height of the polished,
			uncoated substrate of the optic.  {\bf Top right:}
			The absorption of the coated mirror at 1064\,nm.  {\bf Bottom left:}
			Transmission of light through the coated optic at 1064\,nm.
			{\bf Bottom right:} Average scatter from the coated mirror surface at 1064\,nm.
			The surface heigh map was measured by Zygo, with the
			absorption, transmission and scatter maps provided by the
			vendor, Laboratoire des Mat\'{e}riaux Avanc\'{e}s (LMA). 			
			}
\label{fig:maps}
\end{figure}

Realistic mirrors differ from their ideal form in that their optical
and geometric properties are not uniform over the optic.
A possible categorisation of mirror imperfections in interferometers are:
\begin{itemize}
\item misalignment and curvature mismatch, i.e. a mismatch between the
  position, orientation and shape of the optics with respect to the
  laser beam
\item non-uniform mirror phase effects, distorted surfaces and
  substrates will change the phase distribution of a reflected or
  transmitted beam
\item non-uniform amplitude effects: dirty or distorted optics can
  cause non-uniform absorption and reflection
\item apertures created by the finite size of the optics
\end{itemize}
The mirrors can be characterised in detail before installation, see
for example the presentation~\cite{billingsley2012}.
Figure~\ref{fig:maps} shows plots detailing some  measured
properties from an Advanced LIGO mirror~\cite{Billingsley}.
These \emph{mirror maps} (see Section~\ref{sec:characterising}) can be used
in simulations of detectors for a more accurate comparison with
experimental results or for the purposes of producing design requirements
for the mirrors.
In following we discuss the effects of misalignment, mode mismatch
and mirror surface distortions, as these are expected to be the dominant source
of spatial beam distortions.  Similar methods can be used to model non-uniform
absorption or reflection properties.

\subsection{Coupling between higher-order modes}
\label{sec:HOMcoupling}

Any paraxial beam can be expressed as a sum of Hermite or
Laguerre-Gauss modes.   An expansion, in terms of Hermite-Gauss
modes, of the arbitrary field $u(x,y,z)$, can be written as~\cite{siegman}:
\begin{equation}
u(x,y,z) = \sum_{n,m} k_{nm} u_{nm}(x,y,z),
\end{equation}
where $k_{nm}$ refer to coefficients which describe the amplitude and phase of
each Gaussian mode in the field $u(x,y,z)$.
The Hermite-Gauss (and Lagurre-Gauss) modes are orthonormal and
the coefficients can be calculated from an inner product with the relevant HG (or LG)
mode:
\begin{equation}
k_{nm} = \int_{-\infty}^{\infty} \int_{-\infty}^{\infty} u(x,y,z) \ u_{nm}^*(x,y,z) \ \mathrm{d}x \ \mathrm{d}y.
\label{eq:cnm}
\end{equation}
The integral for a generic distortion from an input mode $u_{n'm'}$ to an output
mode $u_{nm}$ due to some distortion to the input beam described
by the complex function $A(x,y)$ is:
\begin{equation}
k_{n,m,n',m'}(q, q', A) = \iint_{-\infty}^{\infty}{u_{nm}(x,y;\,q)\, A(x,y)\, u^{\ast}_{n'm'}(x,y;\,q')}\, dx\,dy.
\end{equation}
In the special case when $A(x,y)$ presently only misalignment
\begin{equation}
A(x,y) \Rightarrow A(x,y,\gamma_x, \gamma_y) = e^{\I 2kz' (\sin^2(\gamma_x/2)+\sin^2(\gamma_y/2))e^{\I k}},
\end{equation}
the
integral can be simplified to the Bayer-Helms coupling equation \ref{eq:tem_conv} as described
in Section~\ref{sec:kmnmn}.
In general, however, the integral cannot be solved analytically.
To model realistic mirror surfaces and how they couple higher-order-modes,
numerical metrology data is used directly in the coupling coefficient integral.
In the generic case $A(x,y)$ represents a complex valued function that
interpolates the measured data. For example, for the coupling in
reflection from a mirror surface  we have
\begin{equation}
k^{\mathrm{refl.}}_{nm,n'm'} = \int_{-\infty}^{\infty} \int_{-\infty}^{\infty} u_{nm}(x,y,z) \exp{(2\I k n_1 z(x,y))} u^*_{n'm'}(x,y,z)
\ \mathrm{d}x \ \mathrm{d}y ,
\label{eq:krefl}
\end{equation}
where $z(x,y)$ describes the distorted surface height and $n_1$ is the index of
refraction for the incident and reflected fields.
Similarly, for transmission through a distorted surface we have
\begin{equation}
k^{\mathrm{trans.}}_{nm,n'm'} = \int_{-\infty}^{\infty} \int_{-\infty}^{\infty} u_{nm}(x,y,z) \exp{(\I k(n_2-n_1)z(x,y))} u^*_{n'm'}(x,y,z) ,
\ \mathrm{d}x \ \mathrm{d}y ,
\label{eq:ktrans}
\end{equation}
where $n_1$ is the index of refraction for the incident beam and $n_2$
is the index for the transmitted beam.
This process of distorting the beam is refereed to as coupling into other
modes, as the action of reflection from a distorted surface creates modes
other than those contained in the incoming beam.

In interferometer simulations such as \Finesse that use modes to
describe the beam shape, a maximum order
of the modes  included $O_{\rm max}$ must be defined for each model.
The coupling between all modes with an order less than the maximum
order of modes is calculated in reflection and transmission of a distorted optic.
This is represented as a coupling coefficient matrix, as described
in Section~\ref{sec:kmnmn}, which computes the transformation of the
incident light field as it interacts with the distorted optic.
These coupling matrices are inserted into the matrix describing
the interferometer behaviour, as described in Section~\ref{sec:coupling_matrices},
giving the higher-order mode content at any position
within the simulated setup.  For well behaved optics, such as those
installed in gravitational wave detectors, we can accurately model
realistic distortions of the beam with a finite number of
modes, as long as we chose a good Gaussian basis (eigenmode)
to work in.  The further from the ideal
eigenmode the more modes you will require to converge to the correct result.
It has been our experience that the best eigenmode is most often that
of the optical cavity, as given by the mirror curvatures and positions.


\subsection{Simulation methods}

Optical simulation tools inherently involve approximations in order
to provide meaningful results within practical computation times.
The most common approximation when modelling laser
interferometers for gravitational wave detectors is assuming a paraxial beam,
allowing the use of a small angle approximation.
There are two distinct simulation methods based on the paraxial
approximation:
\begin{enumerate}
\item Modal decomposition with light fields expressed as linear combinations
of Gaussian modes (solutions to the paraxial wave equation).
\item Fast Fourier Transform (FFT) methods where the light fields are represented
as finite numerical grids which are propagated through an optical system
in the Fourier domain.
\end{enumerate}
The most important aspects of performing simulations with modal
models are: 1) to use the correct Gaussian basis for the higher-order
mode expansion; and 2) to use enough higher-order modes to recreate
distortions of the wavefront. A good choice of Gaussian basis means a
small number of modes should be sufficient to reproduce the
distortions we expect in gravitational wave interferometers. In this
review we make extensive use of the modal simulation \Finesse, see
Appendix~\ref{sec:finesse}.
Other modelling tools for laser interferometers
are based on, or are using, Gaussian
modes~\cite{optickle, MIST}.
The FFT method formed the beginning
of optical modelling in the gravitational wave community and has been
used extensively since~\cite{Vinet1992,
  Bochner03, Day14, OSCAR}.
Both methods contain further approximation, in the addition
to assuming paraxial behaviour.  In the case of modal models this arises
from the finite number of modes. In FFT codes the finite grid size and
resolution restricts the accuracy.  A balance between accuracy and efficiency
often determines the number of modes and grid dimensions used
in these simulations.  Some powerful tools have been developed which
can use modal and FFT based methods internally~\cite{Siesta-paper, E2E}.


\subsection{Mirror surface maps}
\label{sec:characterising}

In order to analyse the effects of mirror surface distortions
we require numerical descriptions of actual mirrors.
In this section we discuss several methods for representing
mirror surface data, with some methods more suited to use
in numerical simulations, whilst others allow an analytic analysis.

A powerful way to implement mirror surface distortions in modal models
is by using a numerical grid representing the surface height of the real
mirrors, known as \emph{mirror maps}.
This is how mirror surface effects are implemented in the interferometer simulation
\Finesse~\cite{finesse_webpage,Freise04}.
The surface data is given as a function over the $x$-$y$ surface of the optic
and can either be measured data from a real mirror or data generated from
mathematical functions, for example, describing the expected thermal distortion
of a mirror.  Mirror map data can be produced for surface height,
reflectivity, transmissivity or absorption over the surface
of the optic.  Unless otherwise notes we use the generic term mirror
map referring to surface height. Figure~\ref{fig:etm08} shows an example of a
mirror map depicting the surface of an end test mass produced for
Advanced LIGO (shown here with any curvature, tilt and offset removed)
to illustrate
the kinds of distortions of the mirrors we can expect.
Note the nano-metre scale of the graph, which is typical for mirrors in
such high-precision interferometers.  We can also see that the central region of the
mirror exhibits less surface height variation.  Again this is expected, as the
requirements on the polishing of the mirror are much more stringent
in the centre of the optic where the beam is most intense.

\epubtkImage{etm08.png}{%
  \begin{figure}[htb]
    \centerline{\includegraphics[scale=0.6]{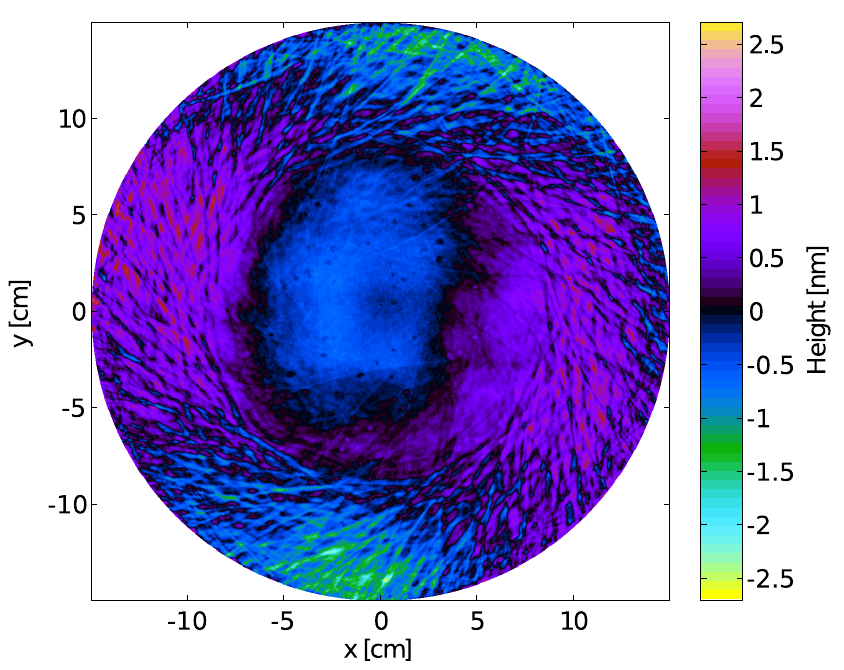}}
    \caption[]{Mirror map describing the measured surface height of an Advanced LIGO
    		end test mass, before coating.  The curvature, tilt and average offset of
		the surface are removed from the data, to clearly show the
    higher-order distortions of the mirror surface.}
    \label{fig:etm08}
\end{figure}}

\begin{figure}[htb]
\centering
	\includegraphics[scale=0.63]{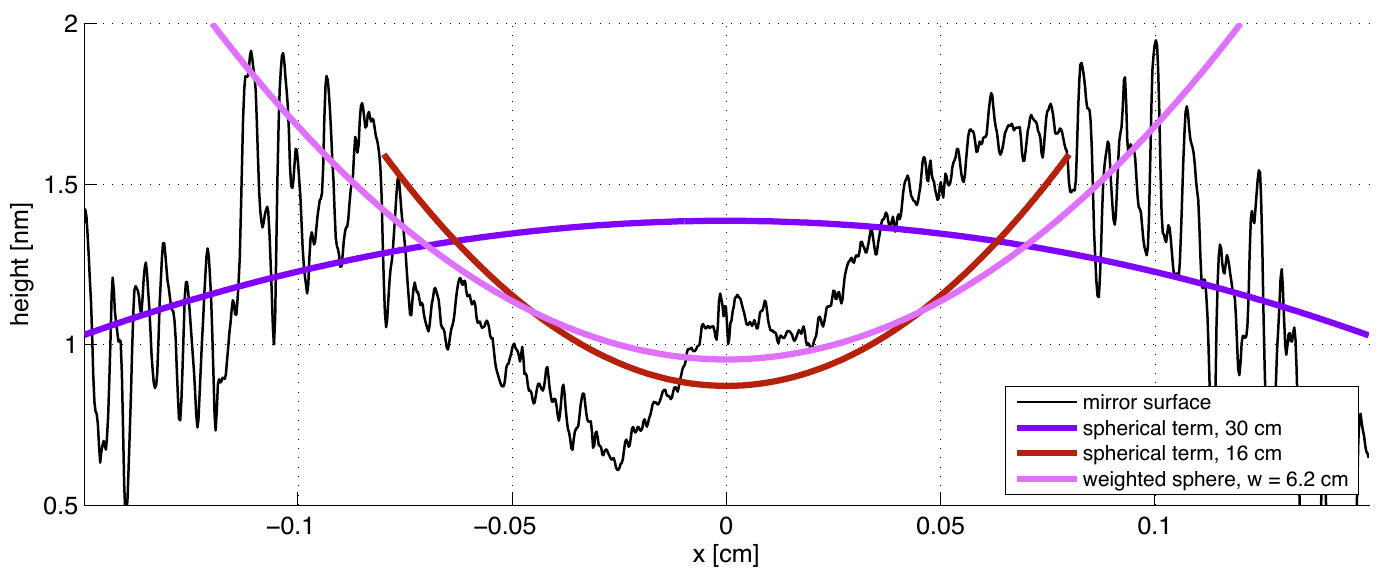}
\caption[Comparing different methods for measuring the curvature of a distorted mirror]
	{Estimates of mirror curvature over different radii and with different weightings
		for a distorted mirror surface.  3 different estimates for the
		curvature of the surface are shown, 1) the spherical
		term over the whole 30\,cm; 2) the spherical term over the central
		16\,cm region; and 3) the Gaussian weighted curvature of the surface, using
		a weighting beam size of $w=6.2$\,cm.
		}
\label{fig:RC_weighting}
\end{figure}

Essentially mirror maps characterise the surfaces by their deviation
from a perfect sphere. The terms for any piston, tilt and
curvature are then expressed by individual numbers (amplitude, angle and radius of
curvature respectively).  This raises the problem of how to optimally
define these low-order features for a distorted surface.  For example, measureing
the curvature of a real mirror is done by fitting a spherical function
to the measured surface data, minimising the difference between
our reference function, the spherical surface $Z_{sphere}$, and our data, the mirror map
$Z_{map}$.
This is represented by minimising the function
\begin{equation}
f = \int_0^{2\pi} \int_0^R  [Z_{map} - Z_{sphere}]^2 \ r \ \mathrm{d}r \ \mathrm{d}\theta ,
\end{equation}
where $R$ is the radius over which we are measuring.
For a typical distorted surface the result can vary greatly with $R$.
We could chose $R$ to be the radius of the mirror, taking a
measure of the curvature over the whole surface.
However, if we consider the part of the mirror over which the Gaussian
beams interact we can measure the curvature the beam `sees' more effectively.
Therefore, it makes sense
to weight our fitting routine using a Gaussian function:
\begin{equation}
f = \int_0^{2\pi} \int_0^R W(r,\theta) [Z_{map} - Z_{sphere}]^2 \ r \ \mathrm{d}r \ \mathrm{d}\theta ,
\end{equation}
where $W(r,\theta)$ is the weighting function, in most cases given byy
the intensity distribution of
the fundamental Gaussian beam and $R$ is the radius of the mirror.
The plots in figure~\ref{fig:RC_weighting} show different
estimates for the curvature of a mirror surface measured over
different regions and using a weighted
fitting function.
There is a significant difference in curvature depending on the area
or weighting used and we must take care to use the correct measurement
for accurate models.
It is especially important for modal simulations that the correct curvature is
measured, as this will determine the cavity eigenmodes,
and the basis of our calculations.  In most cases working in the cavity
eienmodes ensures efficient simulations: accurate results using the
least higher order modes.  The offset and tilt can be measured using
similar methods, specifying the area or weighting with which to measure the defect.


\subsection{Spectrum of surface distortions}

\begin{figure}[htb]
\centering
	\includegraphics[width=\textwidth]{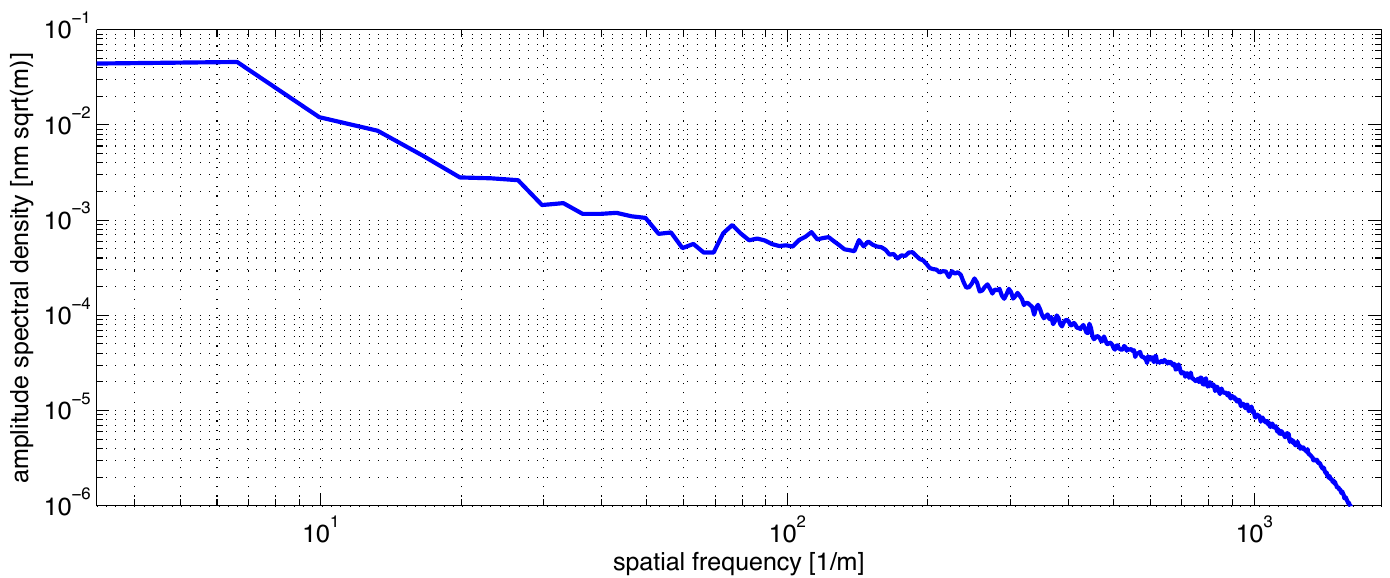}
	\caption[Spectrum of mirror spatial frequencies]
		{Amplitude spectral density of the different spatial frequencies present in the
			Advanced LIGO mirror map ETM08, calculated using
			a 2D FFT and computing a radial average.  The offset, tilt
			and common curvature terms have been removed
			prior to this analysis.}
\label{fig:etm08_spectrum}
\end{figure}

It is desirable to have an analytic description of a mirror surface,
not just numeric data, for example, to categorise specific
types of distortion.
A commonly used method for describing surfaces is to use
a spectrum over spatial frequencies or wavelengths.
The distortion of a surface along the $x$-axis at a specific spatial
frequency, $F$, can be written as:
\begin{equation}
Z(x) = A\cos{\left( 2\pi F x + \phi\right)} ,
\end{equation}
where $A$ is the amplitude of the distortion and $\phi$
is the initial phase of the distortion.  For a purely cosine distortion $\phi=0$ and for
purely sine $\phi=\frac{\pi}{2}$.  A generic distortion
can be described by a sum of sines and cosines
at different frequencies, $F_n$:
\begin{equation}
Z(x) = \sum_n A_n \cos(2\pi F_n + \phi_n) .
\end{equation}
The coefficients and phases can be extracted from a discrete Fourier transform
of measured surface data $z(x)$, calculated using a
Fast Fourier Transform (FFT) algorithm:
\begin{equation}
Z(k) = \sum_{n=0}^{N-1} z(n) \exp{\left(-\frac{\I 2 \pi (k-1)(n-1)}{N}\right)} ,
\end{equation}
where $N$ is the number of elements in $z$, and $n$ and $k$ are integer
indices related  to the spatial coordinate $x$ and $k$ and spatial frequency $F$
respectively.
This method can be adapted to a 2 dimensional surface,
for example, by averaging a 2D Fourier transform into a single
1D amplitude spectrum, similar to the root mean squared ($rms$) for each
spatial frequency.  In figure~\ref{fig:etm08_spectrum} this analysis is shown
for an Advanced LIGO mirror. From such an analysis we can identify
what spatial frequencies are present or dominant in the mirror surfaces distortions.
Low spatial
frequency distortions correspond to the overall mirror shape,
higher spatial frequencies refer to the \emph{roughness}
of the mirror. The amplitude of the lower
spatial frequencies is significantly higher, as expected.
Higher spatial frequencies occur naturally with smaller amplitudes but
are also required to be very small in gravitational wave mirrors
to reduce wide angle scattering out of the beam path.

\subsection{Surface description with Zernike polynomials}
\begin{figure}[htb]

\begin{minipage}[l]{0.4\textwidth}
\begin{tabular}{|c|c|c|}
\hline
$n$ & $m$ & Common name \\
\hline
0 & 0 & Offset \\
1 & $\pm1$ & Tilt in $x$/$y$ direction \\
2 & 0 & Curvature \\
2 & $\pm2$ & Astigmatism \\
3 & $\pm 1$ & Coma along $x$/$y$ axis \\
\hline
\end{tabular}
\end{minipage}
\begin{minipage}[r]{0.45\textwidth}
\begin{minipage}[c]{1\textwidth}
\centering
\includegraphics[scale=0.1]{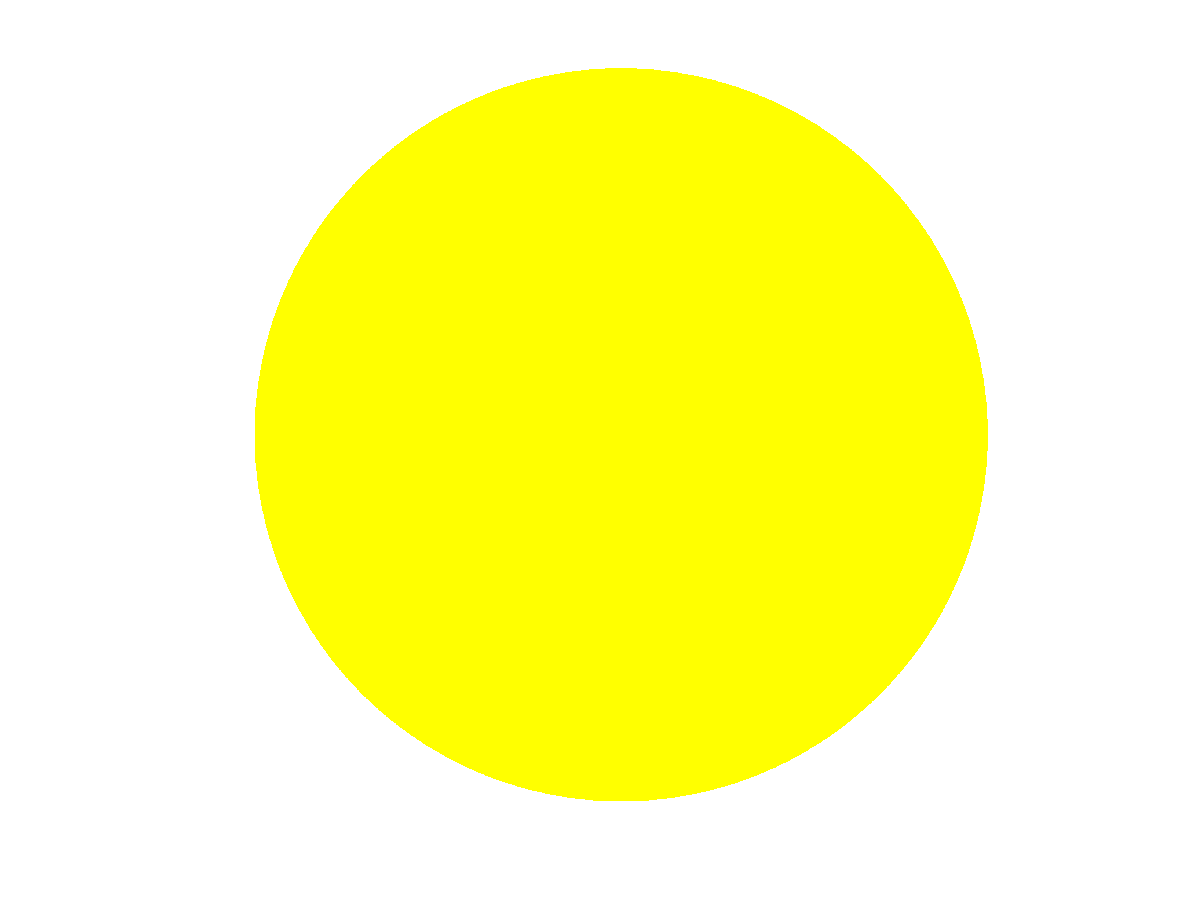}
\\
$n=0$
\end{minipage}
\begin{minipage}[c]{1\textwidth}
\end{minipage}
\begin{minipage}[c]{1\textwidth}
\centering
\includegraphics[scale=0.1]{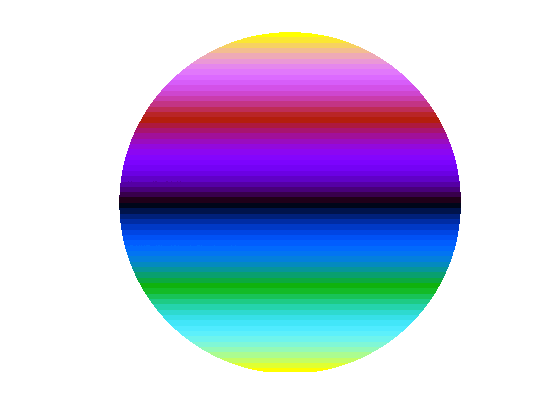}
\includegraphics[scale=0.1]{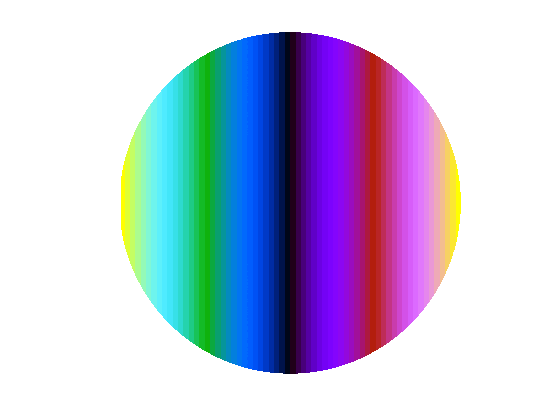}
\\
$n=1$
\end{minipage}
\begin{minipage}[c]{1\textwidth}
\end{minipage}
\begin{minipage}[c]{1\textwidth}
\centering
\includegraphics[scale=0.1]{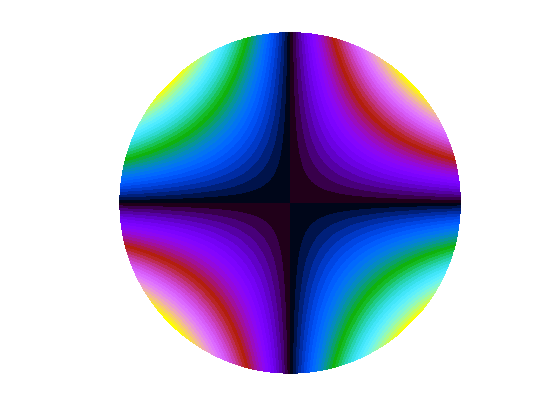}
\includegraphics[scale=0.1]{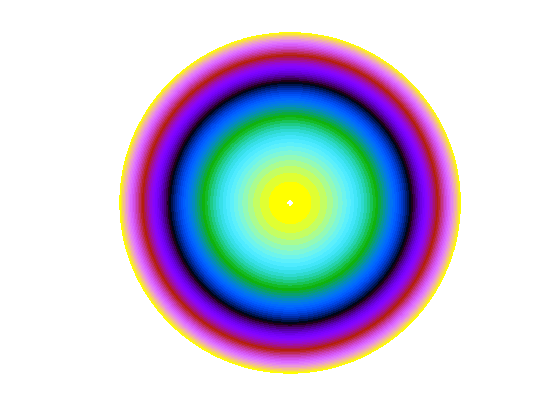}
\includegraphics[scale=0.1]{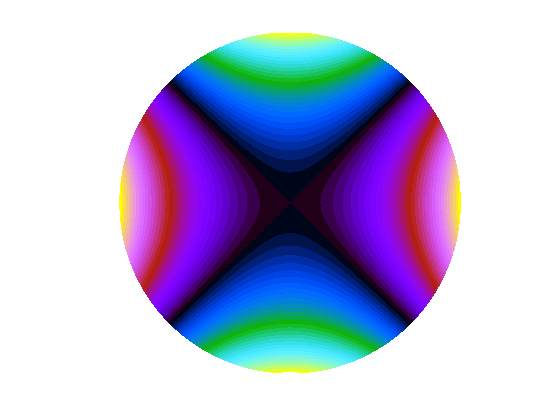}
\\
$n=2$
\end{minipage}
\begin{minipage}[c]{1\textwidth}
\end{minipage}
\begin{minipage}[c]{1\textwidth}
\centering
\includegraphics[scale=0.1]{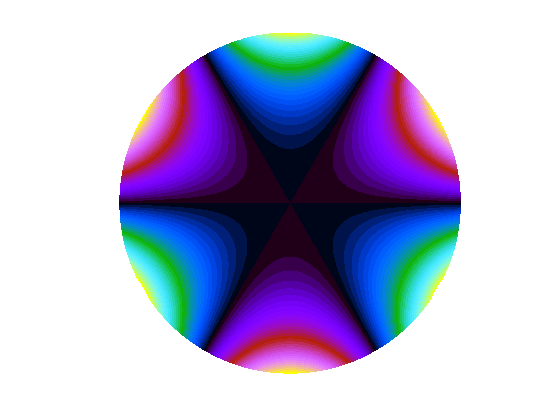}
\includegraphics[scale=0.1]{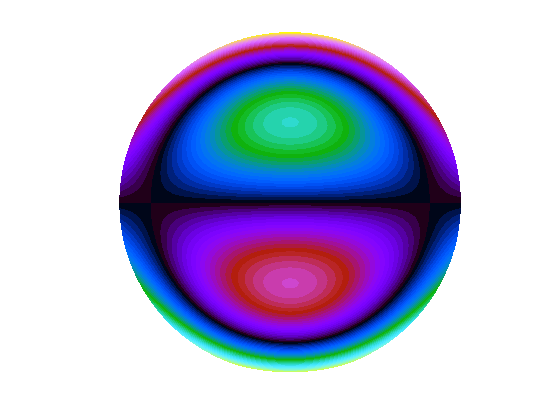}
\includegraphics[scale=0.1]{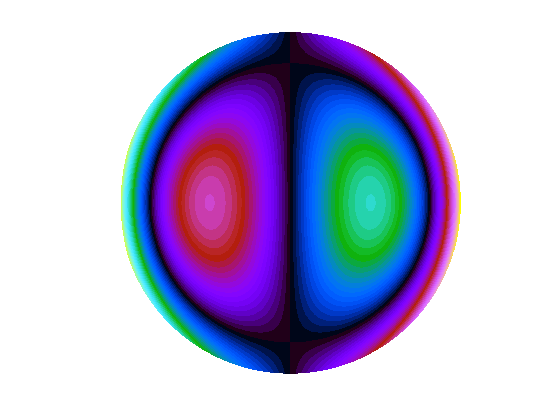}
\includegraphics[scale=0.1]{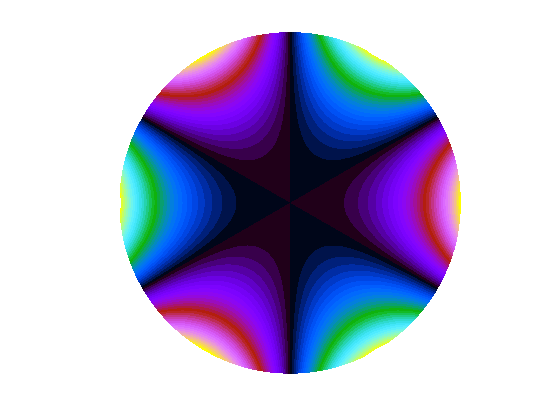}
\\
$n=3$
\end{minipage}
\end{minipage}
\caption{Plots of all the non-zero Zernike polynomials from $n=0$ to
  $n=3$.  They go from odd polynomials with $m=-n$ on the far left to
  even polynomials with $m=n$ on the far right, in steps of 2.  The
  colour scale represents negative surface heights with greens and
  blues, zero with black and positive surface heights with reds and
  purples.}
\label{fig:zern}
\end{figure}

A convenient model for describing the overall shape and low spatial frequency
distortion of a mirror surface are Zernike polynomials.
Zernike polynomials are a complete set of functions which are orthogonal over
the unit disc and defined by radial index, $n$, and azimuthal index, $m$,
with $m \leq n$.  For any index $m$ we have
\begin{equation}
   \begin{split}
       Z_{n}^{+m}(\rho,\phi)= {}& \cos(m\phi)R_{n}^{m}(\rho) \quad \quad \mbox{ the even polynomial} \\
       \\
         Z_{n}^{-m}(\rho,\phi)={}& \sin(m\phi)R_{n}^{m}(\rho) \quad \quad \  \mbox{the odd polynomial}\\
   \end{split}
\end{equation}
with $\rho$ the normalised radius, $\phi$ the azimuthal angle and $R_{n}^{m}(\rho)$
the radial function
\begin{equation}
R_{n}^{m}(\rho)= \left\{
\begin{array}{l l}
   \sum_{h=0}^{\frac{1}{2}(n-m)} \frac{(-1)^{h}(n-h)!}{h! \left(\frac{1}{2}(n+m)-h\right)! \left(\frac{1}{2}(n-m)-h\right)!} \rho^{n-2h} & \quad \mbox{for even $n-m$}\\
  0 & \quad \mbox{for odd $n-m$}\\ \end{array} \right. \
\end{equation}
This gives $n+1$ non-zero Zernike polynomials for each value of $n$ (for $m=0$
the odd polynomial is zero).
Some common optical features are described by the low order Zernike
polynomials, as shown in Figure~\ref{fig:zern}.
The simplest polynomials represent effects we are familiar with:
offset (longitudinal tuning), tilt (misalignment)
and curvature (mode mismatch).  The higher $n$ polynomials represent
higher spatial frequencies.

Odd and even Zernike polynomials describe the same shape for given
$n$ and $m$, with a rotation of $\frac{90^{\circ}}{m}$ with respect
to each other.  A combination of the odd and even polynomials
result in the same shape rotated by a given angle with an
amplitude:
\begin{equation}
A_n^m = \sqrt{(A_n^{-m})^2 +(A_n^{+m})^2} .
\end{equation}
Any surface defined over a disc can be described as
a sum of Zernike polynomials, in the same way any beam shape
can be described as a sum of Gaussian modes,
making these function suitable for the purposes of
describing mirror surface distortions.
Mirror surface data, $Z_{\rm map}$, can
be expressed as
\begin{equation}
Z_{\rm map} = \sum_{n,m} A_n^m Z_n^m ,
\end{equation}
where $A_n^m$ is the amplitude of the relevant Zernike
polynomial in the surface.  In the approach taken here this amplitude
has the same units as the the map data.
We can analyse the surface data contained in mirror maps
by decomposing the surface into Zernike polynomials,
calculating the Zernike coefficients using an
inner product and exploiting the orthogonal nature of the polynomials
\begin{equation}
\int_S Z_{\rm map}\,  \left( N_n^m \right)^2 Z_n^m {\rm d} S = A_n^m \left(N_n^m\right)^2
\int_S Z_n^m  \, Z_n^m {\rm d} S = A_n^m .
\end{equation}
Here $N_n^m$ is a normalisation factor which gives
$\int_S (N_n^m Z_n^m) (N_n^m Z_n^m) {\rm d}S = 1$
and has the form
\begin{equation}
N_n^m = \sqrt{\frac{2(n+1)}{(1+\delta_{m,0})\pi}} .
\end{equation}

\begin{figure}[htb]
\centering
	\includegraphics[scale=0.34]{etm08}
	\includegraphics[scale=0.34]{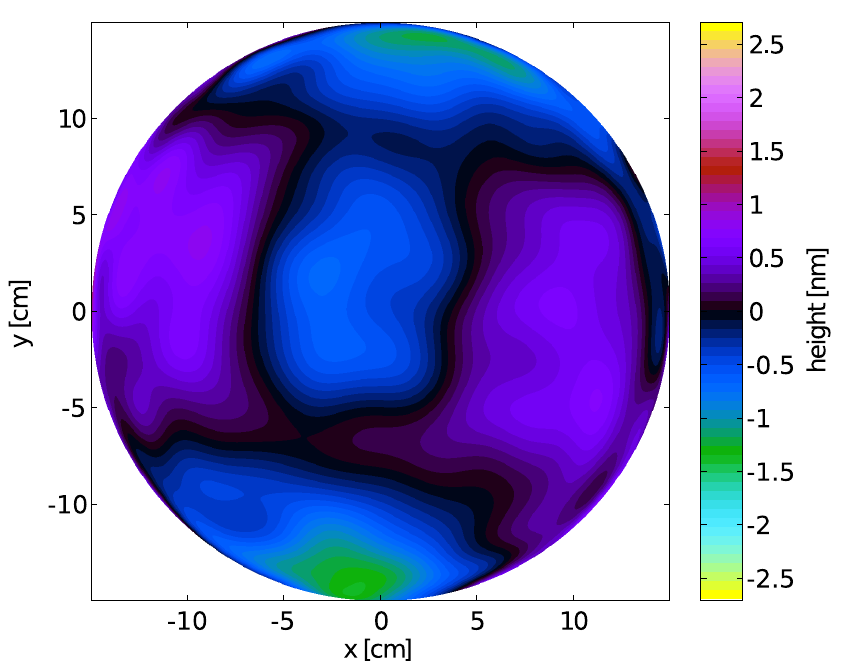}
	\includegraphics[scale=0.34]{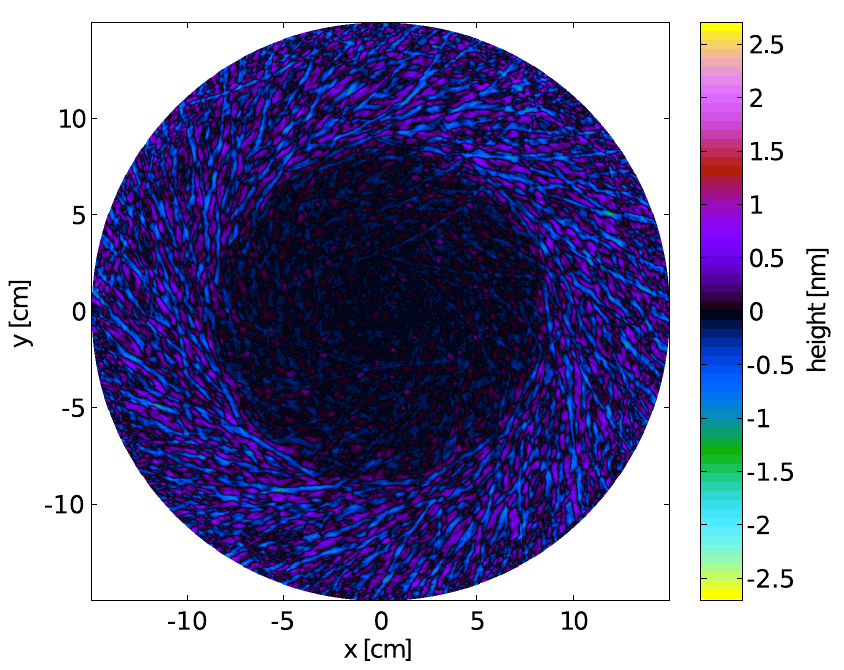}
	\caption
		{Representations of an Advanced LIGO mirror surface.
			{\bf Left:} Original surface map over 30\,cm
			region, with offset, tilt and curvature (Z$_2^0$)
			removed.  {\bf Centre:} Map recreated from Zernike
			polynomials with $n\leq20$, representing the overall
			shape of the mirror.  {\bf Right:} Residual surface after
			the Zernike map is removed, showing the higher spatial
			frequencies.
			}
\label{fig:etm08_zmaps}
\end{figure}

\begin{figure}[htb]
\centering
	\includegraphics[scale=0.63]{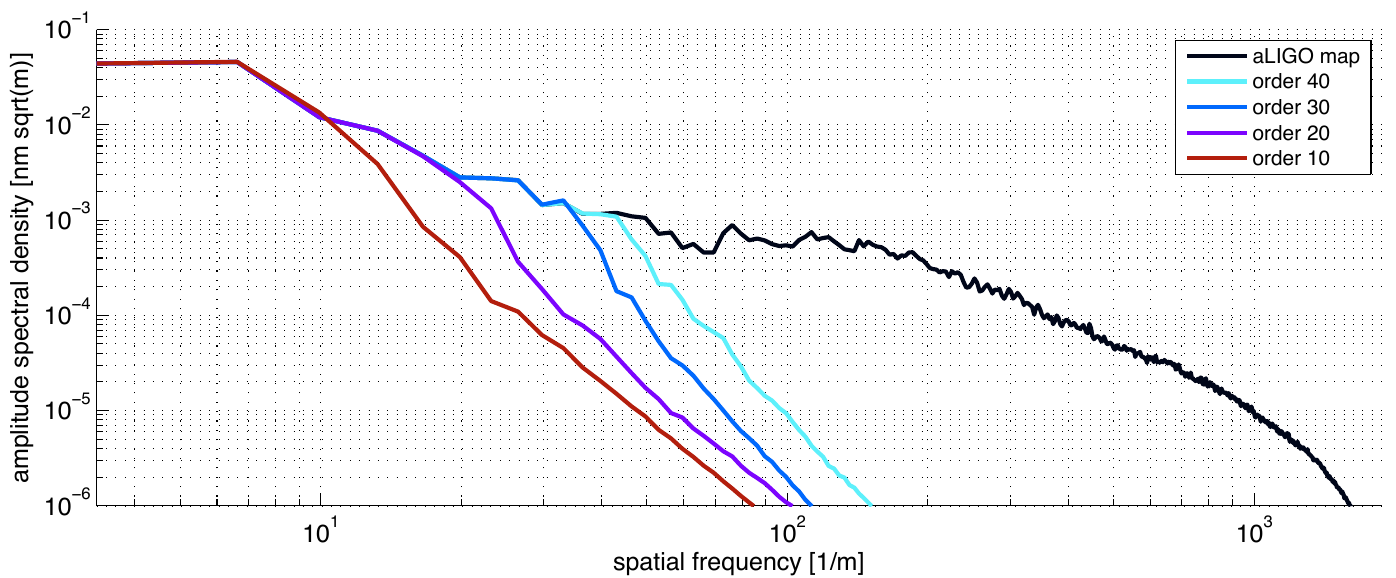}
	\caption[Spatial frequency spectra of Zernike mirror surfaces]
		{Spectra of spatial frequencies in different ETM08 maps.
			The spectrum for the original map is shown, as well as those
			for maps created from Zernike polynomials up to a given order.
			As more polynomials are added to the model the spectra tend
			to the original result.
			}
\label{fig:zmap_spectra}
\end{figure}

Using numerical integration routines real surface data can be
represented as a sum of Zernike polynomials.  This is illustrated
in figure~\ref{fig:etm08_zmaps}, where an Advanced LIGO mirror map
is recreated using low order Zernike polynomials ($n\leq20$).
The overall shape of the Zernike surface looks very similar to the original
map, but lacks the high spatial frequencies.  These are shown in the
residual map
which also illustrates the high polishing requirements for
the central 16\,cm region.
Although high spatial frequencies can be represented by Zernike polynomials
it is often convenient for mirror surface analysis
to consider only the low order Zernike polynomials, with the
rest of the mirror description contained in spectra of spatial
frequencies.  In figure~\ref{fig:zmap_spectra} the spectrum of
an Advanced LIGO mirror map
is shown, as well as the spectra for Zernike maps recreated using
polynomials up to a given order, illustrating how low
order polynomials correspond to low spatial frequencies.  Including
more polynomials in our model tends towards the original map.

\subsection{Mode coupling due to mirror surfaces defects}
In Section~\ref{sec:HOMcoupling} the method for calculating coupling coefficients
numerically, for a generic surface distortion, was discussed.
For design of new laser interferometers we want tools to predict which types of distortions
will couple light into which higher-order Gaussian modes. Such a tool
would allows us to compute specific  requirements for
the distortions in optics for future detectors.
For example, in~\cite{Bond11} the proposal of a new input laser mode, LG$_{33}$,
is analysed in terms of the performance of such a high-order mode with the current
mirrors.  This involves an analysis of the mirror shapes which will couple
between LG$_{33}$ and other modes of the same order, as these modes have the potential
to seriously degrade the performance.  In such a case an analytic approach to coupling,
where the distortions are described by functions such as the Zernike polynomials,
is highly desirable.

\subsubsection*{Scattering into HOMs}

In~\cite{Winkler94} Winkler presents an analytic approach to the
scattering of light in the modal picture,
taking the approach of describing a mirror surface using spatial
frequencies and considering
the x and y spatial components separately:
\begin{equation}
Z(x) = h_0 \cos{\left(\frac{2\pi}{\Lambda}x + \phi \right)} ,
\end{equation}
where $h_0$ is the amplitude of the distortion and $\Lambda$
is the spatial wavelength of the distortion.
For symmetric distortions $\phi = 0$, for anti-symmetric $\phi = \frac{\pi}{2}$.
The coupling from a particular spatial wavelength can be calculated
using Hermite-Gauss modes and splitting them into separate $x$ and $y$
components (see equation ~\ref{eq:knmnm}).  For the coupling
in $x$ we have
\begin{equation}
k_{n,n'} = \int_{-\infty}^{+\infty} U_n \exp{(2\I k Z) } U_{n'}^* \mbox{d}x .
\end{equation}
Winkler took this approach, and an assumption of relatively small
mirror surface distortions, to derive coupling coefficients for an incident
HG$_{00}$ mode.  Here we present this result and expand it for the case
of a generic incident mode.

$U_n$ is the $x$ component of a Hermite-Gauss mode in
the incident beam and $U_{n'}$ is the $x$ component of a
mode in the reflected beam.
\begin{equation}
U_n U_{n'}^* = \frac{1}{w} \sqrt{\frac{2}{\pi}}  \frac{\exp{(\I (n-n') \Psi)}}{\sqrt{2^{n+n'} n! n'!}}
 			H_n\left(\frac{\sqrt{2}x}{w}\right)
      H_{n'}\left(\frac{\sqrt{2}x}{w}\right)
      \exp{\left(-\frac{2x^2}{w^2}\right)} .
\end{equation}
Assuming the distortion of the surface is small compared to the
wavelength of the light, a valid assumption when considering the mirrors
of gravitational wave detectors, we can approximate
\begin{equation}
\exp{(2 \I k Z)} = 1 + 2\I kZ .
\end{equation}
The integral then becomes
\begin{equation}
\begin{split}
	I {}& = C \int_{-\infty}^{\infty} H_n(v) H_{n'}(v) \exp{(-v^2)}
	\cos{\left(\frac{\sqrt{2}\pi w}{\Lambda}v + \phi\right)} \mbox{d} v \\
	 {}& = C \cos{(\phi)} \int_{-\infty}^{\infty} H_n(v) H_{n'}(v) \exp{(-v^2)}
	 \cos{\left(\frac{\sqrt{2}\pi w}{\Lambda}v\right)} \mbox{d}v	\\
	 {}& - C \sin{(\phi)} \int_{-\infty}^{\infty} H_n(v) H_{n'}(v) \exp{(-v^2)}
	 \sin{\left(\frac{\sqrt{2}\pi w}{\Lambda}v\right)} \mbox{d}v		
\end{split}
\end{equation}
where
\begin{equation}
\begin{array}{ccc}
C = \frac{2\I k h_0 }{\sqrt{\pi}} \frac{\exp{(\I (n-n')\Psi)}}{\sqrt{2^{n+n'}n!n'!}} &
\mbox{and} & v = \frac{\sqrt{2}x}{w} .
\end{array}
\end{equation}
These two integrals can be solved using these identities~\cite{Gradstein8511}
\begin{equation}
\label{eq:int_sol}
\begin{split}
 {}& \int_0^{\infty} e^{-x^2} \sin{(bx)} H_p(x) H_{p+2m+1}(x) \mbox{d}x
=
2^{p-1} (-1)^m \sqrt{\pi} \ p! b^{2m+1} \ \exp{\left(-\frac{b^2}{4}\right)} L_p^{2m+1}\left(\frac{b^2}{2}\right) \\
{}& \int_0^{\infty} e^{-x^2} \cos{(bx)} H_p(x) H_{p+2m}(x) \mbox{d}x
=
2^{p-1}(-1)^m\sqrt{\pi} \ p!  b^{2m} \ \exp{\left(-\frac{b^2}{4}\right)} L_p^{2m}\left(\frac{b^2}{2}\right) .
\end{split}
\end{equation}
for $b>0$ and where $L(x)$ refer to the Laguerre polynomials.  The first integral
will refer to coupling where $n-n'$ is odd from asymmetric distortions (sine term)
and the second refers to even $n-n'$ couplings from symmetric distortions (cosine term).
The integral solutions look very similar to the amplitude
of the Laguerre-Gauss modes:
\begin{equation}
|U_{p,l}| =
\frac{1}{W} \sqrt{\frac{2p!}{\pi(|l|+p)!}} \exp{\left(-\frac{r^2}{W^2}\right)}
\left(\frac{\sqrt{2}r}{W}\right)^{|l|} \left|L_p^{|l|}\left(\frac{2r^2}{W^2}\right)\right| .
\end{equation}
Note that $r$ and $W$ are not the radial coordinate and beam spot size as in the
definition of an LG mode, but related to the ratio of the beam size to the
spatial wavelength, $\frac{w}{\Lambda}$.
For a complete solution with the correct phase the sign of the Laguerre-polynomial
should be included, as this disappears when taking the amplitude of the LG mode.
Using these identities we have
\begin{equation}
\begin{split}
I {}& = C 2^p \pi W \sqrt{\frac{p!(|l|+p)!}{2}} (\sqrt{2})^{|l|} \left|U_{p,l}\right| [\cos{(\phi)}
\cos{\left(|l|\tfrac{\pi}{2}\right)} - \sin{(\phi)}\sin{\left(|l|\tfrac{\pi}{2}\right)}] \\
{}& = C 2^p \pi W \sqrt{\frac{p!(|l|+p)!}{2}} (\sqrt{2})^{|l|} \left|U_{p,l}\right|
\cos{\left(\phi + |l|\tfrac{\pi}{2}\right)} ,
\end{split}
\end{equation}
where
\begin{equation}
\begin{array}{ccc}
p = \min{(n,n')} & l = n-n' & \frac{r}{W} = \frac{\pi}{\sqrt{2}}\frac{w}{\Lambda} .
\end{array}
\end{equation}
The factors $\sin{(|l|\frac{\pi}{2})}$ and $\cos{(|l|\frac{\pi}{2})}$ come from
the combination of
the factors $(-1)^{|l|/2}$, $(-1)^{(|l|-1)/2}$ and the fact that the integral
including the sine term is 0 for even $n-n'$ and integral including
the cosine term is 0 for odd $n-n'$.  For simplicity we set $W=\frac{\sqrt{2}}{\pi}$,
which gives $r=\frac{w}{\Lambda}$, the ratio of the beam spot size to
the wavelength of the spatial distortion.  Finally substituting in the values
for $C$ and using $p+|l|=\max{(n,n')}$ and $n+n'=2p+|l|$ gives
\begin{equation}
k^1_{n,n'} = \delta_{n,n'} +  \mbox{sign}\left(L_p^{|l|}(\pi^2 r^2)\right)\frac{2 \I k \ h_0}{\sqrt{\pi}} \exp{(il\Psi)} |U_{p,l}(W=\tfrac{\sqrt{2}}{\pi})| \cos{\left(\phi + |l|\tfrac{\pi}{2}\right)} , 
\end{equation}
adding in the sign of the Lagurre-polynomial to get the correct phase.
The coupling between Hermite-Gauss modes of different orders
is well expressed by this first order approximation, where
the coupling is described with Laguerre-Gauss modes of order $n+n'$.
This is illustrated in figure~\ref{fig:scat_test} where the first order analytical
coupling is compared with the numerical solution of the coupling integral for
$k_{2,6}$ over a range of spatial frequencies.

\epubtkImage{knn_approx_test.png}{%
  \begin{figure}[htb]
    \centerline{\includegraphics[scale=0.63]{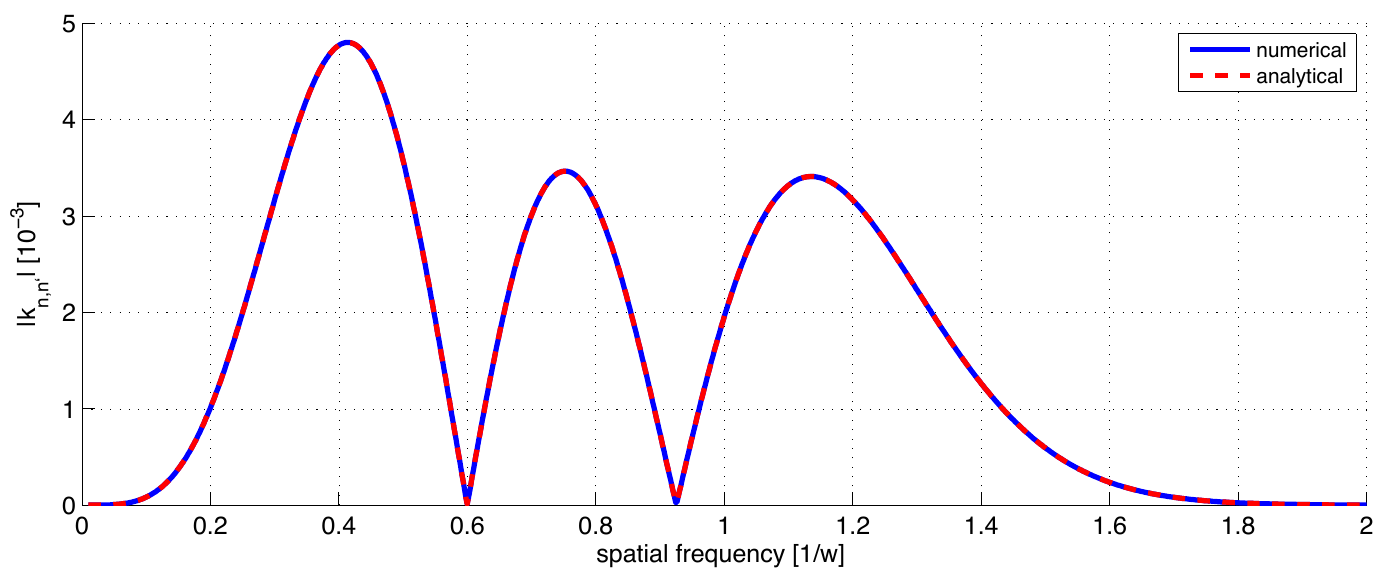}}
    \caption{Comparison of the first order coupling  approximation (analytical)
    with the numerical solution of the coupling between the $x$ components
    of the Hermite-Gauss modes, $U_2$ and $U_6$.}
    \label{fig:scat_test}
\end{figure}}

For coupling back into the same order ($n=n'$) we require up to second order,
derived using the same method as described above.  We have
\begin{equation}
k_{n,n'} \approx k^1_{n,n'} -  k^2h_0^2\delta_{n,n'} -\mbox{sign}\left(L_p^{|l|}( 4\pi^2r^2)\right)
\frac{k^2h_0^2}{\sqrt{\pi}}
|U_{p,l}(W=\tfrac{1}{\sqrt{2}\pi})| \cos{\left(2\phi + |l|\tfrac{\pi}{2}\right)} , 
\end{equation}
where the second order corrections are also described by Laguerre-Gauss
modes of order $n+n'$ but with a beam spot parameter half the size of the that of
the first order coupling.

Figure~\ref{fig:scatHOMs} illustrates the scattering into a range of higher
order modes for different spatial frequency mirror distortions.  Two
examples are given, an incident mode with $n=0$ and an incident
mode with $n=3$.
For low frequency spatial distortions the coupling occurs mostly into low orders.
For higher spatial frequencies, where the wavelength of the spatial distortion
is smaller than the beam spot size, coupling occurs into a vast number of
higher order modes.  In practice the amplitude of spatial distortions
is not constant across the spectrum of spatial wavelengths, as is illustrated here
($h_0=1$\,nm), but decreases with spatial frequency.  In realistic simulations
and in experiments the low order modes dominate, so much so that we can
model gravitational wave interferometers well with a finite number of modes.

 \epubtkImage{scattering_into_HOMs.png}{%
  \begin{figure}[!b]
    \centerline{\includegraphics[scale=0.43]{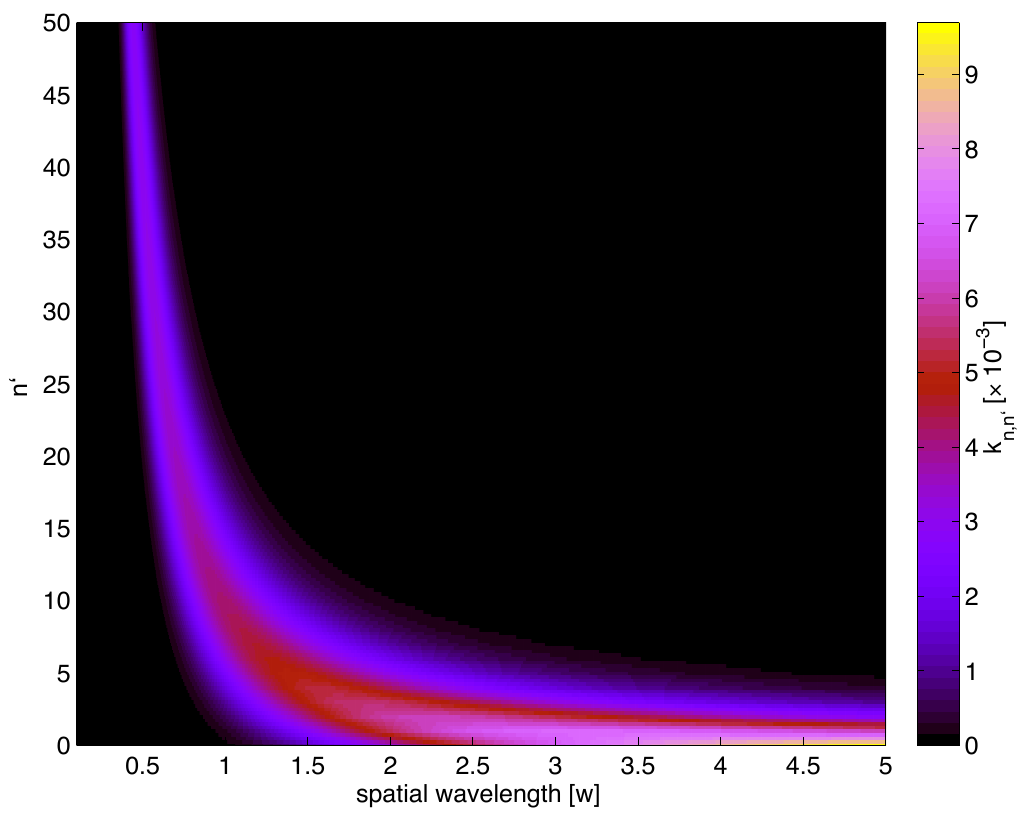}
    		\includegraphics[scale=0.43]{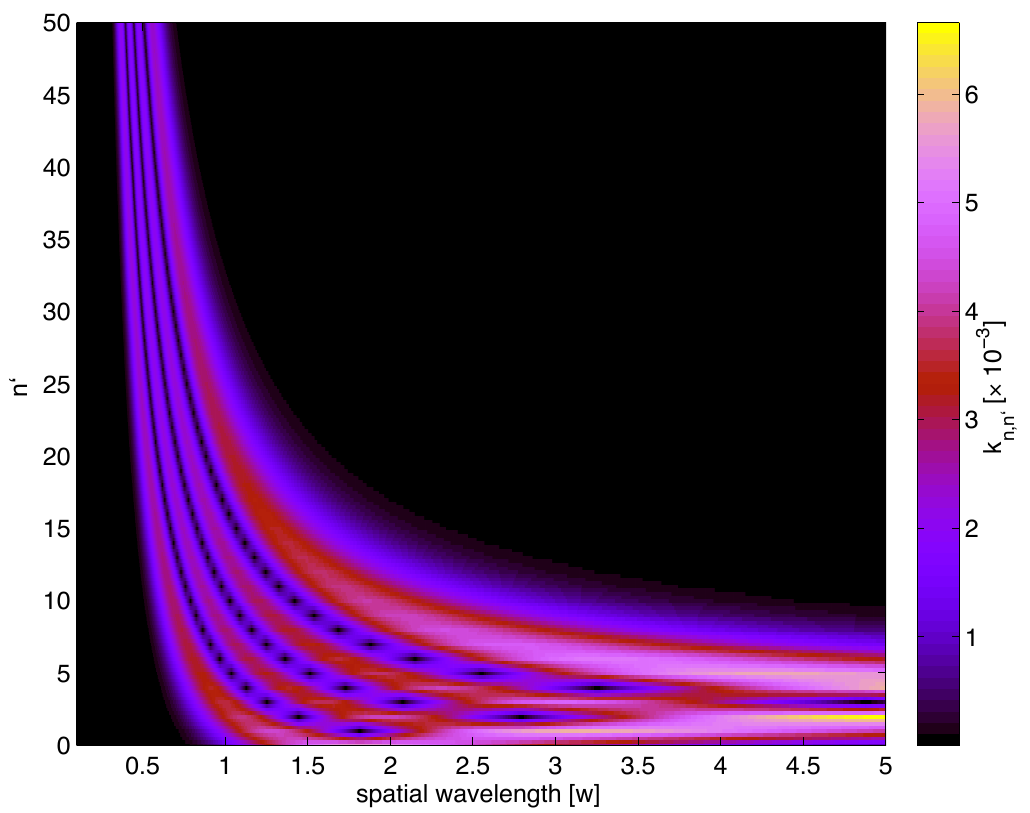}
    }
    \caption{The scattering into higher order modes across a range of
    	spatial distortions (frequencies) for an incident $U_0$ mode (left) and an incident
	$u_3$ mode (right).}
    \label{fig:scatHOMs}
\end{figure}}

Typically individual $k_{n.n'}$ (where $n\neq n'$) are of the order $10^{-3}$
for 1\,nm distortions.   For coupling back into the same mode $k_{n,n}\approx 1$
for small distortions.  We would therefore expect coupling from HG$_{n,m}$ to
HG$_{n,m'}$ or HG$_{n',m}$ would be significantly larger than coupling where
both indices change, as these are of the order $10^{-3}$ rather than $10^{-6}$.

\subsubsection*{Zernike coupling}

Common analysis using spatial frequencies involves taking a statistical approach:
performing numerous simulations using randomly generated realisations of mirror
surfaces
to determine the higher order mode behaviour for a mirror conforming to a particular
spectrum
of spatial frequencies. Such an approach for an LG$_{33}$ mode investigation is
detailed in~\cite{Hong11}.
Here we present an analytic approach which aims to identify the exact
shapes which couple between different modes.

An analysis of mirror surface distortions in terms of
Zernike polynomials is complementary to the approach of
describing the shape of the beam in terms of Gaussian modes.
Both methods deal with the overall shape of the
beam/mirror by expressing them as sums of orthonormal
functions.  Now we have an analytic
representation of mirror surface distortions we
can try and formulate a relation between particular
mirror shapes and their impact on the beam.
The coupling on reflection from a mirror described
by an individual Zernike polynomial, $Z_n^m$, is
\begin{equation}
k^{n,m}_{p,l,p',l'}=\int_{S}U_{p,l}\exp{\left(2\I kZ_{n}^{m}\right)}U^{*}_{p',l'} {\rm d}S ,
\end{equation}
where $U_{p,l}$ is a mode in the incident beam and
$U_{p',l'}$ is a mode in the reflected beam.
The Laguerre-Gauss modes are most suited for this
analysis as they, like the Zernike polynomials,
are naturally described in cylindrical coordinates.
Assuming the distortions are small enough that the
beam parameter remains unchanged, the
product of the two LG fields is
\begin{equation}
    \begin{split}
       U_{p,l}  U^{*}_{p',l'} = {}&  \frac{1}{w^{2}}\frac{2}{\pi}\sqrt{\frac{p!p'!}{(|l|+p)!(|l'|+p')!}} \exp{\left(\I \left(2p+|l|-2p'-|l'|\right)\Psi\right)} \\
          {}& \left(\frac{\sqrt{2}r}{w}\right)^{|l|+|l'|} L^{|l|}_{p}\left(\frac{2r^{2}}{w^{2}}\right) L^{|l'|}_{p'}\left(\frac{2r^{2}}{w^{2}}\right)\exp{\left(-\frac{2r^{2}}{w^{2}}\right)} \exp{\left(\I \phi\left(l-l'\right)\right)} .  \\ 
    \end{split}
 \end{equation}
To simplify the integral we can simplify the expansion
of $\exp{(2\I kZ)}$.  Assuming $2kZ$ is small we can use the approximation
 \begin{equation}
 \exp{(2\I kZ)} \approx 1+2\I kZ .
 \end{equation}
 This is a valid approximation for advanced gravitational wave interferometers,
 where the scale of distortions is not expected to exceed the order of 1\,nm.
$Z=10$\,nm and a wavelength of 1064\,nm gives $2kZ\approx 0.1$.
Making this approximation the coupling coefficients are simplified
\begin{equation}
\begin{split}
k^{n,m}_{p,l,p',l'} {}& = \int_{S}U_{p,l}U^{*}_{p',l'}(1+2\I kZ_{n}^{m}) \\
			  {}& = \delta_{p,p'} \delta_{l,l'} + \int_{0}^{2\pi} \int_{0}^{R} U_{p,l}U^*_{p',l'}(2\I kZ_n^m) \ r  \mbox{d}r \ \mbox{d}\phi ,
\end{split}
\end{equation}
where $\delta_{p,p'}\delta_{l,l'}$ refers to coupling back
into the same mode.
The integral to calculate is over the
surface with $r\rightarrow \infty$, as Laguerre-Gauss
modes are orthogonal over this range.  But since the
integrand is proportional to the Zernike polynomial,
$S$ becomes the Zernike surface as
$Z_{n}^{m}(\frac{r}{R} > 1)= 0$, with $R$ the Zernike radius.

Both Zernike polynomials and Laguerre-Gauss modes
can be separated into there angular and radial parts.  The
angular integrand is
\begin{equation*}
\exp{(\I\phi(l-l'))}
\begin{array}{lr}
	 \cos{(m\phi)} & \mbox{   even } Z_n^m \\
	\sin{(m\phi)}  &  \mbox{   odd } Z_n^m
\end{array}
\end{equation*}
Considering the even Zernike polynomial
 the angular integral becomes
 \begin{equation}
I_{\phi} =  \int_{0}^{2\pi}e^{\I\phi(l-l')} \frac{e^{\I m\phi}+e^{-\I m\phi}}{2} d\phi = \left[ \frac{e^{\I\phi(l-l'+m)}}{2\I (l-l'+m)}+\frac{e^{\I \phi(l-l'-m)}}{2\I (l-l'-m)} \right]^{2\pi}_{0} .
 \end{equation}
As $e^{\I0}=e^{\I N\times2\pi}=1$, for integer $N$,
the integral is equal to 0.  The only combination of
Zernike polynomials and Laguerre-Gauss modes
to give a non-zero result occurs when one of the
exponentials disappears before the integration
takes place.  This occurs for $l-l'+m=0$ or
$l-l'-m=0$.
These same conditions also give the only non-zero
results for the odd Zernike polynomials.
This forms a \emph{coupling condition} between the azimuthal
indices of the shape of the mirror, the Zernike polynomial ($m$),
and the LG modes such a mirror couples between ($l$/$l'$).
This is summarised as
\begin{equation}
m=|l-l'|  .
\label{eq:m_condition}
\end{equation}
Unless this condition is satisfied the coupling between
modes $l$ and $l?$ is 0, to first order.
This condition allows quick identification of the modes
which are created from certain mirror shapes.  It also agrees
with previous work on misalignment and mode-mismatch.
For example, the Zernike polynomial Z$_1^1$ corresponds
to misalignment and couples from the fundamental mode
into LG$_{0,\pm1}$, the order 1 Laguerre-Gauss modes.
Similarly the curvature polynomial, Z$_{2}^{0}$, couples
from LG$_{0,0}$ to the order 2 mode LG$_{0,1}$.

Using this condition we can integrate with respect to $\phi$:
\begin{equation}
I_{\phi} =
\begin{array}{rcl}
0 				&	& m \neq |l-l'| \\
\pi				&	& \mbox{even } Z_n^m \\
\pm \I \pi 			&	& \mbox{odd } Z_n^m \\
2\pi				&	& m = |l-l'| = 0
\end{array}
\end{equation}

The next step is to solve the radial integration.
Making the variable substitution $x = \frac{2r^2}{w^2}$ the
coefficient becomes
\begin{equation}
   \begin{split}
     k^{n,m}_{p,l,p',l'}    = {}& \delta_{p,p'} \delta_{l,l'} + \I k  \frac{I_{\phi}}{\pi} \sqrt{\frac{p!p'!}{(|l|+p)!(|l'|+p')!}}
					\exp{(i \Delta o \Psi)} \\
	{}& \int_0^X x^{\frac{|l|+|l'|}{2}} L_p^{|l|}(x) L_{p'}^{|l'|}(x) \	 \exp{(-x)} Z_n^m \left(\sqrt{\frac{x}{2}}w \right)\mbox{d}x ,
   \end{split}
\end{equation}
with $X = \frac{2R^2}{w^2}$ and $\Delta o = 2p + |l| - 2p' - |l'|$, the difference
in order between the incident and reflected modes.  The integrand is
in the form of $f(x)g(x)$, where $f(x)$ is a polynomial of $x$ whose order
depends on the mode and Zernike indices, and $g(x)=\exp(-x)$.
This integration is solved using the incomplete gamma function,
$\gamma(n,x)=\int_{0}^{x} t^{n-1}e^{-t}dt$~\cite{Gradstein8511},
which for integer $n = 1, 2, \dots$
is
\begin{equation}
\gamma(n,x) = (n-1)! \left[ 1 - e^{-x} \sum_{m=0}^{n-1}\frac{x^m}{m!}\right]
\end{equation}
Substituting in this solution to the integral we have the final solution to
this coupling approximation as
\begin{equation}
\label{eq:kplpl}
   \begin{split}
      k_{p,l,p',l'}^{n,m} = {}& \delta_{p,p'} \delta_{l,l'} + A_n^m \I k \frac{I_{\phi}}{\pi} \sqrt{p!p'!(p+|l|)!(p'+|l'|)!}
      \\
         {}&  \times  \sum_{i=0}^{p} \sum_{j=0}^{p'} \sum_{h=0}^{\frac{1}{2}(n-m)}
         \frac{(-1)^{i+j+h}} {(p-i)!(p'-j)!(|l|+i)!(|l'|+j)! i! j!}  \frac{1}{X^{\frac{1}{2}(n-2h)}}  \\
         {}& \times \frac{(n-h)!}{\left(\frac{1}{2}(n+m)-h\right)! \left(\frac{1}{2}(n-m)-h\right)! h!}
         \ \gamma (i+j-h+\frac{1}{2}(|l|+|l'|+n)+1,X) \  . \\
   \end{split}
\end{equation}
It is worth noting that the first order direct coupling described
here is proportional to the amplitude of the Zernike polynomial, $A_n^m$.

As with the Winkler scattering approximation detailed above we require
up to second order in the exponential expansion to accurately calculate the
coupling back into the incident mode.  We have:
\begin{equation}
\begin{split}
	k_{p,l,p',l'}^{n,m} {}& = \int_S u_{p,l} \exp{(2\I k Z_n^m)} u^*_{p',l'} \ \mbox{d}S \\
		{}& \approx \delta_{p,p'} \delta_{l,l'} + k^{n,m,1}_{p,l,p',l'} + k^{n,m,2}_{p,l,p',l'},
\end{split}
\end{equation}
where $k^{n,m,1}_{p,l,p',l'}$ is the first order coupling as given by equation~\ref{eq:kplpl}
and $k^{n,m,2}_{p,l,p',l'}$ is the second order coupling, given by:
\begin{equation}
k^{n,m,2}_{p,l,p',l'} = \int_{S} u_{p,l} \ u^*_{p',l'} (-2k^2(Z_n^m)^2) \ \mbox{d}S .
\end{equation}
As with the first order coupling we can split the integration into the radial and
angular parts.  The angular integration is:
\begin{equation}
I_{\phi} = \int_0^{2\pi}
\begin{smallmatrix}
 \cos^2{(m\phi)}    \\
 \sin^2{(m\phi)}
\end{smallmatrix}
\exp{(\I \phi(l-l'))} \ \mathrm{d}\phi .
\end{equation}
Taking the even Zernike polynomial we have:
\begin{equation}
\int_0^{2\pi} \frac{1}{4}\left(e^{\I m\phi} + e^{-\I m\phi}\right)^2 e^{\I \phi(l-l')} \ \mathrm{d}\phi
= \int_0^{2\pi} \frac{1}{4} \left[e^{\I \phi(l-l'+2m)}+e^{\I\phi(l-l'-2m)} + 2e^{\I\phi(l-l')}\right]  \ \mathrm{d}\phi .
\end{equation}
As with the angular integration for the first order coupling, a non-zero
value is only achieved when the exponentials disappear before the integration.
We therefore have conditions for non-zero second order coupling:
\begin{equation}
\begin{split}
	{}& 2m = |l-l'| \\
	{}& \mathrm{or} \\
	{}& l = l' .
\end{split}
\end{equation}
Integrating with respect to $\phi$ we have:
\begin{equation}
I_{\phi} =
\begin{array}{rcl}
  0 					&  	& 2m \neq |l-l'|, \ \ l\neq l' \\
  \frac{\displaystyle \pi}{2}  &	& 2m=|l-l'|, \ \ \mathrm{even}\ Z_n^m  \\
  - \frac{\displaystyle \pi}{2} &	& 2m=|l-l'|, \ \ \mathrm{odd}\ Z_n^m \\
  \pi					&	& l=l', \ \ m\neq 0 \\
  2\pi					&	& l=l', \ \ m=0 . \\
\end{array}
\end{equation}

For the radial integration we make the variable substitution $x = \frac{2r^2}{w^2}$
which gives:
\begin{equation}
\begin{split}
k_{p,l,p',l'}^{n,m,2} = {}& -2 k^2  \frac{1}{w^2} \frac{2 I_{\phi}}{\pi} \sqrt{\frac{p! p'!}{(|l|+p)!(|l'|+p')!}} \exp{(\I \Delta o \ \psi)} \\
	{}& \int_0^X x^{\frac{|l|+|l'|}{2}} L_p^{|l|}(x)L_{p'}^{|l'|}(x)\exp{(-x)} \left[ R_n^m\left(\sqrt{\frac{x}{2}}
	\frac{w}{R}\right)\right]^2 \sqrt{\frac{x}{2}}w \ \mathrm{d}x ,
\end{split}
\end{equation}
where $\Delta o$ is the difference in order between the incident and coupled mode
and $X=\frac{2R^2}{w^2}$ is the limit of the exponential.
As with the first order coupling we use the lower incomplete gamma function,
$\gamma(a,x)=\int_0^x t^{a-1}e^{-t}\mathrm{d}t$ to get the final solution:
\begin{equation}
\begin{split}
k_{p,l,p',l'}^{n,m,2} = {}& -\frac{I_{\phi}}{\pi} k^2 A^2 \sqrt{p! p'!(p+|l|)!(p'+|l'|)!} \exp{(\I \Delta o \ \psi)} \\
	{}&	\sum_{i=0}^p \sum_{j=0}^{p'} \sum_{h=0}^{\frac{1}{2}(n-m)} \sum_{g=0}^{\frac{1}{2}(n-m)}
		\frac{(-1)^{i+j+h+g} (n-h)! (n-g)! X^{h+g-n}}{(p-i)!(p'-j)!(|l|+i)!(|l'|+j)!i!j!h!g!} \\
	{}&	 \frac{\gamma(i+j+n-h-g+\frac{1}{2}(|l|+|l'|)+1,X)}{(\frac{1}{2}(n+m)-h)!(\frac{1}{2}(n+m)-g)!(\frac{1}{2}(n-m)-h)!(\frac{1}{2}(n-m)-g)!} .
\end{split}
\end{equation}
Using this derivation of the second order term, combined with our previous derivation
of the first order coupling, the amplitude/ power coupled back into the incident
mode can be calculated.
In the left panel figure~\ref{fig:Zcoupling} the power scattered out of an LG$_{00}$ mode incident
on an mode-mismatched mirror is plotted against the relative beam size.  The larger
the beam size the more of the distortion the beam `see' and hence the more
power is scattered into higher order modes.  In the right panel shows the power coupled
from an LG$_{33}$ mode incident on an astigmatic mirror into 2 other order
9 modes, LG$_{41}$ and LG$_{25}$.  In both cases the analytic coupling approximation and numerical results
agree.

 \epubtkImage{zernike_coupling.png}{%
  \begin{figure}[!b]
    \centerline{\includegraphics[scale=0.6]{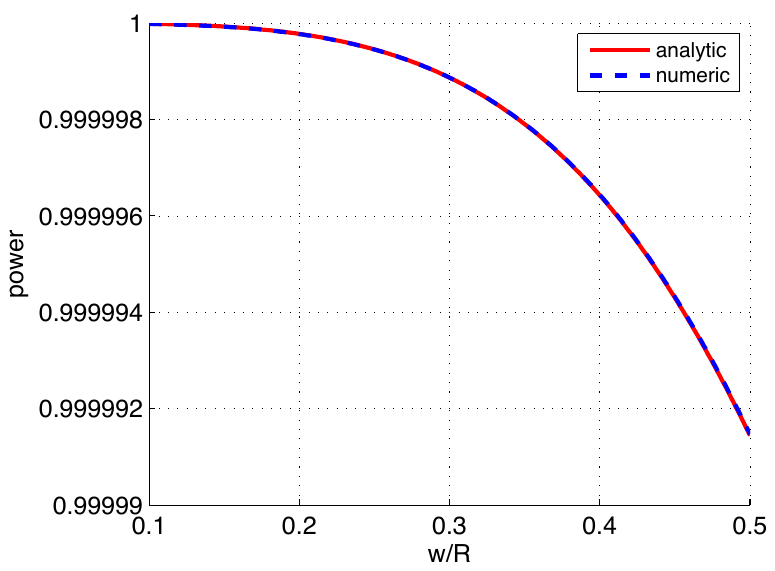}
    		\includegraphics[scale=0.6]{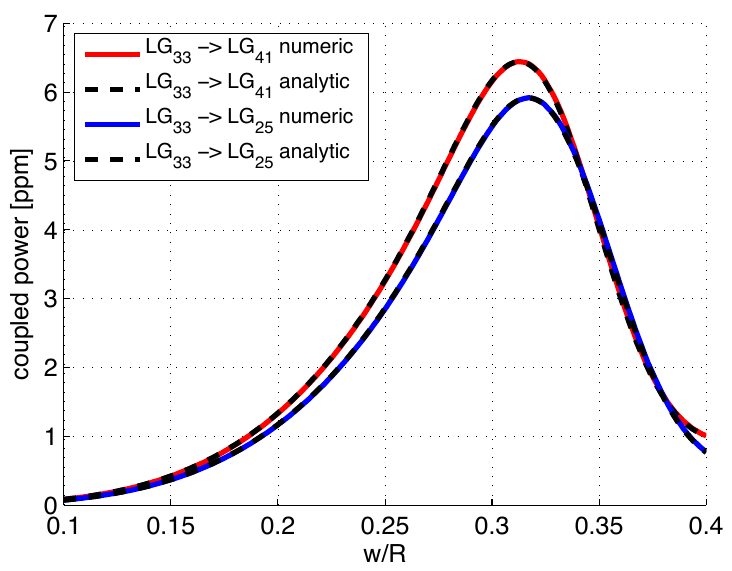}
    }
    \caption{Plots illustrating relationship between power scattered into higher order modes
    		by a curvature mismatched mirror and the
		relative beam size.  {\bf Left:} Power scattered out of
		an incident 00 mode.  {\bf Right:} Power scattered from LG$_{33}$ into 2 other order
		9 modes, LG$_{41}$ and LG$_{25}$.  For both plots both the analytical coupling
		approximation and numerical results are shown.}
    \label{fig:Zcoupling}
\end{figure}}

This coupling approximation has proved particularly useful as it allows
for quick identification of the sources of coupled modes.
For example, this approximation was used in the case of an investigation into the compatibility of
the LG$_{33}$ mode with the current advanced detector mirrors~\cite{Bond11}.
The main cause of worse performance, compared to the fundamental
mode, is coupling into modes of the same order (in the case of LG$_{33}$ order
9).  This approximation means we can quickly identify which mirror distortions
will couple between LG$_{33}$ and other order 9 modes, and hence
which mirror shapes need stricter requirements to produce LG$_{33}$ compatible
mirrors.  Table~\ref{table:LG33} summarise the shapes which couple (to first order) into
each order 9 mode from LG$_{33}$.  Only shapes with specific azimuthal structures
cause coupling between specific modes, and here we see only even azimuthal
indices ($m$) will cause problems for the LG$_{33}$ mode.  Using this approximation
mirror requirements were derived to produce an equivalent performance between
an injected LG$_{33}$ laser mode and a fundamental mode.

\begin{table}[htp]
\begin{center}
\begin{tabular}{l|c|c|c|c|c|c|c|c|c}
mode ($p$,$l$) 	 & 4,1 & 2, 5	& 4, -1 	& 1, 7	& 3, -3	& 0, 9 	& 2, -5	& 1, -7	& 0, -9 \\
\hline
$m$		         & 2 	  & 2		& 4		& 4		& 6		& 6		& 8		& 10		& 12		 \\
\end{tabular}
\end{center}
\caption{Azimuthal index ($m$) of the Zernike shapes required to cause first order coupling
		from an incident LG$_{33}$ mode into each of the other order 9 modes.}
\label{table:LG33}
\end{table}%

\subsection{Efficient coupling matrix computations with multiple distortions}

Evaluating coupling coefficients numerically is a computationally expensive
task if an analytic solution is not known for a particular distortion to the
beam shape. Analytic solutions such as those from
Bayer-Helms~\cite{bayer-helms} for mode-mismatches and misalignments
(see Section~\ref{sec:kmnmn}) provide a fast way to compute the
matrices for such effects. However, if a surface defect or some other
distortion is also applied to a mirror this can require full numerical
integration which is very slow, especially if the simulation varies
the mode-mismatch or alignments. Different distortions can
mathematically be separated into multiple coupling coefficient
matrices, allowing a fast method to solve one which varies often,
like mode-mismatch, and a slow numerical integration which often
need only be performed once.

Consider two general distortions to the beam $A$ and $B$, these
could tilts, apertures, surface defects, etc.
\begin{equation}
k_{MN} = \iint_{-\infty}^{\infty}{U_N(x,y,q_1) A(x,y) B(x,y) U^{\ast}_M(x,y,q_2)} dx\,dy.
\end{equation}
What we want to be able to do is separate the effects as the
coupling caused by the distortion $A$ might be analytically
solveable and variable and whereas $B$ may take along time to
recompute and is constant, therefore we only want to compute it
once. This is the typical scenario when considering simulating
varying mode-mismatches and static surface distortions or
apertures on a mirror for example.

Such a coupling coefficient computation can be represented as
vectors in a Hermite-Gaussian polynomial basis - for convenience
we write for shorthand $|U_N(x,y,q_1)\rangle \rightarrow |N,q_1\rangle$.
\begin{eqnarray}
k_{MN} &=& \langle N,q_1| A(x,y) B(x,y) |M,q_2\rangle , \\
&& \nonumber \\
| N,q \rangle &=&
\begin{bmatrix}
0 \\ \vdots \\ U^\ast_N(x,y,q)  \\ \vdots \\ 0
\end{bmatrix}, \,\,
\langle N,q| =
\begin{bmatrix}
0 & \hdots & U_N(x,y,q) & \hdots & 0
\end{bmatrix}, \\
&& \nonumber \\
\langle N,q_1 | M,q_2 \rangle &=& \iint_{-\infty}^{\infty} U_N(x,y,q_1)U^\ast_M(x,y,q_2) dx\,dy \\  \nonumber \\
\hat{I}_{MN} &=& \sum^\infty_{M}\sum^\infty_{N} | N,q \rangle \langle M,q |   \\ && = \nonumber \sum^\infty_{N,M=N} \left( \iint_{-\infty}^{\infty} U_N(x,y,q)U^*_N(x,y,q) dx\,dy \right) = 1 .\\ \nonumber
\end{eqnarray}
We then define two new vectors and the inner product between them
\begin{eqnarray}
k_{MN} &=& \langle v_N | v_{M} \rangle \\
|v_M,q_2 \rangle &=&
\begin{bmatrix}
0 \\ \vdots \\ U^\ast_M(x,y,q_2)B(x,y)  \\ \vdots \\ 0
\end{bmatrix}, \,\,
\langle v_N,q_1| =
\begin{bmatrix}
0 & \hdots & A(x,y) U^\ast_N(x,y,q_1) & \hdots & 0
\end{bmatrix},
\end{eqnarray}
We then insert the identity matrix, formed by a complete orthonormal basis set of modes,
\begin{eqnarray}
k_{MN} &=& \langle v_N, q_1 | \hat{I} | v_{M},q_2 \rangle \\
 &=& \langle v_N q_1 | \left(\sum^\infty_{L} | L,q_l \rangle\langle L,q_l | \right) | v_{M},q_2 \rangle \\
&=& \sum^\infty_{L} \langle v_N, q_1 | L,q_l \rangle \langle L,q_l | v_{M}, q_2 \rangle \\
&=& \sum^\infty_{L} \langle N, q_1 | A(x,y) | L,q_l \rangle \langle L,q_l | B(x,y) | M,q_2 \rangle \\
\end{eqnarray}
From this we can see that we now have two separate inner products and a sum over the infinite number of Hermite basis functions. In practice
this is limited to a certain number of modes of interest. The last line is identical to a matrix multiplication where each inner product
represents the element of a matrix,
\begin{eqnarray}
k_{MN} &=& \sum_{L} \langle N | A(x,y) | L \rangle \langle L | B(x,y) | M \rangle \\
&=& \sum_{L} \hat{A}_{NL} \hat{B}_{LM} = (\hat{A}\hat{B})_{MN} ,
\end{eqnarray}
where each matrix is now coupling coefficients for each distortion.

The expansion beam parameter $q_l$ can in theory be set to any value;
however the computational requirements can be reduced if it is chosen
sensibly. Remember that a mode-mismatch is present if $q_1 \neq q_2$,
so if $q_l$  is chosen to be either $q_1$ or $q_2$ the mode-mismatch
is present in only one of the matrices. This is beneficial as coupling
coefficient matrices are Hermitian if there is no mode-mismatch. Thus
only one half of the matrix elements need to be computed -- when solving
via numerical integration this can save a great deal of time. There is
also the issue of matrix commutation, $A(x,y)$ and $B(x,y)$ are
interchangeable in the derivation thus it appears $[\hat{A}, \hat{B}] = 0$,
which is a surprising result seeing as the functions can be any arbritary
values. In practice it is found that commutation errors are only present
if the functions are not described using enough higer-order-modes. If
significant amount of information is lost in modes that are not considered,
commutation errors are likely to occur.

\subsection{Clipping by finite apertures}

Another spatial affect present in real interferometers is the finite size of
the optics.  Often in simulations with Gaussian modes or plane waves
there is some intrinsic assumption that the optics are infinite.  In reality
the size of the optics is carefully chosen, optimising between large optics
to contain the power of the incident beams and smaller optics to reduce
the impact of thermal noise.

A finite aperture in the path of a laser beam will produce higher-order modes.
However, in the case of well designed interferometers, such as gravitational
wave detectors, the effect can often be modelled as just a loss of
power in the fundamental mode, so called \emph{clipping loss}.
In such interferometers the size of the optics are chosen such that
they are large enough, compared to the beam
size, that very little power is lost over the edges.
For an LG mode this loss is given by:
\begin{equation}
l_{clip} = 1 - \int_{S}|u_{p,l}|^2 \ \mbox{d} S .
\end{equation}
The integral represents the normalised power reflected by a
mirror with a finite aperture.  For a large mirror  the loss is
effectively 0.  The loss for LG modes is derived as:
\begin{equation}
\begin{split}
l_{clip} ={}& 1-p!(p+|l|)! \sum_{m=0}^p \sum_{n=0}^p \frac{(-1)^{n+m}}{(p-n)!(p-m)!} \\
	       {}& \times	\frac{1}{(|l|+n)!(|l|+m)!n!m!} \gamma(|l|+n+m+1,X) ,
\end{split}
\end{equation}
where $X=\frac{2R^2}{w^2}$ and $\gamma$ is the lower incomplete gamma function.
Generally a clipping loss of the order $1$\,ppm ($10^{-6}$) is desirable.

Generic analytic coupling coefficients describing clipping
at a circular aperture is available in~\cite{VBP2}.

\subsection{Cavity modes of many shapes}

A realistic cavity has an resonate mode which deviates slightly from that of
a pure Gaussian eigenmode: i.e. it is a perfect resonator for a slightly distorted
Gaussian beam, as described by the distorted cavity mirrors.
As an example we consider the case of an astigmatic cavity and the Laguerre-Gauss
modes.  An astigmatic cavity has differing curvatures along the $x$ and $y$ axes.
The Laguerre-Gauss modes, with their cylindrically symmetric properties,
are not eigenmodes of such a system.  In cases with a large
astigmatism \emph{frequency splitting} can be observed,
where an injected Laguerre-Gauss mode is broken down into the eigenmodes
of the cavity.  For an astigmatic cavity these eigenmodes are astigmatic
Hermite-Gauss modes, which can be separated in $x$ and $y$
\begin{equation}
u_{n,m}(x,y,z) = u_n(x,z,R_{C,x})u_m(y,z,R_{C,y}) ,
\end{equation}
with different curvature associated with the $x$ and $y$.
The frequency splitting
phenomenon has its cause in the difference in Gouy phase accumulated in $x$ and $y$
for the different modes. The total Gouy phase is:
\begin{equation}
\varphi_{n,m} = (n+\tfrac{1}{2}) \psi_x(z) + (m+\tfrac{1}{2})\psi_y(z) .
\end{equation}
This results in slightly different resonance frequencies for different
Hermite-Gauss modes of the same order.  Consider for instance the Laguerre-Gauss
mode LG$_{33}$, an order 9 mode.  As described in Section~\ref{sec:LGmodes}
this mode can be described as a sum of order 9 Hermite-Gauss
modes.  In the case where this mode is injected into an astigmatic cavity,
instead of a single clean resonance peak at the order 9 resonance, we see
a spread of resonances corresponding to the different Hermite-Gauss
modes.  This is illustrated in
figures~\ref{fig:LG33_freq_split} and~\ref{fig:LG33_freq_split2}
where the individual higher-order resonances are split across
the resonances of the astigmatic Hermite-Gauss modes: the eigenmodes
of such a cavity.

\epubtkImage{LG33_freq_split_glasgow_sim.png}{%
  \begin{figure}[htb]
    \centerline{\includegraphics[scale=0.63]{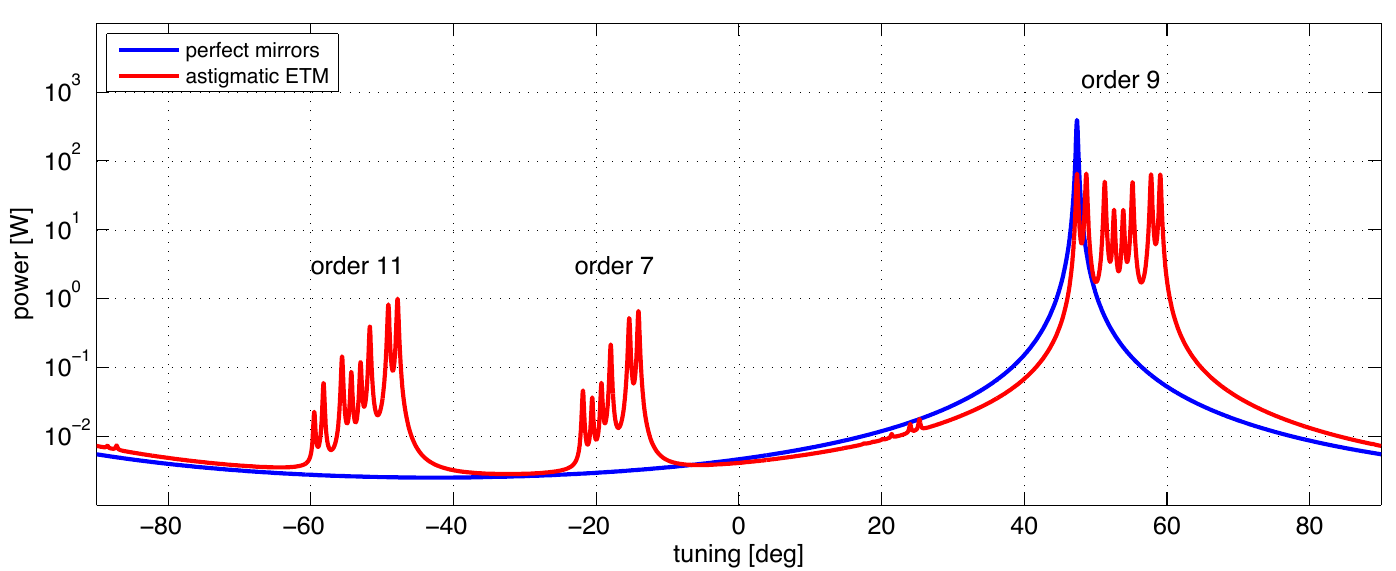}
    }
    \caption{Scan of a cavity injected with an LG$_{33}$ mode.  The results
    		for a cavity with perfectly spherical mirrors show a single resonance
		for the order 9 mode.  For an astigmatic cavity the high order resonances
		are split into the resonances of the astigmatic Hermite-Gauss modes, the
		eigenmodes of the cavity.  Coupling into orders 7 and 11 caused by
		the astigmatism also display this frequency splitting.}
    \label{fig:LG33_freq_split}
\end{figure}}

\epubtkImage{LG33_freq_split_glasgow_sim.png}{%
  \begin{figure}[htb]
    \centerline{\includegraphics[scale=0.63]{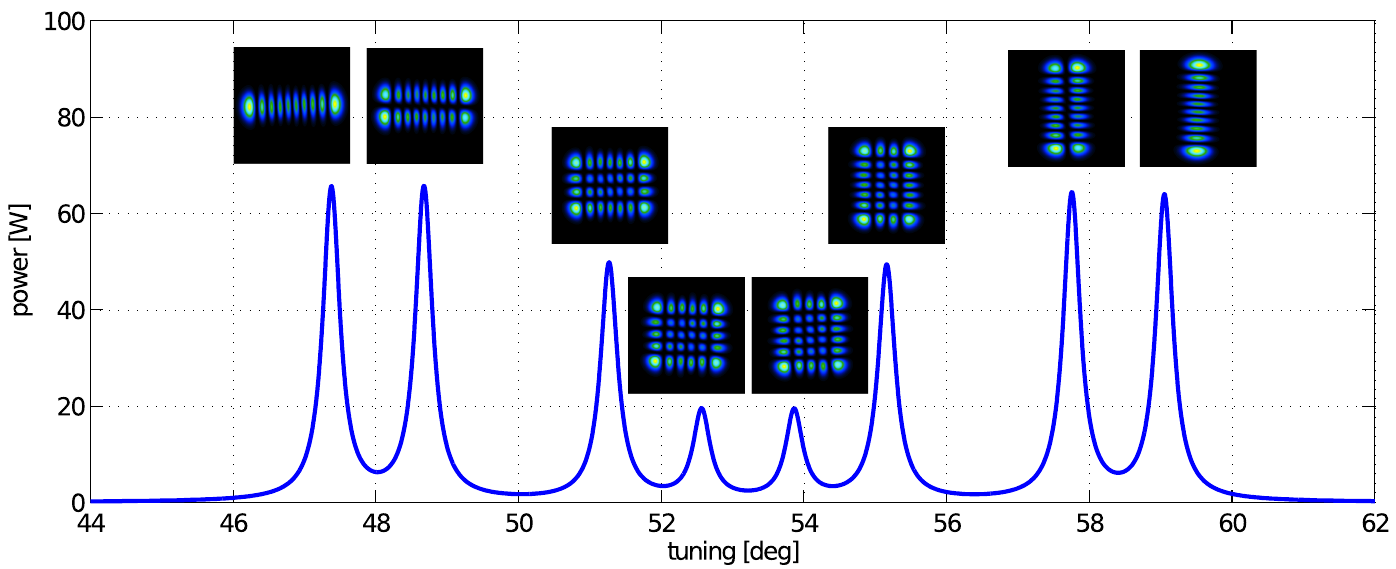}
    }
    \caption{Scan over the order 9 peak in an astigmatic cavity injected
    		with an LG$_{33}$ mode.  The peak is split into the
		resonances of the order 9 Hermite-Gauss modes, the
		eigenmodes of the cavity.}
    \label{fig:LG33_freq_split2}
\end{figure}}

This example is illustrative of a more general effect: the fact that
the resonant modes of a distorted cavity will differ from a perfect Gaussian
mode.  We also note that the finite size of the cavity mirrors
makes the situation more complex, for example it affects the
orthogonality of the cavity eigenmodes~\cite{Siegman1979369}.

\newpage

\section{Acknowledgements}
\label{section:acknowledgements}

We would like to thank our colleagues in the GEO\,600 project and in
the LIGO Scientific Collaboration for many useful discussions over the years.
We'd like to thank GariLynn Billingsley for providing us with data and advice
regarding Advanced LIGO mirror maps.
We thank Miguel Dovale \'Alvarez, Anna Green, Daniel T\"oyr\"a and Haixing
Miao for their time to read many of the new sections in this article
and for the helpful discussions.
AF, CB and DB acknowledge support from the
University of Birmingham. KS acknowledges support from the University
of Glasgow and the Albert Einstein Institute, Hannover. Some of the
illustrations have been prepared using the component library by
Alexander Franzen.
This document has been assigned the LIGO Laboratory document number LIGO-P1500233.

\newpage
\appendix

\section{The Interferometer Simulation \Finesse}
\label{sec:finesse}

Throughout this document we have provided a number
examples using the interferometer simulation
\Finesse~\cite{finesse_webpage, Freise04}. We encourage the reader to obtain \Finesse and to learn its basic
usage by running the included example files (and by making use of its
extensive manual).
The program has been designed
to allow the analysis of arbitrary, user-defined optical setups. In
addition, it is easy to install and use. Therefore \Finesse is well
suited to study basic optical properties.
The \Finesse input files provided in this article
are in most cases very simple and illustrate single concepts in
interferometry. We believe that even a \Finesse novice should be able
to use them as starting points to play and explore freely, for example,
by changing parameters, or by adding further optical components. This
type of `numerical experimentation' can provide insights similar to
real experiments, supplementing the understanding through a
mathematical analysis with experience and intuitions.

\subsubsection*{Free software:}%
\Finesse is a free software package developed and maintained by
some of us (AF and DB): we provide free downloads of binaries
for Linux, Windows and Macintosh computers
online at: \url{http://www.gwoptics.org/finesse/}.
The code is available under the GPLv2 license and does
not require any commercial software (such as Matlab) to
compile and run.

\subsubsection*{Development:}%
\Finesse is a numerical simulation written in the C language;
it is actively developed to fix bugs as they are found and to add
new features required to simulate new interferometric systems. The
recent updates allow modelling of suspended optics that are effected
by radiation pressure of the laser light as well as quantum
effects such as squeezing and vacuum noise.
We are further developing \textsc{Pykat}~\cite{pykat_webpage}, a
Python based toolbox to extend the use of \Finesse, in particular for
automating tasks and data post-processing.

\subsubsection*{Gravitational wave research:}%
The \Finesse development started in 1997 to support the design and
commissioning of the gravitational wave detector GEO\,600 and
\Finesse has remained the standard simulation software of that
project. Since then \Finesse has been developed continuously to meet
the new challenges of the interferometric gravitational wave detectors
world wide such as LIGO, Virgo and KAGRA\footnote{A list of scientific papers
and reports citing \Finesse is provided in~\cite{finesse_webpage}.}.
\Finesse provides a fast and versatile tool that is optimised
for the type of low-noise, steady state laser interferometry
used by gravitational-wave detectors.

\newpage

\section{Advanced LIGO optical layout}
\label{sec:ALOL}

In this review the specific example of Advanced LIGO
has been used to illustrate the operation of an advanced gravitational wave
detector.  In this section the layout and parameters
of Advanced LIGO are presented, with a brief description of the motivation
for some of the key parameters.

 \epubtkImage{aLIGO.png}{%
  \begin{figure}[htb]
    \centerline{\includegraphics*[width=0.85\textwidth, viewport= 38 100 460 480]{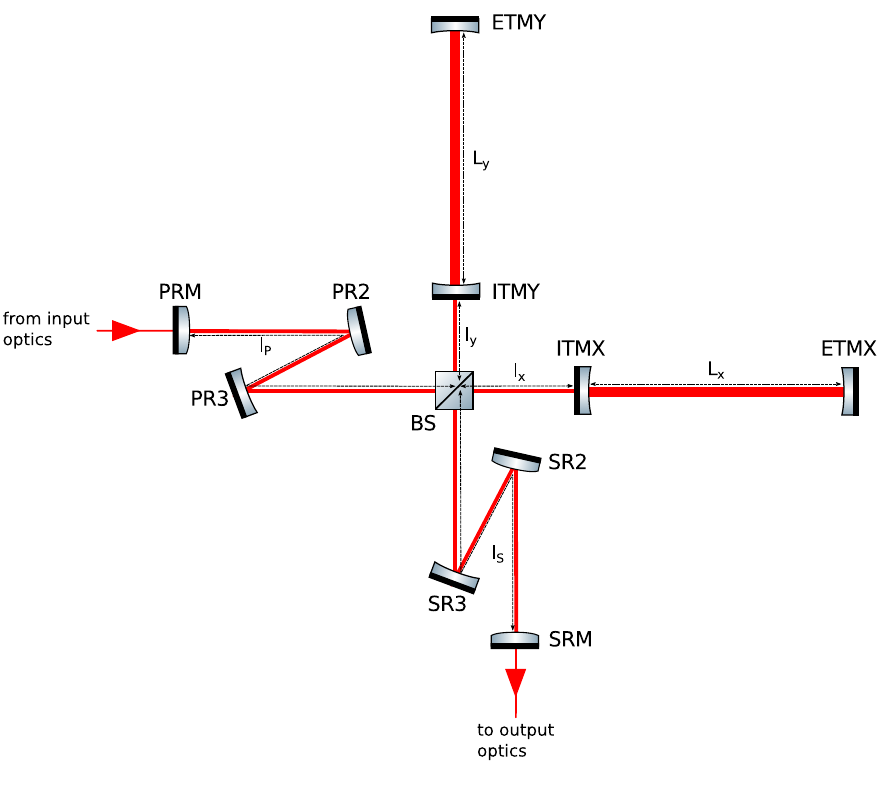}}
    \caption{Optical layout of the central interferometer of Advanced LIGO, a dual recycled Michelson
    		interferometer with Fabry-Perot arm cavities.  The input beam comes from an input
		mode cleaner, which filters the beam of unwanted spatial and frequency
		components.  The output beam is sent to an output mode cleaner before
		it reaches the final detection photodiode.  The arm cavities are characterised
		by the input and end test masses (ITMs and ETMs).  Each recycling cavity consists
		of 3 mirrors, the power/signal recycling mirror (PRM/SRM) and two folding
		mirrors (PR2/SR2 and PR3/SR3).
		The distances between the key optics are labelled:
		the distance between the recycling mirrors and the beam splitter
		($l_P$ and $l_S$), the short Michelson
		arms ($l_x$ and $l_y$) and the arm cavity lengths ($L_x$ and $L_y$).
		}
    \label{fig:aLIGO}
\end{figure}}

\subsection{Advanced LIGO configuration}
\label{sec:aLIGO_params}

Figure~\ref{fig:aLIGO} shows the configuration of the central interferometer
of Advanced LIGO, a dual recycled Michelson interferometer with Fabry-Perot arm
cavities.
The interferometer is characterised by the parameters and performance of the different
subsystems of the interferometer: the arm cavities, power recycling cavity,
signal recycling cavity and the Michelson.


Table~\ref{table:aLIGO_optical_params} summarises the design optical parameters
of the key components of Advanced LIGO: the input and end test masses (ITM and ETM)
and the power and signal recycling mirrors (PRM and SRM).  A 50:50 beam splitter is used and
the design assumes a loss per optic of 37.5\,ppm.  The additional mirrors which
make up the recycling cavities are highly reflective with a transmittance on the order of 1\,ppm.

\begin{table}[htb]
\begin{center}
\begin{tabular}{l|l|}
Parameter	& Value		\\
\hline
ITM $T$			& 1.4\%		\\
ETM $T$			& 5\,ppm		\\
PRM $T$			& 3.0\%		\\
SRM $T$			& 20\%		\\
BS $T$			& 50\%		\\
Arm cavity finesse	& 450		\\
Loss per optic		& 37.5\,ppm
\end{tabular}
\end{center}
\caption{Summary of the key optical parameters of the Advanced LIGO design.
		The transmittance, $T$, of the main mirrors is quoted (in \% or parts-per-million, ppm),
		as well as the expected loss per optic and the finesse of the arm cavities.
		The reflectivity of the mirrors is simple calculated as $R=1-T-L$, where $T$ is
		the transmittance and $L$ is the loss for each mirror~\cite{Arain09}.
		Note that these parameters represent the design at a certain
    period in time and not always the exact parameters of the
    currently operating detectors. For example at the time of writing the
    transmittance of the installed SRM mirror is $\approx 35$\%.}
\label{table:aLIGO_optical_params}
\end{table}%

As well as the optical properties the response of an interferometer is also
characterised by it's geometric parameters, such as the lengths of the arm cavities.
Table~\ref{table:aLIGO_lengths} summarises the key lengths of the Advanced LIGO
design: the lengths of the arm and recycling cavities.
Also included is the Schnupp asymmetry length, a small difference in the short
Michelson length of the $x$ and $y$ arms.

\begin{table}[htb]
\begin{center}
\begin{tabular}{l|l|l}
Length					& Symbol							& Value [m]		\\
\hline
Power recycling cavity 		& $L_{\rm{PRC}} = l_P + \frac{l_x+l_y}{2}$& 57.656			\\
Signal recycling cavity		& $L_{\rm{SRC}} = l_S + \frac{l_x+l_y}{2}$& 56.008			\\
Arm cavity					& $L_x=L_y$						& 3994.5			\\
Schnupp asymmetry 		& $l_{\rm{Sch.}}=l_x-l_y$				& 0.050			\\
\end{tabular}
\end{center}
\caption{Key lengths of the Advanced LIGO design.  The length of the recycling
		cavities ($L_{\rm{PRC}}$/$L_{\rm{SRC}}$) is the average distance
		between the recycling mirror (PRM or SRM)
		and the ITMs (using the average of the short Michelson arms for the distance from the
		BS and ITM, $\frac{l_x+l_y}{2}$).  The arm cavity length is the distance between the highly
		reflective surfaces of the ITM and ETM ($L_x$/$L_y$).
		Finally, the Schnupp asymmetry, $l_{\rm{Sch.}}$, is the difference
		between the short Michelson arms $l_x-l_y$~\cite{Arain09}.  Figure~\ref{fig:aLIGO}
		shows a diagram of the Advanced LIGO layout which illustrates these lengths.
		}
\label{table:aLIGO_lengths}
\end{table}%

\subsection{Choice of parameters}

Above the parameters of Advanced LIGO as
used in models throughout this review are summarised.  In this section
we provide a brief description of the motivation behind the choice of specific
parameters. The majority of the parameters and arguments presented in
this section are taken from~\cite{Abbott2010}.

\subsubsection*{Arm cavity finesse}
The first consideration is the arm cavities.  The length of the arms is
set to $\sim 4$\,km, from the existing LIGO infrastructure.  This gives a free spectral
range of 37.5\,kHz.  The choice of the reflectivity of the mirrors, or the finesse of
the arm cavities, will impact other aspects of the design, so is chosen
with care.  The decision to have $\sim800$\,kW
circulating in the arms (during high power operation) was made early on, in order
to reduce shot noise in the
interferometer, the limiting noise source at most frequencies for initial LIGO.
A combination of parameters determine this intra-cavity power: input laser power,
power recycling gain and arm cavity finesse; the finesse was designed
to be 450 ~\cite{Fritschel2007}.

\subsubsection*{PRM transmission}
\begin{figure}[htb]
\centering
	\includegraphics[scale=0.55]{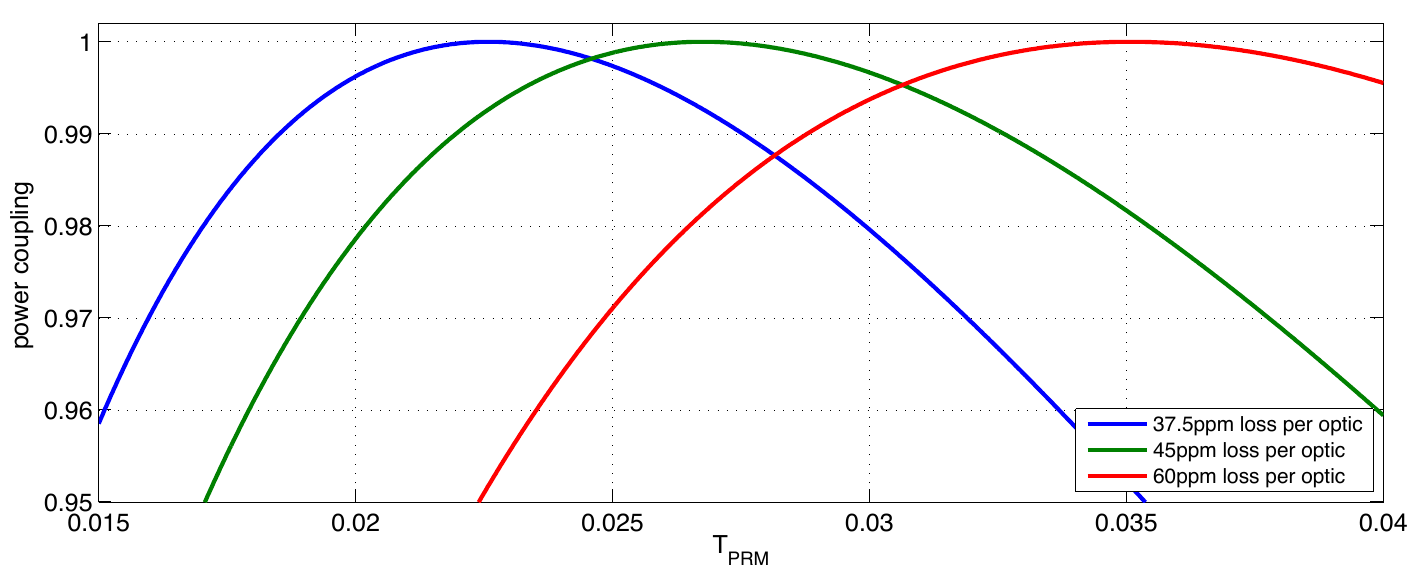}
\caption{Power coupled into an Advanced LIGO interferometer versus
		power recycling mirror transmission.  The power coupled is
		simulated for 3 different values of scattering loss per optic,
		37.5\,ppm (nominal value), 45\,ppm and 60\,ppm.}
\label{fig:Tprm}
\end{figure}

The transmission of the power recycling mirror (PRM) is carefully chosen
to be close to impedance matched with the Fabry-Perot Michelson,
ensuring maximum power coupled into the interferometer and
close to zero reflected.  This requires some knowledge or estimate of the
loss per mirror, which the design states should not be greater than 37.5\,ppm
per optic.  In figure~\ref{fig:Tprm}
the power coupled into the interferometer for different
PRM transmissions is modelled, considering different mirror losses~\cite{Abbott2010}.
A PRM transmission of 3\% was chosen, giving
slight over-coupling for 37.5\,ppm losses but providing robustness against greater losses
and the potential to detect error signals in reflection.
It should be noted that most of the power is `transmitted' through the losses
in the arms, rather than through the end mirrors.

\subsubsection*{Recycling cavity lengths}\label{sec:RC_lengths}
The choice of lengths for the power and signal recycling cavities are closely linked
to the control scheme of Advanced LIGO.  This involves injecting 2 pairs of
control sidebands into the interferometer, at two different frequencies.
The 2 sidebands have (design) frequencies of  $f_1 = 9099471$\,Hz
and $f_2 = 5\times f_1= 45497355$\,Hz, which are chosen to be anti-resonant in the arm
cavities when the carrier is resonant (i.e.~they are reflected by the
arm cavities) and to avoid coinciding with any higher order mode
resonances in the arms.

In the case of Advanced LIGO both sidebands must be resonant in the power
recycling cavity when it is locked to the carrier.
This puts a strict condition on the PRC length:
\begin{equation}
L_{PRC} = \left(N+\frac{1}{2}\right) \frac{c}{2f_1} .
\end{equation}
The factor of $\frac{1}{2}$ is included as the sidebands are $180^{\circ}$
out of phase with the carrier, as the carrier enters the arm cavities whilst
the sidebands do not.
A power recycling cavity length of 57.6557\,m ($N=3$) was chosen
to be compatible with the opto-mechanical layout for a stable recycling
cavity~\cite{Abbott2010}.  The signal recycling cavity length was chosen to be
resonant for $f_2$ but not for $f_1$, i.e.~\cite{Abbott2010}
\begin{equation}
L_{SRC} = M\frac{c}{2f_2} \neq Q\frac{c}{2f_1} ,
\end{equation}
where $M$ and $Q$ are integers.  An SRC length of 56.0084\,m was chosen~\cite{Abbott2010}.

For control of Advanced LIGO one of the sidebands should exit the
Michelson into the signal recycling cavity and eventually the output
port.  However, in a Michelson where the short arms between the
beam splitter and ITMs ($l_x$/$l_y$) are the same length the dark fringe for the
carrier will also be the dark fringe for the sidebands.  In order for
one of the sidebands ($f_2$) to leak into the signal recycling
cavity we need some asymmetry between the two arms, the so-called
\emph{Schnupp asymmetry}, $l_{Sch} = l_x-l_y\neq 0$~\cite{Hild09b}.
In the case of Advanced LIGO a Schnupp asymmetry of
5\,cm was a compromise between maximising the coupling of $f_2$ into
the signal recycling cavity for two different cases: broadband signal recycling
and slightly detuned signal recycling (optimised for neutron
star\,$-$\,neutron star in-spiral (NS\,$-$\,NS)
signals).

\subsubsection*{Mode-matching}
The curvatures of the highly reflective mirror surfaces and distances between
individual mirrors determine the size and shape of the resonating beams.
The individual cavities in the detector need to be well mode matched
to maximise the power build up and avoid high contrast defects.

The eigenmode of the arm cavities is selected to produce large beams at the
ITM (5.3\,cm) and ETM (6.2\,cm) to reduce thermal noise, with slightly smaller beams at the ITM
due to lower thermal noise (fewer coating layers) and
to prevent scattering into the recycling cavities.  The curvatures are also carefully
selected for a specific Gouy phase to avoid higher order modes easily ringing up
in the arms: $R_C=1934$\,m (ITM) and $R_C=2245$\,m (ETM).

In initial LIGO the power recycling cavity was marginally unstable,
enhancing the power in higher order modes in the control
sidebands and causing problems for control~\cite{gretarsson07}.   To avoid this in Advanced
LIGO the interferometer was designed with stable recycling cavities: folded
cavities which do not share the arm cavity eigenmode~\cite{Arain08}, see
section~\ref{sec:recycling_design}.  These cavities consist of 3 mirrors,
the primary recycling mirrors (PRM/SRM) and two additional mirrors
to shape and direct the beam (PR2/SR2 and PR3/SR3).
The curvatures and positions of the mirrors are chosen to ensure good mode
matching into the arm cavities and to achieve a good spacing between
higher order resonances (see section~\ref{sec:recycling_design}).

The mode matching of the beams between the recycling cavities and
arms is complicated by thermal effects, specifically thermal lensing and
the change in mirror curvatures.  Although this will be corrected by
thermal compensation systems~\cite{Willems2009} it was decided to match
the recycling cavities to the arms in the presence of 50\,km
lenses in the ITMs, as expected for an input power of $\sim~12.5$\,W
with coating absorptions of 0.5\,ppm~\cite{Arain09}.  This will potentially
mitigate the use of TCS at low power.  In the end the cavities
were designed for a 50\,km lens inside the substrate (effective 34.5\,km
lens) corresponding to 18\,W input power.

\subsection{Degrees of freedom}

 \epubtkImage{DR-FPMI_fields.png}{%
  \begin{figure}[htb]
    \centerline{\includegraphics[width=0.9\textwidth]{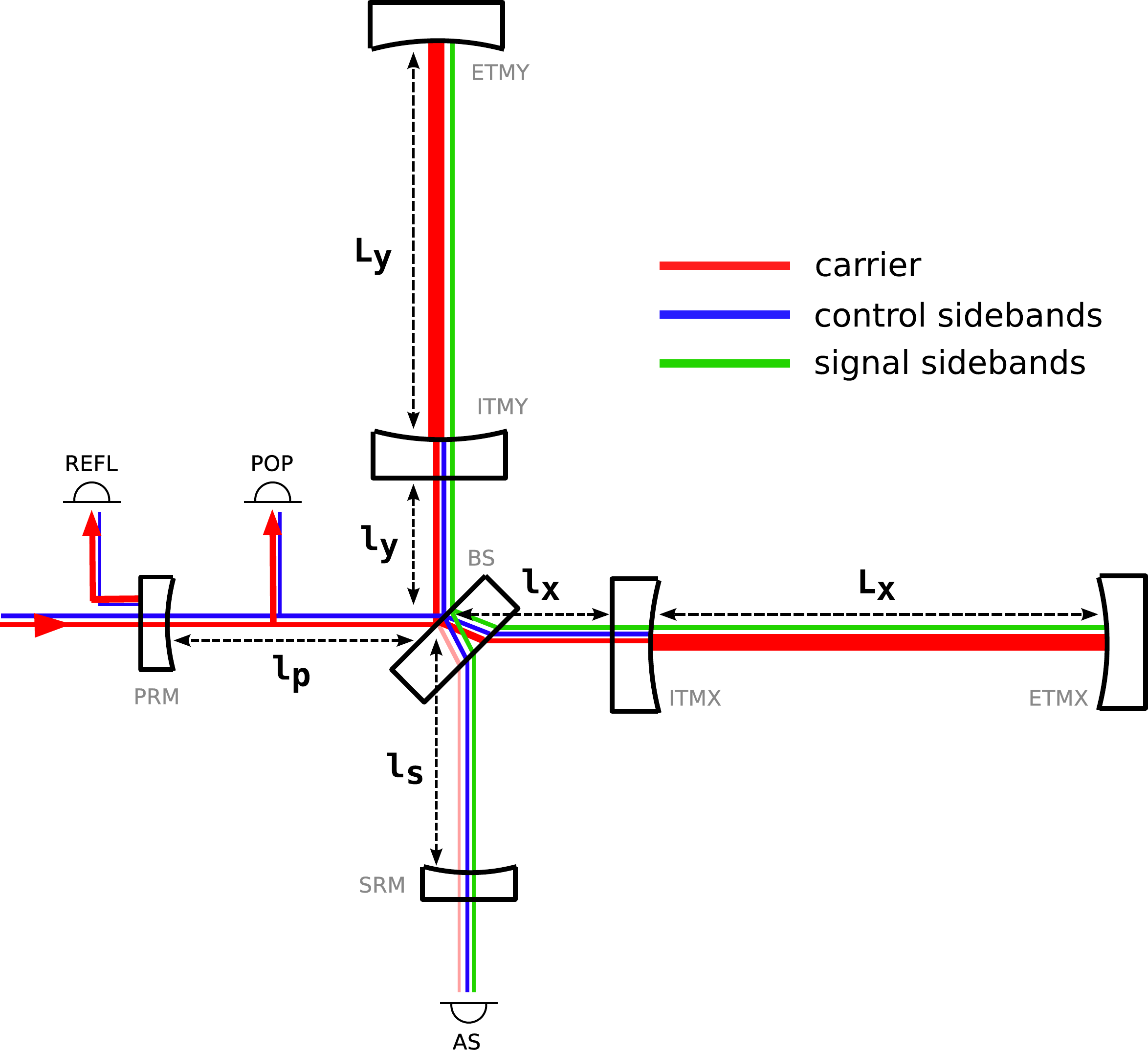}}
    \caption{Simplified layout of the dual recycled Fabry-Perot Michelson of Advanced LIGO,
    illustrating the fields inside the interferometer.  All main mirrors are shown: power and signal recycling
    mirrors (PRM and SRM); input test masses (ITMX and ITMY); end test masses (ETMX and ETMY);
    and the central beam splitter.  Also shown are the locations of various photodiodes: REFL (reflected port),
    POP (power recycling pick off port) and AS (anti-symmetric port).
    The paths of the different light fields in the interferometer
    are shown.  The carrier resonates inside the arm cavities and power recycling cavity, with the Fabry-Perot
    Michelson tuned close to the carrier dark fringe.  A small amount of carrier light leaks into
    the anti-symmetric port due to a small DC offset.  The RF sidebands used for control resonate in the
    power recycling cavity but are reflected from the arms.  The Michelson is not at the dark fringe for
    the control sidebands and they appear in the signal recycling cavity and at the Michelson output.
    The signal sidebands generated by a gravitational wave originate in the arm cavities and
    then exit the Michelson through the anti-symmetric port. }
    \label{fig:DRFPMI_fields}
\end{figure}}

In figure~\ref{fig:DRFPMI_fields} a simplified diagram of the Advanced LIGO
dual recycled configuration is shown, with the lengths which
require control labelled.  The microscopic degrees of freedom associated
with these lengths are:
\begin{itemize}
\item {\bf CARM}  Common-mode arm motion, CARM $=L_x+L_y$.
This tunes the average length
of the arm cavities and is used to keep the arms on resonance.
\item {\bf PRCL} Power recycling cavity length, PRCL $=l_p +\frac{l_x+l_y}{2}$.
The power recycling cavity
is operated on resonance to maximise the power coupled into the central
interferometer.
\item {\bf MICH} Michelson arm length, MICH $=l_x-l_y$.  MICH controls the short
arms of the Michelson (between the ITMs and beam splitter) and determines the fringe
at the output port.  Generally the Michelson is operated on the dark fringe.
\item {\bf DARM} Differential arm motion, DARM $=L_x-L_y$.  This controls the difference
in length of the two arm cavities and is used to get the best interference between the two
arms at the dark port.
\item {\bf SRCL} Signal recycling cavity length, SRCL $=l_s+\frac{l_x+l_y}{2}$.  Used to
control the operation of the signal recycling cavity.  The operating point of SRCL depends
on the mode of operation of the interferometer.  It can be tuned for a particular frequency
of gravitational wave or for broadband operation.
\end{itemize}

\clearpage
\newpage
\bibliography{article}

\end{document}